\documentclass[]{aa}
\pdfoutput=1
\usepackage[citecolor=blue]{hyperref}
\usepackage{graphicx}
\usepackage{enumerate}
\usepackage{epsfig}
\usepackage{color}
\usepackage{txfonts}
\usepackage{natbib}
\usepackage{multirow}
\usepackage{tabularx}
\usepackage[dvipsnames]{xcolor}
\usepackage{braket}
\usepackage{longtable} 
\usepackage[font=small]{caption}

\title{The e-TidalGCs Project}
 \subtitle{Modeling the extra-tidal features generated by Galactic globular clusters}

\titlerunning{The e-TidalGCs Project}

\author{Salvatore Ferrone$^1$, Paola Di Matteo$^1$, Alessandra Mastrobuono-Battisti$^1$,  Misha Haywood $^1$, Owain N.~Snaith$^1$,  Marco Montuori$^2$, Sergey~Khoperskov$^{3,1}$, David~Valls-Gabaud$^4$}
\authorrunning{Ferrone et al}

\institute{GEPI, Observatoire de Paris, PSL Research University, CNRS, Place Jules Janssen, 92195 Meudon Cedex, France \\ \email{salvatore.ferrone@obspm.fr}
\and 
SMC-ISC-CNR and Dipartimento di Fisica, Sapienza University, P.le Aldo Moro 2, 00185, Rome, Italy
\and
Leibniz-Institut f\"ur Astrophysik Potsdam (AIP), An der Sternwarte 16, 14482 Potsdam, Germany
\and
LERMA, CNRS UMR 8122, Observatoire de Paris, PSL, 61 Avenue de l'Observatoire, 75014 Paris, France
}

\date{Received: 29 May 2022; accepted: 27 December 2022}


\abstract{We present the e-TidalGCs Project which aims at modeling and predicting the extra-tidal features surrounding all Galactic globular clusters for which 6D phase space information, masses and sizes are available (currently 159 globular clusters). We focus the analysis and presentation of the results on the distribution of extra-tidal material on the sky, and on the different structures found at different heliocentric distances. We emphasize the wide variety of morphologies found: beyond the canonical tidal tails, our models reveal that the extra-tidal features generated by globular clusters take a wide variety of shapes, from thin and elongated shapes, to thick, and complex halo-like structures. We also compare some of the most well studied stellar streams found around Galactic globular clusters to our model predictions, namely those associated to the clusters NGC~3201, NGC~4590, NGC~5466 and Pal~5. Additionally, we investigate how the distribution and extension in the sky of the simulated streams vary with the Galactic potential by making use of three different models, containing or not a central spheroid, or a stellar bar. Overall, our models predict that the mass lost by the current globular cluster population in the field from the last 5~Gyrs is between $0.3-2.1\times10^{7}M_{\odot}$, an amount comparable between 7--55\% of current mass. Most of this lost mass is found in the inner Galaxy, with the half-mass radius of this population being  between 4--6~kpc. The outputs of the simulations will be publicly available, at a time when the ESA Gaia mission and complementary spectroscopic surveys are delivering exquisite data to which these models can be compared.}

\begin{document}

\maketitle

\section{Introduction}\label{intro}
Globular clusters are the oldest, gravitationally-bound stellar systems in the Galaxy \citep{meylan97}. About 170 are currently known in the Milky Way \citep{vasiliev21}, and this census is possibly still incomplete, especially in the inner regions of the bulge and disk of our Galaxy, where  dust extinction and  high stellar number density limit detections. It is indeed in these regions that new globular cluster candidates have been recently discovered, thanks in particular to the analysis of near-infrared surveys  
\citep{minniti11,monibidin11,minniti17a,minniti17b,minniti18,gran19,garro20,garro2022unveiling,garro2022inspection,garro21c,minniti21a,minniti21b,gran22}.  The current population of globular clusters  possibly represents only the left-over of an initially more numerous and more massive one, depopulated by many disruptive processes \citep{gnedin97, murali97a, murali97b, vesperini97, fall01}. One of the main processes affecting the globular cluster population, its evolution in number, mass and size, is tidal stripping.   

As all stellar systems having a finite size and orbiting the Galaxy, globular clusters are  subject to tidal effects, which arise because the  opposite sides of these systems experience a different gravitational acceleration. The long-term effect of this process strips the system of its most loosely-bound stars, which redistribute themselves onto orbits similar to that of their progenitor, forming so-called ``tidal tails'' or streams around it \citep[see][for some of the earliest studies]{grillmair95, leon00}. Some spectacular tails have been discovered and studied  over the past twenty years around Milky Way globular clusters: from the roughly $30^\circ$ degree long tails departing from the Palomar~5 cluster \citep{odenkirchen01, odenkirchen03, grillmair06, odenkirchen09, thomas16, starkman20,  ibata21} to those of NGC~5466 \citep{belokurov06}, Palomar~14 \citep{sollima11} and to the GD-1 stream, whose parent cluster has still to be discovered \citep[or it has been already completely destroyed, and the stream is the only vestige of its past existence, see][]{grillmair06b, webb19, bonaca20}. These studies have been boosted in the last few years thanks to the publication of the ESA Gaia mission catalogues \citep{gaia16, gaia18, gaia21a, gaia21b} which---by delivering parallaxes, proper motions and magnitudes for about 1.4 billion stars, and radial velocity for several million---is allowing for searches of stars with coherent distances and motions in the Galaxy, revealing the existence of a number of new and spectacular streams, as well as rediscovering already known ones \citep[][]{navarrete17, malhan18a, malhan18b, ibata18, piatti18,  shipp18, ibata19a, ibata19b, kaderali19, bianchini19, malhan19a, malhan19b, palau19, piatti19, caldwell20,  ibata20b, piatti20a, piatti20b, piatti20c, shipp20, thomas20, boldrini21, ibata21, malhan21, jensen21, palau21, piatti21a, piatti21b, yuan2022complexity, zhang22, nie22, piatti22}. \citet{mateu22} provides a recent compilation of known stellar streams.

All these studies are unraveling  a very complex and rich set of stellar structures in the Milky Way, mainly distributed in the halo, where their identification is the easiest because of the low density of the background stellar field.

From the numerical and theoretical point of view, many studies have been focused over the years on the formation and evolution of tidal streams around globular clusters \citep{keenan75, oh92a, oh92b, grillmair98, combes99, ibata02, johnston02, yim02, capuzzo05, dimatteo05, montuori07, siegal08, kupper10, lane10, kupper12, mastrobuono12, sanders13,  bovy14, amorisco16, erkal16, sanders16, pearson17, carlberg18,  thomas18, carlberg20, vitral2022properties}. These studies have contributed to understanding how these structures form and evolve, to what extent they trace the globular cluster orbit and how their shape, extension and morphology depend on the orbital phase, characteristics of the Galactic potential or on the tidal shocks experienced by the cluster itself when it crosses the Galactic disk.

Some works have presented models and simulations for specific streams \citep{dehnen04, mastrobuono12, banik19, bonaca19, banik21, bonaca21}, contributing to understanding their morphology, density variations and extent. From these works, it is clear that the tidal loss of stars from globular clusters and the formation of related structures are important for several aspects: (1) quantifying to what extent  globular clusters have contributed  to the field stellar populations---from the halo to the disk to the bulge---and to what extent they still do, (2) reconstructing the properties (in terms of numbers and masses) of the early Galactic globular clusters, through their current mass loss, and, last but not least, (3) using globular cluster streams as a probe of the Galactic potential and---more generally---of the physical laws governing gravity \citep[see for example][]{thomas18, bianchini19, naik20, banik21, banik21b}.

In this paper, we wish to contribute to the current studies on this matter by presenting the first complete catalogue of simulated extra-tidal features around globular clusters. We emphasize that we talk generically about features and not specifically about tails, or streams, because the  latter are but one  of the morphologies that  extra-tidal material can reveal, as we will show. This project is motivated, on the one hand, by the aforementioned discoveries of many numerous new streams and  tails in the Galaxy, and on the other hand, by the availability of the full 6D phase information and internal parameters (masses and sizes) for more than 150 Galactic globular clusters \citep{baumgardt18, baumgardt21, vasiliev21}. The aims of this project are manyfold: (1) to have a complete view of the expected distribution of globular clusters tidal structures in the sky; (2) to help the interpretation of recent and future discoveries; (3) to support the search for new extra-tidal features in the data; (4) to offer the community a repository of all these models to be compared to other theoretical and numerical predictions, which adopt different Galactic potentials and/or gravity laws.

\section{\bf{Numerical method}}\label{methods}

To model the formation and evolution of extra-tidal features around Galactic globular clusters, we use a set of codes, called GCsTT (Globular Clusters' Tidal Tails), developed by our group. GCsTT comprises two python codes, for the backward and forward integration of a stellar system, made of N test-particles (see Sect.~\ref{numerical}). These codes are separated for data organization and management, the computationally most expensive part---the calculation of the accelerations acting on the N particles and the orbits integration---is realized by means of a Fortran module written by our group. This module is interfaced to python by means of f2py directives from NumPy. The use of test-particle methods for modeling the tidal stripping process is widespread in the literature, where these methods are  usually applied to one or few clusters at a time \citep[see, for example, ][]{lane12, mastrobuono12, palau19, piatti21a, grillmair22}. In this work, we apply a test-particle methodology to the whole set (159) of Galactic globular clusters for which this is currently possible, taking also into account, for each cluster, errors on astrometry, line-of-sight velocities\footnote{Note that the term ``line-of-sight velocities''  adopted in this paper corresponds to the term ``radial velocities'' often used in the literature, as well as in the Gaia catalogues. We prefer the use of the first term, since the second is usually used also to indicate the (Galactocentric) radial velocities and can induce to some ambiguity, especially when different coordinate systems are used. We emphasize that the choice to use the term ``line-of-sight velocity'' is not new \citep[see, for example][]{vasiliev21}.} and distances. In the following of this section, we describe the two main steps of the procedure used by GCsTT  to simulate the tidal stripping process (Sect.~\ref{numerical}), the initial conditions adopted for the clusters' parameters and their mass distribution (Sect.~\ref{initialconds}), as well as the Galactic potentials (Sect.~\ref{galmod}).

\subsection{Simulations of the tidal stripping process: a two-step procedure}\label{numerical}
To model the formation and evolution of extra-tidal features around Galactic globular clusters, and predict their current properties, we proceed as follows:

\begin{enumerate}[\it(i)]
\item \textit{Backward integration---Reconstructing the globular cluster orbit in the last 5~Gyr}.\\ First, for each Galactic globular cluster for which distances from the Sun, proper motions, line-of-sight velocities, and structural parameters are  available (see Sect.~\ref{initialconds}), we determine their current positions and velocities in a Galactocentric reference frame, in which the Sun is at $(x_\odot, y_\odot, z_\odot) = (-8.34, 0., 0.027)$~kpc \citep{chen01, reid14}, a velocity for the local standard of rest, $v_{LSR}= 240$~km/s \citep{reid14}, and a peculiar velocity of the Sun with respect to the LSR, $(U_{\odot}, V_{\odot}, W_{\odot})  = (11.1, 12.24, 7.25)$~km/s \citep{schonrich10}. We then integrate the orbit of a single point mass, representing the cluster barycenter, backwards in time for 5~Gyr, and in this way we retrieve its position and velocity at that time in the chosen Galactic potential (see Sect.~\ref{galmod}). We notice that other choices for the Sun's position or velocity with respect to the Galactocentric frame would have been possible. For example, \citet{piatti21a} adopt the same values as ours for the $v_{LSR}$ and for the peculiar velocity of the Sun, but a different distance to the Galactic center \citep[8.1~kpc in their work, see][]{gravity18}. The difference in the adopted position of the Sun is, however, in general smaller than the uncertainties affecting our knowledge of the distance of Galactic globular clusters to the Sun. For this reason we do not to explore the dependency of the results presented in this paper on these choices. 

\item \textit{Forward integration---Test-particle streams from the past to the present day}. \\Once the positions and velocities of the barycenter of each cluster, 5~Gyr ago, are determined, we build 
the corresponding $N$-body system, with N = 100 000 particles.  The phase-space coordinates of these particles are generated following a Plummer distribution, with total mass and half-mass radius as described in Sect.~\ref{initialconds}. 
The barycenter of this $N$-body cluster is then assigned initial positions and velocities in the Galactic model, as those retrieved at step $(i)$ and the cluster is then integrated forward in time, until the present day. Particles in this $N$-body system are modeled as test-particles, that is they experience the gravitational field exerted by the globular cluster itself (see Sect.~\ref{initialconds}) and by the Galaxy (see Sect.~\ref{galmod}), but not generate any gravitational field themselves. This allows us to maintain a computational time which scales as $O(N)$ and not as $O(N^2)$, as it would be the case for a direct $N$-body self-consistent computation.
\end{enumerate}

In the following, we refer to these simulations, made by using the most probable values on distances, proper motions and line-of-sight velocities, as the ``reference simulations''. In addition, for each globular cluster, we also take into account the errors on its distance, proper motions, and line-of-sight velocity,  assuming Gaussian distributions of the errors, treated as independent, and by generating 50 random realizations of these parameters.  For each of these realizations, we repeat the steps  described above, that is: $(i)$ we determine the associated current positions and velocities in the chosen Galactocentric reference frame, we integrate the orbit of the single-point mass (representing the cluster barycenter) backwards in time, retrieving the corresponding values 5~Gyr ago, $(ii)$ we build a $N$-body cluster containing $N$= 100~000 particles, with total mass and half-mass radius as those used for the reference simulation, and then we integrate the $N$-body cluster forwards in time until the present day position. 

To summarize, for a given Galactic potential, we run $159\times (50+1)=8109$ simulations, where 159 is the total number of clusters for which we currently have both 6D phase-space information and structural parameters. As we will discuss in the following section, the whole set of globular clusters has been evolved in three different Galactic potentials, which implies that a total of 24 327 simulations has been run.

For the orbit integration, a leap-frog algorithm is used, with a fixed time-step, $\Delta t$, and a total number of steps, $N_{steps}$, such that the total simulated time is  $\Delta t \times N_{steps}=5$~Gyr. The choice of the value of $\Delta t$ adopted to simulate each cluster in the Galactic potential has been based on the energy conservation of the corresponding cluster evolved in isolation (that is without the effect of the Galactic gravitational field for 5~Gyr). For the majority of the clusters (109/159) this value has been set to $\Delta t = 10^5$~yr (for a corresponding value of $N_{steps}=50\,000$), for the remaining clusters (50/159) a  $\Delta t = 10^4$~yr (for a corresponding value of $N_{steps}=500\,000$) has been used. We refer the reader to Appendix~\ref{deltat}, and in particular to Table~\ref{tcross-energy}, for additional details on the choice of $\Delta t$ for the whole set of clusters. As for the total simulated time, while globular clusters are much older than 5~Gyr, we chose this time limit because the longer back in time we could go, the less certain we would be of the Galactic environment. In addition, the last significant mergers in the Galaxy happened between 9 and 11~Gyr ago  \citep[see][]{belokurov18, helmi18, dimatteo19, gallart19, kruijssen20}, thus well before the time interval simulated in this study. Other more recent interactions, such as the accretion of Sagittarius and of the Magellanic Clouds, may perturb the Galactic potential as well \citep[see, for example,][]{vasiliev21b} and we plan to investigate their impact on the properties of globular cluster streams in the future.

For each realization, we generate an output file in an hdf5 format\footnote{\url{https://www.hdfgroup.org/solutions/hdf5/}} containing right ascension ($\alpha$), declination ($\delta$), distance from the Sun ($D$); the components for proper motion in the equatorial coordinate system ($\rm \mu_{\alpha}\cos(\delta)$ and $\rm \mu_\delta$), the line-of-sight velocity ($\rm v_{\ell os}$), longitude ($\ell$), latitude ($ b$); the components for proper motion in the Galactic coordinate system ($\rm \mu_{\ell}\cos(\mathit{b})$ and $\rm \mu_b$), and the Galactocentric positions ($x, y, z$), velocities ($v_x, v_y, v_z$) and energy, $E$, of each particle in the simulated system. We used Astropy \citep{astropy13, astropy18} to convert the Galactocentric positions and velocities in the equatorial and Galactic quantities $\alpha, \delta, D, \rm \mu_{\alpha}\cos(\delta), \rm \mu_\delta, \rm v_{\ell os}, \ell, b, \rm \mu_{\ell}\cos(\mathit{b})$ and $\rm \mu_b$.

For each particle, we also save its escape time $t_{\rm esc}$,  defined as the time at which the particle escapes from the cluster, that is the time $t$ at which the particle satisfies the relation\footnote{If the particle is gravitationally bound to the cluster until the end of the simulation, $t_{\rm esc}$ is set equal to $-9999$.}:
\begin{equation}
E_{GC}= 0.5 \times \left( (v_x-v_{x, GC})^2+(v_y-v_{y,GC})^2+(v_z-v_{z,GC})^2\right)+\Phi_{GC} > 0
\end{equation}
$E_{GC}$ being the total specific energy of the particle relative to the cluster, that is the sum of the potential energy, $\Phi_{GC}$, due to the gravitational field of the cluster (see Eq.~\ref{gcpot}), and of the kinetic energy, relative to the cluster barycenter, $T_{GC}=0.5 \times \left( \left(v_x-v_{x, GC}\right)^2+\left(vy-v_{y,GC}\right)^2+\left(vz-v_{z,GC}\right)^2\right)$, where $v_x, v_y$, and $v_z$ are its velocity components at time $t$, and $v_{x,GC}, v_{y,GC}$ and $v_{z,GC}$ those of the cluster barycenter at the same time. A positive value of $E_{GC}$ implies that the particle is no longer gravitationally bound to the cluster, and hence it is lost in the field.\\

Overall, the total volume of the whole set of  24 327 simulations, saved in hdf5 format, amounts to about 370 Gb.

\subsection{Simulations of the tidal stripping process: Globular clusters' current and initial conditions and their gravitational potential}\label{initialconds}

To be executed, steps $(i)$ and $(ii)$ described in the previous section require some input conditions. The current distances from the Sun, proper motions, and line-of-sight velocities, and the related uncertainties, of all 159 globular clusters considered in this study are taken respectively from \citet{baumgardt21} and \citet{vasiliev21}. These values are then converted into Galactocentric positions and velocities by making use of Astropy, and used as initial conditions to execute step $(i)$. 

Step $(ii)$ requires generating an $N$-body system, representing the globular cluster, whose initial total mass and half-mass radius are assigned on the basis of their current values, as given by \citet{baumgardt18}\footnote{In particular, the adopted  values have been taken from the edition available at \url{https://people.smp.uq.edu.au/HolgerBaumgardt/globular/parameter.html}, as to January 14th 2022.} and reported in Table~\ref{TableIC}. As anticipated at step $(ii)$ in Sect.~\ref{numerical}, the phase-space coordinates of each $N$-body cluster are generated by assuming  a Plummer distribution of total mass $M_{GC}$ and half-mass radius $r_{h}$, for which the corresponding potential is:
\begin{equation}\label{gcpot}
\Phi_{GC}(r) = -\frac{GM_{GC}}{\sqrt{r^2+{r_c}^2}},
\end{equation}
where $r_c$ is the cluster scale radius, and is related to the half-mass radius $r_{h}$ through $r_{h} \simeq 1.305 r_c$ \citep{heggie03}. The variable $r$ here indicates the distance of the test particle from the center of the cluster. For each cluster, the same Plummer distribution used to generate the $N$-body system is also used to calculate the accelerations exerted on each particle as the system moves through time. The Plummer sphere, representing the cluster potential, moves indeed through the Galaxy along the orbit retrieved at step (i), traveling this time in the opposite direction, from 5 Gyr ago to the present day.

The reader might note that this implies that the globular cluster density profile and its internal parameters (total mass and characteristic radius) are constant in time in these models. This is of course a crude approximation, because in reality both the internal parameters and the density profile itself can change over time. We consider these assumptions to be acceptable within the scope of our work given that we are primarily interested in the distribution of extra-tidal stars, which once escaped from the cluster have dynamics primarily dictated by the Galactic potential rather than the globular cluster itself. Of course,  the density of stars along the extra-tidal structures, as well as the total mass lost, depend on these assumptions (that is, if the mass of the cluster was not assumed constant in time, but could decrease, the gravitational attraction exerted by the cluster itself on its stars would be weaker, and this would lead to an increasing mass loss and density along the tails). We could have proceeded diminishing the mass over time, based on some assumptions on the temporal behavior of this relation, but we did not find this approach satisfying. In this way we would have taken into account a temporal evolution of the mass, but not of the size of the cluster, adding supplementary hypothesis to the problem. For these reasons, we decided to maintain the simplest approach. We emphasize that other groups followed the same methodology, maintaining masses and sizes constant with time \citep[see, for example,][]{palau19}.

The summary tables giving both the current internal parameters of the clusters (total mass and half-mass radius), their astrometric quantities of relevance for this study and the line-of-sight velocities are publicly available\footnote{All data can be found here \url{https://people.smp.uq.edu.au/HolgerBaumgardt/globular}.}. We have made use of these tables for our work, and we report them in a unique table in our paper for completeness and self-consistency of the data used (see Table~\ref{TableIC}).

\subsection{Simulations of the tidal stripping process: Galactic potentials}\label{galmod}

\begin{table*}
\centering
\caption{The parameters of the Galactic mass models adopted in this work. Masses are in units of $2.32\times10^7M_{\odot}$, distances in units of kpc.
\label{PII}}
\tiny
  \begin{tabular}{  l c  c  c  c  c  c  c  c  c  c  c  c  c   c } \hline
   Parameters & $M_{bulge}$ &  $M_{bar}$ & $M_{thin}$ &  $M_{thick}$  & $M_{halo}$ & \ $b_{bulge}$  & $a_{bar}$ & $b_{bar}$ & $c_{bar}$ & $a_{thin}$ & $b_{thin}$  & $a_{thick}$ &  $b_{thick}$ & $a_{halo}$ \\  \hline \hline \\
   PI & 460.0 & 0.0 & 1700.0 & 1700.0  & 6000.0  & 0.3 & -- & -- & -- & 5.3000 & 0.25 & 2.6 & 0.8 & 14.0 \\  \hline    
    PII & 0.0 & 0.0 & 1600.0 & 1700.0  & 9000.0  & -- & -- & -- & -- & 4.8000 & 0.25 & 2.0 &  0.8 & 14.0 \\    \hline    
    PII-0.3-SLOW & 0 & 990.0 & 1120.0 & 1190.0  & 9000.0  & -- & 4.0 & 1. & 0.5 & 4.8000 &0.25 & 2.0 &  0.8 & 14.0 \\ \hline 
  \end{tabular} 
  \normalsize
\end{table*}

As for the Galactic mass distribution, we make use of the two axisymmetric Galactic mass models presented in \citet{pouliasis17},  and of an asymmetric mass model, containing a central stellar bar, that we present here for the first time. We recall below the main properties of the two models of  \citet{pouliasis17}, and we describe in more detail the asymmetric Galactic mass model, which is presented here for the first time.

\subsubsection{Model I by \citet{pouliasis17}: an axisymmetric mass model for the Galaxy including a spherical bulge}

Model I by \citet{pouliasis17} (abbreviated name: PI) consists of four components: two disks (thin and thick) both described by Miyamoto \& Nagai potentials, a dark matter halo, and a central bulge. Its total potential is:
\begin{equation}
\Phi_{tot}(R, z) = \Phi_{thin}(R, z) + \Phi_{thick}(R, z) + \Phi_{halo}(r)+  \Phi_{bulge}(r)
\end{equation}
with $r=\sqrt{R^2 + z^2}$,
\begin{eqnarray}
\Phi_{thin}(R,z)&=&\frac{-GM_{thin}}{\left(R^2+\left[a_{thin}+\sqrt{z^2+b_{thin}^2}\right]^2\right)^{1/2}}\\
\Phi_{thick}(R,z)&=&\frac{-GM_{thick}}{\left(R^2+\left[a_{thick}+\sqrt{z^2+b_{thick}^2}\right]^2\right)^{1/2}}
\end{eqnarray}
\begin{equation}
\begin{split}
\Phi_{halo}(r)=&\frac{-GM_{halo}}{r}-\frac{M_{halo}}{1.02a_{halo}}\times\\
& \Bigg[\frac{-1.02}{1+\left(\frac{r}{a_{halo}}\right)^{1.02}}+ln{(1+\left(\frac{r}{a_{halo}}\right)^{1.02})}\Bigg]_R^{100},
\end{split}
\end{equation}
and 
\begin{equation}
\Phi_{bulge}(r) = -\frac{GM_{bulge}}{\sqrt{r^2+{b_{bulge}^2}}},
\end{equation}
where $M_{thin}, M_{thick}$, $M_{halo}$, and $M_{bulge}$ are the masses of the disks, halo and bulge and: $a_{thin}, b_{thin},  a_{thick}, b_{thick}, a_{halo}, b_{bulge}$ are the characteristic scale lengths of the thin and thick disks, the halo and the central bulge, respectively (see Table~\ref{PII}).

This model is a modification of the classical \citet{allen91} model, made to include also the presence of a thick disk. As it has been discussed in detail by \citet{pouliasis17}, the choice to include  a massive spheroid in this model---as well as in the original \citet{allen91} model---is dictated by the need to reproduce CO/HI-based velocity curves, as those provided by \citet{sofue12}, which show a rise and then a sudden decrease of the velocity curve in the inner Galactic regions ($R \le 2-3$~kpc). In an axisymmetric model, such rise can be reproduced only if a central spheroidal component, with a typical mass greater than 10\% of that of the disk(s), is added. However, as shown by \citet{chemin15}, the central rise observed in the rotation of the molecular gas in the inner Galaxy may be an effect of non circular motions generated by large scale asymmetries like the bar. Moreover, this feature is not reported in all the observational studies \citep[see, for example][on which model PII is based]{reid14}. In other words, if we do not assume that the mass distribution of the inner Galaxy is axisymmetric, the need for a massive spheroidal component to reproduce velocity curves such as those by \citet{sofue12} no longer persists. In addition to that, in the last decade, a number of works  have shown that if a spheroidal bulge exists in the central regions of our Galaxy, it has to be small \citep[few percents of the mass of the disk at the most, see among others][]{shen10, kunder12, dimatteo15, gomez18}. All these arguments suggest to employ this model \citep[as well as all models including a massive central spheroid; see, for example,][]{irrgang13}, with care when dealing with the central parts of the Galaxy. Since models with a massive central spheroid, however, are still used in the literature, we have included model PI here, as a term of comparison.

\subsubsection{Model II by \citet{pouliasis17}: an axisymmetric, bulge-less, mass model for the Galaxy}

Model II by \citet{pouliasis17} (abbreviated name: PII) consists of a spherical dark matter halo, with the same functional form adopted in the \citet{allen91} model, and of two disk components (a thin and a thick disk) with same functional form as PI. This model does not include any central spheroid (i.e. it is a bulge-less model) and thus its total potential is the sum of three components only:
\begin{equation}
\Phi_{tot}(R, z) = \Phi_{thin}(R, z) + \Phi_{thick}(R, z) + \Phi_{halo}(r)
\end{equation}
with the thin, thick disks and dark matter halo having the same functional forms adopted in PI.

As it has been shown in \citet{pouliasis17}, this model satisfies a number of observational constraints, such as the stellar density at the solar vicinity, thin and thick disc scale lengths and heights, the rotation curve as provided by \citet{reid14} , and the absolute value of the perpendicular force $K_z$ as a function of distance to the Galactic centre \citep[see Sect.~2.5 in][]{pouliasis17}. Being however, an axisymmetric model, it fails describing accurately the inner few kpc of the Galaxy, where the stellar mass distribution has been shown to be asymmetric.

\subsubsection{Model~II with a massive, slowly rotating, stellar bar}
The third mass model (abbreviated name: PII-0.3-SLOW)  that we use in this paper is a version of PII by \citet{pouliasis17} modified to include a rotating stellar bar, whose mass has been assigned to be 30\% of the (thin+thick) disk mass of PII. We assume that the bar rotates with a constant pattern speed of $\Omega_{bar}=38\rm km~s^{-1}kpc^{-1}$, and that it is currently inclined of $25^\circ$ with respect to the Sun-Galactic center direction \citep[see][]{blandhawthorn16}. We model it as a triaxial distribution, whose gravitational potential is given by  \citet{long92}:
\begin{equation}
\Phi_{bar}(x,y,z)=\frac{GM_{bar}}{2a_{bar}}ln\left( \frac{x-a_{bar}+T_{-}}{x+a_{bar}+T_{+}}   \right)
\end{equation}
with $T_{\pm}=\left[ (a_{bar}\pm x)^2+ y^2 +  (b_{bar}+ \sqrt{c_{bar}^2+z^2})^2 \right]^{1/2}$
and $a_{bar}, b_{bar}, c_{bar}$ the characteristic bar parameters. The total gravitational potential generated by this model has thus the form:
\begin{equation}
\Phi_{tot}(x, y, z) = \Phi_{thin}(R, z) + \Phi_{thick}(R, z) + \Phi_{halo}(r)+  \Phi_{bar}(x, y, z)
\end{equation}
(see Table~\ref{PII} for all characteristic values). Practically, to include the bar we have reduced the mass of the disks in such a way to maintain the total stellar mass of this model as that of PII.  \citet{long92} provide the formulas of the accelerations generated by this triaxial distribution in the reference frame of the bar. To calculate and add them to the accelerations generated by the disks and dark matter halo, at each time step we convert the positions of all particles  in the rotating, non-inertial, reference frame of the bar, compute the corresponding accelerations on each particle, and then transform these accelerations back in the inertial reference frame described in Sect.~\ref{numerical}. In this way, the accelerations due to the bar are added to those generated by the other terms of the Galactic mass distribution. 

We emphasize that we do not consider this model as the best possible representation of the Galactic mass distribution, especially in the central region. It can, however, provide a first indication on how the inclusion of a rotating asymmetric component in the inner Galaxy can affect the globular cluster streams, near and far from the Galactic center. 

Moreover, since the exact characteristics of the Milky Way bar are still subject to debate \citep[see, for example,][]{blandhawthorn16}, it is important to explore how varying the parameters adopted in this paper, such as the pattern speed, the mass or the length of the bar, can affect the characteristics of the whole set of streams.  More complex shapes for the bar can also be explored, for example substituting the inner parts of the triaxial bar with a boxy/peanut-shaped morphology, which has been shown to characterize the inner Milky Way  \citep[see, for example][]{wegg13, wegg15}. These topics are, however, beyond the scope of the present paper. In sum, given the uncertainties on the bar's physical extent and how it can change over the time span investigated here, its affect on the streams presented here are purely indicative.

\section{Results}\label{results}

\begin{figure*}[h!]
\begin{center}
\includegraphics[clip=true, trim = 0mm 20mm 0mm 20mm, width=0.9\columnwidth]{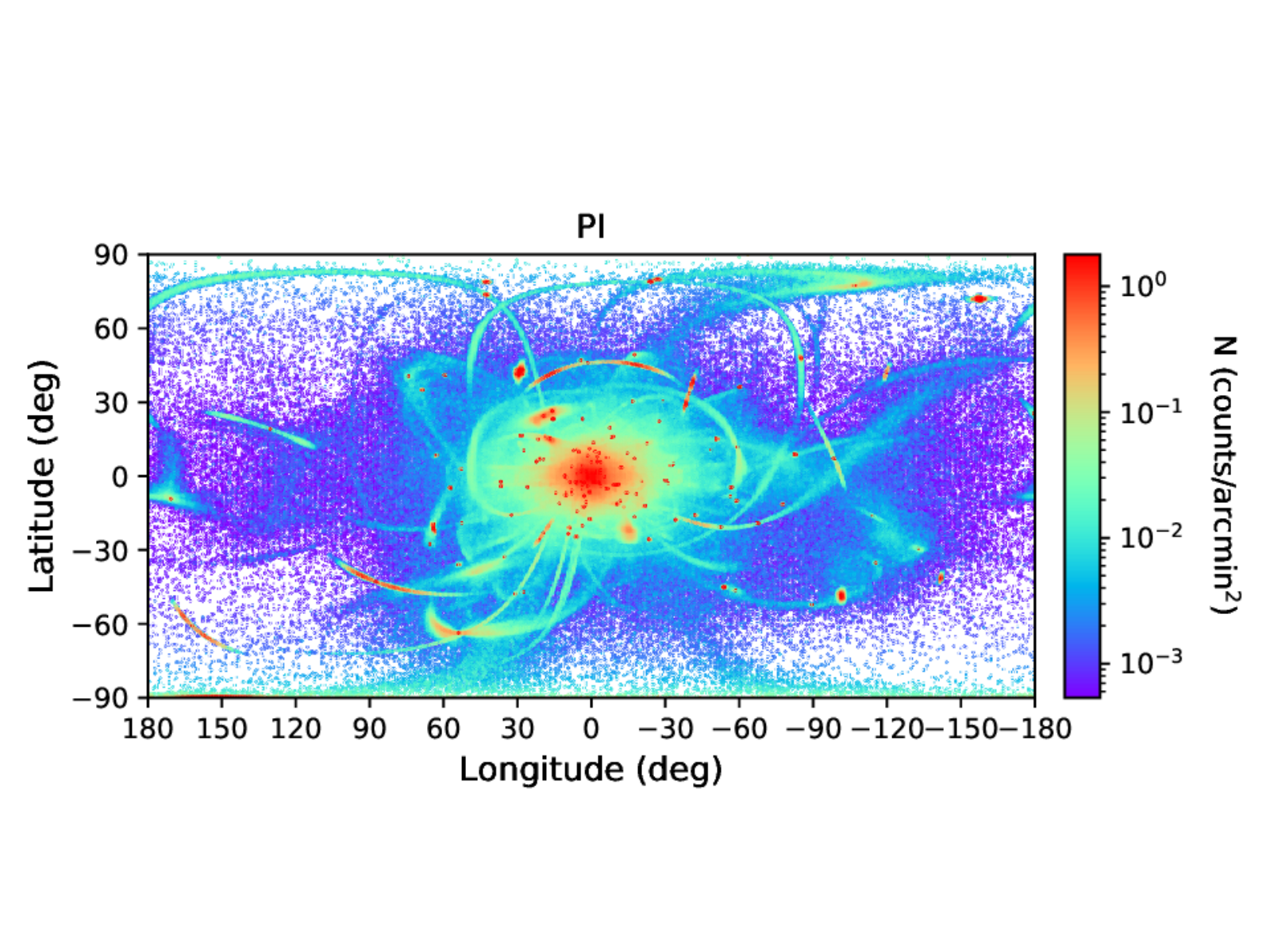}
\includegraphics[clip=true, trim = 0mm 20mm 0mm 20mm, width=0.9\columnwidth]{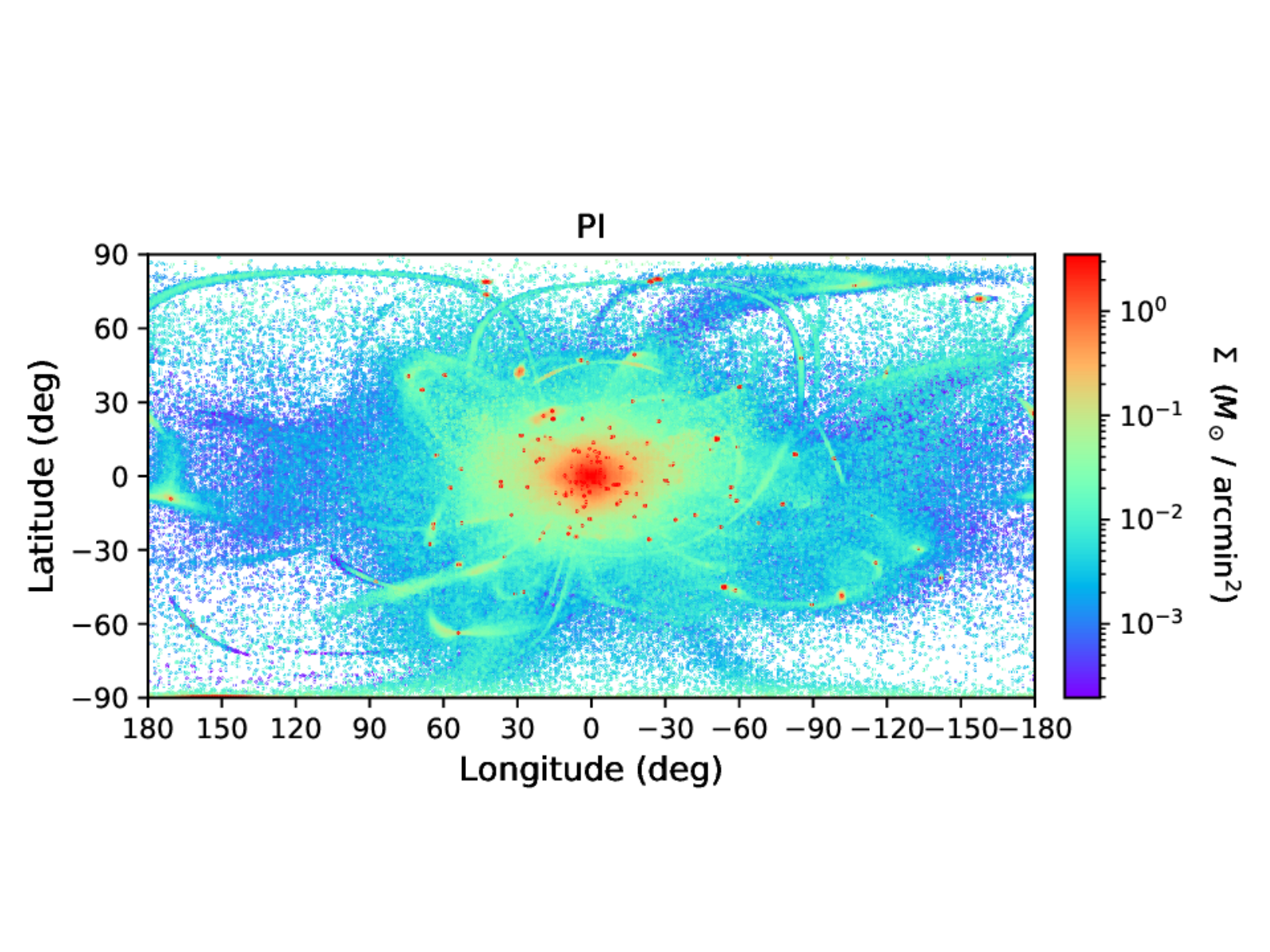}
  
\includegraphics[clip=true, trim = 0mm 20mm 0mm 20mm, width=0.9\columnwidth]{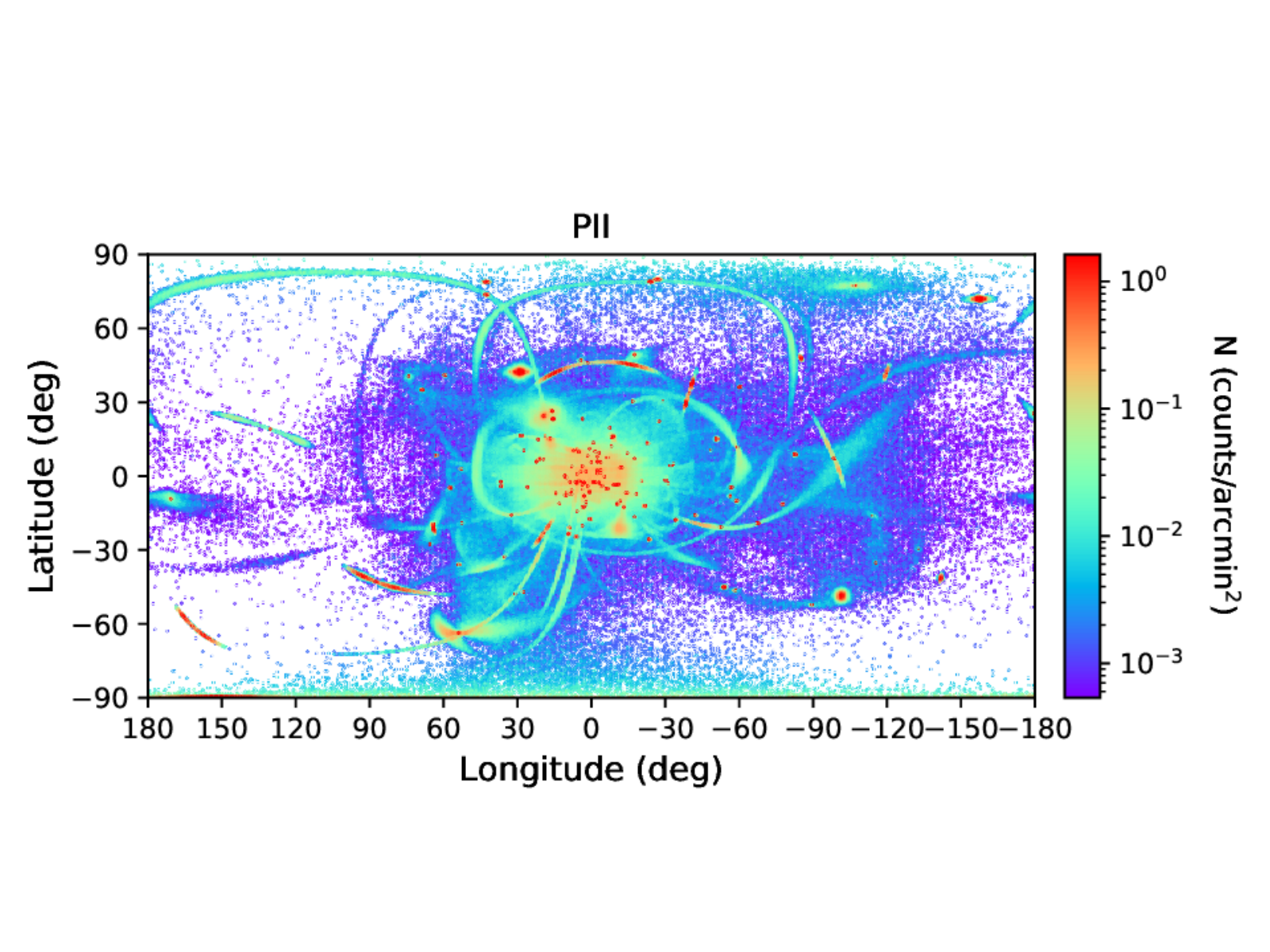}
\includegraphics[clip=true, trim = 0mm 20mm 0mm 20mm, width=0.9\columnwidth]{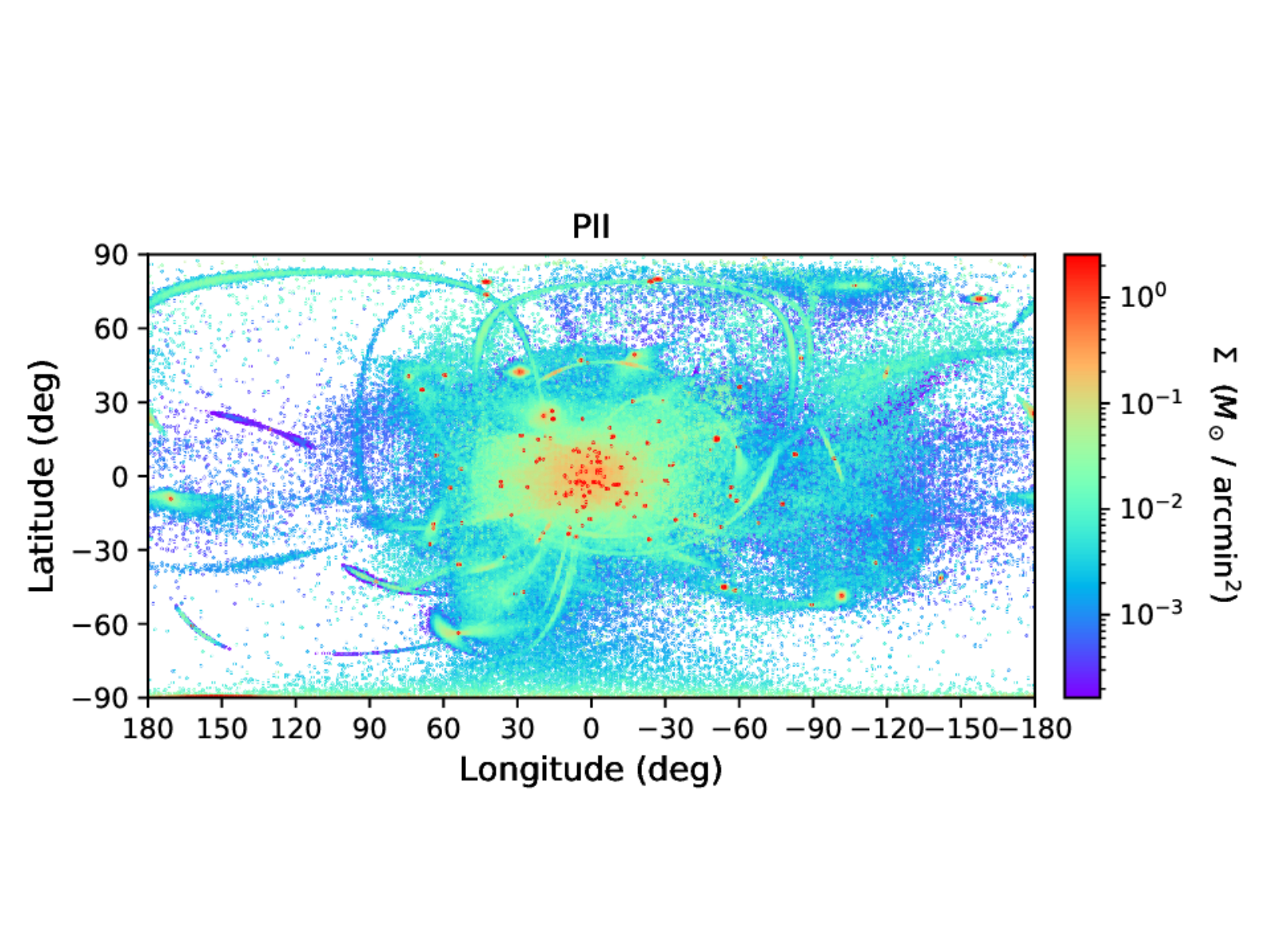}

\includegraphics[clip=true, trim = 0mm 20mm 0mm 20mm, width=0.9\columnwidth]{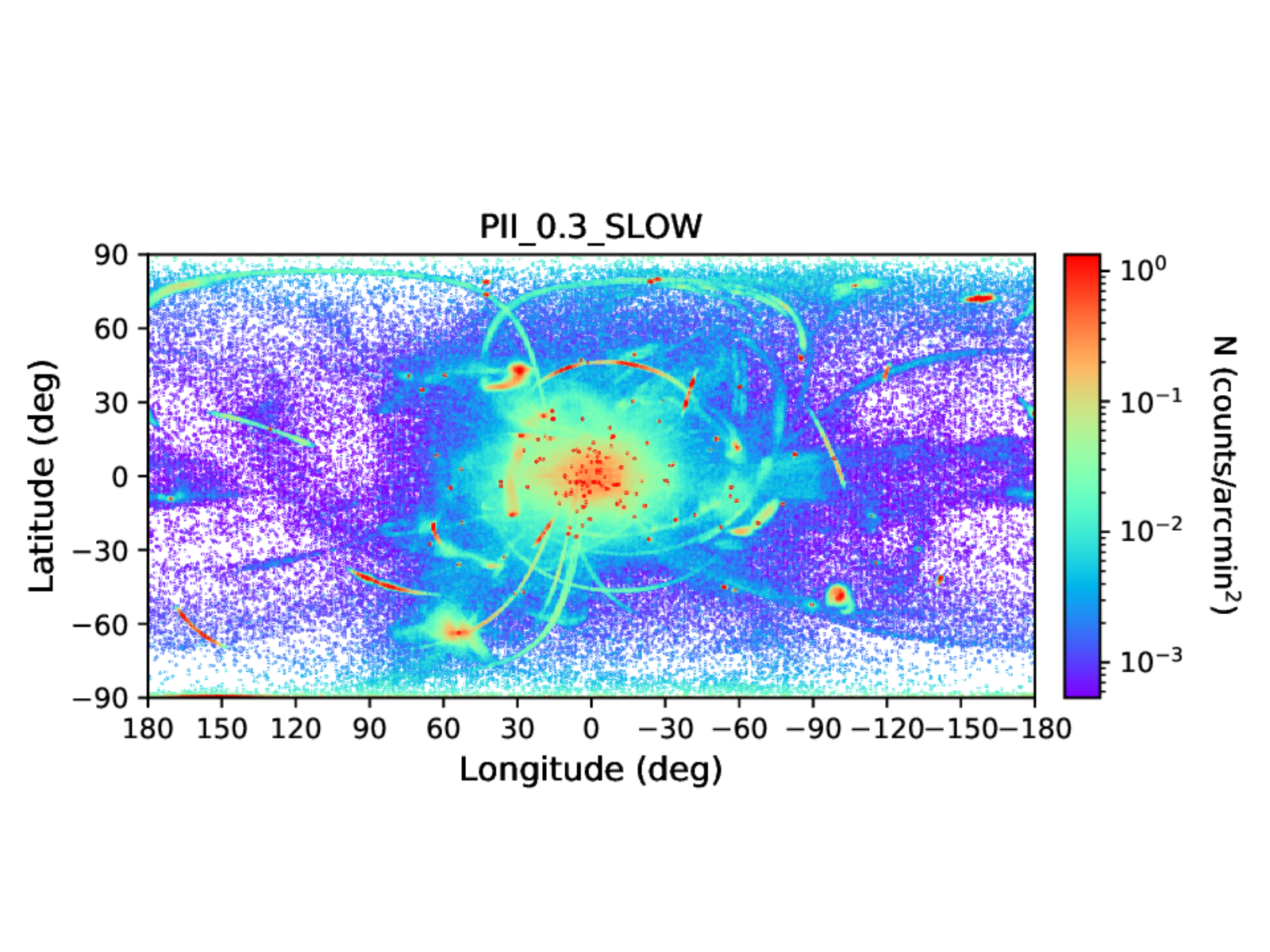}
\includegraphics[clip=true, trim = 0mm 20mm 0mm 20mm, width=0.9\columnwidth]{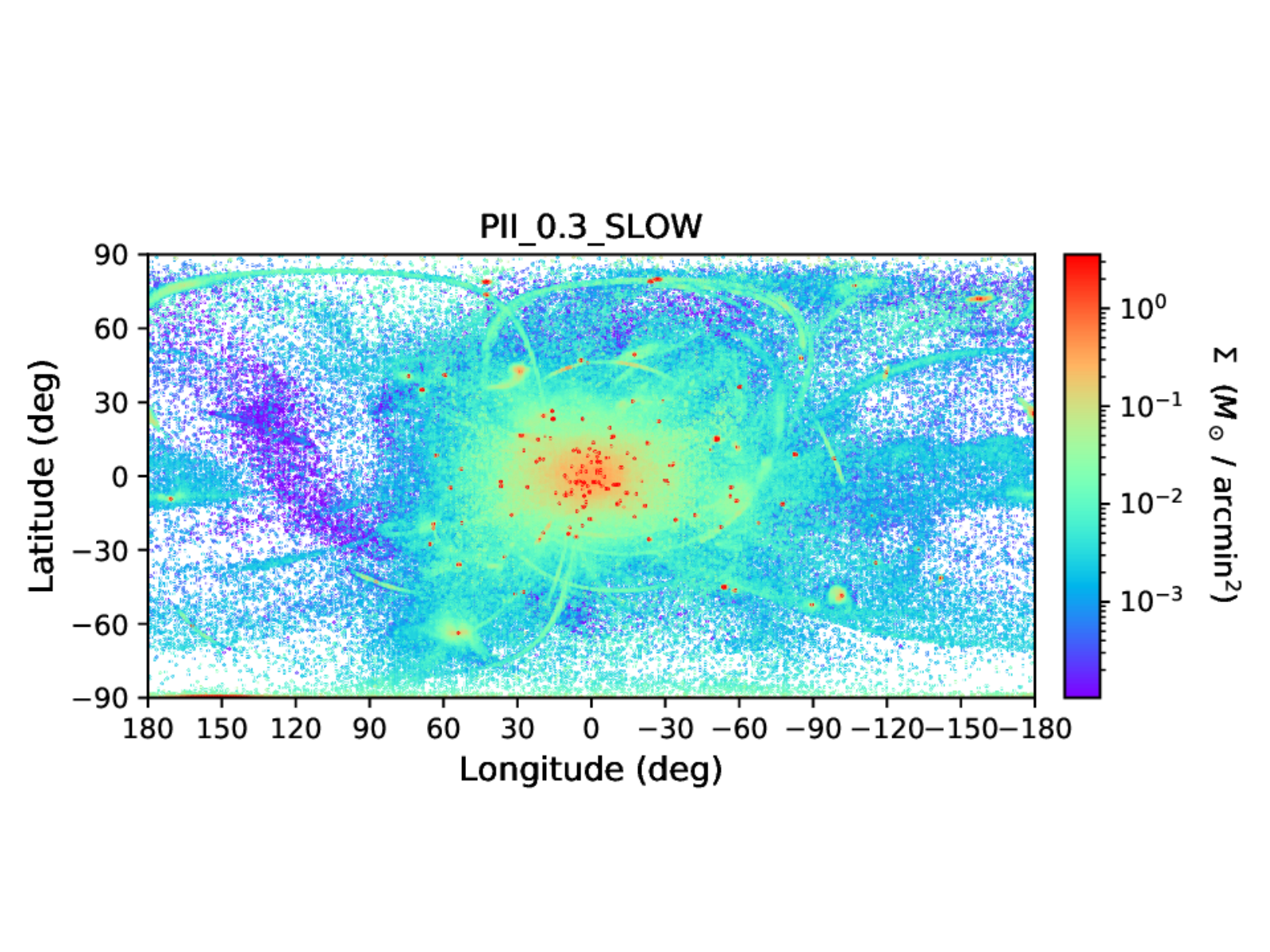}

\caption{(\textit{Top}) \textit{Left}: Surface number density distribution in $(\ell, b)$ plane of the \textit{ensemble} of extra-tidal features around the entire population of Galactic globular clusters at the current time, as predicted by our models. \textit{Right}: Surface mass density, as predicted by our models. Top row corresponds to model PI, the middle to model PII, and the bottom column to model PII-0.3-SLOW, as indicated. All densities are expressed in a logarithmic scale. The red point-like density maxima correspond to the current positions of the globular clusters. Values of higher density are over-plotted. Thus in the case of mass density, diffuse tidal debris of more massive globular clusters covers the entire $(\ell,b)$ space and occults delicate tidal features, which are more visible when considering number density counts. In all panels, only the reference simulations are shown, for better clarity.} \label{galleryLB}
\end{center}
\end{figure*}

\paragraph{A few premises}
Given the large number of simulations carried out and the wealth of information contained in them, it is not possible to exhaust all possible applications of this simulation database in this paper. We have therefore chosen to proceed as follows.

In Section~\ref{results1} we present an overview of the distribution of all streams in  Galactic coordinates. This coordinate space will be the one used in the remainder of the entire article. This first section allows us to show qualitatively how the global distribution of streams varies, depending on the Galactic potential used.

We  then move on (Section~\ref{streamvsD}) to present the global system of streams as a function of their distance from the Sun. In this Section, we also show the kinematic properties of the streams, such as proper motions and line-of-sight velocities, that can be directly compared to Gaia data or other astrometric and spectroscopic surveys. This section also allows us to show the variety of morphologies that the stars which escaped from globular clusters can take. In Section~\ref{sec:morphologies}, we explore this issue in more detail, showing how these morphologies depend primarily on the orbital characteristics of the clusters, and their distance from the Galactic center. For the most interesting cases, we compare the tidal structures predicted by our simulations with streams found in observational data. For this purpose, we make use of the \emph{galstreams} library of stellar streams in the Milky Way \citep{mateu22}, which constitutes a unique and public database summarizing angular positions, distances, proper motions and line-of-sight velocity tracks for nearly a hundred Galactic stellar streams. Any stream not in this library will not be compared to our simulations, in the context of this paper.

For the interested reader, the tidal features associated with each of the 159 simulated clusters are presented in Appendix~\ref{allstreams}. In order not to make this appendix too long, the above tidal features are presented only in the case of the potential PII. However, all the data will be made available to the community at a dedicated site\footnote{\url{http://etidal-project.obspm.fr/}}, and thus, the interested reader will be able to see how the characteristics of these streams change, for any cluster, in the three chosen potentials.

\subsection{A sky full of streams}\label{results1}

Fig.~\ref{galleryLB} shows the number and mass density distributions of the whole set of simulated globular clusters and their extra-tidal features in Galactic longitude and latitude, and for the three Galactic potentials.

For all Galactic mass models adopted, a striking characteristics of the plots in Figs.~\ref{galleryLB} is the variety of features that our models predict, which are reminiscent of the tidal tails, stellar streams, and shells, produced by interacting and merging galaxies in the process of mass assembly \citep[see, for example, ][]{mancillas19}. Some clusters have very thin and elongated streams, which describe arcs that can extend up to $180^\circ$ in longitude, or tens of kpc in physical space. In some other cases, extra-tidal features appear shorter (a few up to ten degrees), and sometimes also thicker (about $10^\circ$ in the sky) than others. Finally, in some cases, clusters are surrounded by extended structures, such as halos, rather than coherent and thin streams. 

This variety of properties depends on several factors: the distance of the stream to the Sun (due to projection; i.e. for a given physical thickness, the closer the stream is to the Sun, the more extended it appears in the $(\ell, b)$ plane), its orbital phase (towards the peri-center or the apo-center of the orbit) and the orbital properties of the parent globular cluster. We also see from these figures that stellar particles stripped from their parent clusters do not only redistribute in coherent structures, but in some cases can also contribute to a more diffuse density distribution. 

Because of the large number of simulated clusters, we have chosen not to present the corresponding extra-tidal features one by one in the main part of this paper, but have rather decided to describe these extra-tidal features with a global approach, by first adopting a criterion based on the distance of these features to the Sun (see Sect.~\ref{streamvsD}), and then discussing the types of distributions tidal debris can have and how these depend on the cluster orbital parameters (see Sect.~\ref{sec:morphologies}). All the extra-tidal structures generated by the 159 globular clusters simulated in this paper, and their corresponding uncertainties, are reported in Appendix~\ref{allstreams}. Among them, the reader will find clusters with thin and elongated tails---as IC~4499, NGC~3201, NGC~4590, NGC~5024, NGC~5053, Pal~5, to cite only a few---clusters like AM~1, Pal~14, Pal~4, and Pal~15 whose extra-tidal material shows a halo-like configuration, and clusters like NGC~1261, NGC~4147, NGC~6356, UKS~1, whose stripped stars show a remarkable diffuse distribution in the field. 

Finally, even if our models are not tailored to accurately reproduce the mass loss from globular clusters---since, as described in Sect.~\ref{initialconds}, we adopt a test-particle approach with a time-constant globular cluster potential---it is however tempting to estimate, to first order, the total mass associated to the tidally stripped population, and compare it to the current mass. By calculating the mass lost in the field in the past 5~Gyr as the sum of the mass of all particles\footnote{\label{footnote:mass}To estimate the mass of particles in each cluster we have quantified the number of particles, $N_{bound}$, bound to the cluster at the end of the simulation and calculated the corresponding particle mass as $m_p=M_{GC}/N_{bound}$, where $M_{GC}$ is the current mass of a cluster given in Table~\ref{TableIC}.} which have escaped the cluster ($t_{\rm esc}> 0$), we find that the PI model sheds $2.1\times 10^{7} M_{\odot}$, which is 55\% of the GC population's current mass. The PII model shed $2.7\times 10^{6} M_{\odot}$, which is 7\% of the current mass. Similarly, PII-0.3-SLOW model lost $3.7\times 10^{6} M_{\odot}$, which is 10\% of the current mass and gives a half mass radius of 6.3~kpc.

This mass roughly constitutes one-hundredth to one-tenth of the total stellar halo mass \citep{blandhawthorn16}, and it is probably only a lower limit to the mass of escaped stars in the field, since a number of clusters initially in the Galaxy must have been destroyed over time (see Introduction), and thus are not identifiable any longer as globular clusters today. It is also interesting to note that escaped stars are mostly redistributed in the inner Galaxy, the half mass radius of the PI model being 4.0~kpc and that of the PII and PII-0.3-SLOW models being 6.3~kpc. 

The total mass lost from the clusters, as well as its spatial distribution, hence depends on the Galactic potential adopted: the variations between the PI model and both the PII and PII-0.3-SLOW models are of course caused by the PI's inclusion of the bulge, which leads to larger tidal forces in the center of the galaxy and subsequently drives larger mass loss. 

Despite the differences in the modeling approach, it is interesting to compare our results to those of \citet{baumgardt2017global}. Briefly, our experiments differ in the following ways: they employ N-body simulations while we use our test-particle approach; they have an integration time of 12 Gyrs compared to our 5 Gyrs; and lastly their clusters have circular orbits in a Galactic potential modeled as an isothermal sphere as compared to the more realistic orbits and Galactic potentials considered in this work. Interestingly, the authors find that over 12 Gyrs their population of globular clusters loose 2/3 of their initial mass. This is roughly consistent with our usage of the PI model, whose globular clusters shed 35\% of their initial mass in 5 Gyrs, which is roughly half the mass found by \citet{baumgardt2017global} in a period that is also about half as long.

\subsection{From the nearest to the furthest extra-tidal structures}\label{streamvsD}
\begin{figure*}[h!]
\begin{center}
\includegraphics[clip=true, trim = 0mm 15mm 0mm 20mm, width=0.9\columnwidth]{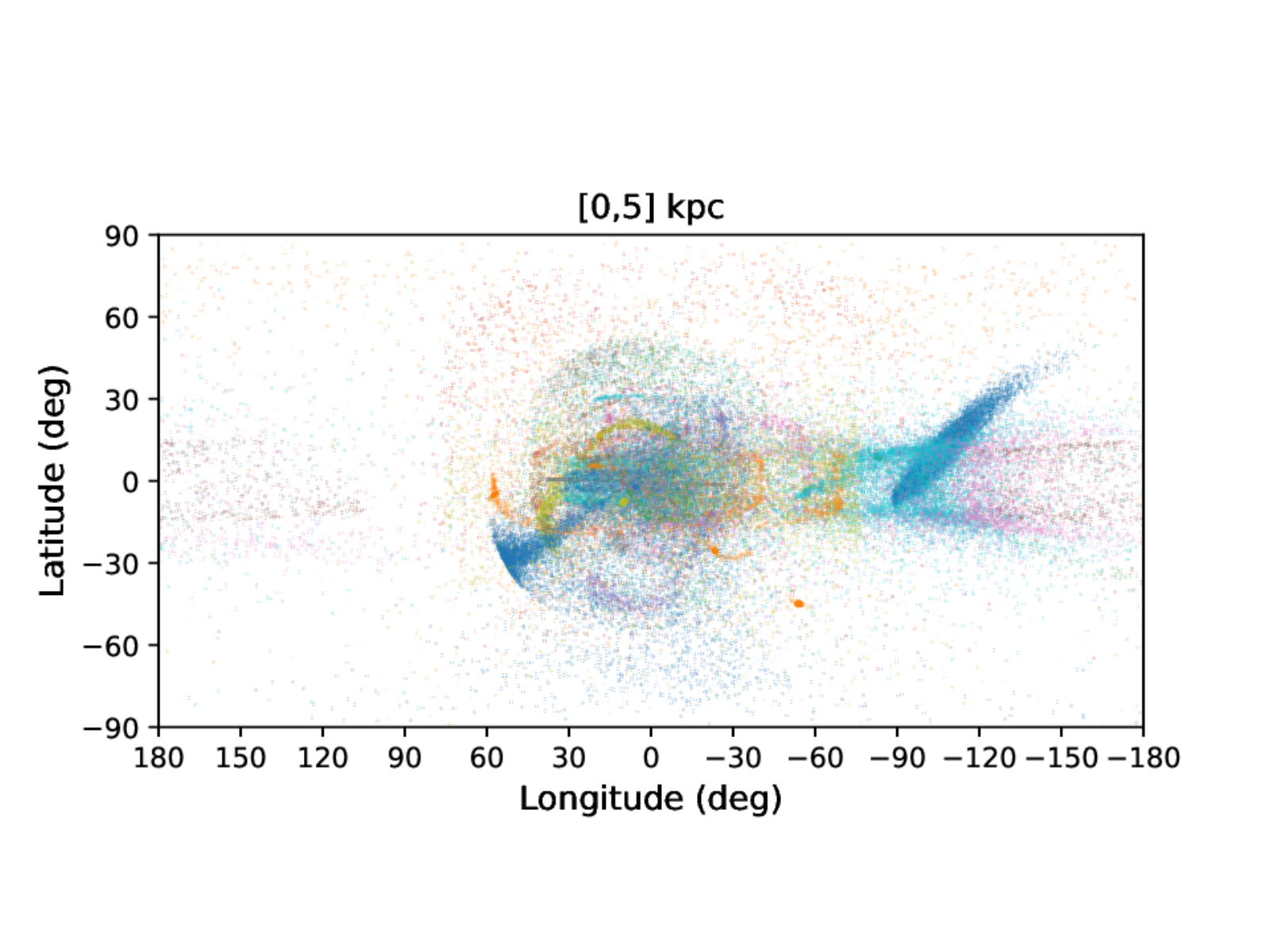}
\includegraphics[clip=true, trim = 0mm 15mm 0mm 20mm, width=0.9\columnwidth]{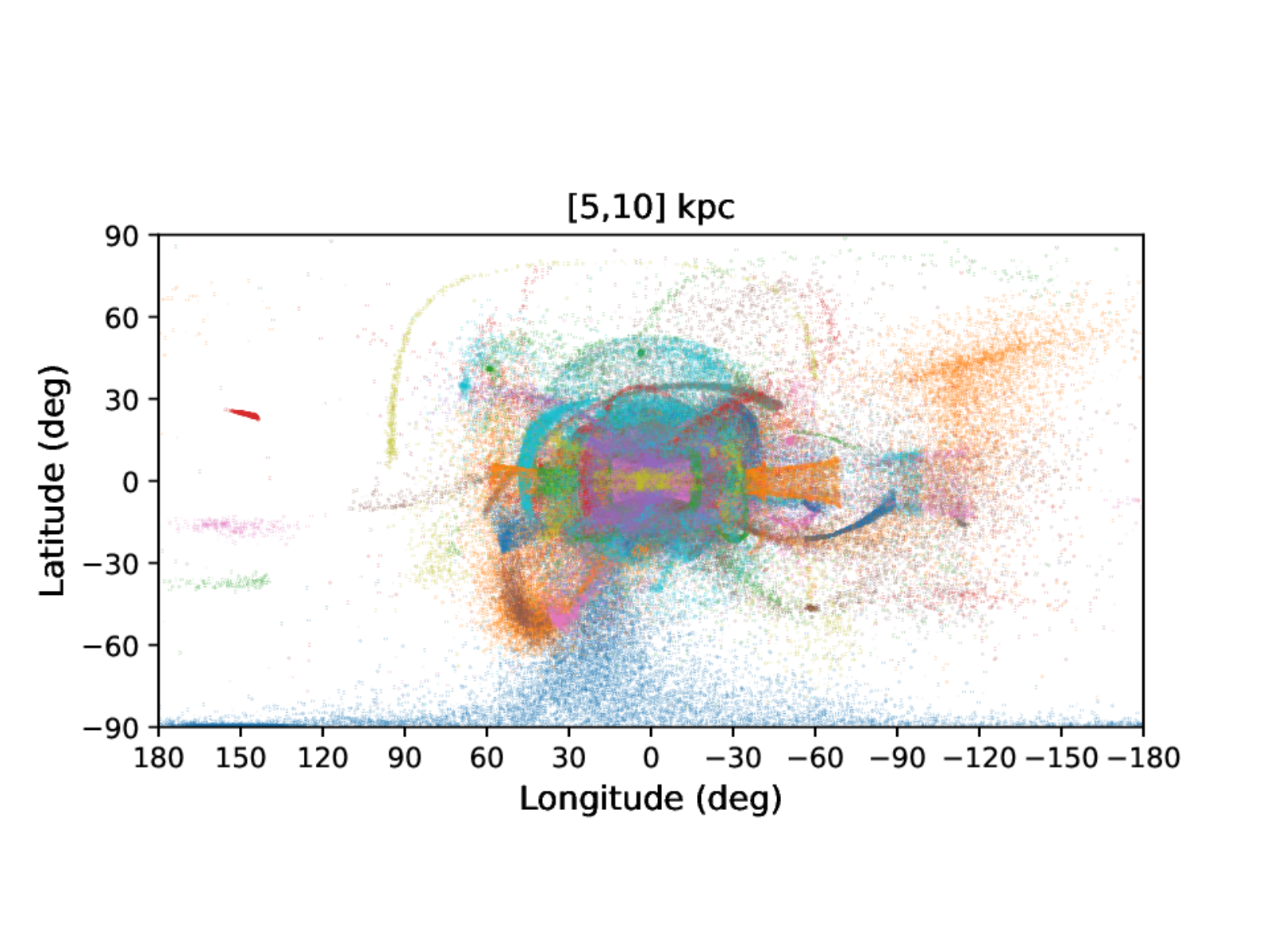}

\includegraphics[clip=true, trim = 0mm 20mm 0mm 20mm, width=0.9\columnwidth]{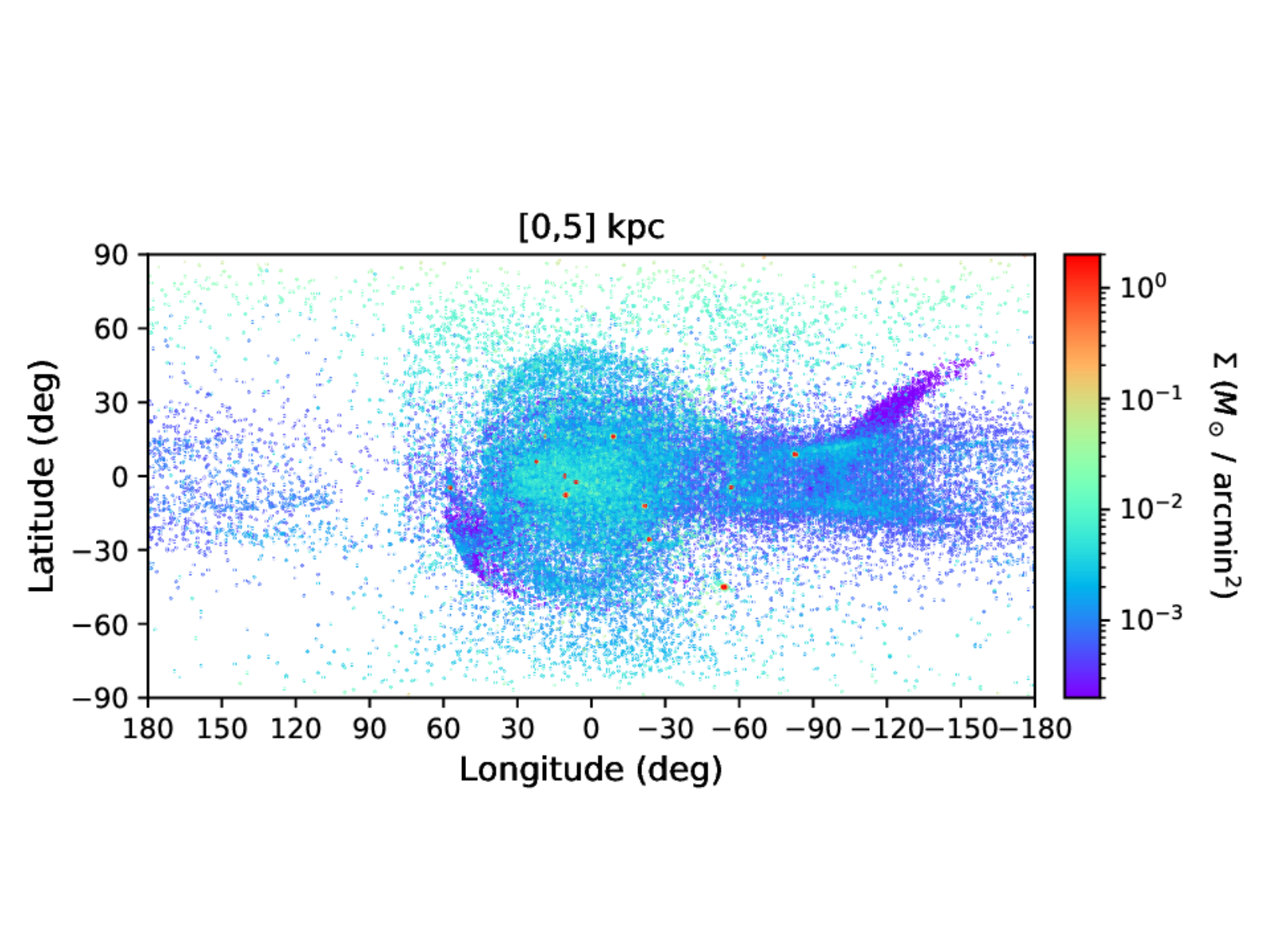}
\includegraphics[clip=true, trim = 0mm 20mm 0mm 20mm, width=0.9\columnwidth]{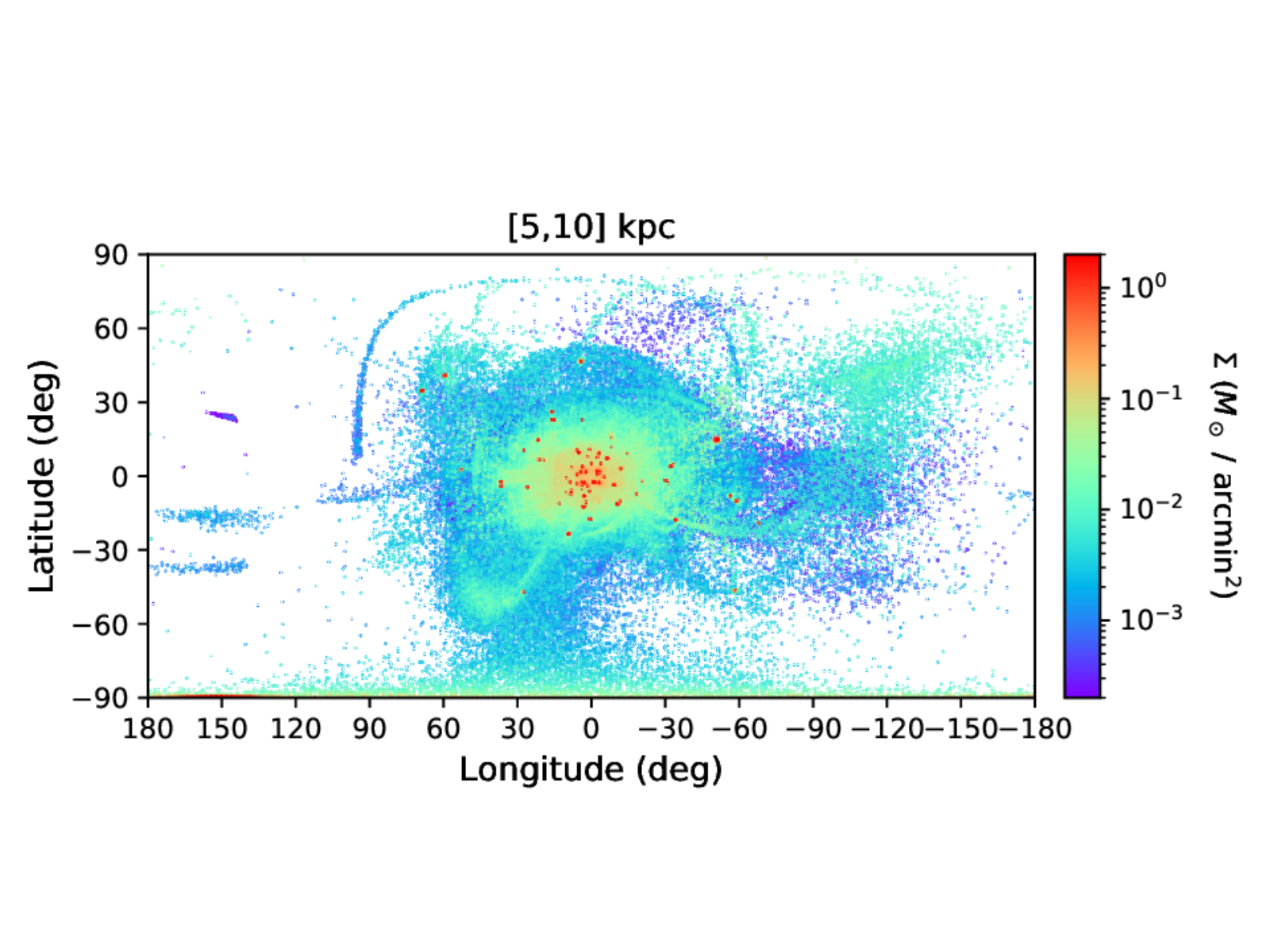}

\includegraphics[clip=true, trim = 0mm 20mm 0mm 20mm, width=0.9\columnwidth]{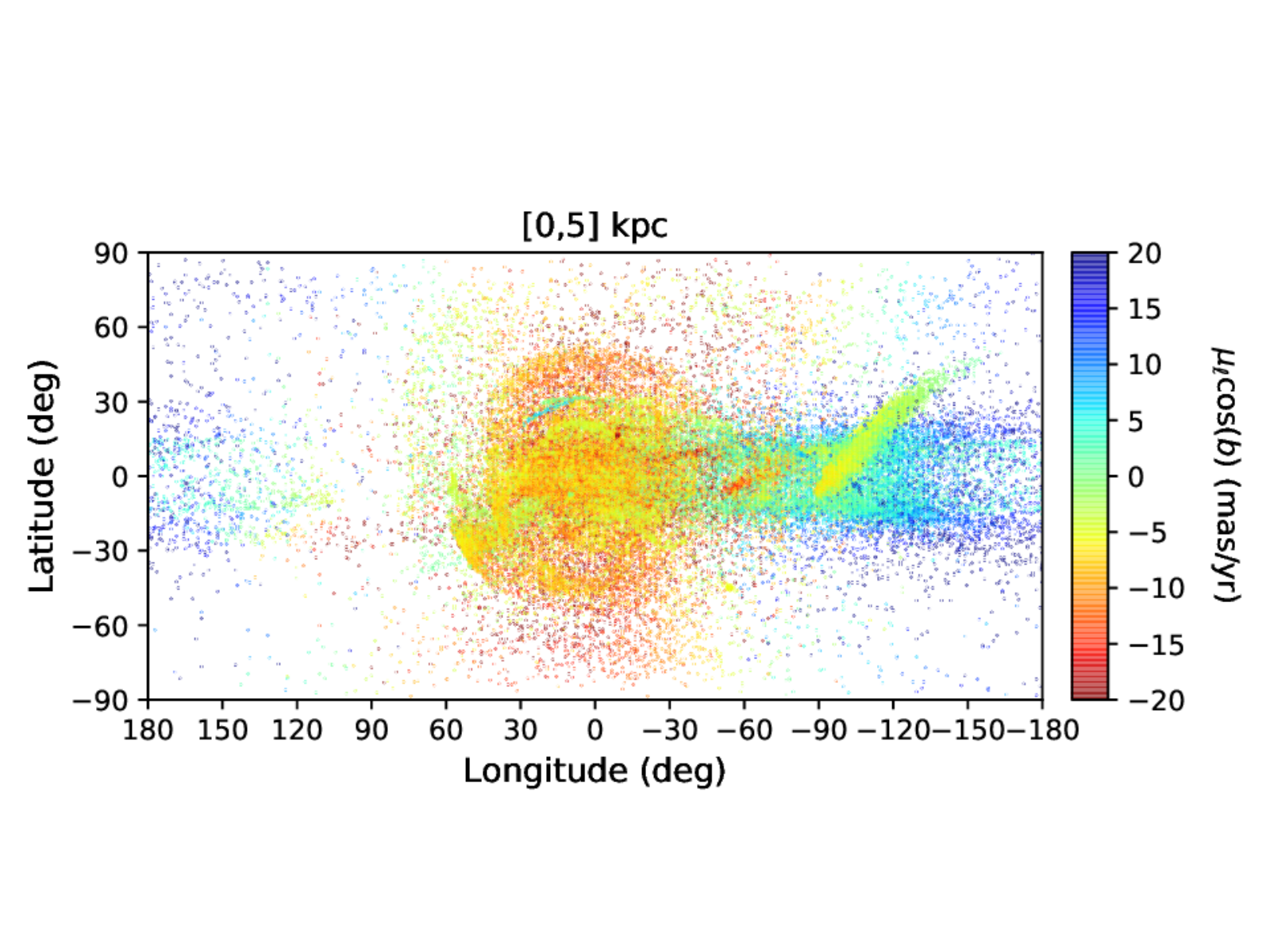}
\includegraphics[clip=true, trim = 0mm 20mm 0mm 20mm, width=0.9\columnwidth]{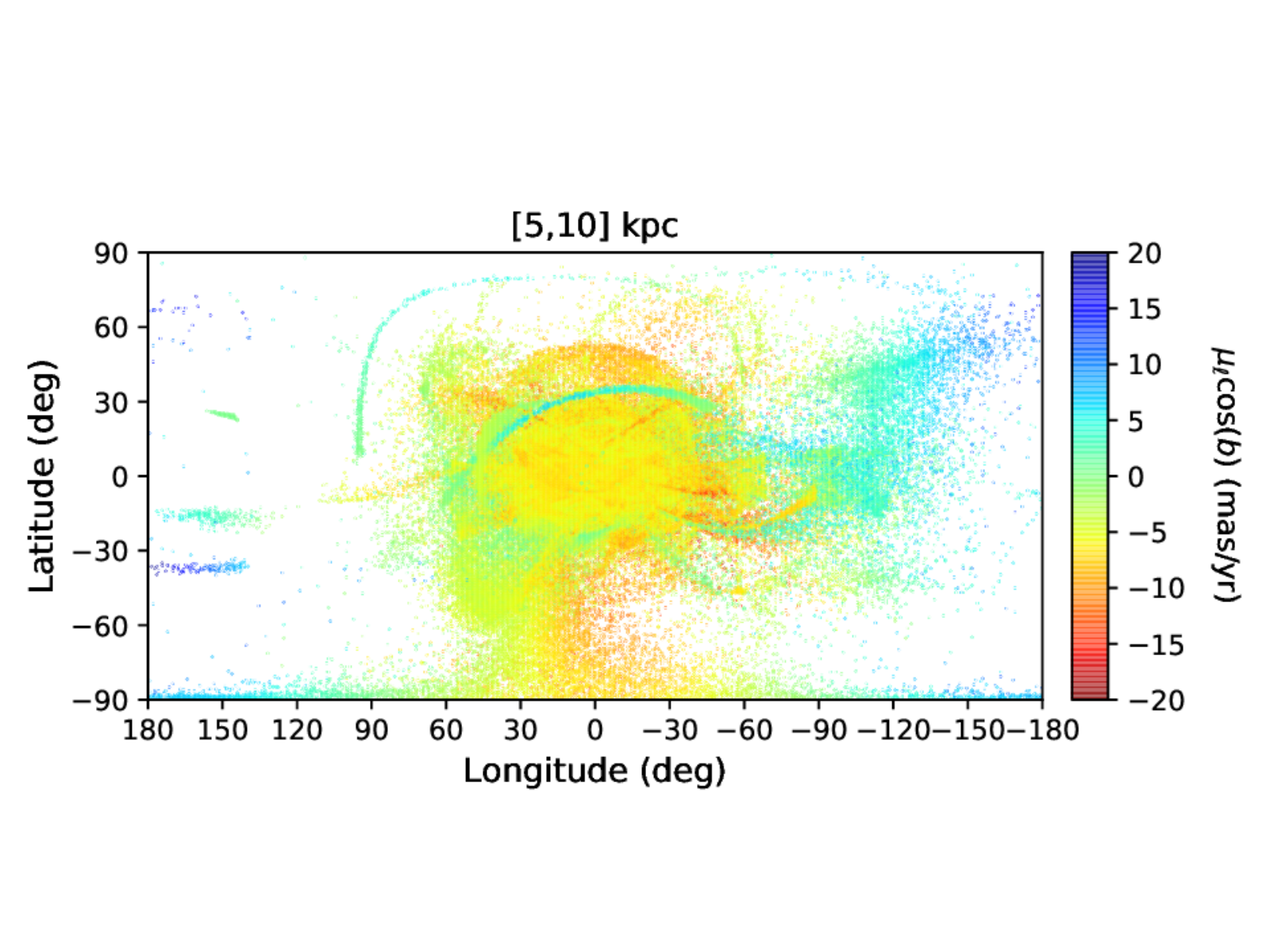}

\includegraphics[clip=true, trim = 0mm 20mm 0mm 20mm, width=0.9\columnwidth]{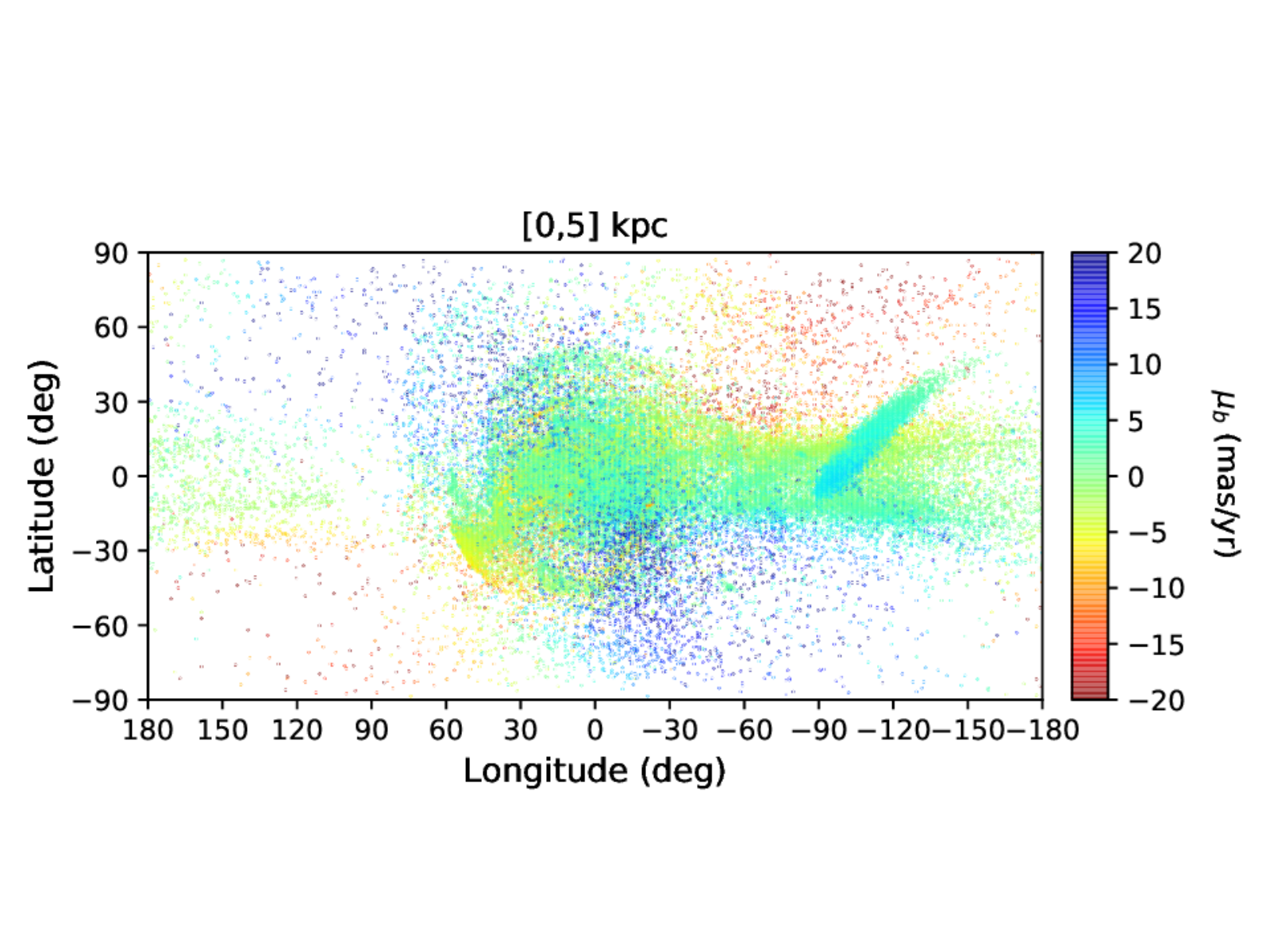}
\includegraphics[clip=true, trim = 0mm 20mm 0mm 20mm, width=0.9\columnwidth]{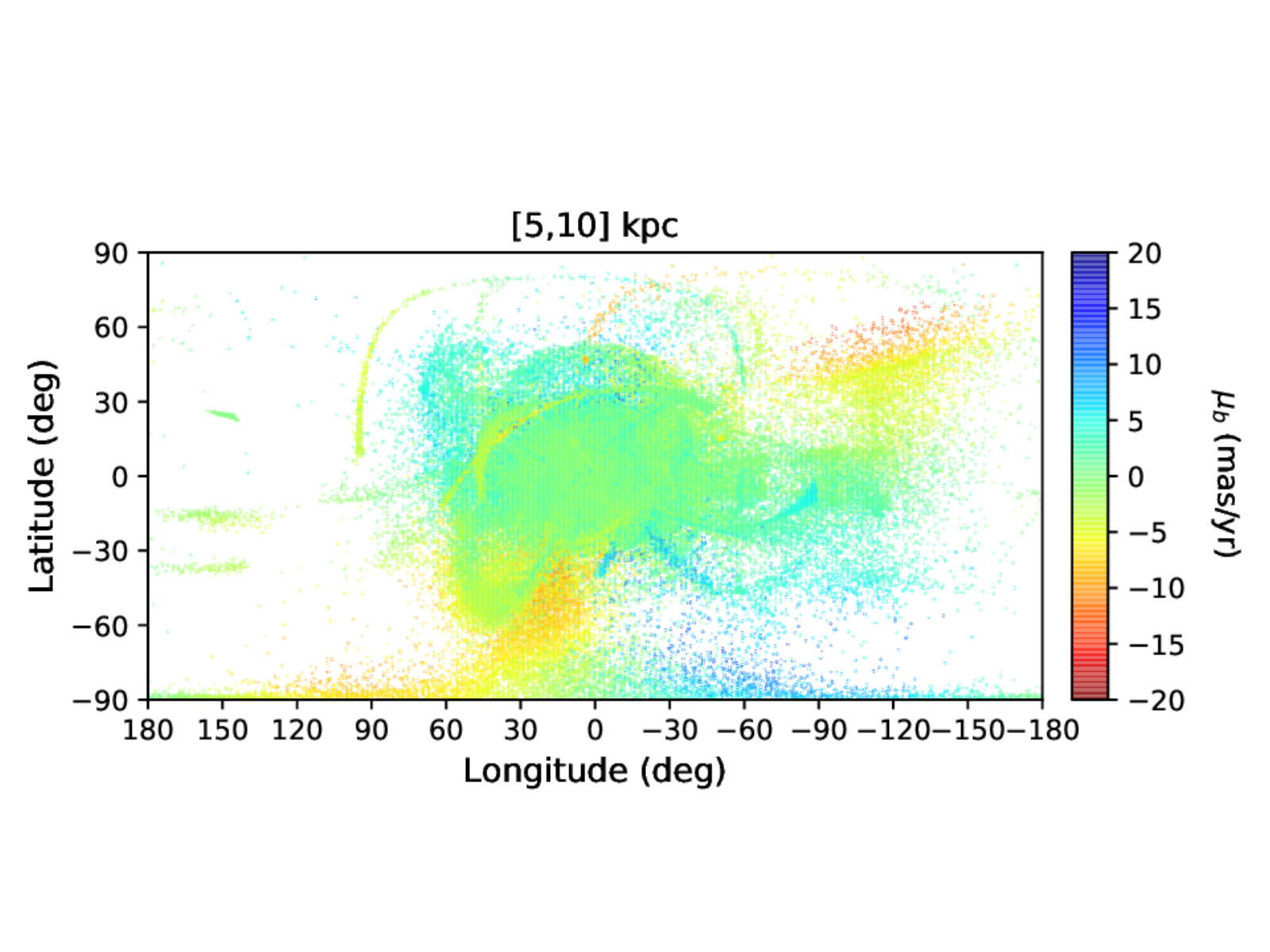}

\includegraphics[clip=true, trim = 0mm 20mm 0mm 20mm, width=0.9\columnwidth]{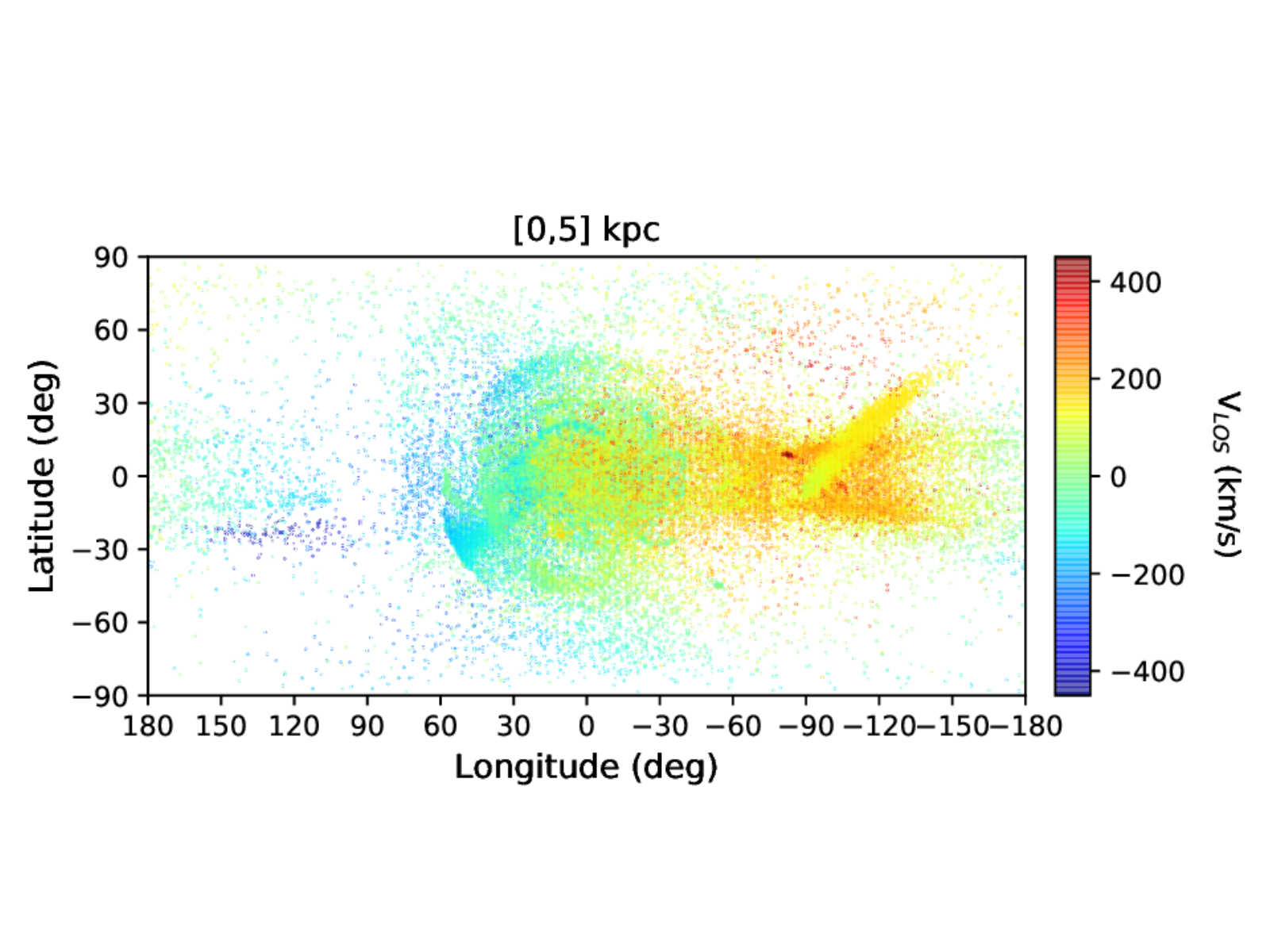}
\includegraphics[clip=true, trim = 0mm 20mm 0mm 20mm, width=0.9\columnwidth]{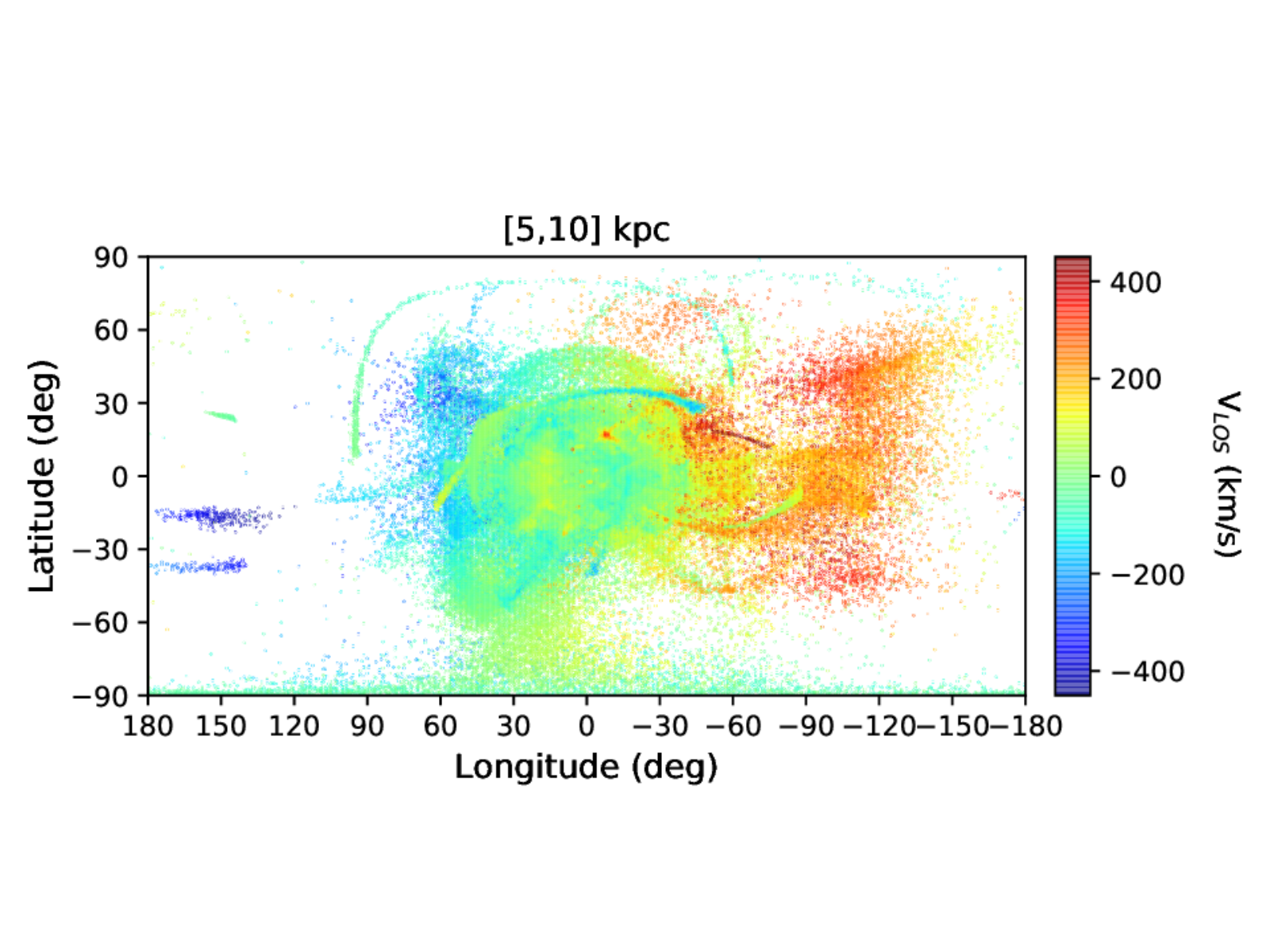}
\end{center}

\caption{(\emph{Left column}): Extra-tidal features found at a distance of [0,5]~kpc from the Sun and projected in the $(\ell, b)$ plane. \emph{Top row:} Scatter plot, with different colors indicating different progenitor clusters;  \emph{Second row:} Mass density map in logarithmic scale; \emph{Third row:} Map color-coated by proper motions in longitudinal direction;    \emph{Fourth row:} Map color-coated by proper motions in latitudinal direction;  \emph{Bottom row:} Map color-coded by line-of-sight velocities. (\emph{Right column}): same as left column, but for the tidal features found at a distance of [5, 10]~kpc from the Sun. \textit{Note} that the 10 colors used in the top panels are recycled between the 159 clusters. Thus, all particles from the same cluster share one color, but a color is not unique to a cluster. This figure shows streams, and their corresponding properties, as found for model PII only. In all panels, only the reference simulations are shown, for better clarity.}\label{D0-10}
\end{figure*}

\begin{figure*}[h!]
\begin{center}
\includegraphics[clip=true, trim = 0mm 15mm 0mm 20mm, width=0.9\columnwidth]{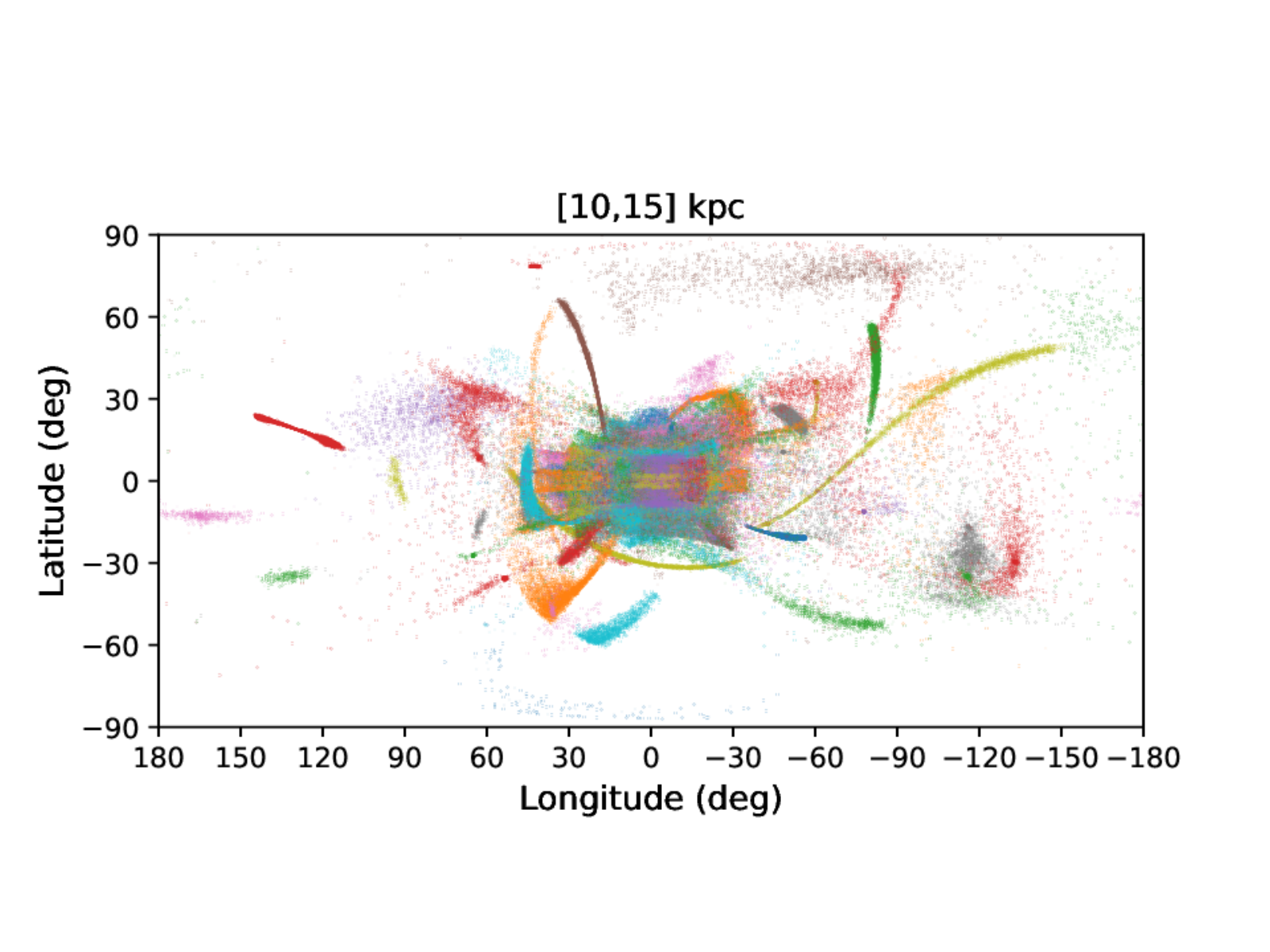}
\includegraphics[clip=true, trim = 0mm 15mm 0mm 20mm, width=0.9\columnwidth]{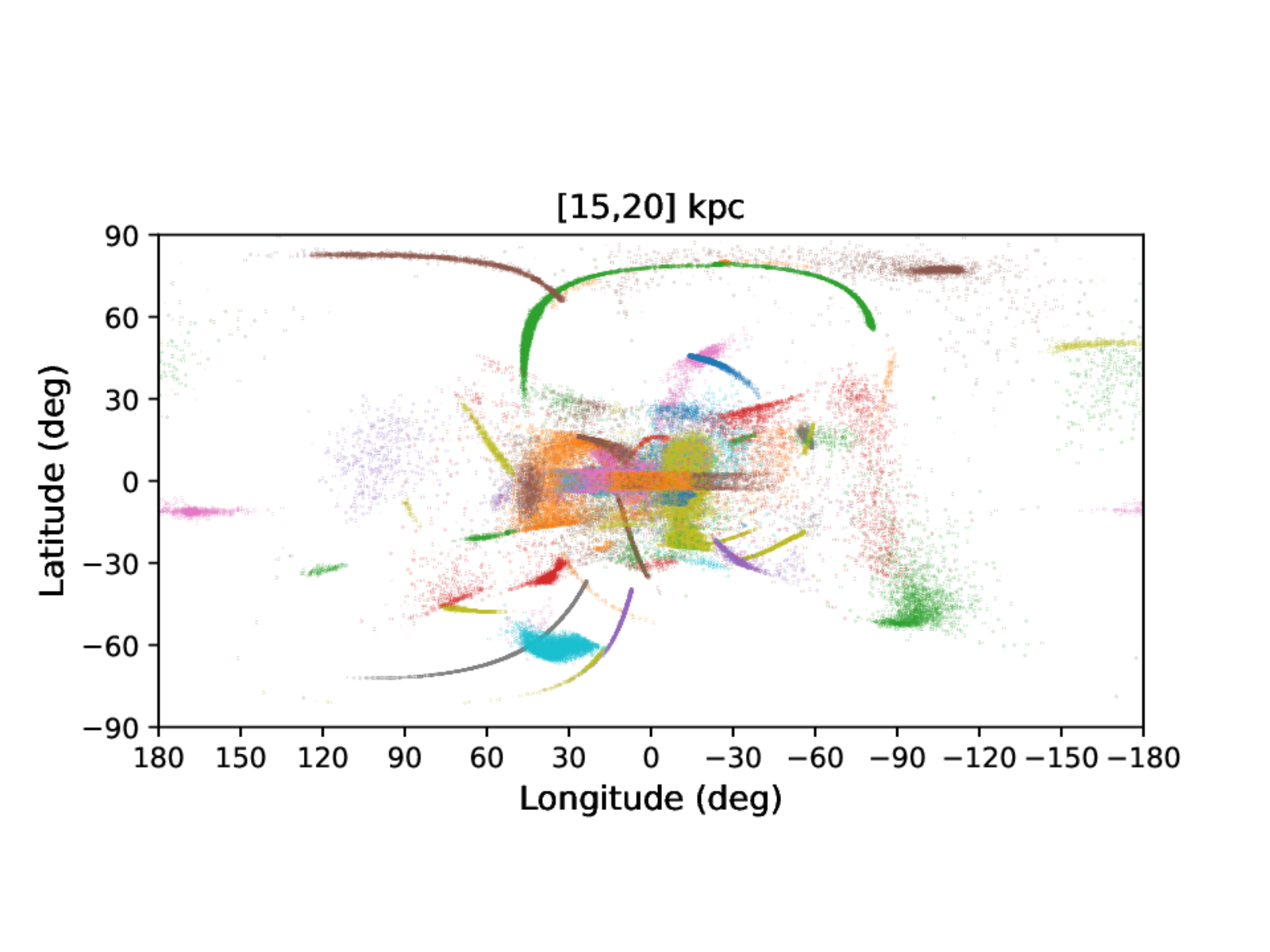}

\includegraphics[clip=true, trim = 0mm 20mm 0mm 20mm, width=0.9\columnwidth]{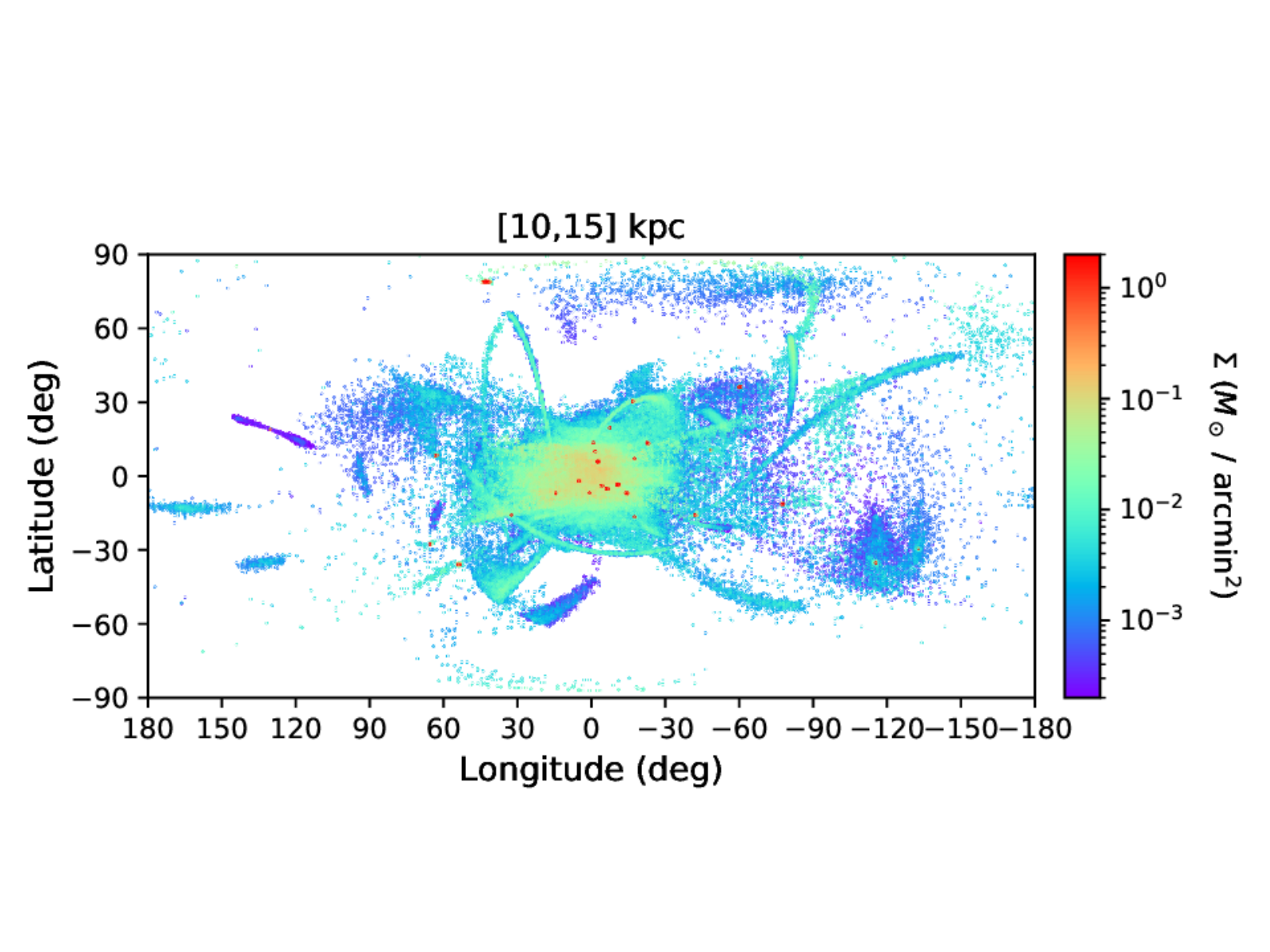}
\includegraphics[clip=true, trim = 0mm 20mm 0mm 20mm, width=0.9\columnwidth]{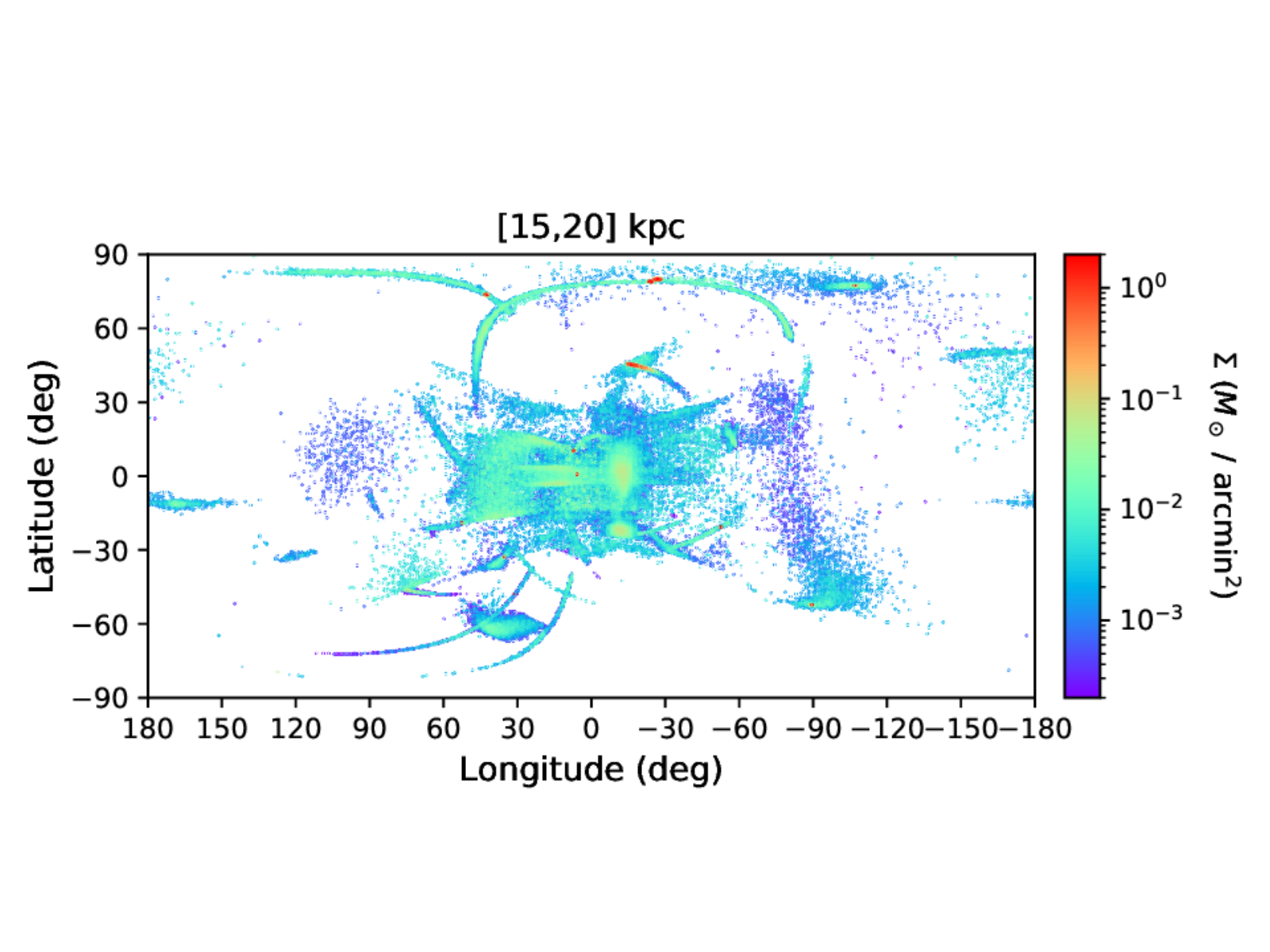}

\includegraphics[clip=true, trim = 0mm 20mm 0mm 20mm, width=0.9\columnwidth]{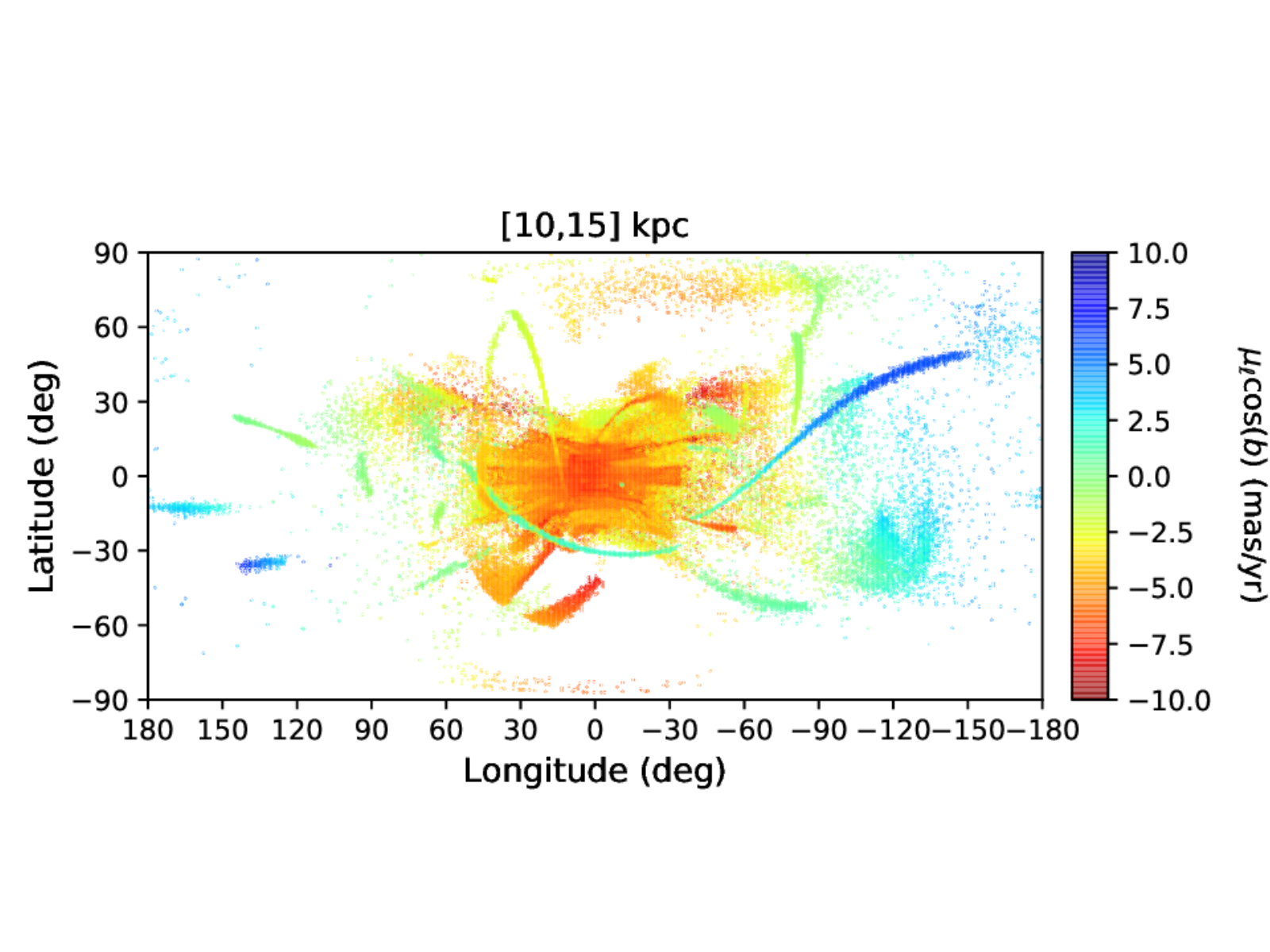}
\includegraphics[clip=true, trim = 0mm 20mm 0mm 20mm, width=0.9\columnwidth]{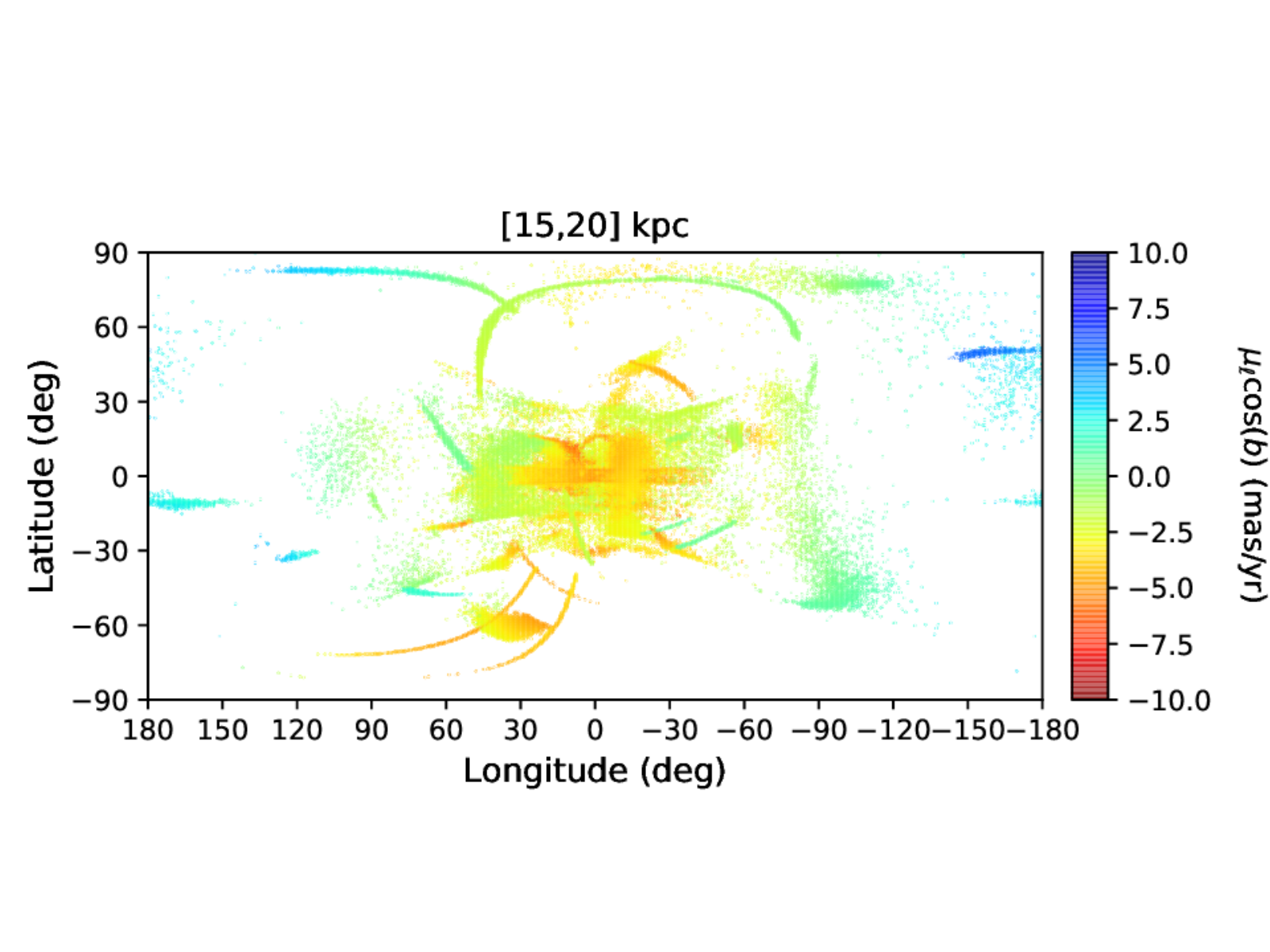}

\includegraphics[clip=true, trim = 0mm 20mm 0mm 20mm, width=0.9\columnwidth]{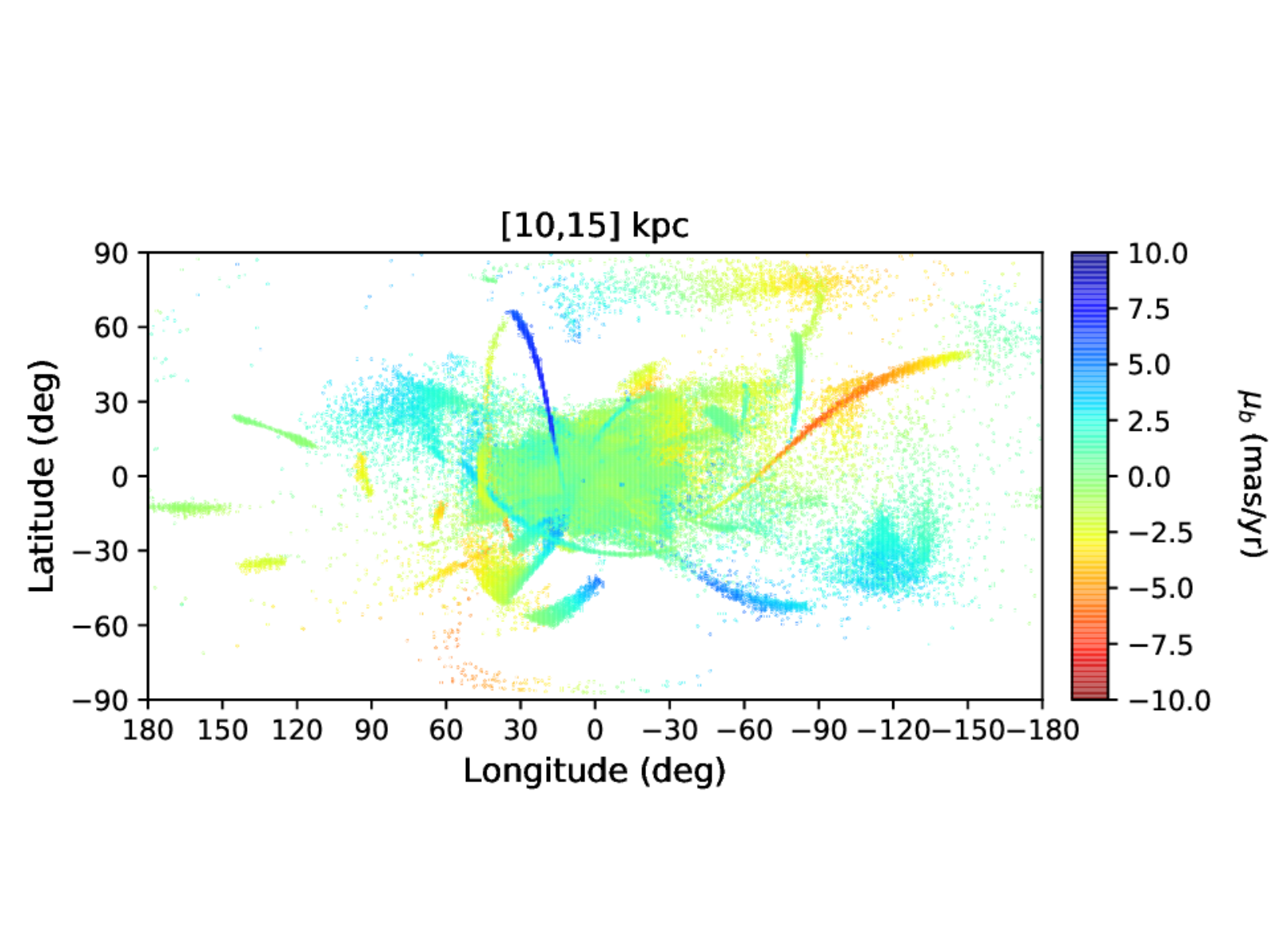}
\includegraphics[clip=true, trim = 0mm 20mm 0mm 20mm, width=0.9\columnwidth]{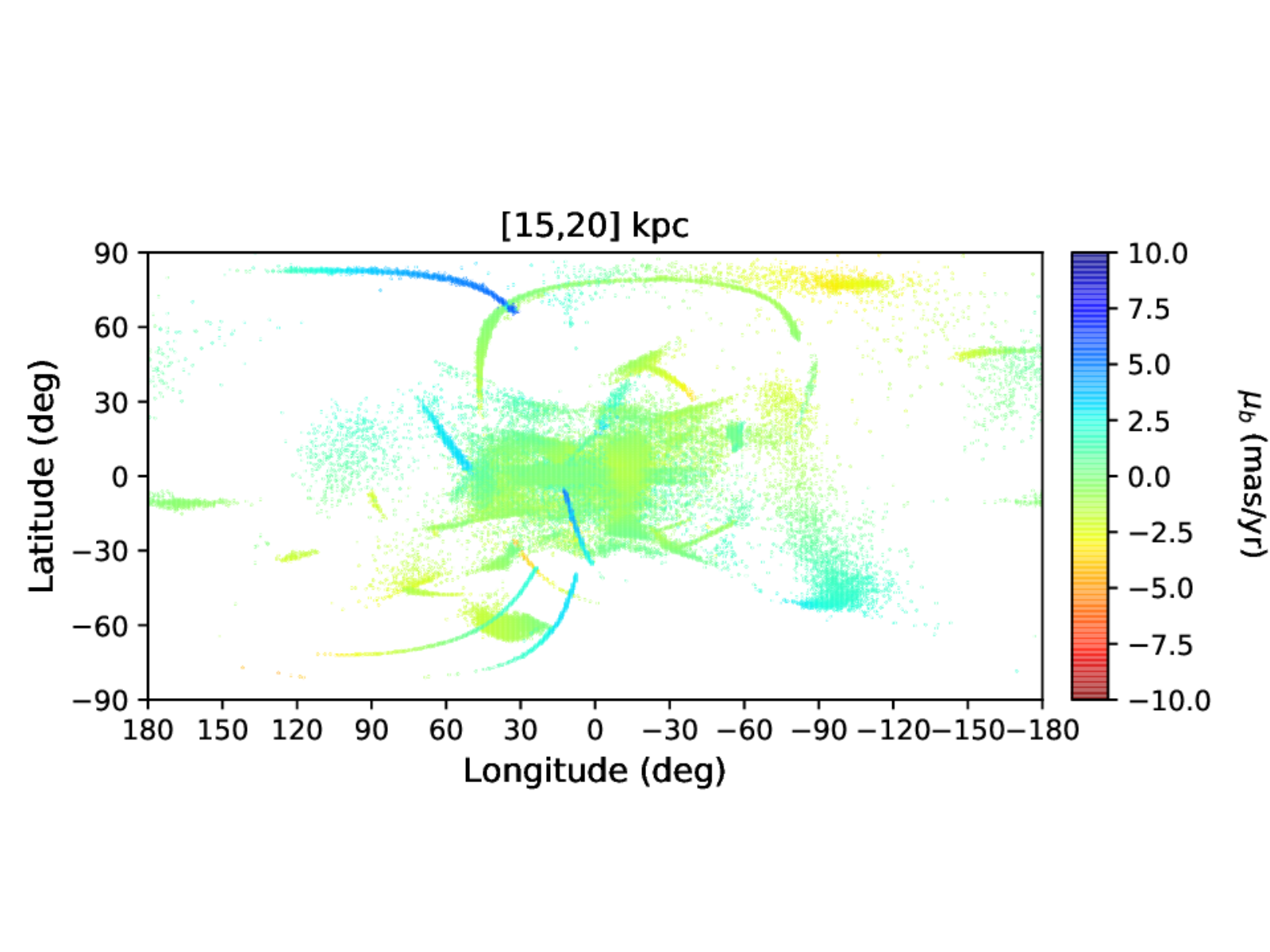}

\includegraphics[clip=true, trim = 0mm 20mm 0mm 20mm, width=0.9\columnwidth]{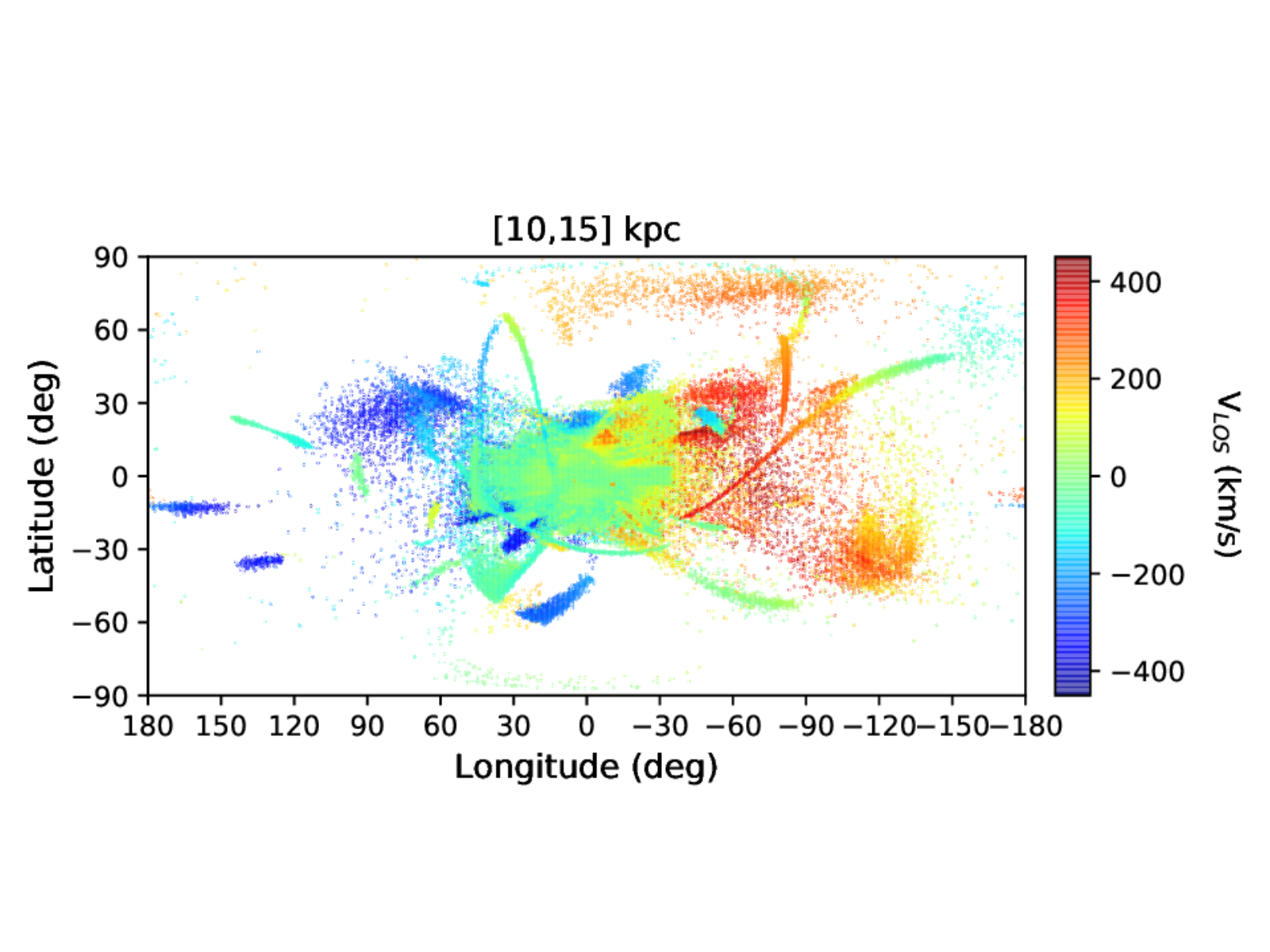}
\includegraphics[clip=true, trim = 0mm 20mm 0mm 20mm, width=0.9\columnwidth]{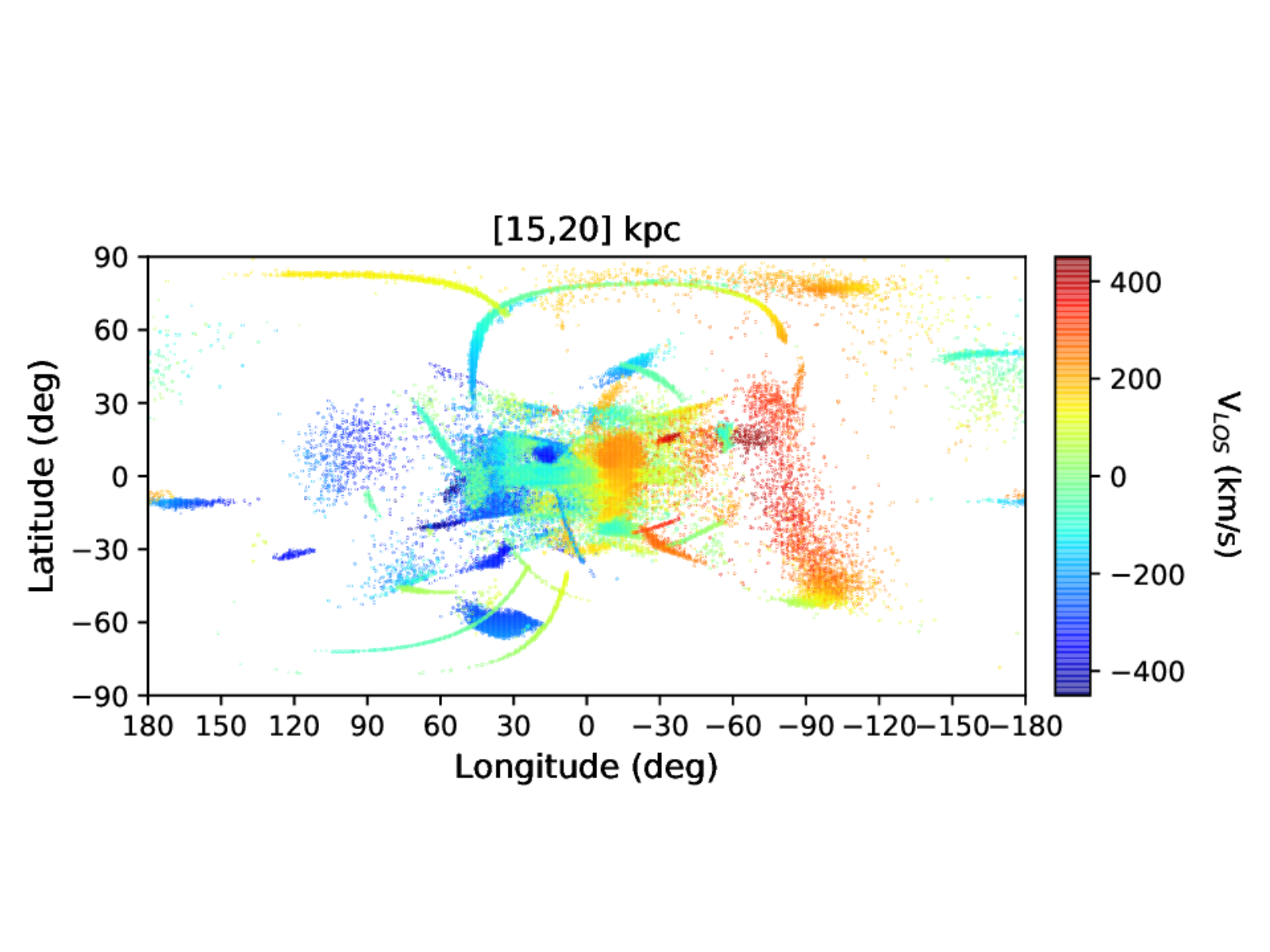}
\end{center}
\caption{As for  Fig.~\ref{D0-10}, but for two different distance bins: [10, 15]~kpc from the Sun (\emph{left column}) and  [15, 20]~kpc from the Sun (\emph{right column}). . This figure shows streams, and their corresponding properties, as found for model PII only. In all panels, only the reference simulations are shown, for better clarity.}\label{D10-20}
\end{figure*}

\begin{figure*}[h!]
\begin{center}
\includegraphics[clip=true, trim = 0mm 15mm 0mm 20mm, width=0.9\columnwidth]{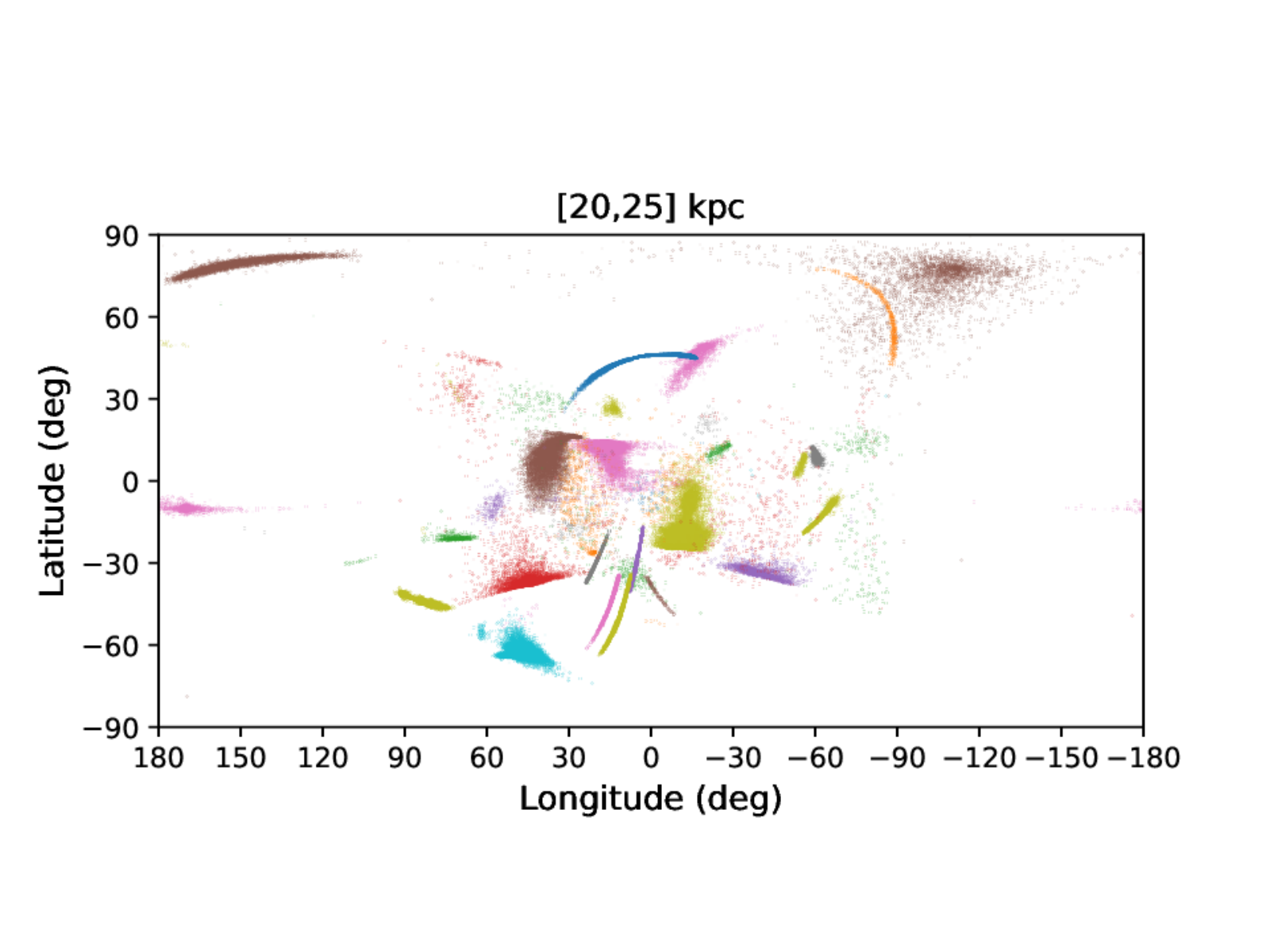}
\includegraphics[clip=true, trim = 0mm 15mm 0mm 20mm, width=0.9\columnwidth]{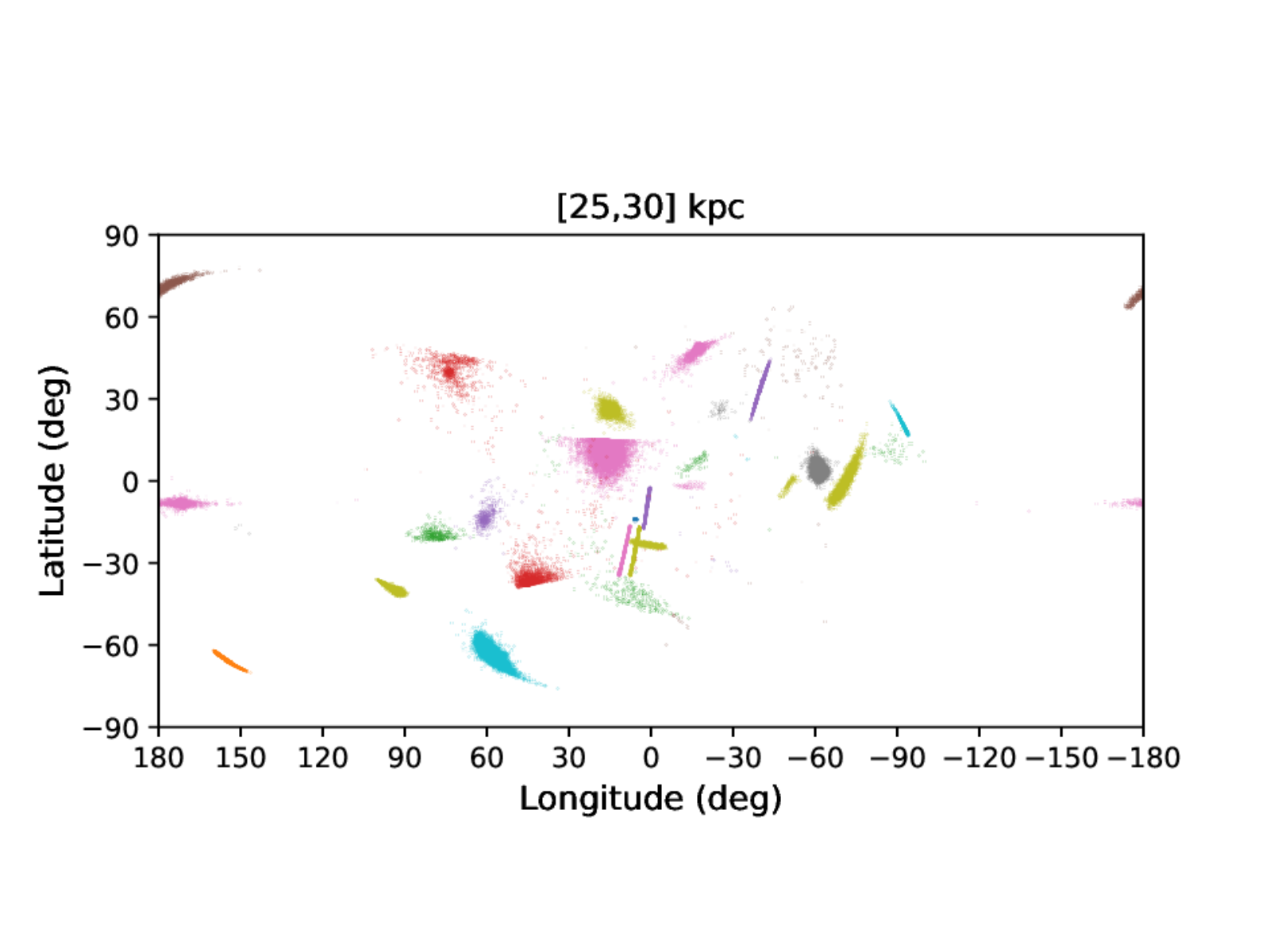}

\includegraphics[clip=true, trim = 0mm 20mm 0mm 20mm, width=0.9\columnwidth]{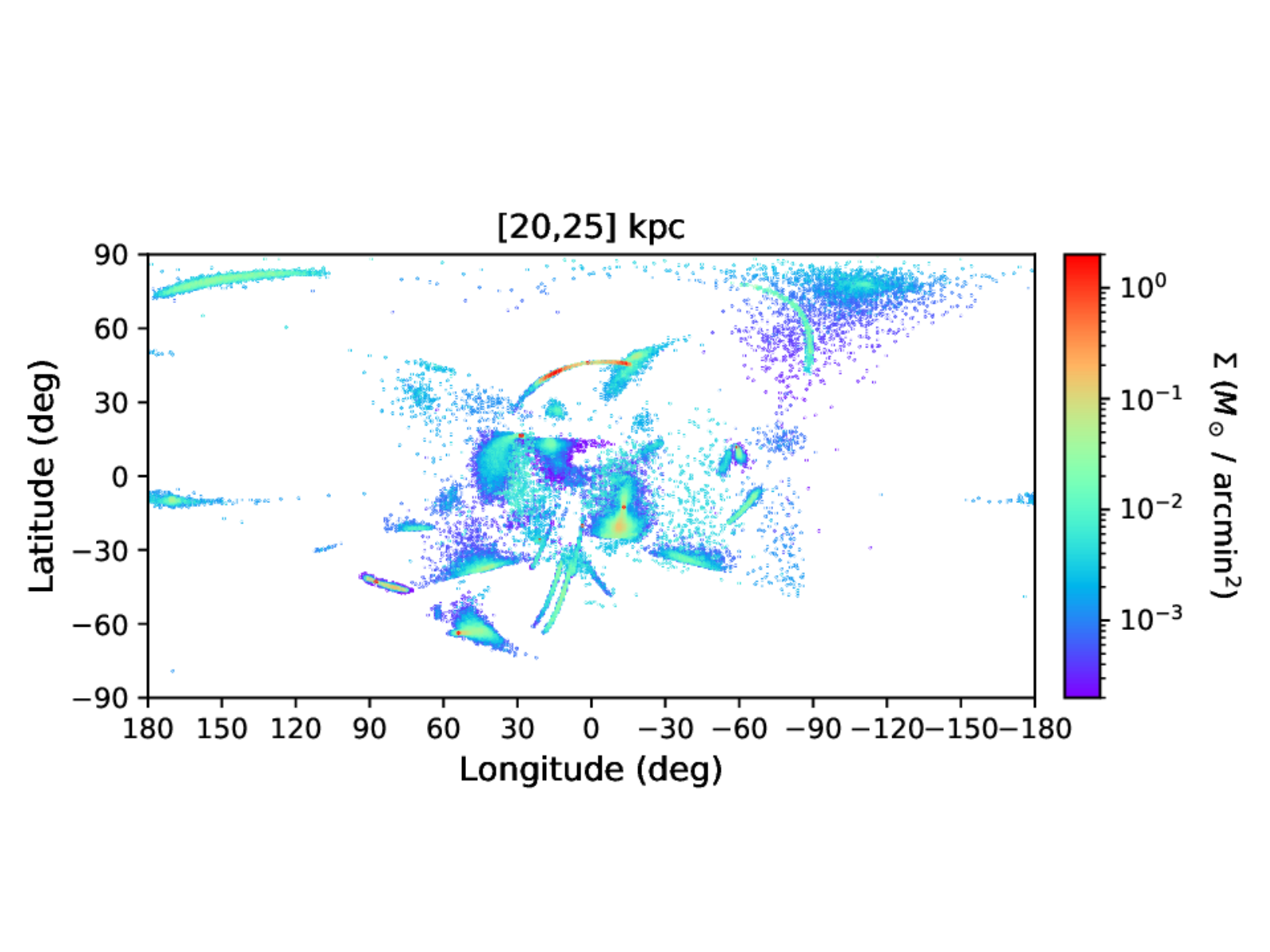}
\includegraphics[clip=true, trim = 0mm 20mm 0mm 20mm, width=0.9\columnwidth]{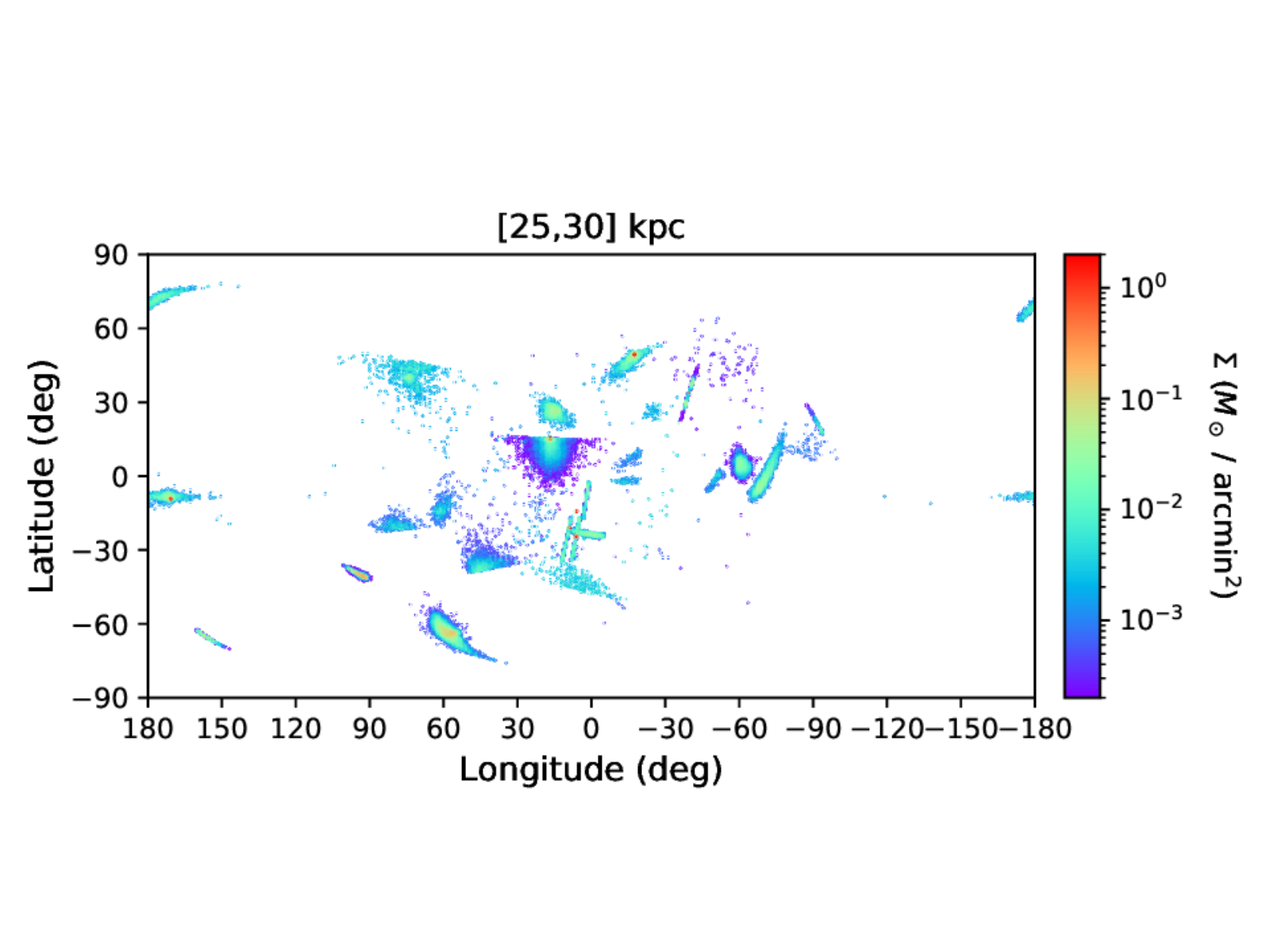}

\includegraphics[clip=true, trim = 0mm 20mm 0mm 20mm, width=0.9\columnwidth]{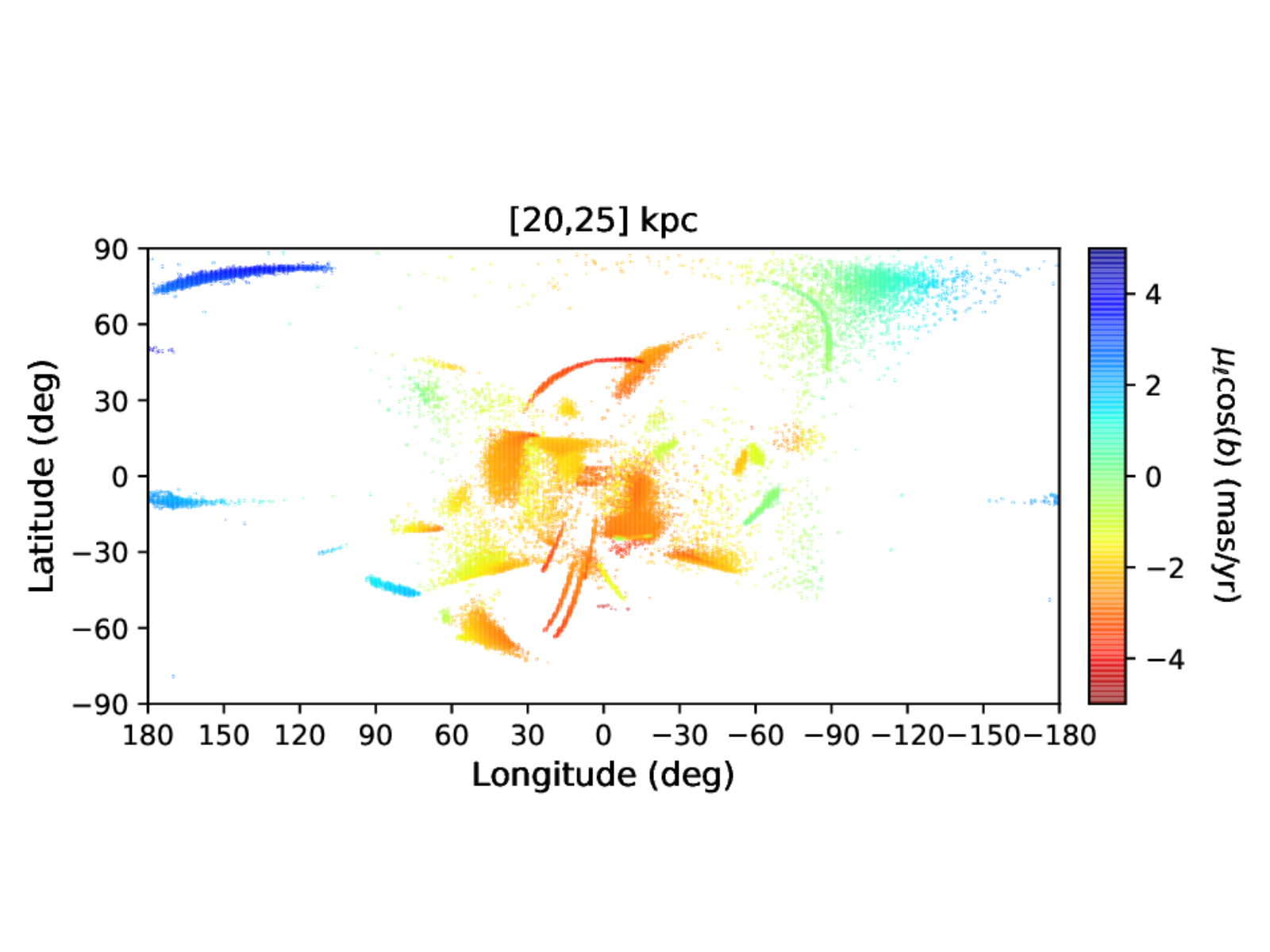}
\includegraphics[clip=true, trim = 0mm 20mm 0mm 20mm, width=0.9\columnwidth]{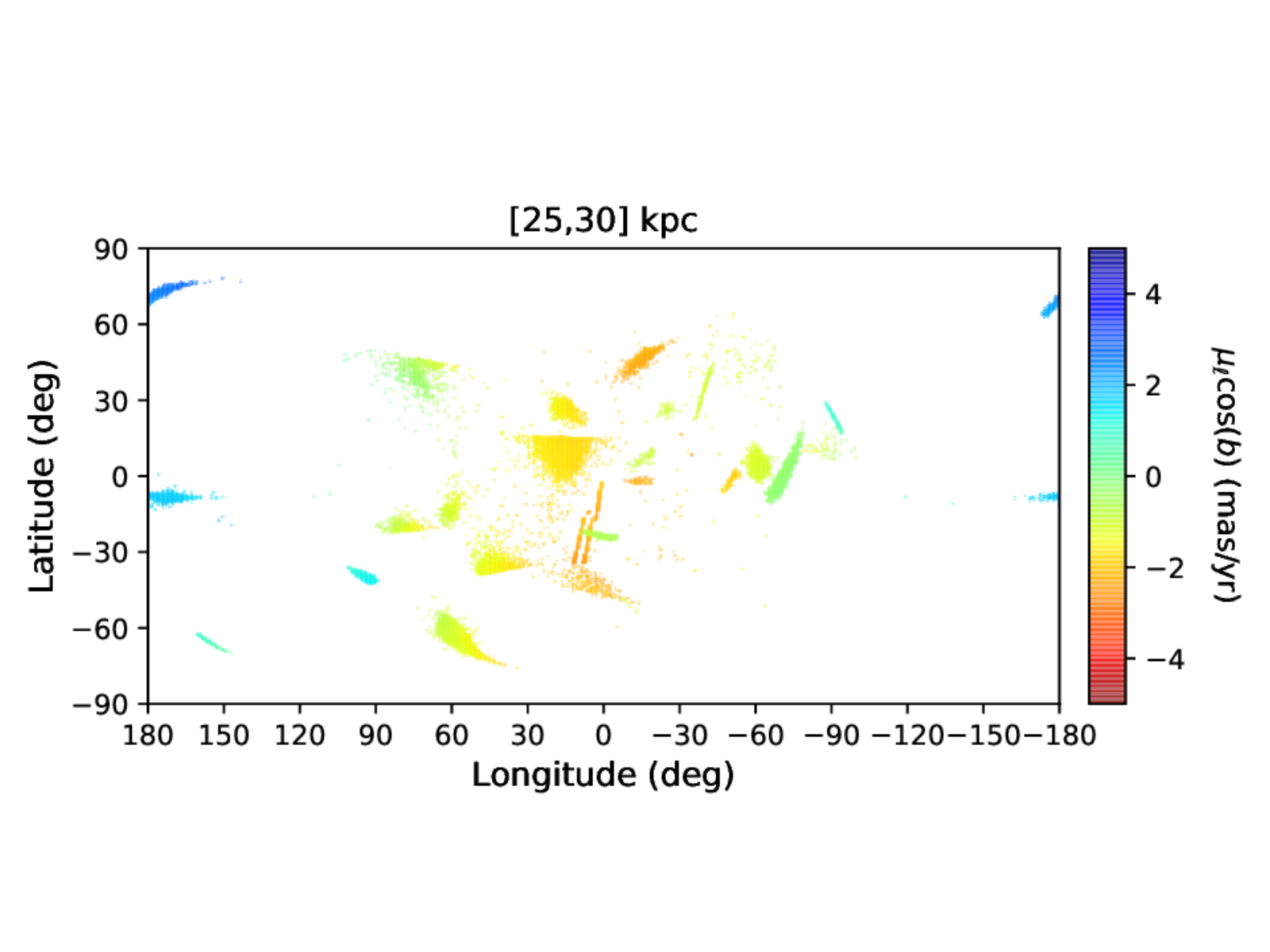}

\includegraphics[clip=true, trim = 0mm 20mm 0mm 20mm, width=0.9\columnwidth]{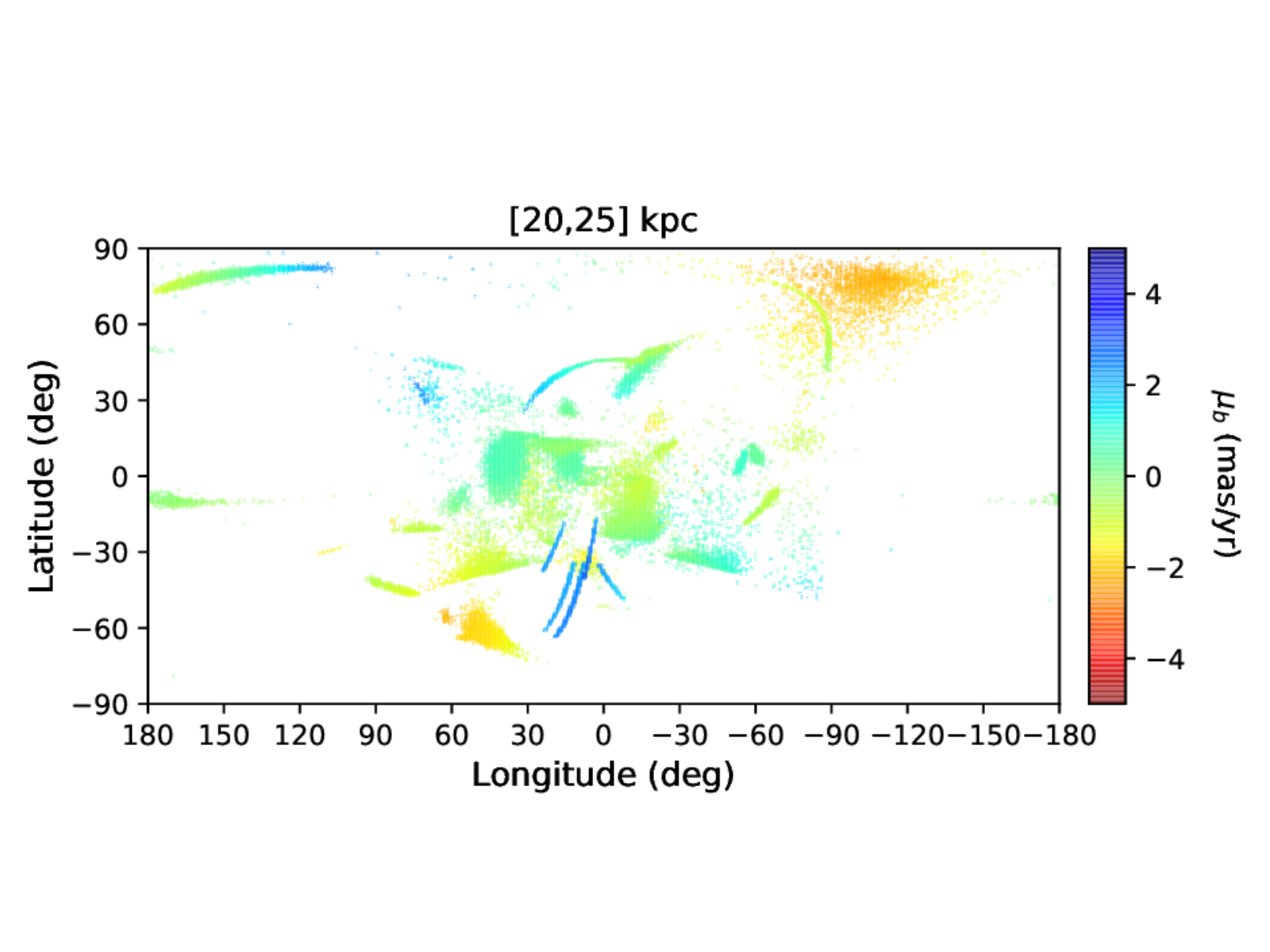}
\includegraphics[clip=true, trim = 0mm 20mm 0mm 20mm, width=0.9\columnwidth]{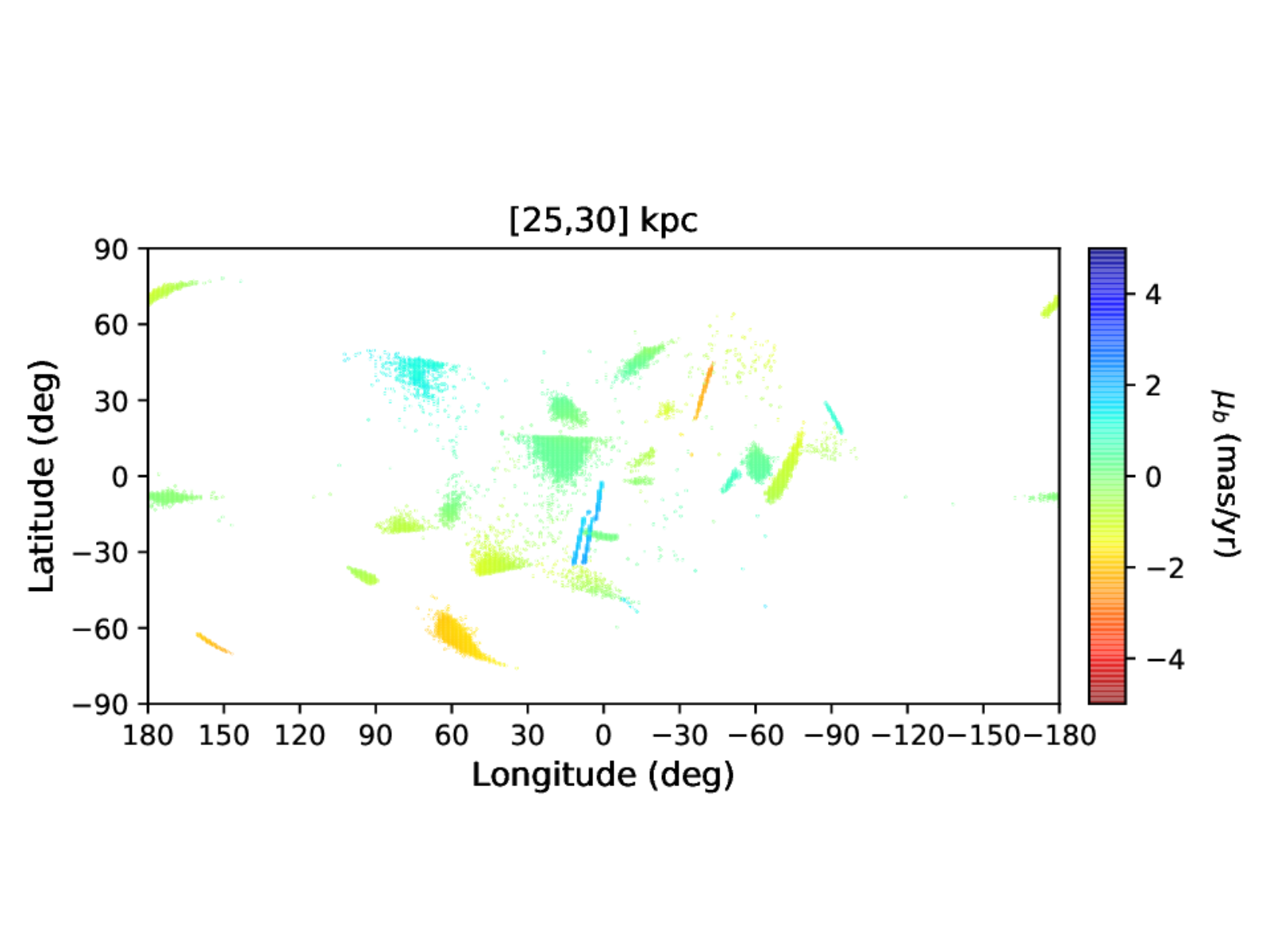}

\includegraphics[clip=true, trim = 0mm 20mm 0mm 20mm, width=0.9\columnwidth]{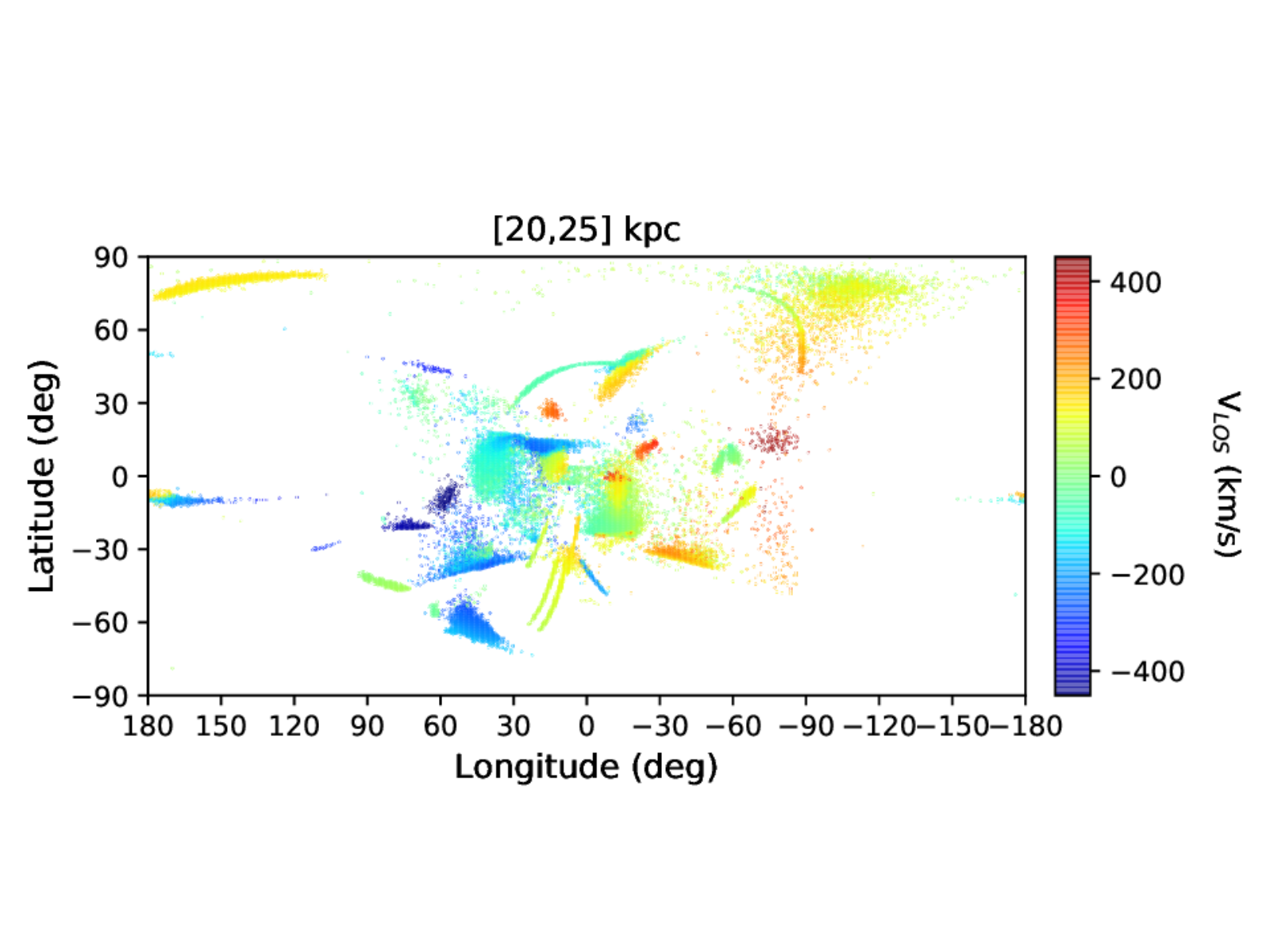}
\includegraphics[clip=true, trim = 0mm 20mm 0mm 20mm, width=0.9\columnwidth]{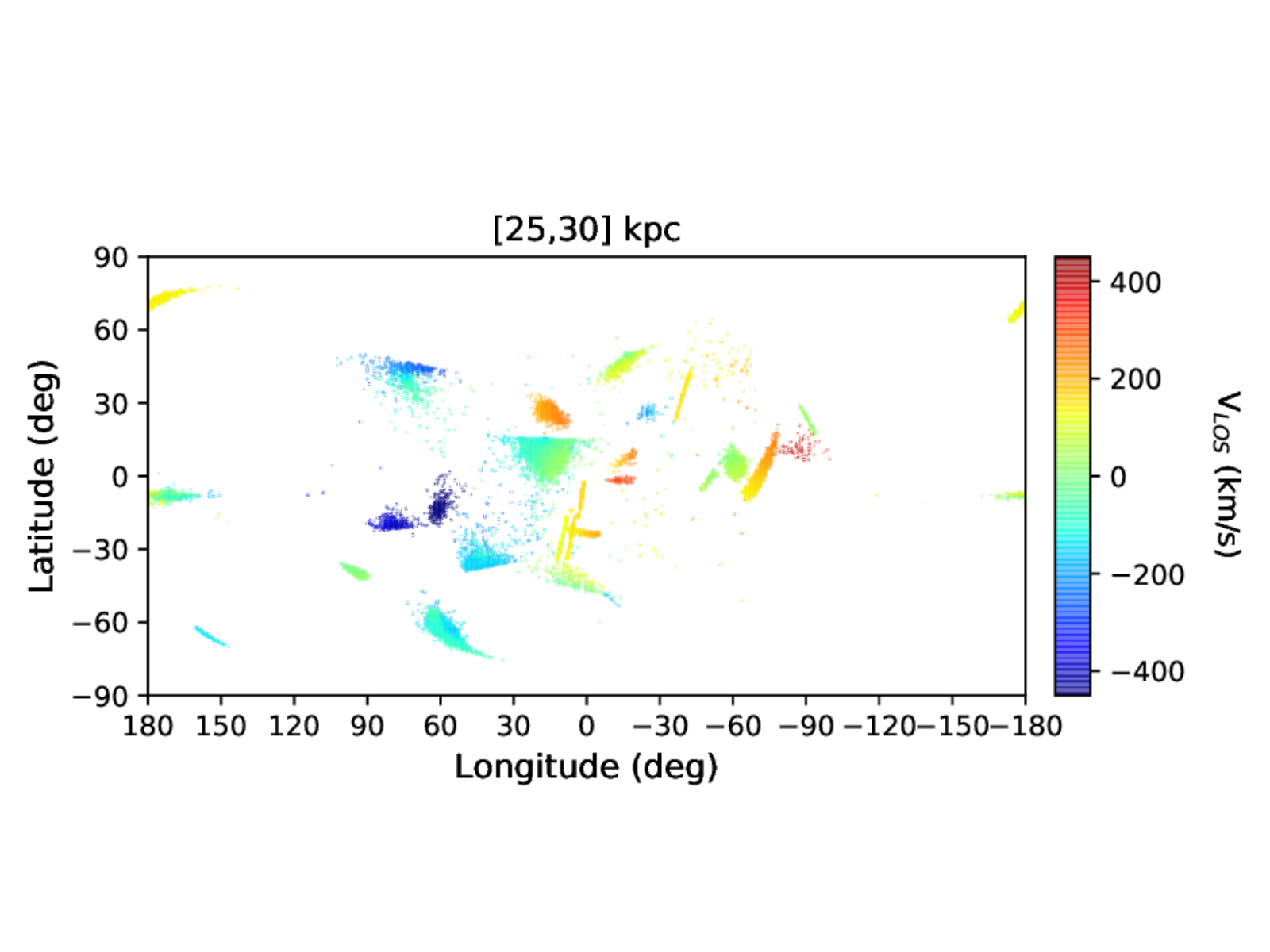}
\end{center}
\caption{As for  Fig.~\ref{D0-10}, but for two different distance bins: [20, 25]~kpc from the Sun (\emph{left column}) and  [25, 30]~kpc from the Sun (\emph{right column}). This figure shows streams, and their corresponding properties, as found for model PII only. In all panels, only the reference simulations are shown, for better clarity.}\label{D20-30}
\end{figure*}

\begin{figure*}[h!]
\begin{center}
\includegraphics[clip=true, trim = 0mm 15mm 0mm 20mm, width=0.9\columnwidth]{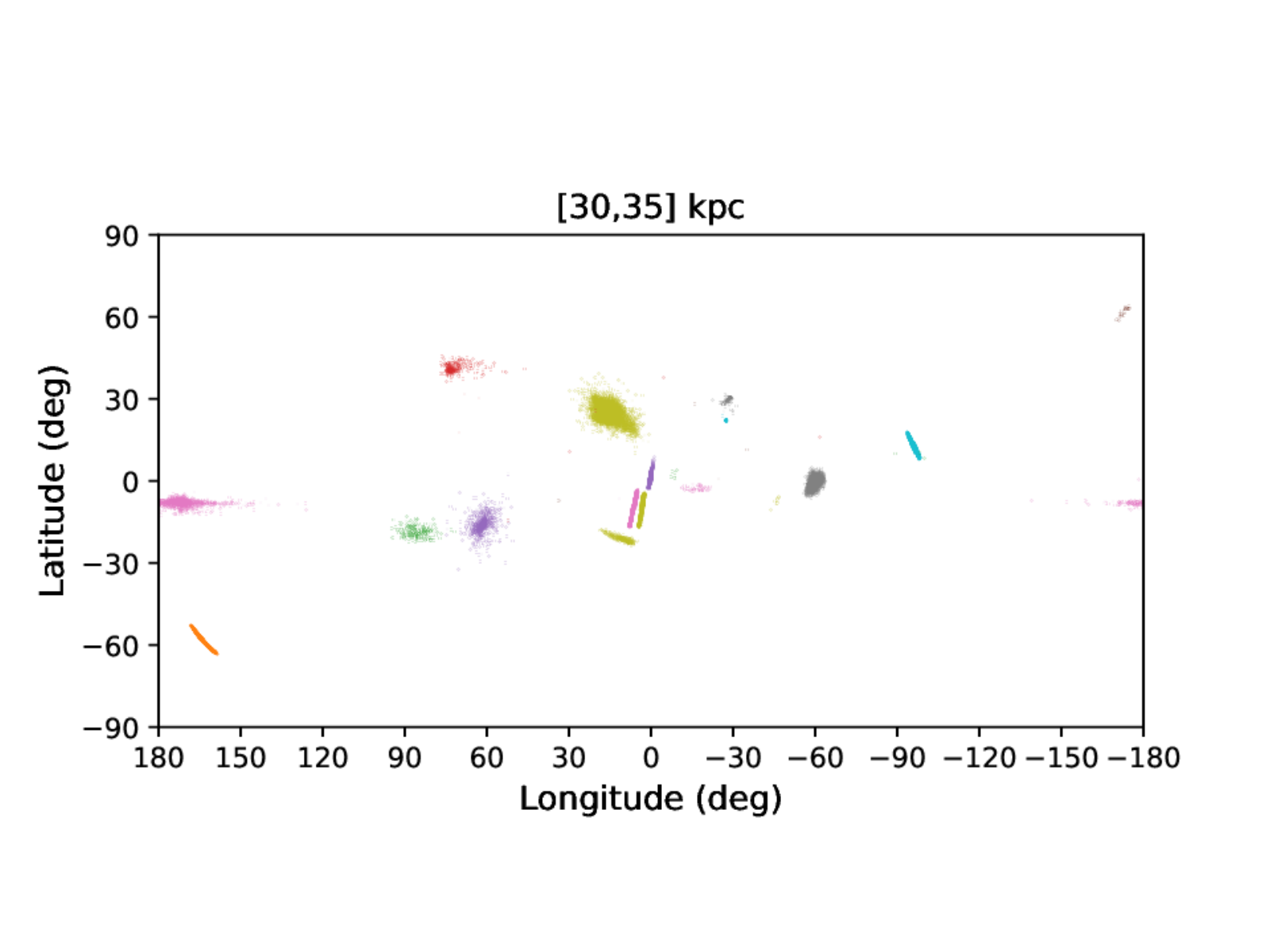}
\includegraphics[clip=true, trim = 0mm 15mm 0mm 20mm, width=0.9\columnwidth]{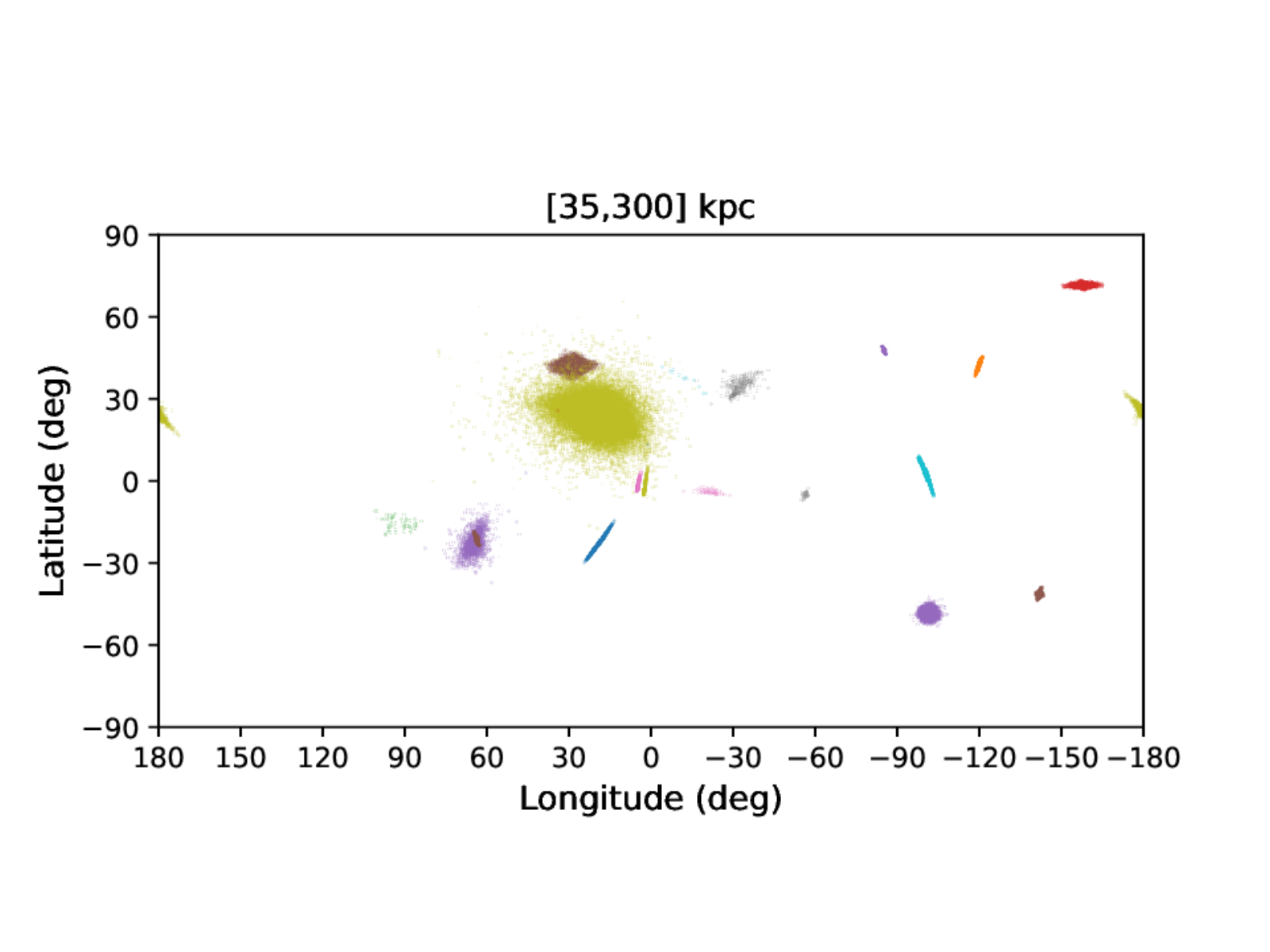}

\includegraphics[clip=true, trim = 0mm 15mm 0mm 20mm, width=0.9\columnwidth]{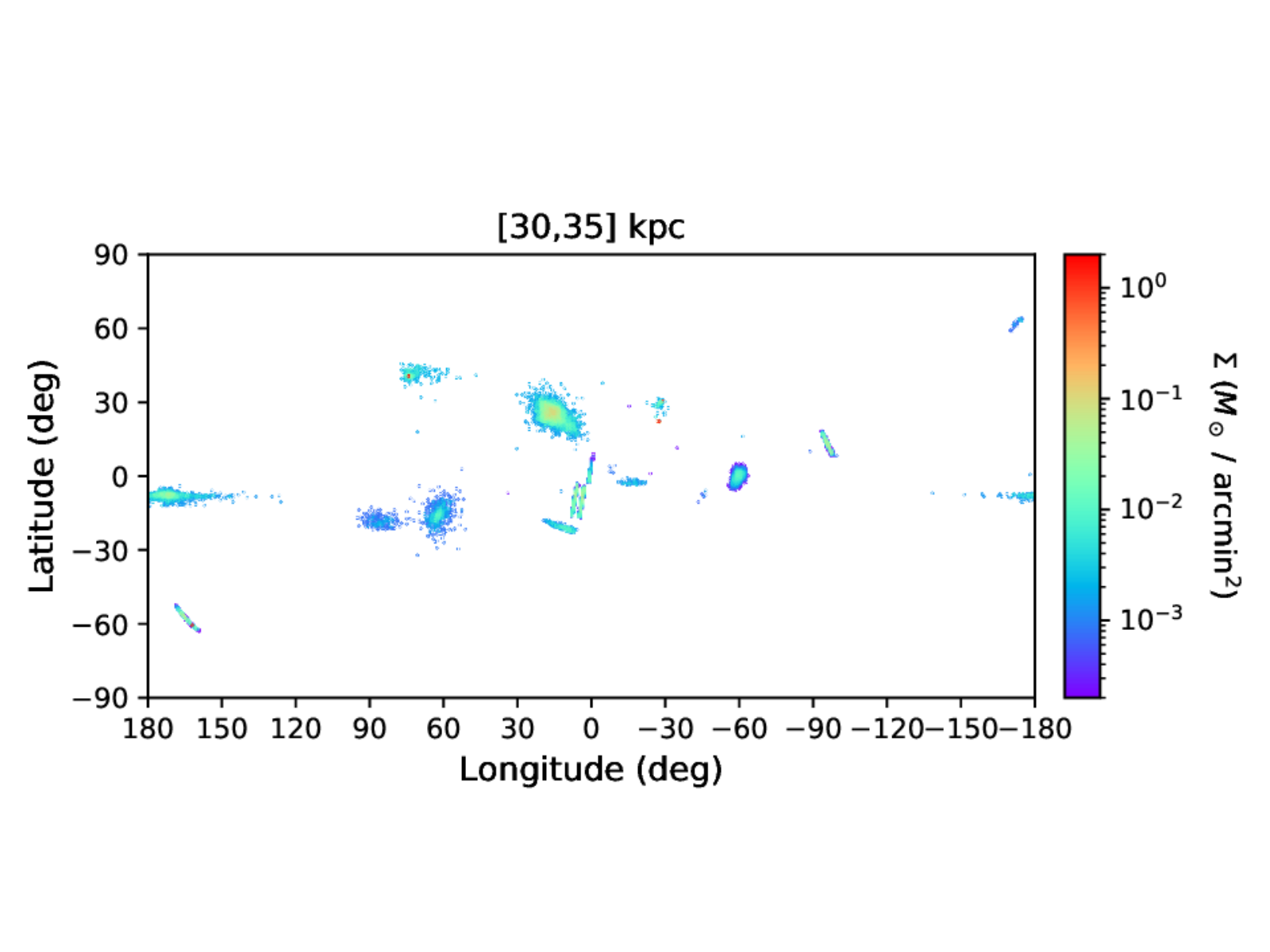}
\includegraphics[clip=true, trim = 0mm 15mm 0mm 20mm, width=0.9\columnwidth]{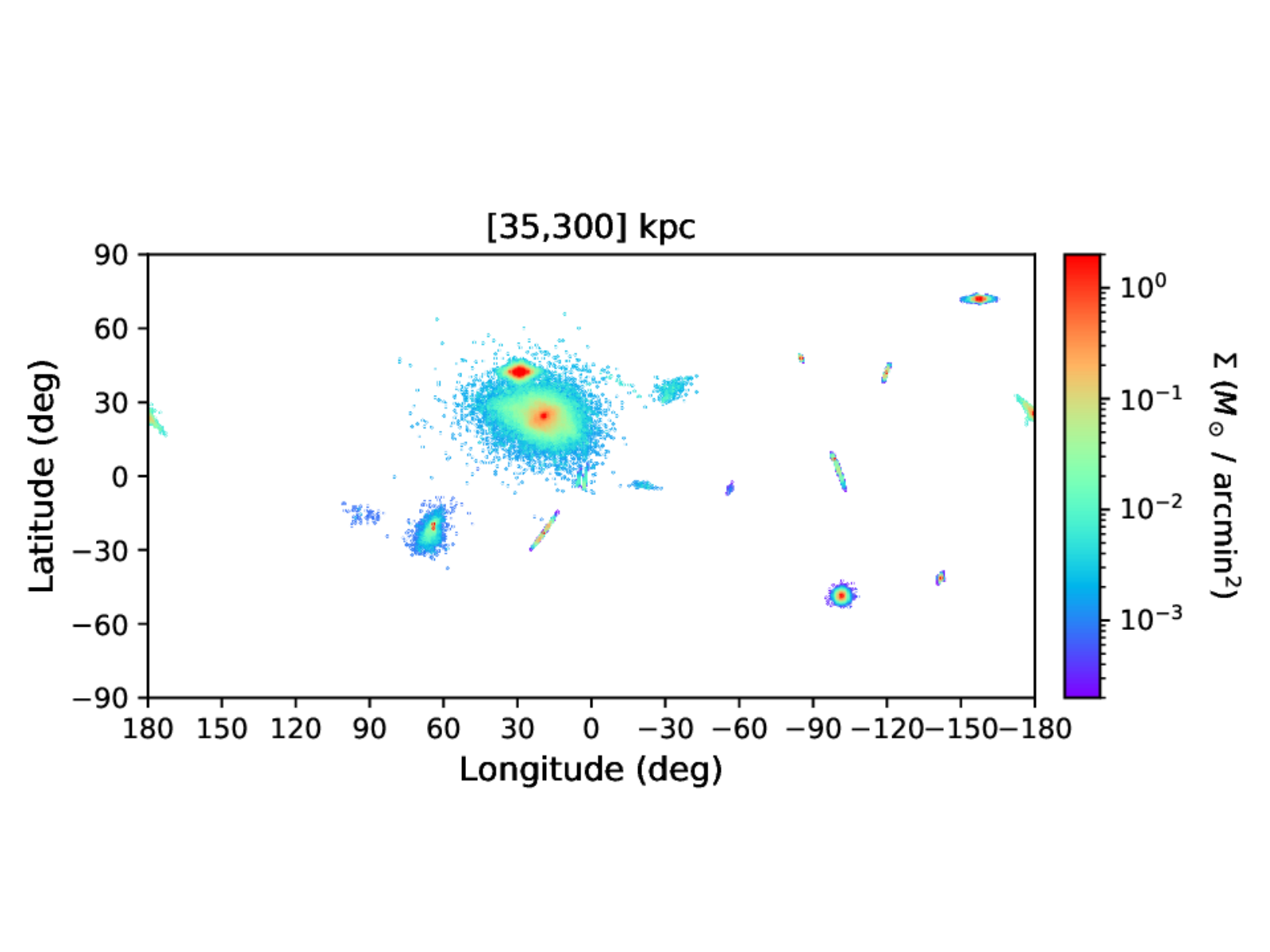}

\includegraphics[clip=true, trim = 0mm 20mm 0mm 20mm, width=0.9\columnwidth]{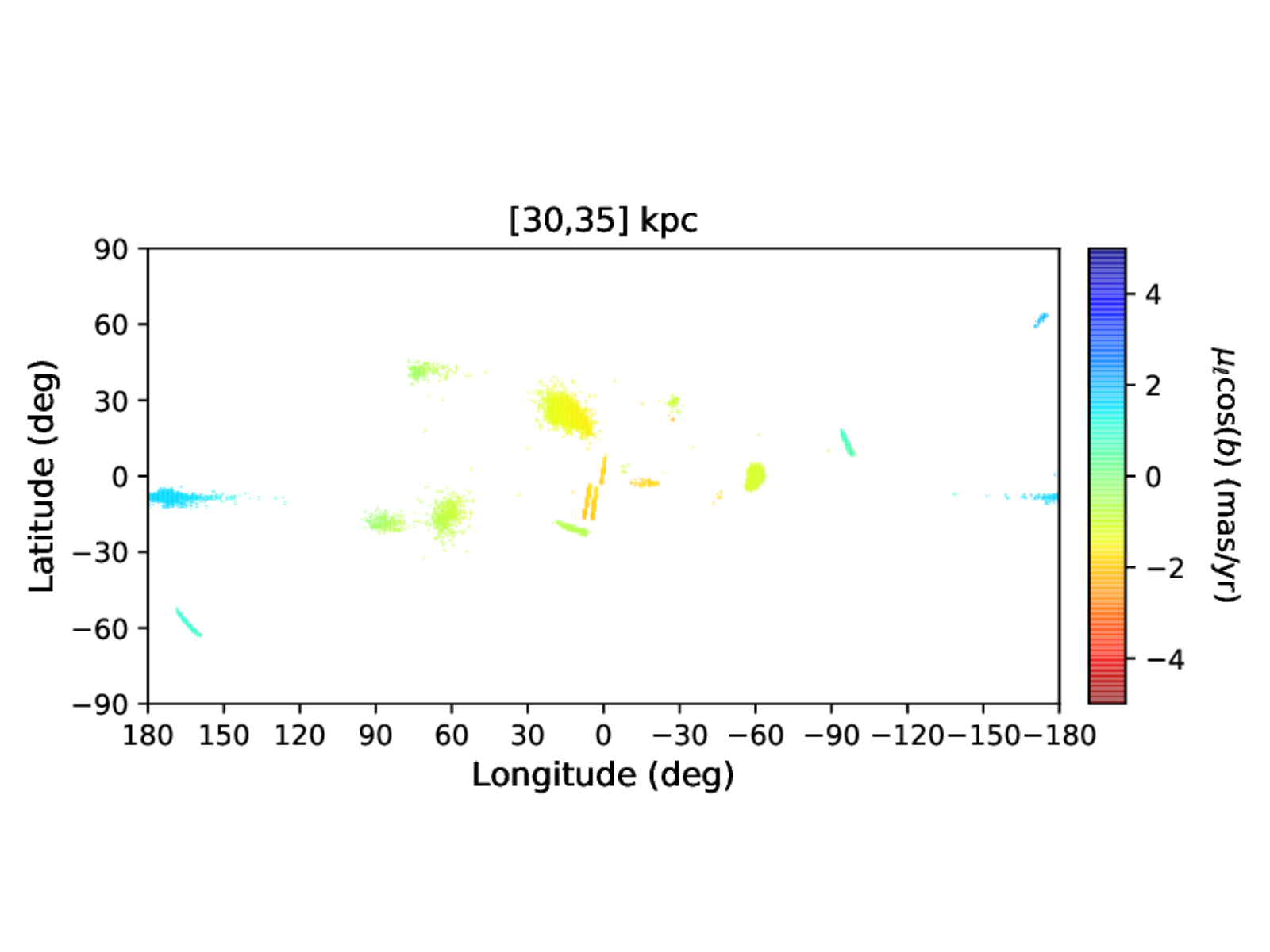}
\includegraphics[clip=true, trim = 0mm 20mm 0mm 20mm, width=0.9\columnwidth]{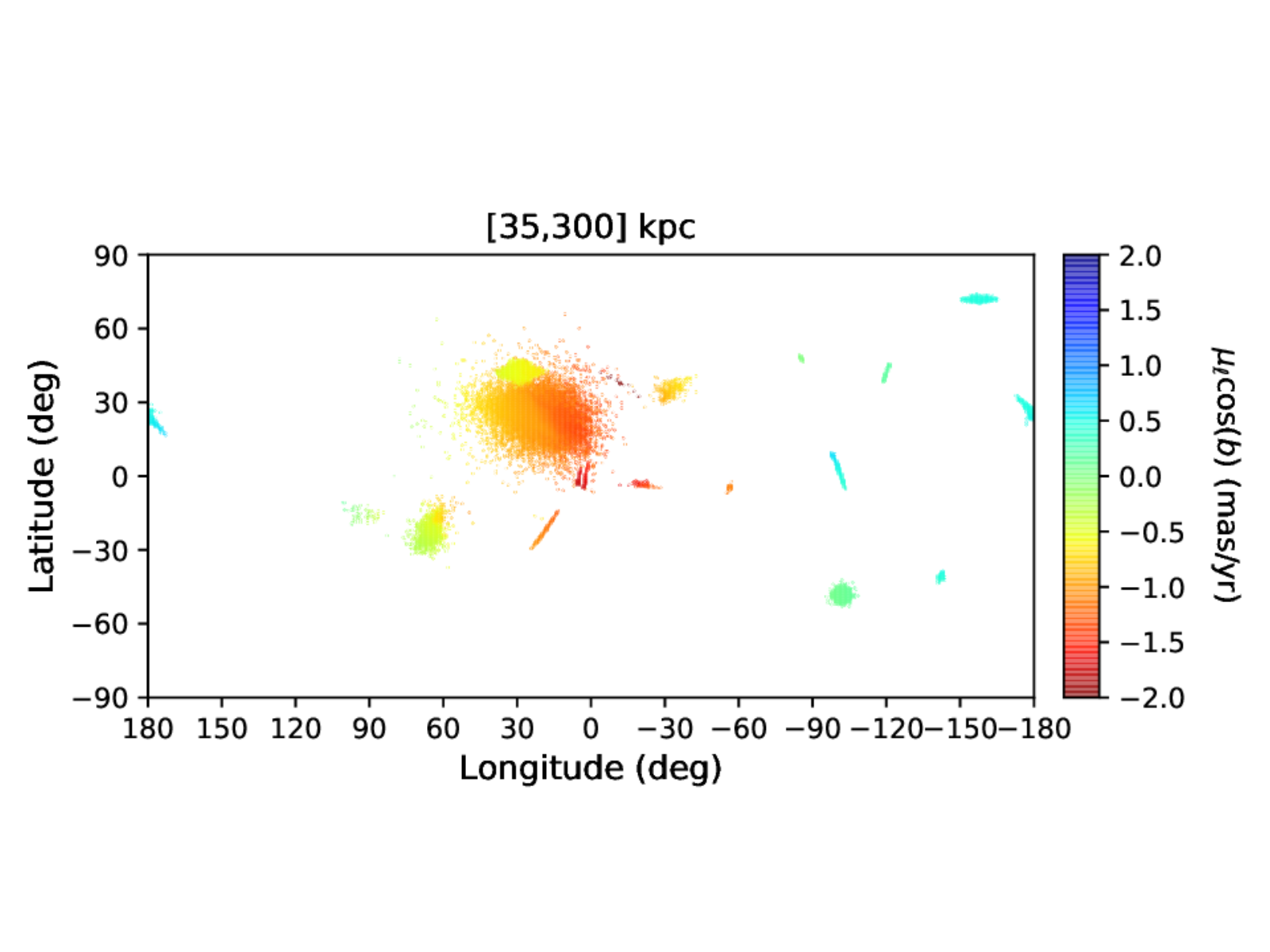}

\includegraphics[clip=true, trim = 0mm 20mm 0mm 20mm, width=0.9\columnwidth]{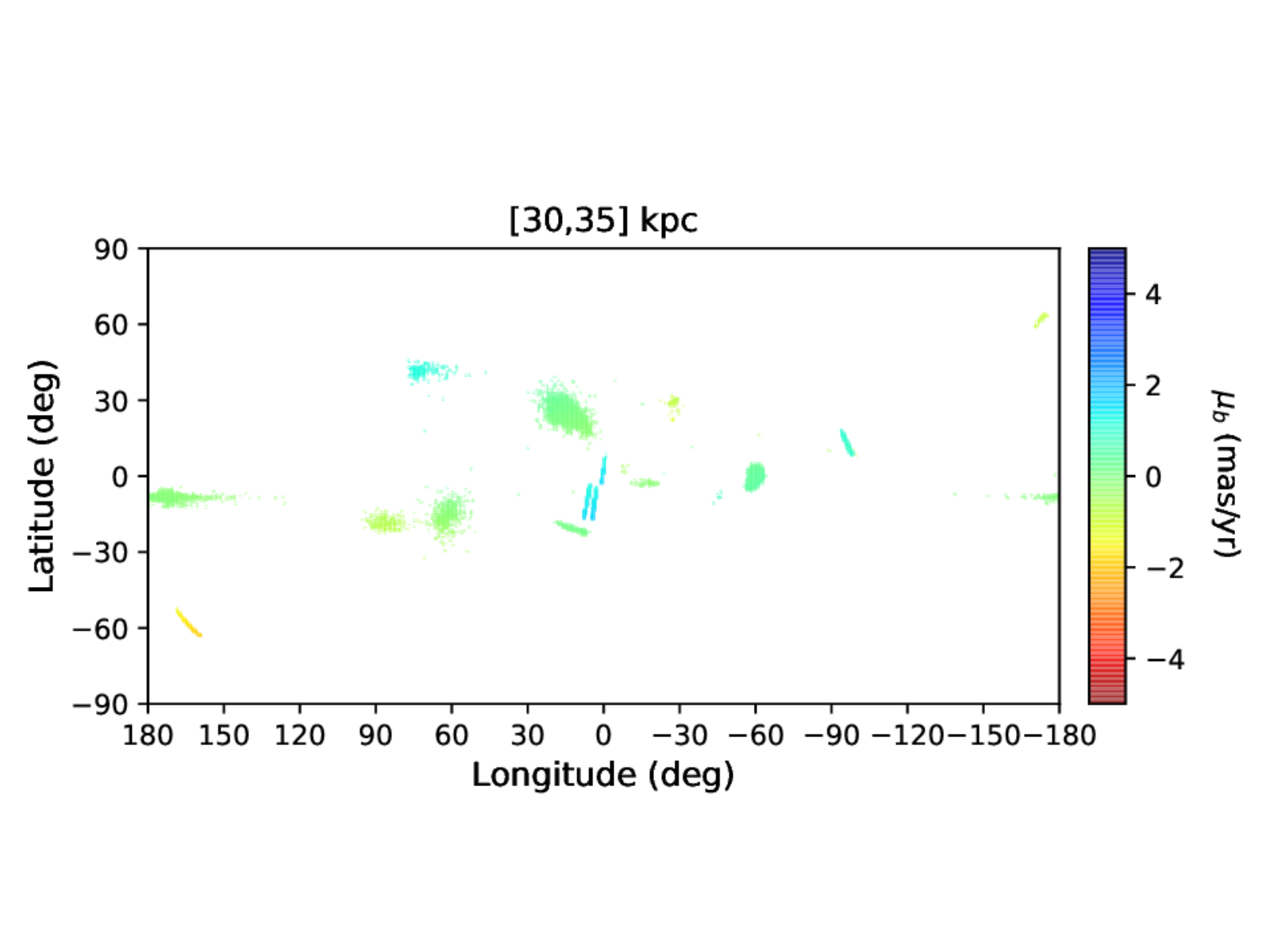}
\includegraphics[clip=true, trim = 0mm 20mm 0mm 20mm, width=0.9\columnwidth]{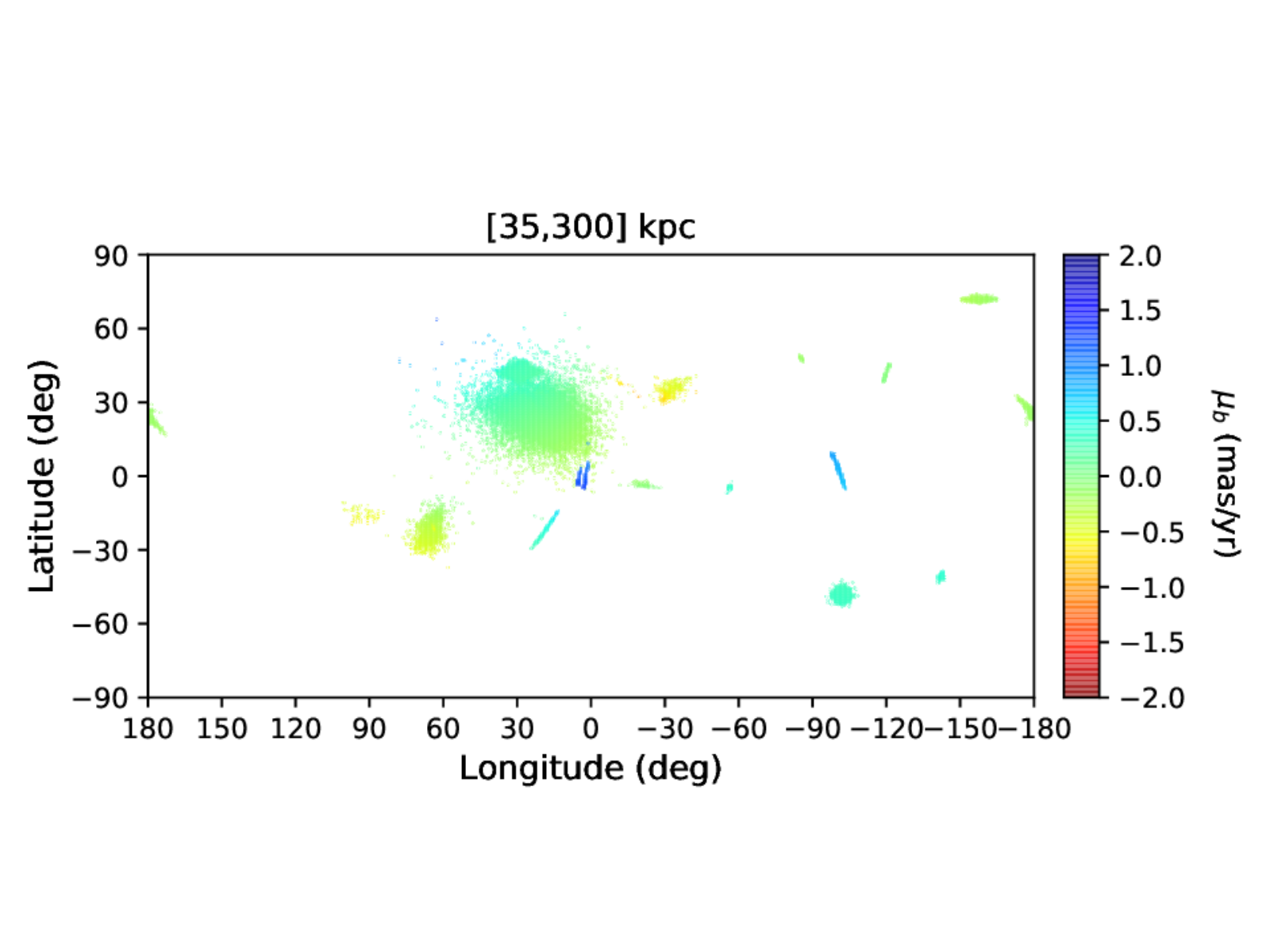}

\includegraphics[clip=true, trim = 0mm 20mm 0mm 20mm, width=0.9\columnwidth]{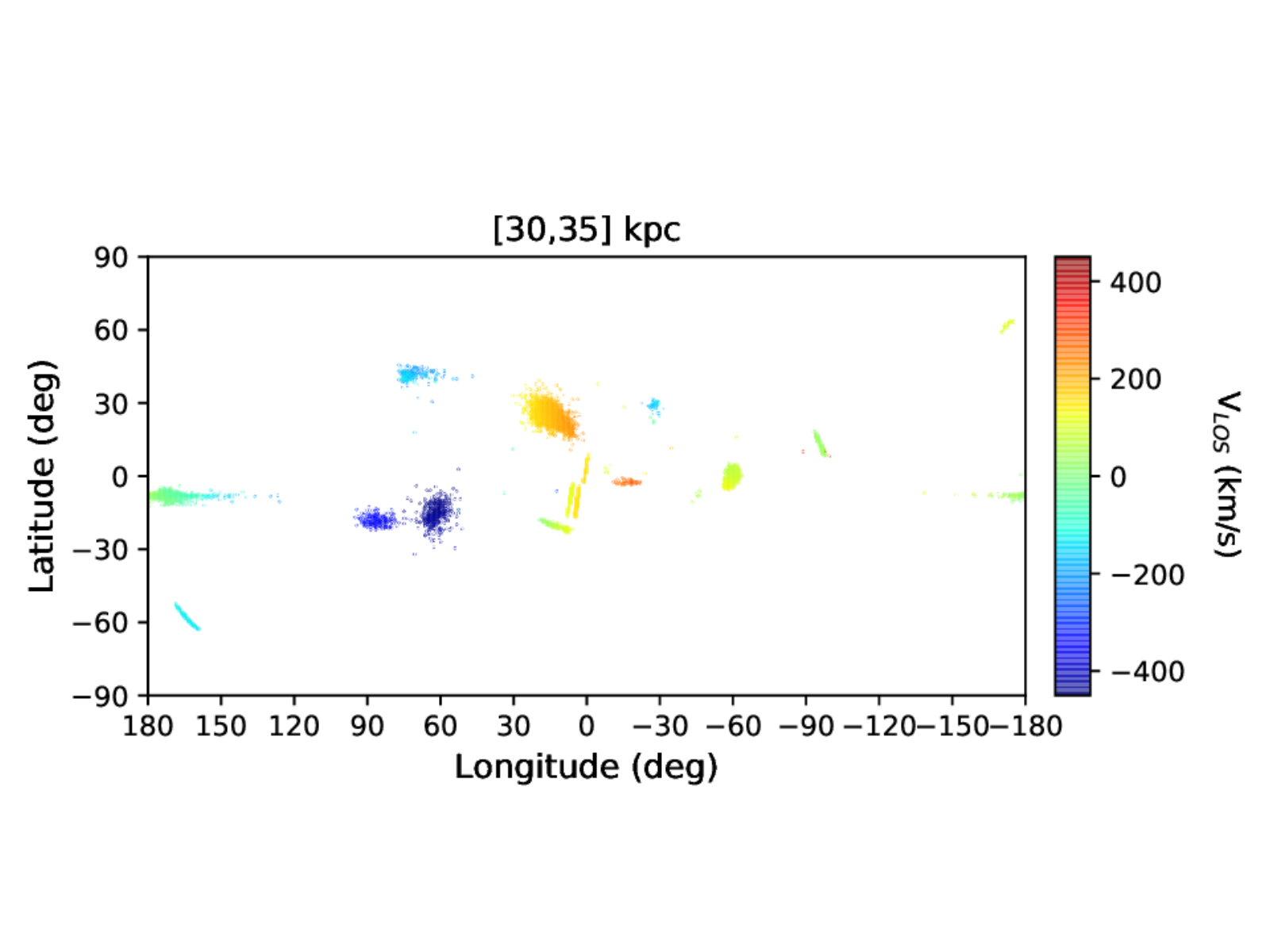}
\includegraphics[clip=true, trim = 0mm 20mm 0mm 20mm, width=0.9\columnwidth]{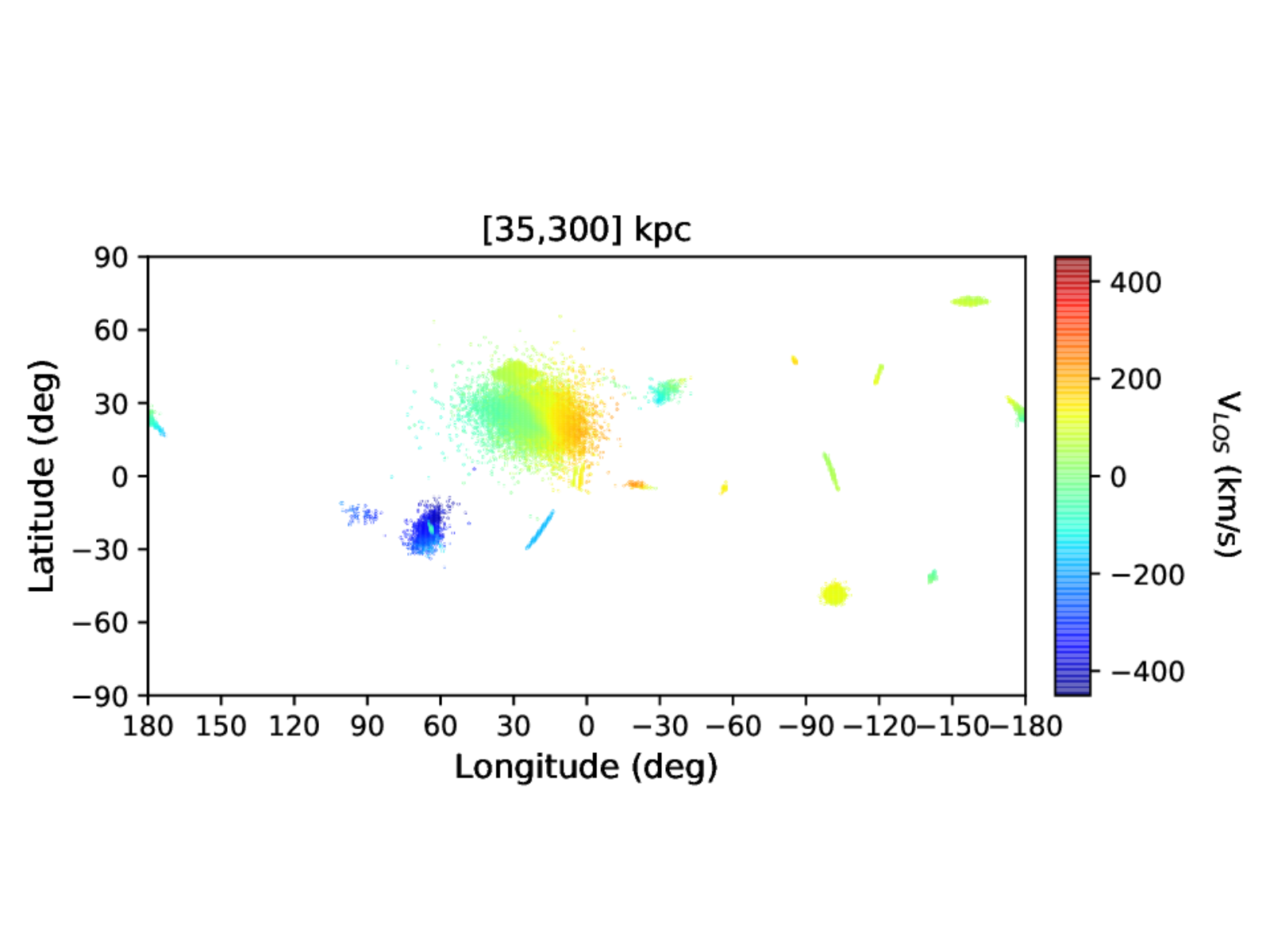}
\end{center}
\caption{As for  Fig.~\ref{D0-10}, but for two different distance bins: [30, 35]~kpc from the Sun (\emph{left column}) and  [35, 300]~kpc from the Sun (\emph{right column}). This figure shows streams, and their corresponding properties, as found for model PII only. In all panels, only the reference simulations are shown, for better clarity.}\label{D30-300}
\end{figure*}

The analysis presented in this Section, as well as the corresponding Figures~\ref{D0-10},~\ref{D10-20},~\ref{D20-30},~and~\ref{D30-300}, are restricted to the PII model.

In this section we analyze structures located at different distances from the Sun. We identify structures as main or secondary on the basis of the fraction of stars they represent. If, for example, a cluster contributes more than 10\% of its stripped stars in a given distance range, we consider the associated structure to be significant and we define it as a main structure in this distance range. If, on the other hand, the fraction of stripped stars from a cluster is between 1\% and 10\%, we define the structure associated as secondary. Structures which constitute less than 1\% of the cluster mass are considered insignificant in that range. Main and secondary structures for each distance bin are reported in Table~\ref{summarySTREAMS}, where they are named by their progenitor cluster. Below we describe the structures encountered at different distances from the Sun, from the closest to the most distant. For the following discussions,  we refer to Figs.~\ref{D0-10} to \ref{D30-300}, where we report the different streams in a given distance range (top row in each figure), the corresponding mass density generated by all the structures found in that bin (second row), the longitudinal and latitudinal proper motions map (third and fourth rows), and the line-of-sight velocities (bottom row).

\paragraph{The [0--2]~kpc distance range: }
Among the 159 simulated clusters, only  NGC~6121 is currently at a distance of less than 2~kpc from the Sun. In addition to stars stripped from this cluster within this distance-to-the-Sun range, we have identified very diffuse and low density extra-tidal stars associated to five other clusters, which are currently several kpc away from the Sun, as reported in Table~\ref{summarySTREAMS}. Of these, only UKS~1 and NGC~6121 have a significant fraction of their stripped mass in this distance range. All the others contribute with only a few percent. None of these clusters seem to show well defined, stream-like features in the ($\ell,b$) plane. This extra-tidal material appears indeed quite uniformly redistributed, in a latitude range mostly inside $-30^\circ$ to $30^\circ$, and mostly at negative longitudes. Because in this interval range no remarkable extra-tidal structure is found on the sky, we have decided to plot this distance bin together with the [2--5]~kpc bin (see Fig.~\ref{D0-10}, left column).

\paragraph{The [2--5]~kpc distance range: }
As the distance from the Sun increases, many more tidal structures are intercepted. Eleven globular clusters are found at a distance between 2 and 5~kpc from the Sun; and together with structures emanating from these clusters we also find extra-tidal material associated to other 48 clusters, 19 of which are significant, as listed in Table~\ref{summarySTREAMS}. Some extra-tidal structures are clearly identifiable, even when over-plotted with all the others: it is the case, for example, of the tidal material associated to the E~3 cluster, which appears as an extended thick stream, as shown of Fig.~\ref{D0-10} at approximately $-100^{\circ}$ longitude; and of the complex tidal structure associated to BH~140, which appears like a ribbon in the sky, with a bifurcation at negative longitudes whose edges extend to about $-30^\circ$ and $30^\circ$ latitude. In addition to these, there are a series of circular halos concentric about the Galactic center, with abrupt drop-offs in density tracing the furthest extent of diffuse debris emanating from a variety of clusters. Overall, few streams are immediately recognizable in this range, though plenty of debris is present. This is expected given that this range samples a bite of the Sun-side of the Galactic disk, and that this range has a relatively small volume. In this, as in the following distance ranges, the $v_{\ell os}$ maps show that the tidal material at negative Galactic longitudes has, on average, positive $v_{\ell os}$ while material at positive Galactic longitudes has, on average, negative  $v_{\ell os}$, which is due to the solar reflex velocity. This is the same trend observed for the whole set of radial (i.e. line-of-sight) velocities in Gaia DR3---for instance see the bottom panel of Fig.~5 from \citet{katz22}, although less extreme velocities are reported in their plot since their data is dominated by disk stars whereas our maps have a high proportional contribution of halo stars (additionally, they use median values in their bins while we use an average). In order to quantify the net rotation of the system of streams, we calculate the mean angular momentum about the Galactic pole as $\braket{L_z} =\sum_i L_{z,i} m_{p,i} / \sum_i m_{p,i} $, where $m_{p,i}$ is the mass of each star particle indexed by $i$ as discussed in footnote~\ref{footnote:mass} and $L_{z,i}$ is the corresponding particle's angular momentum. The mean angular momentum is found to be $\braket{L_z}=-300$~kpc~km~s$^{-1}$, which shows a slight co-rotation of the system of streams with the disk though there is much dispersion about this value as shown in bottom right panel of Fig.~\ref{orbparam} in the Appendix.

\paragraph{The [5--10]~kpc distance range: }
The [5--10]~kpc distance range, which includes the Galactic center, contains much more material. A total of 79 clusters are found in this range, together with tidal structures associated to 131 different progenitors redistributed among main and secondary structures. Some tiny streams are visible in the density maps as well as in proper motions and line-of-sight velocity spaces (see Fig.~\ref{D0-10}, right column): the trailing portion of the tail of the globular cluster Pal~1, at $(\ell, b)\sim(140^\circ, 25^\circ)$; the most extreme portion of the trailing tail of NGC~3201 at $(\ell, b)\sim(150^\circ, -37^\circ)$; the waterfall-like shape of NGC~288, particularly evident at $b \lesssim -60^{\circ}$; the thin inverted U-shape of NGC~4590 at positive latitudes spanning a large longitude extent from $\ell \simeq -60^\circ$ to $100^\circ$; the portion of the E~3 tails the closest to the cluster at $(\ell, b)\sim(-75^{\circ},-15^{\circ})$, which continues from the more easily recognizable portion in the [0--5]~kpc bin.

\paragraph{The [10--15]~kpc distance range: }
In the [10--15]~kpc range, we find tidal structures associated to 134 progenitors, as listed in Table~\ref{summarySTREAMS} and reported in Fig.~\ref{D10-20} (left column), 27 of which are related to globular clusters that are also found in this distance bin. We note that, at these distances from the Sun, the distribution of tidal features in directions towards the Galactic center, from $-30^{\circ}$ to $30^{\circ}$ longitude, appears less fuzzy than the one characterizing the [0--5] and [5--10]~kpc distance bins. Tidal features here are beyond the Galactic center, and are mostly associated to disk or halo clusters. 

Among the thinnest structures, we find the stream associated to Pal~1, at $(\ell, b)\sim(120^\circ, 15^\circ)$ which is also visible in the distance bin [5--10]~kpc (see previous discussion), but which is even more elongated here. NGC~6101 shows the nearest portion of its long thin diagonal tidal tail that spans negative longitudes and ranges from $-15^{\circ}$ to $45^\circ$ latitude. Additionally this stream is also unique against its counter parts in proper motion space. NGC~5053's nearest portion appears as a vertical tidal tail at $-80^{\circ}$ longitude. Similarly, NGC~5466 is shown vertically at $25^{\circ}$ longitude.

Among the thickest structures, we can recognize general diffuse and bow-tie like shapes. There are also spoke-like structures departing radially from the Galactic center. For instance, the extra-tidal material associated to: NGC~7078 at $(\ell, b)\sim(60^\circ,-28^{\circ})$; NGC~7089, nearly parallel to the previous structure, but at lower latitudes at $(\ell, b)\sim(50^\circ,-40^{\circ})$.

\paragraph{The [15--20]~kpc distance range: }
At larger distances ($[15--20]$~kpc range, see Fig.~\ref{D10-20}), some of the most striking features are associated to the clusters NGC~5024 and NGC~5053, whose long thin tails essentially overlap in this distance, with the latter covering the former, and appearing at high latitudes spanning a longitudinal range from $\ell \sim -90^{\circ}$ to $\ell \sim 45^{\circ}$; again, the long thin stream of NGC~5466 appears in this range (as will be the case for the next) and is at high latitudes at roughly $85^{\circ}$ and positive longitudes; the thicker extended structure of NGC~4147 whose diffuseness emanates from about $(\ell, b)\sim(-100^{\circ},80^{\circ})$.

In this distance bin, we find long tidal tails emanating from globular clusters associated to the Sagittarius dwarf galaxy, which are particularly visible at negative longitudes: a long thin stream is associated to Pal~12 at positive longitudes and latitudes $b \le-15^{\circ}$, as well as two overlapping structures at $0^\circ \lesssim \ell \lesssim 30^\circ$ longitude, i.e. Ter~7 (and Ter~8 in the next distance bin). A word of caution is needed here: the mass loss from these clusters may be incorrect, since we do not include the presence of the Sagittarius dwarf galaxy itself. The potential well associated to this latter could change the tidal effects experienced by clusters associated to Sagittarius, especially in the case of NGC~6715, which sits at the center of this dwarf galaxy. The inclusion of the Sagittarius dwarf will be the subject of future investigations. Overall, in this distance bins, we find 11 clusters  and 61 streams, all listed in Table~\ref{summarySTREAMS}.

\paragraph{The [20--25] and [25--30]~kpc distance ranges: }
In the following distance bins ([20--25]~kpc and [25--30], see Fig.~\ref{D20-30}), globular clusters and extra-tidal structures become less numerous, though some are still visible, for instance Pal~5 at $(\ell, b)\sim(0^{\circ},45^{\circ})$. In more detail, in the [20--25]~kpc bin we find tidal features associated to 37 different progenitors, 8 of which are associated to globular clusters whose current positions are in the same distance bin; in the [25--30]~kpc bin 7 clusters are found, together with tidal features associated to 30 other progenitor clusters which do not lie in this same distance range. In both bins, the streams emanating from globular clusters associated to the Sagittarius dwarf galaxy are still visible, as well as the most extreme portion of the tail associated to NGC~5466. 

\paragraph{The [30--35] and [35--300]~kpc distance ranges: }
Finally, in the last distance bins (see Fig.~\ref{D30-300}), thin streams become rare. Some small streams are visible: Pyxis at $(\ell, b)\sim(-100^{\circ},0^{\circ})$; NGC~2419 at $(\ell, b)\sim(-180^{\circ},30^{\circ})$; Pal~4 at $(\ell, b)\sim(-160^{\circ},75^{\circ})$; Pal~3 at $(\ell, b)\sim(-120^{\circ},45^{\circ})$. Many more have a diffuse and halo-like structure. For instance, the blob associated to Pal~15, centered at $(\ell, b)\sim (15^\circ, 20^\circ$); AM~1 at $(\ell, b)\sim(-100^{\circ},-55^{\circ})$; Eridanus at $(\ell, b)\sim(-140^{\circ},-45^{\circ})$; Pal~14 at $(\ell, b)\sim(30^{\circ},45^{\circ})$; Laevens~3 at $(\ell, b)\sim(65^\circ,-20^\circ)$, which is completely enveloped by NGC~7006. In total, in these two distance bins we find 4 and 12 clusters, and their associated streams, together with extra-tidal material associated to 14 and 19 progenitors total, respectively.

\begin{table*}
\tiny
\centering                                      
\caption{List of tidal structures found in different intervals of distance to the Sun. The tidal structures are named by their progenitor clusters. If the parent cluster is also in the distance range under consideration, the name of the tidal structure is shown in bold. Second column reports the main structures found in a given distance bin. The third column list secondary structures. The numbers in parenthesis in the second column  and third column (numbers with normal font) correspond to the total number of main and secondary structures found in a given distance range. The number of clusters in each distance bin is also reported in the second column (bold numbers in parenthesis). }\label{summarySTREAMS}
\begin{tabularx}{\textwidth}{l X X }          
\hline
Distance (kpc) &   Main tidal structures & Secondary tidal structures\\ 
\hline   \\
\tiny
\vspace{0.1cm}
[0-2] & (\textbf{1},2) \textbf{NGC6121}, UKS1 & (4) BH140, Djor1, NGC6333, NGC6356\\ 
\vspace{0.1cm}

[2-5] & (\textbf{11},30) \textbf{NGC6397}, \textbf{NGC6544}, \textbf{NGC3201}, \textbf{BH140}, \textbf{NGC104}, \textbf{NGC6838}, \textbf{NGC6366}, \textbf{NGC6752}, \textbf{IC1276}, \textbf{NGC6656}, \textbf{2MASS-GC01}, NGC6284, NGC6356, NGC6287, VVV-CL001, NGC6254, NGC5927, E3, NGC6121, VVV-CL001, Djor1, UKS1, Ter10, 2MASS-GC02, Pal10, NGC5139, NGC6333, NGC6441, NGC6541, NGC288 & (29) FSR1716, FSR1758, NGC1851, NGC1904, NGC2298, NGC2808, NGC362, NGC4372, NGC4833, NGC5897, NGC5986, NGC6205, NGC6218, NGC6235, NGC6273, NGC6316, NGC6352, NGC6388, NGC6496, NGC6681, NGC6749, NGC6760, NGC6809, NGC6864, NGC7078, Pal2, Pal8, Ter12, Ton2\\ 
\vspace{0.1cm}

[5-10] & (\textbf{79},124) \textbf{VVV-CL001}, \textbf{NGC7099}, \textbf{NGC6362}, \textbf{Ton2}, \textbf{Djor1}, \textbf{VVV-CL001}, \textbf{NGC6496}, \textbf{Djor2}, \textbf{NGC6535}, \textbf{NGC6528}, \textbf{NGC6539}, \textbf{NGC6540}, \textbf{NGC6553}, \textbf{2MASS-GC02}, \textbf{Ter12}, \textbf{BH261}, \textbf{Ter9}, \textbf{NGC6712}, \textbf{NGC6717}, \textbf{NGC6723}, \textbf{NGC6749}, \textbf{NGC6760}, \textbf{Pal10}, \textbf{HP1}, \textbf{Ter4}, \textbf{Ter2}, \textbf{Ter3}, \textbf{NGC2298}, \textbf{E3}, \textbf{NGC4372}, \textbf{NGC4833}, \textbf{NGC5904}, \textbf{NGC5927}, \textbf{FSR1716}, \textbf{Lynga7}, \textbf{NGC6144}, \textbf{NGC6171}, \textbf{NGC6352}, \textbf{ESO452-SC11}, \textbf{NGC6218}, \textbf{FSR1735}, \textbf{NGC6254}, \textbf{NGC6256}, \textbf{NGC6287}, \textbf{NGC6293}, \textbf{NGC6304}, \textbf{NGC6355}, \textbf{NGC6809}, \textbf{NGC6637}, \textbf{NGC6402}, \textbf{NGC6325}, \textbf{NGC6341}, \textbf{NGC6342}, \textbf{NGC6380}, \textbf{NGC6401}, \textbf{NGC6440}, \textbf{NGC6517}, \textbf{NGC6522}, \textbf{NGC6541}, \textbf{NGC6558}, \textbf{NGC6624}, \textbf{NGC6626}, \textbf{NGC6638}, \textbf{NGC6642}, \textbf{NGC6652}, \textbf{NGC6681}, \textbf{Pal6}, \textbf{Ter1}, \textbf{Ter5}, \textbf{Ter6}, \textbf{NGC6333}, \textbf{NGC288}, \textbf{NGC362}, \textbf{NGC6273}, \textbf{NGC6266}, \textbf{NGC6205}, \textbf{NGC5139}, \textbf{Liller1}, \textbf{NGC5946}, NGC5897, NGC1904, NGC6752, NGC6656, NGC6121, NGC1851, NGC6864, NGC7078, NGC7089, NGC6316, NGC5272, NGC6779, NGC2808, NGC4590, Rup106, NGC104, NGC4147, NGC3201, Pal11, Pal1, NGC1261, BH140, NGC6235, Ter10, NGC6569, UKS1, NGC6453, NGC6139, NGC6426, NGC6397, NGC6093, NGC6388, FSR1758, IC1276, NGC6838, NGC6366, NGC5986, NGC6441, NGC6356, NGC6584, NGC5286, NGC6544, NGC6284, Pal8, NGC6981 & (7) 2MASS-GC01, IC1257, NGC5634, NGC5694, NGC6229, NGC7006, Pal2\\ 
\vspace{0.1cm}

[10-15] & (\textbf{27},115) \textbf{NGC6453}, \textbf{NGC5272}, \textbf{NGC6584}, \textbf{Pal8}, \textbf{NGC6316}, \textbf{NGC5897}, \textbf{NGC6284}, \textbf{NGC6139}, \textbf{NGC6093}, \textbf{NGC5986}, \textbf{NGC5286}, \textbf{NGC6235}, \textbf{NGC2808}, \textbf{NGC1904}, \textbf{NGC1851}, \textbf{NGC6101}, \textbf{NGC6779}, \textbf{NGC6388}, \textbf{Pal11}, \textbf{Pal1}, \textbf{Ter10}, \textbf{NGC4590}, \textbf{NGC7089}, \textbf{NGC6569}, \textbf{NGC7078}, \textbf{FSR1758}, \textbf{NGC6441}, NGC6304, NGC6254, FSR1735, NGC6426, NGC6397, NGC6362, Ton2, NGC6256, NGC6366, NGC6287, NGC6352, NGC6355, NGC6356, NGC6218, NGC6293, VVV-CL001, NGC6171, NGC5024, NGC1261, NGC2298, E3, NGC3201, NGC4147, NGC4372, Rup106, BH140, NGC4833, NGC5053, Ter3, NGC5466, NGC5634, IC4499, NGC5904, NGC5927, FSR1716, UKS1, NGC6121, NGC6144, Lynga7, NGC6553, VVV-CL001, NGC6496, NGC5946, NGC6205, NGC6266, NGC6273, NGC6333, NGC6341, NGC6342, NGC6401, NGC6402, NGC6517, NGC6541, NGC6544, NGC6558, NGC6626, NGC6652, NGC6656, NGC6681, NGC6864, Pal6, Ter1, Ter5, NGC5139, NGC362, NGC104, Ter12, Djor2, NGC6535, NGC6528, NGC6539, NGC6540, 2MASS-GC01, Ter9, 2MASS-GC02, IC1276, BH261, NGC7099, NGC6712, NGC6723, NGC6749, NGC6752, NGC6760, NGC6809, NGC6838, NGC6934, NGC6981, NGC288 & (18) Djor1, ESO280-SC06, ESO452-SC11, HP1, IC1257, NGC5694, NGC5824, NGC6229, NGC6325, NGC6380, NGC6638, NGC6642, NGC6717, NGC7006, NGC7492, Pal10, Pal2, Ter6\\ 
\vspace{0.1cm}

[15-20] & (\textbf{11},46) \textbf{NGC6356}, \textbf{NGC5466}, \textbf{NGC1261}, \textbf{UKS1}, \textbf{NGC4147}, \textbf{IC4499}, \textbf{NGC5024}, \textbf{NGC5053}, \textbf{Pal12}, \textbf{NGC6981}, \textbf{NGC6934}, NGC6101, NGC5904, Pal5, NGC5824, NGC5634, NGC7089, FSR1758, NGC4833, BH140, NGC4590, Rup106, NGC3201, Pal2, NGC5272, Djor1, NGC6426, NGC7078, NGC6864, NGC6715, NGC6656, NGC6341, NGC6333, NGC5286, NGC362, NGC2808, NGC1851, NGC104, NGC7492, Pal10, Ter7, NGC6779, NGC6584, IC1276, ESO280-SC06, NGC288 & (15) 2MASS-GC02, IC1257, NGC1904, NGC2298, NGC4372, NGC5139, NGC5694, NGC6121, NGC6205, NGC6229, NGC6838, NGC7006, NGC7099, Pal13, Ter8\\ 
\vspace{0.1cm}

[20-25] & (\textbf{8},29) \textbf{Pal5}, \textbf{NGC7492}, \textbf{Pal13}, \textbf{Rup106}, \textbf{NGC6864}, \textbf{NGC6426}, \textbf{ESO280-SC06}, \textbf{Ter7}, NGC5466, IC4499, NGC5634, NGC7089, NGC5824, NGC5024, NGC4590, NGC4147, NGC3201, NGC5272, IC1257, NGC5904, NGC6101, NGC6584, Arp2, Ter8, NGC6934, NGC6981, Pal12, NGC6229, Pal2 & (9) Djor1, FSR1758, NGC1261, NGC1851, NGC1904, NGC2298, NGC2808, NGC5694, NGC7006\\ 
\vspace{0.1cm}

[25-30] & (\textbf{7},20) \textbf{NGC6715}, \textbf{Pal2}, \textbf{AM4}, \textbf{NGC5634}, \textbf{Ter8}, \textbf{Arp2}, \textbf{IC1257}, NGC5824, Rup106, NGC5694, IC4499, NGC6101, NGC5904, NGC6229, Ter7, NGC6934, NGC6981, Pal13, NGC7492, Whiting1 & (10) NGC1851, NGC1904, NGC3201, NGC4147, NGC4590, NGC5466, NGC7006, NGC7089, Pal15, Pyxis\\ 
\vspace{0.1cm}

[30-35] & (\textbf{4},11) \textbf{NGC6229}, \textbf{NGC5824}, \textbf{NGC5694}, \textbf{Whiting1}, NGC7006, Ter8, Arp2, Ter7, Rup106, Pyxis, Pal2 & (3) NGC6101, NGC6934, Pal15\\ 
\vspace{0.1cm}

[35-300] & (\textbf{12},15) \textbf{Laevens3}, \textbf{NGC7006}, \textbf{SagittariusII}, \textbf{Pal15}, \textbf{Pal14}, \textbf{Crater}, \textbf{Pal4}, \textbf{Pal3}, \textbf{Pyxis}, \textbf{NGC2419}, \textbf{Eridanus}, \textbf{AM1}, NGC6715, NGC5824, NGC5694 & (4) Arp2, NGC6934, Pal2, Ter8
\end{tabularx}
\normalsize
\end{table*}

\subsection{Disk, inner and outer halo clusters: a variety of morphologies and shapes for extra-tidal structures}\label{sec:morphologies}

The analysis presented in the previous section allows us to appreciate the variety of morphologies found for extra tidal structures, from padlocks to ``Easter eggs'', disks, ribbons, and canonical streams. Moreover, some structures are limited in latitude and longitude, while some others fill nearly the entire sky. 

To more easily capture the similarity and differences in the morphology of the extra-tidal features surrounding Galactic globular clusters, we can group the latter on the basis of their orbital parameters\footnote{We caution the reader that the classification of disk, inner and outer clusters made in this Section is based on the orbital parameters of the clusters, as found when their orbits are integrated in model PII. This classification may slightly change if model PI or model PII-0.3-SLOW were adopted.}(see Appendix~\ref{class} for more details), as follows:

\begin{enumerate}
\item \emph{disk clusters}: a cluster is classified as a disk cluster if  arctan($z_{max}/R_{max}$) $\le 10^\circ$, where $z_{max}$ and $R_{max}$ are, respectively, the maximum height above or below the Galactic plane reached by its orbit in the last 5~Gyr, and its maximum in-plane distance from the Galactic center.
\item \emph{inner clusters}: all clusters with $r_{max} \le R_{\odot}$ which are not classified as disk clusters enter this group. Contrary to $R_{max}$, which is the maximum in-plane distance that a cluster reaches from the Galactic center, $r_{max}$ is the maximum 3D distance, that is $r_{max} = \textrm{max}(\sqrt{R^2+z^2})$ with the maximum calculated over the whole cluster orbit. 
\item \emph{outer clusters}: all clusters with $r_{max} > R_{\odot}$ which are not classified as disk clusters enter this group. 
\end{enumerate}
By using the orbital radius of the Sun as the criterion for inner and outer clusters, debris from outer clusters can span the whole sky while inner clusters must be restricted in longitude and latitude. With these definitions, 21 clusters are disk clusters, 71 are inner clusters, and 67 are outer clusters (see  Table~\ref{classification} in Appendix~\ref{class}). We emphasize that this classification does not aim to suggest any specific origin for these systems \citep[for example whether they are in-situ or accreted, see][]{massari19}, but it is uniquely based on their current orbital characteristics, and helps capturing some of the properties in the extension (projected in to the sky) and shape of their extra-tidal material, as we discuss in the following.

\subsubsection{Extra-tidal features originating from disk clusters: ribbons in the Galactic plane}

Disk clusters are defined on the basis of the flatness of their orbits (i.e. on their $z_{max}/R_{max}$ ratio). As a result, they typically are restricted to low latitudes, though the exact distribution depends on the relationship of their orbit to the solar radius. To specify, clusters whose $R_{max}$ are interior to the Solar radius generate tidal debris in a limited range in longitude and latitude. For instance, the material associated to clusters as Ter~1, Ter~5, Ter~6, and Ter~9 has a disk-like shape and is completely confined to $|\ell| <30^{\circ}$ and $|b|<10^{\circ}$. If $R_{max}$ is greater than the solar radius, material can cover the full longitude space, and most of the material will still appear at low latitudes. This is the case, for example, of BH~140, whose escaped stars diffusely occupy all longitudes and most of them are found at $|b| \le 30^\circ$, Pal~2 and Pal~10  whose extra-tidal stars have a very limited latitudinal extension and appear as ribbons in the sky. In the following, we discuss some of the structures associated to NGC~6121, Pal~2, and Pal~10. We refer the reader to Appendix~\ref{allstreams} for the tidal structures generated by the whole set of disk clusters.  

\begin{figure}
\begin{center}
\includegraphics[clip=true, trim = 0mm 2mm 0mm 0mm, width=0.9\columnwidth]{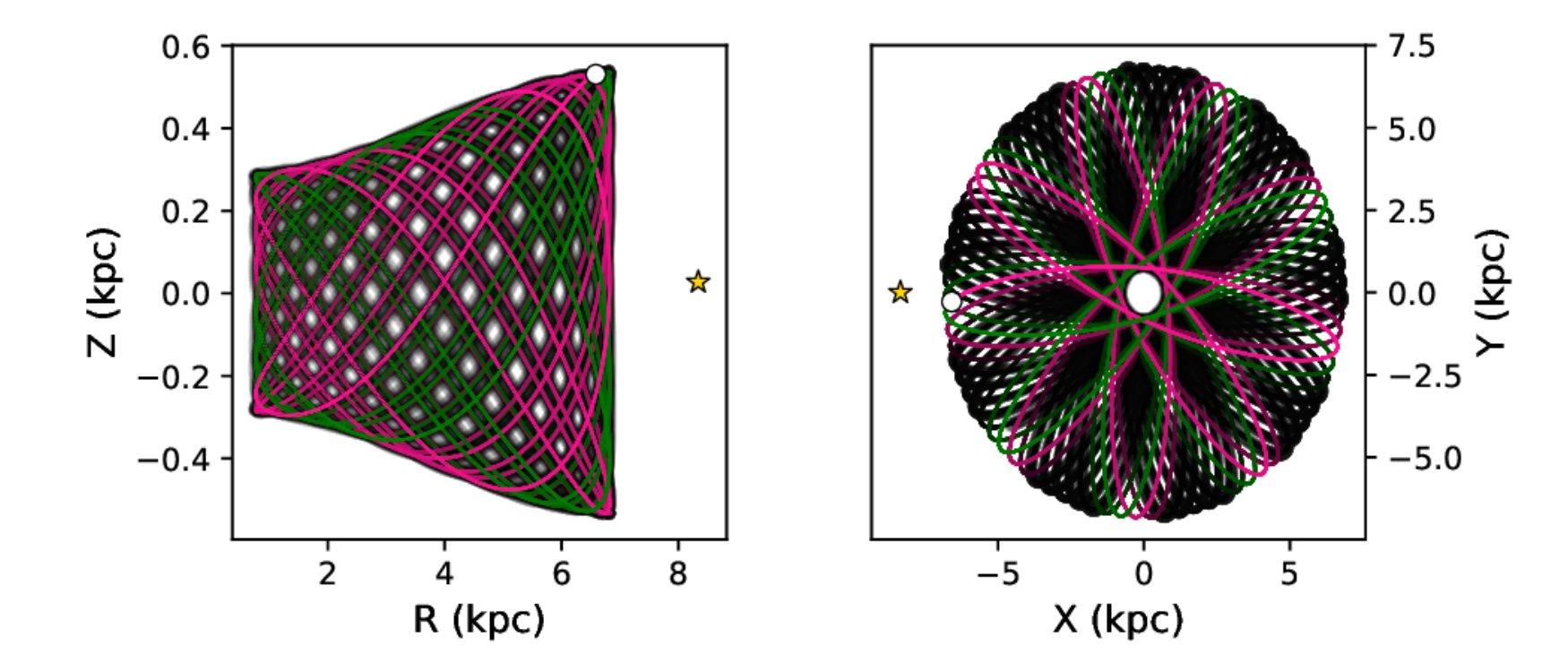}

\includegraphics[clip=true, trim = 0mm 20mm 0mm 10mm, width=0.9\columnwidth]{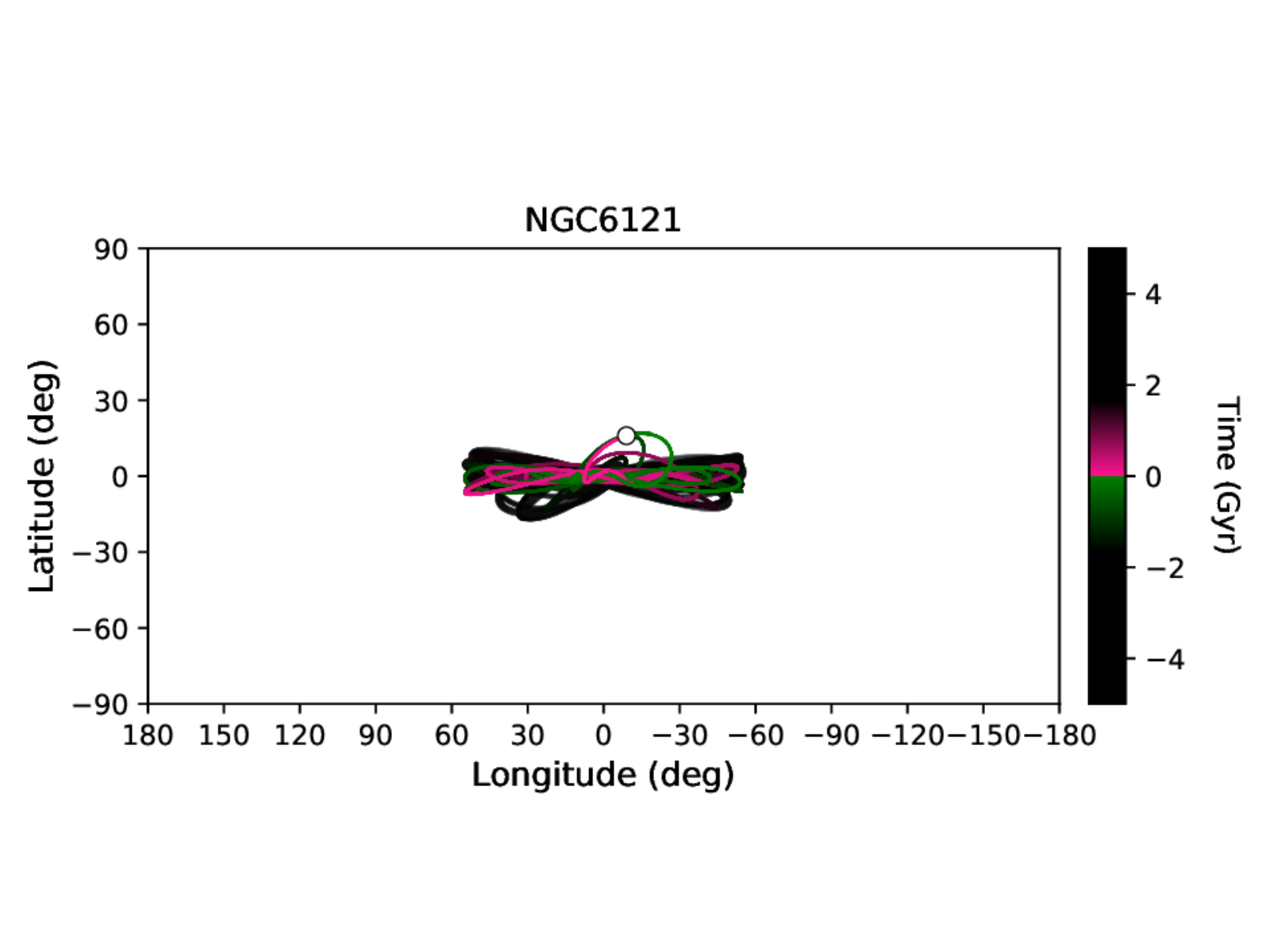}

\includegraphics[clip=true, trim = 0mm 20mm 0mm 10mm, width=0.9\columnwidth]{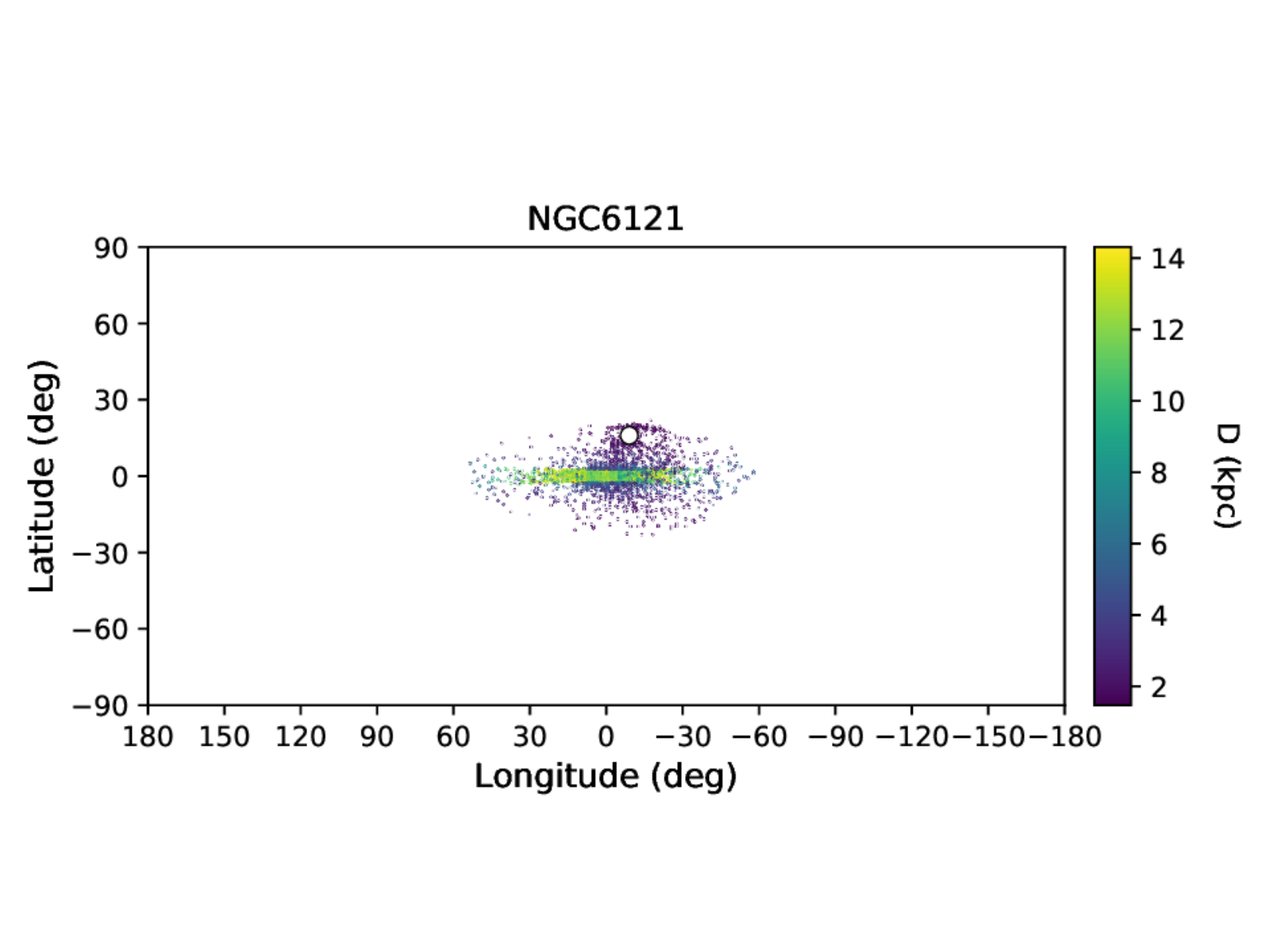}

\includegraphics[clip=true, trim = 0mm 20mm 0mm 10mm, width=0.9\columnwidth]{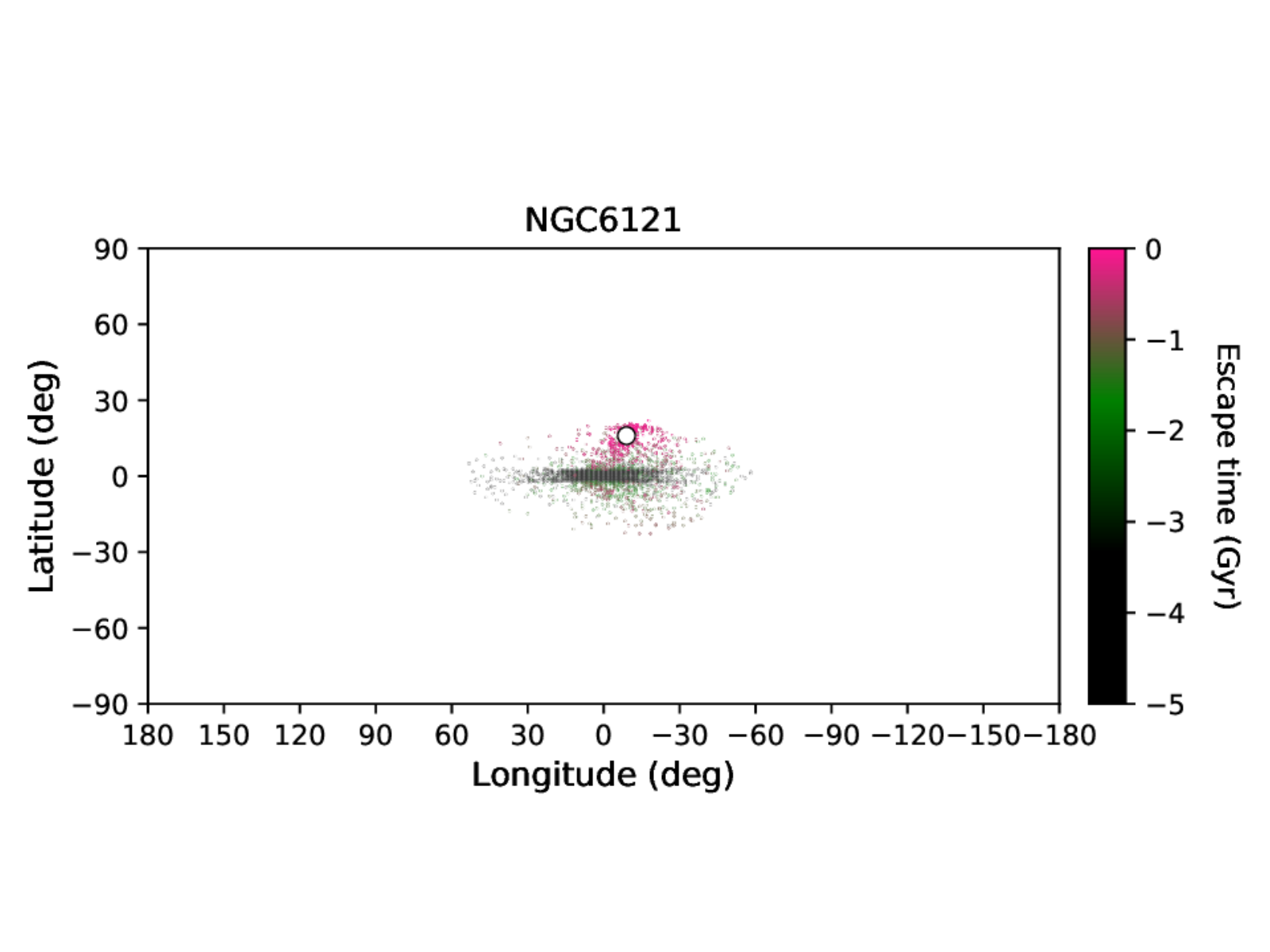}
\end{center}
\caption{ \emph{(Top-left panel):} Projection of the orbit of NGC~6121 in the meridional $\rm R-z$ plane. Colors trace time, from 5~Gyr ago (negative values) to 5~Gyr forward in time (positive values). \emph{(Top-right panel):} Projection of the orbit of NGC~6121 in the Galactic $x-y$ plane.  \emph{(Second row):} Projection of the NGC~6121 orbit for the past and future 5~Gyr, in the longitude-latitude plane.   \emph{(Third row):} Projection in the longitude-latitude plane of the extra-tidal material lost by NGC~6121. Colors indicate the average distance of the stripped material from the Sun.  \emph{(Bottom panel):} Projection in the longitude-latitude plane of the extra-tidal material lost by NGC~6121. Colors indicate the average time at which stellar particles become gravitationally unbound to the cluster, from 5~Gyr ago (negative time), to the current time ($\textrm{escape time} = 0$). In the bottom and middle panels, only the reference simulation without errors is shown, for better clarity. In all plots, the current position of the cluster is given by the white circle with a black edge color. The yellow star, when present, indicates the position of the Sun. }\label{ngc6121_stream}
\end{figure}

\paragraph{NGC~6121: }

With a current position at $x=-6.58$, $y=-0.28$ and $z=0.53$~kpc, NGC~6121 is the closest globular cluster to the Sun in our list. This cluster has a  remarkably planar orbit, with a maximal vertical excursion from the Galactic plane of only 0.5~kpc (see  Fig.~\ref{ngc6121_stream}), and an eccentricity $e=0.80 \pm 0.01$, which makes it oscillate between an apo-center at $R_{\rm max}=6.81 \pm 0.02$~kpc and a peri-center at $R_{min}=0.76 \pm 0.04$~kpc. Because this cluster lies inside the solar circle, its orbit is limited to a longitude interval from $-60^{\circ}$ to $60^{\circ}$; because the cluster currently lies very close to the Sun, and is at its highest height above the Galactic plane, the orbit forms a hook-like pattern in longitude-latitude space. This hook-like portion of the orbit, nearest to the cluster, is traced by the recently stripped tidal material (see Fig.~\ref{ngc6121_stream}), with a leading tail oriented mostly in a vertical direction in the $(\ell,b)$ plane, from the current cluster location, up to approximately $0^\circ$ latitude. This portion of the stripped material lies at less than 2~kpc from the Sun, and it constitutes the nearest stream found in our simulations.

To our knowledge, no extra-tidal structure has been discovered yet around NGC~6121. Recently, \citet{kundu19} have used RR-Lyrae stars to trace the extra-tidal material around NGC~6121, without finding any clear evidence of structures. The current position of the cluster in the sky, at a latitude of roughly $20^\circ$ and at a longitude close to $0^\circ$, makes this search  difficult due to the strong contamination of field disk stars, despite the fact that this portion of the stream should be very close to the Sun.

\paragraph{Pal~10 and Pal~2: }

\begin{figure}
\includegraphics[clip=true, trim = 0mm 2mm 0mm 0mm, width=0.9\columnwidth]{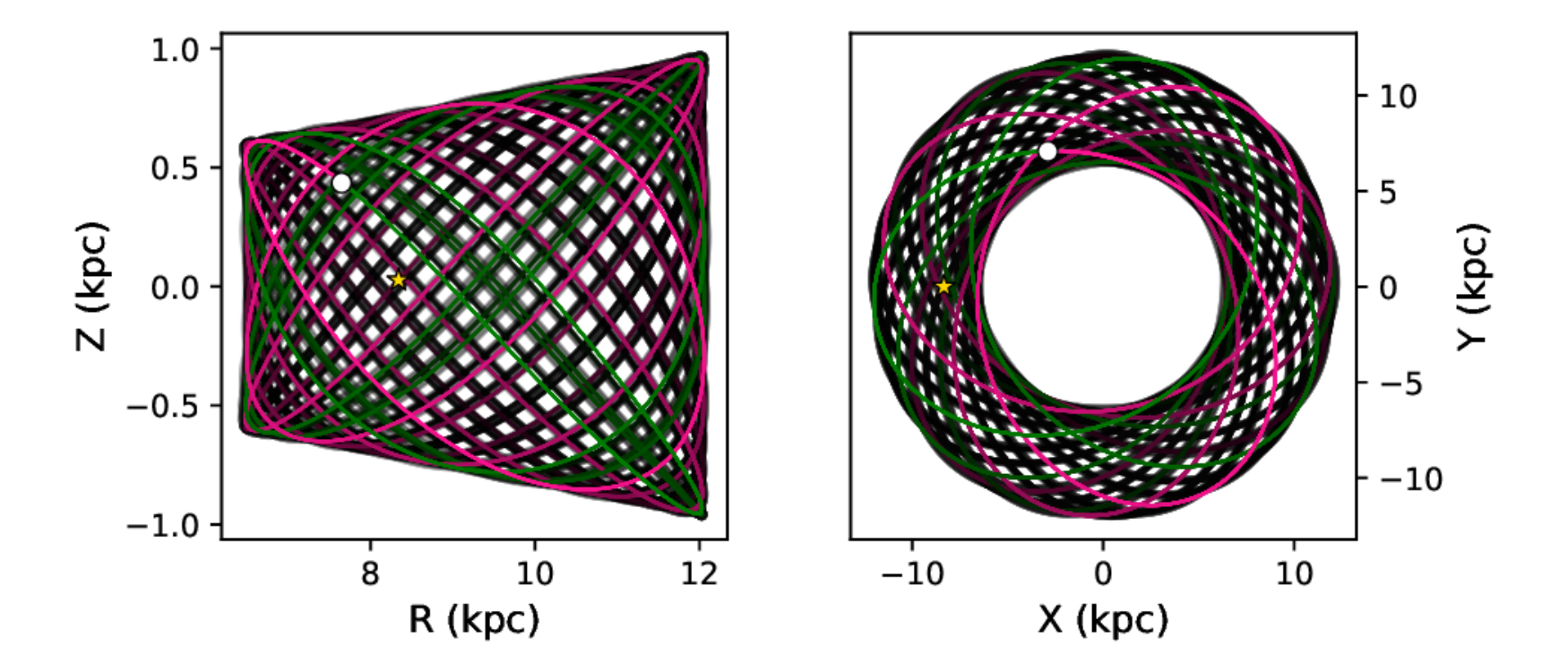}

\includegraphics[clip=true, trim = 0mm 20mm 0mm 10mm, width=0.9\columnwidth]{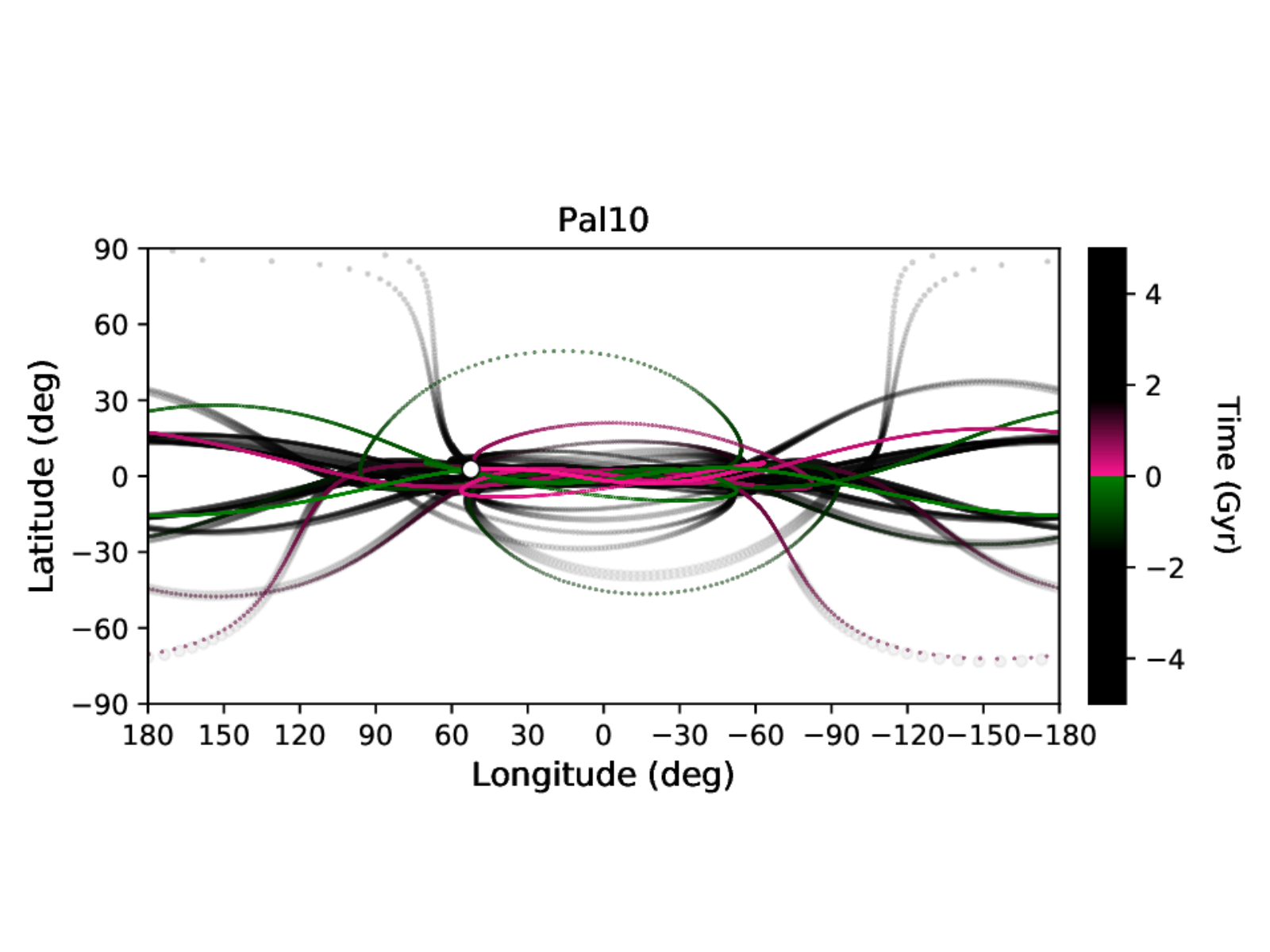}

\includegraphics[clip=true, trim = 0mm 20mm 0mm 10mm, width=0.9\columnwidth]{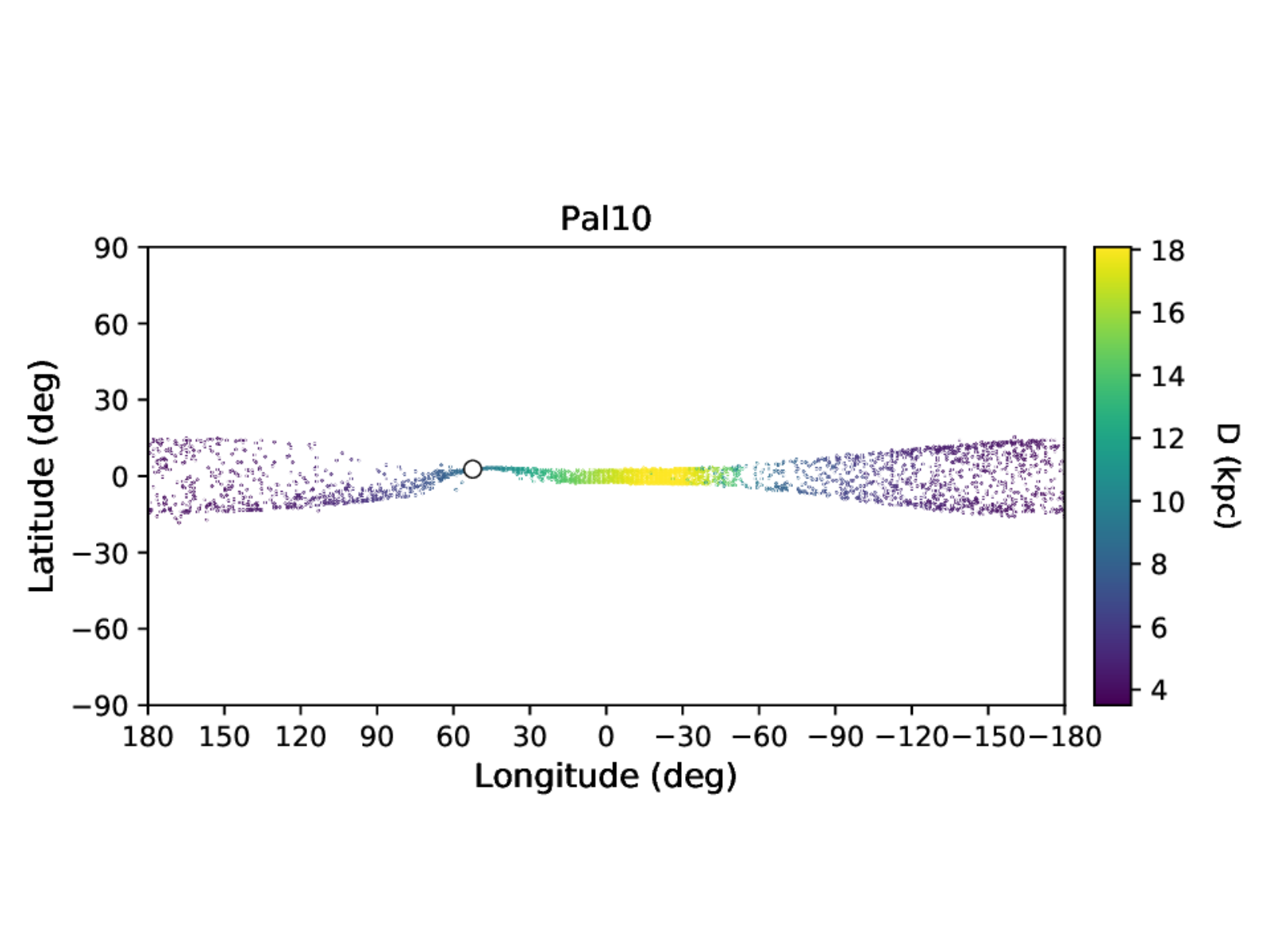}

\includegraphics[clip=true, trim = 0mm 20mm 0mm 10mm, width=0.9\columnwidth]{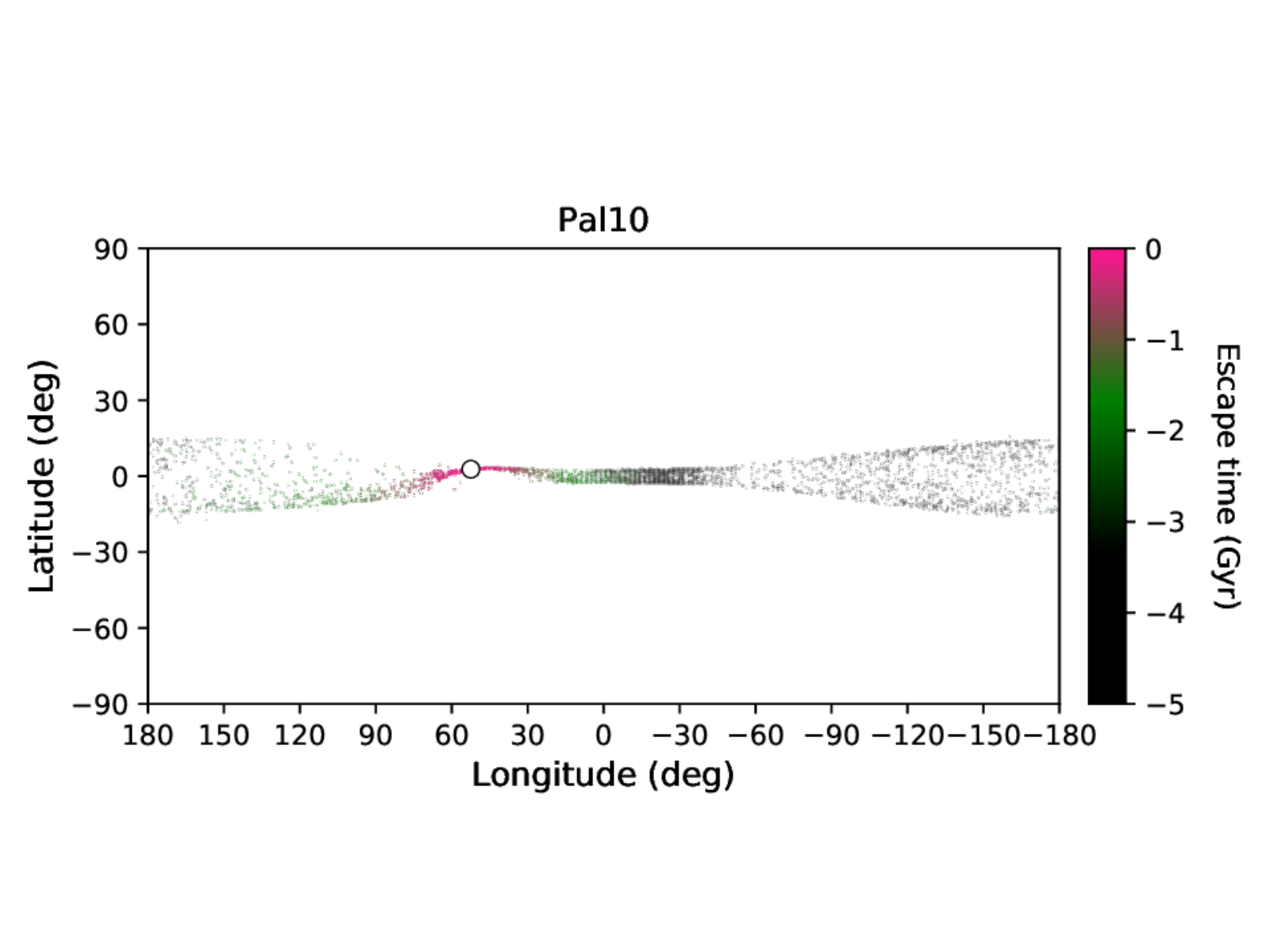}
\caption{Same as Fig.~\ref{ngc6121_stream}, but for the cluster Pal~10. \label{pal10_stream}}
\end{figure}

Pal~10 and Pal~2 are two disk clusters whose orbit crosses the solar radius. While for Pal~10, the maximal in-plane distance $R_{\rm max}$ is approximately $12$~kpc, in the case of Pal~2, the orbit can reach about 40~kpc from the Galactic center. The fact that both these clusters have a radial excursion of the orbit which is beyond the solar radius implies that their stripped stars can redistribute over the whole longitudinal range, thus also in the anti-center direction. The fact that both clusters have orbits confined close to the disk plane implies that the escaped material redistributes in very thin structures (i.e. confined in a limited latitude interval), which look like typical ``ribbons" in the sky.  Our models predict that both clusters are surrounded by a long stream of tidal material, which is however probably very difficult to identify because in both cases these extra-tidal stars are confined close to the Galactic plane. No tidal streams emanating from these two clusters, to our knowledge, have been identified in the observational data so far.  Because the tidal structures associated to these two clusters have similar properties, in Fig.~\ref{pal10_stream} we report only the case of Pal~10. 

\subsubsection{Extra-tidal features originating from inner clusters: bow-ties and more complex shapes}

We have defined inner globular clusters as systems  which are not disk clusters (their orbit is not confined close to the Galactic plane), but which are confined inside the solar radius. Seventy-one clusters are found in this category (see  Table~\ref{classification}). We discuss some of them in the following.

\paragraph{NGC~5946 \& NGC~5986:}

\begin{figure}
 \begin{center}
\includegraphics[clip=true, trim = 0mm 2mm 0mm 0mm, width=0.9\columnwidth]{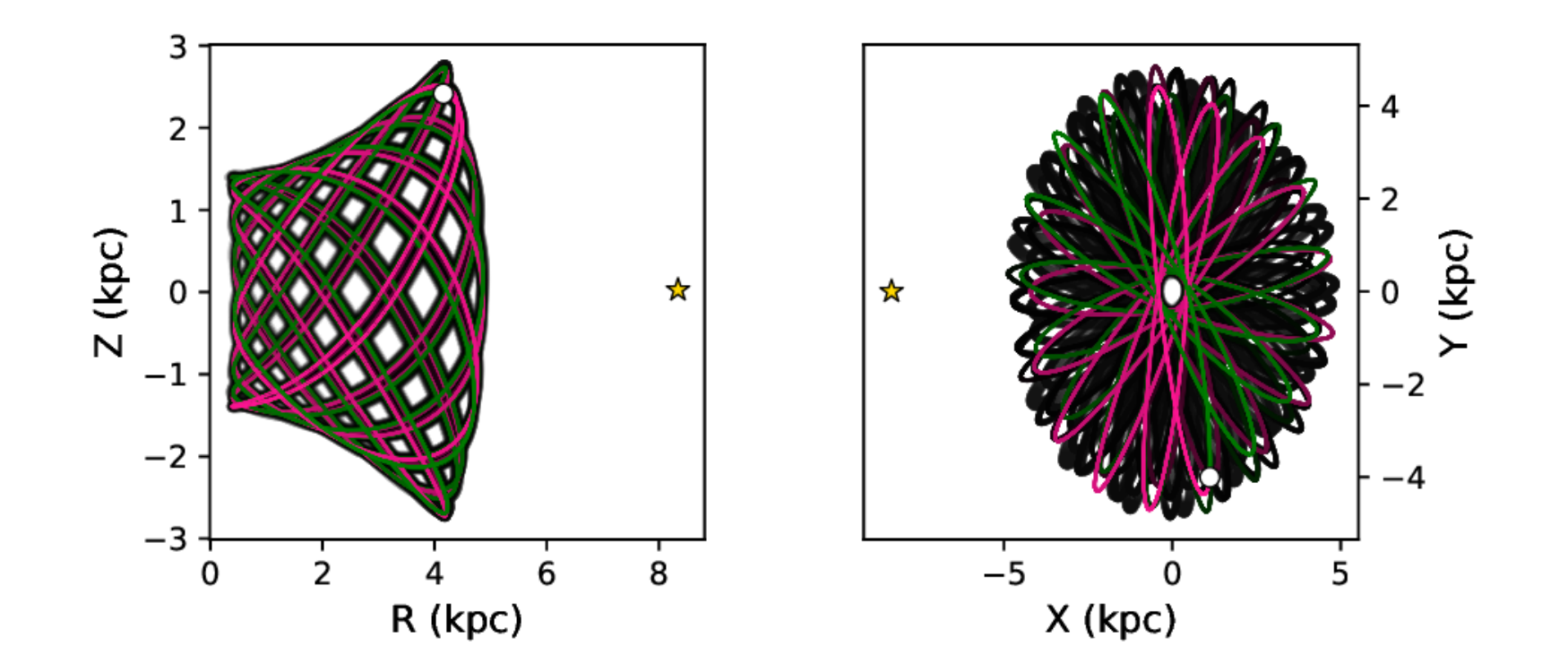}
\includegraphics[clip=true, trim = 0mm 20mm 0mm 10mm, width=0.9\columnwidth]{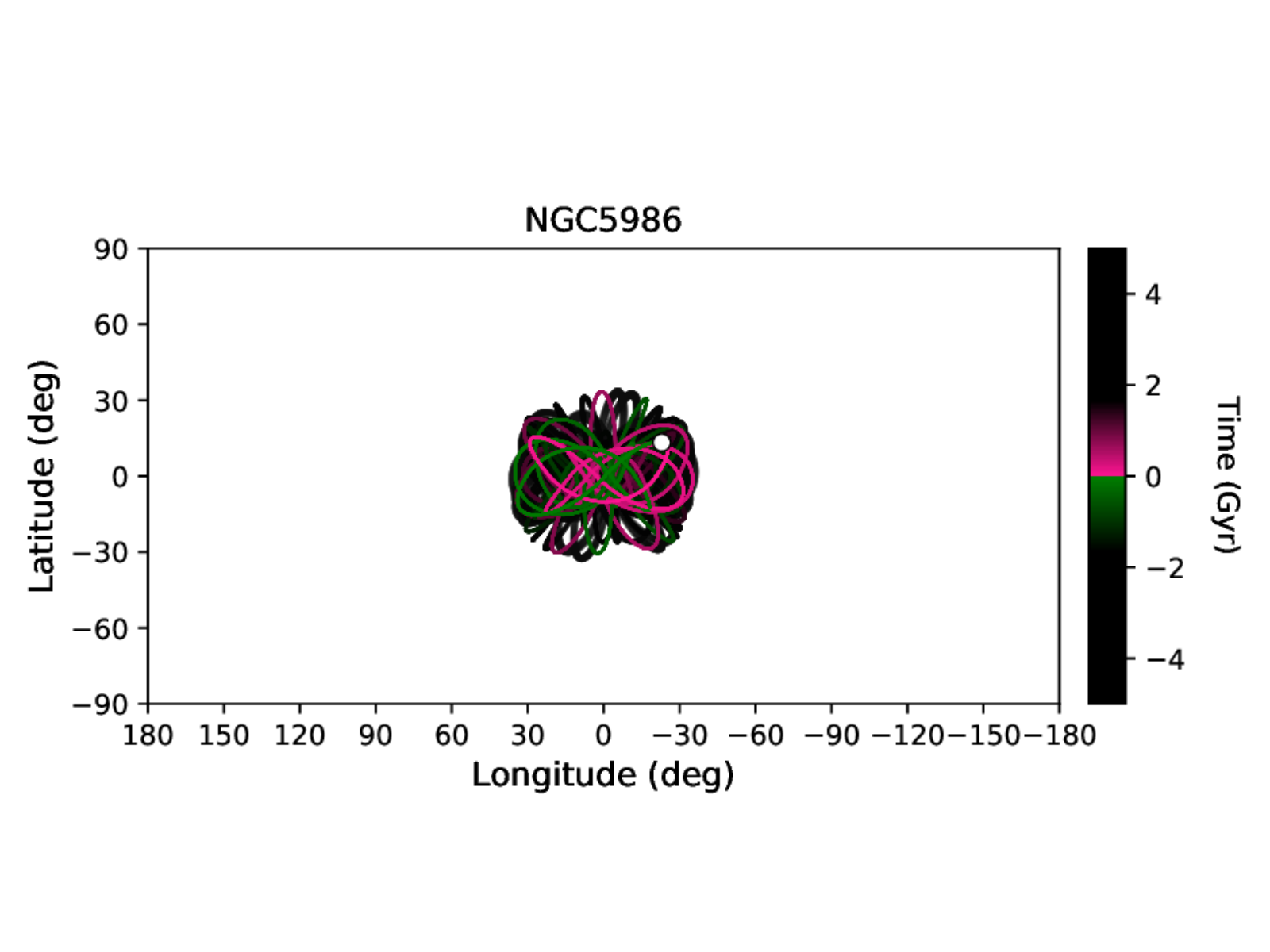}
\includegraphics[clip=true, trim = 0mm 20mm 0mm 10mm, width=0.9\columnwidth]{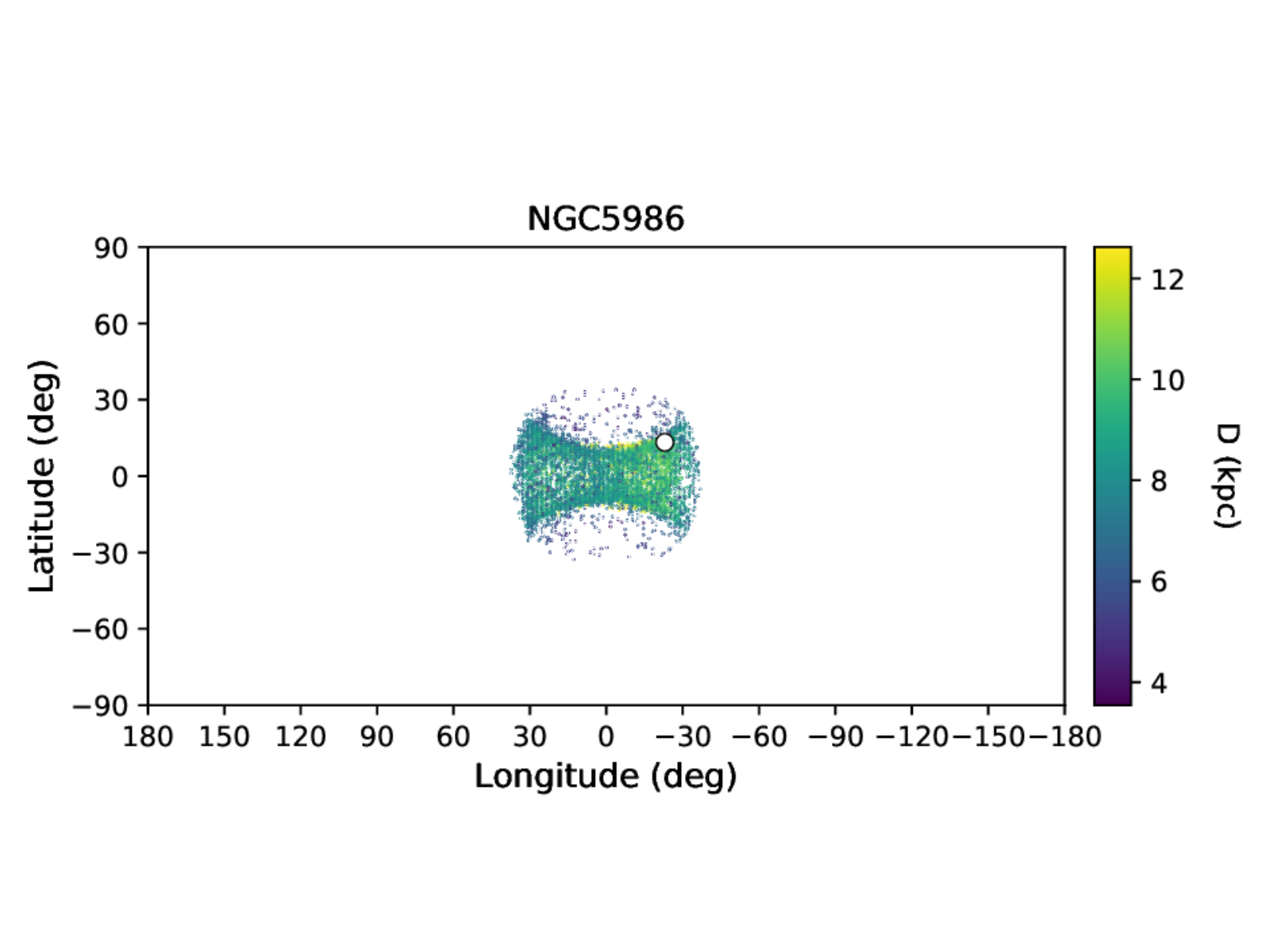}
\includegraphics[clip=true, trim = 0mm 20mm 0mm 10mm, width=0.9\columnwidth]{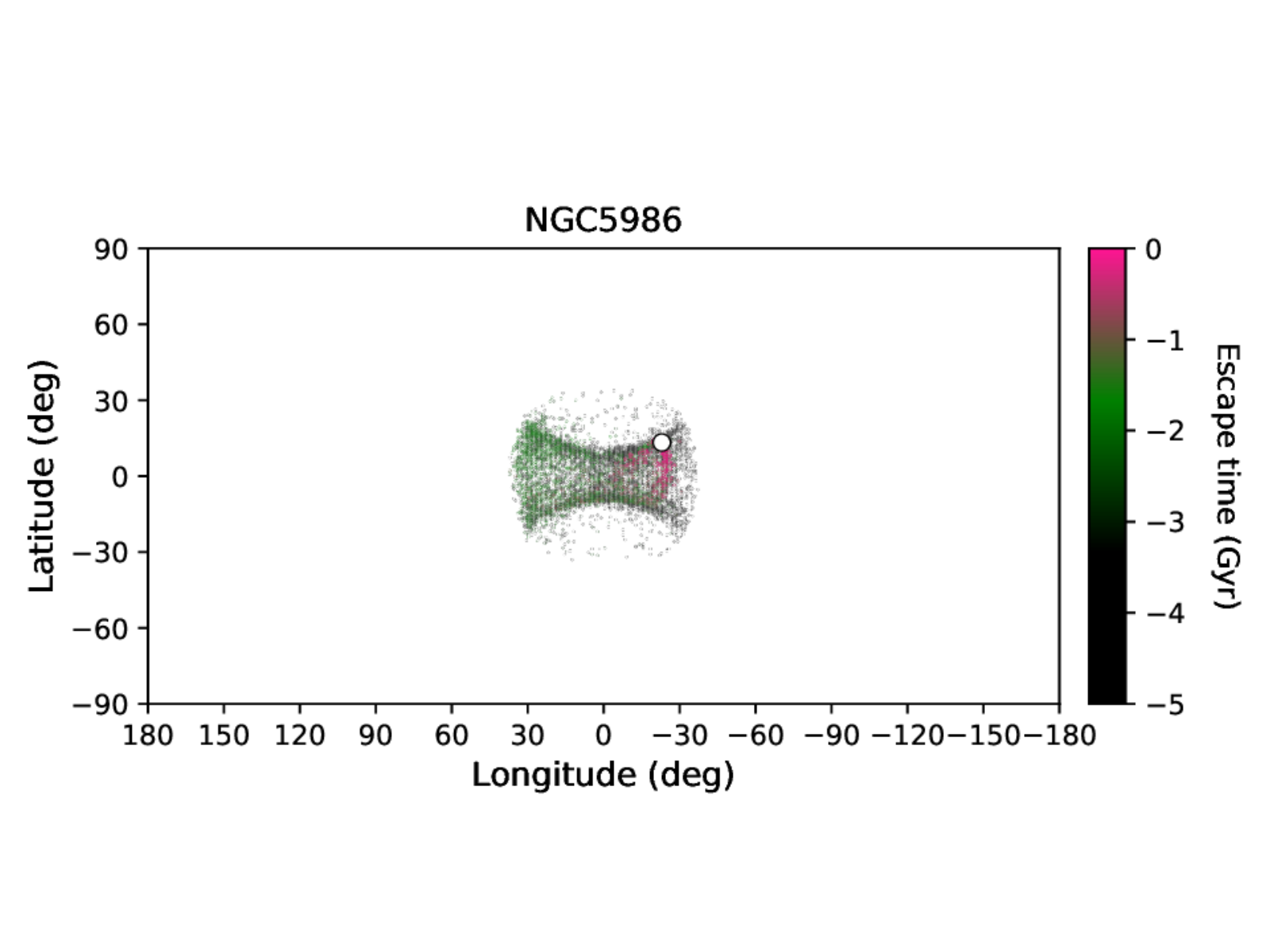}
  \end{center}
 \caption{Same as Fig.~\ref{ngc6121_stream}, but for the cluster NGC~5986. \label{ngc5986_stream}}
 \end{figure}

These are inner-non disk clusters whose escaped stars redistribute in a characteristic ``bow-tie" shape. These stars are all confined in a relatively narrow longitudinal range (typically within $-30^\circ$ to $30^\circ$). Towards the edges of the longitude interval, the distribution of extra-tidal stars tends to flare, whereas it instead shrinks at zero longitude. These trends can be explained as an effect of the projection of the orbits of these clusters in the $(\ell, b)$ plane.  Moreover, because these clusters always stay in the inner region of the Galaxy, where the dynamical timescales are short, their orbit - and consequently their stripped stars - can experience many disk crossings over the whole duration of the simulation, filling the whole $(\ell, b)$ space allowed by their orbital parameters. An example of such a distribution is given in Fig.~\ref{ngc5986_stream} for the extra-tidal material associated to the cluster NGC~5986.

\paragraph{NGC~104: }

 \begin{figure}
  \begin{center}
  \includegraphics[clip=true, trim = 0mm 2mm 0mm 0mm, width=0.9\columnwidth]{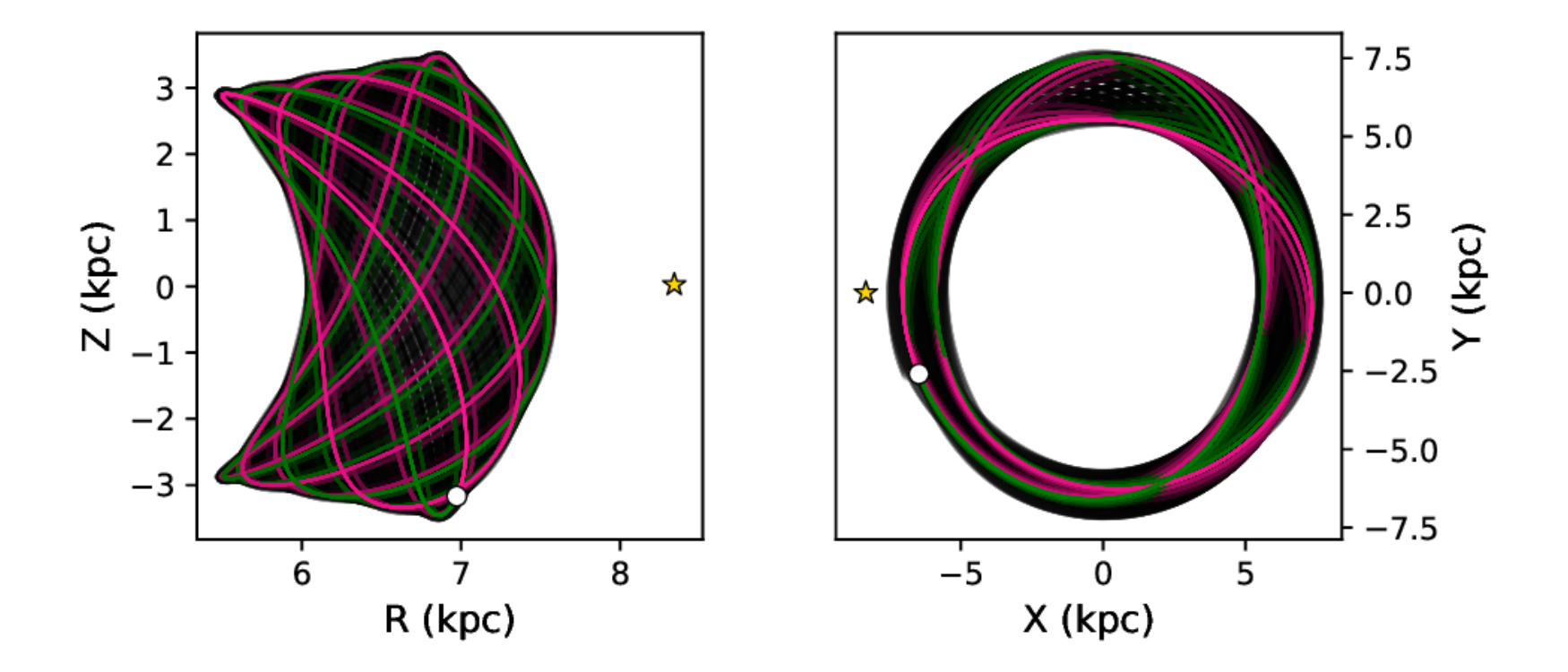}
  \includegraphics[clip=true, trim = 0mm 20mm 0mm 10mm, width=0.9\columnwidth]{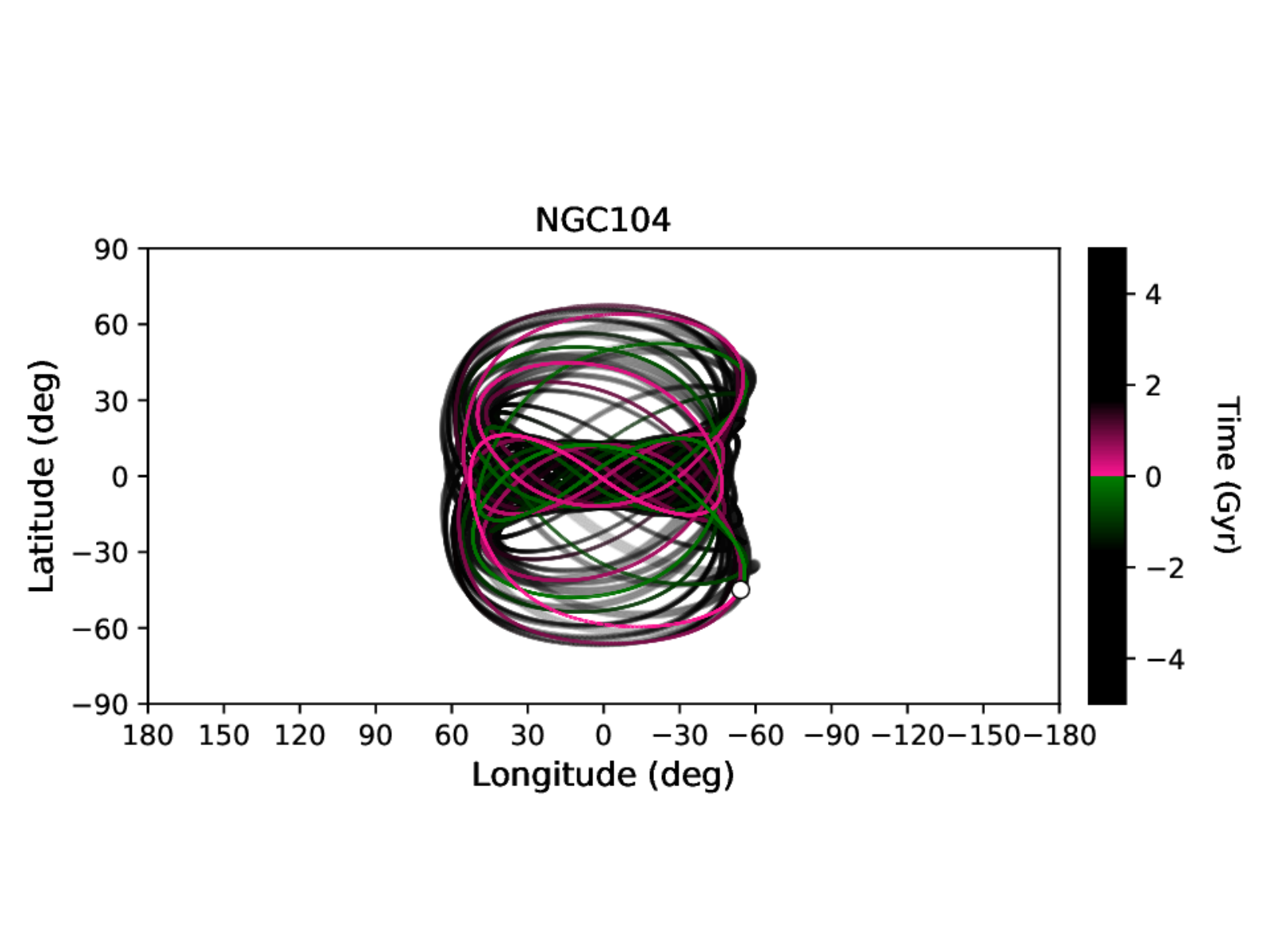}
  \includegraphics[clip=true, trim = 0mm 20mm 0mm 10mm, width=0.9\columnwidth]{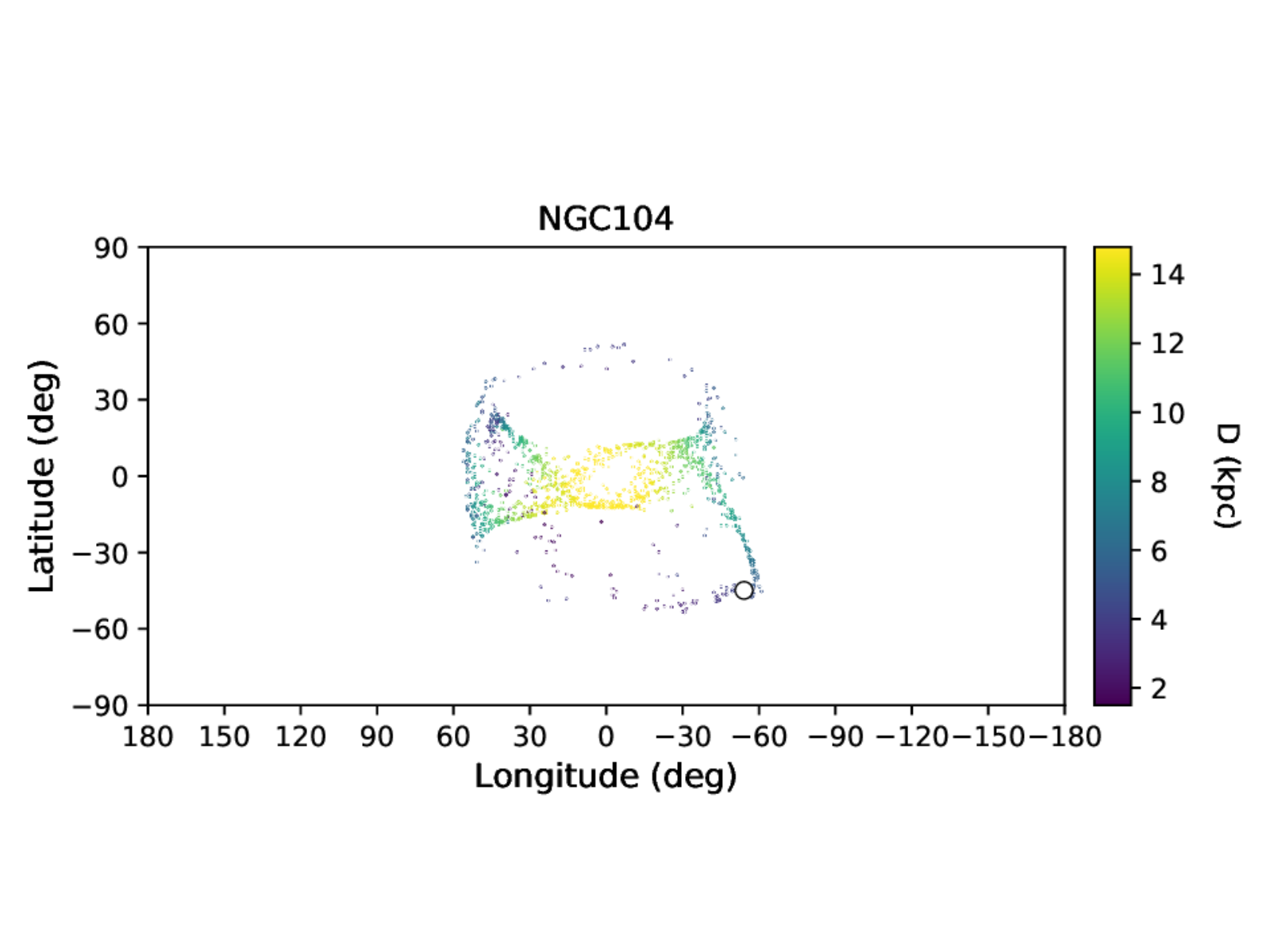}
  \includegraphics[clip=true, trim = 0mm 20mm 0mm 10mm, width=0.9\columnwidth]{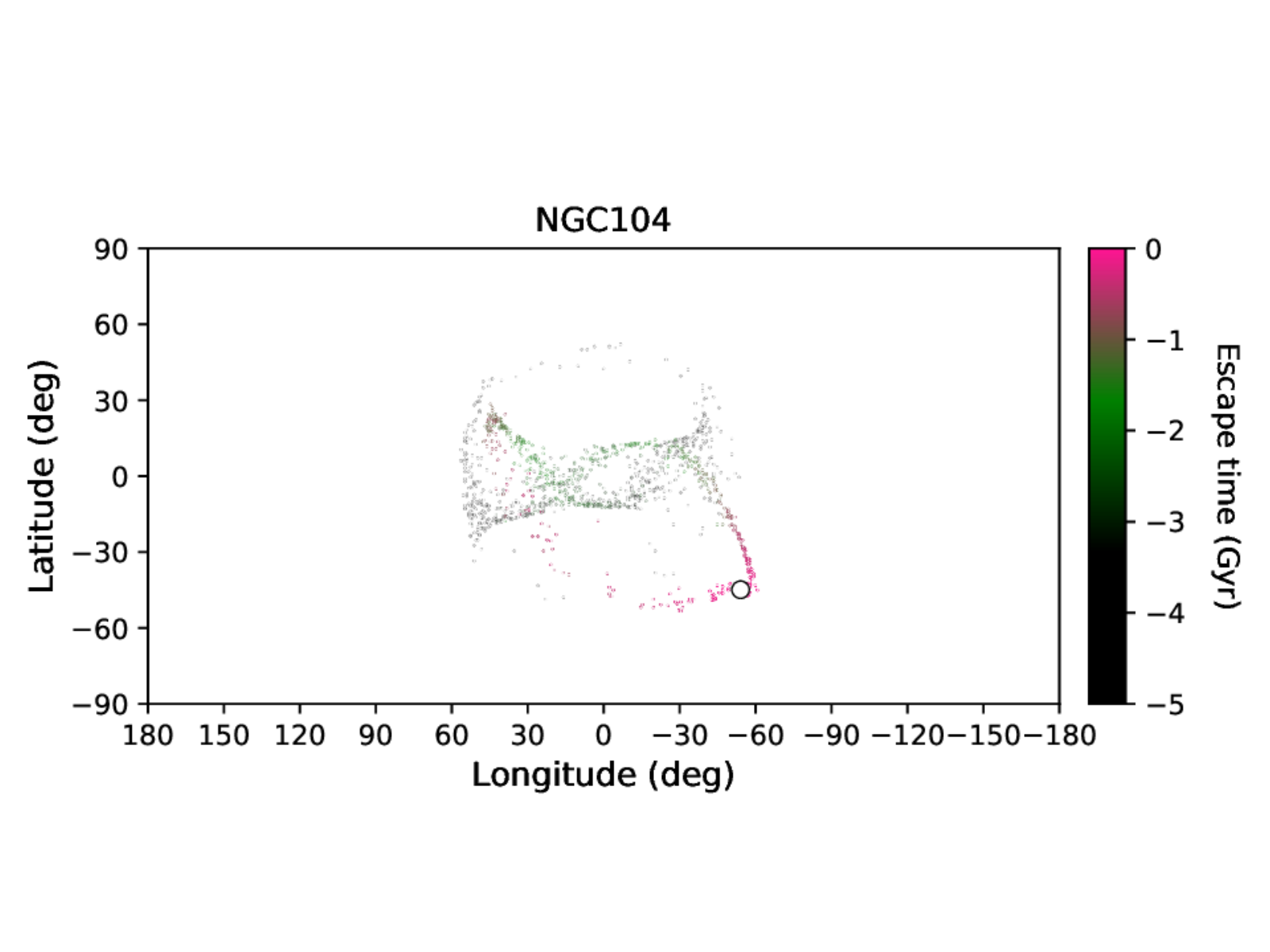}
  \end{center}
 \caption{Same as Fig.~\ref{ngc6121_stream}, but for the cluster NGC~104. \label{ngc104_stream}}
 \end{figure}

Then there are clusters like NGC~104 (47~Tuc) for which the morphology of the extra-tidal material takes more complex shapes. This cluster has an orbit confined inside the solar radius, but which, at its apocenter, can reach a distance of less than 1~kpc from the Sun (see Fig.~\ref{ngc104_stream}). The projection of the past and future orbit in the ($\ell, b$) plane gives rise to a kind of figure-of-eight shape, with the stars stripped more recently from the cluster tracing the portion of this shape which extends to negative latitudes. Our model of NGC~104 agrees with the conclusions reached by \citet{lane12} who suggested that two clear tidal tails should emanate from this cluster, however, the orientation of these tails in the ($\ell, b$) plane differs slightly in the two works. \citet{lane12}'s models suggest indeed that the leading tail of NGC~104 should extend a bit beyond $\ell < -70^\circ$ (see Fig.~4 in their paper, as an example), while our model predicts a minimum longitude of about $-60^\circ$. The exact comparison between these two works is however difficult, since the model of the Galactic potential, the distance to the Sun, proper motions and line-of-sight velocities of NGC~104 used in their study are different from ours. To date, clear tails around NGC~104 have not been found yet, with the most recent observational works pointing to the possibility of the presence of a diffuse extended halo-like structure around this cluster \citep[see][]{piatti17}. We note that we only find a more diffuse distribution of extra-tidal material in the case of the PII-0.3-SLOW model. Additional work for comparing the current observational data with simulations will be needed to resolve this apparent discrepancy between theoretical predictions and observational findings. 

Other shapes found in this category include ``Easter eggs", which are generated by clusters whose orbits are confined to the innermost kpc of the Galaxy, and show significant variations in the z-coordinates---at least as large as those found in the radial direction. Among clusters whose extra-tidal material shows these peculiar shapes we find HP1, NGC~6093, NGC~6273, NGC~6293, NGC~6723, and NGC~6809.

\subsubsection{Extra-tidal features originating from outer clusters: ``canonical" tidal tails}

\textit{Caveat}: In this group, there are some elongated streams emanating from clusters as NGC~6715, Pal~12, Ter~7 and Ter~8, which are  associated to the Sagittarius dwarf galaxy. We caution the reader that for these clusters the elongation and shape of the streams may be severely modified if the gravitational potential generated by the Sagittarius dwarf galaxy itself was included in the model. For completeness, we have decided to include these streams in the paper, and to report them in Appendix~\ref{allstreams}. In future works, we plan to investigate how the inclusion of the Sagittarius dwarf may alter these streams, and possibly affect also those of other clusters, not necessarily associated to this dwarf galaxy.\\

Among the extra-tidal structures emanating from outer globular clusters, we find some of the most beautiful and elongated tidal tails, of which those associated to Pal~5 were the first to be discovered \citep{odenkirchen01}.  In this category, we note the stream associated to the E~3 cluster that extends about  $120^\circ$  in longitude based on our models prediction; the thin stream emanating from IC~4499, which we predict to have an extension of about $150^\circ$ in longitude. To list them, the finest and thinnest stellar streams are predicted from: AM~4, Arp~2, IC~4499, NGC~1261, NGC~3201, NGC~4590, NGC~5024, NGC~5053, NGC~5272, NGC~5466, NGC~5694, NGC~5824, NGC~5904, NGC~6101, NGC~6426, NGC~6584, NGC~6934, Pal~1, Pal~5, Pyxis, Rup~106, Sagittarius~II, Ter~7, Ter~8, and Whiting~1.

Many of the above cited streams have been discovered and found also in observational data, but in many cases the extent of the tails, as predicted by our models, is larger. Of these observed streams, many tracks are available in the \textit{galstreams} \citep{mateu22} library. In the following, we compare our model predictions to some of these tracks in the three different Galactic potentials adopted in this paper. More specifically, we compare the projected density distribution of the simulated stream to observations in the $(\ell, b)$, $(\ell,  \rm \mu_{\ell}cos(\mathit{b}))$, $(\ell, \rm \mu_b)$ and $(\ell, \rm D)$ planes. As for the projected density distributions derived from the models, we calculate them by taking into account all the 51 simulations realized for each cluster,  that is both the reference simulation, and those realized by a Monte-Carlo sampling of the uncertainties. Overall, the agreement between models and observational data is excellent, in all Galactic models used. By looking at more extended regions in the sky than those covered by current observations, it should be however possible to better constrain the streams, and favor/disfavor some of these models.

\begin{figure*}[h!]
  \begin{center}
    \includegraphics[clip=true, trim = 0mm 0mm 0mm 0mm, width=0.65\columnwidth]{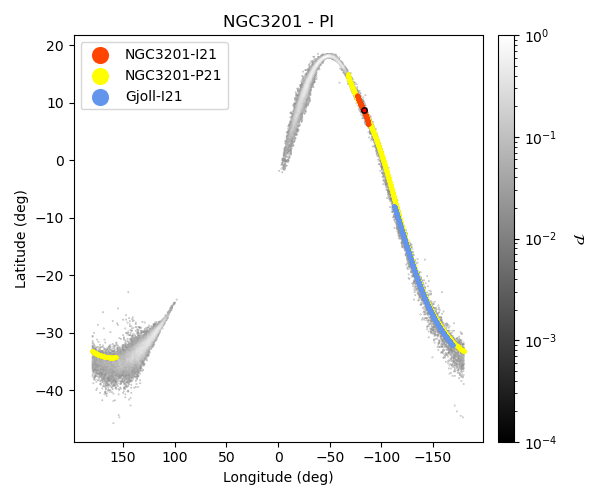}
    \includegraphics[clip=true, trim = 0mm 0mm 0mm 0mm, width=0.65\columnwidth]{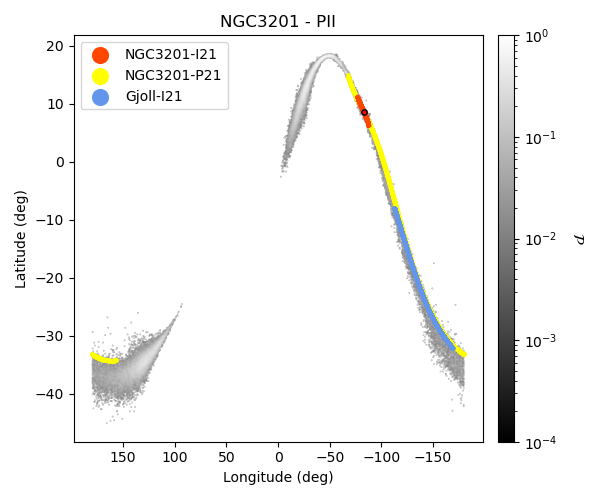}
    \includegraphics[clip=true, trim = 0mm 0mm 0mm 0mm, width=0.65\columnwidth]{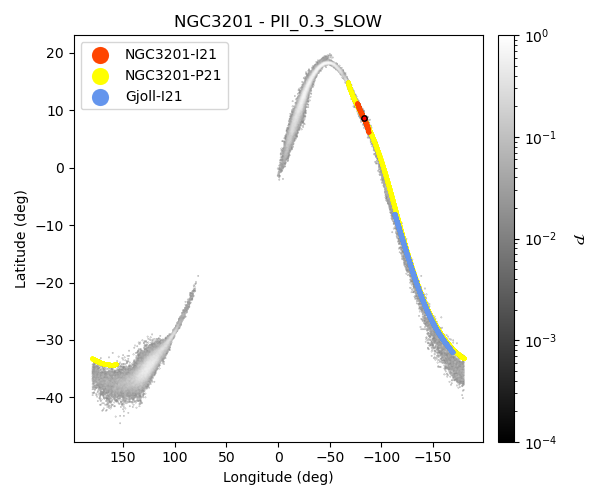}\\
   
    \includegraphics[clip=true, trim = 0mm 0mm 0mm 0mm, width=0.65\columnwidth]{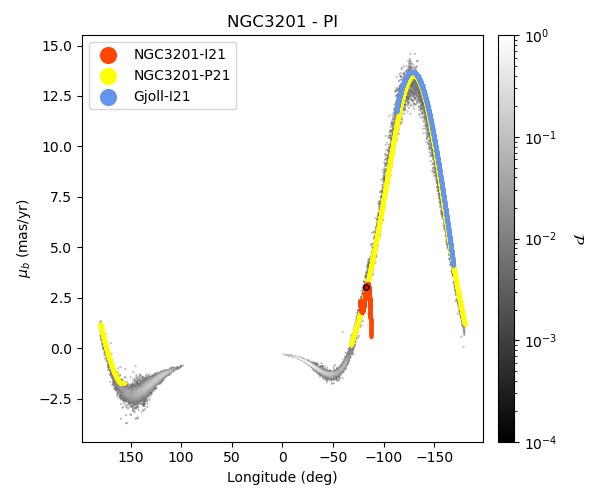}
    \includegraphics[clip=true, trim = 0mm 0mm 0mm 0mm, width=0.65\columnwidth]{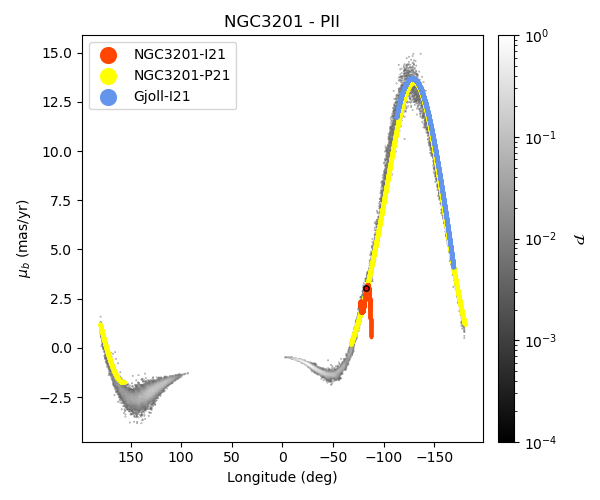}
    \includegraphics[clip=true, trim = 0mm 0mm 0mm 0mm, width=0.65\columnwidth]{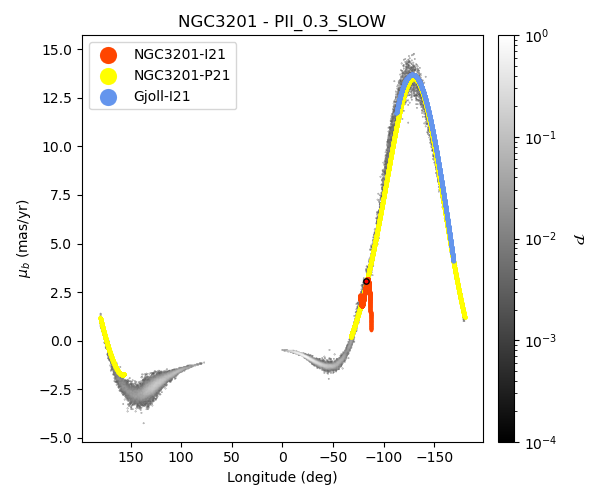}\\
    
    \includegraphics[clip=true, trim = 0mm 0mm 0mm 0mm, width=0.65\columnwidth]{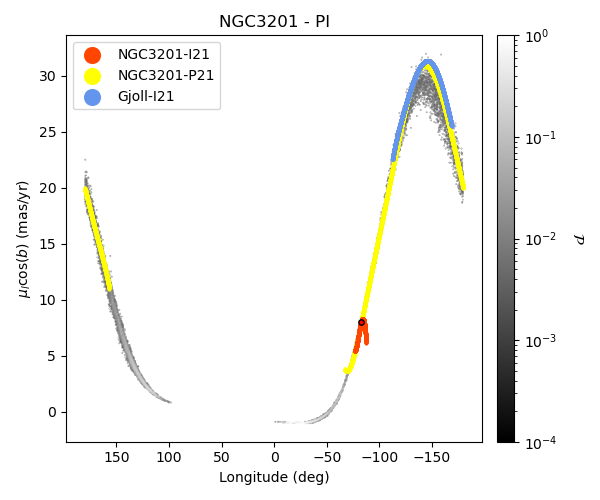}
    \includegraphics[clip=true, trim = 0mm 0mm 0mm 0mm, width=0.65\columnwidth]{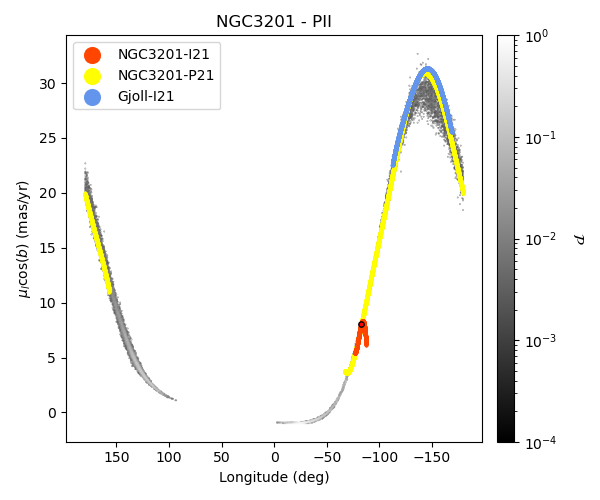}
    \includegraphics[clip=true, trim = 0mm 0mm 0mm 0mm, width=0.65\columnwidth]{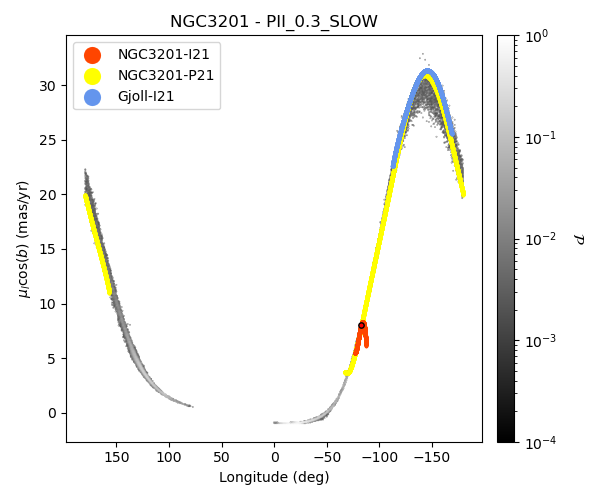}\\
     
    \includegraphics[clip=true, trim = 0mm 0mm 0mm 0mm, width=0.65\columnwidth]{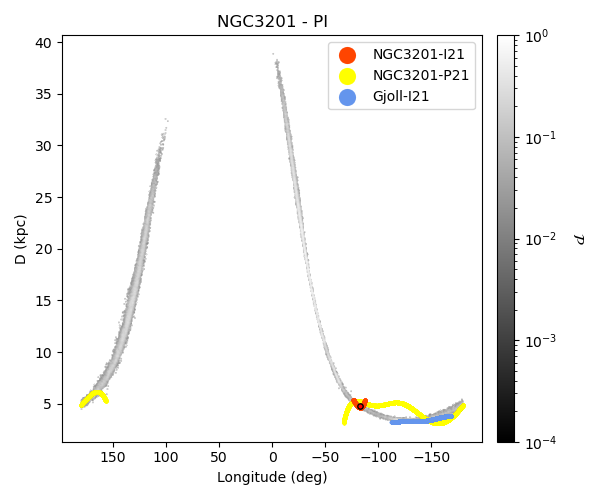}
    \includegraphics[clip=true, trim = 0mm 0mm 0mm 0mm, width=0.65\columnwidth]{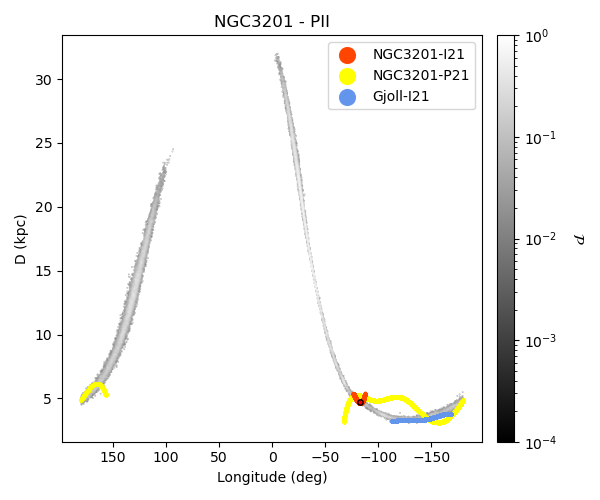}
    \includegraphics[clip=true, trim = 0mm 0mm 0mm 0mm, width=0.65\columnwidth]{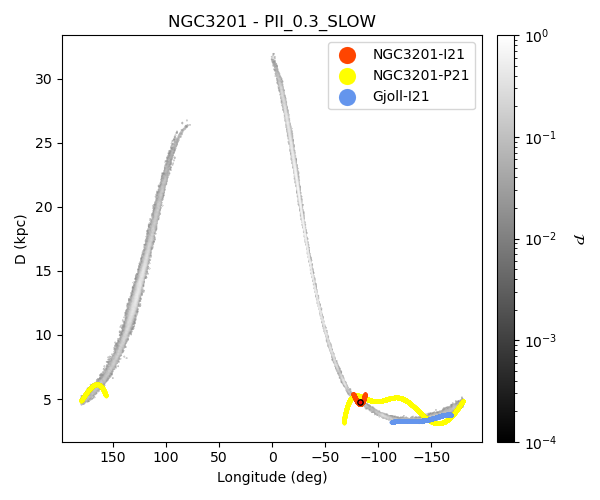}\\
  \end{center}
 \caption{\textit{From top to bottom:} Projected density distribution of the NGC~3201 stream, as predicted by our simulations, in the $(\ell,b)$, $(\ell, \rm \mu_b)$,  $(\ell,  \rm \mu_{\ell}cos(\mathit{b}))$ and $(\ell, \rm D)$ planes. The model predictions are shown for the three Galactic potentials  PI (\textit{left column}), PII (\textit{middle column}), and PII-0.3-SLOW (\textit{right column}) and are compared to the tracks available in the \textit{galstreams} library for this cluster, and which are taken from \citet{ibata21} (NGC3201-I21, red lines), \citet{palau21}  (NGC3201-P21, yellow lines) and from \citet{ibata21}, as for the Gj\"{o}ll stream  (Gj\"{o}ll-I21, blue lines). For each panel, the color-bar indicates the two dimensional probability density quantified by taking into account all the particles from the 51 realizations which is then normalized to its maximum value. In all panels, the current position of the cluster is indicated by a red dot.}\label{NGC3201_comp}
 \end{figure*}

\paragraph{NGC~3201: } NGC~3201 is an outer cluster  at a distance of about 4.7~kpc from the Sun. It has received much attention in the last couple of years,  since the suggestion by \citet{riley20} that part of its tidal tails could be associated to the Gj\"{o}ll stream, discovered by \citet{ibata19b}. The association of Gj\"{o}ll and NGC~3201 stream has been further confirmed by \citet{hansen20}, on the basis of the similarity of the chemical abundances of stars in the Gj\"{o}ll stream and in NGC~3201. More recently, \citet{palau21} have conducted an extensive study of the tidal tails emanating from this cluster, discovering a long stream, which an overall length of $140^\circ$ in the sky. Our models suggest that, in fact, the tails emanating from NGC~3201 may be even more extended than those found by \citet{palau21}. To further illustrate this point,  in Fig.~\ref{NGC3201_comp} we compare our model predictions to the tracks available for this stream in the \textit{galstreams} library, and which are taken from \citet{ibata21}, from \citet{palau21}  and from the Gj\"{o}ll stream, as reported by \citet{ibata21}. All models represent very well the portion of the stream discovered so far, both in distribution of the stream in the sky, distance and proper motion spaces. Our models predict indeed that, for this cluster, the differences in the stream properties change very little with the mass model adopted. The most striking difference is found for the elongation of the stream at $\ell > 0$: in the case of the barred potential, the stream associated to NGC~3201 extends indeed to smaller values of $\ell$ (up to $\ell \sim 70^\circ$) which are not reached in the case of the axisymmetric models. This portion of the stream is expected to be at distances greater than 20~kpc from the Sun (see bottom panels in Fig.~\ref{NGC3201_comp}).

\begin{figure*}
  \begin{center}
    \includegraphics[clip=true, trim = 0mm 0mm 0mm 0mm, width=0.65\columnwidth]{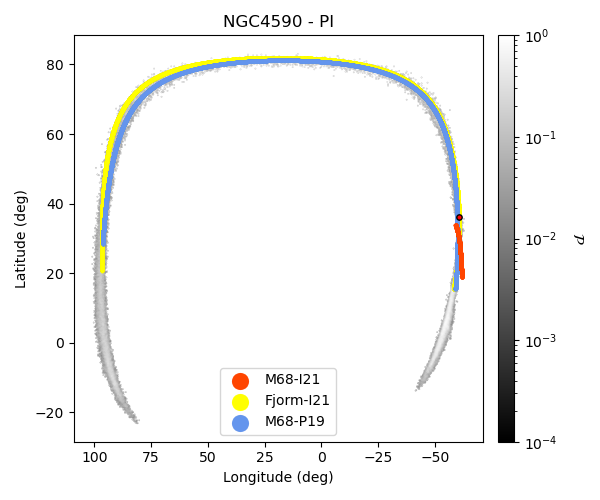}
    \includegraphics[clip=true, trim = 0mm 0mm 0mm 0mm, width=0.65\columnwidth]{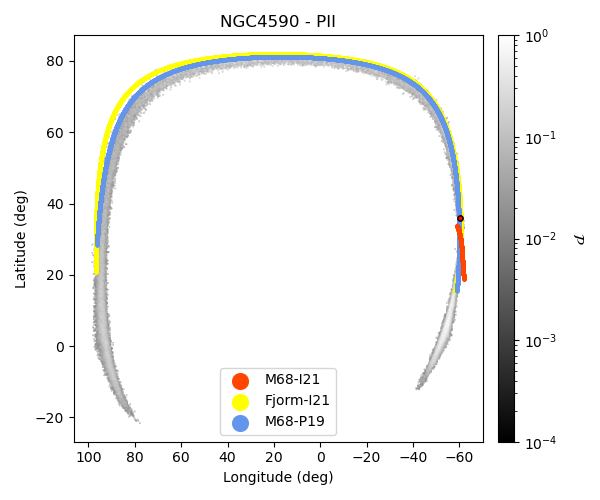}
    \includegraphics[clip=true, trim = 0mm 0mm 0mm 0mm, width=0.65\columnwidth]{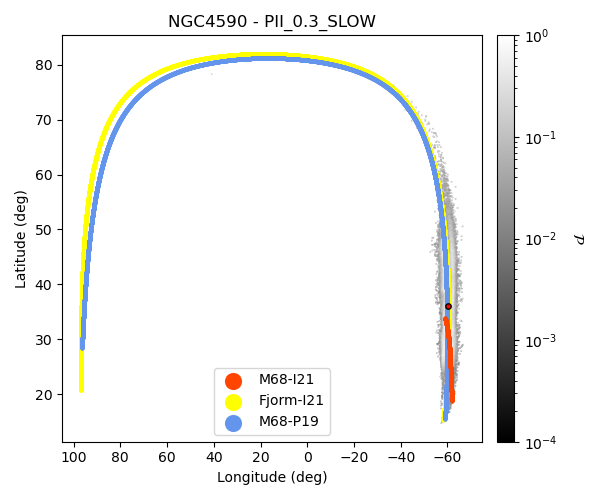}\\
   
    \includegraphics[clip=true, trim = 0mm 0mm 0mm 0mm, width=0.65\columnwidth]{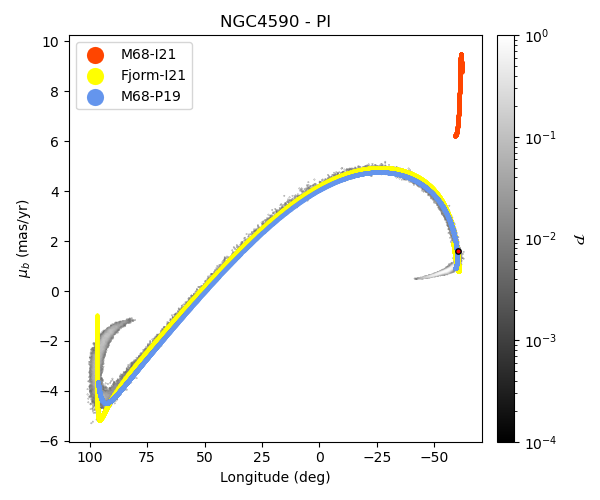}
    \includegraphics[clip=true, trim = 0mm 0mm 0mm 0mm, width=0.65\columnwidth]{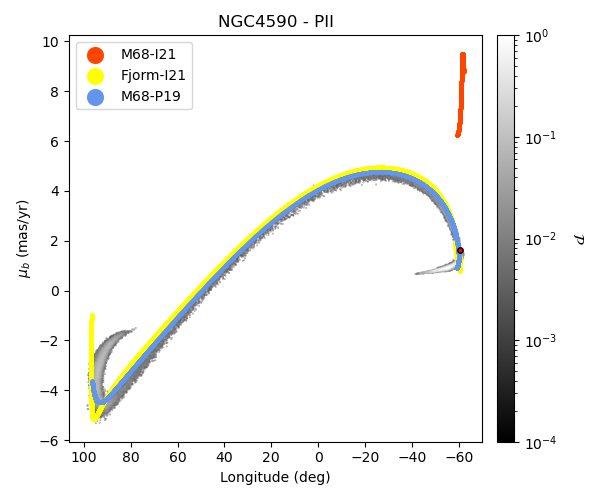}
    \includegraphics[clip=true, trim = 0mm 0mm 0mm 0mm, width=0.65\columnwidth]{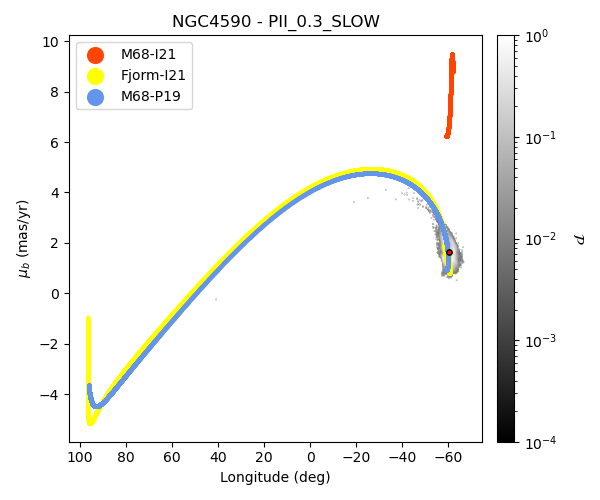}\\
    
    \includegraphics[clip=true, trim = 0mm 0mm 0mm 0mm, width=0.65\columnwidth]{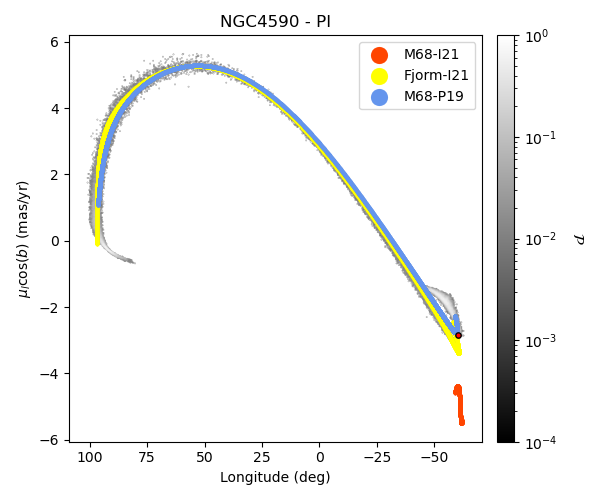}
    \includegraphics[clip=true, trim = 0mm 0mm 0mm 0mm, width=0.65\columnwidth]{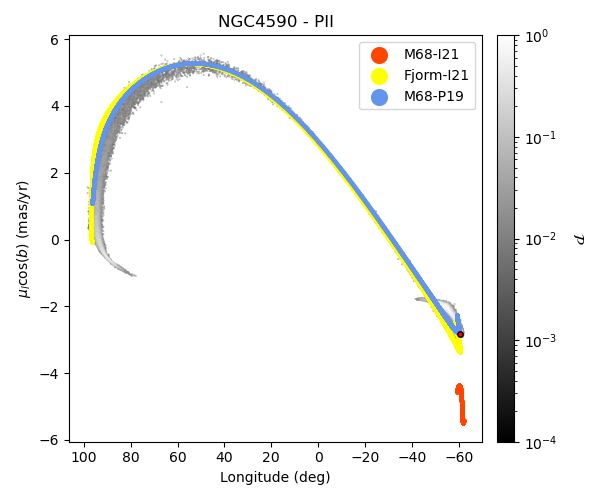}
    \includegraphics[clip=true, trim = 0mm 0mm 0mm 0mm, width=0.65\columnwidth]{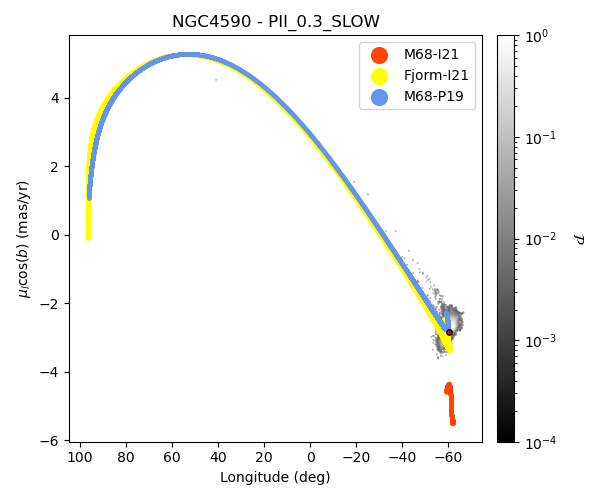}\\
     
    \includegraphics[clip=true, trim = 0mm 0mm 0mm 0mm, width=0.65\columnwidth]{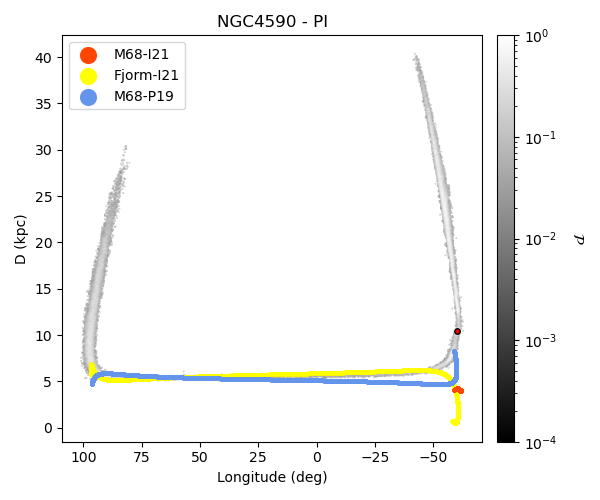}
    \includegraphics[clip=true, trim = 0mm 0mm 0mm 0mm, width=0.65\columnwidth]{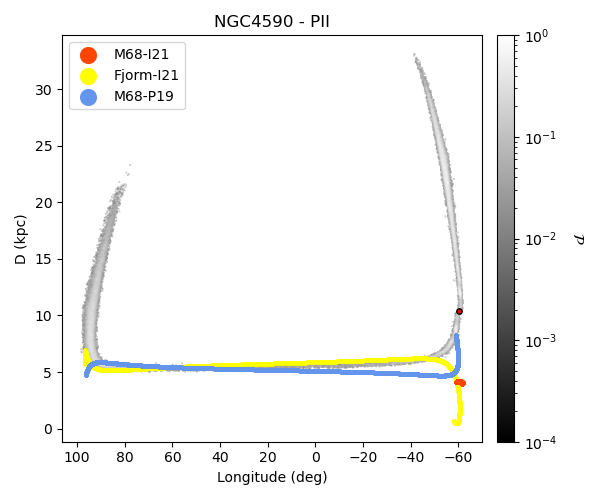}
    \includegraphics[clip=true, trim = 0mm 0mm 0mm 0mm, width=0.65\columnwidth]{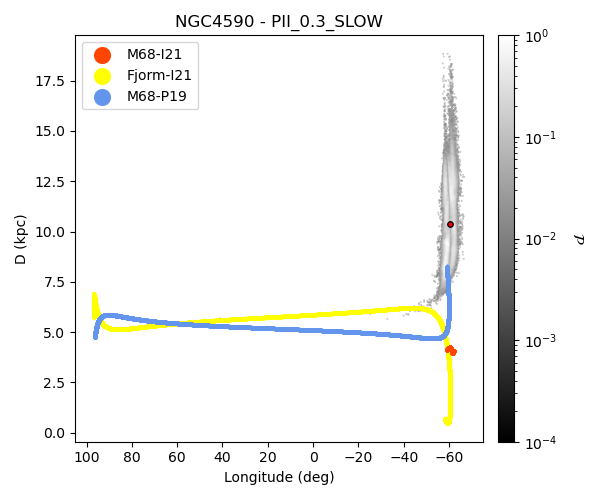}\\
  \end{center}
 \caption{Same as Fig.~\ref{NGC3201_comp}, for the NGC~4590 cluster. The tracks from come from the work by  \citet{palau19} (M68-P19, blue lines), \citet{ibata21} (M68-I21, red lines) and from \citet{ibata21} as for the Fj\"{o}rm stream (Fjorm-I21, yellow lines).} \label{NGC4590_comp}
 \end{figure*}

\paragraph{NGC~4590: } NGC~4590 (M~68) is an outer cluster at a current distance of about 10~kpc from the Sun. This cluster is surrounded by a very extended stream \citep{palau19, ibata21}, a long portion of which is represented by the Fj\"{o}rm stream discovered by \citet{ibata19b}. The comparison between our model predictions and the tracks available in \textit{galstreams} is shown in Fig.~\ref{NGC4590_comp}. Interestingly, and differently from the case of NGC~3201, not all the Galactic potentials adopted in this paper seem to represent equally well the stream distribution in the sky. While the axisymmetric models PI and PII predict generally a good match---with model PI describing the stream at positive longitudes even more accurately than model PII---the model PII-0.3-SLOW fails in reproducing the stream in its observed extension: the modeled stream in this case appears quite thick in the ($\ell, b$) plane, and moreover much shorter than the stream found in the observational data. Interestingly, the axisymmetric models capture very well also the proper motions and distance-to-the-Sun trends as a function of longitude, as found by \citet{ibata21} (for  Fj\"{o}rm) and by \citet{palau19}, while the NGC~4590 as reported by \citet{ibata21} (and named "M68-I21" in Fig.~\ref{NGC4590_comp}) tends to be off in all models, and also off when compared to the other observational tracks. As suggested by \citet{mateu22}, it is possible that the stellar stream associated by  \citet{ibata21} to NGC~4590   has indeed a different progenitor.  Finally, the failure of the barred model to reproduce the extension of the NGC~4590 stream, and in particular Fj\"{o}rm,  can be due to the choice of the pattern speed adopted, or of the bar length. We will explore these topics in future work. Here we simply note that given the sensitivity of NGC~4590 stream to the choice of the barred potential, this stream is potentially very interesting for determining the parameters of this latter.

\begin{figure*}
  \begin{center}
    \includegraphics[clip=true, trim = 0mm 0mm 0mm 0mm, width=0.65\columnwidth]{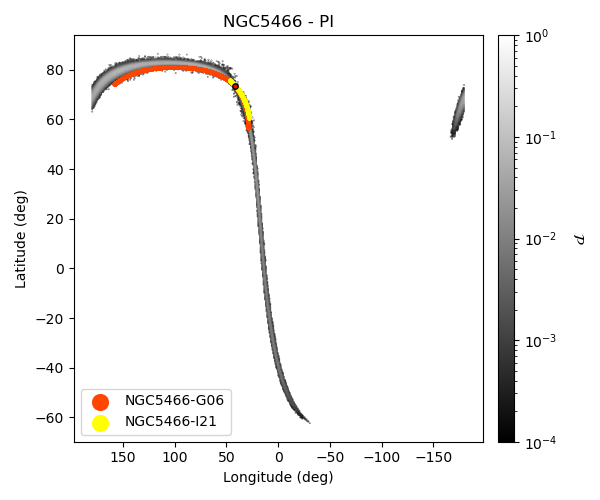}
    \includegraphics[clip=true, trim = 0mm 0mm 0mm 0mm, width=0.65\columnwidth]{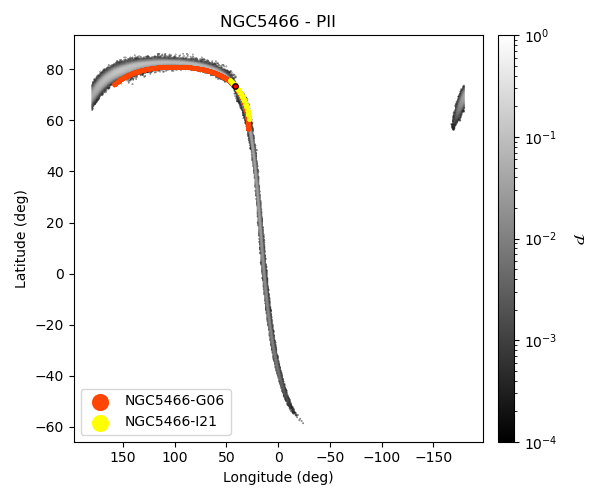}
    \includegraphics[clip=true, trim = 0mm 0mm 0mm 0mm, width=0.65\columnwidth]{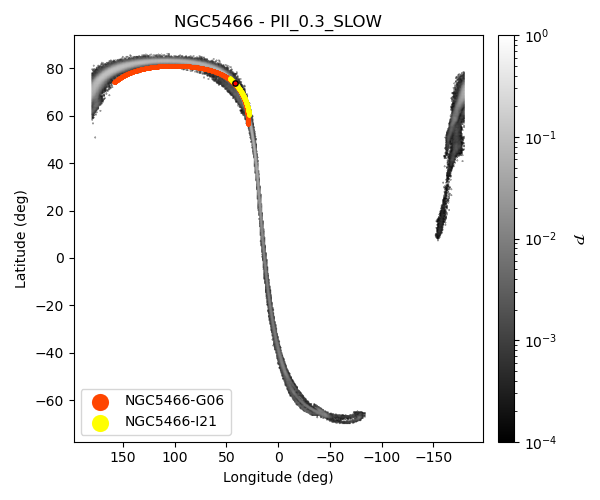}\\
   
    \includegraphics[clip=true, trim = 0mm 0mm 0mm 0mm, width=0.65\columnwidth]{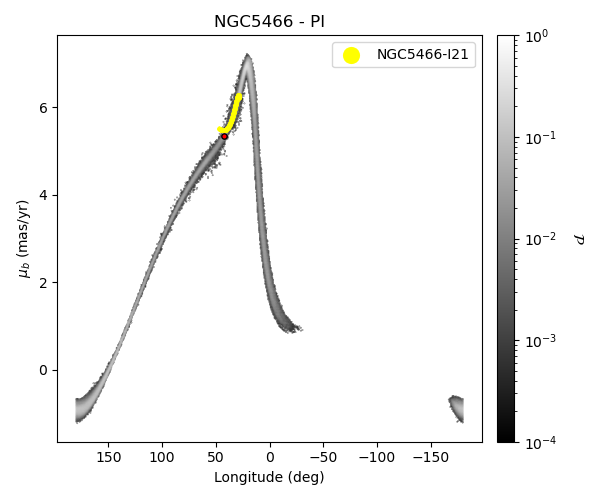}
    \includegraphics[clip=true, trim = 0mm 0mm 0mm 0mm, width=0.65\columnwidth]{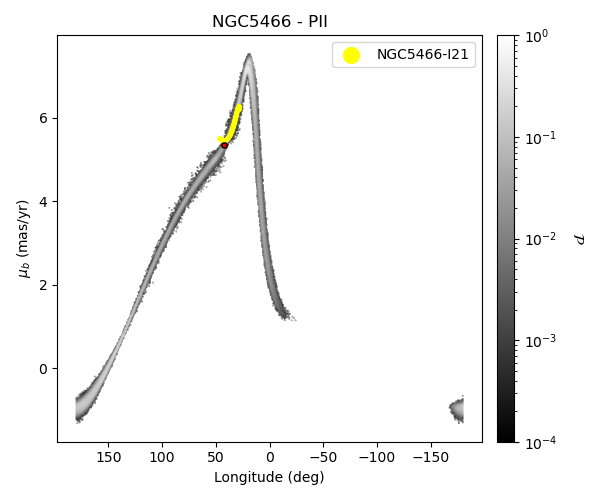}
    \includegraphics[clip=true, trim = 0mm 0mm 0mm 0mm, width=0.65\columnwidth]{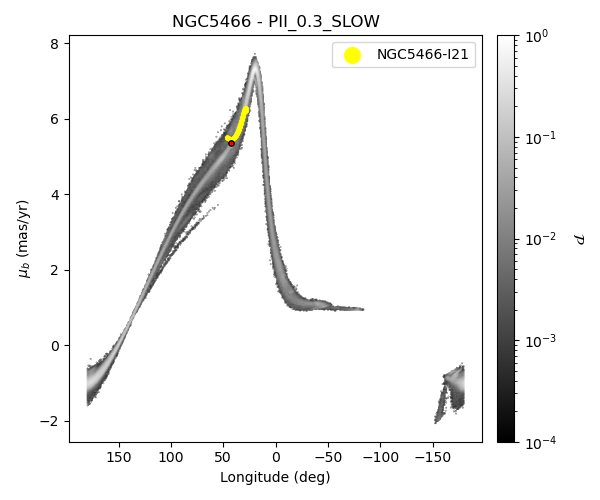}\\
    
    \includegraphics[clip=true, trim = 0mm 0mm 0mm 0mm, width=0.65\columnwidth]{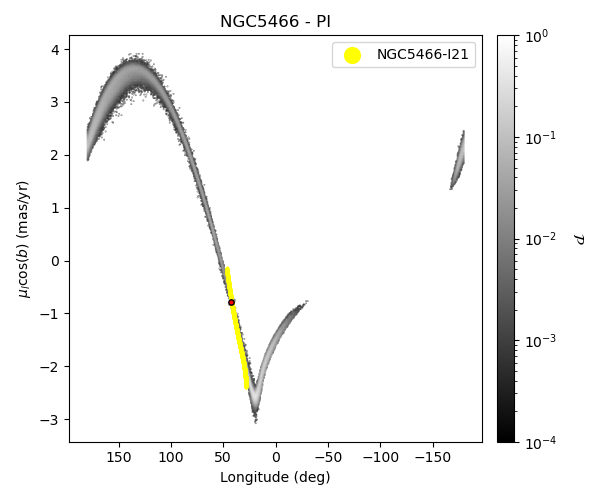}
    \includegraphics[clip=true, trim = 0mm 0mm 0mm 0mm, width=0.65\columnwidth]{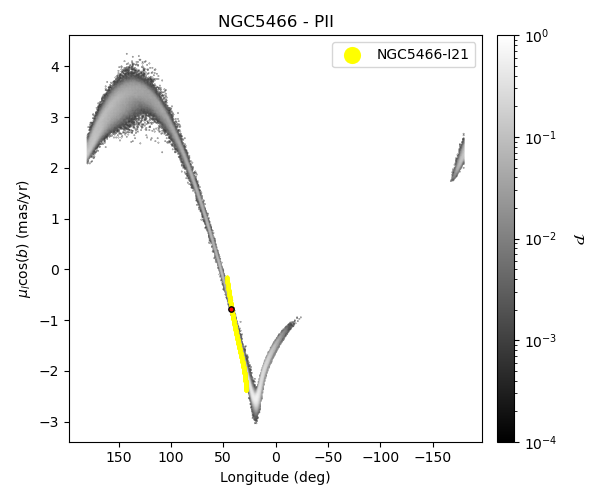}
    \includegraphics[clip=true, trim = 0mm 0mm 0mm 0mm, width=0.65\columnwidth]{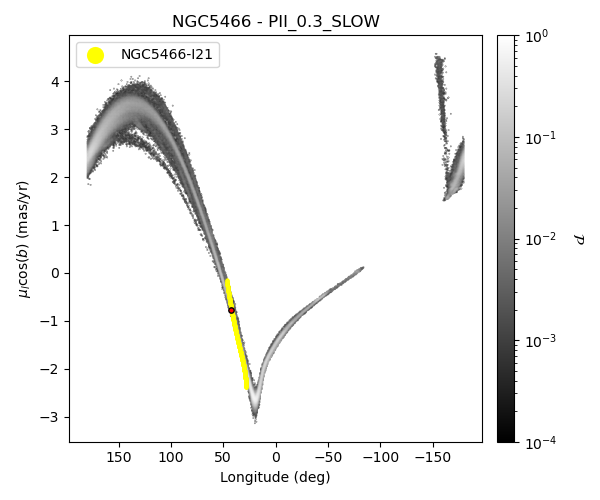}\\
     
    \includegraphics[clip=true, trim = 0mm 0mm 0mm 0mm, width=0.65\columnwidth]{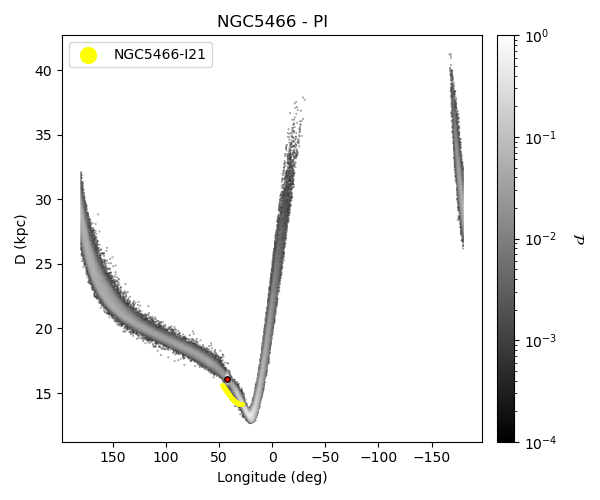}
    \includegraphics[clip=true, trim = 0mm 0mm 0mm 0mm, width=0.65\columnwidth]{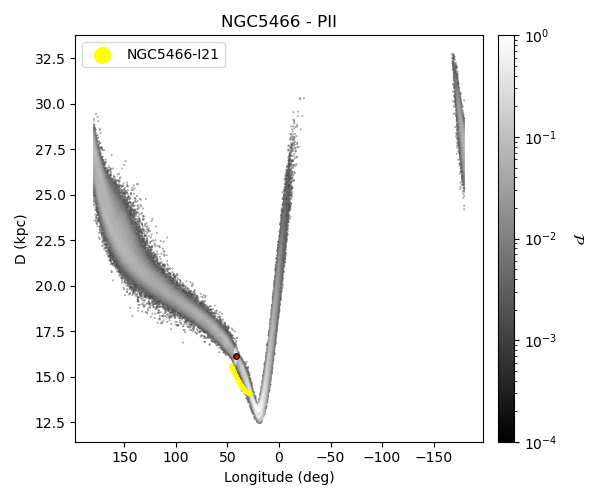}
    \includegraphics[clip=true, trim = 0mm 0mm 0mm 0mm, width=0.65\columnwidth]{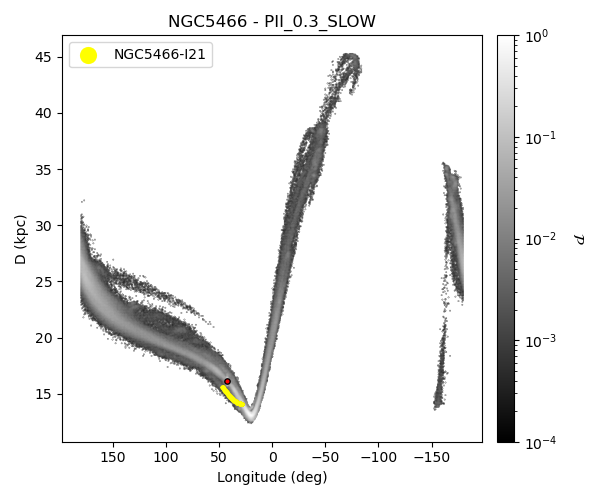}\\
     \end{center}
 \caption{Same as Figs.~\ref{NGC3201_comp} and \ref{NGC4590_comp}, for the case of the NGC~5466 cluster. Model predictions are compared to the tracks available in the \textit{galstreams} library, and which come from the work by  \citet{grillmair06c} (red lines) and \citet{ibata21} (yellow lines). Note that proper motions and distances are available only for the NGC5466-I21 track. Notice that the variations in the streams, specifically the different stripes, originate from the errors considered in the simulations. \label{NGC5466_comp}}
 \end{figure*}

\paragraph{NGC~5466: } NGC~5466 is another cluster known to be surrounded by a thin and very extended stream, as shown by \citet{grillmair06c} and \citet{belokurov06}.  More recently,  \citet{jensen21} have used Gaia~DR2 data to study the stream, finding an extension of about 30 degrees on the sky and a somewhat different orientation than that suggested by \citet{grillmair06c}.  Also \citet{ibata21} confirmed the existence of an elongated stream around this cluster, even if less extended than that found by \citet{grillmair06c}. In Fig.~\ref{NGC5466_comp} we show the comparison of our model predictions to the tracks available in \textit{galstreams}, taken from the works by \citet{grillmair06c}  and \citet{ibata21}. The comparison is excellent for all the Galactic potentials used in this paper. We note that there is a slight offset between the modeled streams and the track by \citet{ibata21}  in the ($\ell-D$) plane (of about 1~kpc at a fixed longitude) which is maybe less evident for the case of the PII-0.3-SLOW potential than in the axisymmetric cases.  We emphasize that all our models predict that the stream should be more extended than that discovered so far in the observations: in particular, there is a portion of the leading tail at $0 < \ell < 42^\circ$ not yet discovered, which lies at distances from the Sun closer or similar to that of the known stream.

\begin{figure*}[h!]
  \begin{center}
 \includegraphics[clip=true, trim = 0mm 0mm 0mm 0mm, width=0.65\columnwidth]{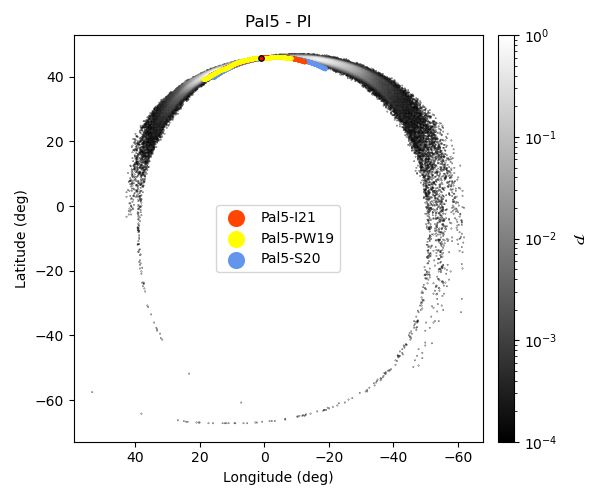}
  \includegraphics[clip=true, trim = 0mm 0mm 0mm 0mm, width=0.65\columnwidth]{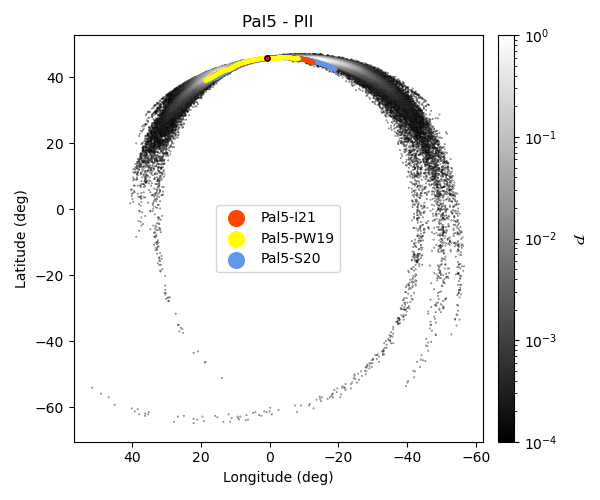}
   \includegraphics[clip=true, trim = 0mm 0mm 0mm 0mm, width=0.65\columnwidth]{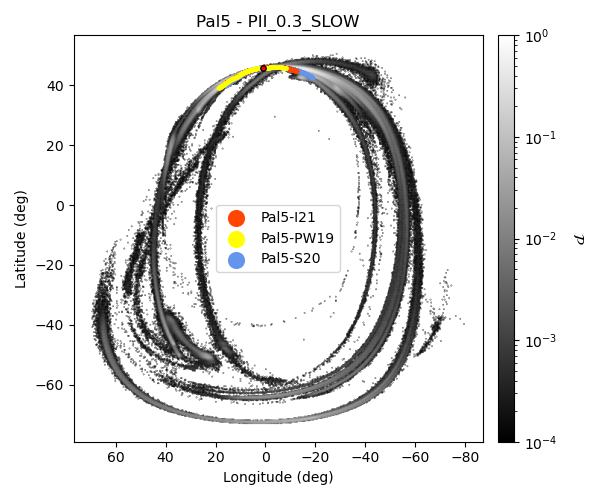}\\
   
  \includegraphics[clip=true, trim = 0mm 0mm 0mm 0mm, width=0.65\columnwidth]{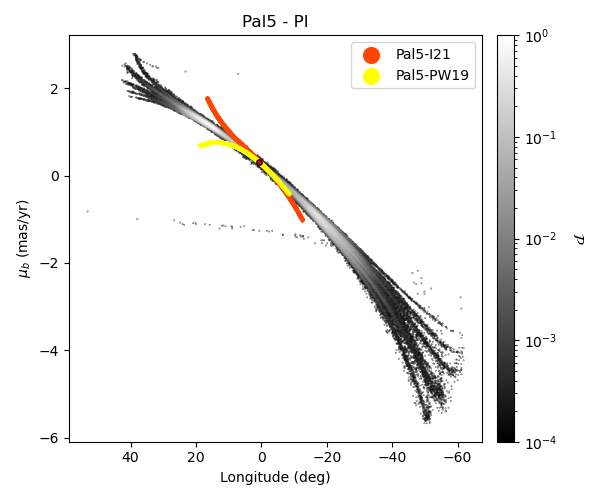}
  \includegraphics[clip=true, trim = 0mm 0mm 0mm 0mm, width=0.65\columnwidth]{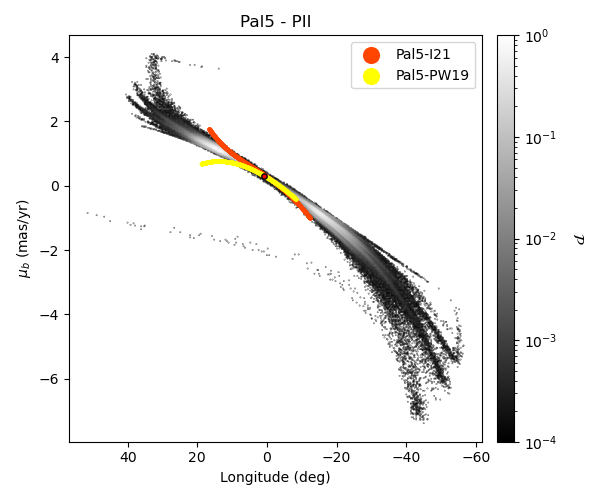}
  \includegraphics[clip=true, trim = 0mm 0mm 0mm 0mm, width=0.65\columnwidth]{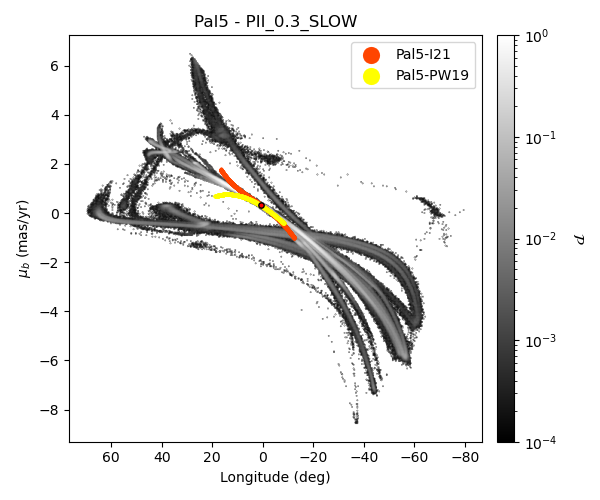}\\
   
  \includegraphics[clip=true, trim = 0mm 0mm 0mm 0mm, width=0.65\columnwidth]{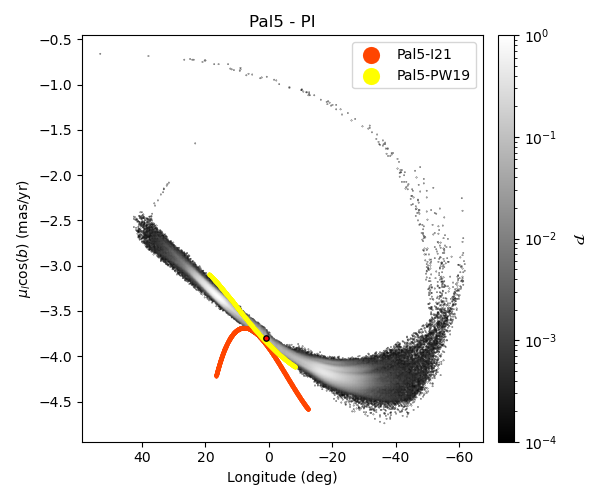}
  \includegraphics[clip=true, trim = 0mm 0mm 0mm 0mm, width=0.65\columnwidth]{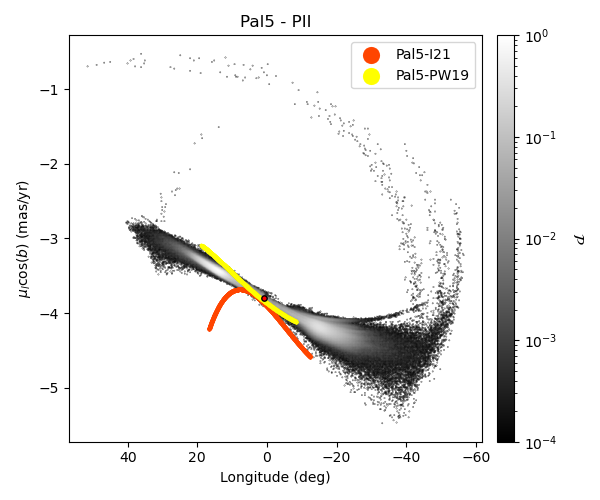}
  \includegraphics[clip=true, trim = 0mm 0mm 0mm 0mm, width=0.65\columnwidth]{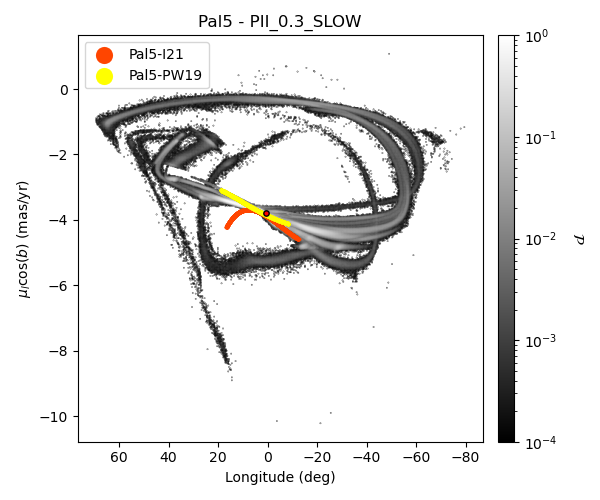}\\
   
  \includegraphics[clip=true, trim = 0mm 0mm 0mm 0mm, width=0.65\columnwidth]{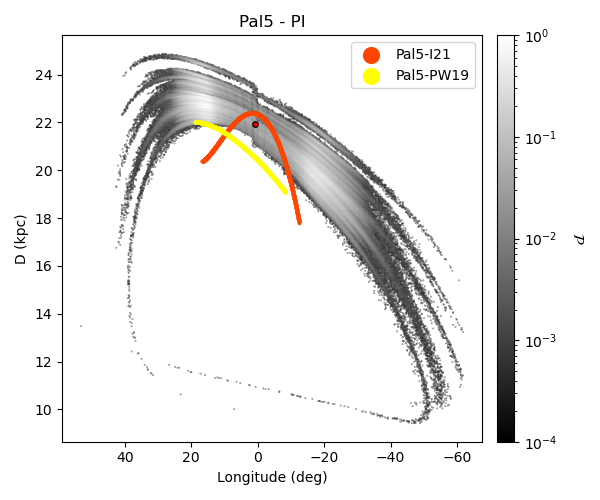}
  \includegraphics[clip=true, trim = 0mm 0mm 0mm 0mm, width=0.65\columnwidth]{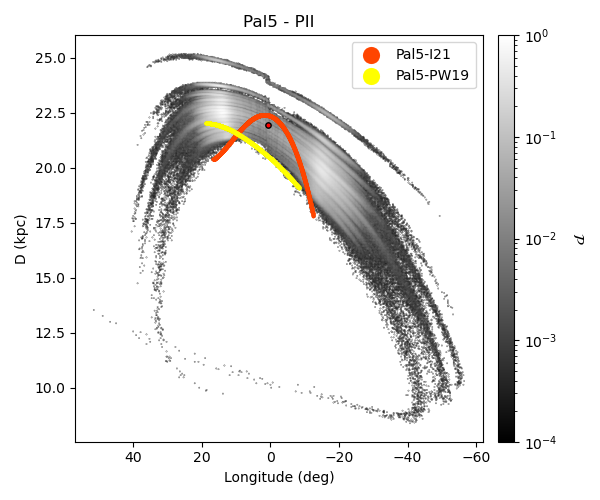}
  \includegraphics[clip=true, trim = 0mm 0mm 0mm 0mm, width=0.65\columnwidth]{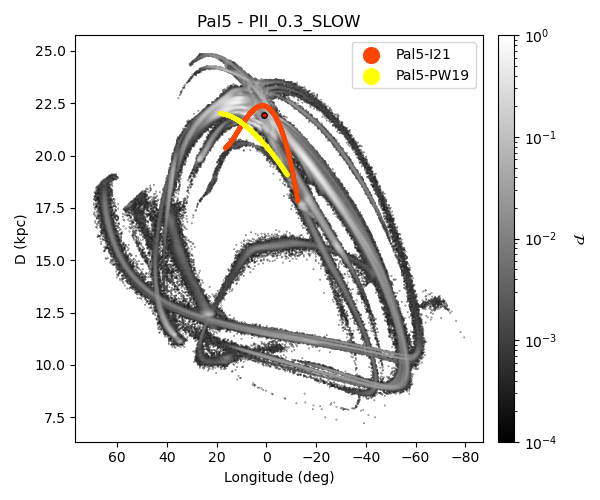}
  \end{center}
 \caption{Same as Figs.~\ref{NGC3201_comp}, \ref{NGC4590_comp} and \ref{NGC5466_comp}, for the case of the Palomar~5 cluster. Model predictions are compared to the tracks available in the \textit{galstreams} library, and which come from the work by \citet{ibata21} (red lines), \citet{pricewhelan19} (yellow-lines) and \citet{starkman20} (blue lines). Note that proper motions and distances are not available for the Pal5-S20 track.\label{Pal5_comp}}
 \end{figure*}

\paragraph{Pal~5: } After about 20 years from the discovery of its tidal tails \citep{odenkirchen01, odenkirchen03}, Pal~5 still represents the prototype cluster surrounded by thin and extended streams of stars. The extension, morphology, kinematics and chemical composition of its tails have been extensively studied, both observationally and numerically \citep{odenkirchen02, rockosi02, dehnen04, koch04, grillmair06, odenkirchen09, mastrobuono12, kupper15, kuzma15, fritz15, ibata16, ishigaki16, thomas16, koch17, ibata17, pearson17, pricewhelan19, starkman20, bonaca20b, ibata21,  philips22, kuzma22}. In Fig.~\ref{Pal5_comp}, we report the comparison of our model predictions to the tracks available for this cluster in \textit{galstreams}, and which come from the work by \citet{pricewhelan19}, \citet{starkman20} and \citet{ibata21}. Projected in the $(\ell, b)$ plane, the observed streams are in excellent agreement with the model predictions, for all the Galactic potentials adopted. Interestingly, all potentials suggest more extended streams than those discovered so far. In the case of the barred potential, we note that the prediction of the stream position in the sky at large angular distances from the cluster center is still highly uncertain. This is due to the uncertainties still affecting the current distance of Pal~5 to the Sun, combined with the impact of the rotating bar, which can be more or less efficient in perturbing the stream depending on the torques experienced by the latter at pericenter---which depend, in turn, on the orbital parameters of the cluster itself. \citet{pearson17} already explored the effect of a rotating bar on the extension and morphology of Pal~5 tails. While we can confirm the \citet{pearson17} findings, that both characteristics are affected by a rotating bar, we also emphasize that both the extension of the leading tail (i.e. portion of the stream at negative longitudes) and its morphology depend on the choice of the bar pattern speed. For example, we do not find a density drop in the leading tail as reported by  \citet{pearson17}. These authors adopted a higher bar pattern speed than the one adopted in this work ($\Omega_b=60$~kms$^{-1}$kpc$^{-1}$ for the example discussed in Fig. 2 of their work, right panel, while $\Omega_b=38$~kms$^{-1}$kpc$^{-1}$ in our PII-0.3-SLOW model). Additionally, taking into account the uncertainties in the cluster distance, proper motions and line-of-sight velocity is also important to have robust predictions on the stream characteristics. Some of our solutions, for example, predict a very extended leading tail, significantly more extended than those found for the axisymmetric potentials.

\section{Conclusion}
In this work, we have presented the first simulated catalogue of \textit{all} Galactic globular clusters for which 6D phase space information, masses and sizes are available. A total of 159~globular clusters has been simulated in three Milky Way-like potentials, modeling  the process of tidal stripping that these clusters have experienced over the past 5~Gyr.  As a result, for all clusters we can predict the distribution of the extra-tidal material in the sky, their proper motions,  distances to the Sun, and line-of-sight velocities. Errors on 6D-phase space  information have been taken into account by generating, for each cluster, 50 complementary simulations, with a Monte-Carlo sampling of the uncertainties.  This catalogue currently contains 24 327 simulations, for a total volume of about 370 Gb. It will be made publicly available\footnote{All data will be available on a dedicated website (\url{http://etidal-project.obspm.fr/})}, and it is intended to provide the community with an instrument to: have a complete view of the expected distribution of globular clusters tidal structures in the sky;  to help the interpretation of recent and future discoveries;  to support the search for new extra-tidal features in the data;  to offer the community a repository of all these models to be compared to other theoretical and numerical predictions, which employ different Galactic potentials and/or gravity laws.

In this first paper, we have presented the distribution in the $(\ell, b)$ plane of all the simulated extra-tidal features. A striking result is the variety of extra-tidal shapes that globular clusters can give rise to. The canonical tidal tails ``\`a la" Palomar~5 are only some of the multiple morphologies these structures can have. Ribbons, bow-ties, padlocks, halo-shapes are also common. This variety of shapes that these stripped stars can show in the sky depends on the characteristics of the cluster orbit, on its current position in the orbit itself and on distance of the extra-tidal material from the Sun. Any search for the left-overs of globular clusters in the field should take into account this richness of distributions and morphologies. 

These simulations have also allowed us to derive an estimate of the expected mass of stars escaped from the clusters in the last 5~Gyr and now in the field. Although approximate, given the limitation of our models, our estimate of the mass lost from the Galactic globular cluster system over the last 5~Gyr is between $2-21\times 10^6 M_{\sun}$ which is comparable to up to half of the total stellar mass found nowadays in the Galactic globular cluster system itself. 

This work is intended to be the first of a series which investigates the properties of globular clusters streams in a variety of realistic Galactic potentials, including the perturbations induced by close dwarf satellites (Sagittarius, Large and Small Magellanic Clouds), as well as more complex and time-varying distributions for the dark matter component. 

\begin{acknowledgements} 

The authors are grateful to Nicolas Leclerc for his help in the construction of the e-TidalGCs Project website. The authors wish to thank H. Baumgardt and E. Vasiliev for making the globular cluster data used in this paper publicly available, as well as C. Mateu for developing the \emph{galstreams} library and making it available to the community. SF and PDM would also like to thank P. Bianchini, P. Bonifacio, R. Ibata, D. Katz, N. Martin, for their interest in this work and their comments. Finally, the authors are grateful to the referee for their suggestions and remarks, which greatly improved the paper and the presentation of the results. 

This work has made use of the computational resources available at the Paris Observatory, as well as those obtained through the DARI grant A1020410154 (PI: P. Di Matteo), and of the \texttt{Astropy}, \texttt{Numpy} and \texttt{Matplotlib} libraries \citep{astropy13, astropy18, harris20array, hunter07}. Salvatore Ferrone and Paola Di Matteo would like to thank the Graduate Program in Astrophysics of  the Paris Sciences et Lettres University for funding this research.
Alessandra Mastrobuono-Battisti acknowledges funding from the European Union’s Horizon 2020 research and innovation program under the Marie Sk\l{}odowska-Curie grant agreement No 895174.

\end{acknowledgements}

\bibliographystyle{aa}
\bibliography{biblio}

\begin{appendix}

\section{Additional Table}
\clearpage
\onecolumn
\tiny
\begin{longtable}{ | l | r | r| r | r | r | r | r | r | r | r | r | r  |} 
\caption{\label{TableIC}Current positions in the sky, proper motions, line-of-sight velocities, distances and relative uncertainties, masses and half-mass radii of all globular clusters analyzed in this study.}\\
\hline\hline
\multicolumn{1}{|c|}{Cluster}& 
\multicolumn{1}{|c|}{$D$} & 
 \multicolumn{1}{|c|}{\rm{err} $D$} & 
 \multicolumn{1}{|c|}{$\alpha$       } & 
 \multicolumn{1}{|c|}{$\delta$       } & 
 \multicolumn{1}{|c|}{$\mu_{\alpha}$} & 
 \multicolumn{1}{|c|}{err $\mu_{\alpha}$} & 
 \multicolumn{1}{|c|}{$\mu_{\delta}$} & 
 \multicolumn{1}{|c|}{err $\mu_{\delta}$} & 
 \multicolumn{1}{|c|}{$\rm v_{\ell os}$  } & 
 \multicolumn{1}{|c|}{err $\rm v_{\ell os}$} & 
 \multicolumn{1}{|c|}{$M_{GC}$         } & 
 \multicolumn{1}{|c|}{$r_{h}$ } \\ 
\multicolumn{1}{|c|}{}& 
\multicolumn{1}{|c|}{kpc       } & 
 \multicolumn{1}{|c|}{kpc       } & 
 \multicolumn{1}{|c|}{degrees   } & 
 \multicolumn{1}{|c|}{degrees   } & 
 \multicolumn{1}{|c|}{mas/yr    } & 
 \multicolumn{1}{|c|}{mas/yr    } & 
 \multicolumn{1}{|c|}{mas/yr    } & 
 \multicolumn{1}{|c|}{mas/yr    } & 
 \multicolumn{1}{|c|}{km/s      } & 
 \multicolumn{1}{|c|}{km/s      } & 
 \multicolumn{1}{|c|}{$M_{\odot}$} & 
 \multicolumn{1}{|c|}{pc        } \\ 
\hline
\endfirsthead
\caption{continued.}\\
\hline\hline
\multicolumn{1}{|c|}{Cluster}& 
\multicolumn{1}{|c|}{$D$} & 
 \multicolumn{1}{|c|}{\rm{err} $D$} & 
 \multicolumn{1}{|c|}{$\alpha$       } & 
 \multicolumn{1}{|c|}{$\delta$       } & 
 \multicolumn{1}{|c|}{$\mu_{\alpha}$} & 
 \multicolumn{1}{|c|}{err $\mu_{\alpha}$} & 
 \multicolumn{1}{|c|}{$\mu_{\delta}$} & 
 \multicolumn{1}{|c|}{err $\mu_{\delta}$} & 
 \multicolumn{1}{|c|}{$\rm v_{\ell os}$  } & 
 \multicolumn{1}{|c|}{err $\rm v_{\ell os}$} & 
 \multicolumn{1}{|c|}{M         } & 
 \multicolumn{1}{|c|}{$r_{h}$ } \\ 
\multicolumn{1}{|c|}{}& 
\multicolumn{1}{|c|}{kpc       } & 
 \multicolumn{1}{|c|}{kpc       } & 
 \multicolumn{1}{|c|}{degrees   } & 
 \multicolumn{1}{|c|}{degrees   } & 
 \multicolumn{1}{|c|}{mas/yr    } & 
 \multicolumn{1}{|c|}{mas/yr    } & 
 \multicolumn{1}{|c|}{mas/yr    } & 
 \multicolumn{1}{|c|}{mas/yr    } & 
 \multicolumn{1}{|c|}{km/s      } & 
 \multicolumn{1}{|c|}{km/s      } & 
 \multicolumn{1}{|c|}{$M_{\odot}$} & 
 \multicolumn{1}{|c|}{pc        } \\ 
\hline
\endhead
\hline
\endfoot
2MASS-GC01    &   3.37 & 0.62 & 272.0909 & -19.8297 &  -1.121 & 0.296 &  -1.881 &  0.235 &  -31.28 &  0.50 &   35100 &  4.70\\ 
2MASS-GC02    &   5.50 & 0.44 & 272.4021 & -20.7789 &   4.000 & 0.900 &  -4.700 &  0.800 & -237.75 & 10.10 &   15800 &  2.85\\ 
AM1           & 118.91 & 3.40 &  58.7596 & -49.6153 &   0.291 & 0.107 &  -0.177 &  0.086 &  118.00 & 14.14 &   19200 & 19.86\\ 
AM4           &  29.01 & 0.94 & 209.0891 & -27.1652 &  -0.291 & 0.445 &  -2.512 &  0.344 &  151.19 &  2.85 &     756 & 15.00\\ 
Arp2          &  28.73 & 0.34 & 292.1838 & -30.3556 &  -2.331 & 0.031 &  -1.475 &  0.029 &  122.64 &  0.29 &   37000 & 18.45\\ 
BH140         &   4.81 & 0.25 & 193.4729 & -67.1773 & -14.848 & 0.024 &   1.224 &  0.024 &   90.30 &  0.35 &   59900 &  9.53\\ 
BH261         &   6.12 & 0.26 & 273.5275 & -28.6350 &   3.566 & 0.043 &  -3.590 &  0.037 &  -45.00 & 15.00 &   22000 &  4.66\\ 
Crater        & 147.23 & 4.27 & 174.0687 & -10.8770 &  -0.059 & 0.125 &  -0.116 &  0.116 &  148.10 &  0.65 &   10800 & 25.74\\ 
Djor1         &   9.88 & 0.65 & 266.8696 & -33.0664 &  -4.693 & 0.046 &  -8.468 &  0.041 & -359.18 &  1.64 &   79700 &  5.57\\ 
Djor2         &   8.76 & 0.18 & 270.4544 & -27.8258 &   0.662 & 0.042 &  -2.983 &  0.037 & -149.75 &  1.10 &  125000 &  5.16\\ 
E3            &   7.88 & 0.25 & 140.2378 & -77.2819 &  -2.727 & 0.027 &   7.083 &  0.027 &   11.71 &  0.34 &    2890 &  6.14\\ 
ESO280-SC06   &  20.95 & 0.65 & 272.2750 & -46.4233 &  -0.688 & 0.039 &  -2.777 &  0.033 &   93.20 &  0.34 &    7800 &  9.65\\ 
ESO452-SC11   &   7.39 & 0.20 & 249.8542 & -28.3992 &  -1.423 & 0.031 &  -6.472 &  0.030 &   16.37 &  0.44 &    8260 &  3.68\\ 
Eridanus      &  84.68 & 2.89 &  66.1856 & -21.1868 &   0.510 & 0.039 &  -0.301 &  0.041 &  -23.15 &  0.73 &   11600 & 17.91\\ 
FSR1716       &   7.43 & 0.27 & 242.6250 & -53.7489 &  -4.354 & 0.033 &  -8.832 &  0.031 &  -30.70 &  0.98 &   64300 &  5.16\\ 
FSR1735       &   9.08 & 0.53 & 253.0442 & -47.0581 &  -4.439 & 0.054 &  -1.534 &  0.048 &  -69.85 &  4.88 &   72300 &  2.97\\ 
FSR1758       &  11.09 & 0.74 & 262.8000 & -39.8080 &  -2.881 & 0.026 &   2.519 &  0.025 &  227.31 &  0.59 &  628000 & 17.04\\ 
HP1           &   7.00 & 0.14 & 262.7717 & -29.9817 &   2.523 & 0.039 & -10.093 &  0.037 &   39.76 &  1.22 &  124000 &  3.74\\ 
IC1257        &  26.59 & 1.43 & 261.7854 &  -7.0931 &  -1.007 & 0.040 &  -1.492 &  0.032 & -137.97 &  2.04 &   18100 &  5.54\\ 
IC1276        &   4.55 & 0.25 & 272.6844 &  -7.2076 &  -2.553 & 0.026 &  -4.568 &  0.026 &  155.06 &  0.69 &   73900 &  5.21\\ 
IC4499        &  18.89 & 0.25 & 225.0772 & -82.2138 &   0.466 & 0.025 &  -0.489 &  0.025 &   38.41 &  0.31 &  155000 & 14.96\\ 
Laevens3      &  61.77 & 1.65 & 316.7267 &  14.9805 &   0.172 & 0.101 &  -0.666 &  0.080 &  -70.30 &  0.82 &    2120 &  9.46\\ 
Liller1       &   8.06 & 0.34 & 263.3523 & -33.3896 &  -5.403 & 0.109 &  -7.431 &  0.077 &   60.36 &  2.44 &  915000 &  2.01\\ 
Lynga7        &   7.90 & 0.16 & 242.7652 & -55.3178 &  -3.851 & 0.027 &  -7.050 &  0.027 &   17.86 &  0.83 &   79600 &  5.16\\ 
NGC104        &   4.52 & 0.03 &   6.0238 & -72.0813 &   5.252 & 0.021 &  -2.551 &  0.021 &  -17.45 &  0.16 &  895000 &  6.30\\ 
NGC1261       &  16.40 & 0.19 &  48.0675 & -55.2162 &   1.596 & 0.025 &  -2.064 &  0.025 &   71.34 &  0.21 &  182000 &  5.23\\ 
NGC1851       &  11.95 & 0.13 &  78.5282 & -40.0466 &   2.145 & 0.024 &  -0.650 &  0.024 &  321.40 &  1.55 &  318000 &  2.90\\ 
NGC1904       &  13.08 & 0.18 &  81.0458 & -24.5244 &   2.469 & 0.025 &  -1.594 &  0.025 &  205.76 &  0.20 &  139000 &  3.21\\ 
NGC2298       &   9.83 & 0.17 & 102.2475 & -36.0053 &   3.320 & 0.025 &  -2.175 &  0.026 &  147.15 &  0.57 &   55800 &  3.31\\ 
NGC2419       &  88.47 & 2.40 & 114.5353 &  38.8819 &   0.007 & 0.028 &  -0.523 &  0.026 &  -21.10 &  0.31 &  971000 & 26.50\\ 
NGC2808       &  10.06 & 0.11 & 138.0129 & -64.8635 &   0.994 & 0.024 &   0.273 &  0.024 &  103.57 &  0.27 &  864000 &  3.89\\ 
NGC288        &   8.99 & 0.09 &  13.1885 & -26.5826 &   4.164 & 0.024 &  -5.705 &  0.024 &  -44.45 &  0.13 &   93400 &  8.37\\ 
NGC3201       &   4.74 & 0.04 & 154.4034 & -46.4125 &   8.348 & 0.022 &  -1.958 &  0.022 &  493.65 &  0.21 &  160000 &  6.78\\ 
NGC362        &   8.83 & 0.10 &  15.8094 & -70.8488 &   6.694 & 0.025 &  -2.535 &  0.024 &  223.12 &  0.28 &  284000 &  3.79\\ 
NGC4147       &  18.54 & 0.21 & 182.5263 &  18.5426 &  -1.707 & 0.027 &  -2.090 &  0.027 &  179.35 &  0.31 &   39000 &  4.03\\ 
NGC4372       &   5.71 & 0.21 & 186.4391 & -72.6591 &  -6.409 & 0.024 &   3.297 &  0.024 &   75.59 &  0.30 &  198000 &  8.53\\ 
NGC4590       &  10.40 & 0.10 & 189.8666 & -26.7441 &  -2.739 & 0.024 &   1.779 &  0.024 &  -93.11 &  0.18 &  122000 &  7.58\\ 
NGC4833       &   6.48 & 0.08 & 194.8913 & -70.8765 &  -8.377 & 0.025 &  -0.963 &  0.025 &  201.99 &  0.40 &  206000 &  4.76\\ 
NGC5024       &  18.50 & 0.18 & 198.2302 &  18.1682 &  -0.133 & 0.024 &  -1.331 &  0.024 &  -63.37 &  0.25 &  455000 & 10.18\\ 
NGC5053       &  17.54 & 0.23 & 199.1129 &  17.7003 &  -0.329 & 0.025 &  -1.213 &  0.025 &   42.82 &  0.25 &   74200 & 17.31\\ 
NGC5139       &   5.43 & 0.05 & 201.6970 & -47.4795 &  -3.250 & 0.022 &  -6.746 &  0.022 &  232.78 &  0.21 & 3640000 & 10.36\\ 
NGC5272       &  10.18 & 0.08 & 205.5484 &  28.3773 &  -0.152 & 0.023 &  -2.670 &  0.022 & -147.20 &  0.27 &  406000 &  6.34\\ 
NGC5286       &  11.10 & 0.14 & 206.6117 & -51.3742 &   0.198 & 0.025 &  -0.153 &  0.025 &   62.38 &  0.40 &  353000 &  3.79\\ 
NGC5466       &  16.12 & 0.16 & 211.3637 &  28.5344 &  -5.342 & 0.025 &  -0.822 &  0.024 &  106.82 &  0.20 &   59800 & 14.03\\ 
NGC5634       &  25.96 & 0.62 & 217.4053 &  -5.9764 &  -1.692 & 0.027 &  -1.478 &  0.026 &  -16.07 &  0.60 &  228000 &  7.39\\ 
NGC5694       &  34.84 & 0.74 & 219.9012 & -26.5388 &  -0.464 & 0.029 &  -1.105 &  0.029 & -139.55 &  0.49 &  317000 &  4.86\\ 
NGC5824       &  31.71 & 0.60 & 225.9942 & -33.0681 &  -1.189 & 0.026 &  -2.234 &  0.026 &  -25.24 &  0.52 &  762000 &  6.51\\ 
NGC5897       &  12.55 & 0.24 & 229.3517 & -21.0101 &  -5.422 & 0.025 &  -3.393 &  0.025 &  101.31 &  0.22 &  157000 & 10.99\\ 
NGC5904       &   7.48 & 0.06 & 229.6384 &   2.0810 &   4.086 & 0.023 &  -9.870 &  0.023 &   53.50 &  0.25 &  394000 &  5.68\\ 
NGC5927       &   8.27 & 0.11 & 232.0029 & -50.6730 &  -5.056 & 0.025 &  -3.217 &  0.025 & -104.09 &  0.28 &  275000 &  5.28\\ 
NGC5946       &   9.64 & 0.51 & 233.8691 & -50.6597 &  -5.331 & 0.028 &  -1.657 &  0.027 &  137.60 &  0.94 &   93100 &  2.59\\ 
NGC5986       &  10.54 & 0.13 & 236.5125 & -37.7864 &  -4.192 & 0.026 &  -4.568 &  0.026 &  101.18 &  0.43 &  334000 &  4.25\\ 
NGC6093       &  10.34 & 0.12 & 244.2600 & -22.9761 &  -2.934 & 0.027 &  -5.578 &  0.026 &   10.93 &  0.39 &  338000 &  2.62\\ 
NGC6101       &  14.45 & 0.19 & 246.4505 & -72.2022 &   1.756 & 0.024 &  -0.258 &  0.025 &  366.33 &  0.32 &  178000 & 14.06\\ 
NGC6121       &   1.85 & 0.02 & 245.8967 & -26.5257 & -12.514 & 0.023 & -19.022 &  0.023 &   71.21 &  0.15 &   87100 &  3.69\\ 
NGC6139       &  10.04 & 0.45 & 246.9185 & -38.8488 &  -6.081 & 0.027 &  -2.711 &  0.026 &   24.41 &  0.95 &  323000 &  2.47\\ 
NGC6144       &   8.15 & 0.13 & 246.8078 & -26.0235 &  -1.744 & 0.026 &  -2.607 &  0.026 &  194.79 &  0.58 &   79200 &  4.91\\ 
NGC6171       &   5.63 & 0.08 & 248.1328 & -13.0538 &  -1.939 & 0.025 &  -5.979 &  0.025 &  -34.71 &  0.18 &   74900 &  3.94\\ 
NGC6205       &   7.42 & 0.08 & 250.4218 &  36.4599 &  -3.149 & 0.023 &  -2.574 &  0.023 & -244.90 &  0.30 &  545000 &  5.26\\ 
NGC6218       &   5.11 & 0.05 & 251.8091 &  -1.9485 &  -0.191 & 0.024 &  -6.802 &  0.024 &  -41.67 &  0.14 &  107000 &  4.05\\ 
NGC6229       &  30.11 & 0.47 & 251.7452 &  47.5278 &  -1.171 & 0.026 &  -0.467 &  0.027 & -137.89 &  0.71 &  286000 &  4.41\\ 
NGC6235       &  11.94 & 0.38 & 253.3557 & -22.1774 &  -3.931 & 0.027 &  -7.587 &  0.027 &  126.68 &  0.33 &  107000 &  4.78\\ 
NGC6254       &   5.07 & 0.06 & 254.2877 &  -4.1003 &  -4.758 & 0.024 &  -6.597 &  0.024 &   74.21 &  0.23 &  205000 &  4.81\\ 
NGC6256       &   7.24 & 0.29 & 254.8861 & -37.1210 &  -3.715 & 0.031 &  -1.637 &  0.030 &  -99.75 &  0.66 &  125000 &  4.82\\ 
NGC6266       &   6.41 & 0.10 & 255.3042 & -30.1134 &  -4.978 & 0.026 &  -2.947 &  0.026 &  -73.98 &  0.67 &  610000 &  2.43\\ 
NGC6273       &   8.34 & 0.16 & 255.6575 & -26.2680 &  -3.249 & 0.026 &   1.660 &  0.025 &  145.54 &  0.59 &  697000 &  4.21\\ 
NGC6284       &  14.21 & 0.42 & 256.1201 & -24.7648 &  -3.200 & 0.029 &  -2.002 &  0.028 &   28.62 &  0.73 &  129000 &  3.78\\ 
NGC6287       &   7.93 & 0.37 & 256.2889 & -22.7080 &  -5.010 & 0.029 &  -1.883 &  0.028 & -294.74 &  1.65 &  145000 &  3.65\\ 
NGC6293       &   9.19 & 0.28 & 257.5425 & -26.5821 &   0.870 & 0.028 &  -4.326 &  0.028 & -143.66 &  0.39 &  205000 &  4.05\\ 
NGC6304       &   6.15 & 0.15 & 258.6344 & -29.4620 &  -4.070 & 0.029 &  -1.088 &  0.028 & -108.62 &  0.39 &  126000 &  4.26\\ 
NGC6316       &  11.15 & 0.39 & 259.1554 & -28.1401 &  -4.969 & 0.031 &  -4.592 &  0.030 &   99.65 &  0.84 &  318000 &  4.77\\ 
NGC6325       &   7.53 & 0.32 & 259.4963 & -23.7677 &  -8.289 & 0.030 &  -9.000 &  0.029 &   29.54 &  0.58 &   58900 &  2.05\\ 
NGC6333       &   8.30 & 0.14 & 259.7991 & -18.5163 &  -2.180 & 0.026 &  -3.222 &  0.026 &  310.75 &  2.12 &  323000 &  4.17\\ 
NGC6341       &   8.50 & 0.07 & 259.2808 &  43.1359 &  -4.935 & 0.024 &  -0.625 &  0.024 & -120.55 &  0.27 &  352000 &  4.49\\ 
NGC6342       &   8.01 & 0.23 & 260.2916 & -19.5877 &  -2.903 & 0.027 &  -7.116 &  0.026 &  115.75 &  0.90 &   42200 &  2.06\\ 
NGC6352       &   5.54 & 0.07 & 261.3713 & -48.4222 &  -2.158 & 0.025 &  -4.447 &  0.025 & -125.63 &  1.01 &   64700 &  4.56\\ 
NGC6355       &   8.65 & 0.22 & 260.9935 & -26.3528 &  -4.738 & 0.031 &  -0.572 &  0.030 & -195.85 &  0.55 &  101000 &  3.55\\ 
NGC6356       &  15.66 & 0.92 & 260.8958 & -17.8130 &  -3.750 & 0.026 &  -3.392 &  0.026 &   48.18 &  1.82 &  600000 &  6.86\\ 
NGC6362       &   7.65 & 0.07 & 262.9791 & -67.0483 &  -5.506 & 0.024 &  -4.763 &  0.024 &  -14.58 &  0.18 &  127000 &  7.23\\ 
NGC6366       &   3.44 & 0.05 & 261.9344 &  -5.0799 &  -0.332 & 0.025 &  -5.160 &  0.024 & -120.65 &  0.19 &   37600 &  5.56\\ 
NGC6380       &   9.61 & 0.30 & 263.6186 & -39.0695 &  -2.183 & 0.031 &  -3.233 &  0.030 &   -1.48 &  0.73 &  334000 &  4.40\\ 
NGC6388       &  11.17 & 0.16 & 264.0718 & -44.7355 &  -1.316 & 0.026 &  -2.709 &  0.026 &   83.11 &  0.45 & 1250000 &  4.34\\ 
NGC6397       &   2.48 & 0.02 & 265.1754 & -53.6743 &   3.260 & 0.023 & -17.664 &  0.022 &   18.51 &  0.08 &   96600 &  3.90\\ 
NGC6401       &   8.06 & 0.24 & 264.6522 & -23.9096 &  -2.748 & 0.035 &   1.444 &  0.034 & -105.44 &  2.50 &  145000 &  3.28\\ 
NGC6402       &   9.14 & 0.25 & 264.4007 &  -3.2459 &  -3.590 & 0.025 &  -5.059 &  0.025 &  -60.71 &  0.45 &  592000 &  5.14\\ 
NGC6426       &  20.71 & 0.35 & 266.2280 &   3.1701 &  -1.828 & 0.026 &  -2.999 &  0.026 & -210.51 &  0.51 &   71700 &  8.00\\ 
NGC6440       &   8.25 & 0.24 & 267.2202 & -20.3604 &  -1.187 & 0.036 &  -4.020 &  0.035 &  -69.39 &  0.93 &  489000 &  2.14\\ 
NGC6441       &  12.73 & 0.16 & 267.5544 & -37.0514 &  -2.551 & 0.028 &  -5.348 &  0.028 &   18.47 &  0.56 & 1320000 &  3.47\\ 
NGC6453       &  10.07 & 0.22 & 267.7155 & -34.5985 &   0.203 & 0.036 &  -5.934 &  0.037 &  -99.23 &  1.24 &  165000 &  3.85\\ 
NGC6496       &   9.64 & 0.15 & 269.7654 & -44.2659 &  -3.060 & 0.027 &  -9.271 &  0.026 & -134.72 &  0.26 &   68900 &  6.42\\ 
NGC6517       &   9.23 & 0.56 & 270.4608 &  -8.9588 &  -1.551 & 0.029 &  -4.470 &  0.028 &  -35.06 &  1.65 &  195000 &  2.29\\ 
NGC6522       &   7.29 & 0.21 & 270.8920 & -30.0340 &   2.566 & 0.039 &  -6.438 &  0.036 &  -15.23 &  0.49 &  211000 &  3.08\\ 
NGC6528       &   7.83 & 0.24 & 271.2067 & -30.0558 &  -2.157 & 0.043 &  -5.649 &  0.039 &  211.86 &  0.43 &   56700 &  2.73\\ 
NGC6535       &   6.36 & 0.12 & 270.9604 &  -0.2976 &  -4.214 & 0.027 &  -2.939 &  0.026 & -214.85 &  0.46 &   21900 &  3.65\\ 
NGC6539       &   8.16 & 0.39 & 271.2073 &  -7.5859 &  -6.896 & 0.026 &  -3.537 &  0.026 &   35.19 &  0.50 &  209000 &  5.18\\ 
NGC6540       &   5.91 & 0.27 & 271.5357 & -27.7653 &  -3.702 & 0.032 &  -2.791 &  0.032 &  -16.50 &  0.78 &   34500 &  5.32\\ 
NGC6541       &   7.61 & 0.10 & 272.0098 & -43.7149 &   0.287 & 0.025 &  -8.847 &  0.025 & -163.97 &  0.46 &  293000 &  4.34\\ 
NGC6544       &   2.58 & 0.06 & 271.8338 & -24.9982 &  -2.304 & 0.031 & -18.604 &  0.030 &  -38.46 &  0.67 &   91400 &  2.07\\ 
NGC6553       &   5.33 & 0.13 & 272.3153 & -25.9078 &   0.344 & 0.030 &  -0.454 &  0.029 &   -0.27 &  0.34 &  285000 &  4.56\\ 
NGC6558       &   7.47 & 0.29 & 272.5740 & -31.7645 &  -1.720 & 0.036 &  -4.144 &  0.034 & -195.12 &  0.73 &   26500 &  1.70\\ 
NGC6569       &  10.53 & 0.26 & 273.4117 & -31.8269 &  -4.125 & 0.028 &  -7.354 &  0.028 &  -49.83 &  0.50 &  236000 &  3.85\\ 
NGC6584       &  13.61 & 0.17 & 274.6566 & -52.2158 &  -0.090 & 0.026 &  -7.202 &  0.025 &  260.64 &  1.58 &  102000 &  5.37\\ 
NGC6624       &   8.02 & 0.11 & 275.9188 & -30.3610 &   0.124 & 0.029 &  -6.936 &  0.029 &   54.79 &  0.40 &  156000 &  3.69\\ 
NGC6626       &   5.37 & 0.10 & 276.1370 & -24.8698 &  -0.278 & 0.028 &  -8.922 &  0.028 &   11.11 &  0.60 &  299000 &  2.26\\ 
NGC6637       &   8.90 & 0.10 & 277.8463 & -32.3481 &  -5.034 & 0.028 &  -5.832 &  0.028 &   47.48 &  1.00 &  155000 &  3.69\\ 
NGC6638       &   9.78 & 0.34 & 277.7337 & -25.4975 &  -2.518 & 0.029 &  -4.076 &  0.029 &    8.63 &  2.00 &  118000 &  2.20\\ 
NGC6642       &   8.05 & 0.20 & 277.9760 & -23.4756 &  -0.173 & 0.030 &  -3.892 &  0.030 &  -60.61 &  1.35 &   34400 &  1.51\\ 
NGC6652       &   9.46 & 0.14 & 278.9401 & -32.9907 &  -5.484 & 0.027 &  -4.274 &  0.027 &  -95.37 &  0.86 &   48100 &  1.96\\ 
NGC6656       &   3.30 & 0.04 & 279.0998 & -23.9047 &   9.851 & 0.023 &  -5.617 &  0.023 & -148.72 &  0.78 &  476000 &  5.29\\ 
NGC6681       &   9.36 & 0.11 & 280.8032 & -32.2921 &   1.431 & 0.027 &  -4.744 &  0.026 &  216.62 &  0.84 &  116000 &  2.89\\ 
NGC6712       &   7.38 & 0.24 & 283.2680 &  -8.7060 &   3.363 & 0.027 &  -4.436 &  0.027 & -107.45 &  0.29 &   96300 &  3.21\\ 
NGC6715       &  26.28 & 0.33 & 283.7639 & -30.4799 &  -2.679 & 0.025 &  -1.387 &  0.025 &  143.13 &  0.43 & 1780000 &  5.20\\ 
NGC6717       &   7.52 & 0.13 & 283.7752 & -22.7015 &  -3.125 & 0.027 &  -5.008 &  0.027 &   30.25 &  0.90 &   35800 &  4.23\\ 
NGC6723       &   8.27 & 0.10 & 284.8881 & -36.6322 &   1.028 & 0.025 &  -2.418 &  0.025 &  -94.39 &  0.26 &  177000 &  5.06\\ 
NGC6749       &   7.59 & 0.21 & 286.3141 &   1.8998 &  -2.829 & 0.028 &  -6.006 &  0.027 &  -58.44 &  0.96 &  211000 &  7.09\\ 
NGC6752       &   4.12 & 0.04 & 287.7171 & -59.9846 &  -3.161 & 0.022 &  -4.027 &  0.022 &  -26.01 &  0.12 &  276000 &  5.27\\ 
NGC6760       &   8.41 & 0.43 & 287.8003 &   1.0305 &  -1.107 & 0.026 &  -3.615 &  0.026 &   -2.37 &  1.27 &  269000 &  5.22\\ 
NGC6779       &  10.43 & 0.14 & 289.1482 &  30.1835 &  -2.018 & 0.025 &   1.618 &  0.025 & -136.97 &  0.45 &  186000 &  4.51\\ 
NGC6809       &   5.35 & 0.05 & 294.9988 & -30.9647 &  -3.432 & 0.024 &  -9.311 &  0.024 &  174.70 &  0.17 &  193000 &  6.95\\ 
NGC6838       &   4.00 & 0.05 & 298.4437 &  18.7792 &  -3.416 & 0.025 &  -2.656 &  0.024 &  -22.72 &  0.20 &   65600 &  6.57\\ 
NGC6864       &  20.52 & 0.45 & 301.5198 & -21.9212 &  -0.598 & 0.026 &  -2.810 &  0.026 & -189.08 &  1.12 &  370000 &  2.96\\ 
NGC6934       &  15.72 & 0.17 & 308.5474 &   7.4045 &  -2.655 & 0.026 &  -4.689 &  0.026 & -406.22 &  0.73 &  136000 &  5.16\\ 
NGC6981       &  16.66 & 0.18 & 313.3654 & -12.5373 &  -1.274 & 0.026 &  -3.361 &  0.026 & -331.39 &  1.47 &   68900 &  5.96\\ 
NGC7006       &  39.32 & 0.56 & 315.3726 &  16.1873 &  -0.128 & 0.027 &  -0.633 &  0.027 & -383.47 &  0.73 &  136000 &  6.99\\ 
NGC7078       &  10.71 & 0.10 & 322.4930 &  12.1670 &  -0.659 & 0.024 &  -3.803 &  0.024 & -106.84 &  0.30 &  633000 &  4.30\\ 
NGC7089       &  11.69 & 0.11 & 323.3626 &  -0.8233 &   3.435 & 0.025 &  -2.159 &  0.024 &   -3.78 &  0.30 &  620000 &  4.77\\ 
NGC7099       &   8.46 & 0.09 & 325.0921 & -23.1799 &  -0.737 & 0.025 &  -7.299 &  0.024 & -185.19 &  0.17 &  143000 &  4.99\\ 
NGC7492       &  24.39 & 0.57 & 347.1112 & -15.6115 &   0.756 & 0.028 &  -2.320 &  0.028 & -176.70 &  0.27 &   26600 &  9.89\\ 
Pal1          &  11.18 & 0.32 &  53.3335 &  79.5811 &  -0.252 & 0.034 &   0.007 &  0.037 &  -75.72 &  0.29 &    1030 &  3.56\\ 
Pal10         &   8.94 & 1.18 & 289.5069 &  18.5790 &  -4.322 & 0.029 &  -7.173 &  0.029 &  -31.70 &  0.23 &  162000 &  6.33\\ 
Pal11         &  14.02 & 0.51 & 296.3100 &  -8.0072 &  -1.766 & 0.030 &  -4.971 &  0.028 &  -67.64 &  0.76 &   11900 &  7.72\\ 
Pal12         &  18.49 & 0.30 & 326.6618 & -21.2526 &  -3.220 & 0.029 &  -3.333 &  0.028 &   27.91 &  0.28 &    6270 & 10.52\\ 
Pal13         &  23.48 & 0.40 & 346.6852 &  12.7715 &   1.748 & 0.049 &   0.104 &  0.047 &   25.30 &  0.22 &    3020 & 16.95\\ 
Pal14         &  73.58 & 1.63 & 242.7525 &  14.9578 &  -0.463 & 0.038 &  -0.413 &  0.038 &   72.30 &  0.14 &   18900 & 36.70\\ 
Pal15         &  44.10 & 1.14 & 254.9626 &  -0.5390 &  -0.592 & 0.037 &  -0.901 &  0.034 &   72.27 &  1.74 &   50900 & 26.86\\ 
Pal2          &  26.17 & 1.28 &  71.5246 &  31.3815 &   1.045 & 0.034 &  -1.522 &  0.031 & -135.97 &  1.55 &  231000 &  8.06\\ 
Pal3          &  94.84 & 3.23 & 151.3816 &   0.0717 &   0.086 & 0.060 &  -0.148 &  0.071 &   94.04 &  0.80 &   18900 & 27.44\\ 
Pal4          & 101.39 & 2.57 & 172.3183 &  28.9734 &  -0.188 & 0.042 &  -0.476 &  0.041 &   72.40 &  0.24 &   12900 & 21.30\\ 
Pal5          &  21.94 & 0.51 & 229.0192 &  -0.1210 &  -2.730 & 0.028 &  -2.654 &  0.027 &  -58.61 &  0.15 &    9980 & 27.64\\ 
Pal6          &   7.05 & 0.45 & 265.9258 & -26.2250 &  -9.222 & 0.038 &  -5.347 &  0.036 &  177.00 &  1.35 &   94500 &  2.89\\ 
Pal8          &  11.32 & 0.63 & 280.3773 & -19.8289 &  -1.987 & 0.027 &  -5.694 &  0.027 &  -31.54 &  0.21 &   67400 &  5.86\\ 
Pyxis         &  36.53 & 0.66 & 136.9869 & -37.2266 &   1.030 & 0.032 &   0.138 &  0.035 &   40.46 &  0.21 &   24600 & 22.83\\ 
Rup106        &  20.71 & 0.36 & 189.6675 & -51.1503 &  -1.254 & 0.026 &   0.401 &  0.026 &  -38.36 &  0.26 &   34200 & 11.57\\ 
SagittariusII &  66.53 & 1.56 & 298.1647 & -22.0653 &  -0.804 & 0.044 &  -0.882 &  0.028 & -175.73 &  0.37 &   18500 & 39.31\\ 
Ter1          &   5.67 & 0.17 & 263.9467 & -30.4818 &  -2.806 & 0.055 &  -4.861 &  0.055 &   56.75 &  1.61 &  150000 &  2.15\\ 
Ter10         &  10.21 & 0.40 & 270.7408 & -26.0669 &  -6.827 & 0.059 &  -2.588 &  0.050 &  211.37 &  2.27 &  302000 &  4.60\\ 
Ter12         &   5.17 & 0.38 & 273.0658 & -22.7419 &  -6.222 & 0.037 &  -3.052 &  0.034 &   95.61 &  1.21 &   87200 &  3.28\\ 
Ter2          &   7.75 & 0.33 & 261.8879 & -30.8023 &  -2.170 & 0.041 &  -6.263 &  0.038 &  134.56 &  0.96 &  136000 &  4.16\\ 
Ter3          &   7.64 & 0.31 & 247.1625 & -35.3398 &  -5.577 & 0.027 &  -1.760 &  0.026 & -135.76 &  0.57 &   40400 &  7.19\\ 
Ter4          &   7.59 & 0.31 & 262.6625 & -31.5955 &  -5.462 & 0.060 &  -3.711 &  0.048 &  -48.96 &  1.57 &  200000 &  6.06\\ 
Ter5          &   6.62 & 0.15 & 267.0202 & -24.7791 &  -1.989 & 0.068 &  -5.243 &  0.066 &  -82.57 &  0.73 &  935000 &  3.77\\ 
Ter6          &   7.27 & 0.35 & 267.6932 & -31.2754 &  -4.979 & 0.048 &  -7.431 &  0.039 &  136.45 &  1.50 &  104000 &  1.33\\ 
Ter7          &  24.28 & 0.49 & 289.4330 & -34.6577 &  -3.002 & 0.029 &  -1.651 &  0.029 &  159.85 &  0.14 &   24000 & 13.21\\ 
Ter8          &  27.54 & 0.42 & 295.4350 & -33.9995 &  -2.496 & 0.027 &  -1.581 &  0.026 &  148.43 &  0.17 &   62100 & 21.53\\ 
Ter9          &   5.77 & 0.34 & 270.4117 & -26.8397 &  -2.121 & 0.052 &  -7.763 &  0.049 &   68.49 &  0.56 &  120000 &  1.90\\ 
Ton2          &   6.99 & 0.34 & 264.0393 & -38.5409 &  -5.904 & 0.031 &  -0.755 &  0.029 & -184.72 &  1.12 &   69100 &  4.60\\ 
UKS1          &  15.58 & 0.56 & 268.6133 & -24.1453 &  -2.040 & 0.095 &  -2.754 &  0.063 &   59.38 &  2.63 &   77000 &  3.84\\ 
VVV-CL001     &   8.08 & 1.48 & 268.6771 & -24.0147 &  -3.487 & 0.144 &  -1.652 &  0.107 & -327.28 &  0.90 &  135000 &  2.94\\ 
Whiting1      &  30.59 & 1.17 &  30.7375 &  -3.2528 &  -0.228 & 0.065 &  -2.046 &  0.056 & -130.41 &  1.79 &    1970 & 15.49\\ 
\end{longtable}
\clearpage
\twocolumn

\section{Choice of the time-step for orbit integration}\label{deltat}

\begin{table*}
\caption{Crossing time, $t_{cross}$, and median error in energy conservation, $m_{err}$, for all 159 clusters evolved in isolation. All ``isolated" simulations have been run with a $\Delta t=10^5$~yr, and for a total of $N_{steps}=50 000$ steps, except for clusters marked with (*), for which a $\Delta t=10^4$~yr and a a total of $N_{steps}=500 000$ steps have been used. }\label{tcross-energy}
\tiny
\begin{center}
\begin{tabular}{l | c | c | l |  c|   c | l | c | c } 
\hline
         Cluster &       $ t_{cross}$ &  $m_{err}$ & Cluster &       $ t_{cross}$ &  $m_{err}$  & Cluster &       $ t_{cross}$ &  $m_{err}$ \\
         \hline \hline

    2MASS-GC01 & $5.6\times10^5$ &         $ 2.1\times10^{-12}$  &
    2MASS-GC02 &  $3.9\times10^5$ &          $4.4\times10^{-12}$ &
           AM1 &  $6.5\times10^6$ &          $5.4\times10^{-13}$ \\
           AM4 &  $2.2\times10^7$ &          $8.6\times10^{-13}$ & 
          Arp2 &  $4.2\times10^6$ &          $6.1\times10^{-13}$ &  
         BH140 &  $1.2\times10^6$ &         $ 5.3\times10^{-13}$  \\
         BH261 &  $6.9\times10^5$ &          $1.7\times10^{-12}$  &
        Crater &  $1.3\times10^7$ &         $ 4.4\times10^{-13}$ &
         Djor1 &  $4.8\times10^5$ &          $1.0\times10^{-12}$  \\
         Djor2 &  $3.4\times10^5$ &          $3.9\times10^{-12}$ &
            E3 &  $2.9\times10^6$ &          $3.3\times10^{-14}$  &
   ESO280-SC06 &  $3.5\times10^6$ &         $ 1.1\times10^{-13}$  \\
   ESO452-SC11 &  $7.9\times10^5$ &          $7.9\times10^{-13}$  &
      Eridanus &  $7.2\times10^6$ &          $1.3\times10^{-12}$ &
       FSR1716 &  $4.7\times10^5$ &         $ 3.4\times10^{-13}$  \\
       FSR1735 &  $1.9\times10^5$ &          $4.9\times10^{-11}$ &
       FSR1758 &  $9.1\times10^5$ &          $7.4\times10^{-13}$  &
           HP1 &  $2.1\times10^5$ &          $4.0\times10^{-11}$  \\
        IC1257 &  $9.9\times10^5$ &         $ 1.5\times10^{-14}$ &
        IC1276 &  $4.5\times10^5$ &          $3.4\times10^{-12}$ &
        IC4499 &  $1.5\times10^6$ &          $9.0\times10^{-13}$  \\
      Laevens3 &  $6.5\times10^6$ &          $4.7\times10^{-13}$  &
       Liller1 (*) &  $3.0\times10^4$ &          $6.0\times10^{-13}$  &
        Lynga7 &  $4.2\times10^5$ &         $ 2.3\times10^{-13}$  \\
        NGC104 (*) &  $1.7\times10^5$ &         $ 2.4\times10^{-13}$  &
       NGC1261 &  $2.96\times10^5$ &        $  1.4\times10^{-11}$  &
       NGC1851 (*) &  $9.0\times10^4$ &          $3.8\times10^{-13}$  \\
       NGC1904 (*) &  $1.6\times10^5$ &          $6.8\times10^{-10}$  &
       NGC2298 &  $2.6\times10^5$ &          $1.2\times10^{-11}$  &
       NGC2419 &  $1.4\times10^6$ &          $1.2\times10^{-12}$  \\
       NGC2808 (*) &  $8.4\times10^4$ &          $1.6\times10^{-12}$  &
        NGC288 &  $8.1\times10^5$ &          $6.5\times10^{-13}$  &
       NGC3201 &  $4.5\times10^5$ &         $ 1.6\times10^{-12}$ \\
        NGC362 (*) &  $1.4\times10^5$ &          $1.9\times10^{-13}$  &
       NGC4147 &  $4.2\times10^5$ &         $ 5.2\times10^{-14}$ &
       NGC4372 &  $5.7\times10^5$ &          $4.8\times10^{-13}$ \\
       NGC4590 &  $6.1\times10^5$ &          $4.3\times10^{-13}$ &
       NGC4833 &  $2.3\times10^5$ &          $1.8\times10^{-12}$  &
       NGC5024 &  $4.9\times10^5$ &          $3.0\times10^{-12}$  \\
       NGC5053 &  $2.7\times10^6$ &          $4.4\times10^{-13}$  &
       NGC5139 (*) &  $1.8\times10^5$ &          $1.1\times10^{-12}$ &
       NGC5272 &  $2.6\times10^5$ &          $2.0\times10^{-12}$  \\
       NGC5286 (*) &  $1.3\times10^5$ &          $1.5\times10^{-13}$ &
       NGC5466 &  $2.2\times10^6$ &          $1.0\times10^{-12}$  &
       NGC5634 &  $4.3\times10^5$ &          $4.6\times10^{-12}$  \\
       NGC5694 &  $1.9\times10^5$ &          $6.9\times10^{-11}$  &
       NGC5824 &  $1.9\times10^5$ &          $5.2\times10^{-11}$  &
       NGC5897 &  $9.4\times10^5$ &          $2.5\times10^{-13}$ \\
       NGC5904 &  $2.2\times10^5$ &          $8.3\times10^{-12}$  &
       NGC5927 &  $2.4\times10^5$ &          $1.9\times10^{-11}$  &
       NGC5946 (*) &  $1.4\times10^5$ &          $1.1\times10^{-12}$  \\
       NGC5986 (*) &  $1.5\times10^5$ &          $7.7\times10^{-13}$  &
       NGC6093 (*) &  $7.5\times10^4$ &          $7.9\times10^{-13}$  &
       NGC6101 &  $1.3\times10^6$ &          $3.5\times10^{-13}$  \\
       NGC6121 &  $2.5\times10^5$ &          $9.1\times10^{-12}$  &
       NGC6139 (*) &  $7.0\times10^4$ &          $3.9\times10^{-13}$  &
       NGC6144 &  $4.0\times10^5$ &          $3.9\times10^{-12}$  \\
       NGC6171 &  $2.9\times10^5$ &          $1.7\times10^{-13}$  &
       NGC6205 (*) &  $1.7\times10^5$ &          $1.7\times10^{-13}$  &
       NGC6218 &  $2.5\times10^5$ &          $3.0\times10^{-12}$  \\
       NGC6229 (*) &  $1.8\times10^5$ &          $2.5\times10^{-14}$ &
       NGC6235 &  $3.3\times10^5$ &          $3.1\times10^{-12}$  &
       NGC6254 &  $2.4\times10^5$ &          $9.0\times10^{-12}$ \\
       NGC6256 &  $3.1\times10^5$ &          $7.1\times10^{-12}$  &
       NGC6266 (*)&  $5.0\times10^4$ &          $1.1\times10^{-13}$  &
       NGC6273 (*) &  $1.1\times10^5$ &          $8.0\times10^{-13}$  \\
       NGC6284 &  $2.1\times10^5$ &          $1.3\times10^{-11}$  &
       NGC6287 &  $1.9\times10^5$ &          $3.1\times10^{-11}$  &
       NGC6293 &  $1.8\times10^5$ &          $9.9\times10^{-11}$ \\
       NGC6304 &  $2.5\times10^5$ &          $5.9\times10^{-12}$  &
       NGC6316 (*) &  $1.9\times10^5$ &          $1.3\times10^{-13}$  &
       NGC6325 (*) &  $1.2\times10^5$ &          $4.4\times10^{-13}$  \\
       NGC6333 (*) &   $1.5\times10^5$ &          $2.2\times10^{-13}$ &
       NGC6341 (*) &  $1.6\times10^5$ &          $5.6\times10^{-13}$  &
       NGC6342 (*) &  $1.5\times10^5$ &          $3.4\times10^{-13}$  \\
       NGC6352 &  $3.9\times10^5$ &          $2.2\times10^{-12}$  &
       NGC6355 &   $2.2\times10^5 $ &          $4.8\times10^{-11}$  &
       NGC6356 &  $2.4\times10^5$ &          $3.0\times10^{-11}$  \\
       NGC6362 &  $5.6\times10^5$ &          $8.1\times10^{-13}$  &
       NGC6366 &  $6.9\times10^5$ &          $4.0\times10^{-13}$ &
       NGC6380 (*) &  $1.6\times10^5$ &          $2.1\times10^{-13}$ \\
       NGC6388 (*) &  $8.3\times10^4$ &          $6.0\times10^{-13}$ &
       NGC6397 &  $2.5\times10^5$ &          $4.1\times10^{-12}$ &
       NGC6401 (*) &  $1.6\times10^5$ &          $6.5\times10^{-13}$  \\
       NGC6402 (*) &  $1.5\times10^5$ &          $9.7\times10^{-13}$  &
       NGC6426 &  $8.6\times10^5$ &          $8.6\times10^{-13}$  &
       NGC6440 (*) &  $4.6\times10^4$ &         $ 1.1\times10^{-13}$  \\
       NGC6441 (*) &  $5.7\times10^4$ &          $2.6\times10^{-14}$  &
       NGC6453 &  $1.9\times10^5$ &          $5.2\times10^{-11}$  &
       NGC6496 &  $6.3\times10^5$ &          $2.7\times10^{-13}$  \\
       NGC6517 (*) &  $8.0\times10^4$ &          $3.8\times10^{-13}$  &
       NGC6522 (*) &  $1.2\times10^5$ &          $5.9\times10^{-13}$  &
       NGC6528 &  $1.9\times10^5$ &          $5.7\times10^{-11}$  \\
       NGC6535 &  $4.8\times10^5$ &        $  2.9\times10^{-12}$ &
       NGC6539 &  $2.6\times10^5$ &          $2.5\times10^{-12}$  &
       NGC6540 &  $6.8\times10^5$ &          $2.0\times10^{-13}$  \\
       NGC6541 (*) &  $1.7\times10^5$ &          $6.1\times10^{-13}$  &
       NGC6544 (*) &  $1.0\times10^5$ &          $6.4\times10^{-13}$  &
       NGC6553 &  $1.9\times10^5$ &          $7.8\times10^{-11}$  \\
       NGC6558 (*) &  $1.4\times10^5$ &          $7.0\times10^{-13}$  &
       NGC6569 (*) &  $1.6\times10^5$ &          $1.5\times10^{-13}$  &
       NGC6584 &  $4.0\times10^5$ &          $4.7\times10^{-13}$  \\
       NGC6624 (*) &  $1.8\times10^5$ &          $4.3\times10^{-13}$  &
       NGC6626 (*) &  $6.4\times10^4$ &          $2.5\times10^{-12}$  &
       NGC6637 &  $1.8\times10^5$ &          $5.4\times10^{-11}$  \\
       NGC6638 (*) &  $9.7\times10^4$ &          $7.8\times10^{-13}$  &
       NGC6642 (*) &  $1.0\times10^5$ &          $1.3\times10^{-12}$ &
       NGC6652 (*)&  $1.3\times10^5$ &          $3.3\times10^{-13}$  \\
       NGC6656 (*) &  $1.8\times10^5$ &          $1.5\times10^{-13}$ &
       NGC6681 (*) &  $1.5\times10^5$ &          $4.3\times10^{-13}$  &
       NGC6712 &  $1.9\times10^5$ &          $4.5\times10^{-11}$  \\
       NGC6715 (*) &  $9.1\times10^4$ &          $1.4\times10^{-12}$&
       NGC6717 &  $4.7\times10^5$ &          $7.4\times10^{-13}$  &
       NGC6723 &  $2.8\times10^5$ &          $2.1\times10^{-11}$  \\
       NGC6749 &  $4.2\times10^5$ &         $7.0\times10^{-13}$  &
       NGC6752 &  $2.4\times10^5$ &          $2.9\times10^{-11}$ &
       NGC6760 &  $2.4\times10^5$ &          $6.7\times10^{-12}$ \\ 
       NGC6779 &  $2.3\times10^5$ &          $1.3\times10^{-11}$  &
       NGC6809 &  $4.3\times10^5$ &          $1.7\times10^{-12}$  &
       NGC6838 &  $6.7\times10^5$ &          $7.0\times10^{-14}$  \\
       NGC6864 (*) &  $8.6\times10^4$ &          $9.1\times10^{-13}$  &
       NGC6934 &  $3.2\times10^5$ &          $6.4\times10^{-12}$  &
       NGC6981 &  $5.7\times10^5$ &          $7.0\times10^{-13}$ \\
       NGC7006 &  $5.1\times10^5$ &          $9.1\times10^{-13}$ &
       NGC7078 (*) &  $1.1\times10^5$ &          $4.1\times10^{-13}$  &
       NGC7089 (*) &  $1.4\times10^5$ &          $1.3\times10^{-13}$ \\
       NGC7099 & $ 3.0\times10^5$ &          $5.2\times10^{-14}$  &
       NGC7492 &  $1.9\times10^6$ &          $6.0\times10^{-13}$ &
          Pal1 &  $2.1\times10^6$ &          $3.0\times10^{-13}$ \\
         Pal10 &  $4.0\times10^5 $&          $4.6\times10^{-13}$  &
         Pal11 &  $2.0\times10^6$ &          $3.5\times10^{-13}$  &
         Pal12 &  $4.4\times10^6$ &          $1.1\times10^{-12}$  \\
         Pal13 &  $1.3\times10^7$ &          $3.0\times10^{-13}$  &
         Pal14 &  $1.7\times10^7$ &          $7.4\times10^{-13}$  &
         Pal15 &  $6.3\times10^6$ &          $9.2\times10^{-14}$  \\
          Pal2 &  $4.9\times10^5$ &          $4.1\times10^{-13}$  &
          Pal3 &  $1.1\times10^7$ &          $8.1\times10^{-13}$  &
          Pal4 &  $8.8\times10^6$ &          $6.6\times10^{-13}$  \\
          Pal5 &  $1.5\times10^7$ &          $6.2\times10^{-13}$  &
          Pal6 (*) &  $1.6\times10^5$ &          $1.0\times10^{-12}$ &
          Pal8 &  $5.6\times10^5$ &          $1.1\times10^{-12}$  \\
         Pyxis &  $7.2\times10^6$ &          $6.0\times10^{-13}$  &
        Rup106 &  $2.2\times10^6 $&          $9.4\times10^{-13}$  &
 SagittariusII &  $1.9\times10^7$ &          $7.3\times10^{-13}$ \\
          Ter1 (*)&  $8.3\times10^4$ &          $7.1\times10^{-13}$  &
         Ter10 &  $1.8\times10^5 $&          $7.3\times10^{-11}$  &
         Ter12 &  $2.1\times10^5$ &          $3.1\times10^{-11}$ \\
          Ter2 &  $2.4\times10^5$ &          $1.2\times10^{-11}$  &
          Ter3 &  $9.8\times10^5$ &          $1.2\times10^{-12}$  &
          Ter4 &  $3.4\times10^5$ &          $5.9\times10^{-12}$ \\
          Ter5 (*) &  $7.7\times10^4$ &          $3.5\times10^{-13}$ &
          Ter6 (*) &  $4.9\times10^4$ &          $1.5\times10^{-12}$ &
          Ter7 &  $3.2\times10^6 $&         $ 7.4\times10^{-13}$  \\
          Ter8 &  $4.1\times10^6$ &          $9.4\times10^{-13}$ &
          Ter9 (*) &  $7.7\times10^4$ &          $4.7\times10^{-13}$  &
          Ton2 &  $3.8\times10^5 $&          $1.1\times10^{-13}$  \\
          UKS1 &  $2.8\times10^5 $&         $ 2.0\times10^{-12}$  &
     VVV-CL001 (*) &  $1.4\times10^5$ &       $   1.1\times10^{-12}$ &
      Whiting1 &  $1.4\times10^7$ &          $1.6\times10^{-13}$ \\

\hline
\end{tabular}
\end{center}

\end{table*}

To choose the optimal time-step $\Delta t$ for the simulations, we quantified the energy conservation of the orbit integration, in the case where the 159 clusters evolve in isolation, that is in the case where each test-particle in a cluster feels the gravitational attraction of the cluster itself, but not that of the Galaxy. Since in the case of an isolated cluster, the gravitational potential is time-independent, the total energy $E_i$ of each particle in the system (sum of the particle kinetic and potential energy) must be conserved. Any departure from energy conservation is thus a check of the quality of the integration and in particular of the choice of the adopted time-step.

For the isolated simulations, we thus proceeded as follows. We modeled each cluster as a set of $N$=100 000 test-particles, subject to the cluster potential, only. To model this latter, for each cluster we adopted the same Plummer sphere distribution, with same characteristic radius $r_c$ and total masse $M_{GC}$, as those adopted for the simulations described in Sect.~2 (see in particular Sect.~2.2).  A  first set of 159 isolated simulations was run adopting a   $\Delta t=10^5$~yr and a total number of steps $N_{steps}=50 000$ for all clusters, for a total simulated time interval of 5~Gyr.  We then quantified the energy conservation by calculating the median error per time-step, $m_{err}$, of $(\Delta E/E)_i=|(E_{fin,i}-E_{ini,i})/E_{ini,i}|$, where $E_{ini,i}$ and $E_{fin,i}$ are, respectively, the initial (time $t=0$) and final (time t=5~Gyr) energy of each particle in the system.
While for most of the simulated clusters (109 over 159), this choice of the time-step was sufficient to guarantee an excellent energy conservation (with $m_{err}$ typically of the order of $10^{-11}-10^{-12}$), for 50 clusters the corresponding $m_{err}$ values were found to be above $10^{-10}$. This was the case for all clusters with crossing times $t_{cross}=\sqrt{{r_c}^3/GM_{GC}}$ lower than $2\times 10^5$ yr, that is about twice the time-step. For these clusters, we hence reduced the $\Delta t$ of a factor 10,  rerunning the simulations with $\Delta t=10^4$~yr and  $N_{steps}=500 000$. With such a choice, the corresponding energy conservation turned out to be excellent (below $10^{-10}$ per step). In Table~\ref{tcross-energy}, we summarize the  result of this study, reporting the cluster name, crossing time, and median error, $m_{err}$, in energy conservation obtained for all isolated cluster simulations. Clusters for which a $\Delta t=10^4$~yr, and a corresponding number $N_{steps}$ have been used, are indicated in the Table with an asterisk. The values of  $\Delta t$ adopted for the isolated simulations, and the associated number of time steps, $N_{steps}$, are also those used to run the simulations of the same clusters orbiting in the gravitational field of the Milky Way.

\section{Extra-tidal features generated by all the simulated clusters}\label{allstreams}

\begin{figure*}
  \includegraphics[clip=true, trim = 0mm 20mm 0mm 10mm, width=1\columnwidth]{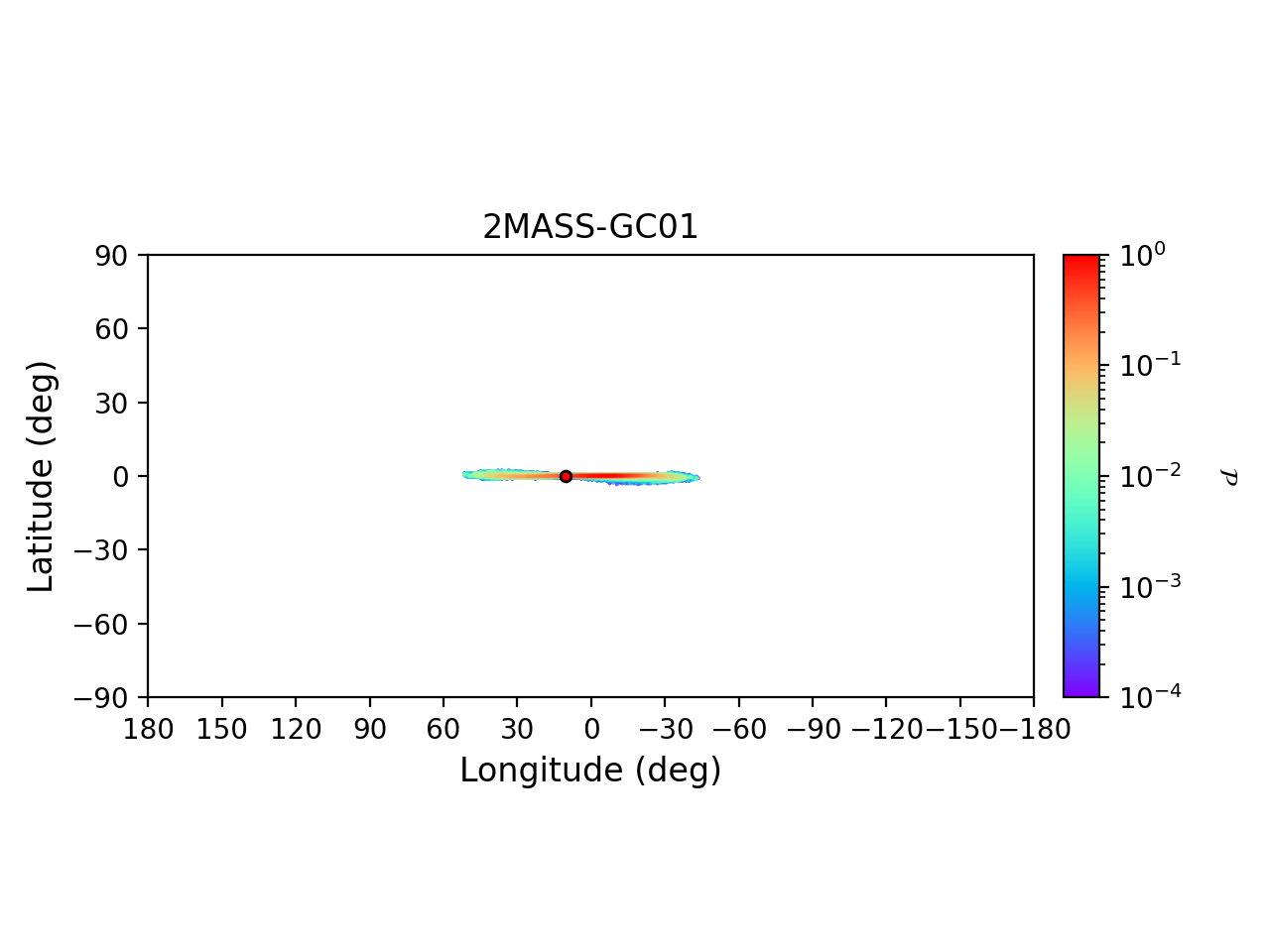}
  \includegraphics[clip=true, trim = 0mm 20mm 0mm 10mm, width=1\columnwidth]{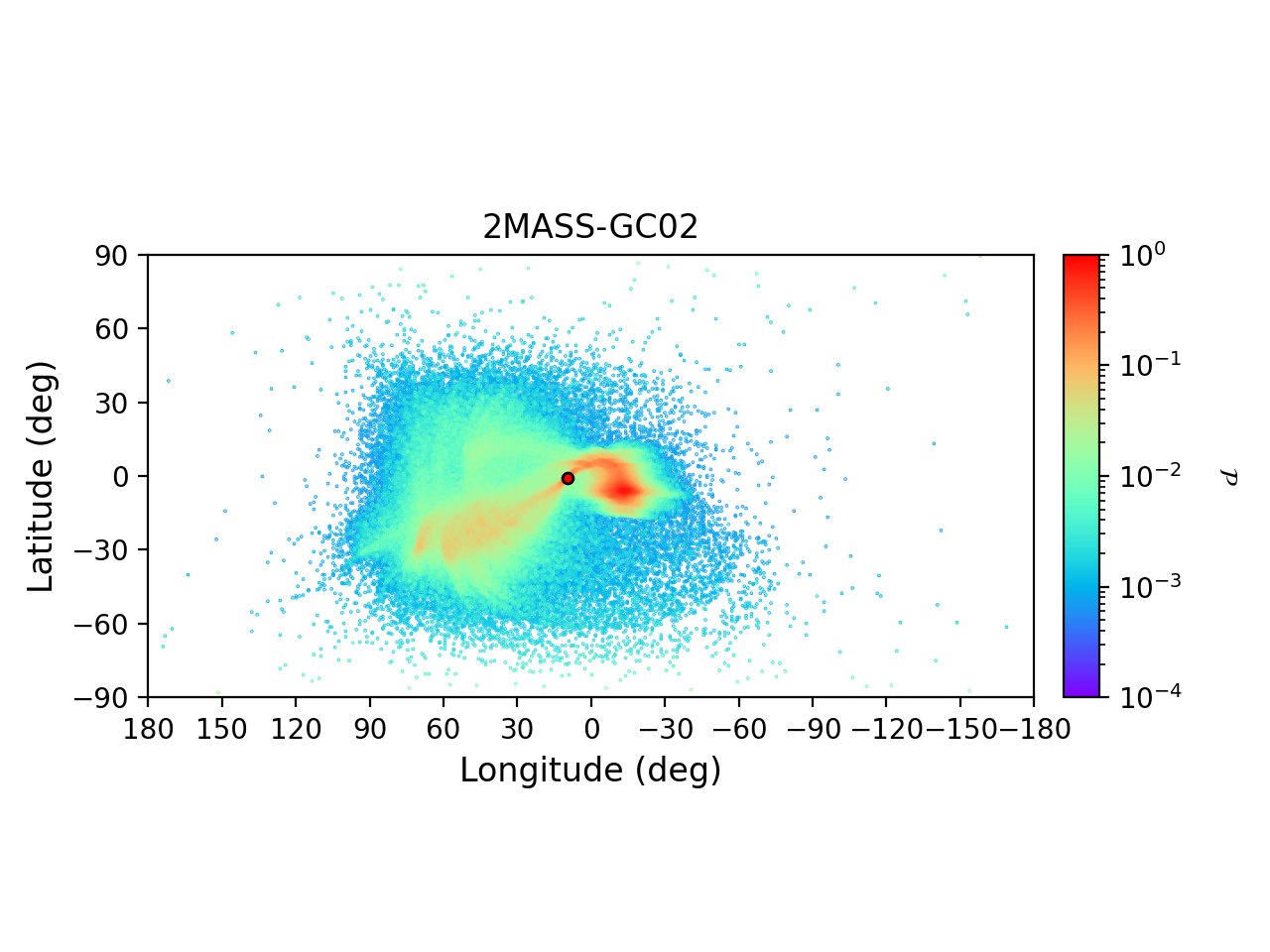}
  \includegraphics[clip=true, trim = 0mm 20mm 0mm 10mm, width=1\columnwidth]{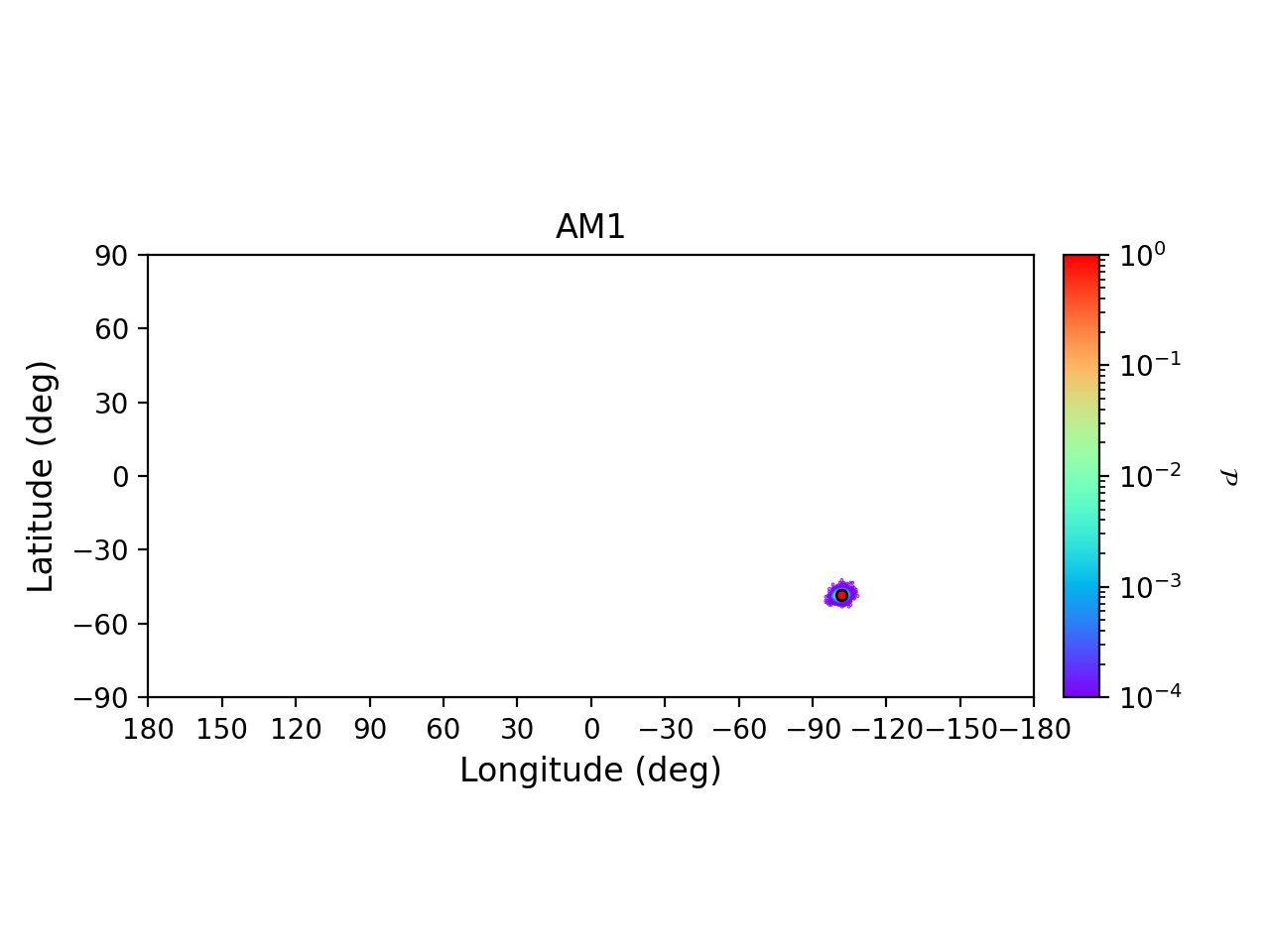}
  \includegraphics[clip=true, trim = 0mm 20mm 0mm 10mm, width=1\columnwidth]{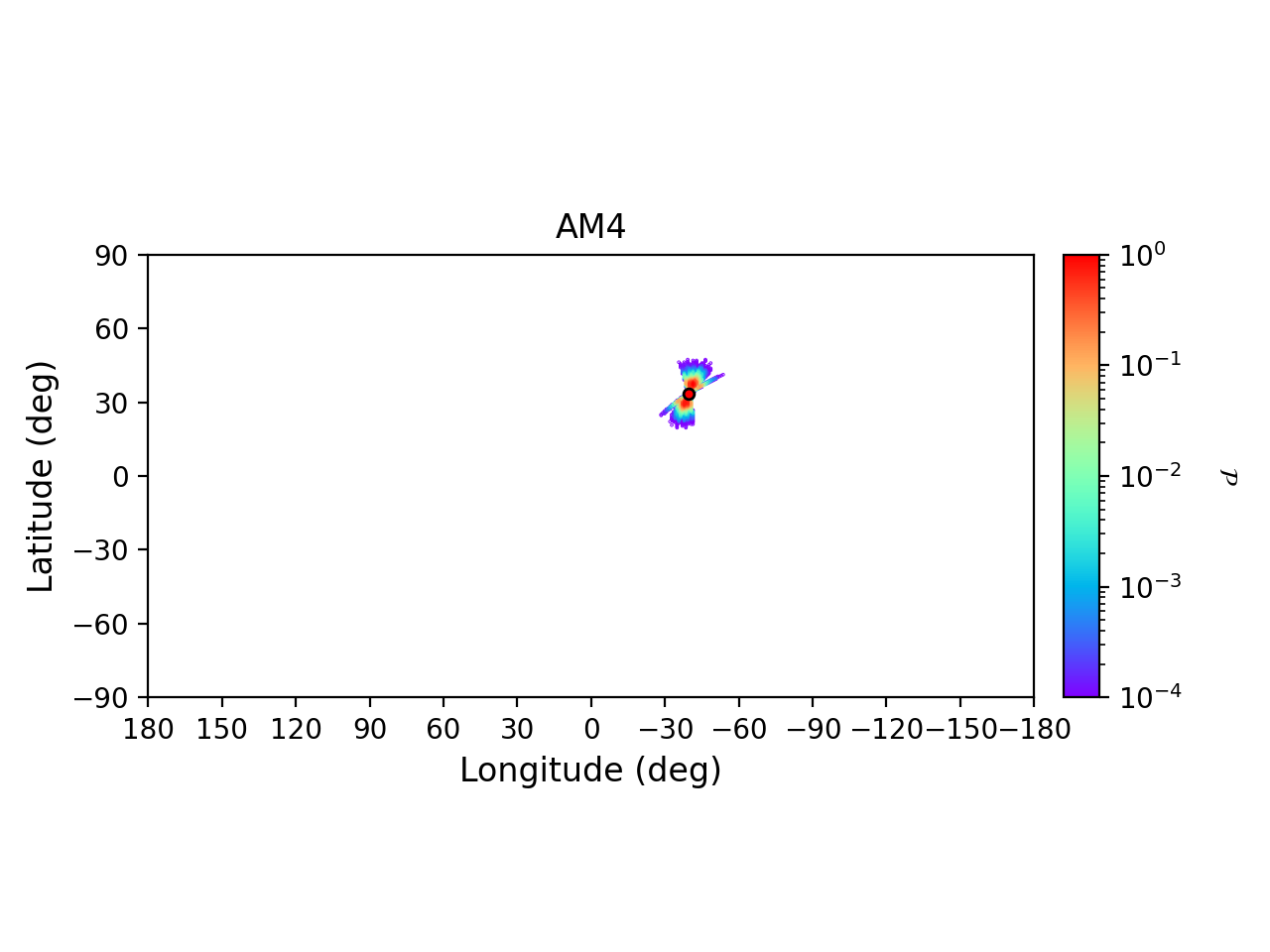}
  \includegraphics[clip=true, trim = 0mm 20mm 0mm 10mm, width=1\columnwidth]{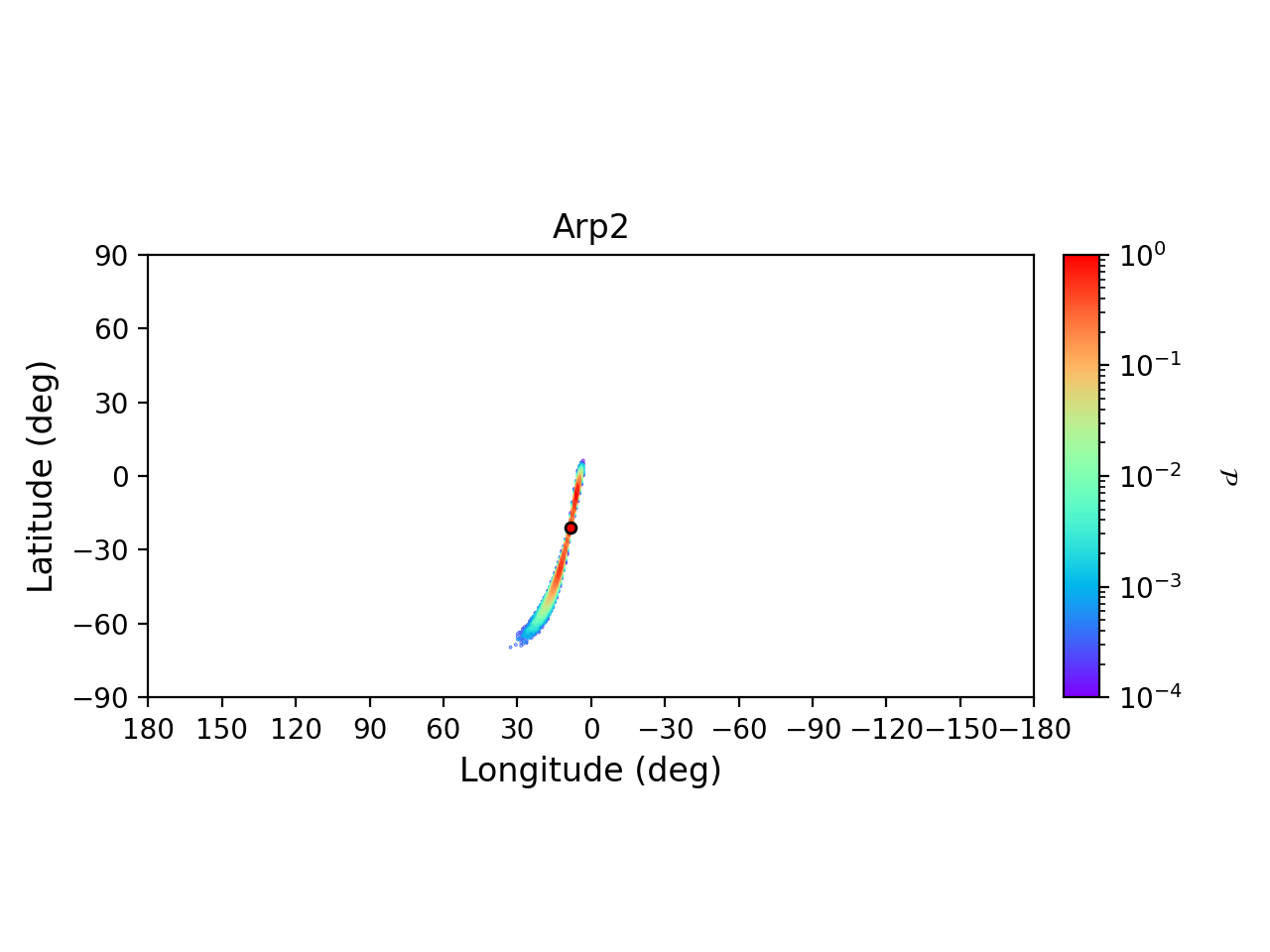}
  \includegraphics[clip=true, trim = 0mm 20mm 0mm 10mm, width=1\columnwidth]{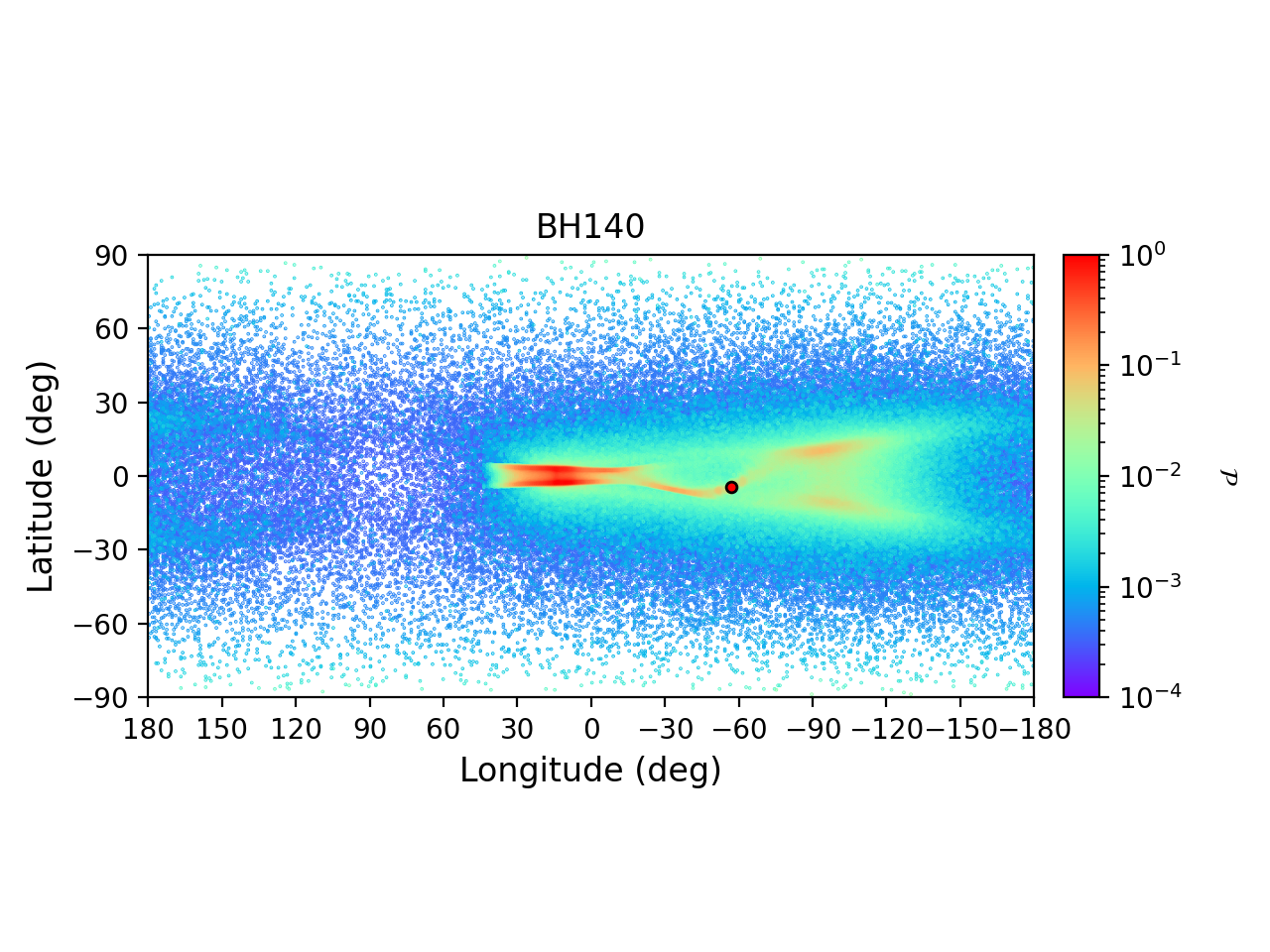}
  \includegraphics[clip=true, trim = 0mm 20mm 0mm 10mm, width=1\columnwidth]{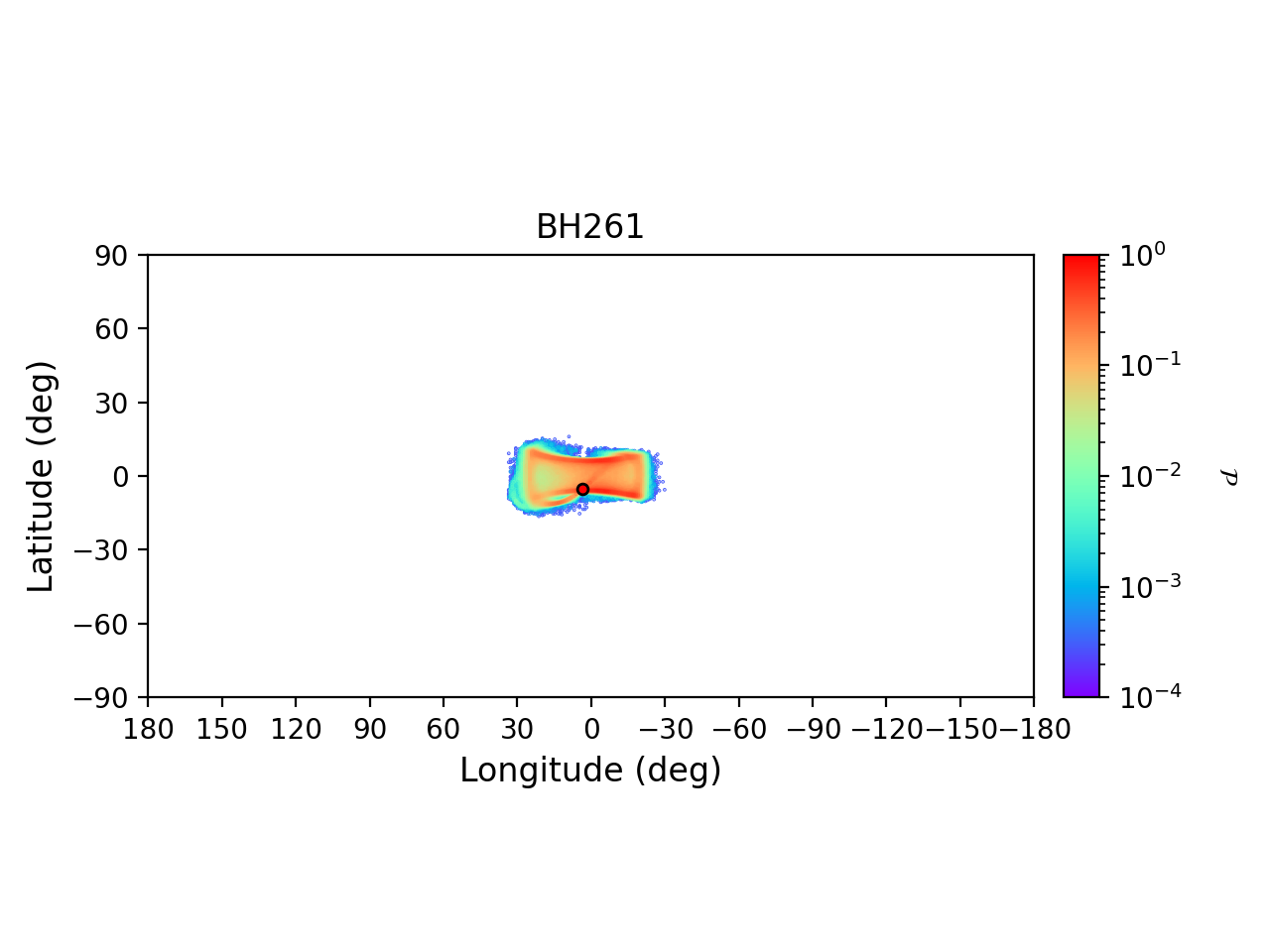}
  \includegraphics[clip=true, trim = 0mm 20mm 0mm 10mm, width=1\columnwidth]{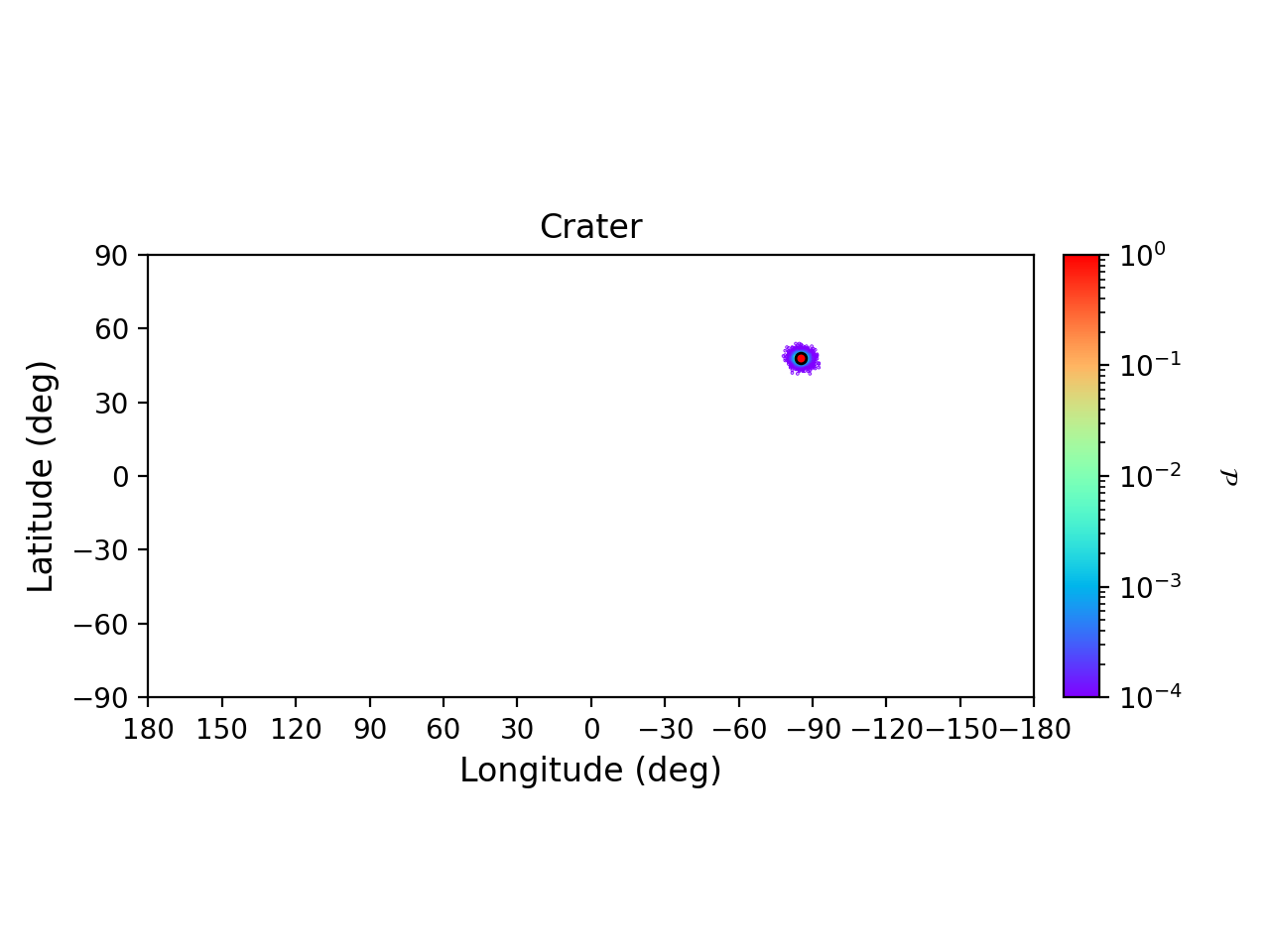}
  \caption{Projected density distribution in the $(\ell, b)$ plane of a subset of simulated globular clusters, as indicated at the top of each panel. In each panel, the red circle indicates the current position of the cluster. The densities have been normalized to their maximum value.}\label{stream1}
  \end{figure*}

\begin{figure*}
\includegraphics[clip=true, trim = 0mm 20mm 0mm 10mm, width=1\columnwidth]{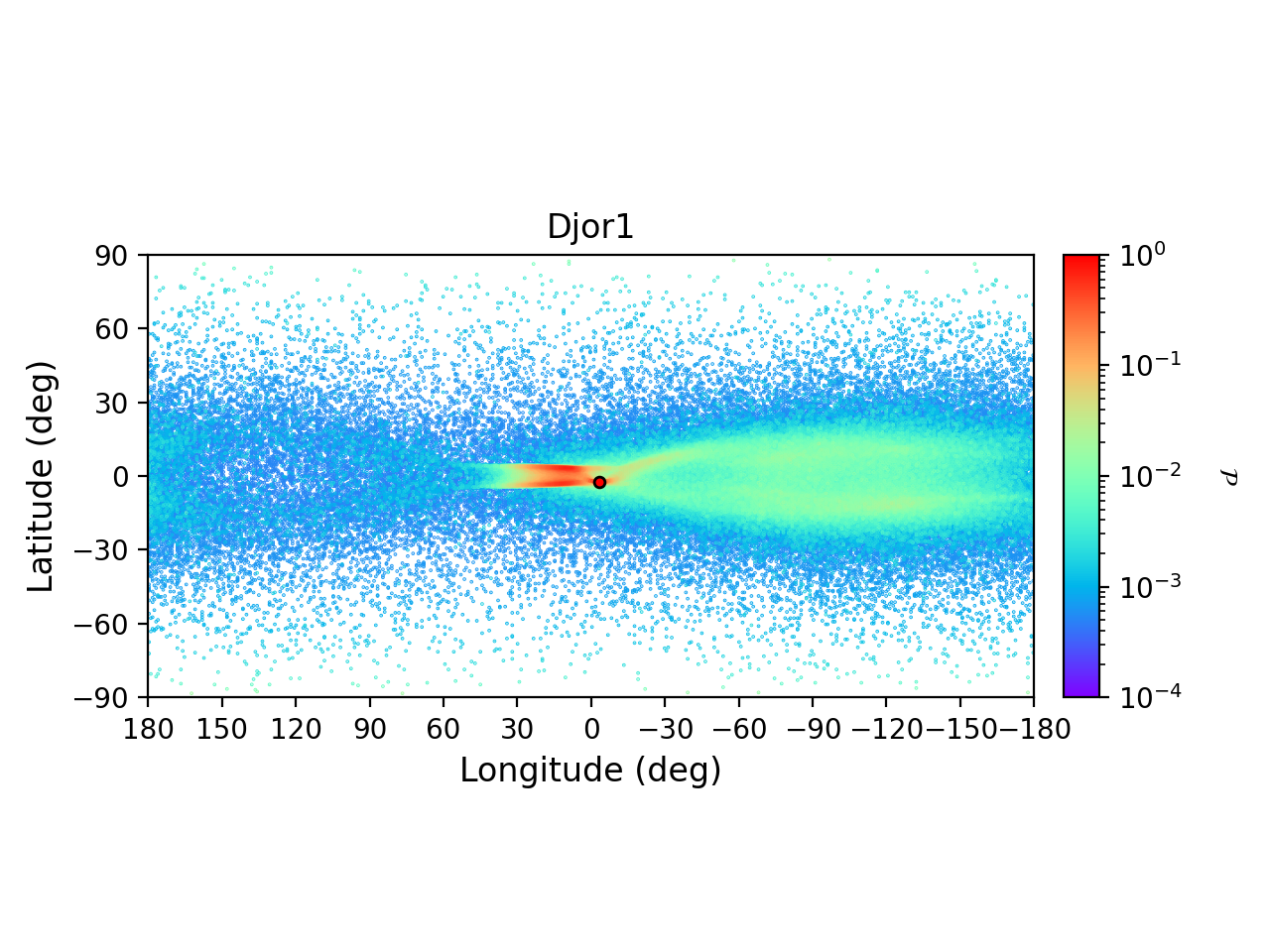}
\includegraphics[clip=true, trim = 0mm 20mm 0mm 10mm, width=1\columnwidth]{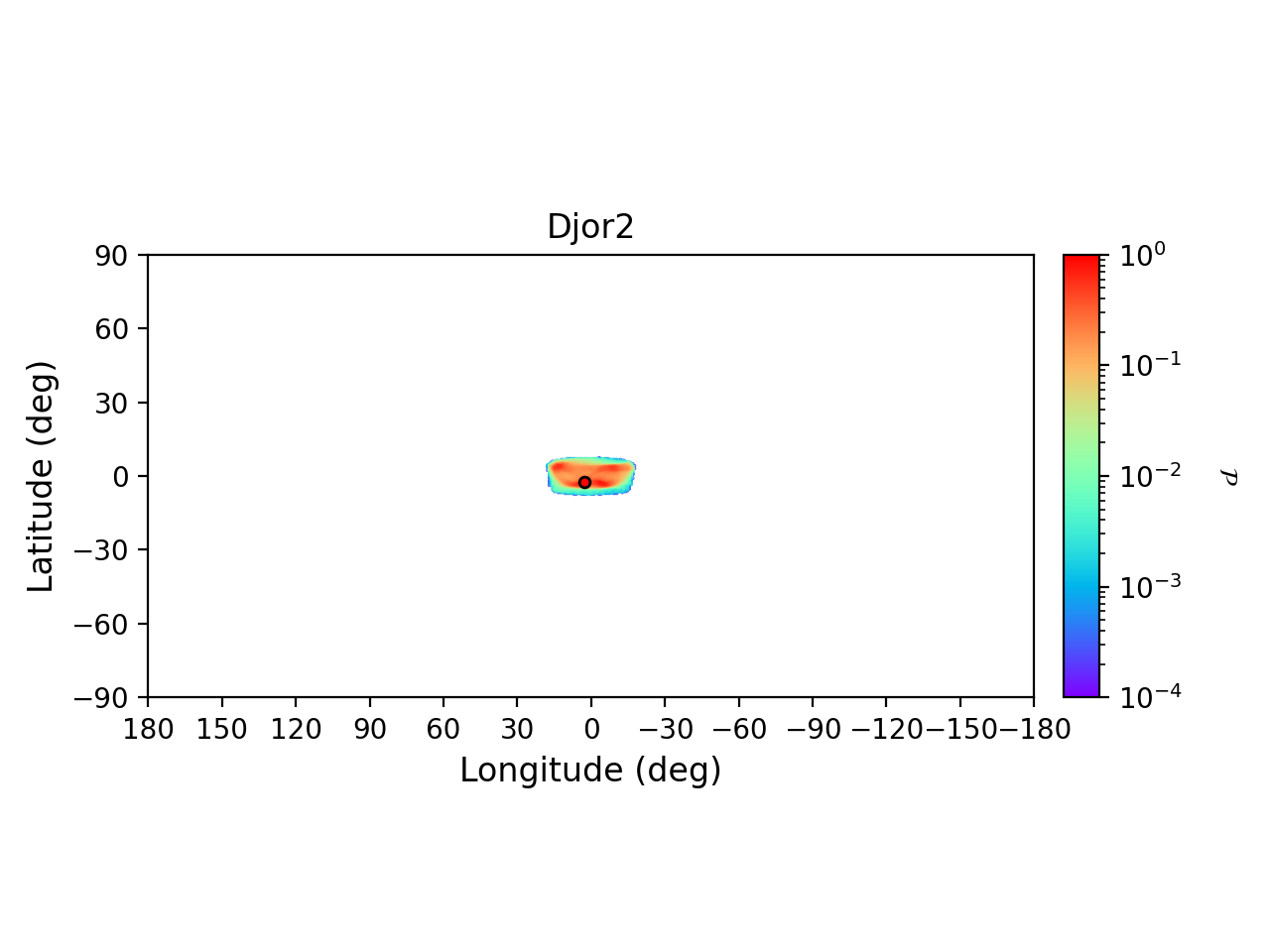}
\includegraphics[clip=true, trim = 0mm 20mm 0mm 10mm, width=1\columnwidth]{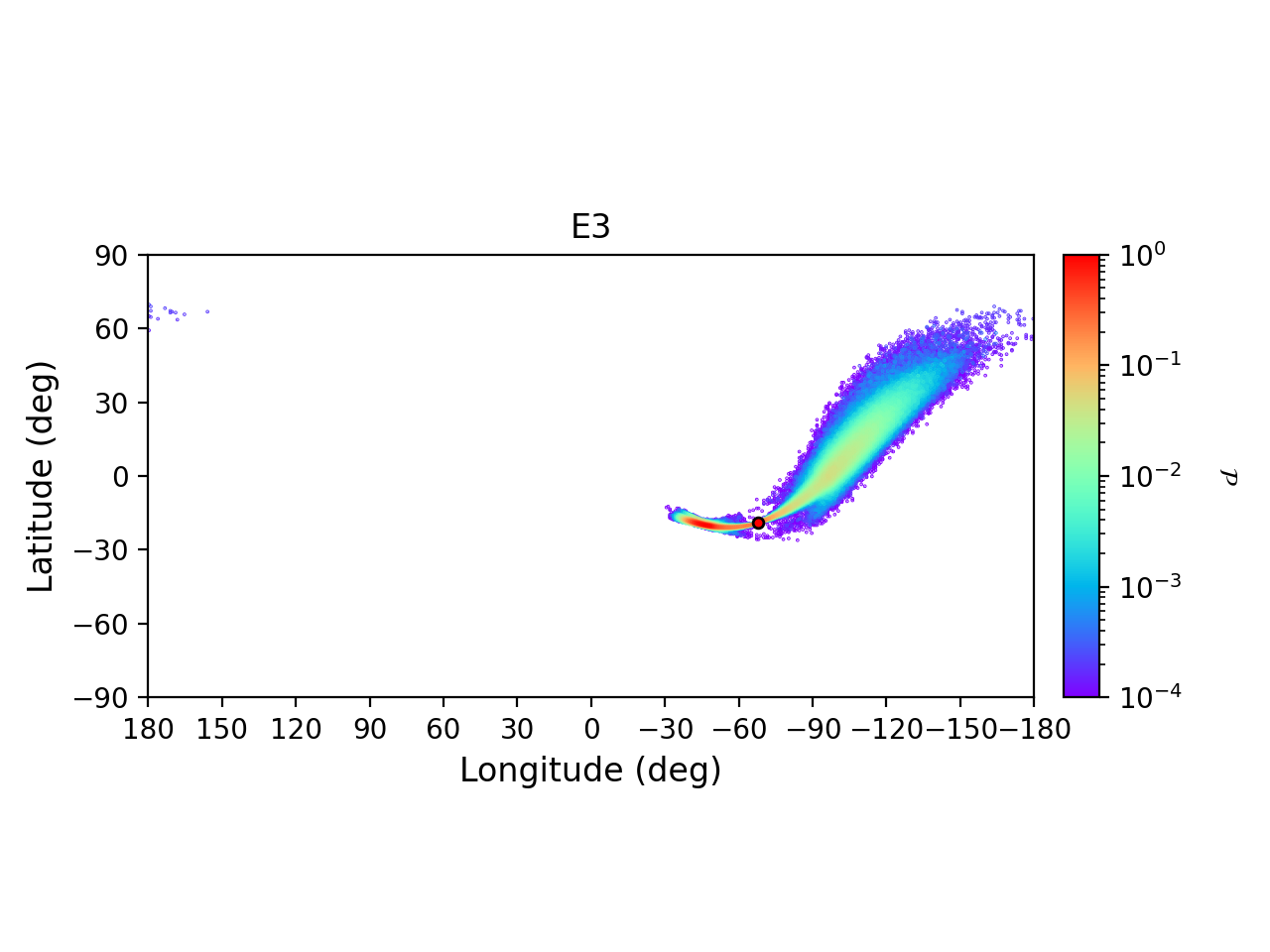}
\includegraphics[clip=true, trim = 0mm 20mm 0mm 10mm, width=1\columnwidth]{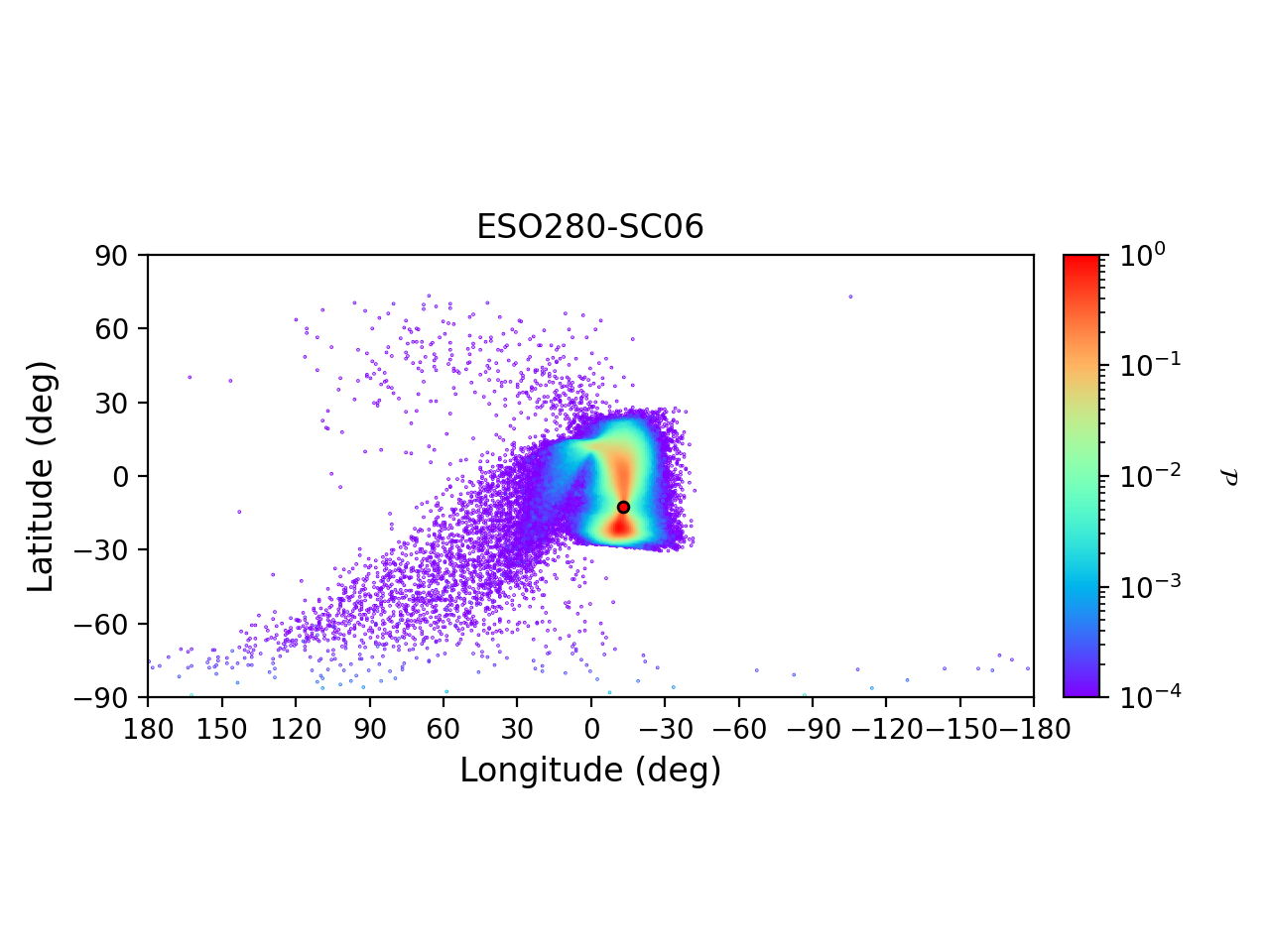}
\includegraphics[clip=true, trim = 0mm 20mm 0mm 10mm, width=1\columnwidth]{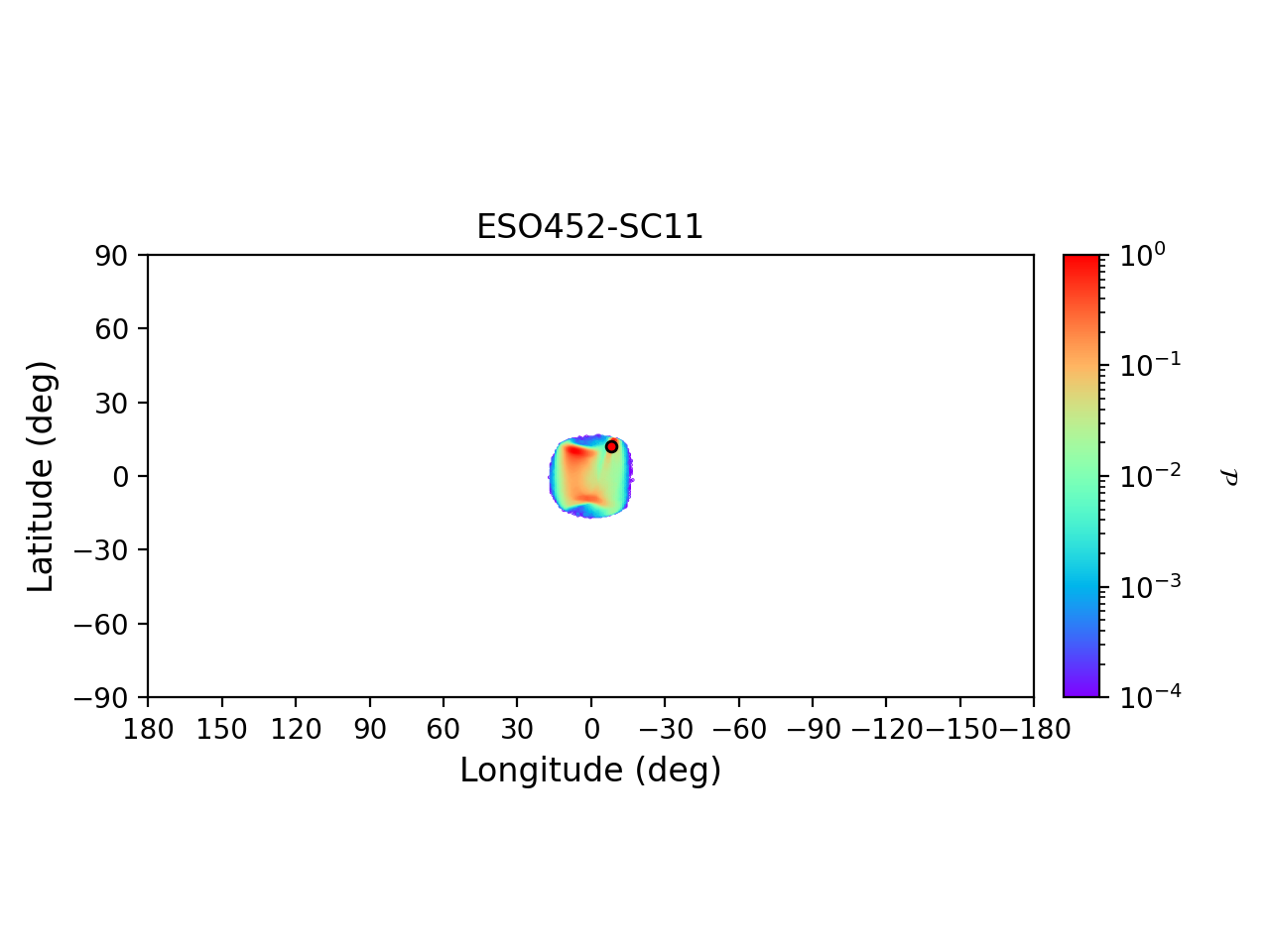}
\includegraphics[clip=true, trim = 0mm 20mm 0mm 10mm, width=1\columnwidth]{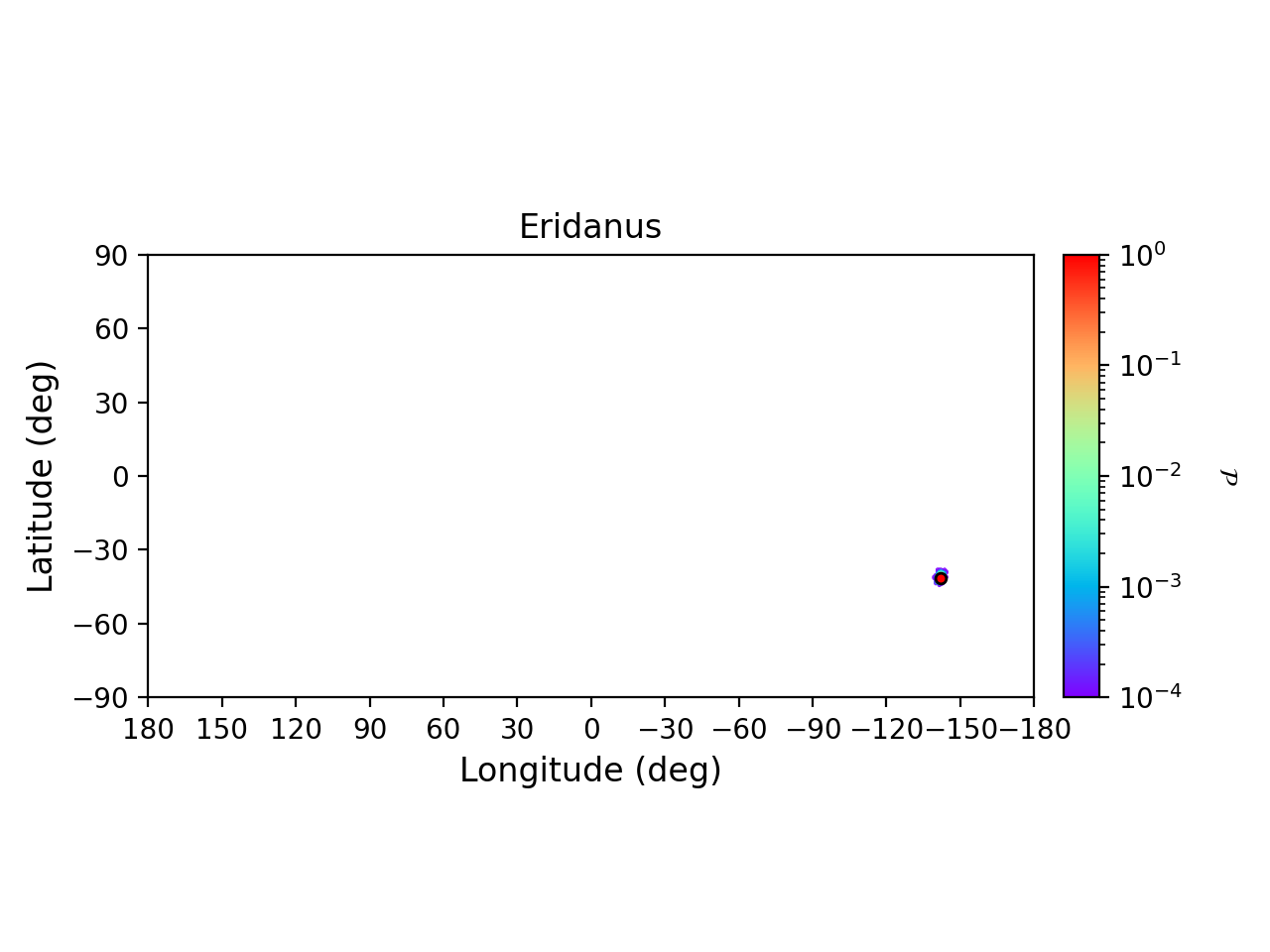}
\includegraphics[clip=true, trim = 0mm 20mm 0mm 10mm, width=1\columnwidth]{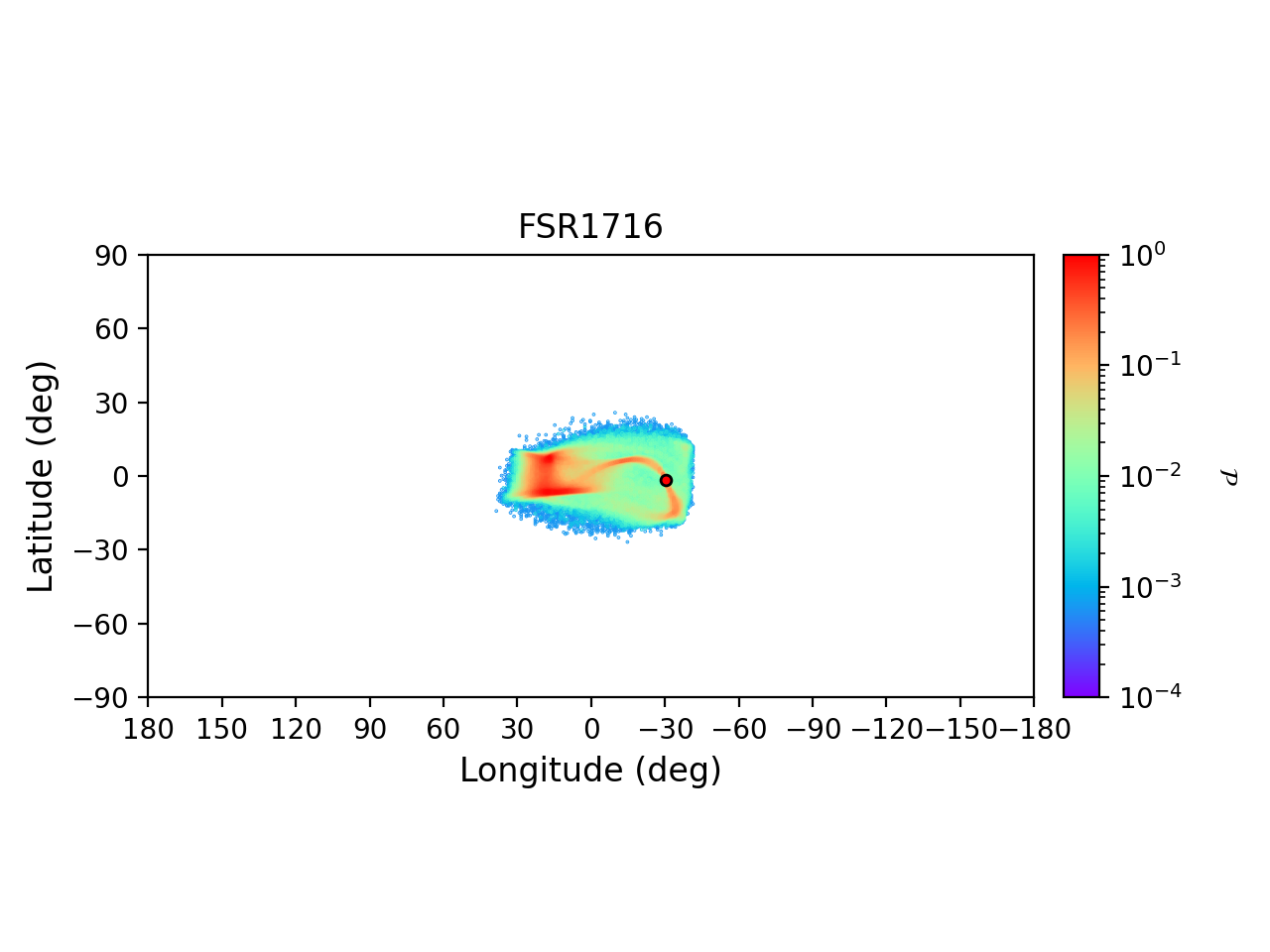}
\includegraphics[clip=true, trim = 0mm 20mm 0mm 10mm, width=1\columnwidth]{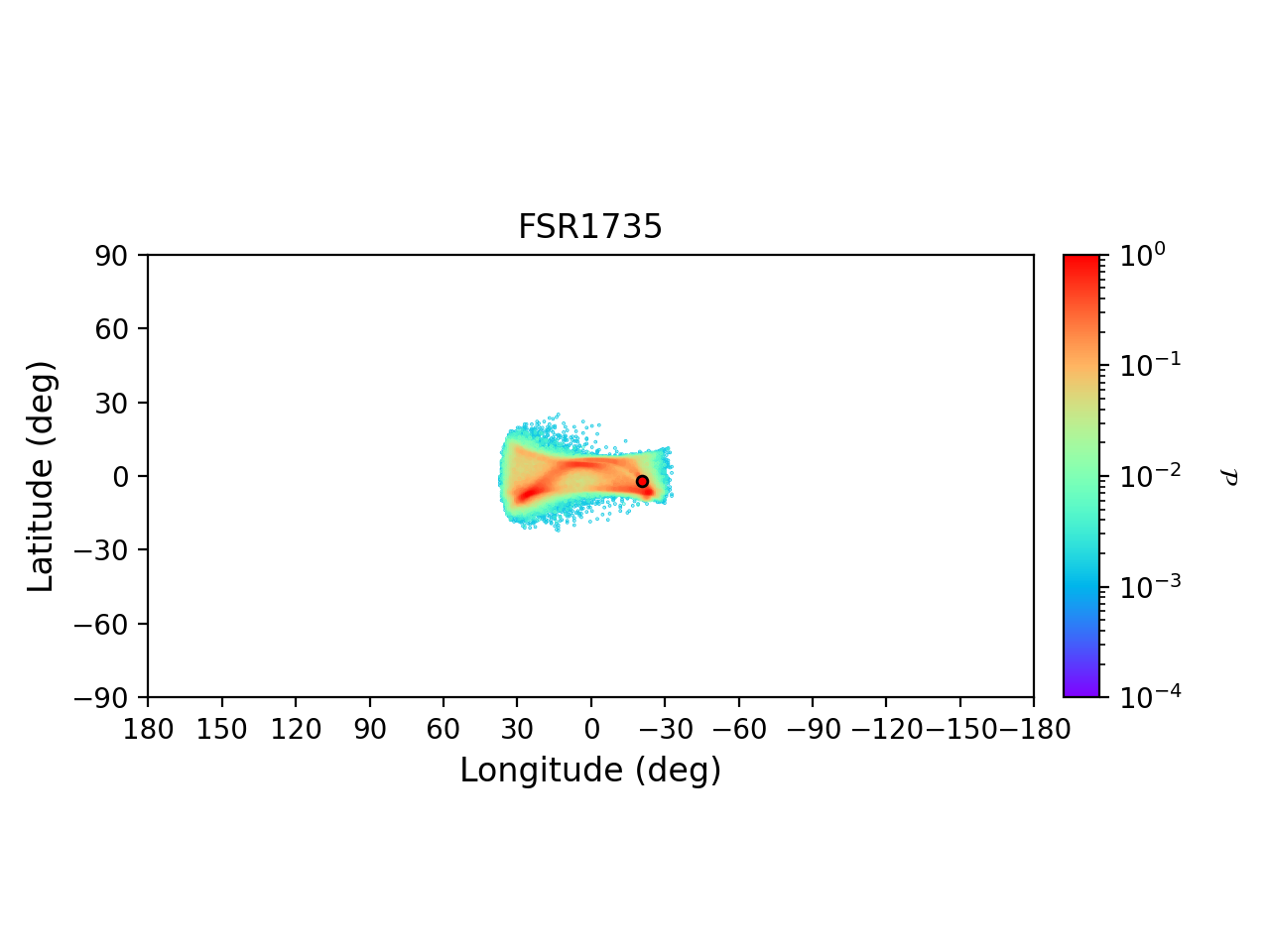}
\caption{Projected density distribution in the $(\ell, b)$ plane of a subset of simulated globular clusters, as indicated at the top of each panel. In each panel, the red circle indicates the current position of the cluster. The densities have been normalized to their maximum value.}\label{stream2}
\end{figure*}

\begin{figure*}
\includegraphics[clip=true, trim = 0mm 20mm 0mm 10mm, width=1\columnwidth]{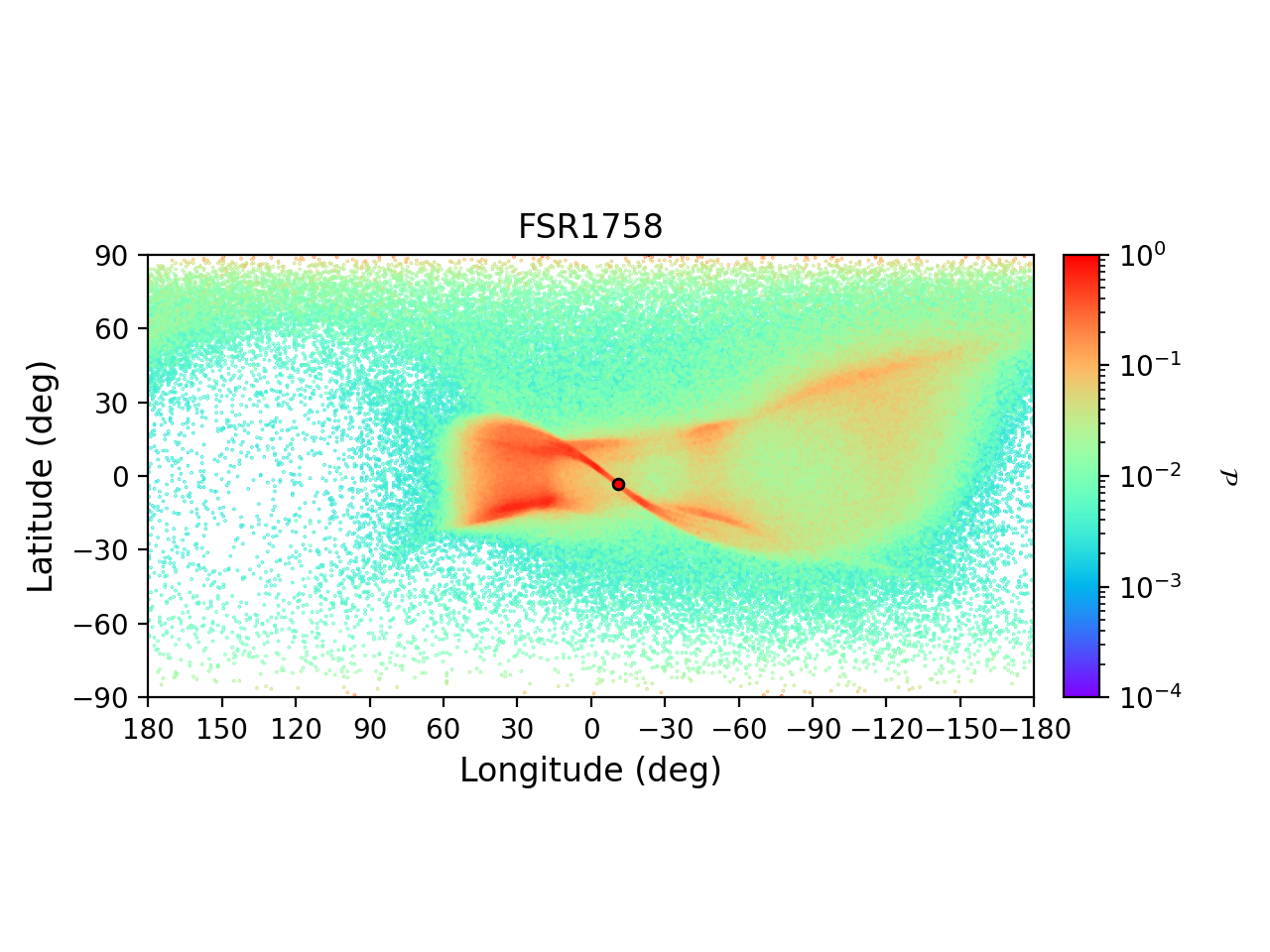}
\includegraphics[clip=true, trim = 0mm 20mm 0mm 10mm, width=1\columnwidth]{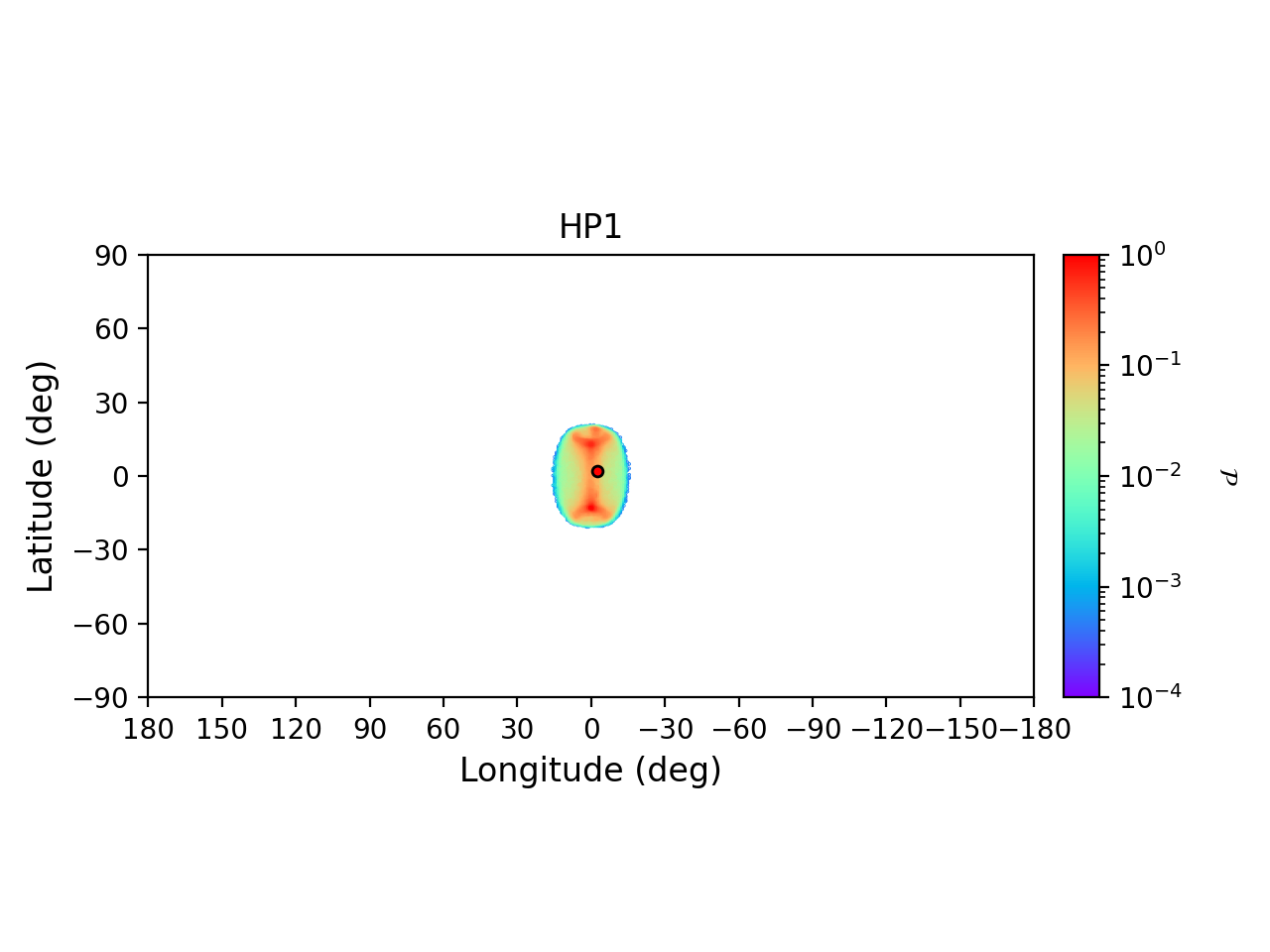}
\includegraphics[clip=true, trim = 0mm 20mm 0mm 10mm, width=1\columnwidth]{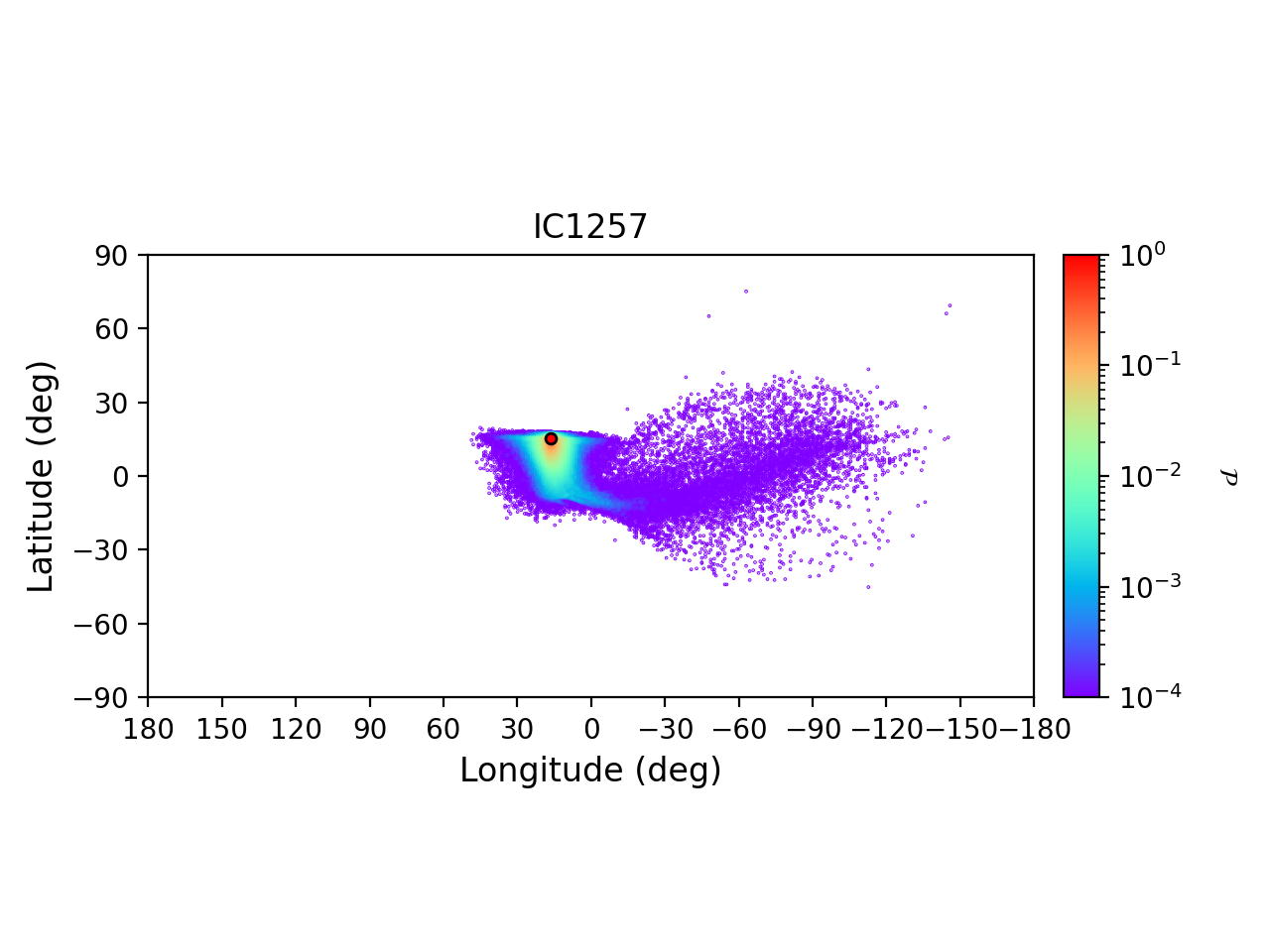}
\includegraphics[clip=true, trim = 0mm 20mm 0mm 10mm, width=1\columnwidth]{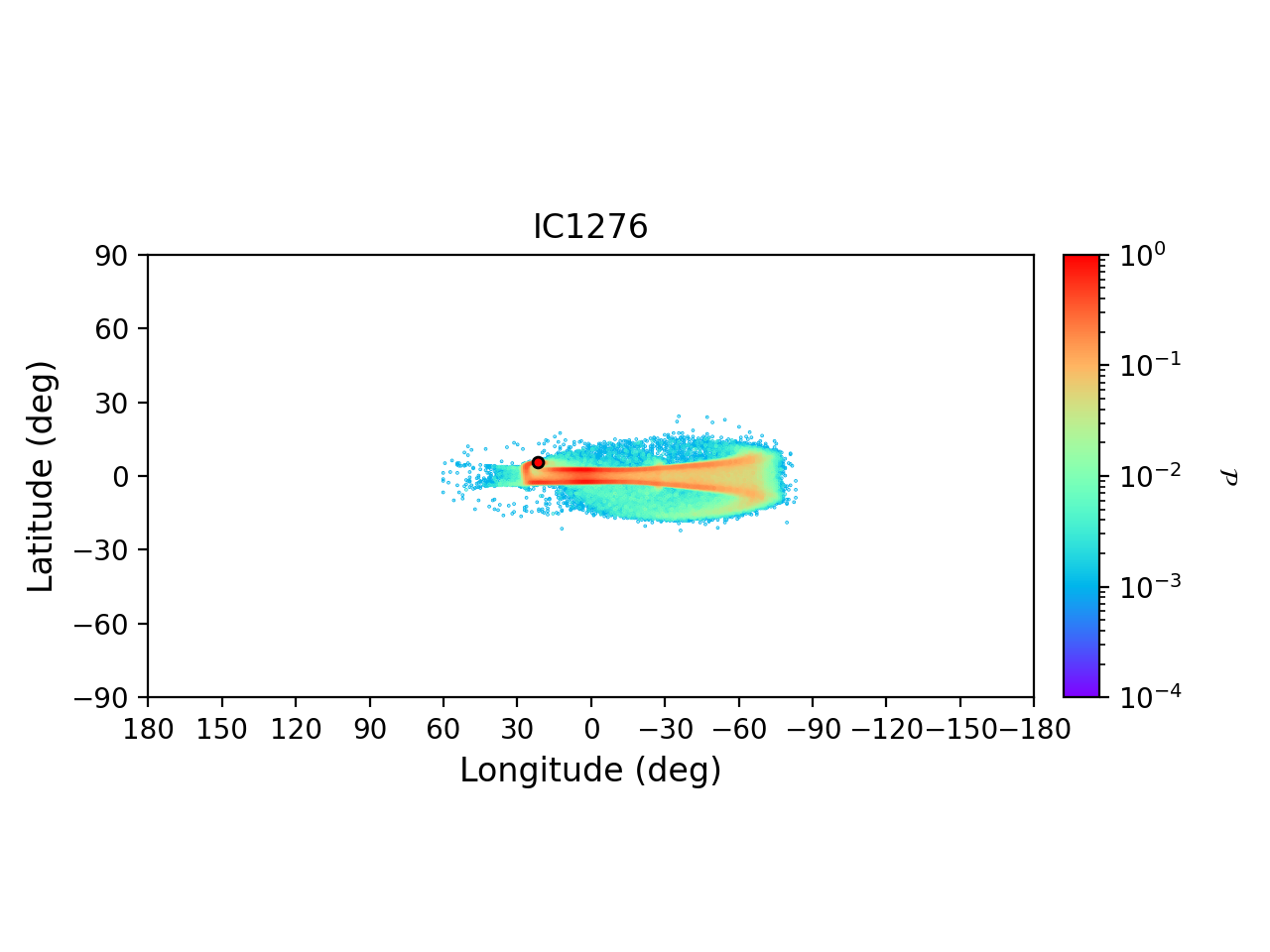}
\includegraphics[clip=true, trim = 0mm 20mm 0mm 10mm, width=1\columnwidth]{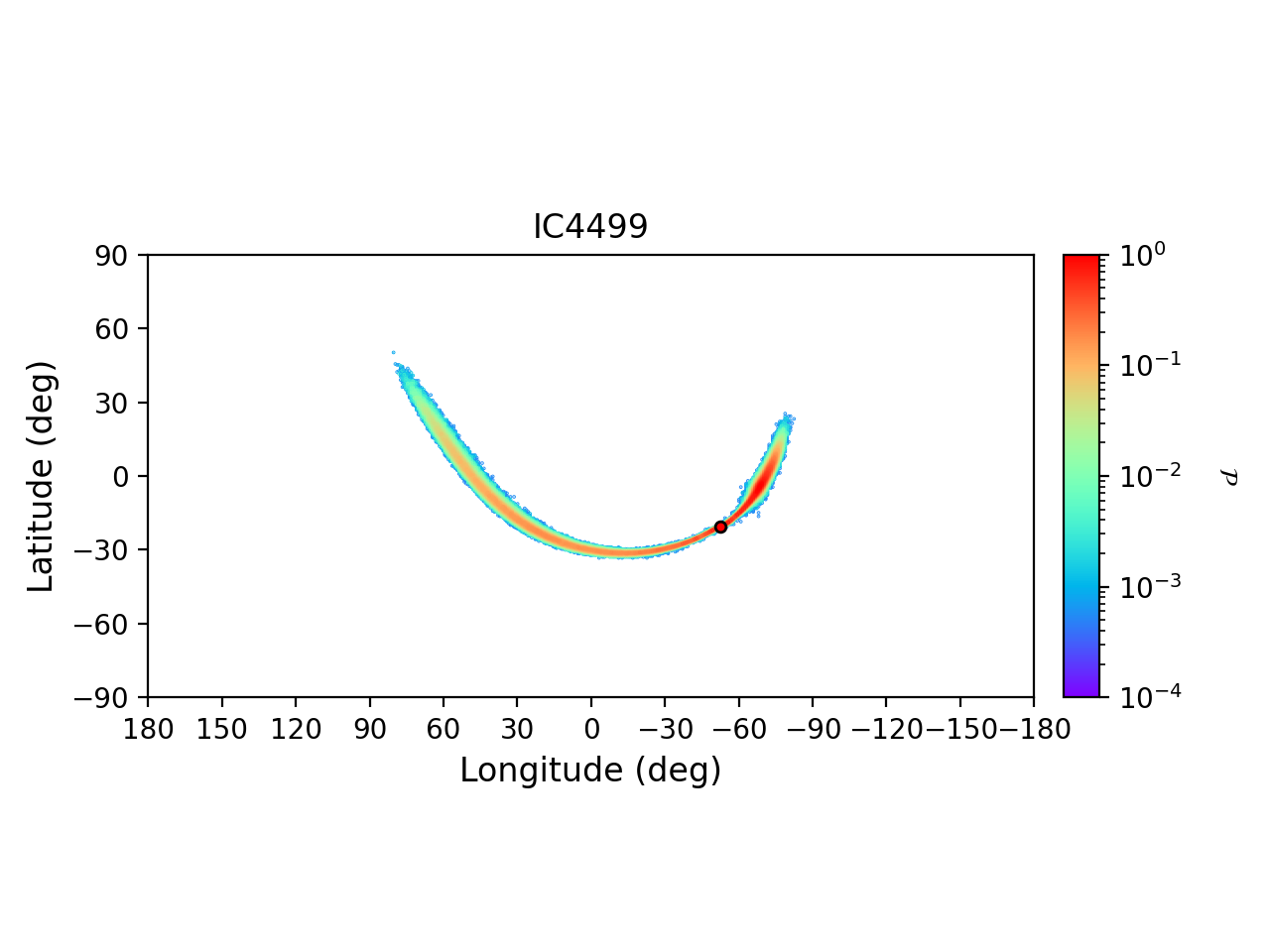}
\includegraphics[clip=true, trim = 0mm 20mm 0mm 10mm, width=1\columnwidth]{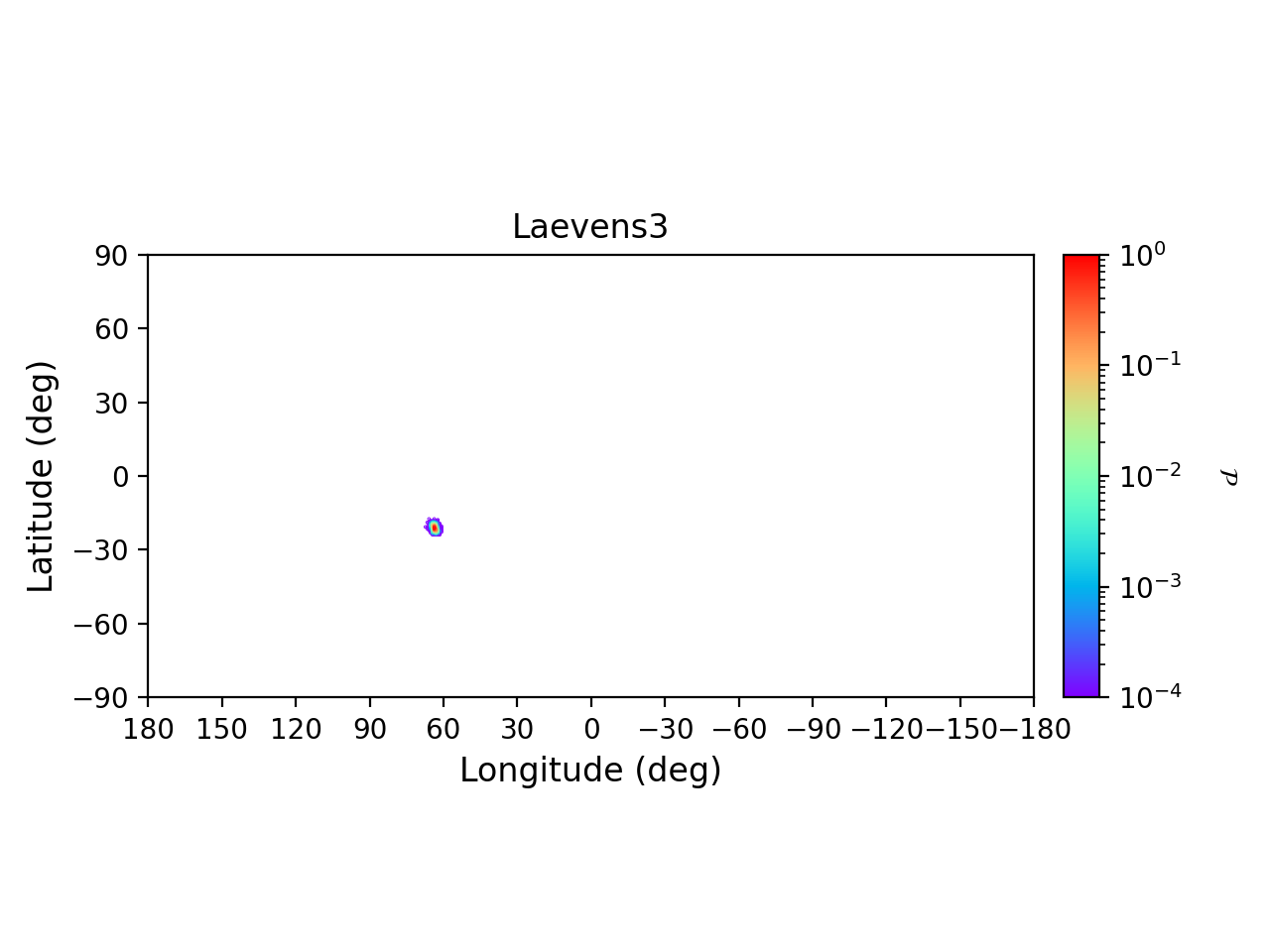}
\includegraphics[clip=true, trim = 0mm 20mm 0mm 10mm, width=1\columnwidth]{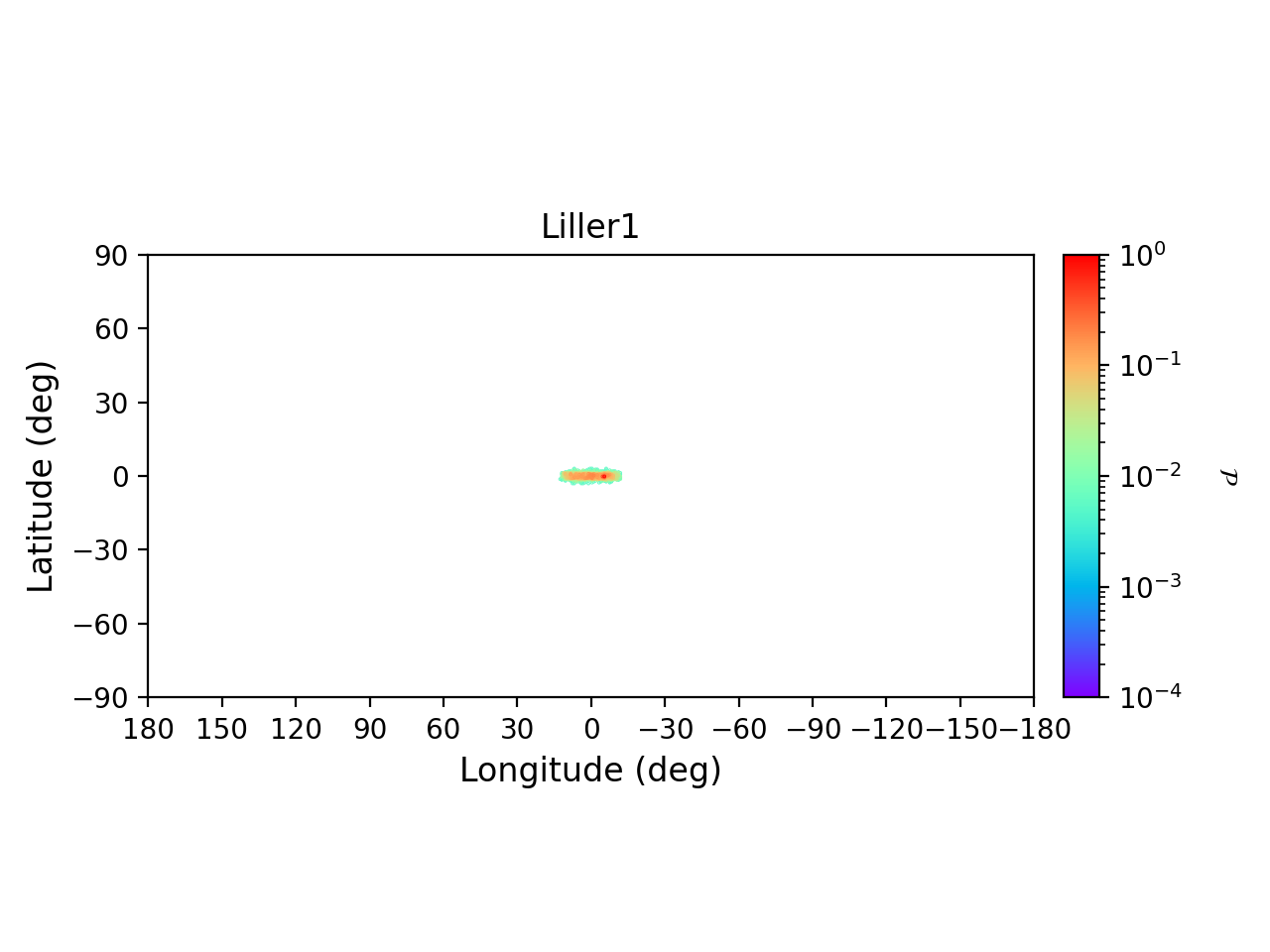}
\includegraphics[clip=true, trim = 0mm 20mm 0mm 10mm, width=1\columnwidth]{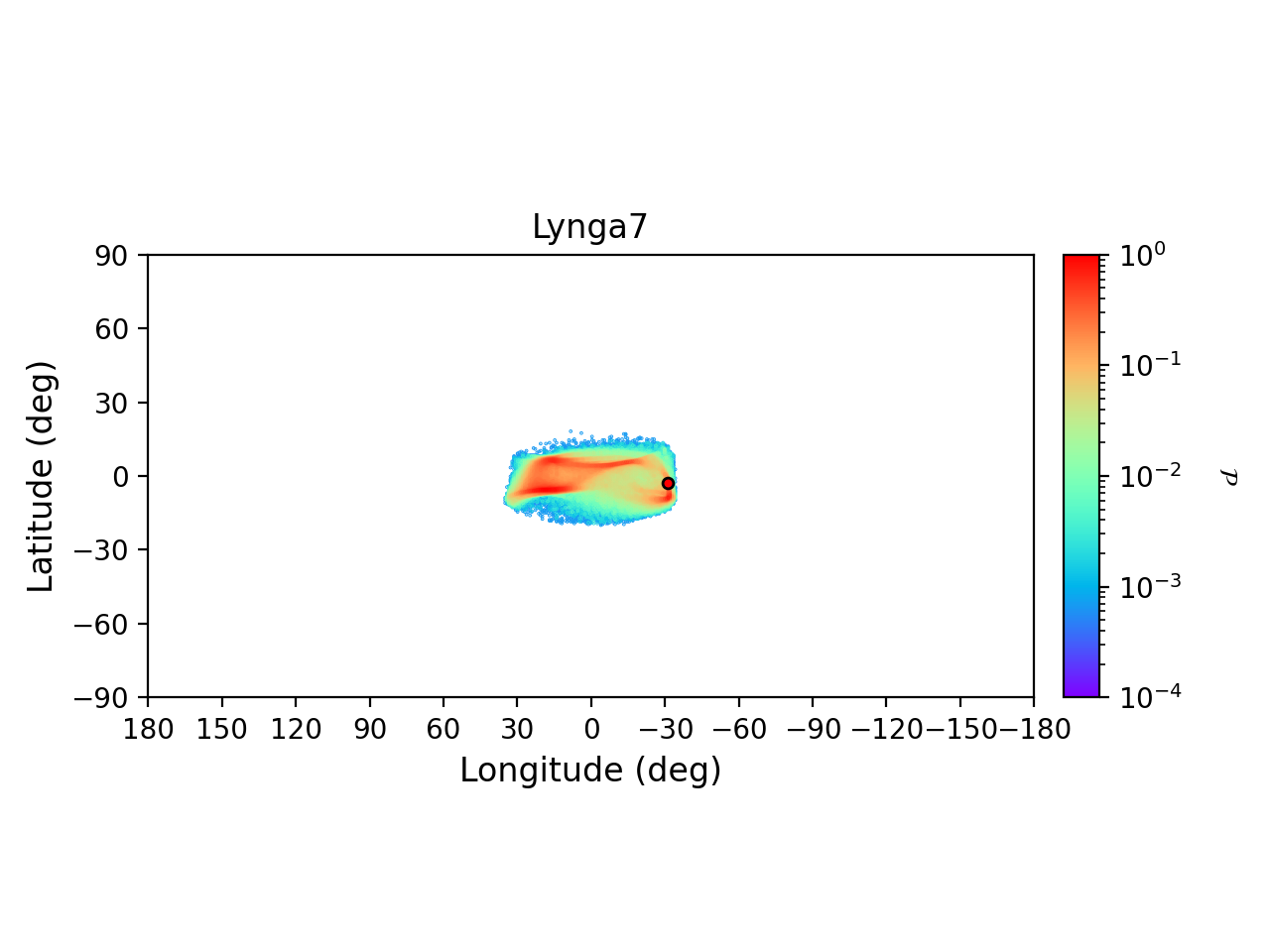}
\caption{Projected density distribution in the $(\ell, b)$ plane of a subset of simulated globular clusters, as indicated at the top of each panel. In each panel, the red circle indicates the current position of the cluster. The densities have been normalized to their maximum value.}\label{stream3}
\end{figure*}

\begin{figure*}
\includegraphics[clip=true, trim = 0mm 20mm 0mm 10mm, width=1\columnwidth]{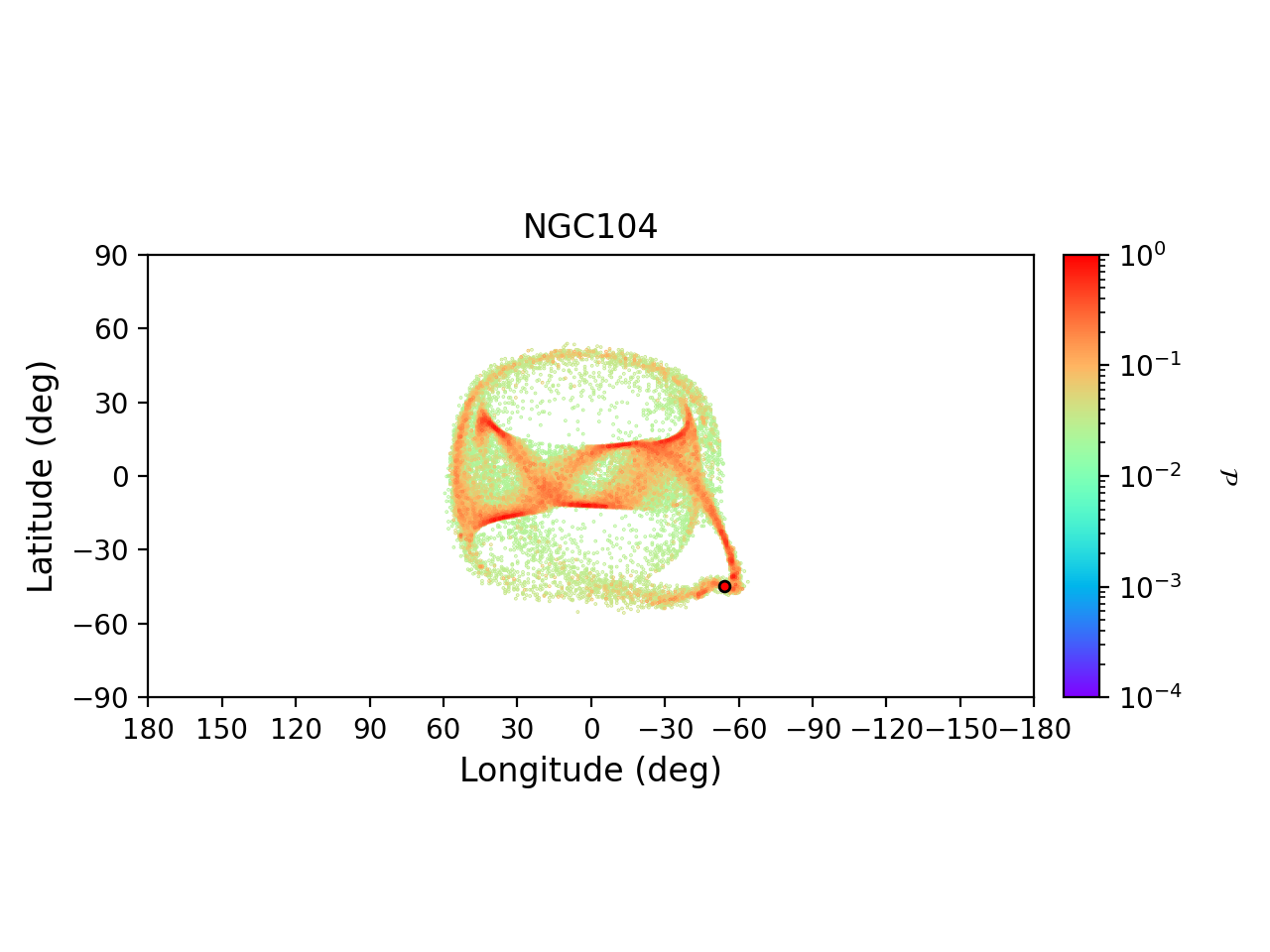}
\includegraphics[clip=true, trim = 0mm 20mm 0mm 10mm, width=1\columnwidth]{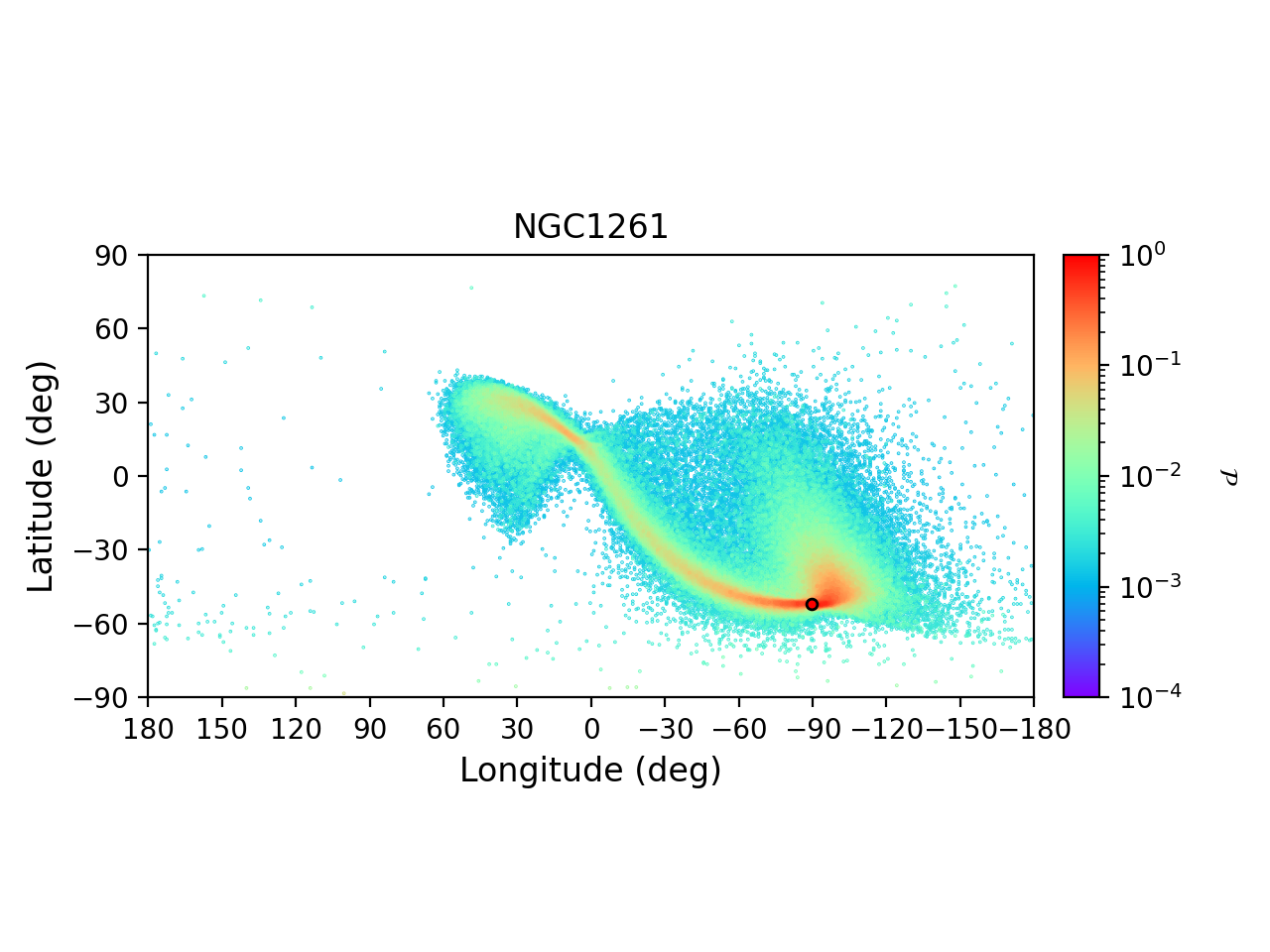}
\includegraphics[clip=true, trim = 0mm 20mm 0mm 10mm, width=1\columnwidth]{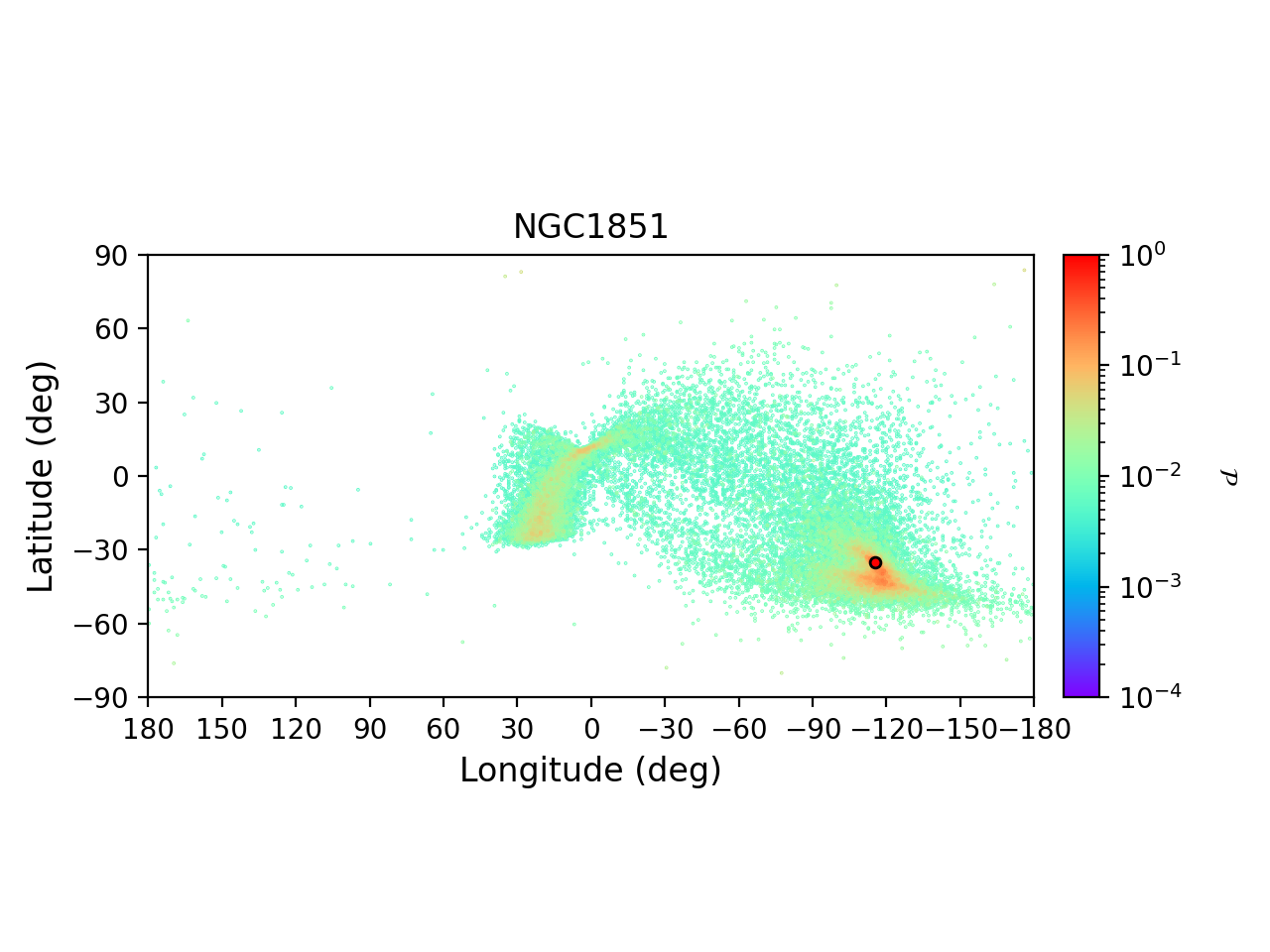}
\includegraphics[clip=true, trim = 0mm 20mm 0mm 10mm, width=1\columnwidth]{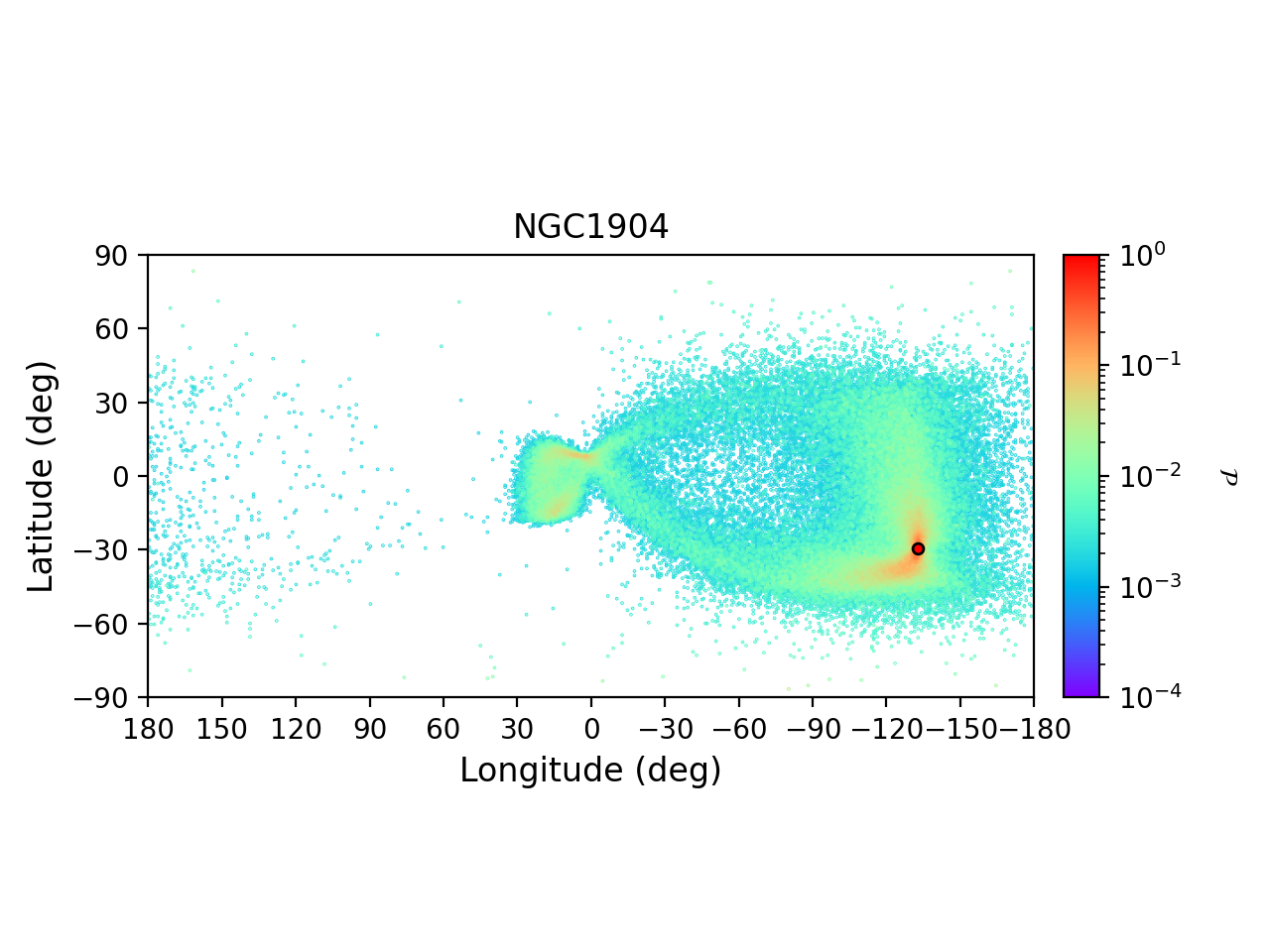}
\includegraphics[clip=true, trim = 0mm 20mm 0mm 10mm, width=1\columnwidth]{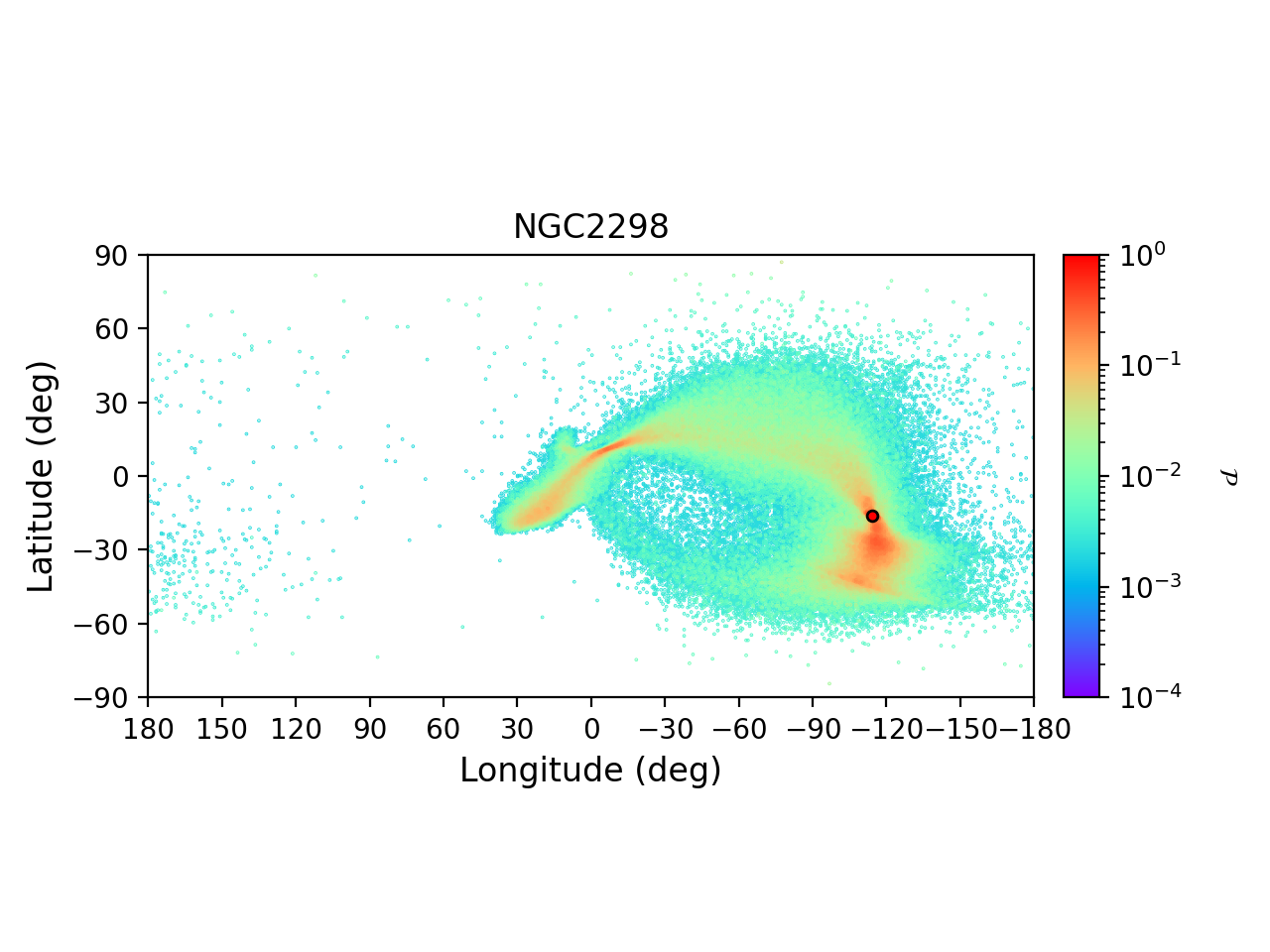}
\includegraphics[clip=true, trim = 0mm 20mm 0mm 10mm, width=1\columnwidth]{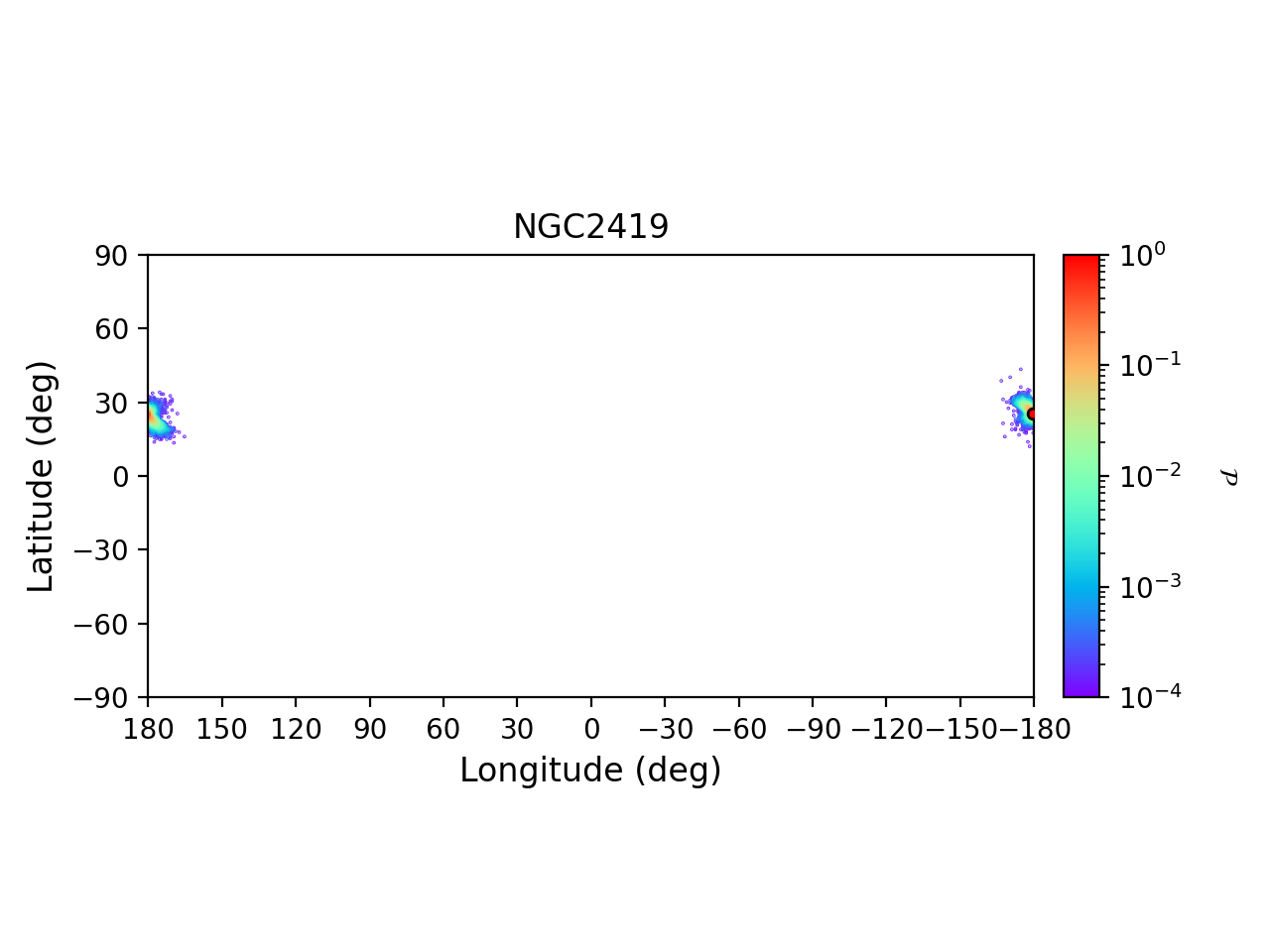}
\includegraphics[clip=true, trim = 0mm 20mm 0mm 10mm, width=1\columnwidth]{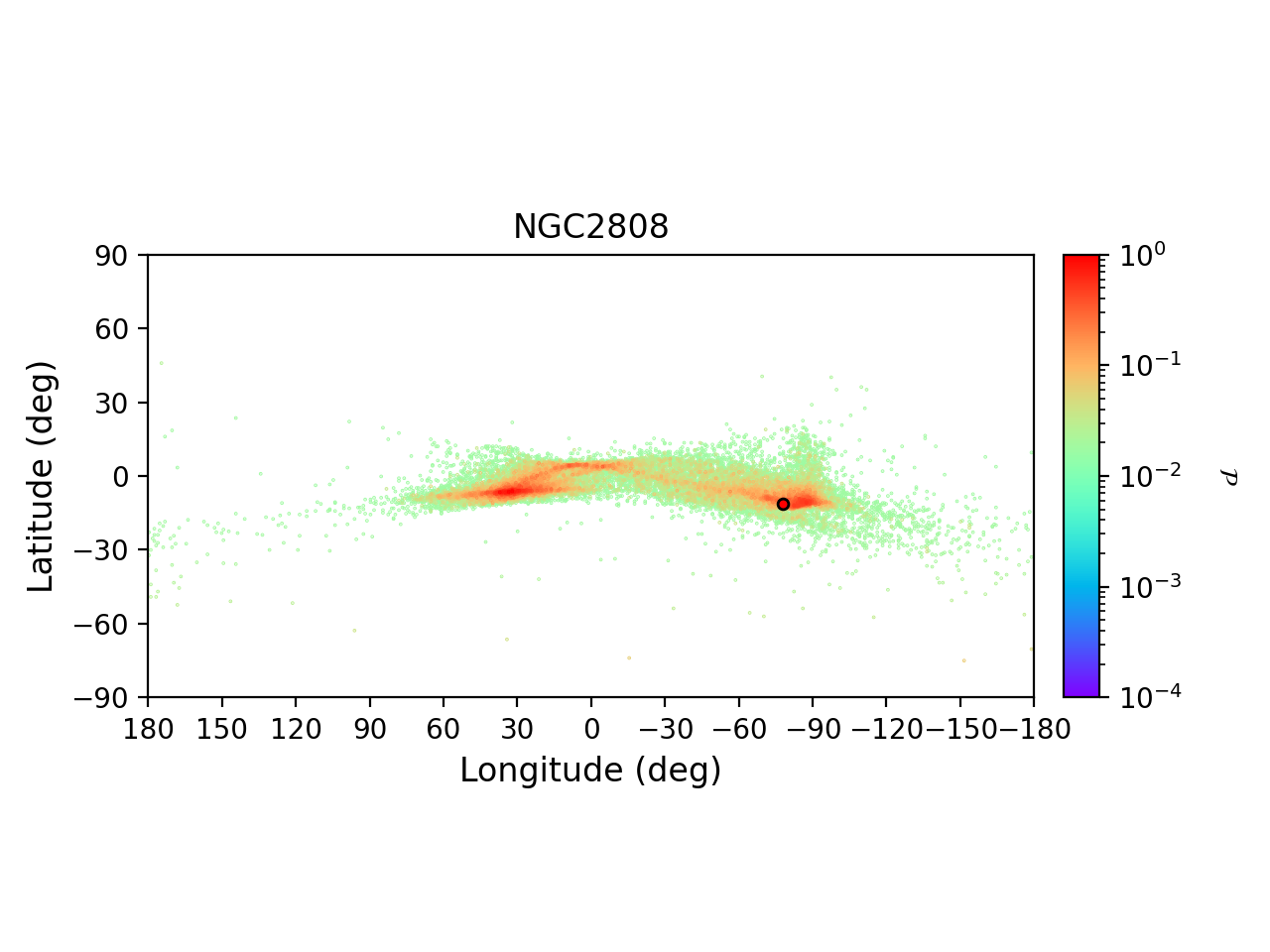}
\includegraphics[clip=true, trim = 0mm 20mm 0mm 10mm, width=1\columnwidth]{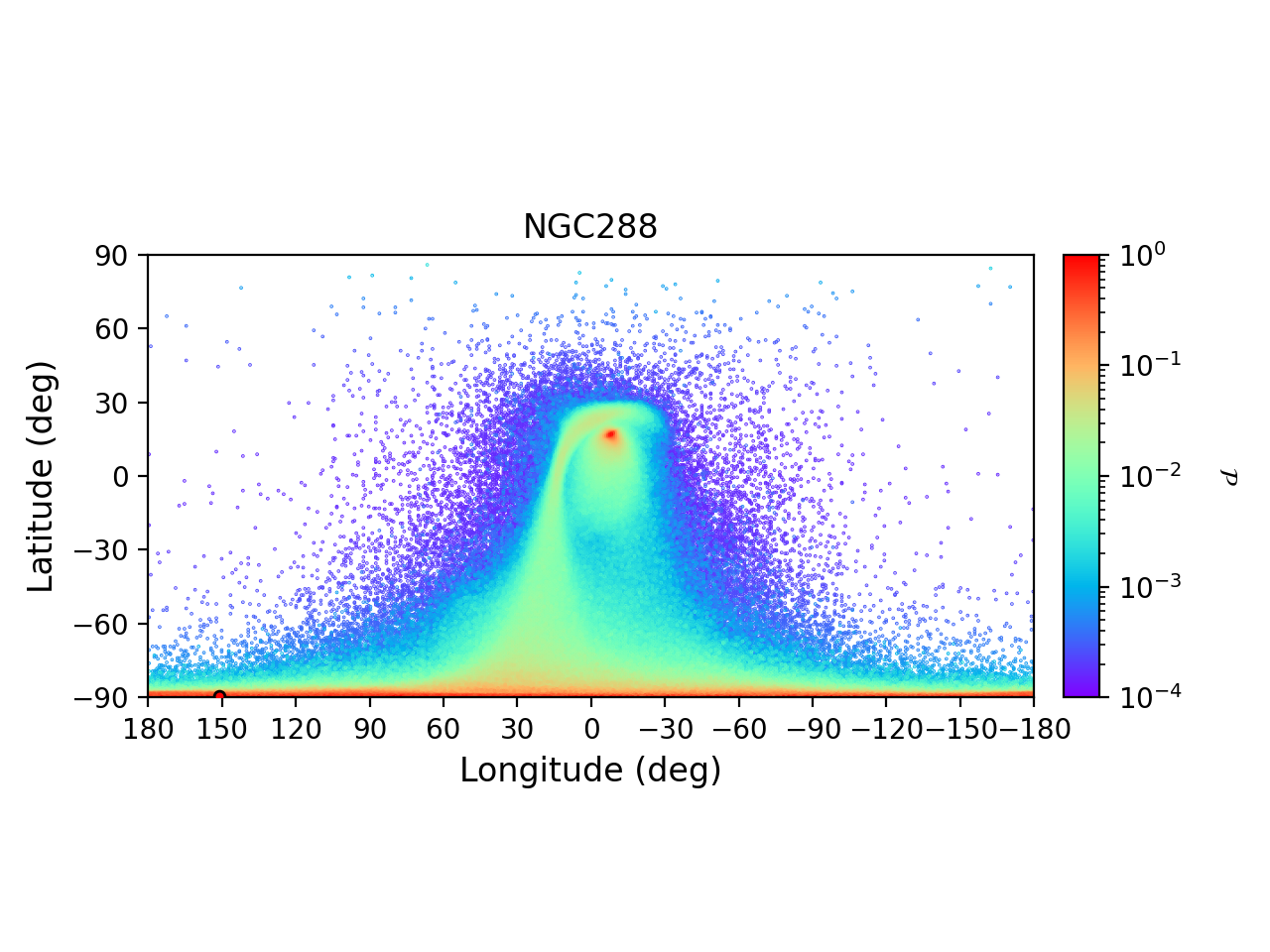}
\caption{Projected density distribution in the $(\ell, b)$ plane of a subset of simulated globular clusters, as indicated at the top of each panel. In each panel, the red circle indicates the current position of the cluster. The densities have been normalized to their maximum value.}\label{stream4}
\end{figure*}

\begin{figure*}
\includegraphics[clip=true, trim = 0mm 20mm 0mm 10mm, width=1\columnwidth]{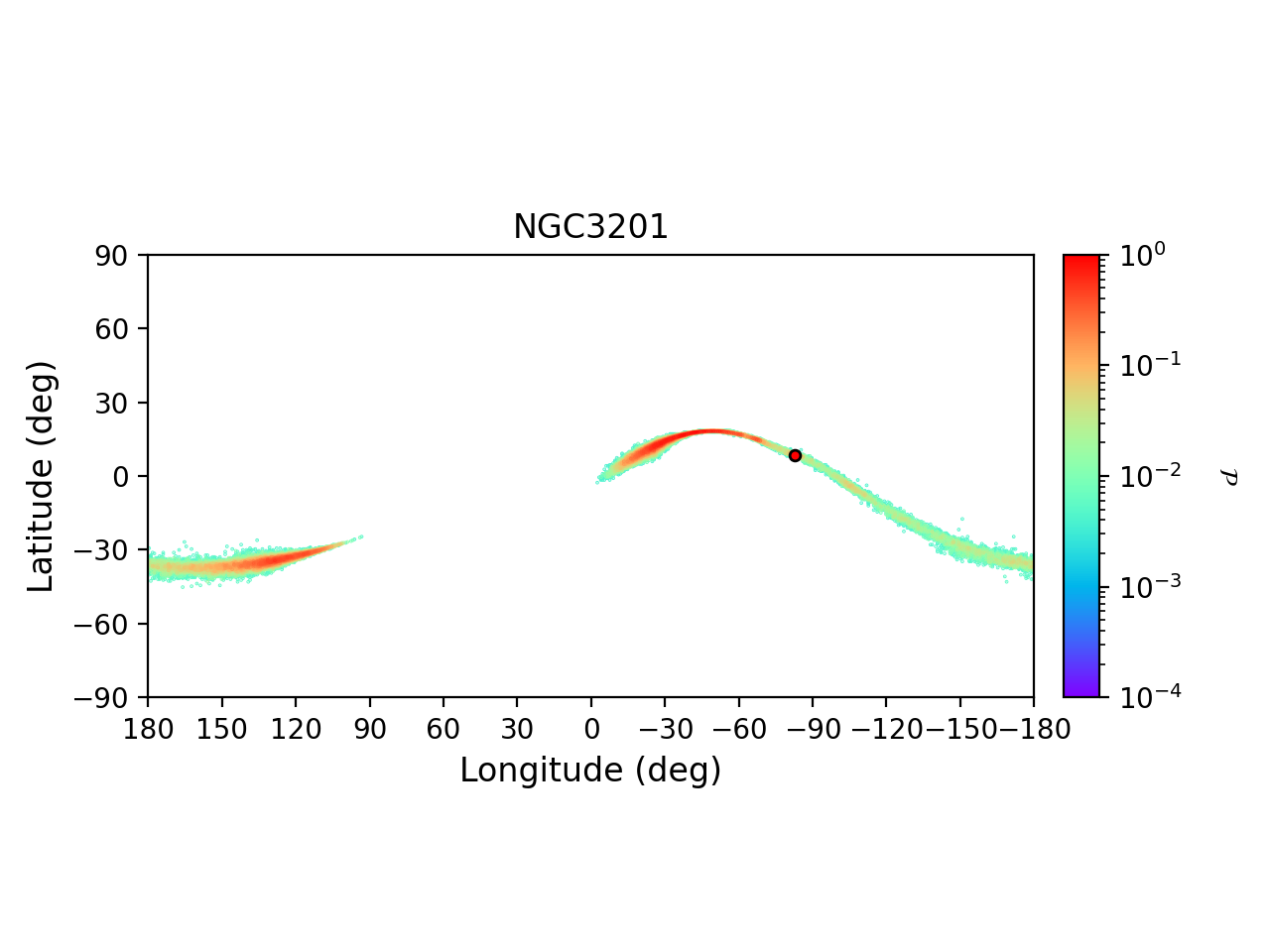}
\includegraphics[clip=true, trim = 0mm 20mm 0mm 10mm, width=1\columnwidth]{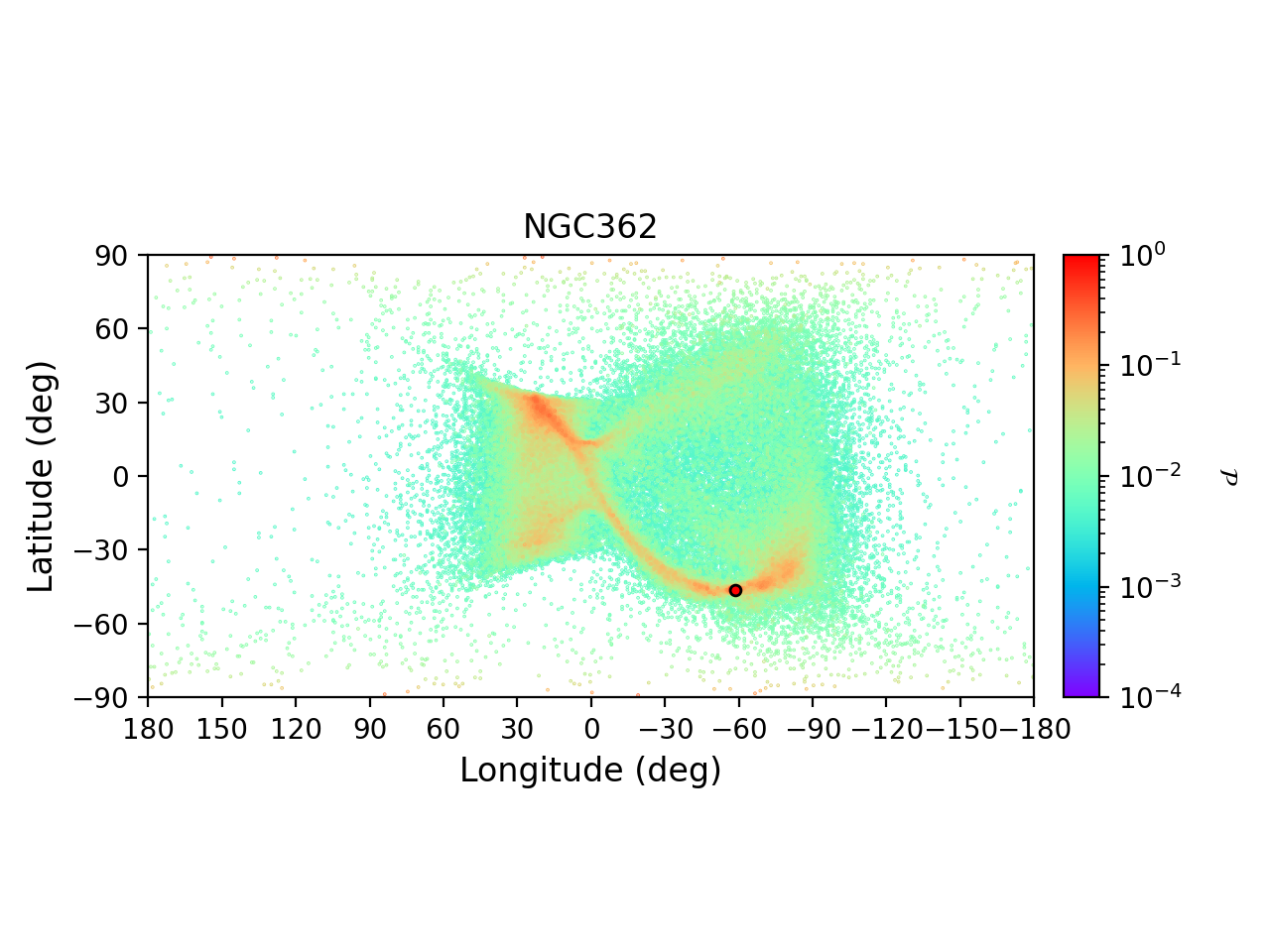}
\includegraphics[clip=true, trim = 0mm 20mm 0mm 10mm, width=1\columnwidth]{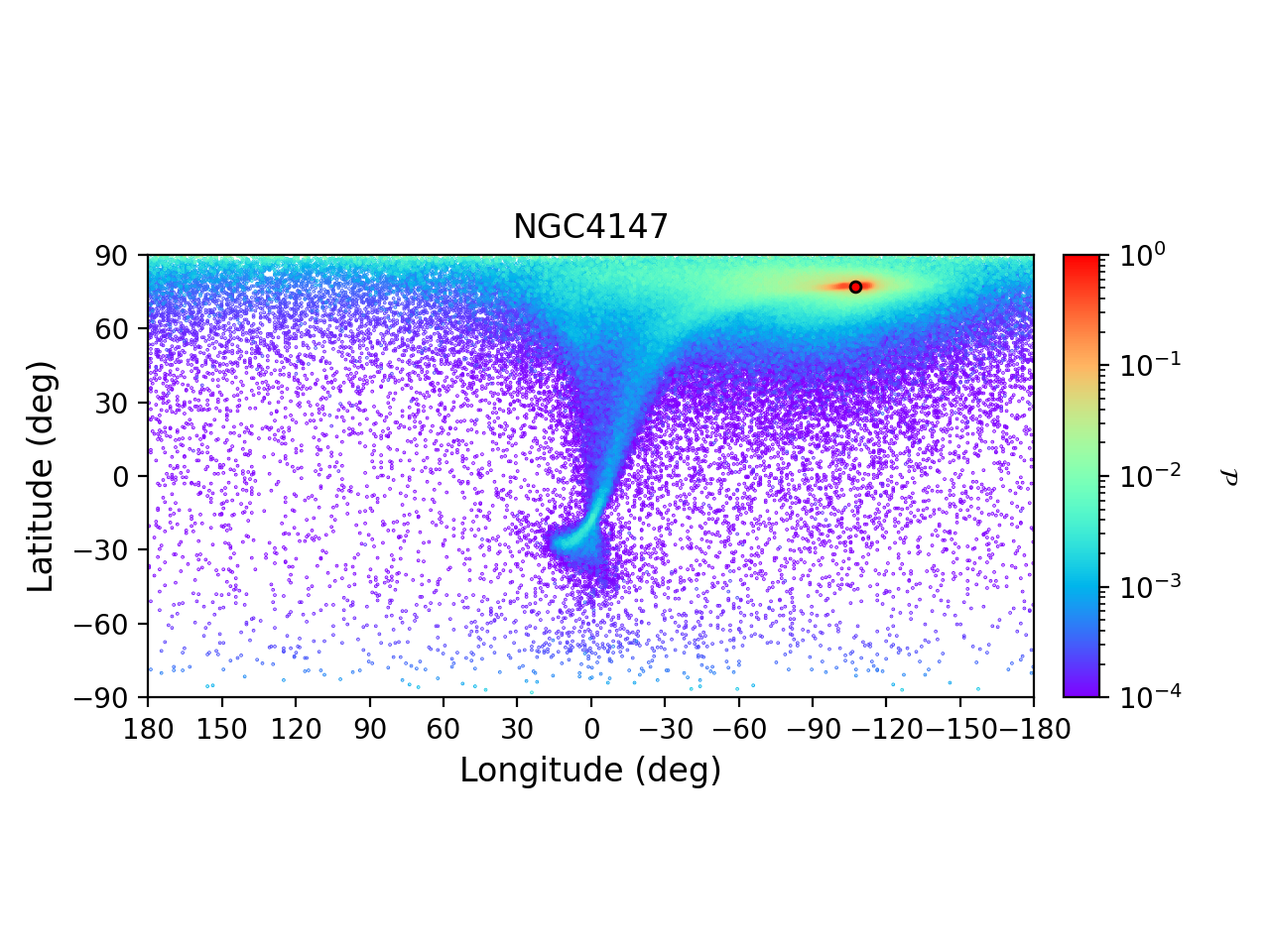}
\includegraphics[clip=true, trim = 0mm 20mm 0mm 10mm, width=1\columnwidth]{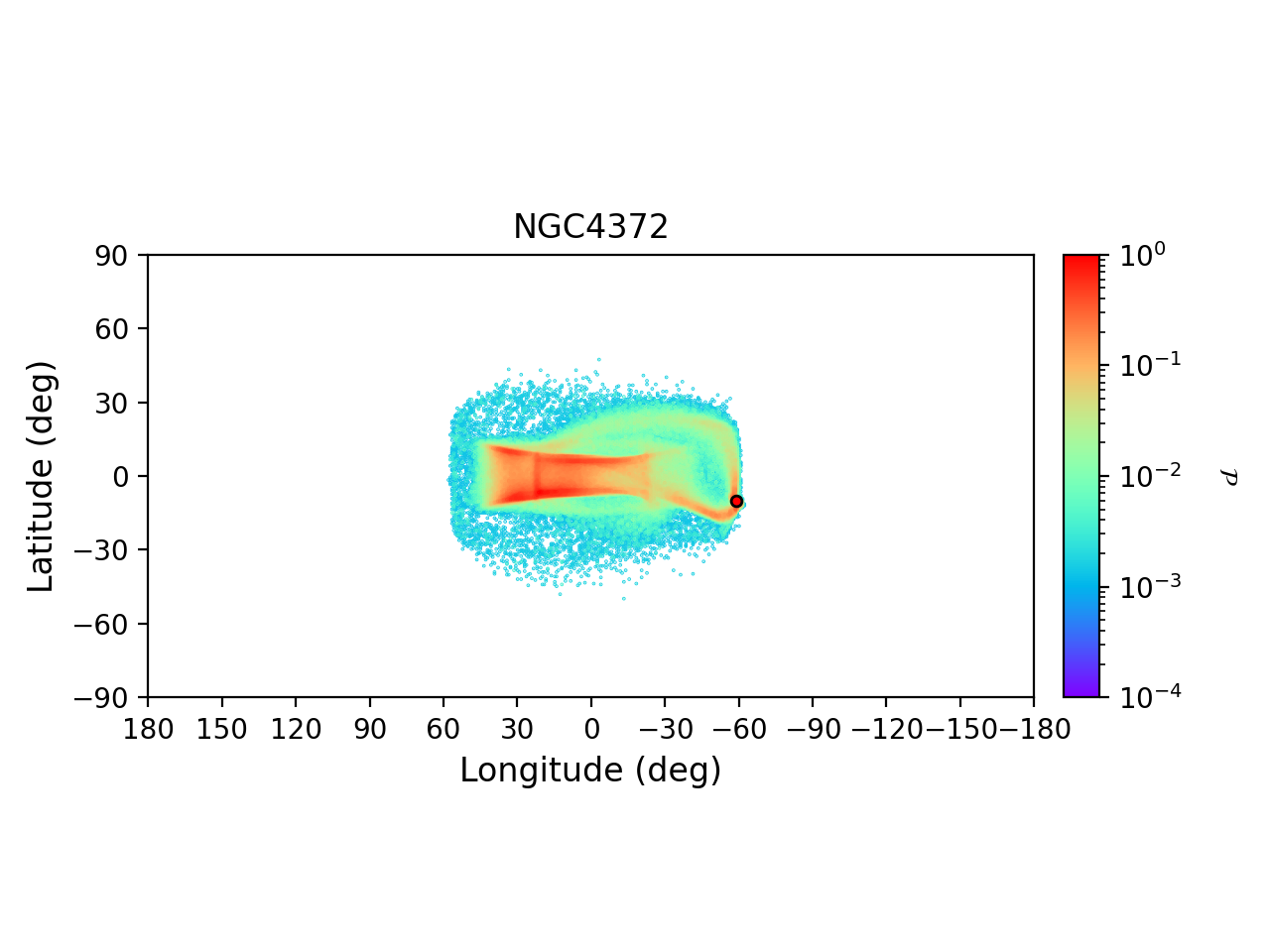}
\includegraphics[clip=true, trim = 0mm 20mm 0mm 10mm, width=1\columnwidth]{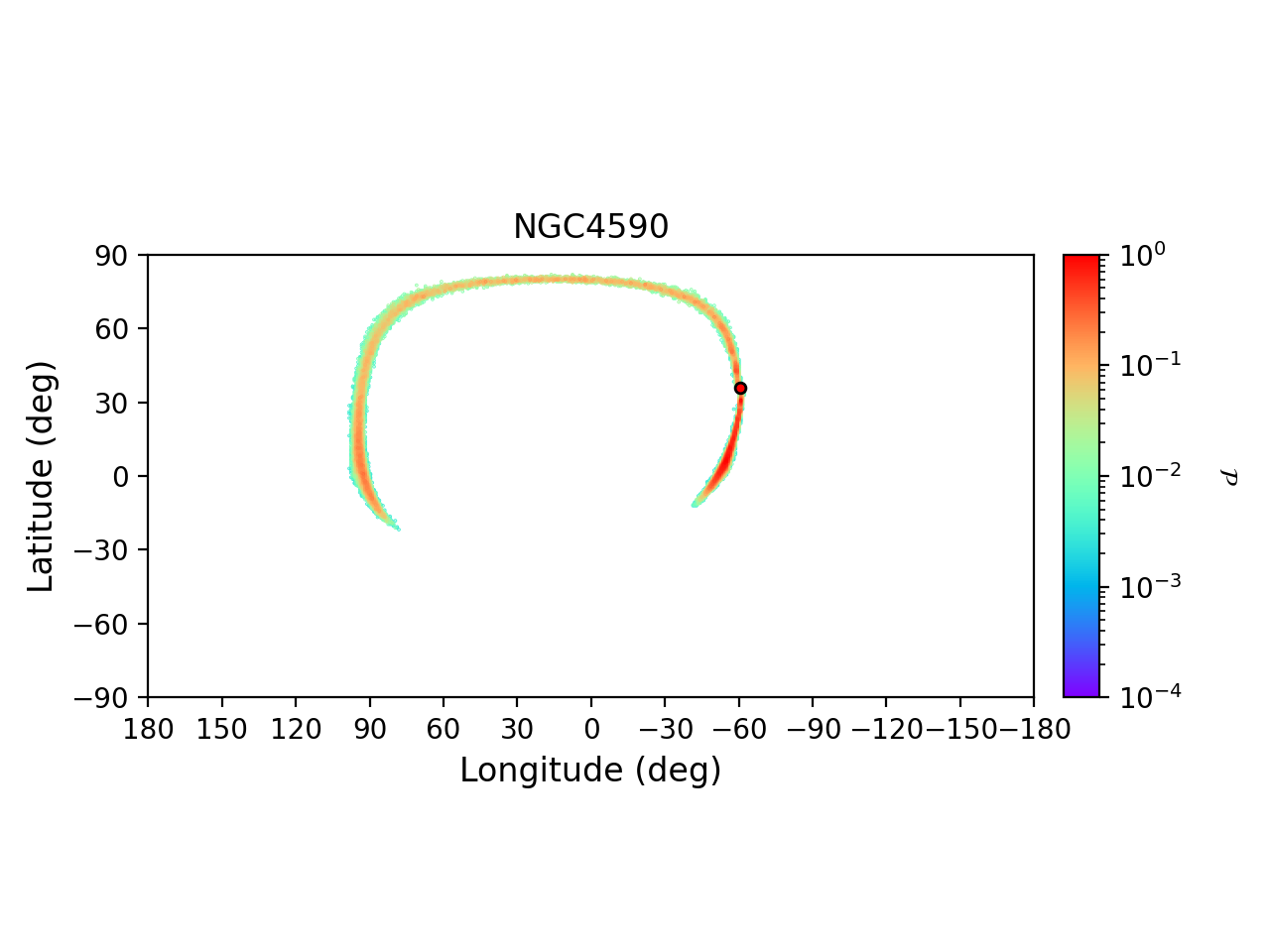}
\includegraphics[clip=true, trim = 0mm 20mm 0mm 10mm, width=1\columnwidth]{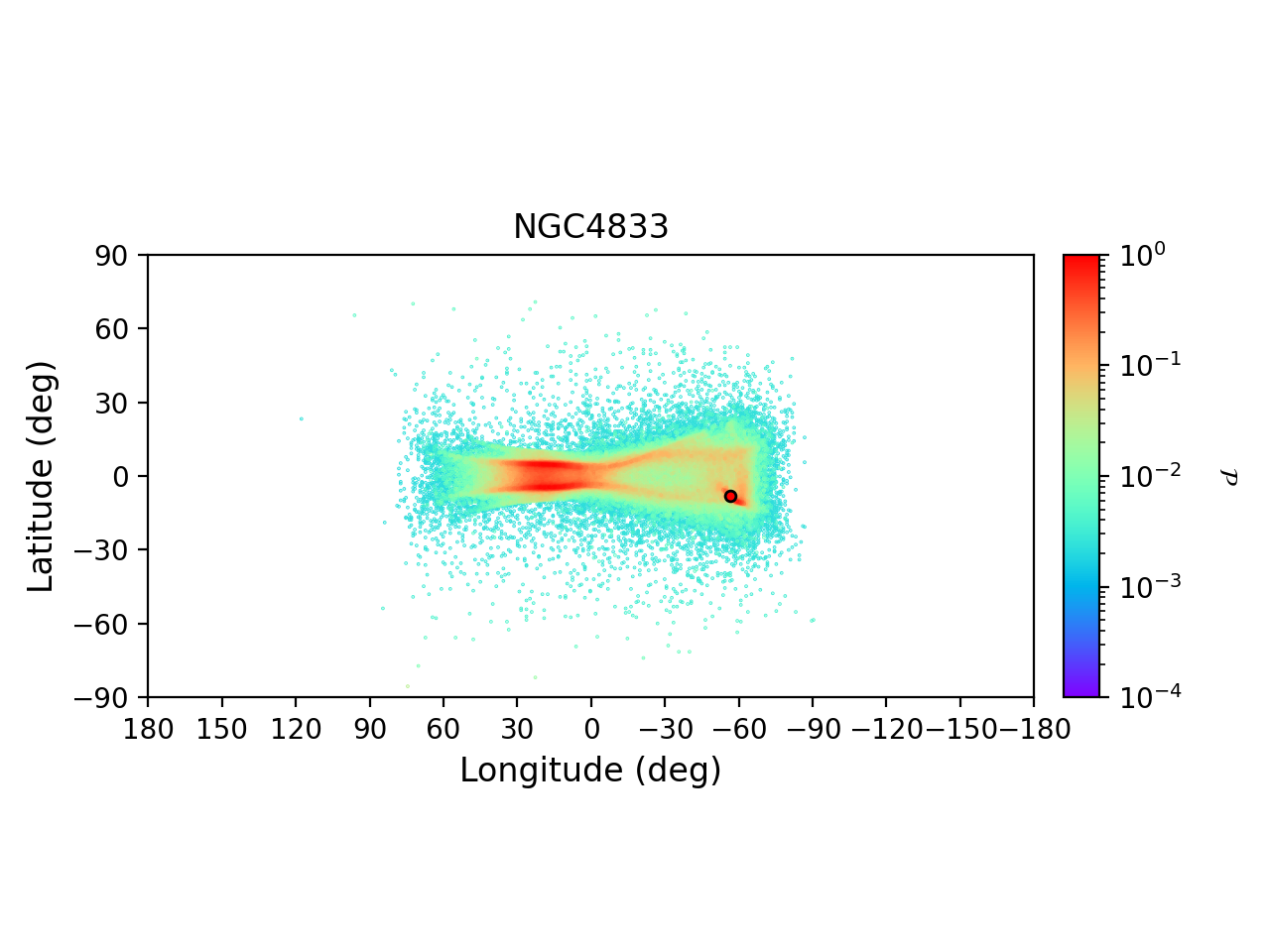}
\includegraphics[clip=true, trim = 0mm 20mm 0mm 10mm, width=1\columnwidth]{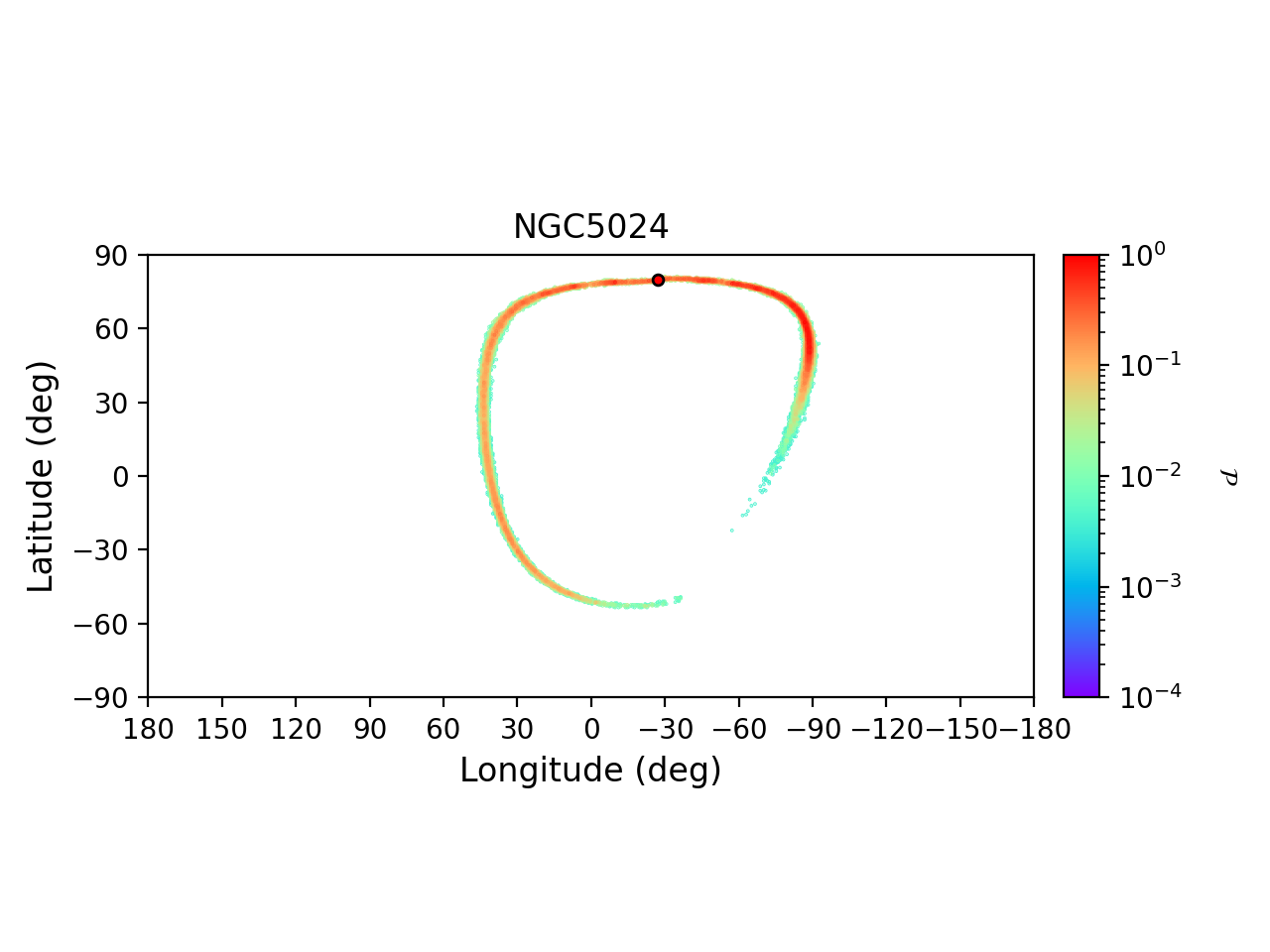}
\includegraphics[clip=true, trim = 0mm 20mm 0mm 10mm, width=1\columnwidth]{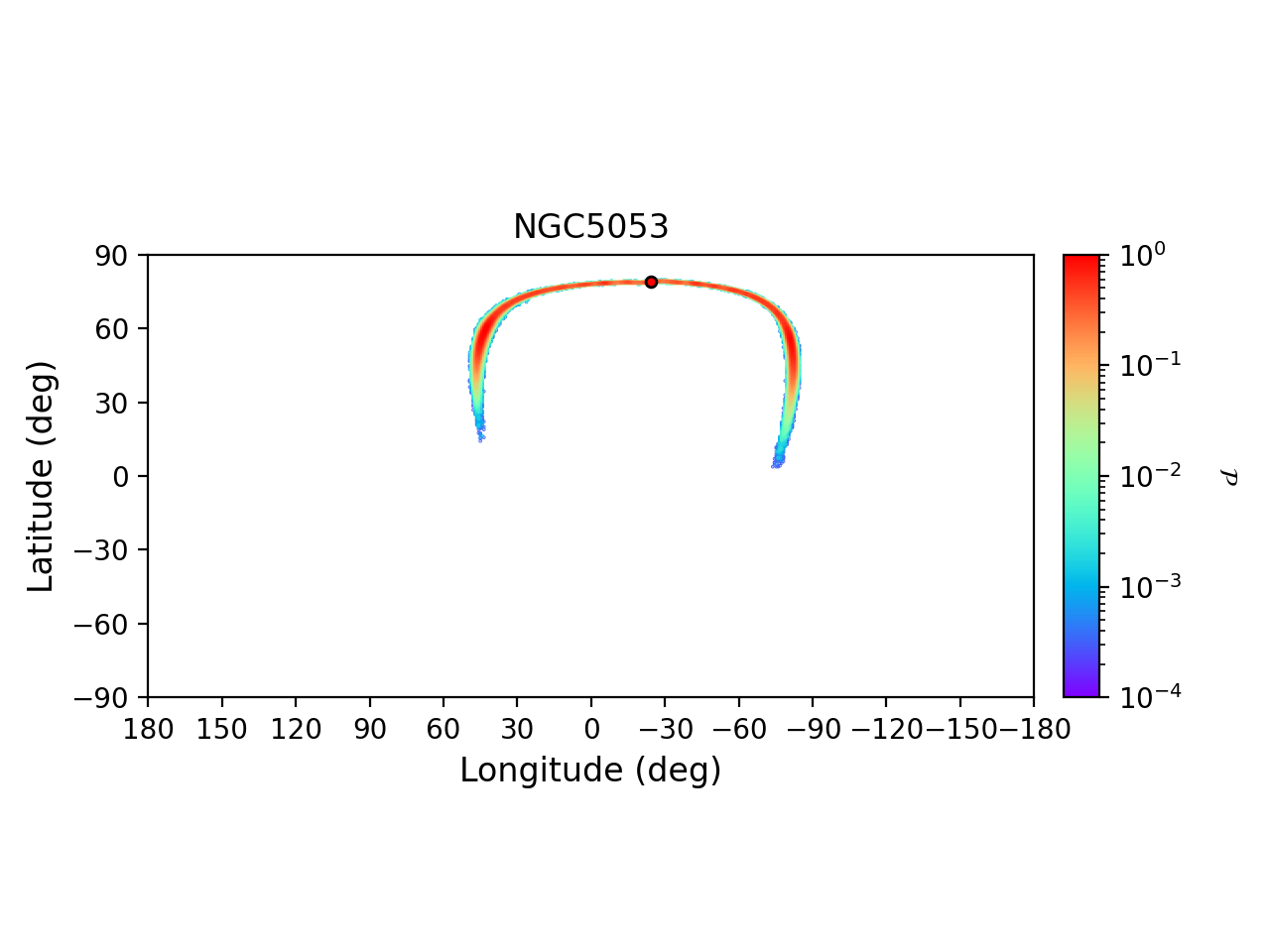}
\caption{Projected density distribution in the $(\ell, b)$ plane of a subset of simulated globular clusters, as indicated at the top of each panel. In each panel, the red circle indicates the current position of the cluster. The densities have been normalized to their maximum value.}\label{stream5}
\end{figure*}

\begin{figure*}
\includegraphics[clip=true, trim = 0mm 20mm 0mm 10mm, width=1\columnwidth]{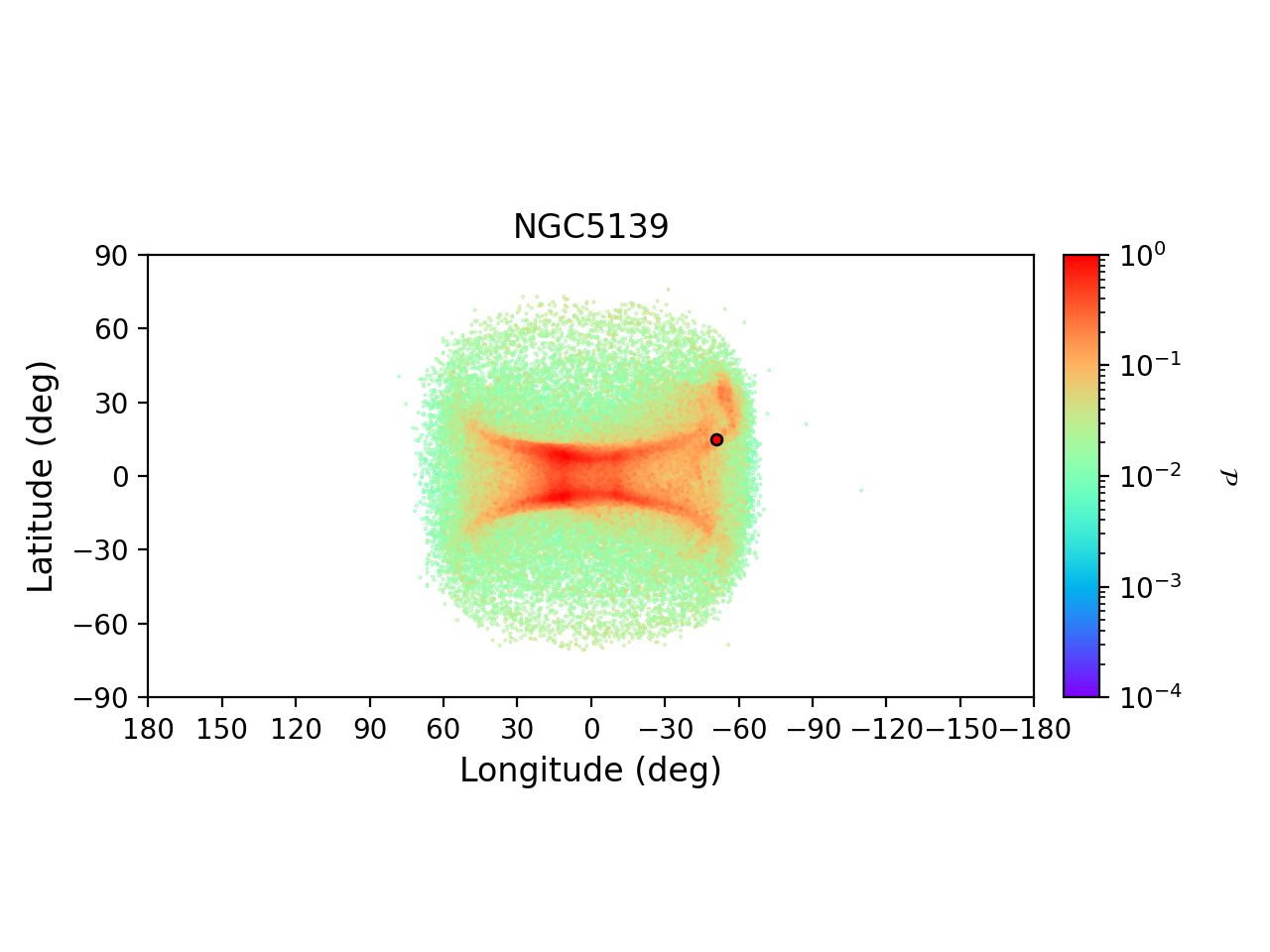}
\includegraphics[clip=true, trim = 0mm 20mm 0mm 10mm, width=1\columnwidth]{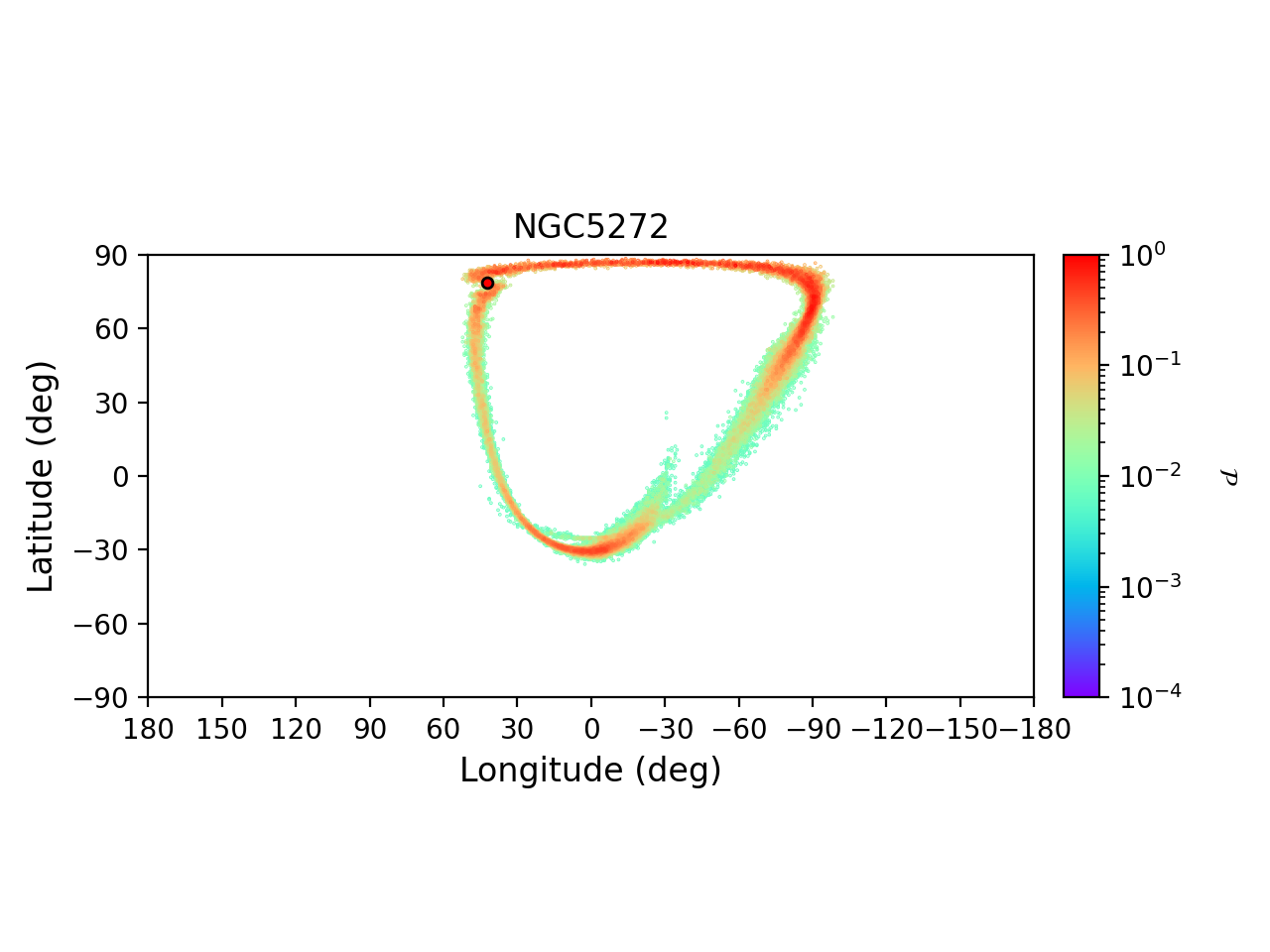}
\includegraphics[clip=true, trim = 0mm 20mm 0mm 10mm, width=1\columnwidth]{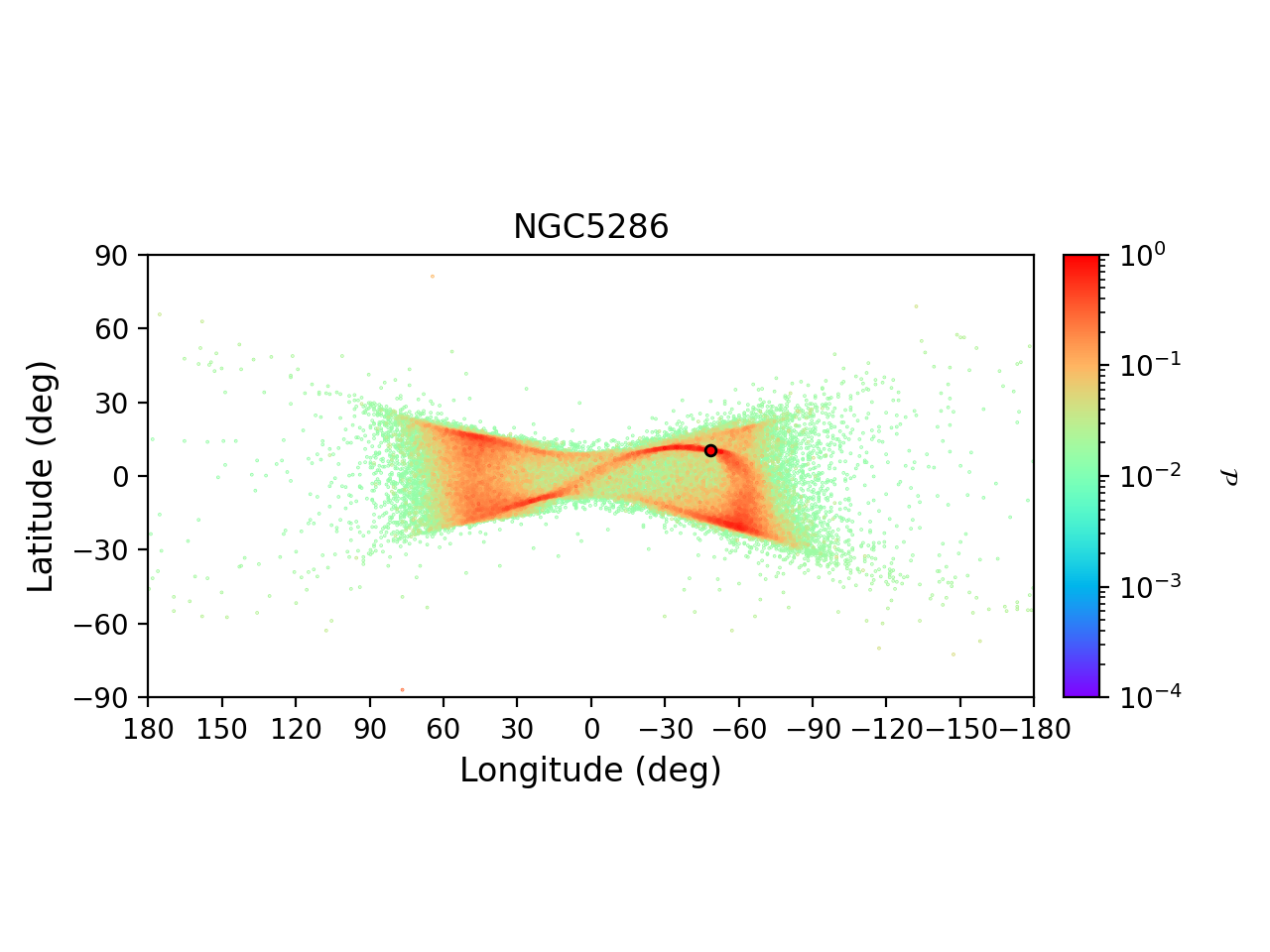}
\includegraphics[clip=true, trim = 0mm 20mm 0mm 10mm, width=1\columnwidth]{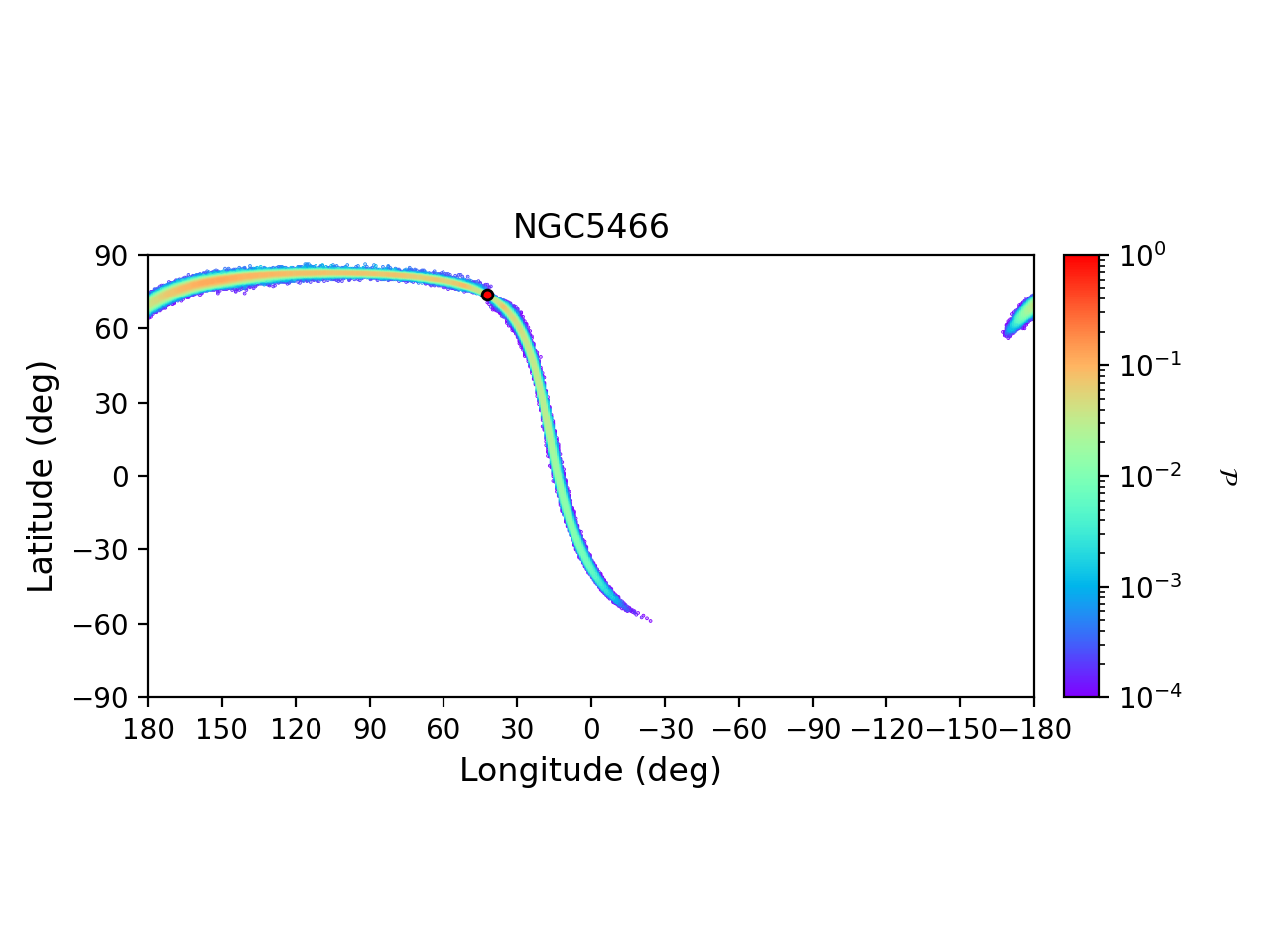}
\includegraphics[clip=true, trim = 0mm 20mm 0mm 10mm, width=1\columnwidth]{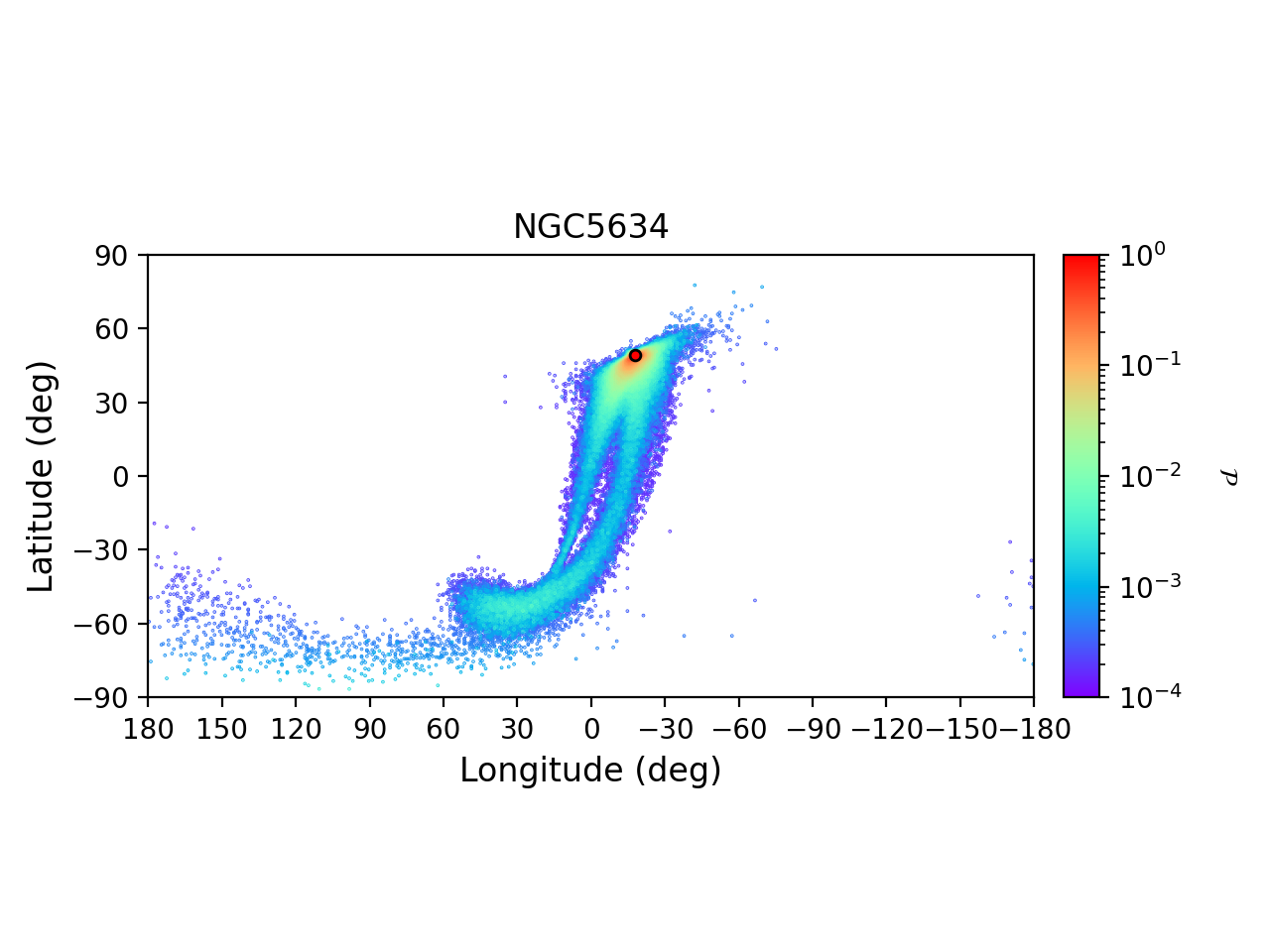}
\includegraphics[clip=true, trim = 0mm 20mm 0mm 10mm, width=1\columnwidth]{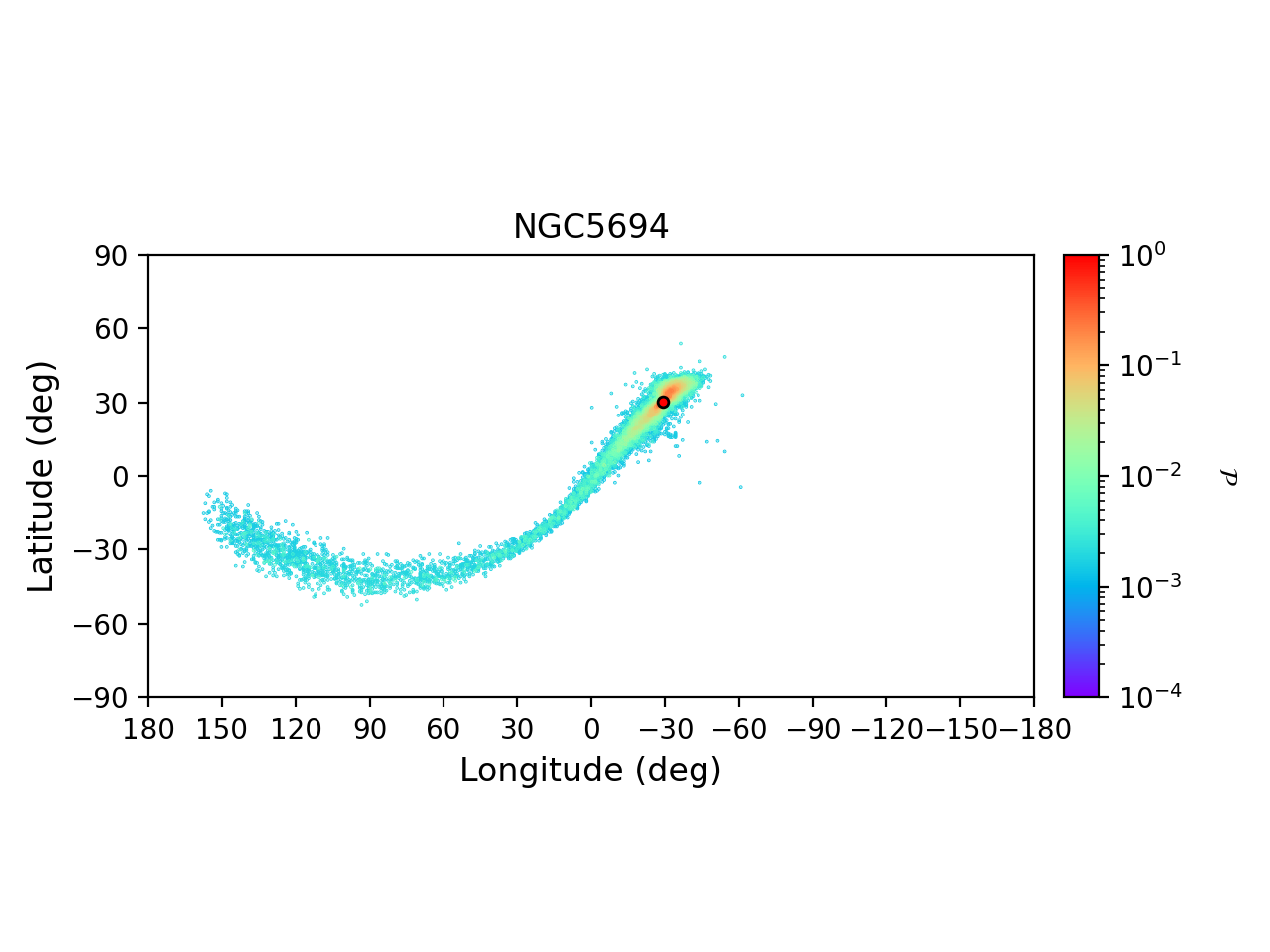}
\includegraphics[clip=true, trim = 0mm 20mm 0mm 10mm, width=1\columnwidth]{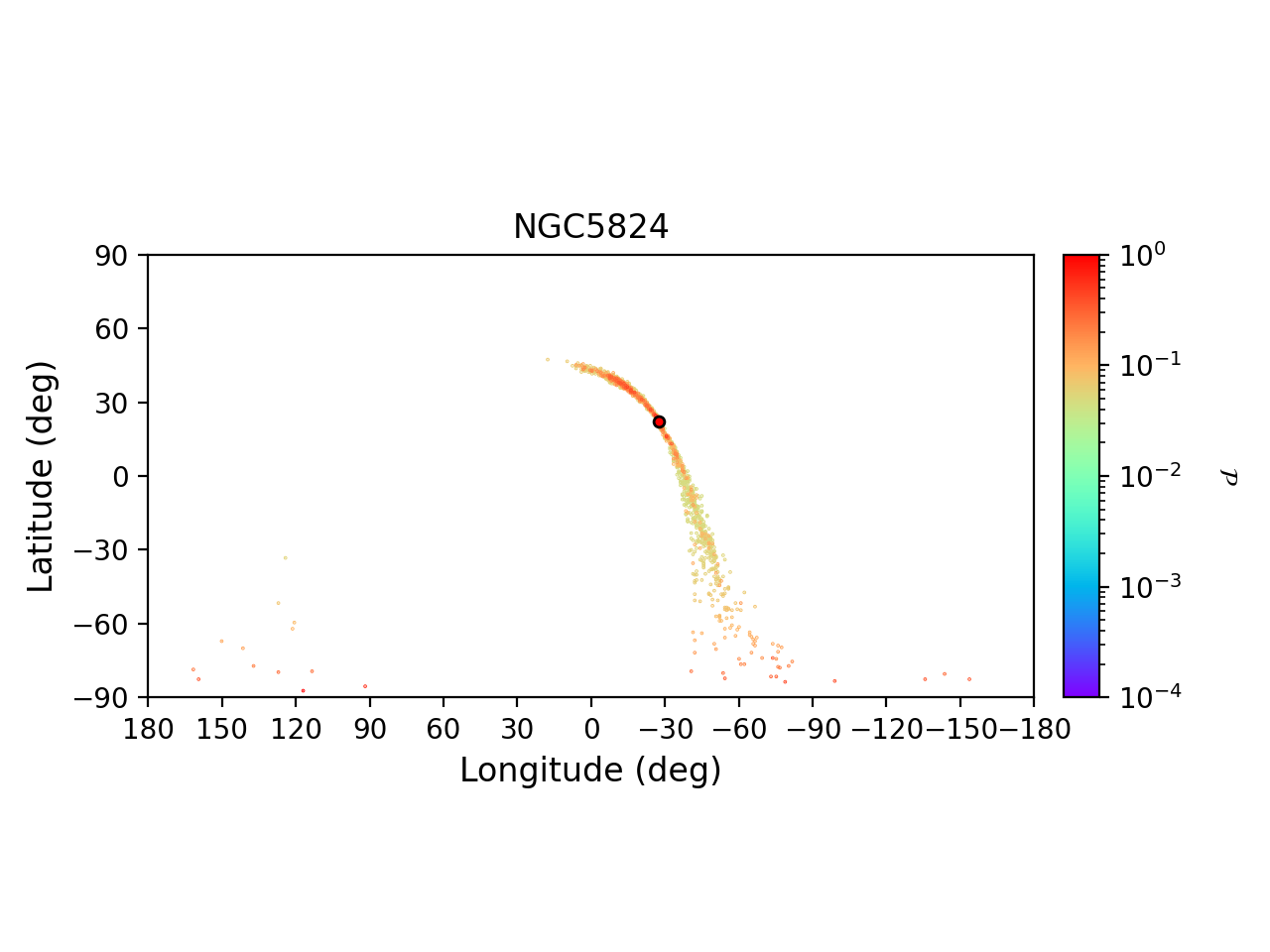}
\includegraphics[clip=true, trim = 0mm 20mm 0mm 10mm, width=1\columnwidth]{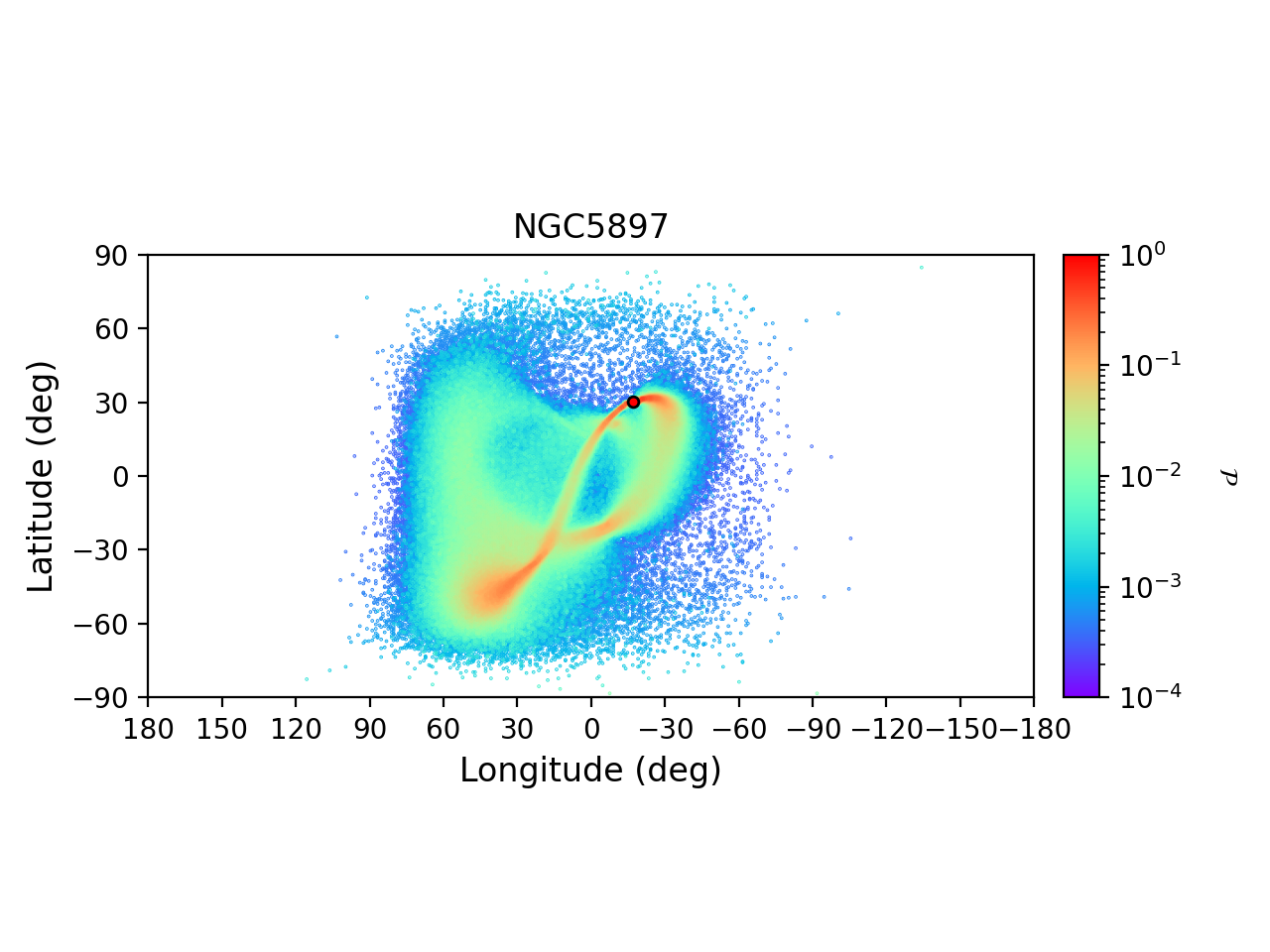}
\caption{Projected density distribution in the $(\ell, b)$ plane of a subset of simulated globular clusters, as indicated at the top of each panel. In each panel, the red circle indicates the current position of the cluster. The densities have been normalized to their maximum value.}\label{stream6}
\end{figure*}

\begin{figure*}
\includegraphics[clip=true, trim = 0mm 20mm 0mm 10mm, width=1\columnwidth]{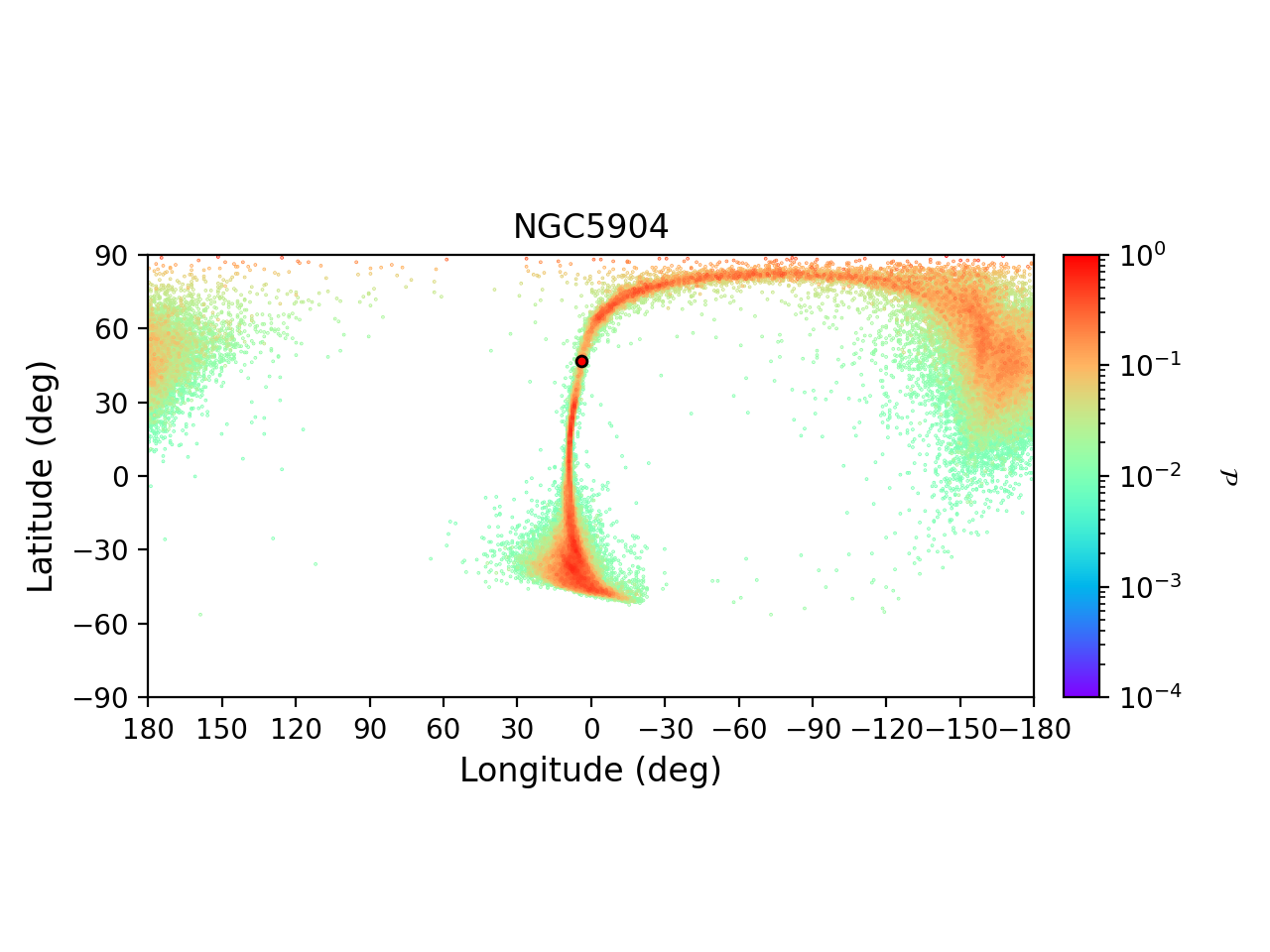}
\includegraphics[clip=true, trim = 0mm 20mm 0mm 10mm, width=1\columnwidth]{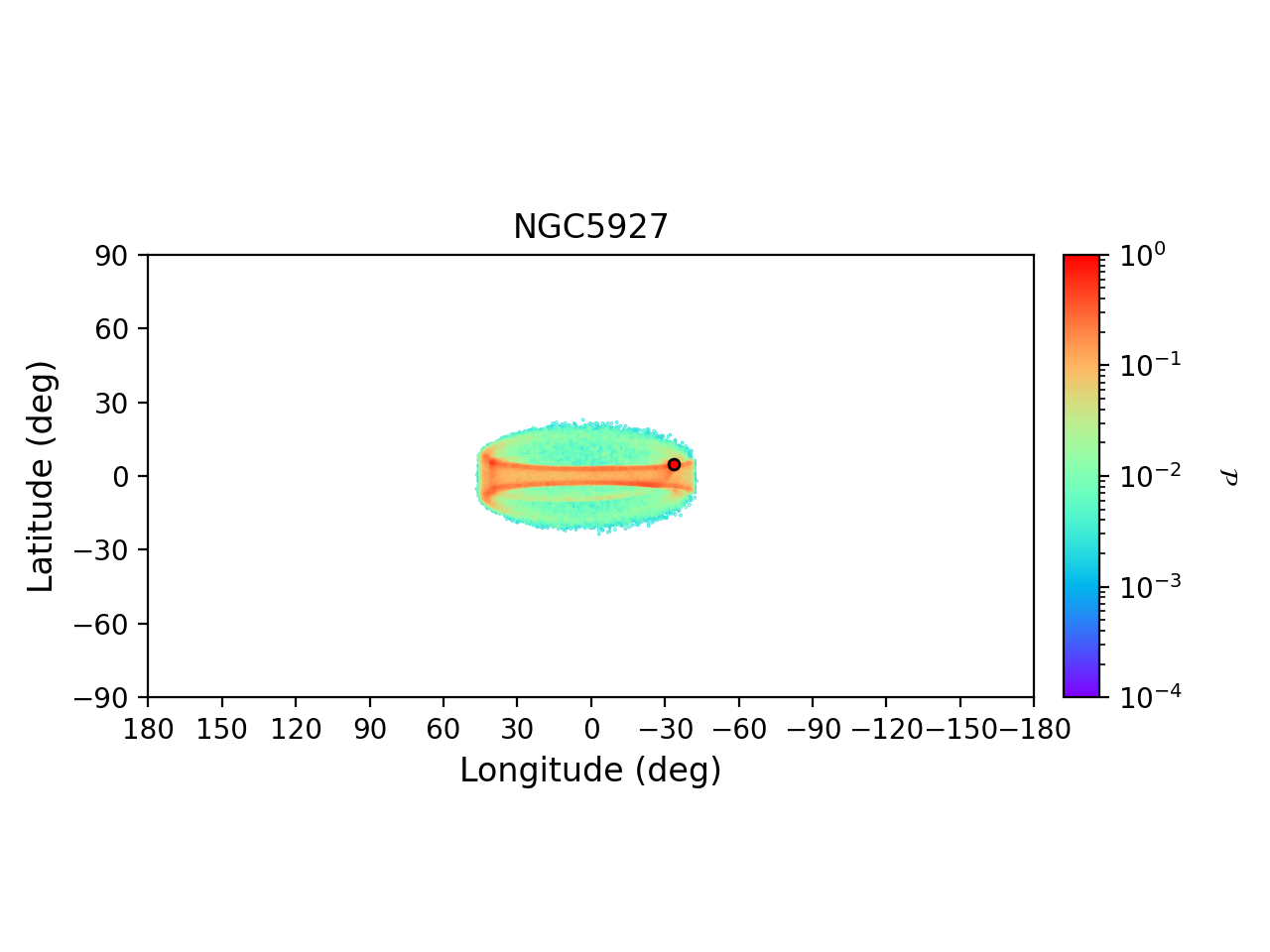}
\includegraphics[clip=true, trim = 0mm 20mm 0mm 10mm, width=1\columnwidth]{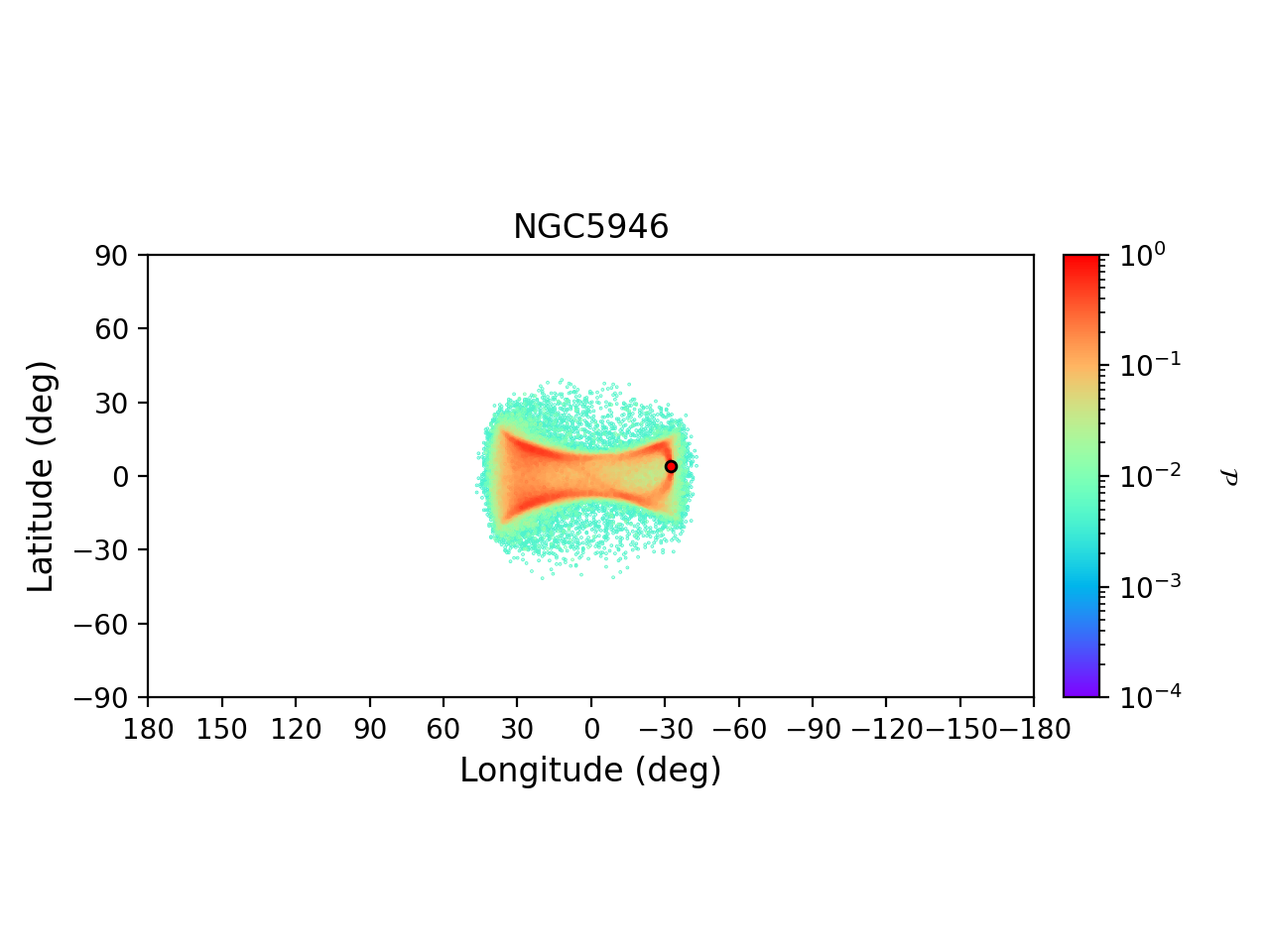}
\includegraphics[clip=true, trim = 0mm 20mm 0mm 10mm, width=1\columnwidth]{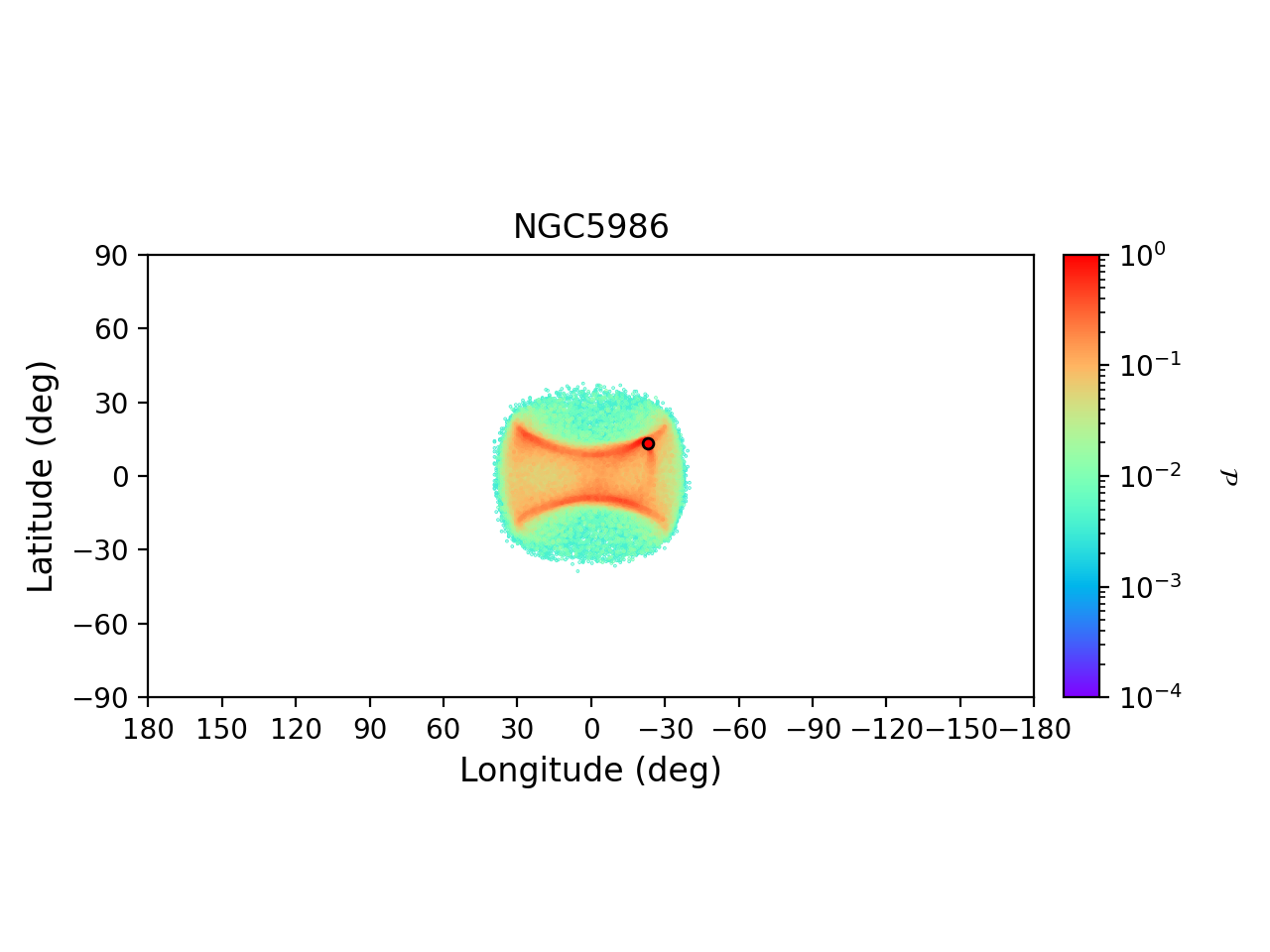}
\includegraphics[clip=true, trim = 0mm 20mm 0mm 10mm, width=1\columnwidth]{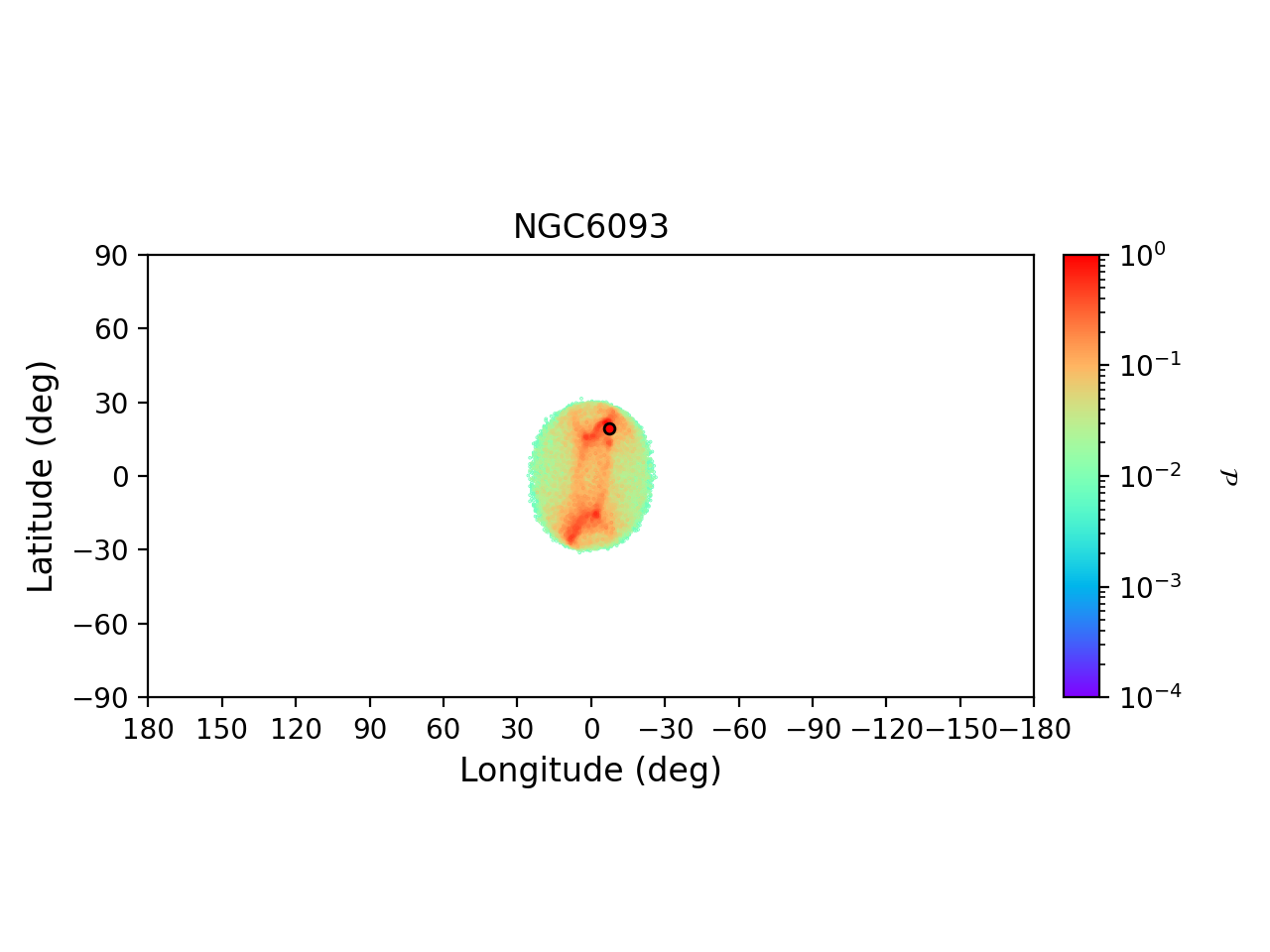}
\includegraphics[clip=true, trim = 0mm 20mm 0mm 10mm, width=1\columnwidth]{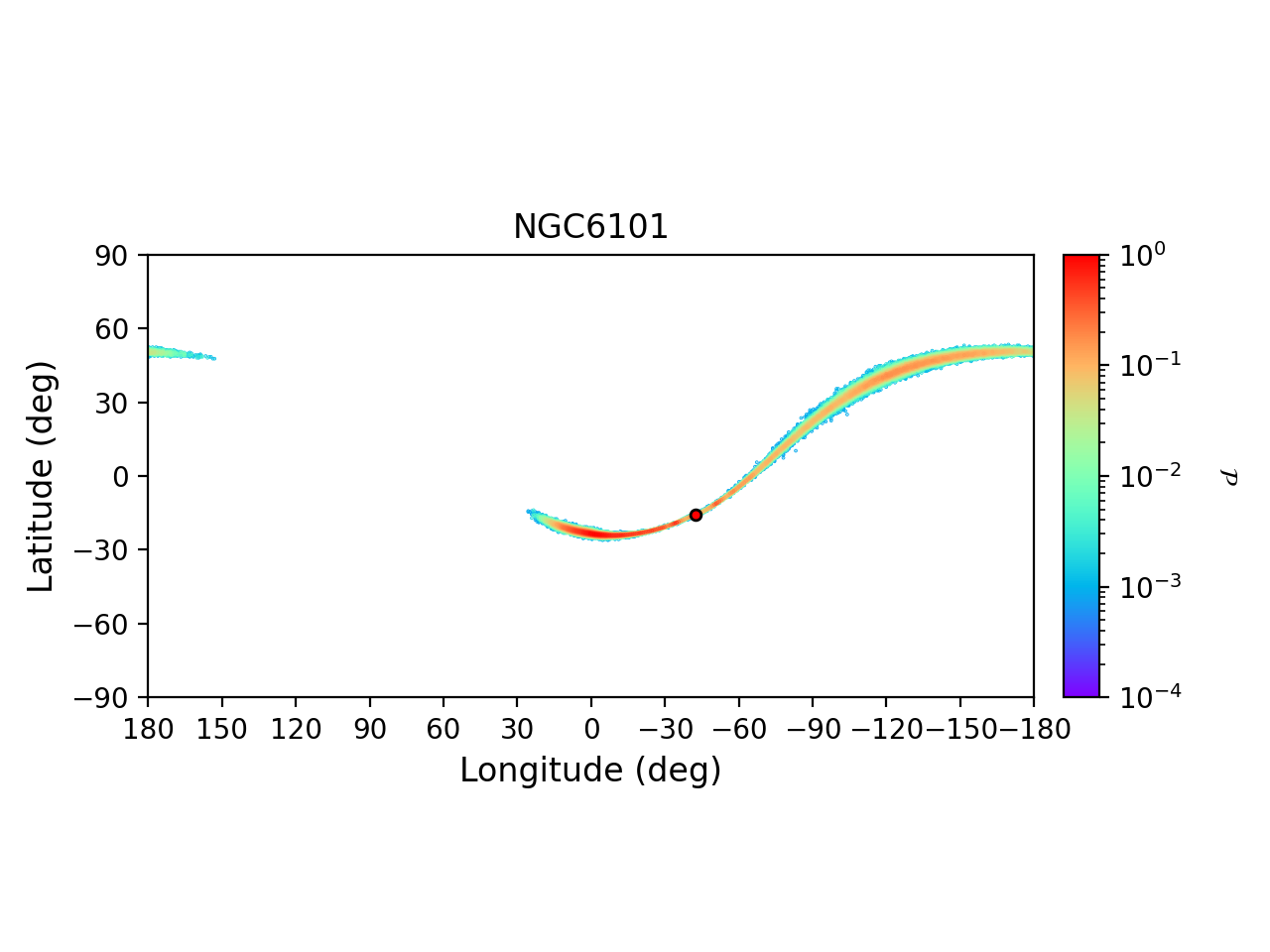}
\includegraphics[clip=true, trim = 0mm 20mm 0mm 10mm, width=1\columnwidth]{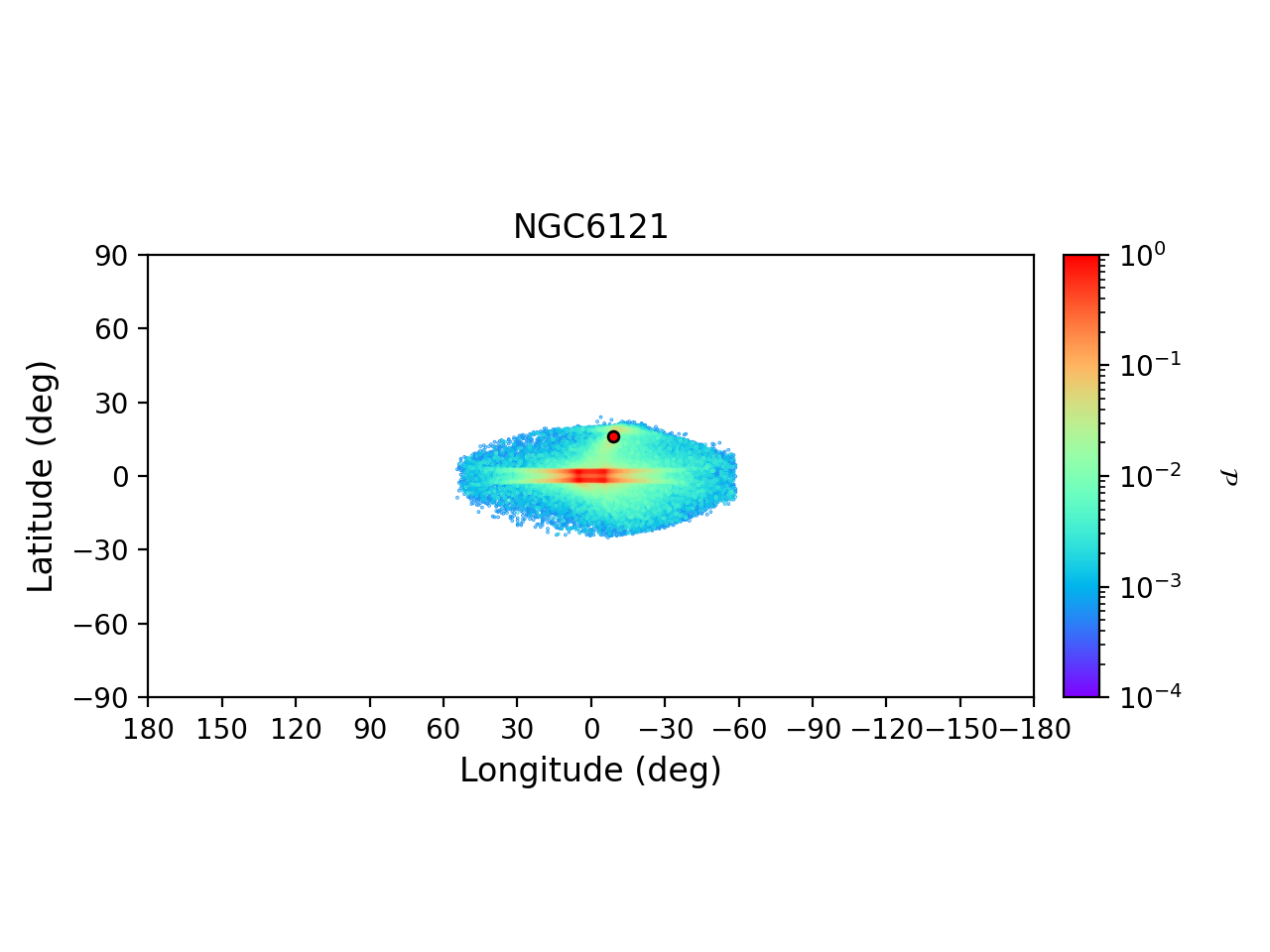}
\includegraphics[clip=true, trim = 0mm 20mm 0mm 10mm, width=1\columnwidth]{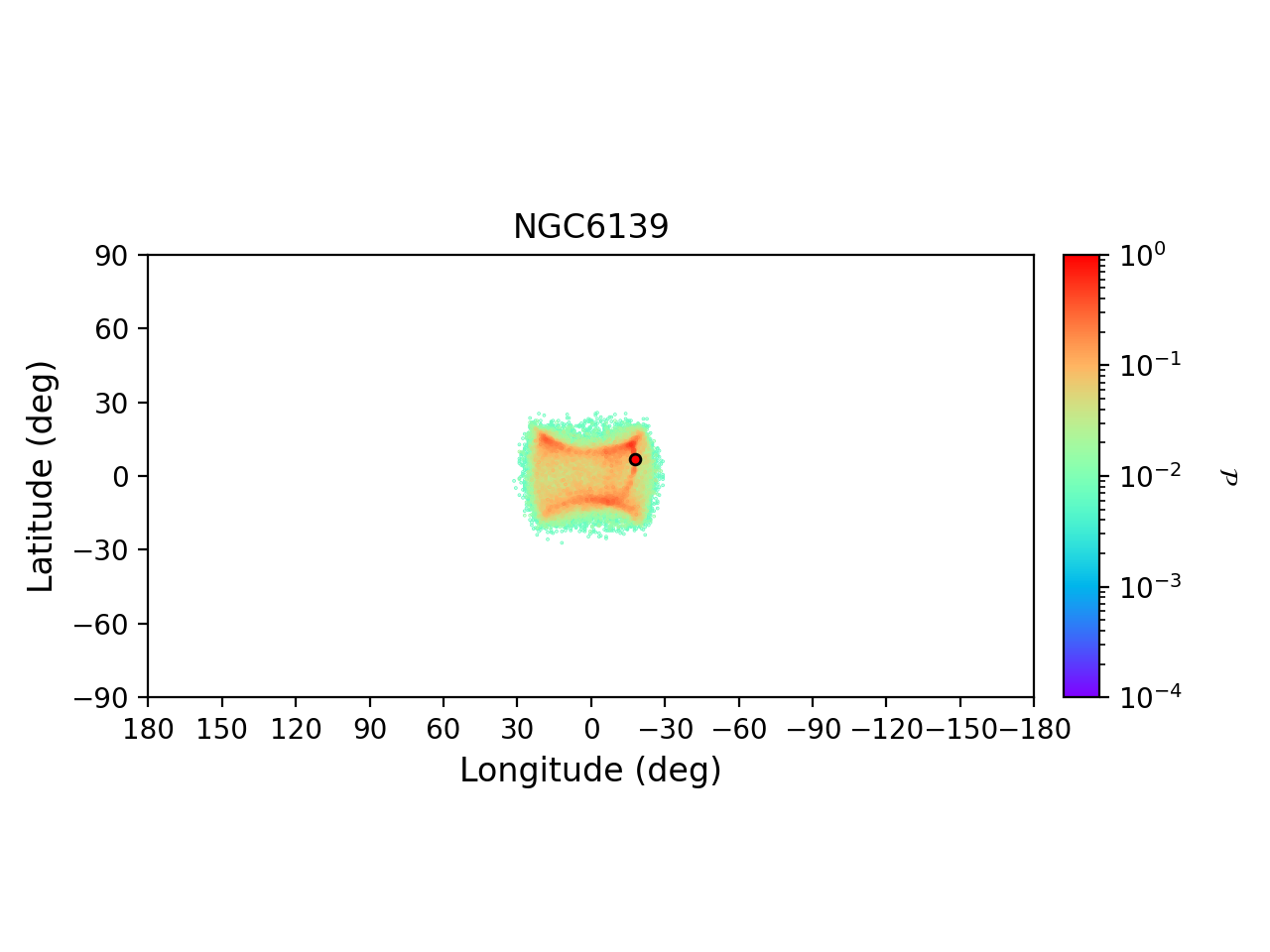}
\caption{Projected density distribution in the $(\ell, b)$ plane of a subset of simulated globular clusters, as indicated at the top of each panel. In each panel, the red circle indicates the current position of the cluster. The densities have been normalized to their maximum value.}\label{stream7}
\end{figure*}

\begin{figure*}
\includegraphics[clip=true, trim = 0mm 20mm 0mm 10mm, width=1\columnwidth]{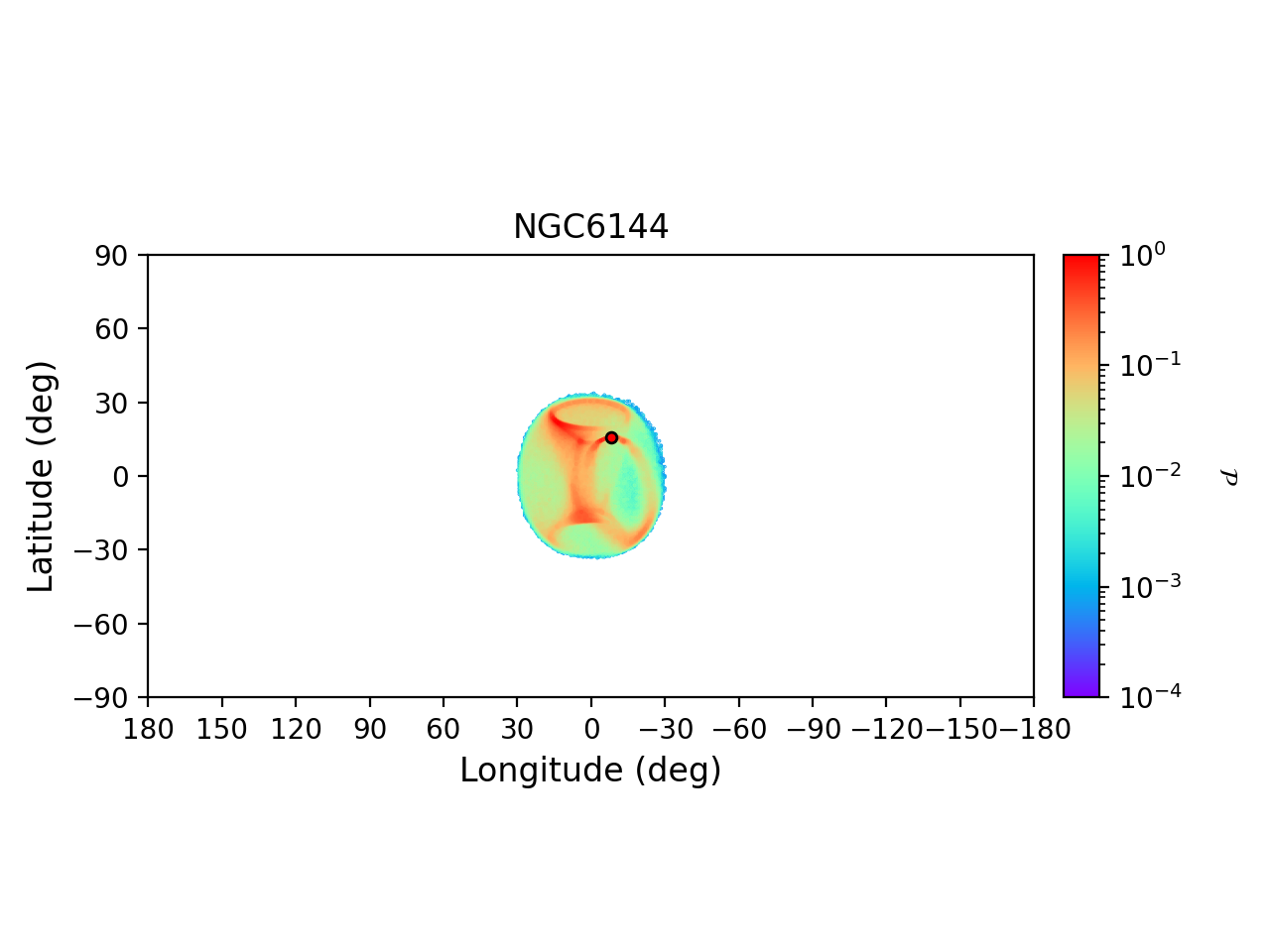}
\includegraphics[clip=true, trim = 0mm 20mm 0mm 10mm, width=1\columnwidth]{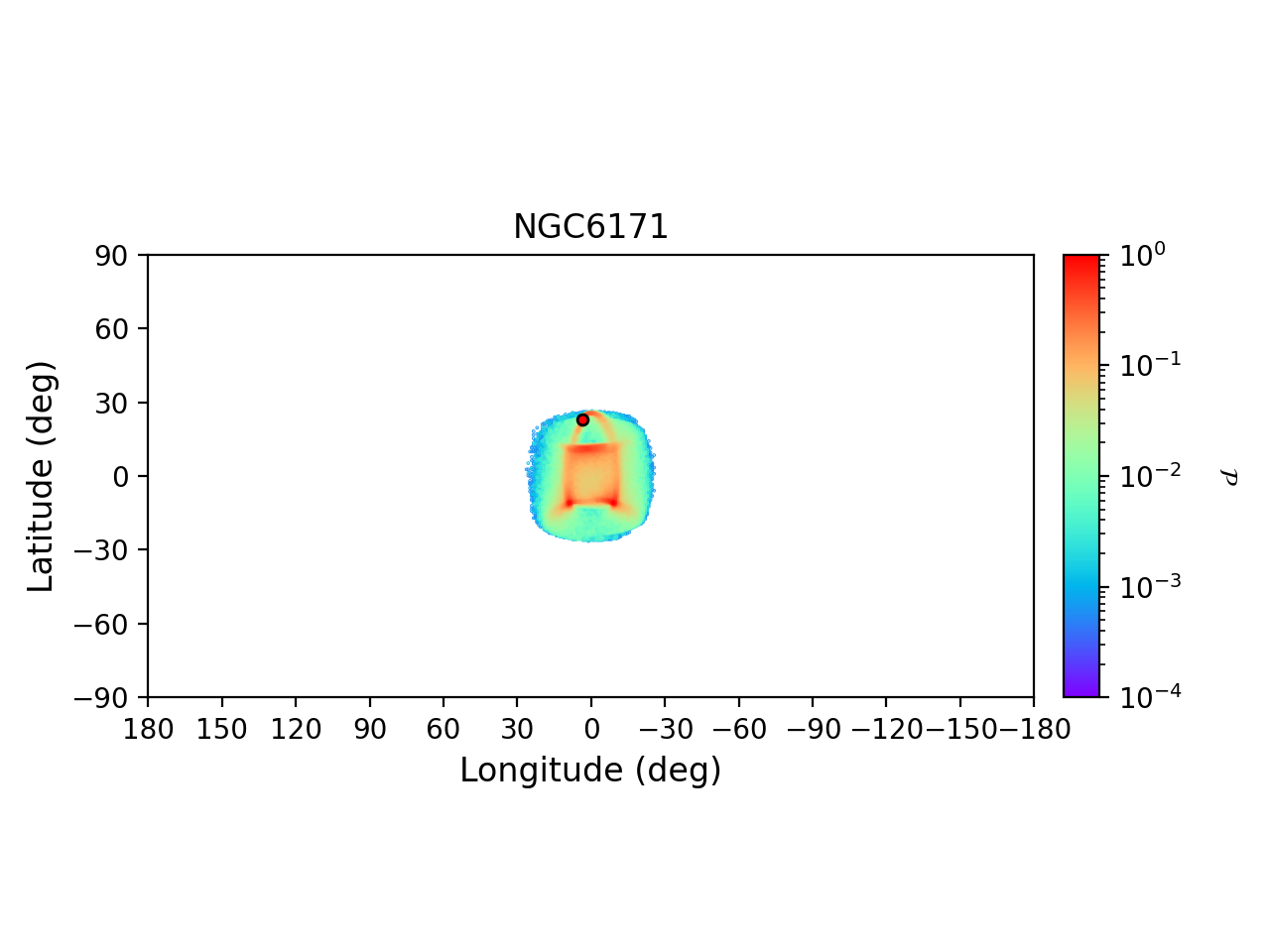}
\includegraphics[clip=true, trim = 0mm 20mm 0mm 10mm, width=1\columnwidth]{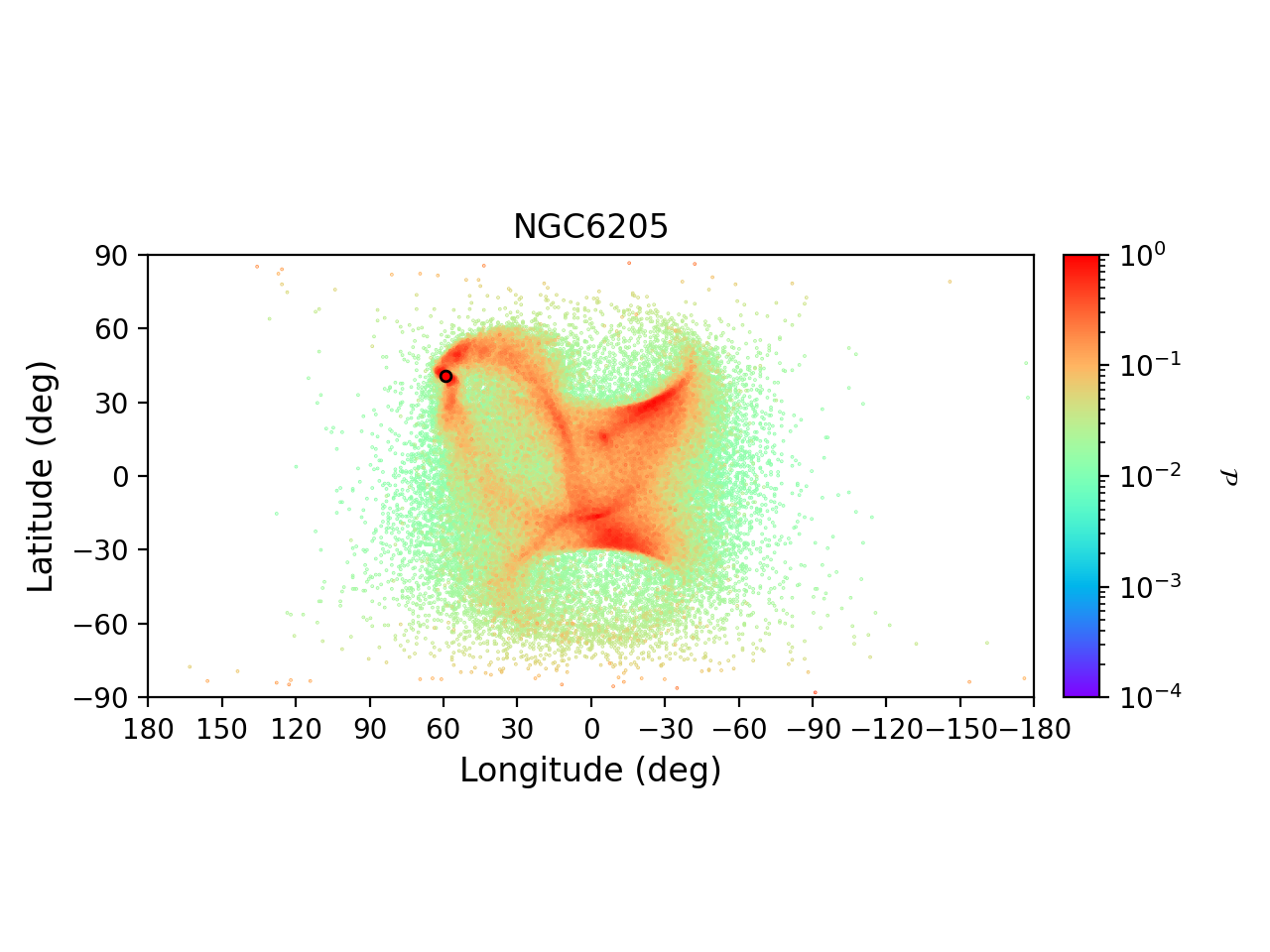}
\includegraphics[clip=true, trim = 0mm 20mm 0mm 10mm, width=1\columnwidth]{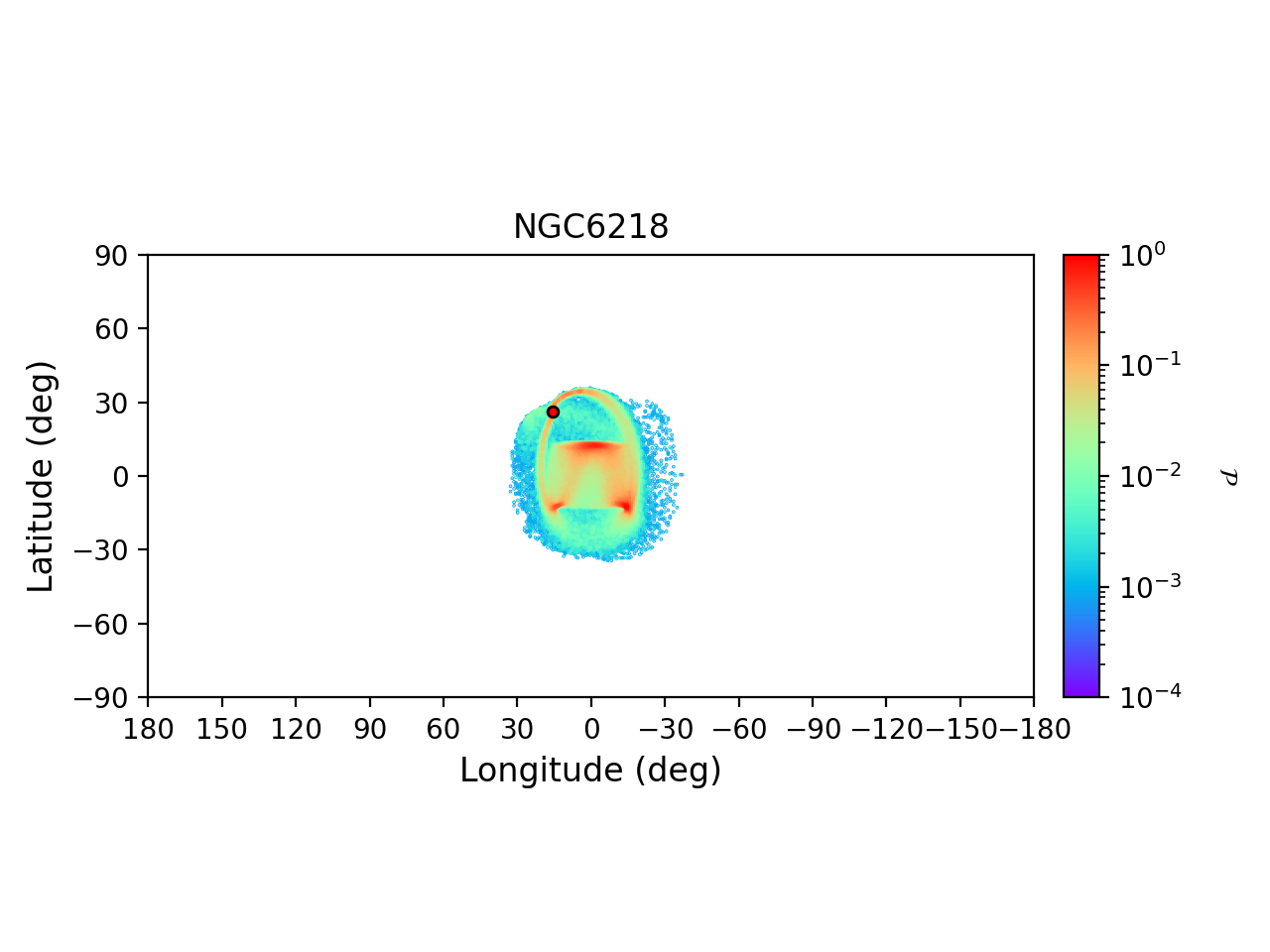}
\includegraphics[clip=true, trim = 0mm 20mm 0mm 10mm, width=1\columnwidth]{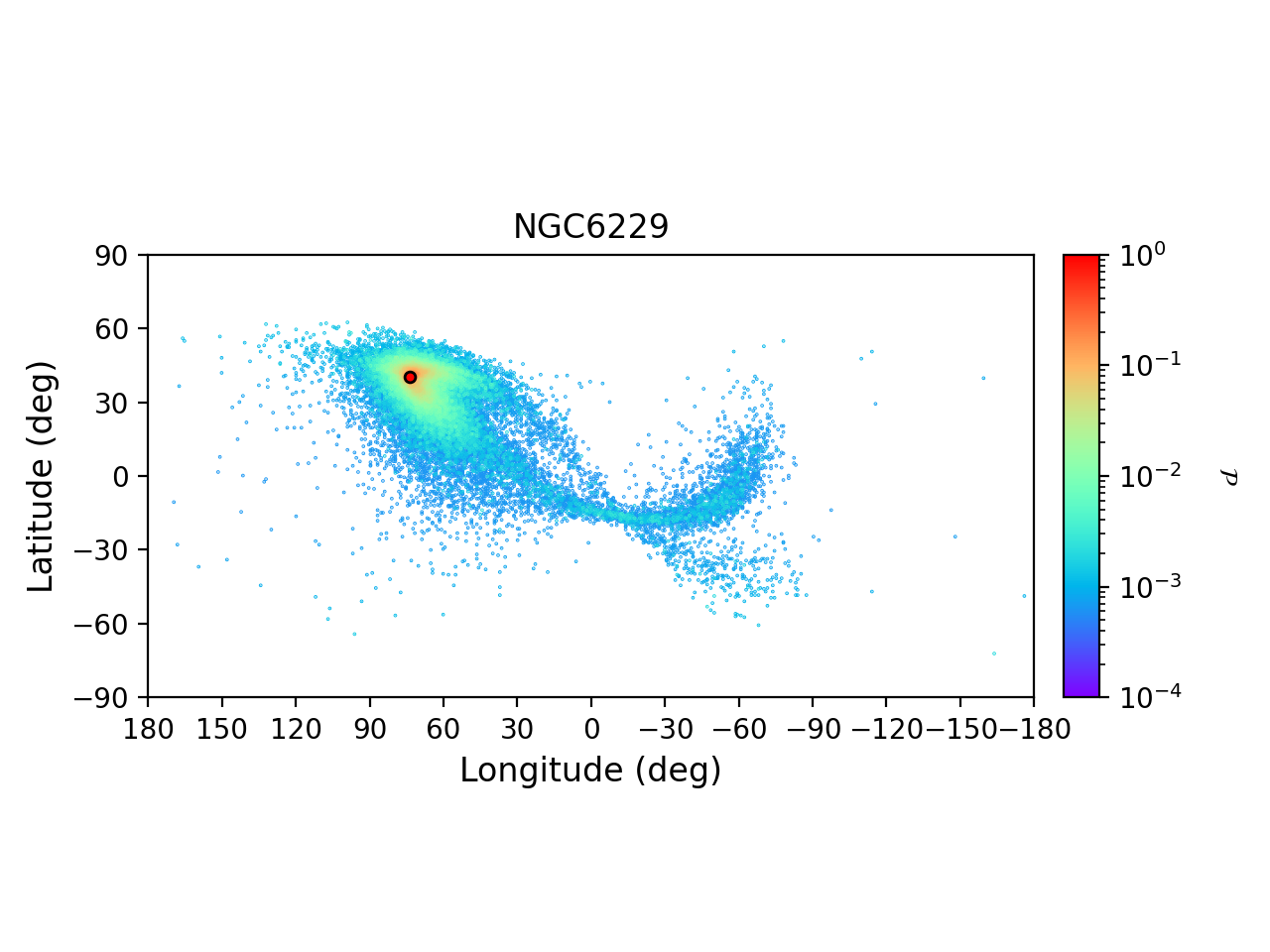}
\includegraphics[clip=true, trim = 0mm 20mm 0mm 10mm, width=1\columnwidth]{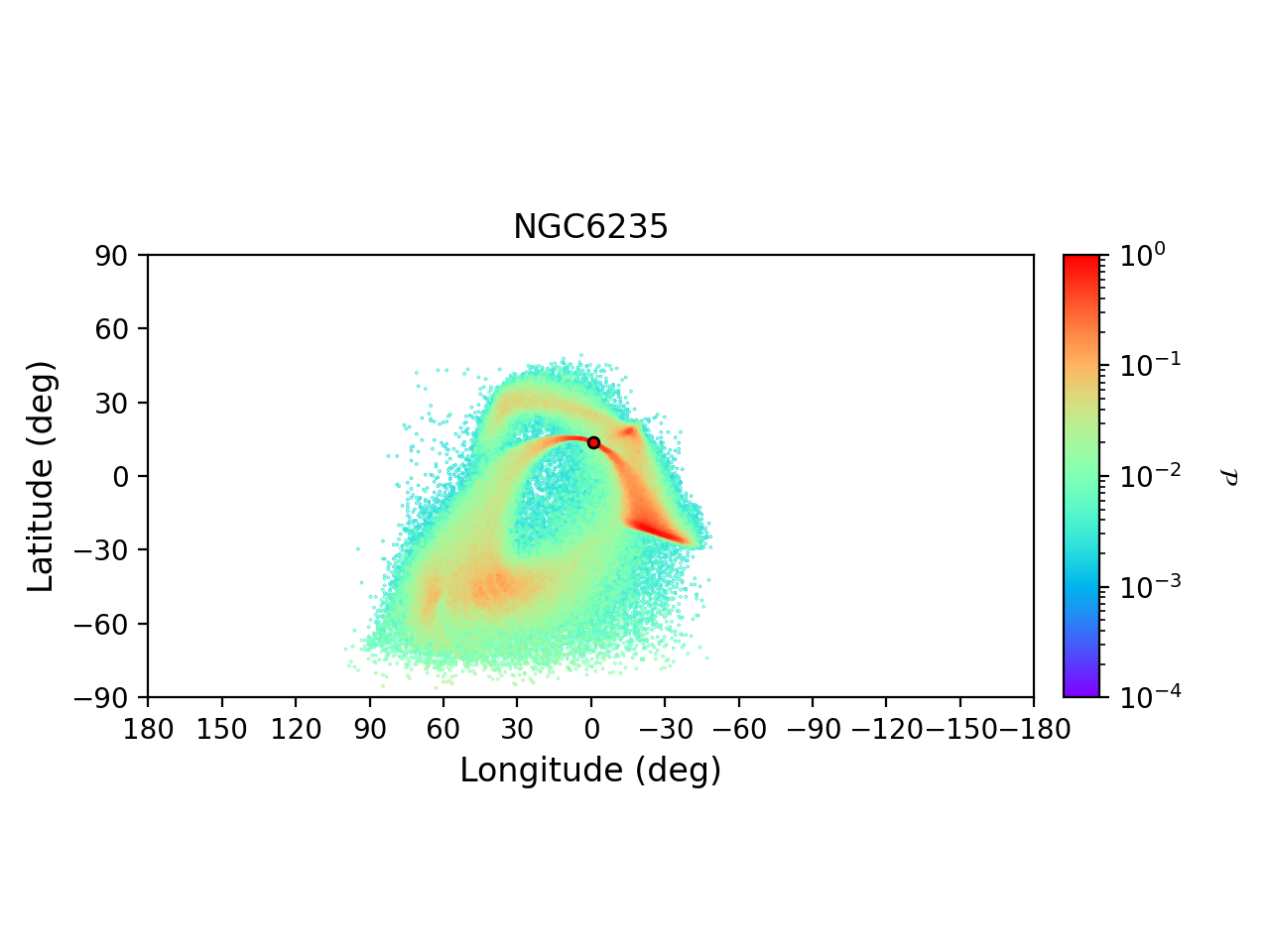}
\includegraphics[clip=true, trim = 0mm 20mm 0mm 10mm, width=1\columnwidth]{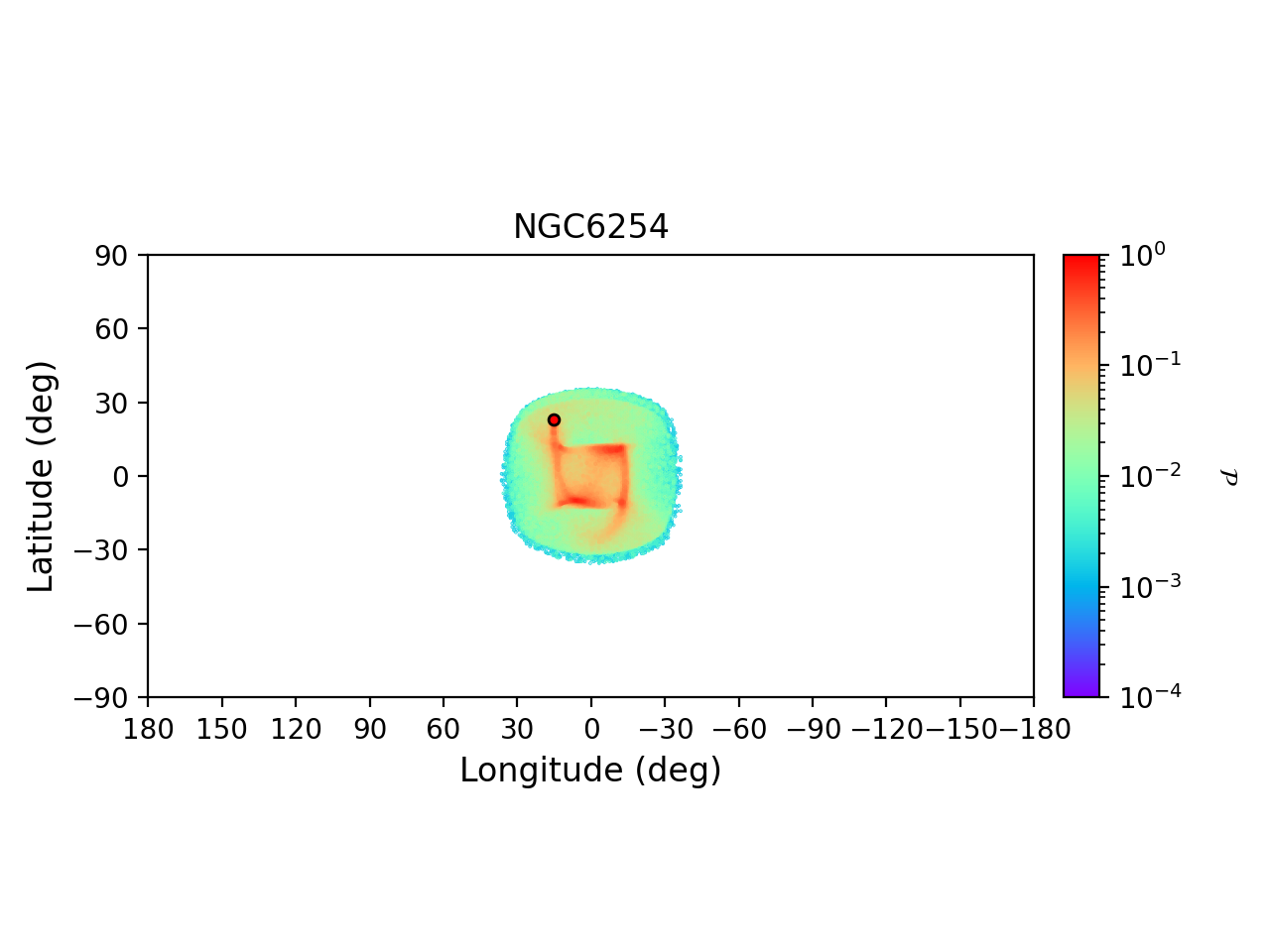}
\includegraphics[clip=true, trim = 0mm 20mm 0mm 10mm, width=1\columnwidth]{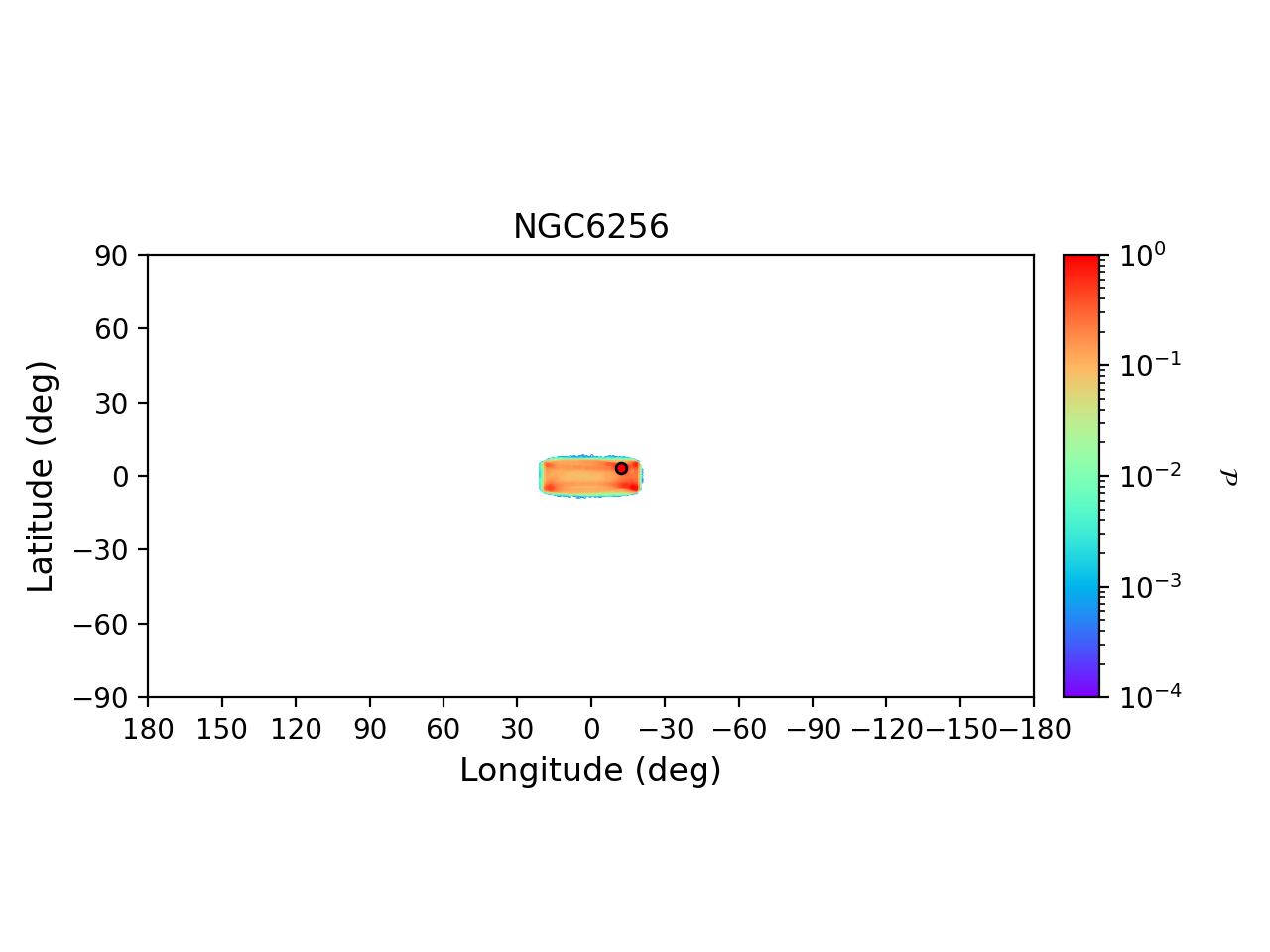}
\caption{Projected density distribution in the $(\ell, b)$ plane of a subset of simulated globular clusters, as indicated at the top of each panel. In each panel, the red circle indicates the current position of the cluster. The densities have been normalized to their maximum value.}\label{stream8}
\end{figure*}

\begin{figure*}
\includegraphics[clip=true, trim = 0mm 20mm 0mm 10mm, width=1\columnwidth]{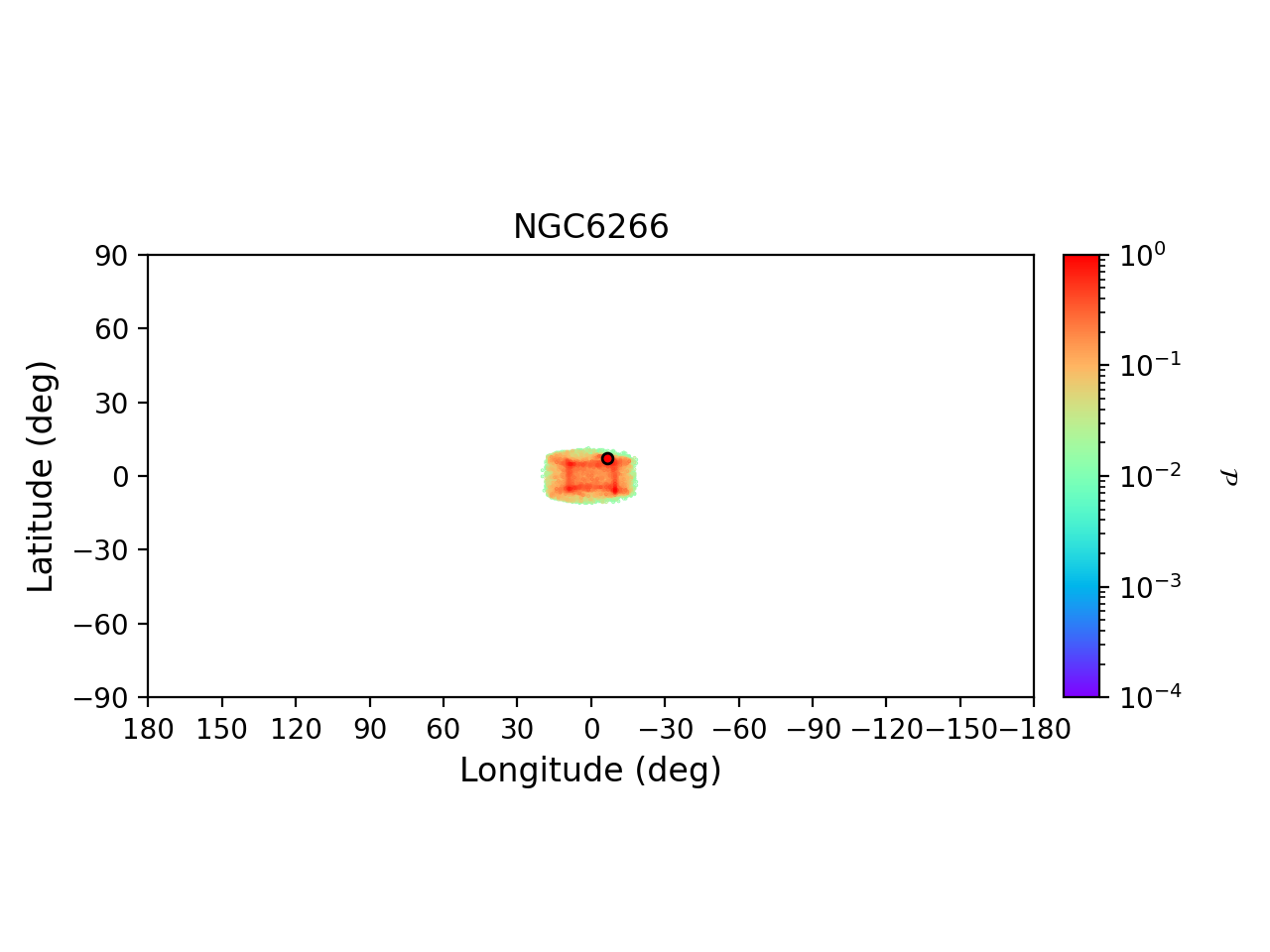}
\includegraphics[clip=true, trim = 0mm 20mm 0mm 10mm, width=1\columnwidth]{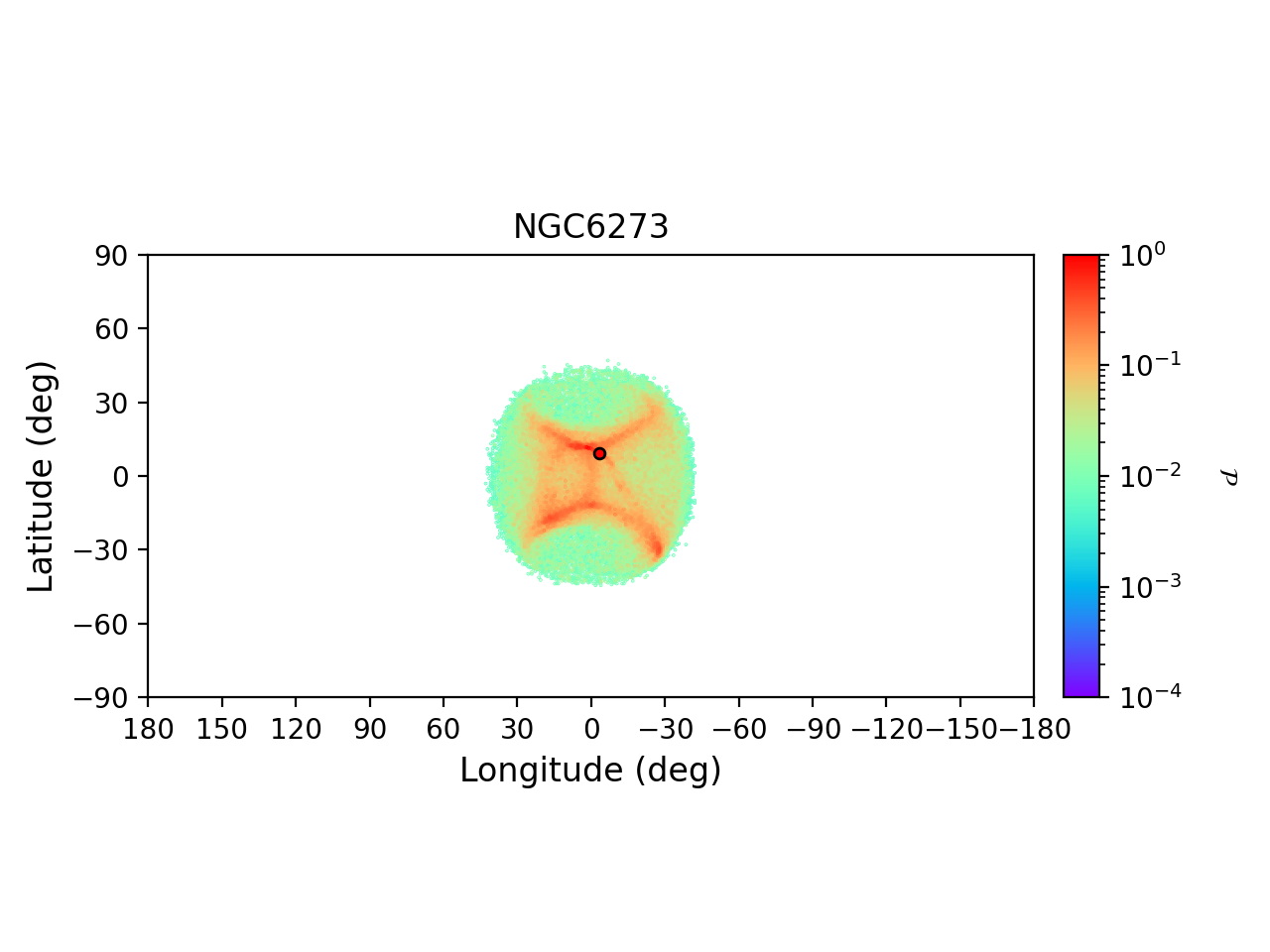}
\includegraphics[clip=true, trim = 0mm 20mm 0mm 10mm, width=1\columnwidth]{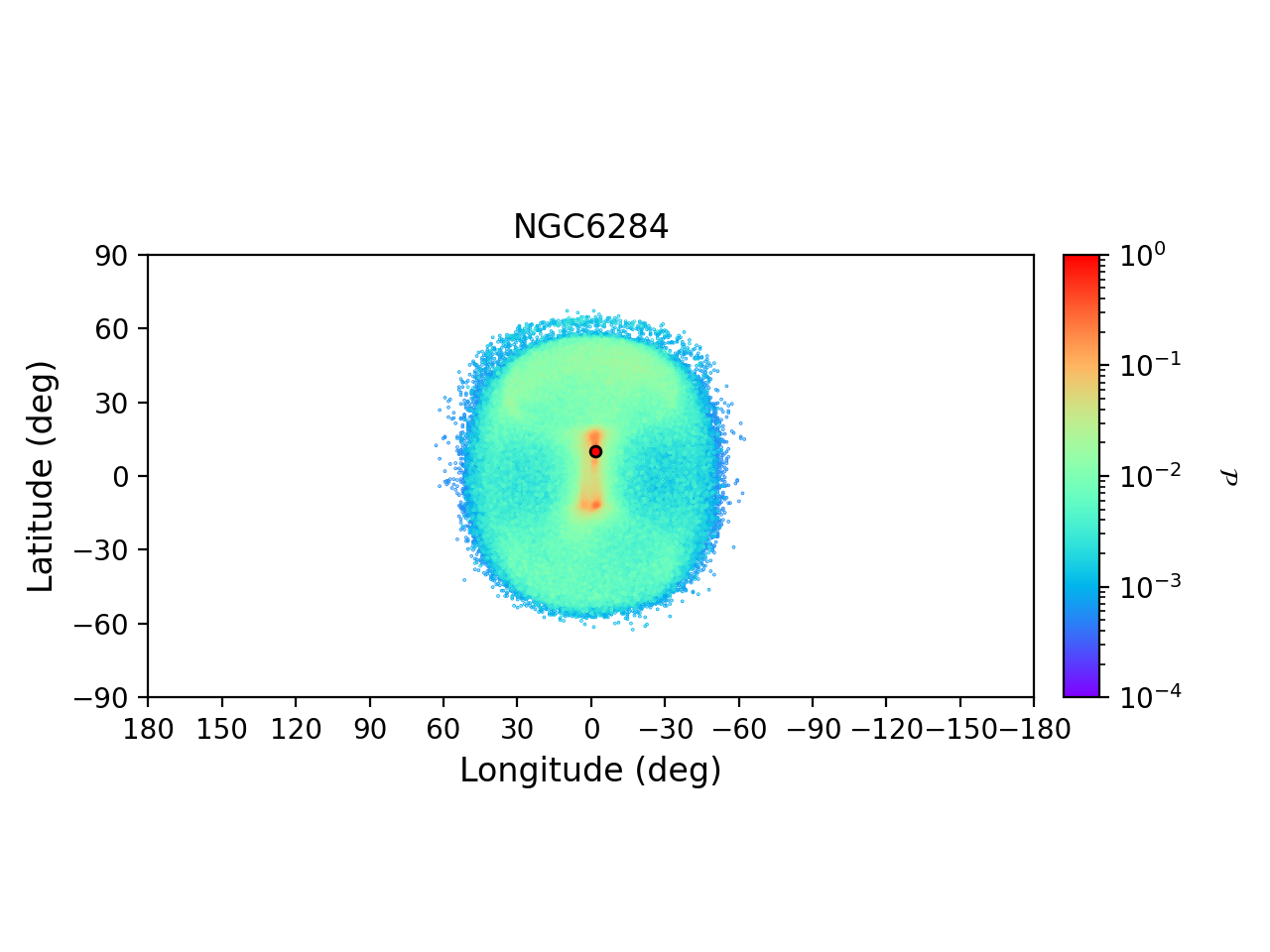}
\includegraphics[clip=true, trim = 0mm 20mm 0mm 10mm, width=1\columnwidth]{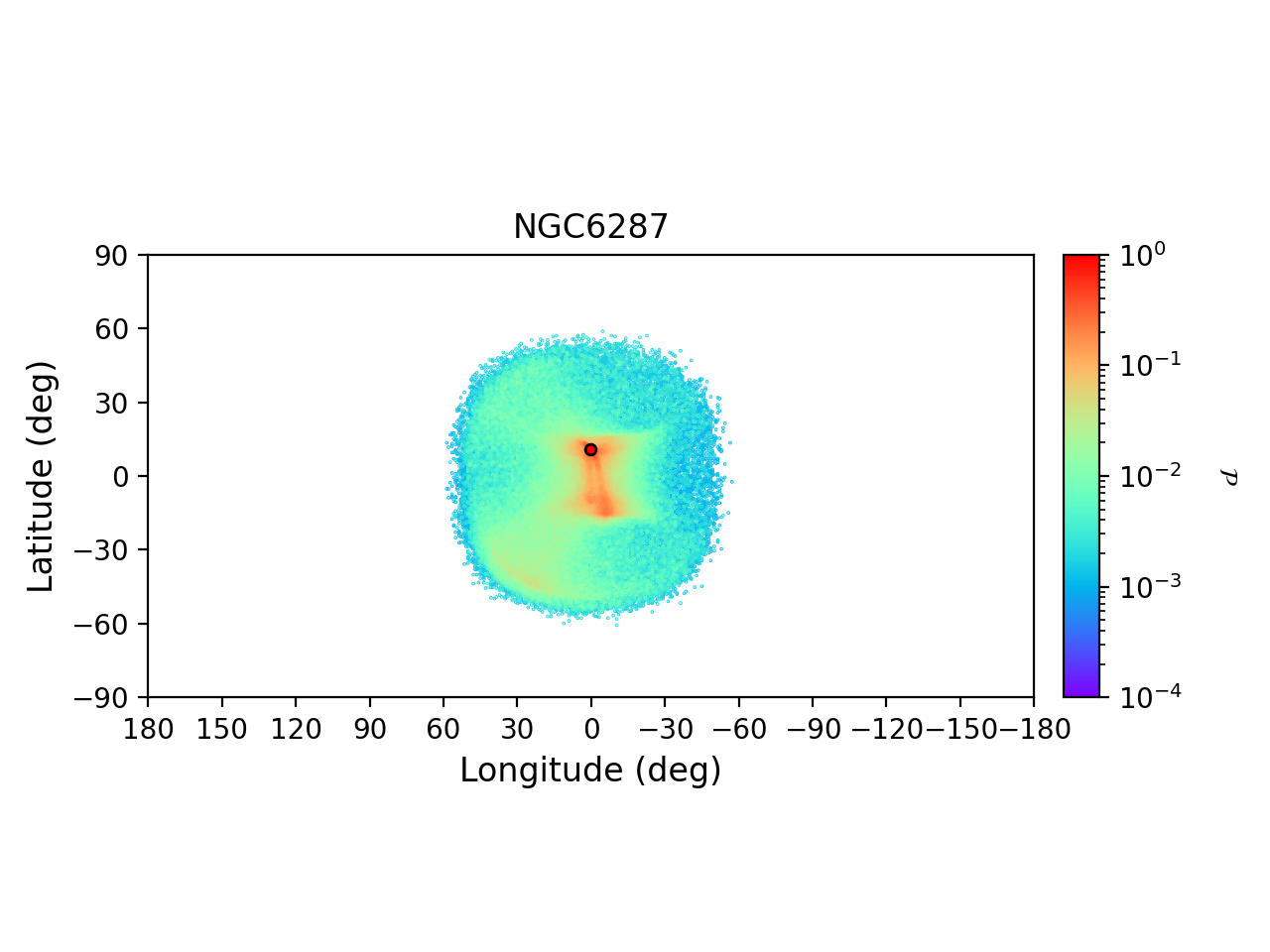}
\includegraphics[clip=true, trim = 0mm 20mm 0mm 10mm, width=1\columnwidth]{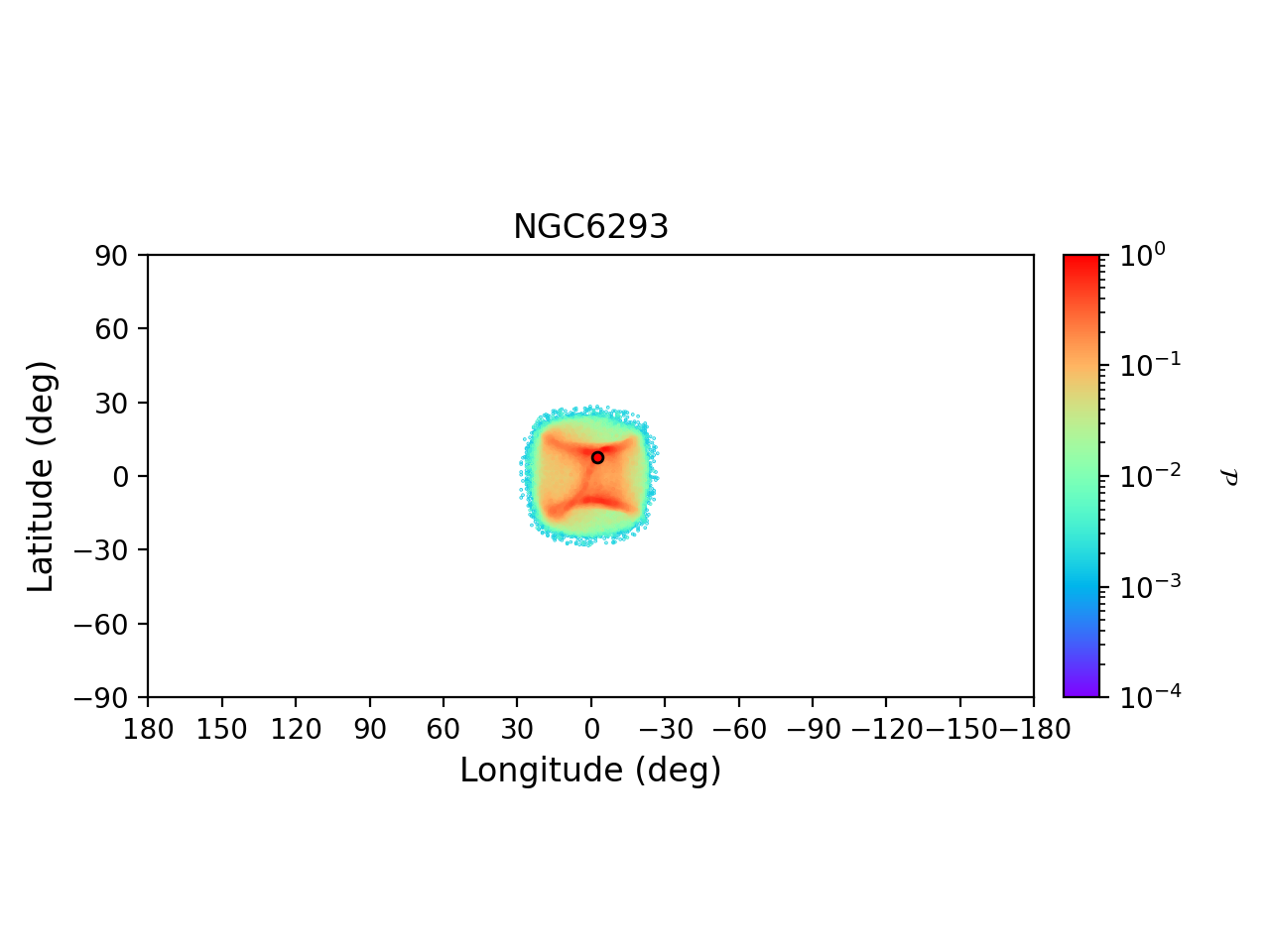}
\includegraphics[clip=true, trim = 0mm 20mm 0mm 10mm, width=1\columnwidth]{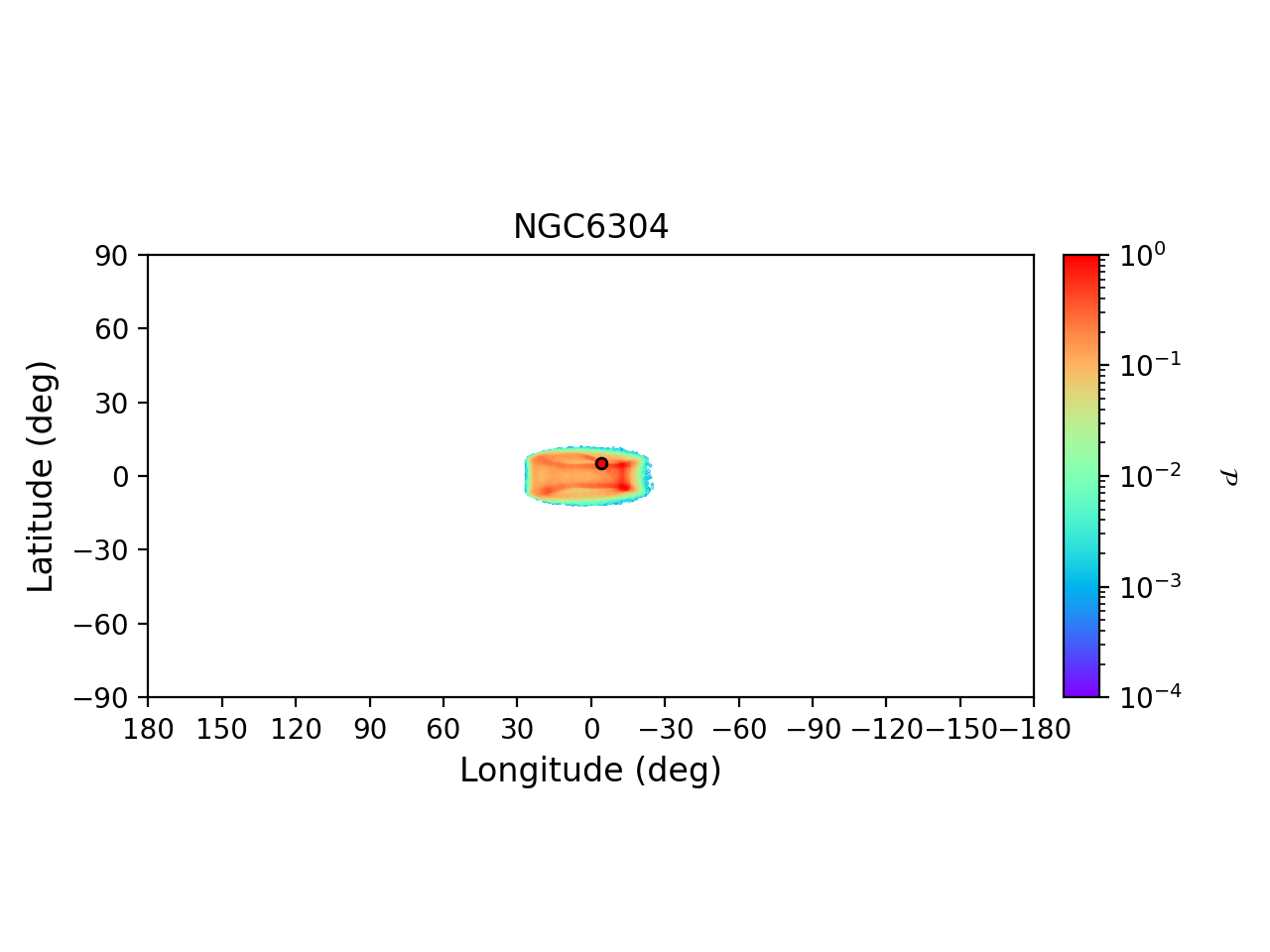}
\includegraphics[clip=true, trim = 0mm 20mm 0mm 10mm, width=1\columnwidth]{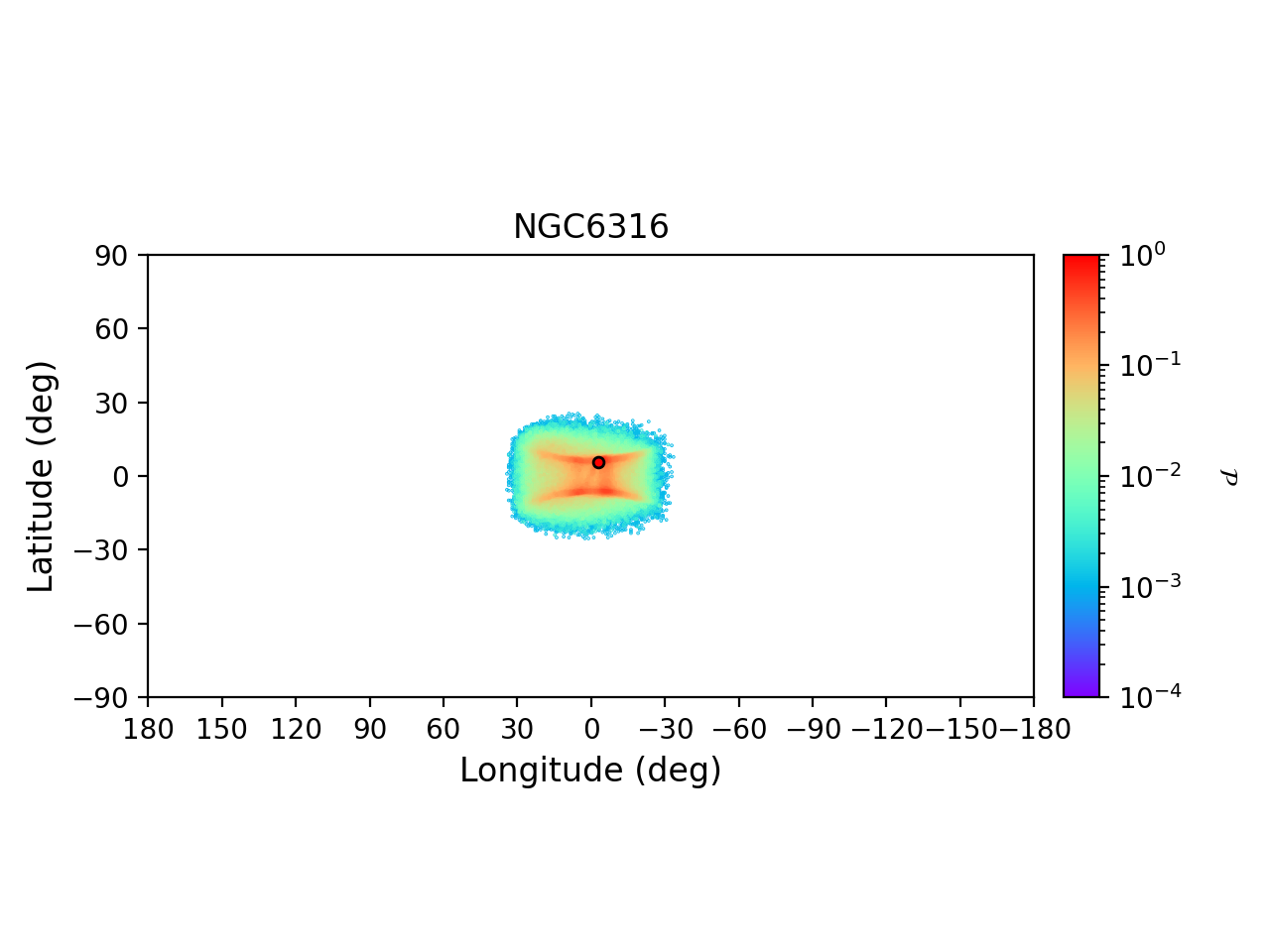}
\includegraphics[clip=true, trim = 0mm 20mm 0mm 10mm, width=1\columnwidth]{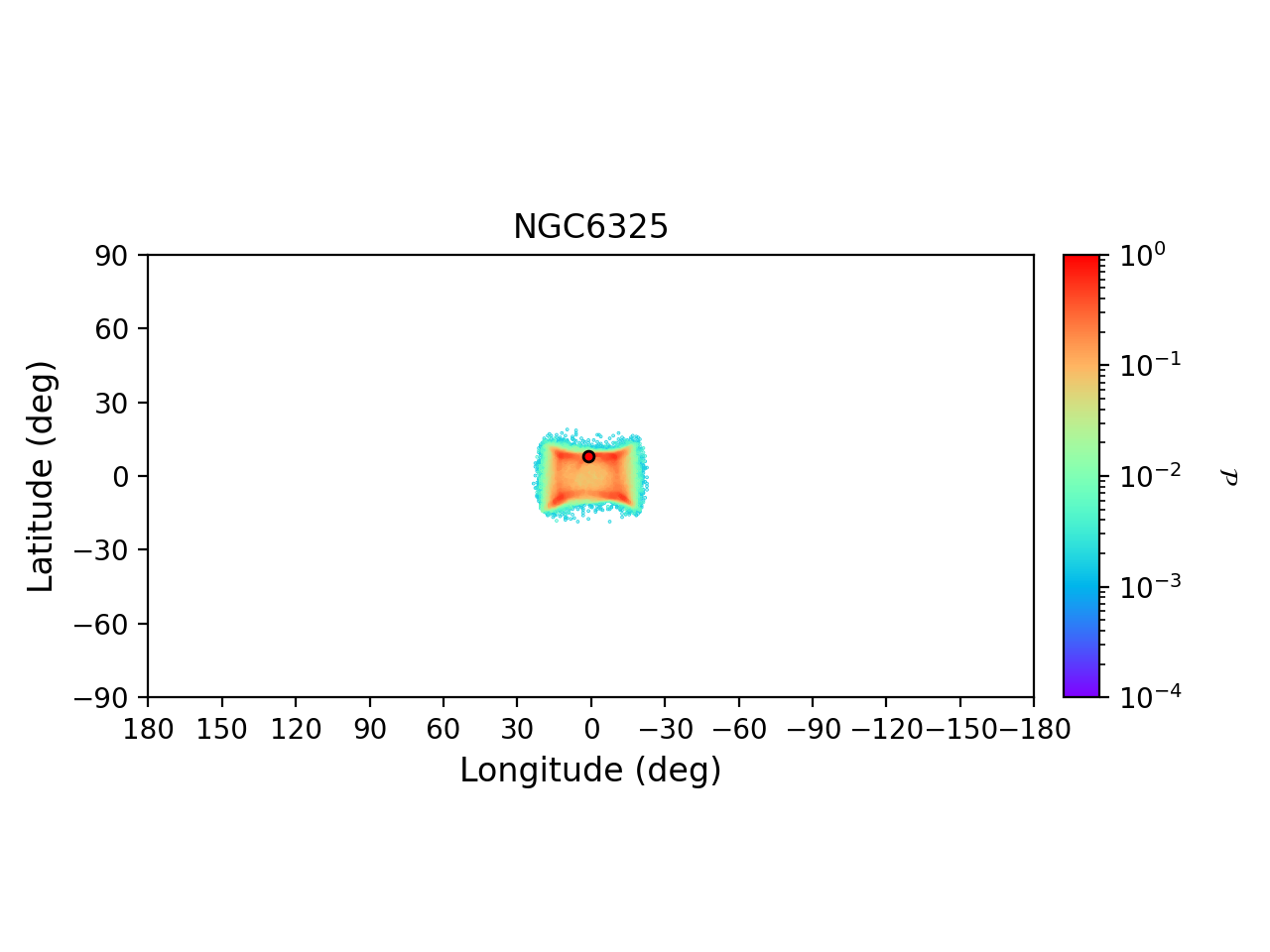}
\caption{Projected density distribution in the $(\ell, b)$ plane of a subset of simulated globular clusters, as indicated at the top of each panel. In each panel, the red circle indicates the current position of the cluster. The densities have been normalized to their maximum value.}\label{stream9}
\end{figure*}

\begin{figure*}
\includegraphics[clip=true, trim = 0mm 20mm 0mm 10mm, width=1\columnwidth]{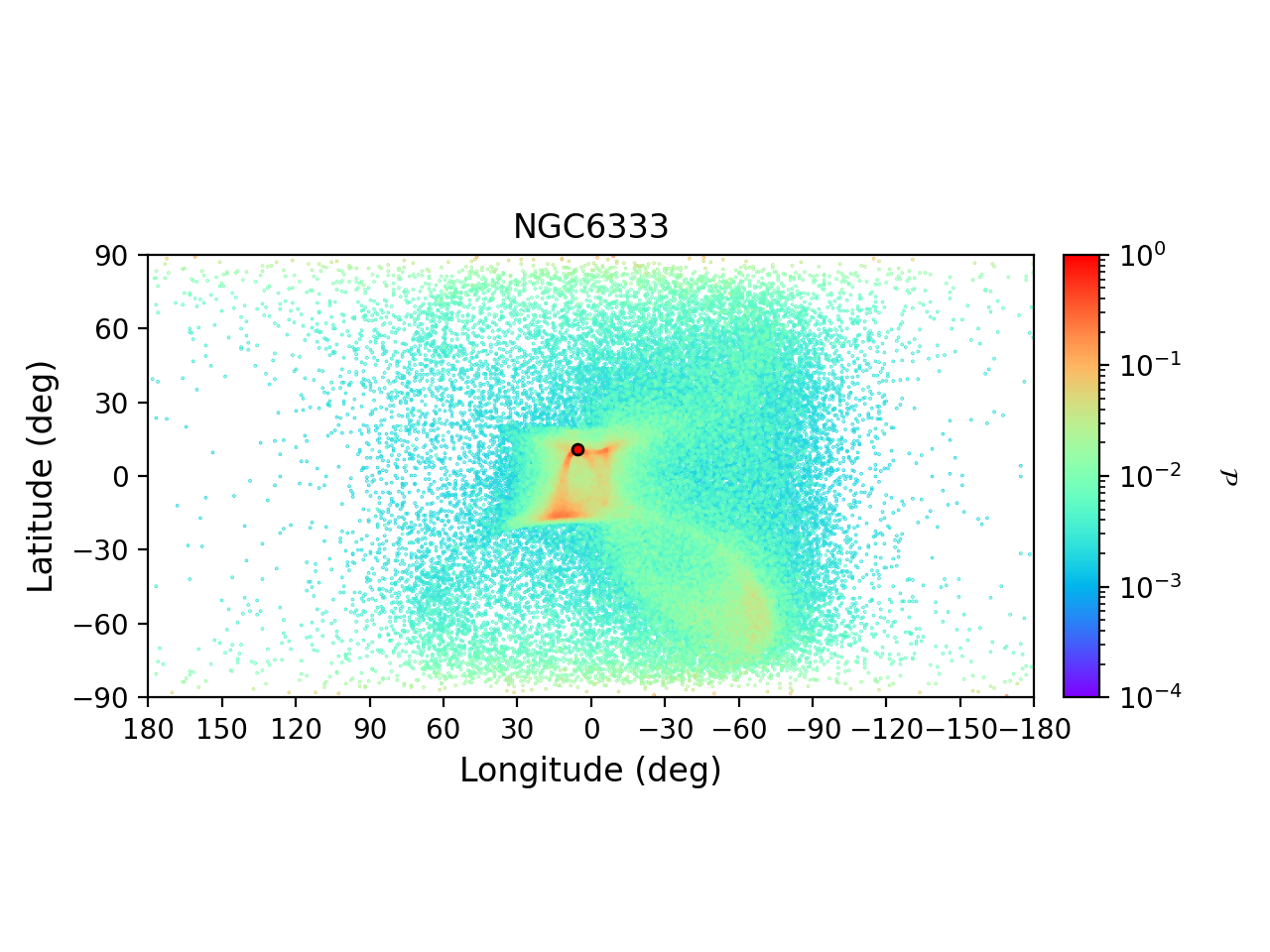}
\includegraphics[clip=true, trim = 0mm 20mm 0mm 10mm, width=1\columnwidth]{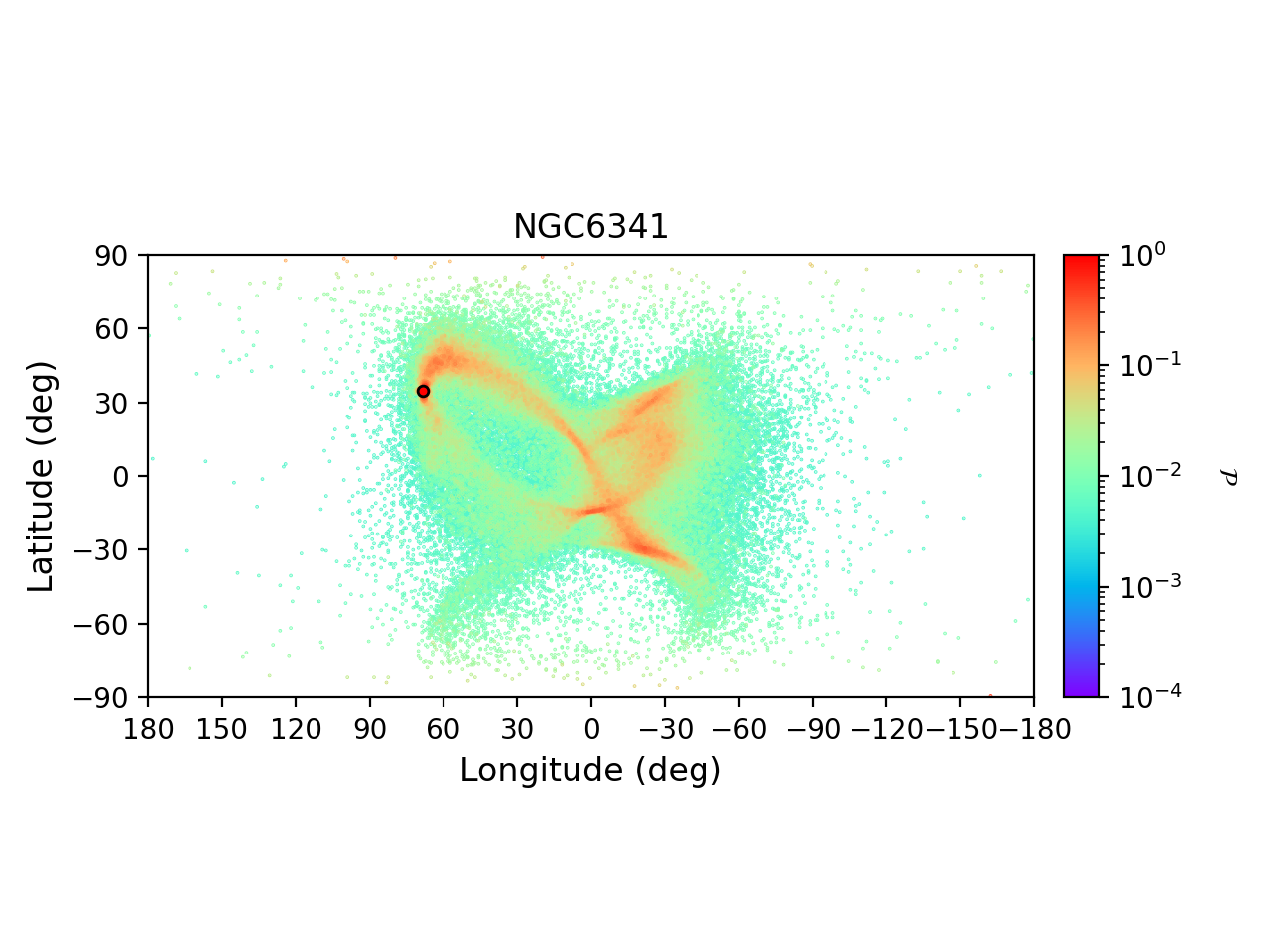}
\includegraphics[clip=true, trim = 0mm 20mm 0mm 10mm, width=1\columnwidth]{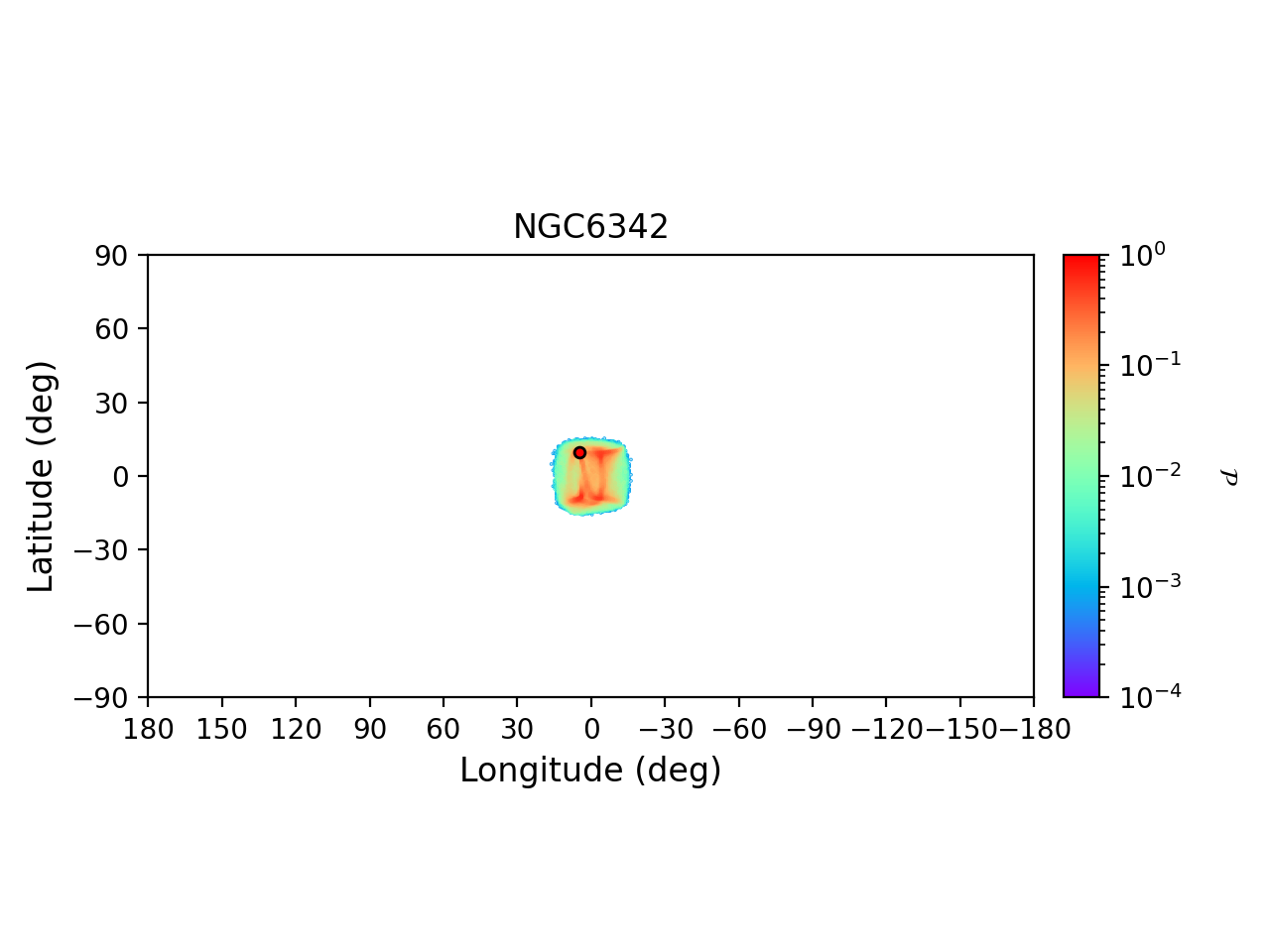}
\includegraphics[clip=true, trim = 0mm 20mm 0mm 10mm, width=1\columnwidth]{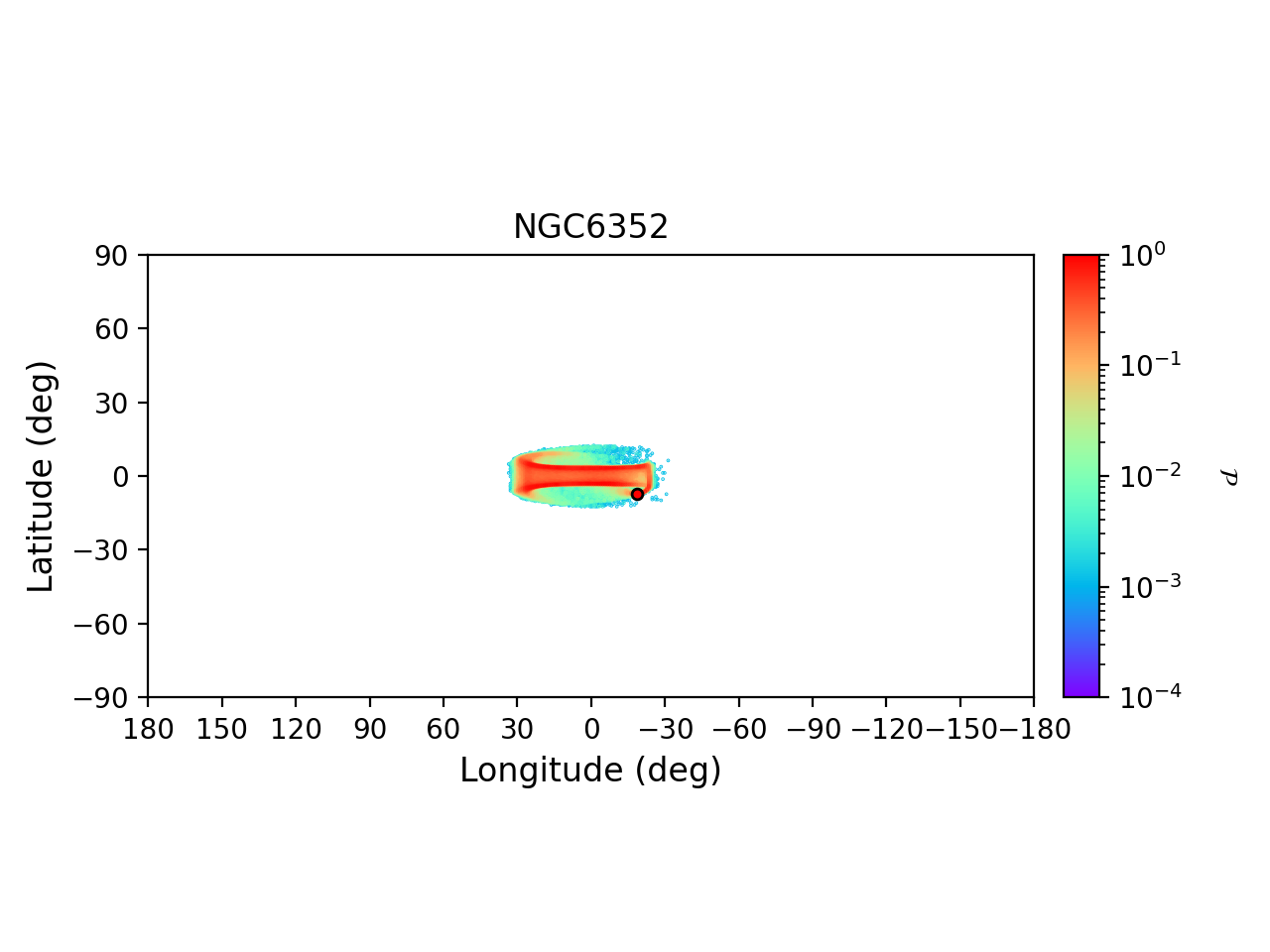}
\includegraphics[clip=true, trim = 0mm 20mm 0mm 10mm, width=1\columnwidth]{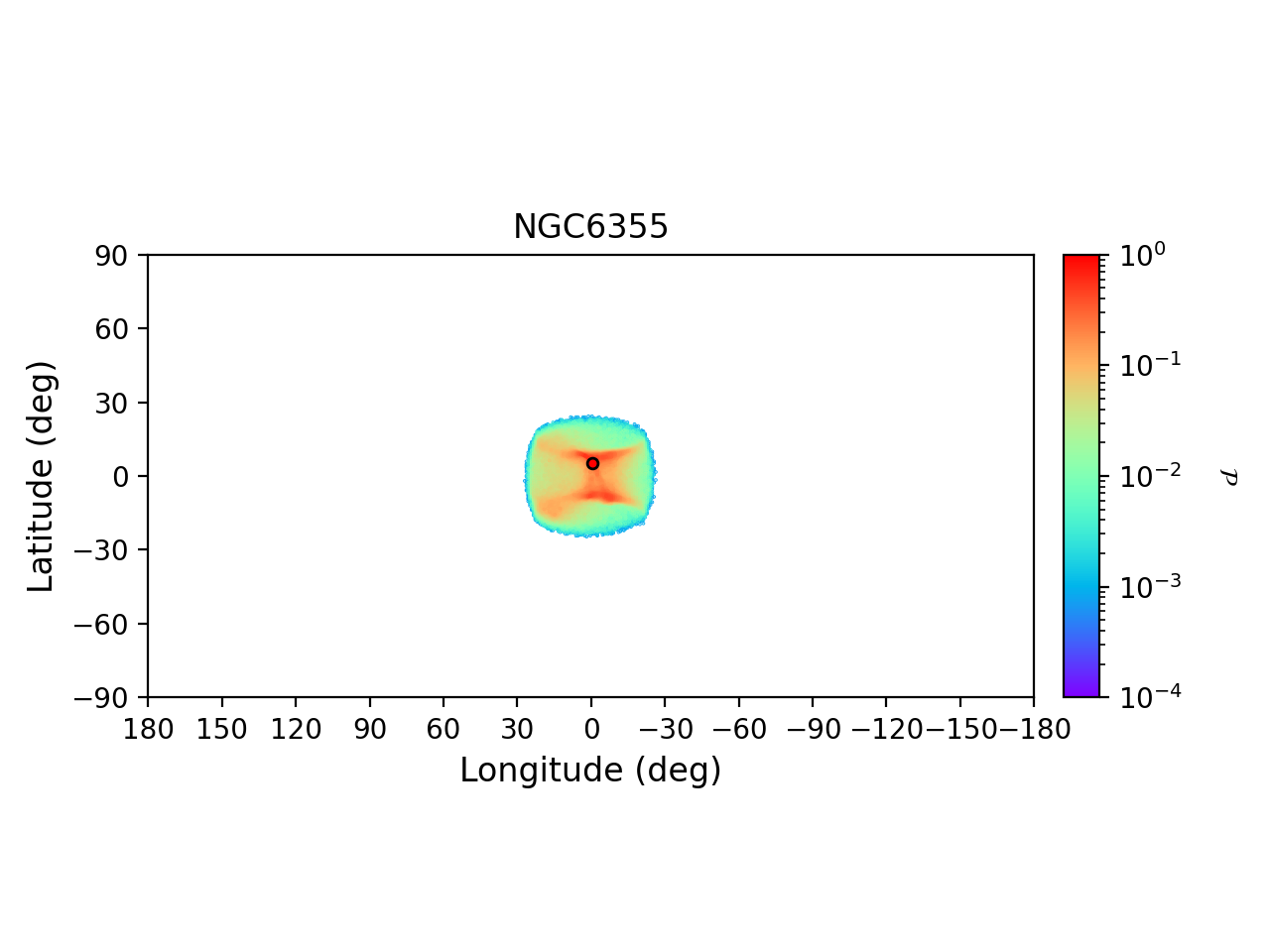}
\includegraphics[clip=true, trim = 0mm 20mm 0mm 10mm, width=1\columnwidth]{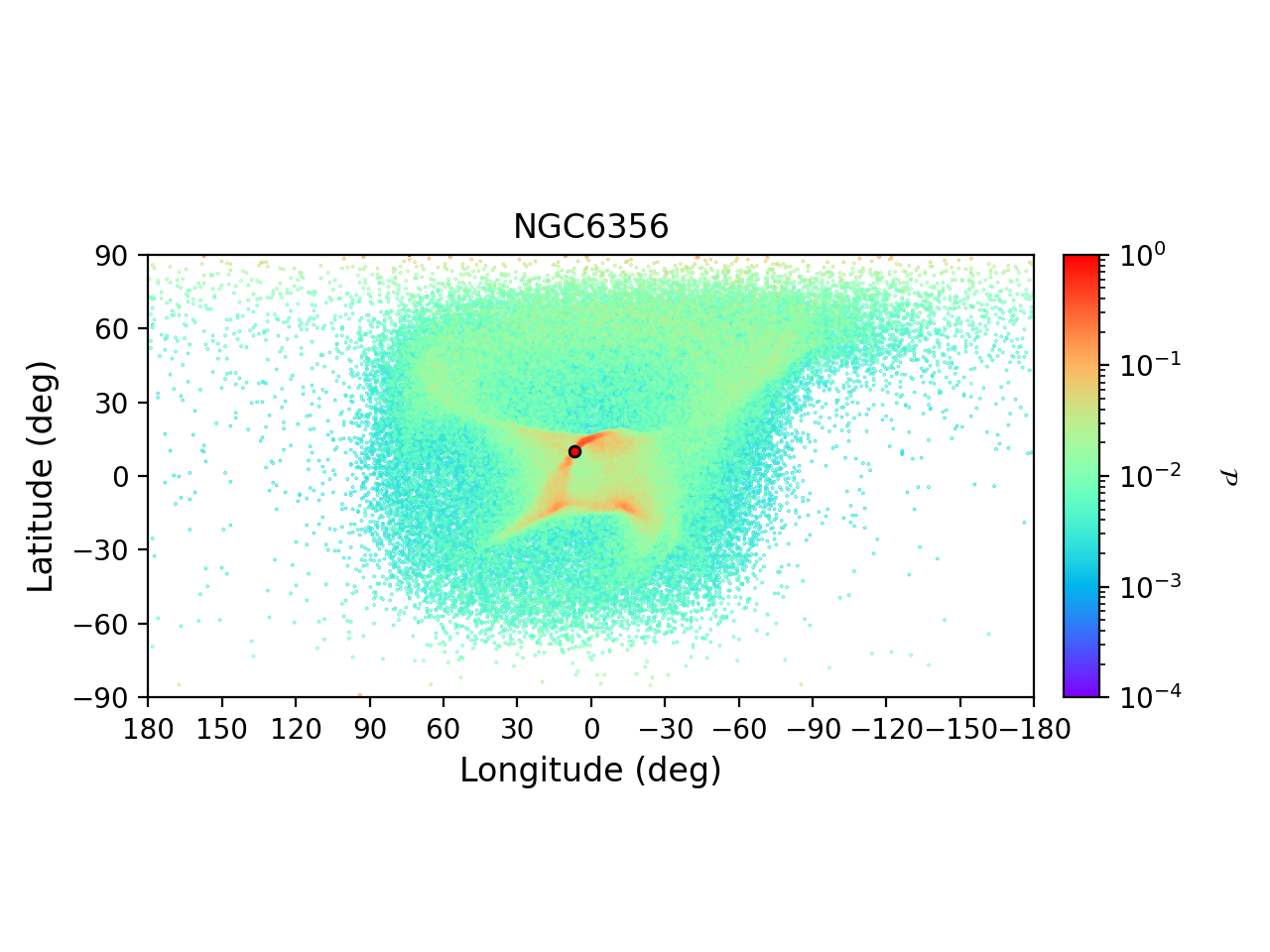}
\includegraphics[clip=true, trim = 0mm 20mm 0mm 10mm, width=1\columnwidth]{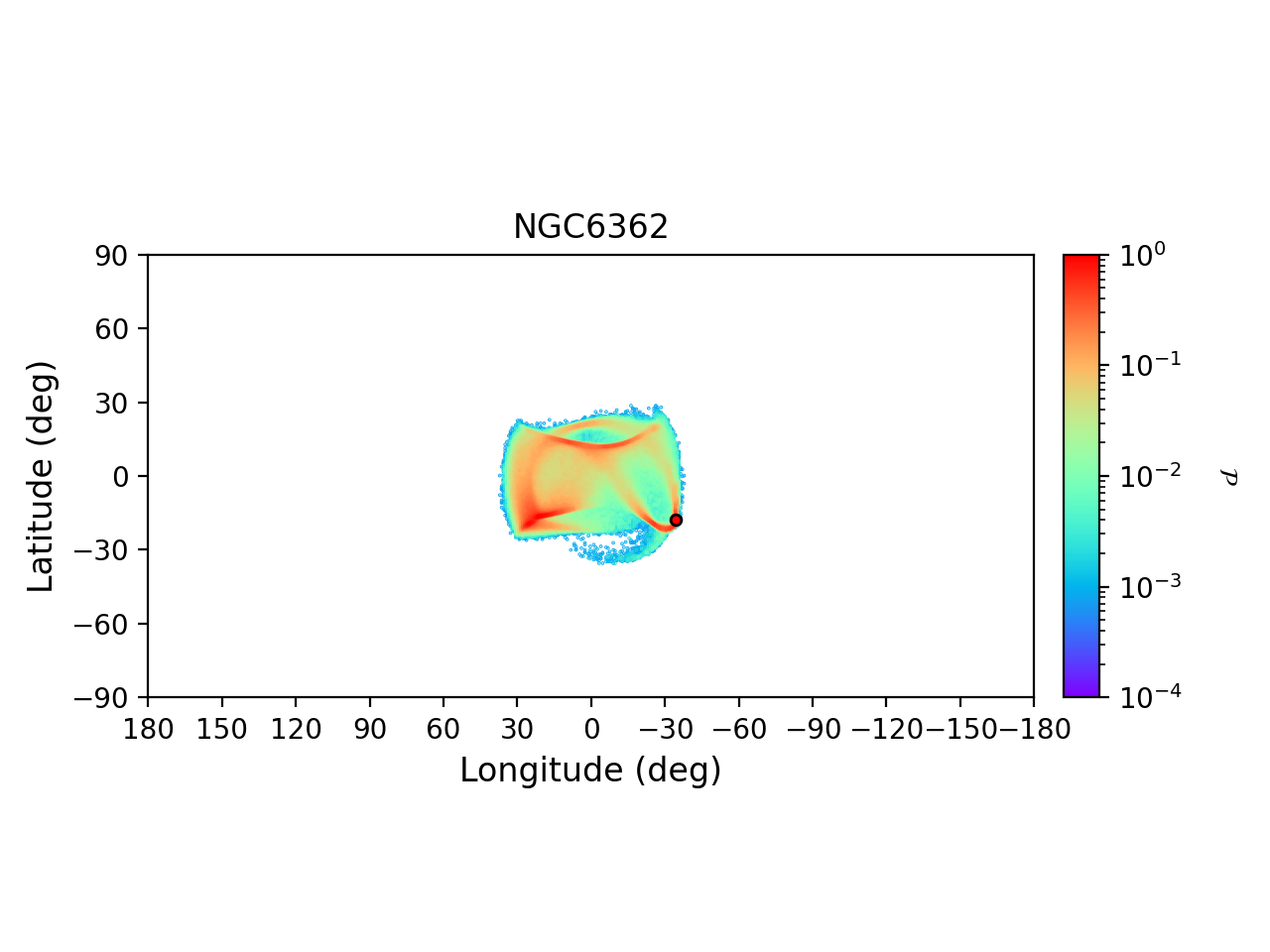}
\includegraphics[clip=true, trim = 0mm 20mm 0mm 10mm, width=1\columnwidth]{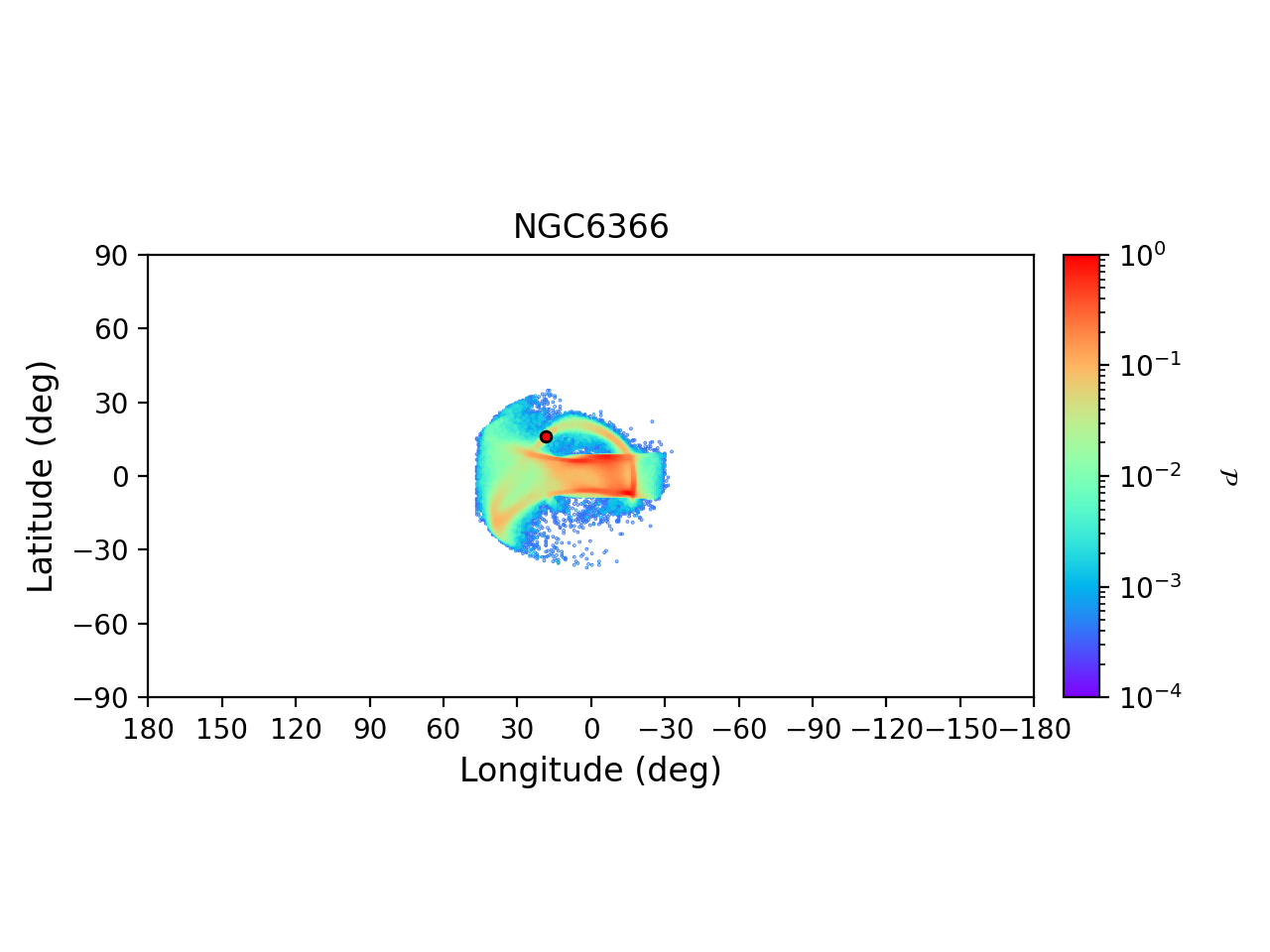}
\caption{Projected density distribution in the $(\ell, b)$ plane of a subset of simulated globular clusters, as indicated at the top of each panel. In each panel, the red circle indicates the current position of the cluster. The densities have been normalized to their maximum value.}\label{stream10}
\end{figure*}

\begin{figure*}
\includegraphics[clip=true, trim = 0mm 20mm 0mm 10mm, width=1\columnwidth]{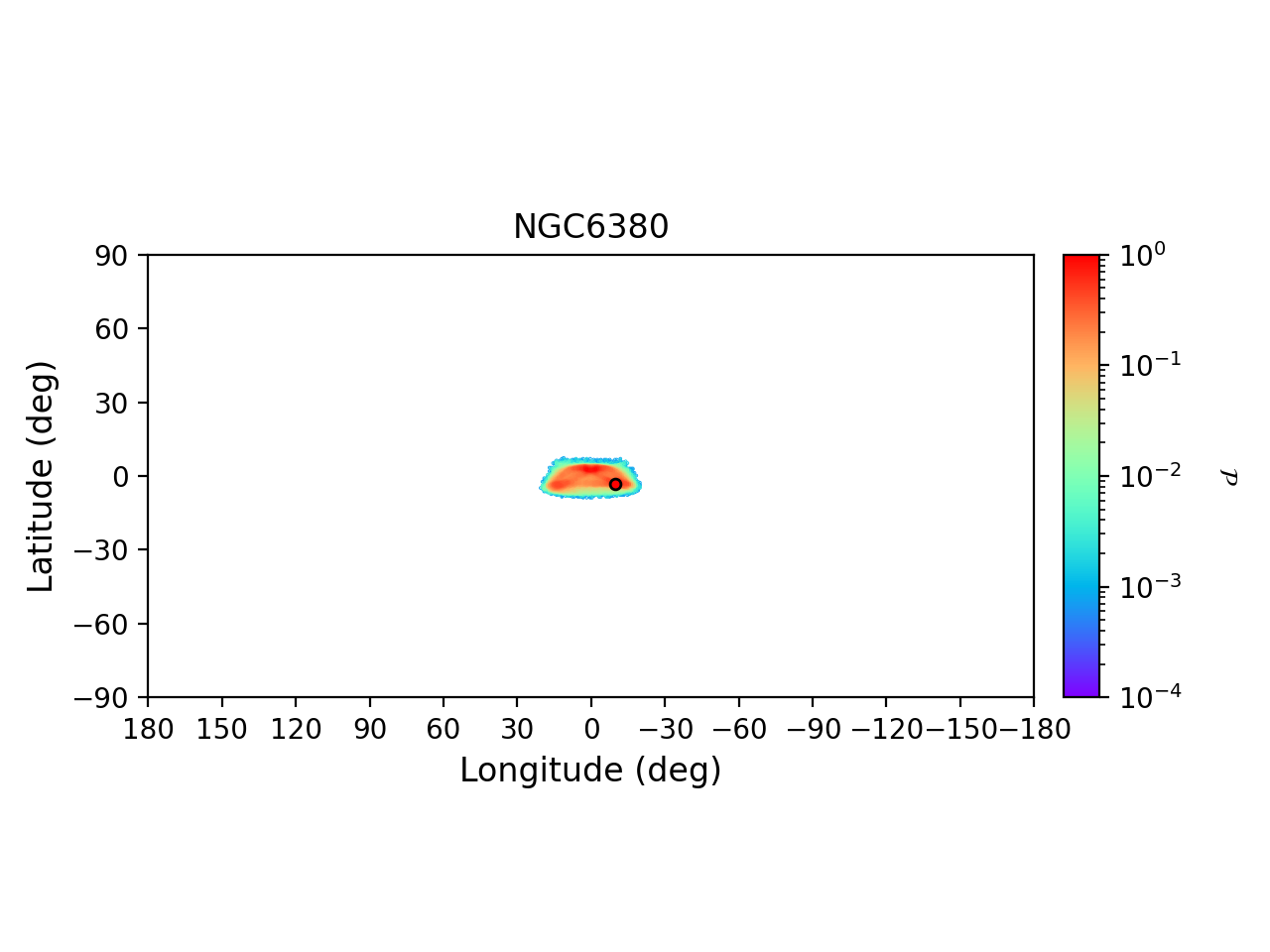}
\includegraphics[clip=true, trim = 0mm 20mm 0mm 10mm, width=1\columnwidth]{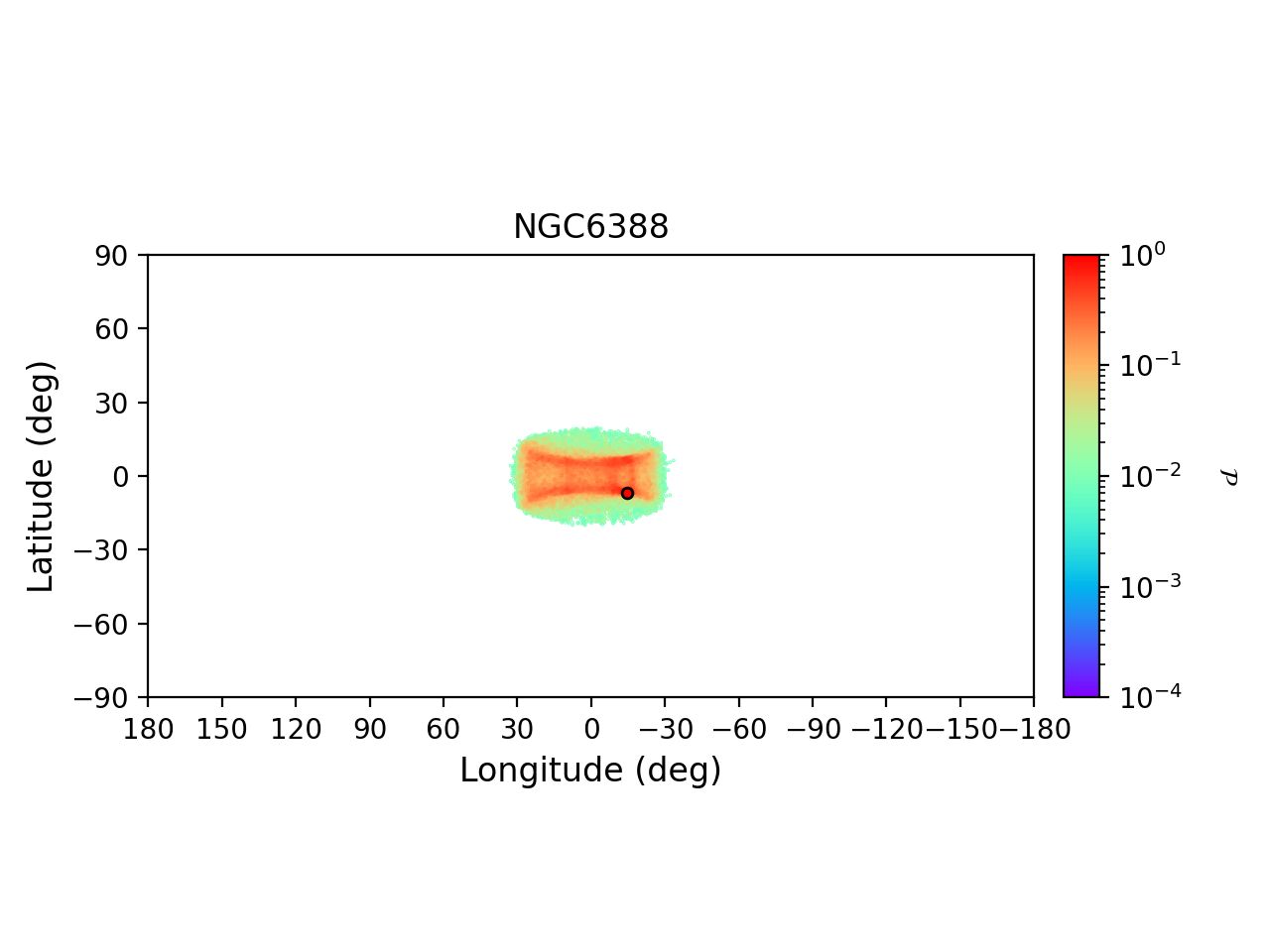}
\includegraphics[clip=true, trim = 0mm 20mm 0mm 10mm, width=1\columnwidth]{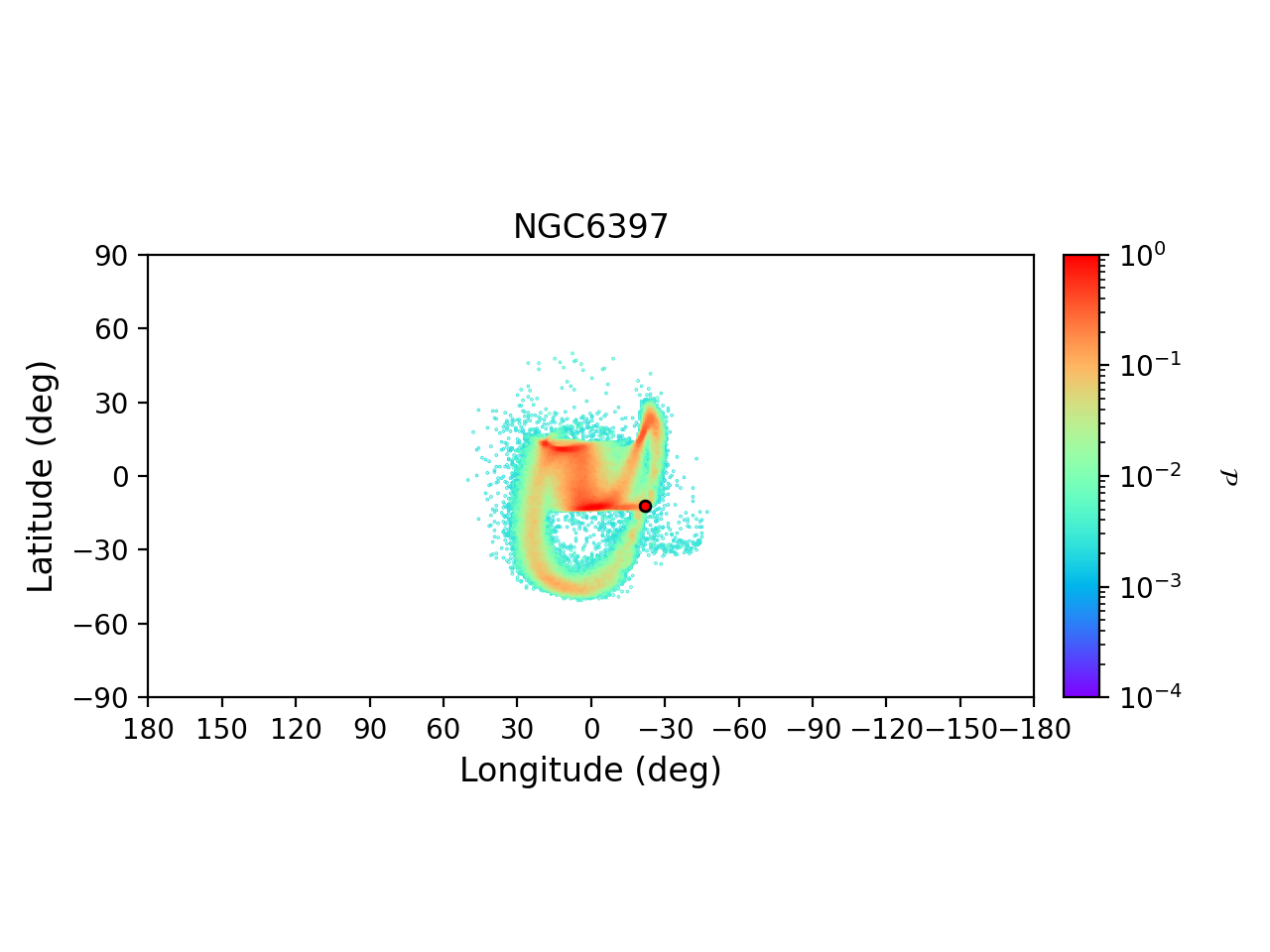}
\includegraphics[clip=true, trim = 0mm 20mm 0mm 10mm, width=1\columnwidth]{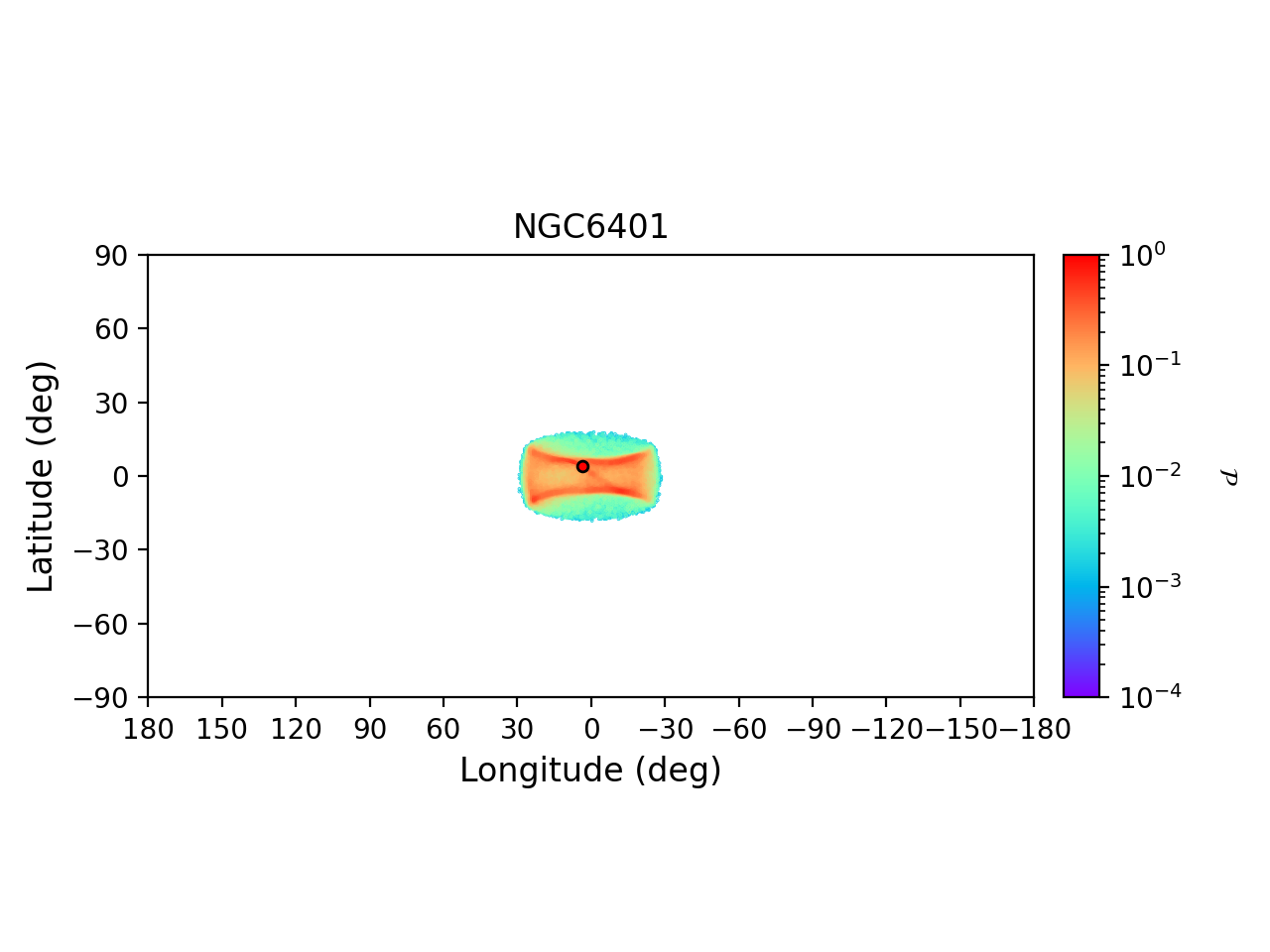}
\includegraphics[clip=true, trim = 0mm 20mm 0mm 10mm, width=1\columnwidth]{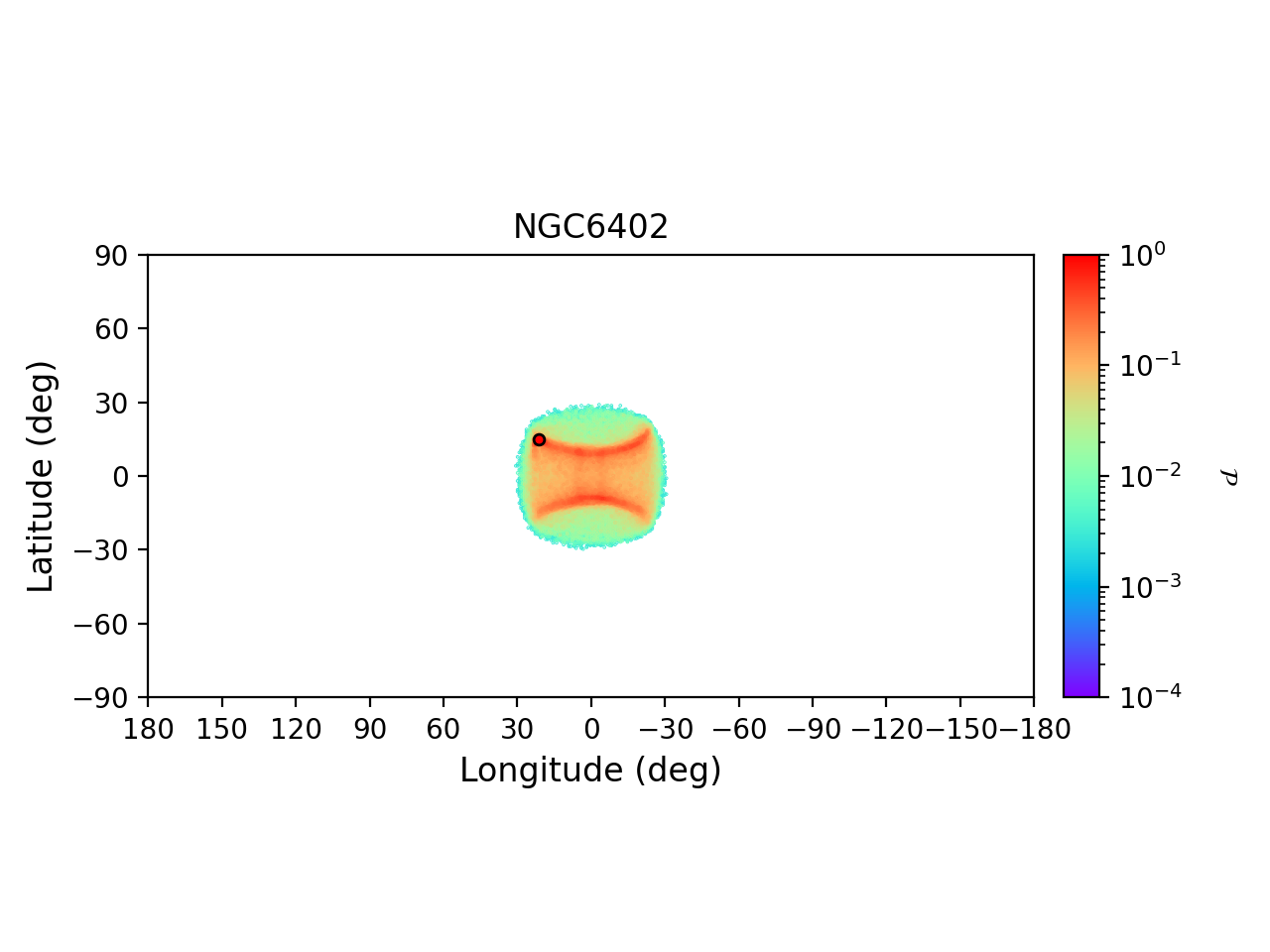}
\includegraphics[clip=true, trim = 0mm 20mm 0mm 10mm, width=1\columnwidth]{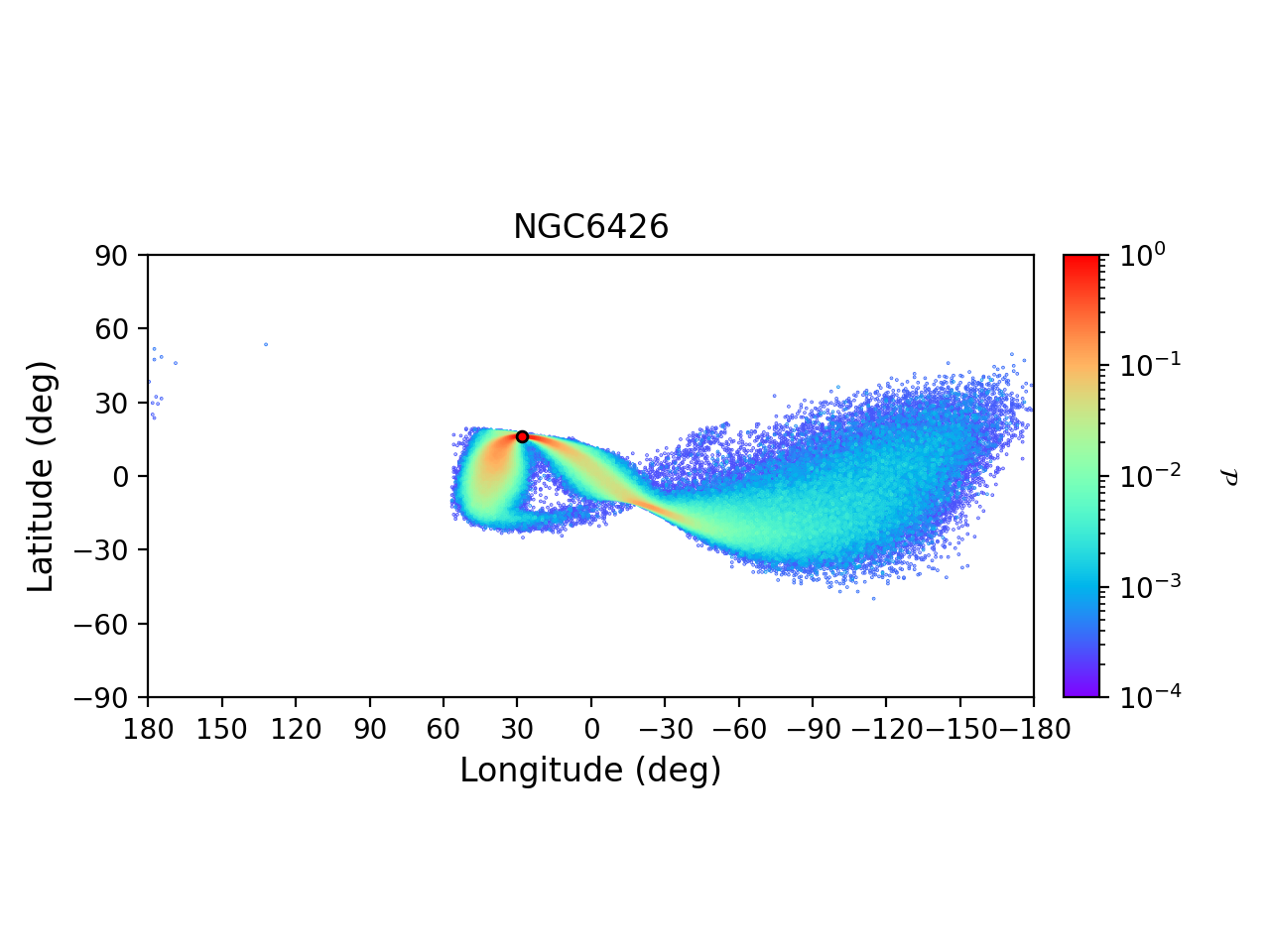}
\includegraphics[clip=true, trim = 0mm 20mm 0mm 10mm, width=1\columnwidth]{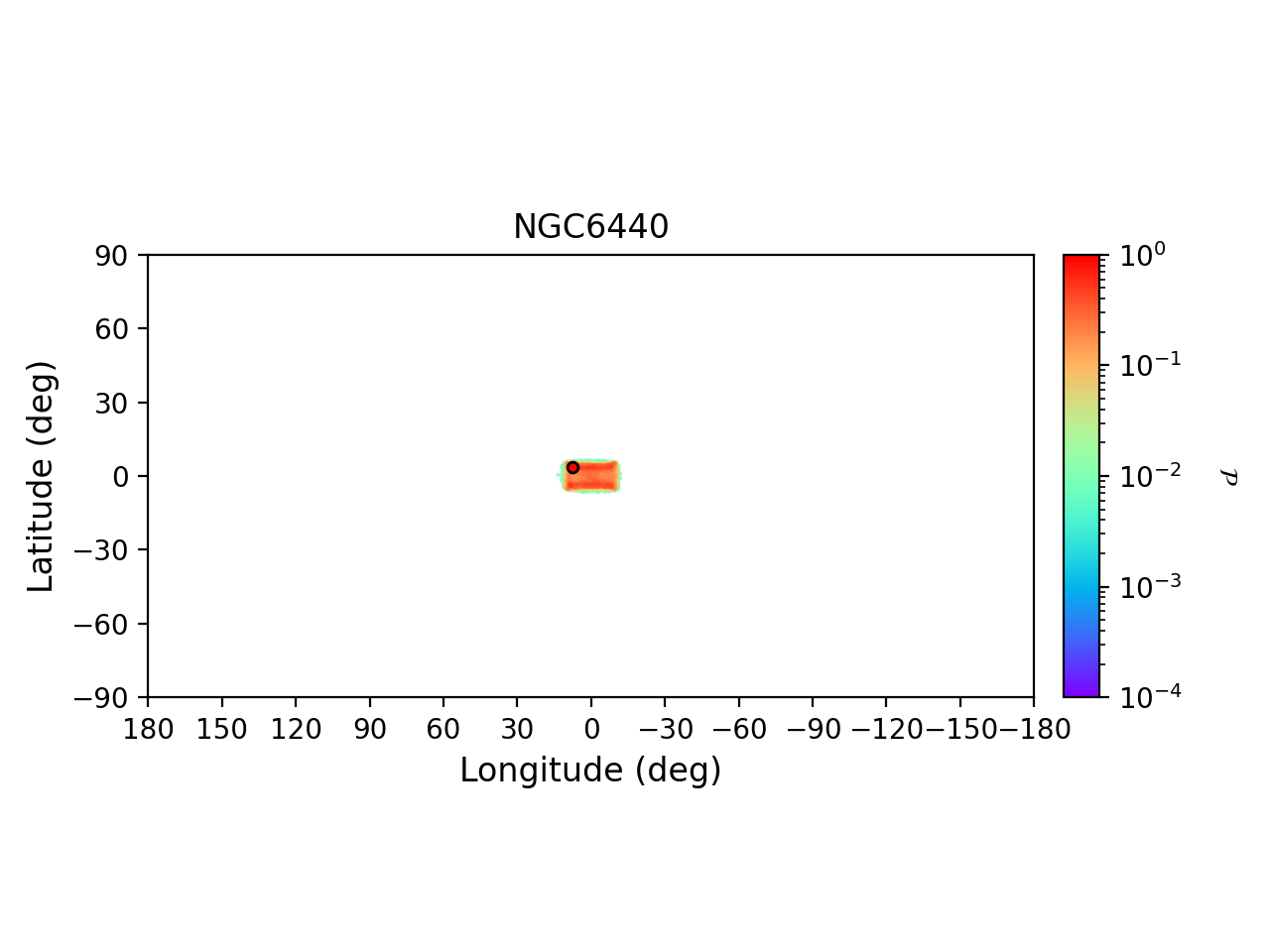}
\includegraphics[clip=true, trim = 0mm 20mm 0mm 10mm, width=1\columnwidth]{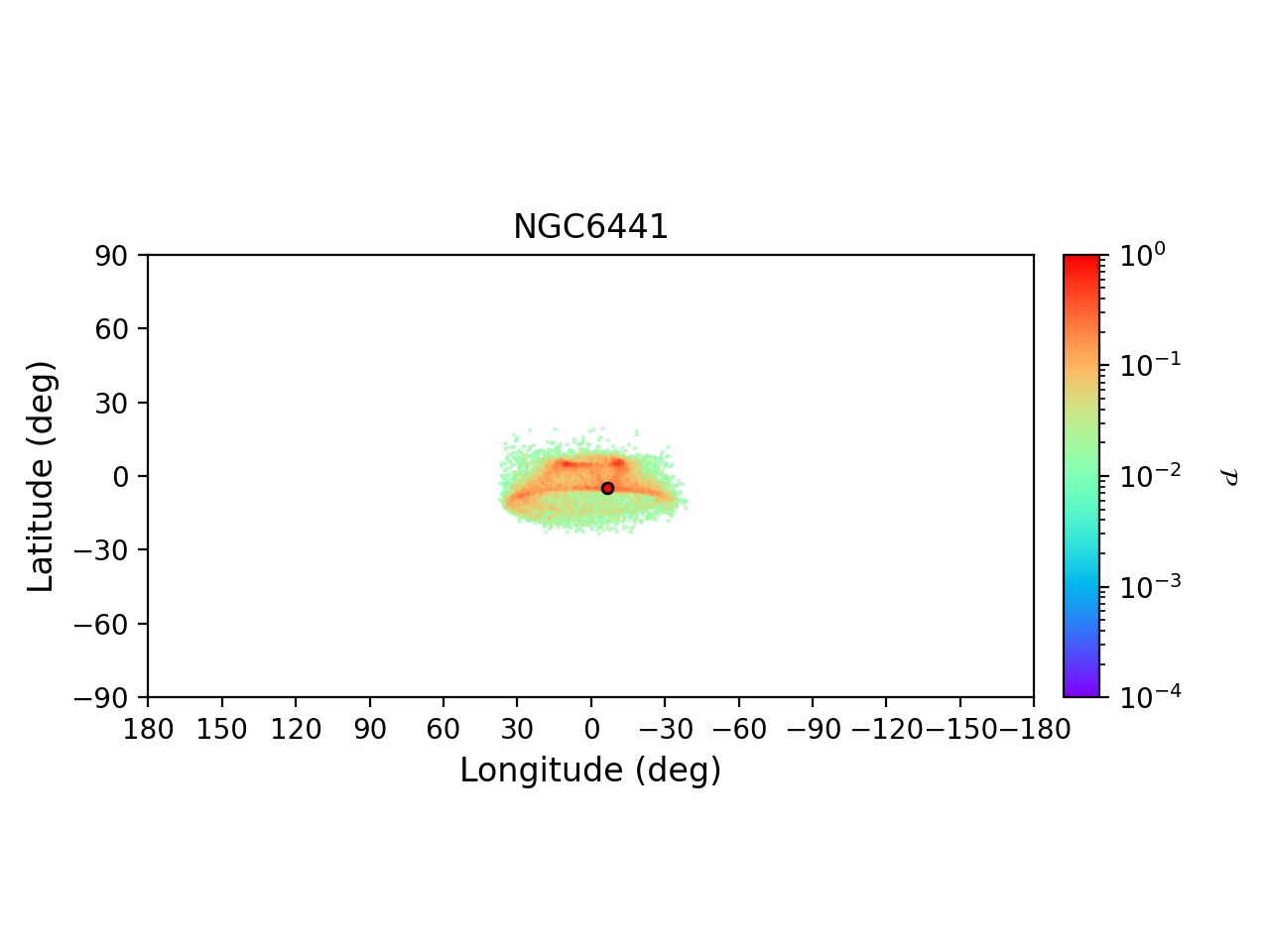}
\caption{Projected density distribution in the $(\ell, b)$ plane of a subset of simulated globular clusters, as indicated at the top of each panel. In each panel, the red circle indicates the current position of the cluster. The densities have been normalized to their maximum value.}\label{stream11}
\end{figure*}

\begin{figure*}
\includegraphics[clip=true, trim = 0mm 20mm 0mm 10mm, width=1\columnwidth]{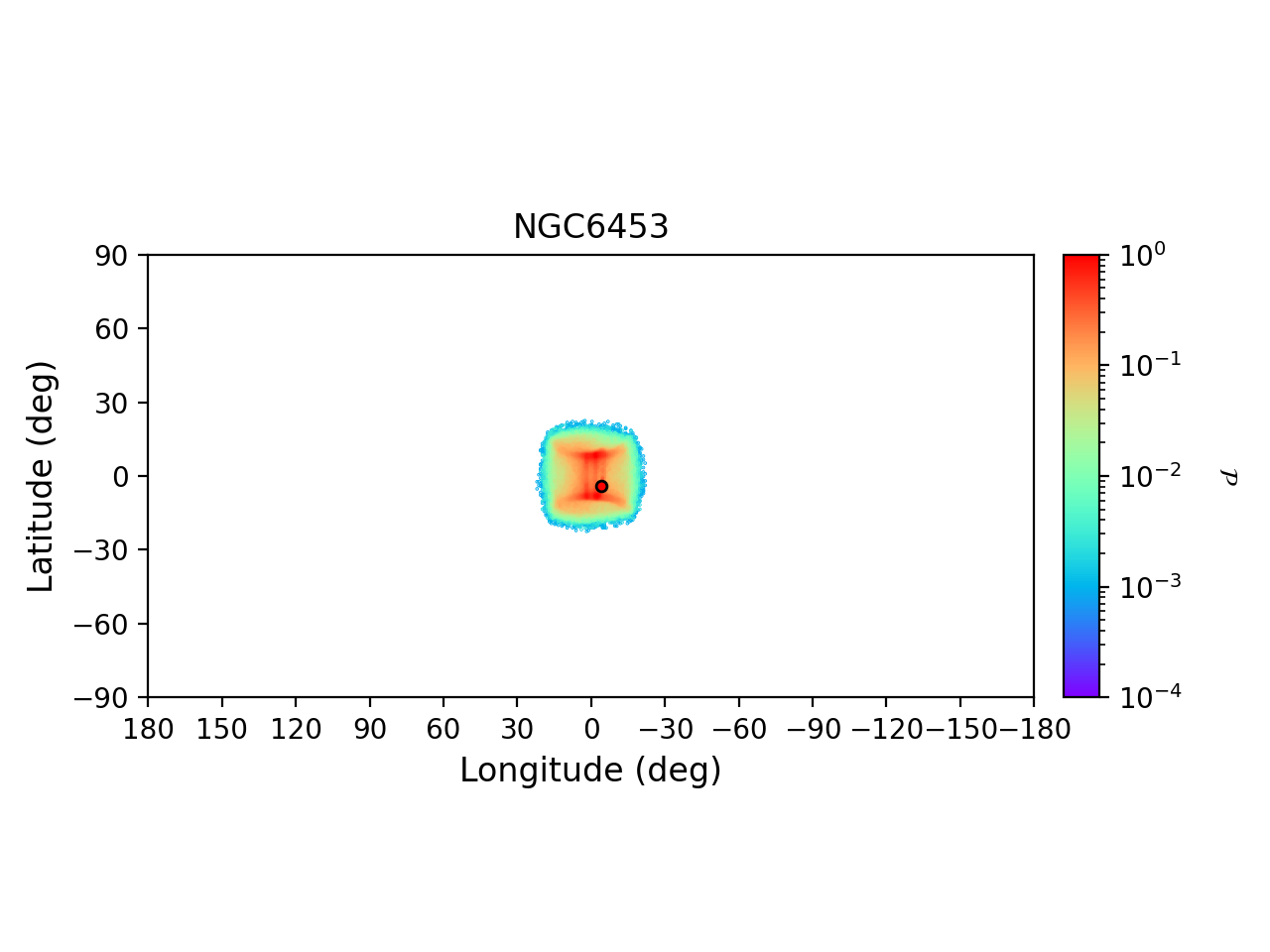}
\includegraphics[clip=true, trim = 0mm 20mm 0mm 10mm, width=1\columnwidth]{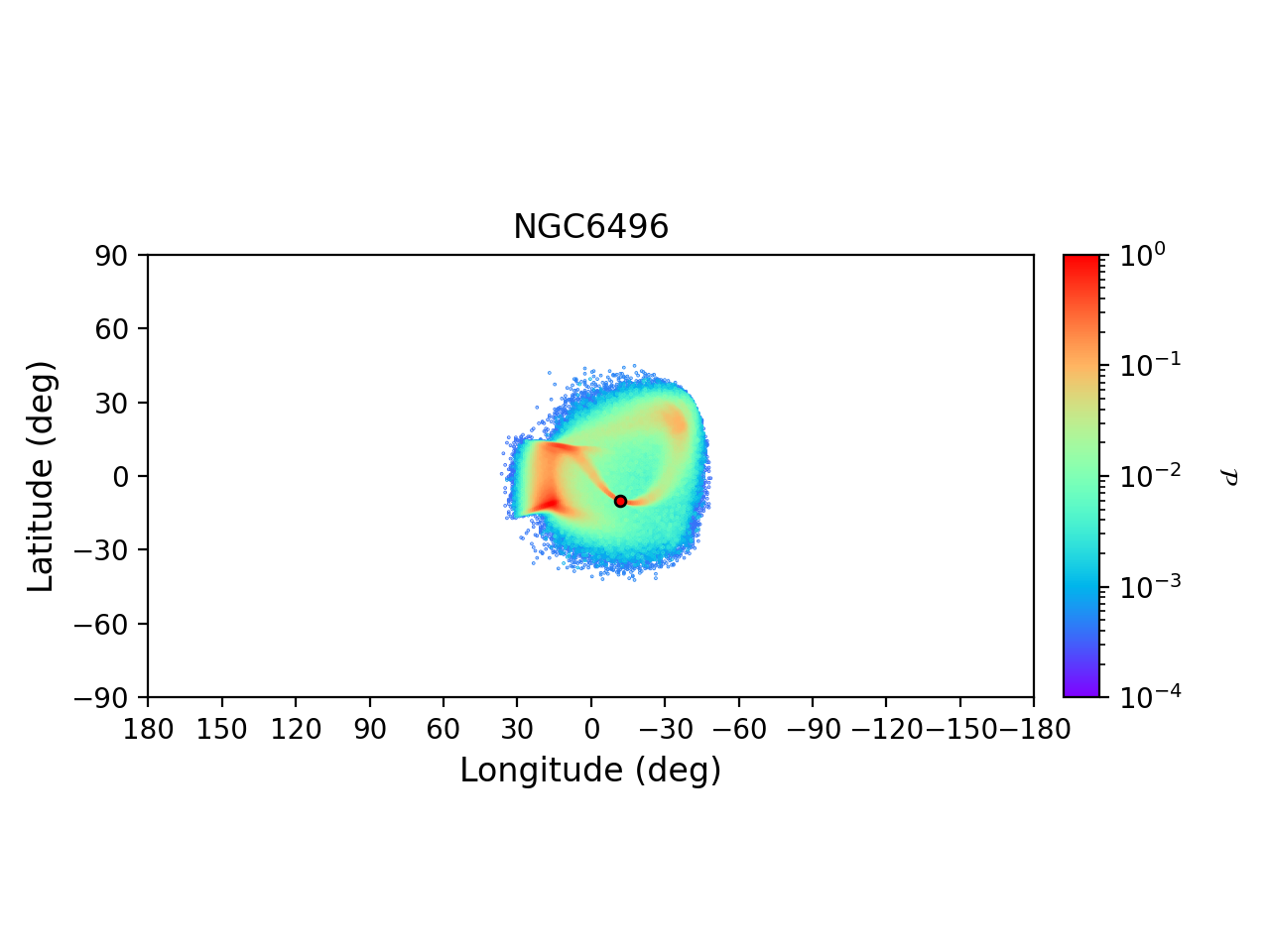}
\includegraphics[clip=true, trim = 0mm 20mm 0mm 10mm, width=1\columnwidth]{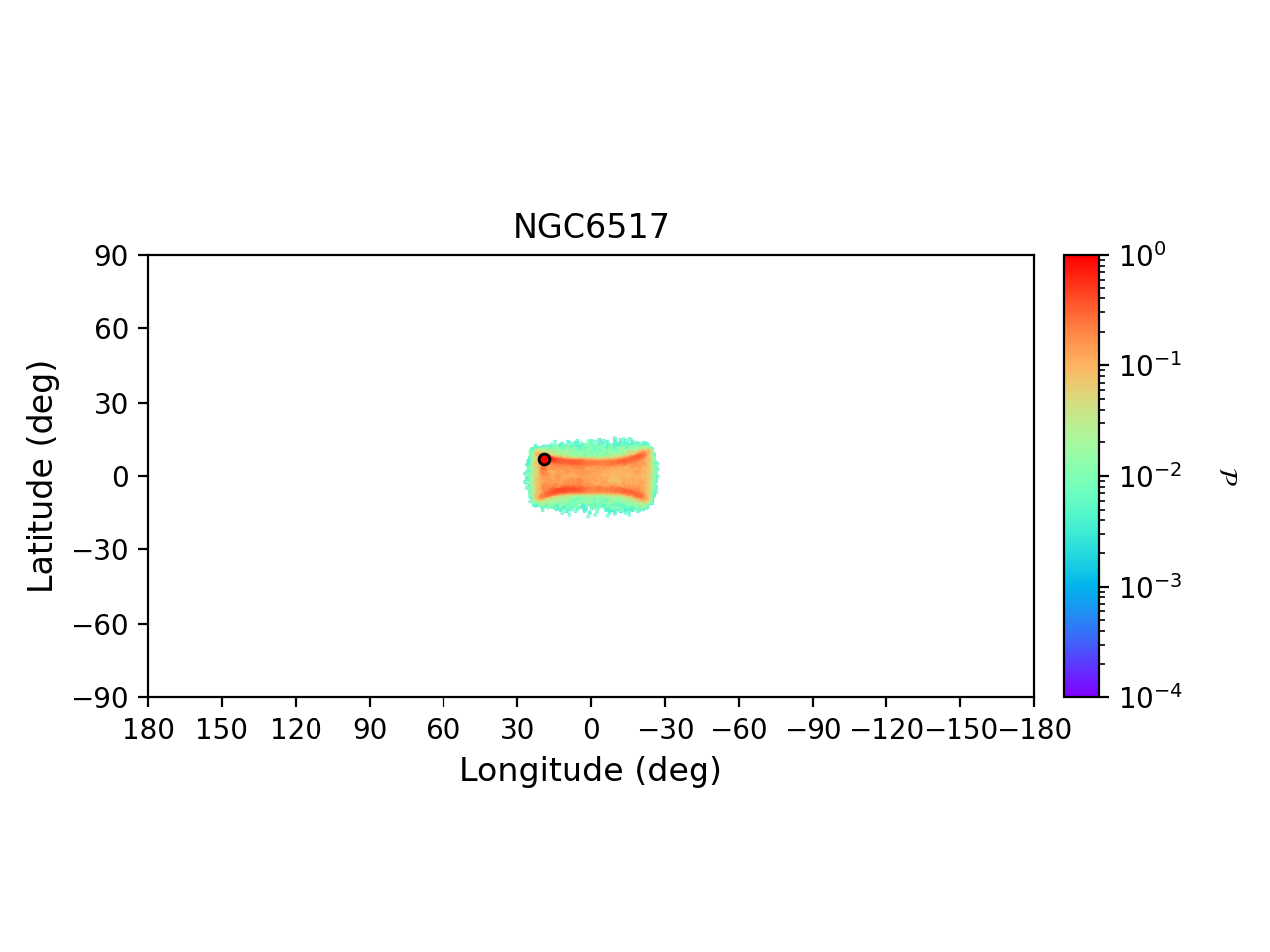}
\includegraphics[clip=true, trim = 0mm 20mm 0mm 10mm, width=1\columnwidth]{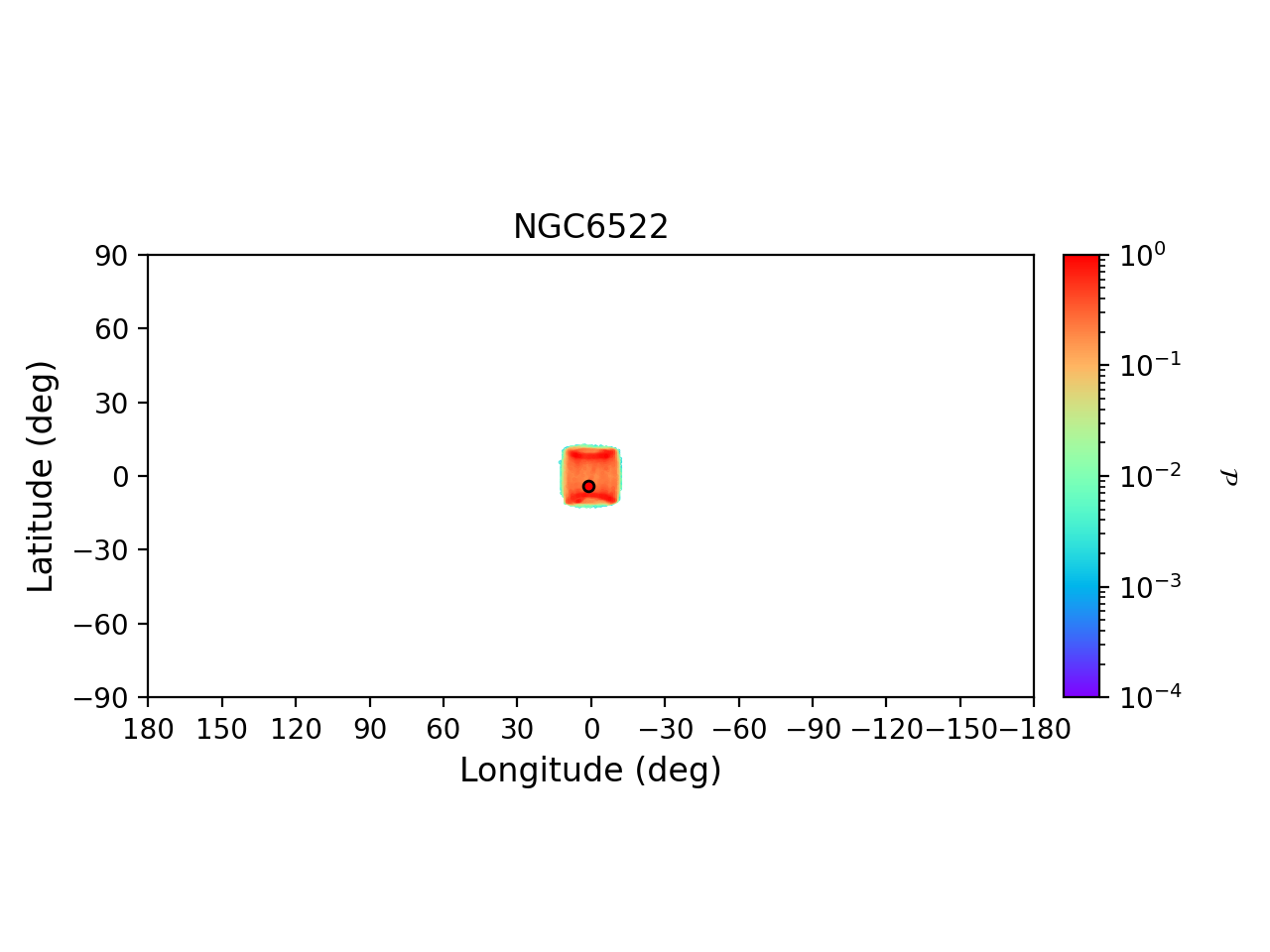}
\includegraphics[clip=true, trim = 0mm 20mm 0mm 10mm, width=1\columnwidth]{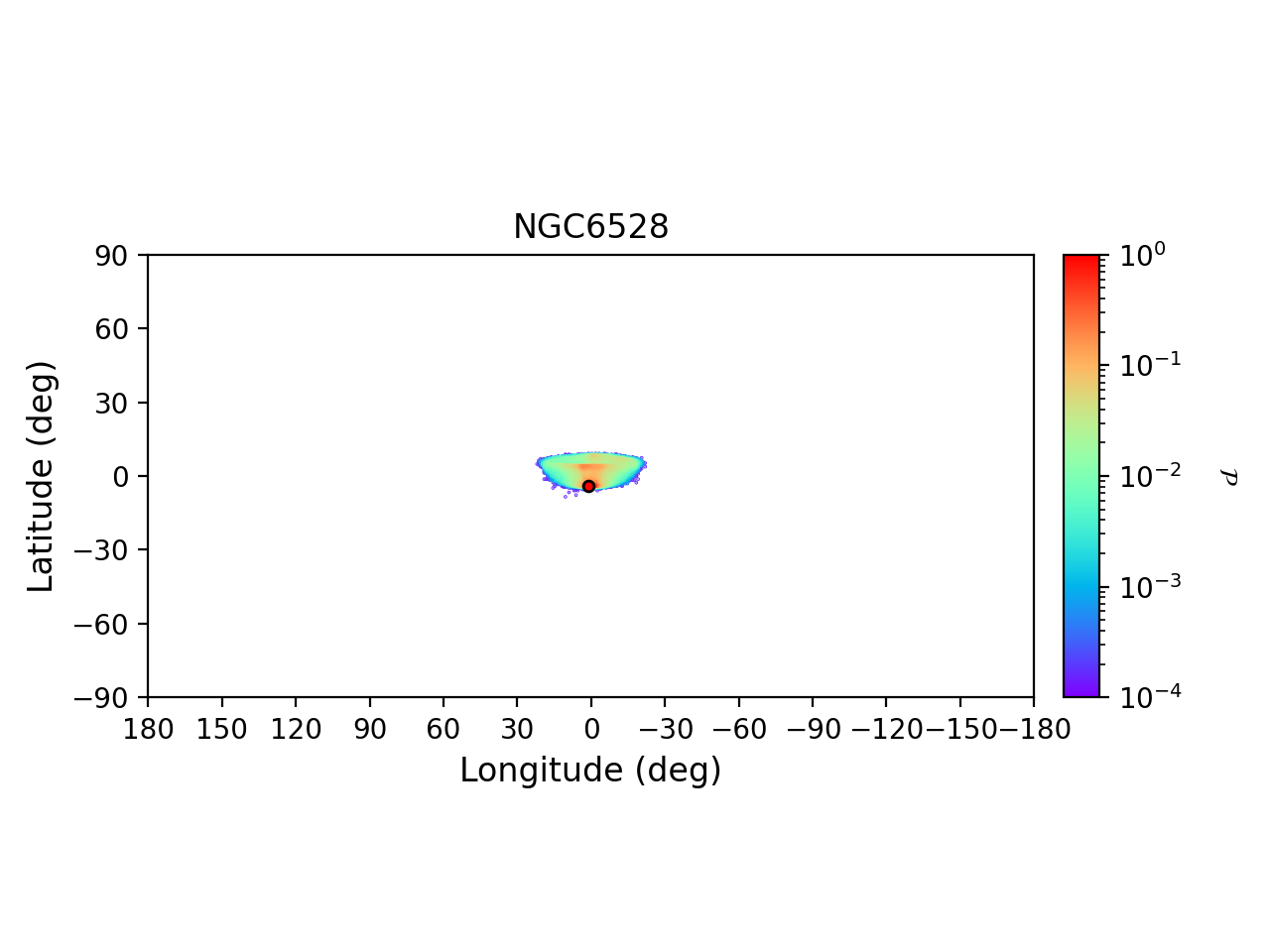}
\includegraphics[clip=true, trim = 0mm 20mm 0mm 10mm, width=1\columnwidth]{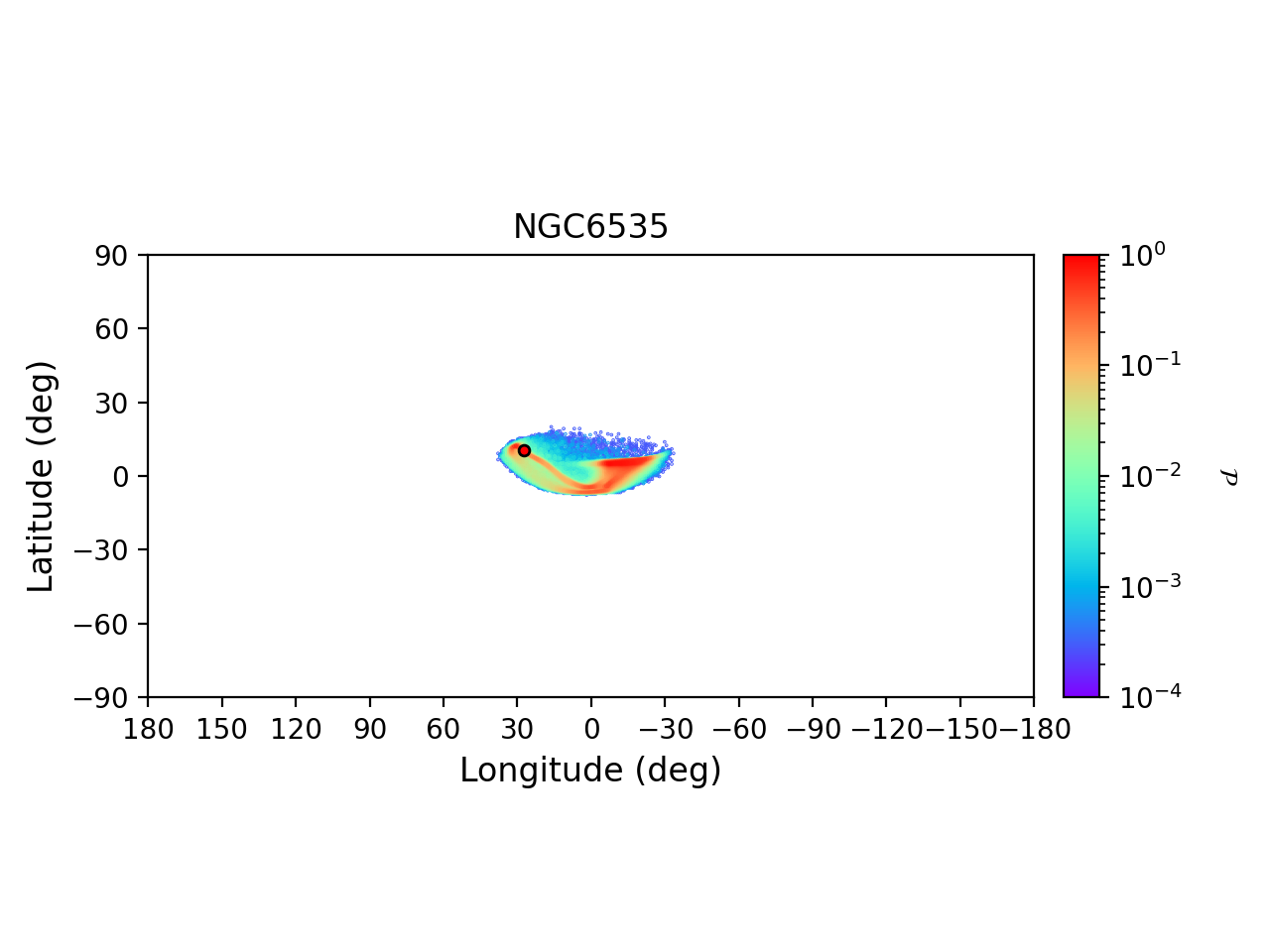}
\includegraphics[clip=true, trim = 0mm 20mm 0mm 10mm, width=1\columnwidth]{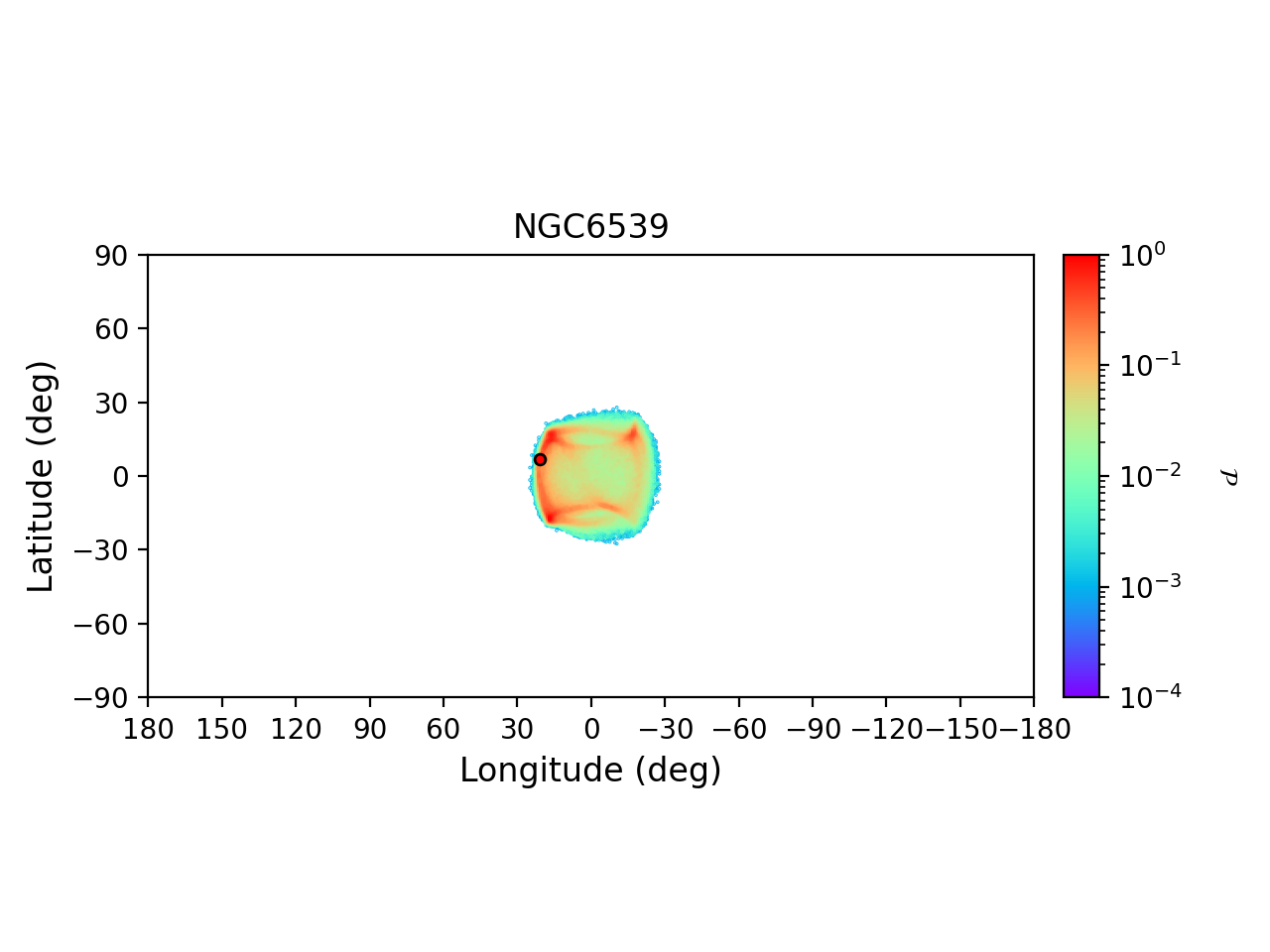}
\includegraphics[clip=true, trim = 0mm 20mm 0mm 10mm, width=1\columnwidth]{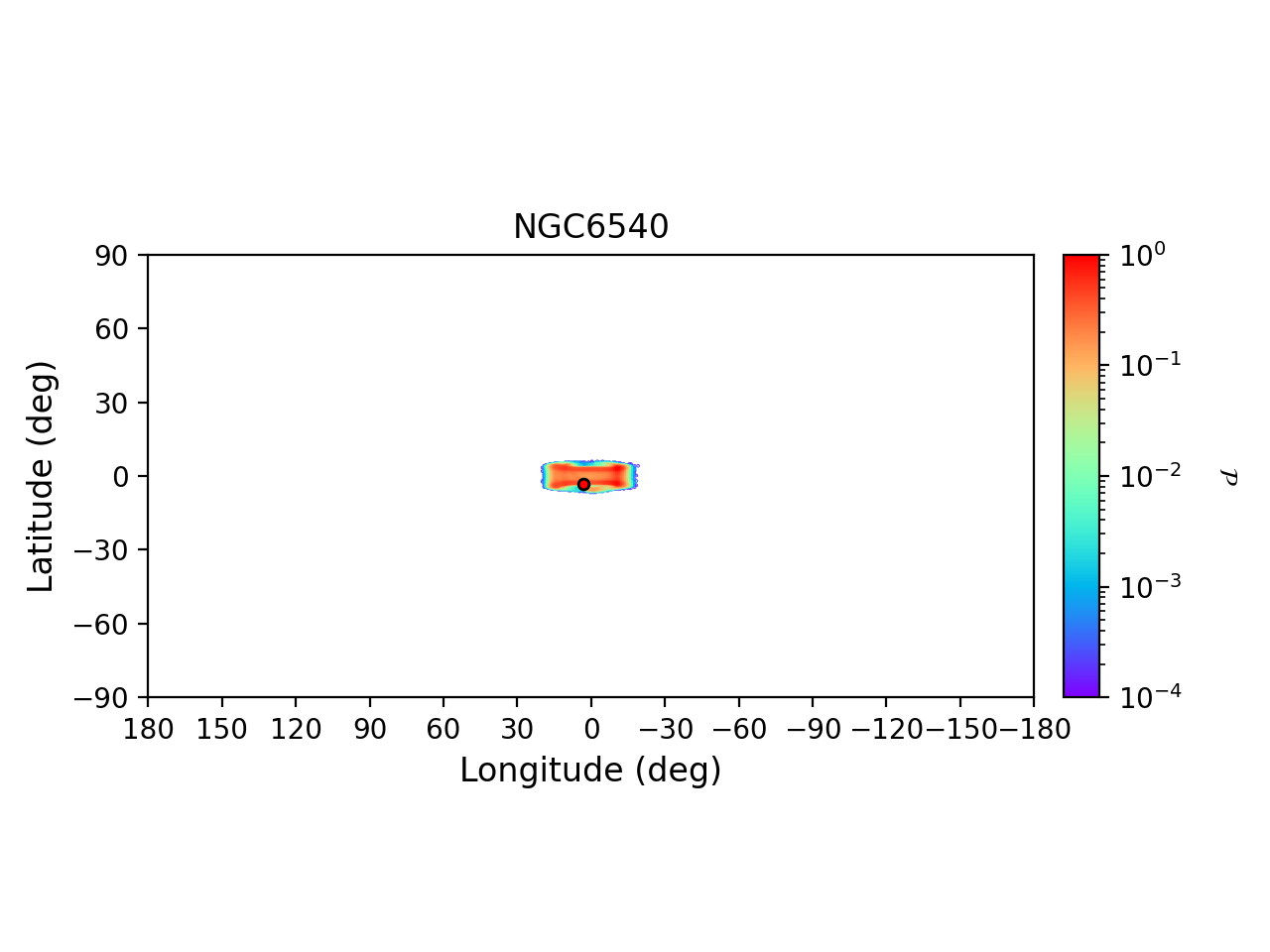}
\caption{Projected density distribution in the $(\ell, b)$ plane of a subset of simulated globular clusters, as indicated at the top of each panel. In each panel, the red circle indicates the current position of the cluster. The densities have been normalized to their maximum value.}\label{stream12}
\end{figure*}

\begin{figure*}
\includegraphics[clip=true, trim = 0mm 20mm 0mm 10mm, width=1\columnwidth]{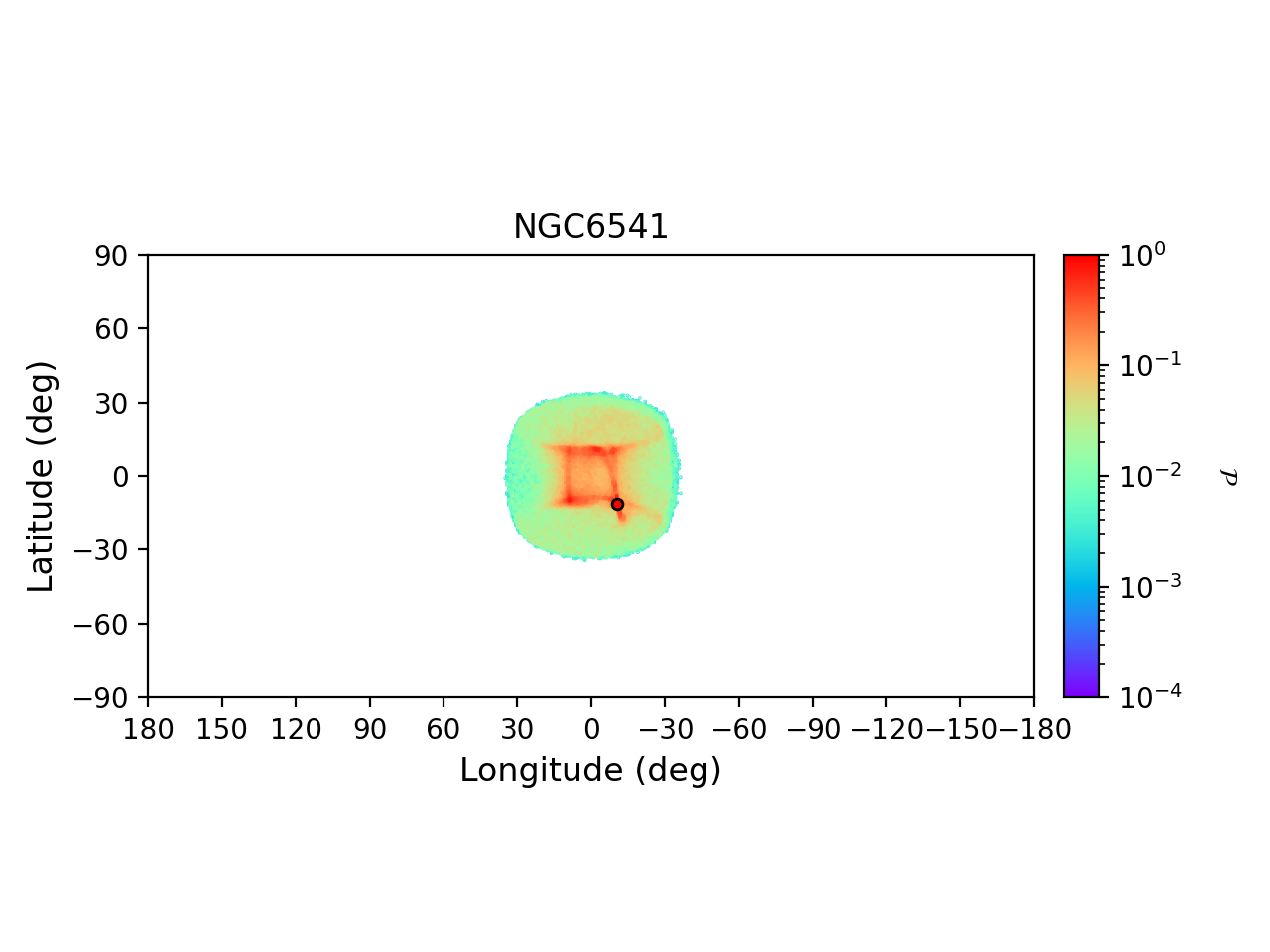}
\includegraphics[clip=true, trim = 0mm 20mm 0mm 10mm, width=1\columnwidth]{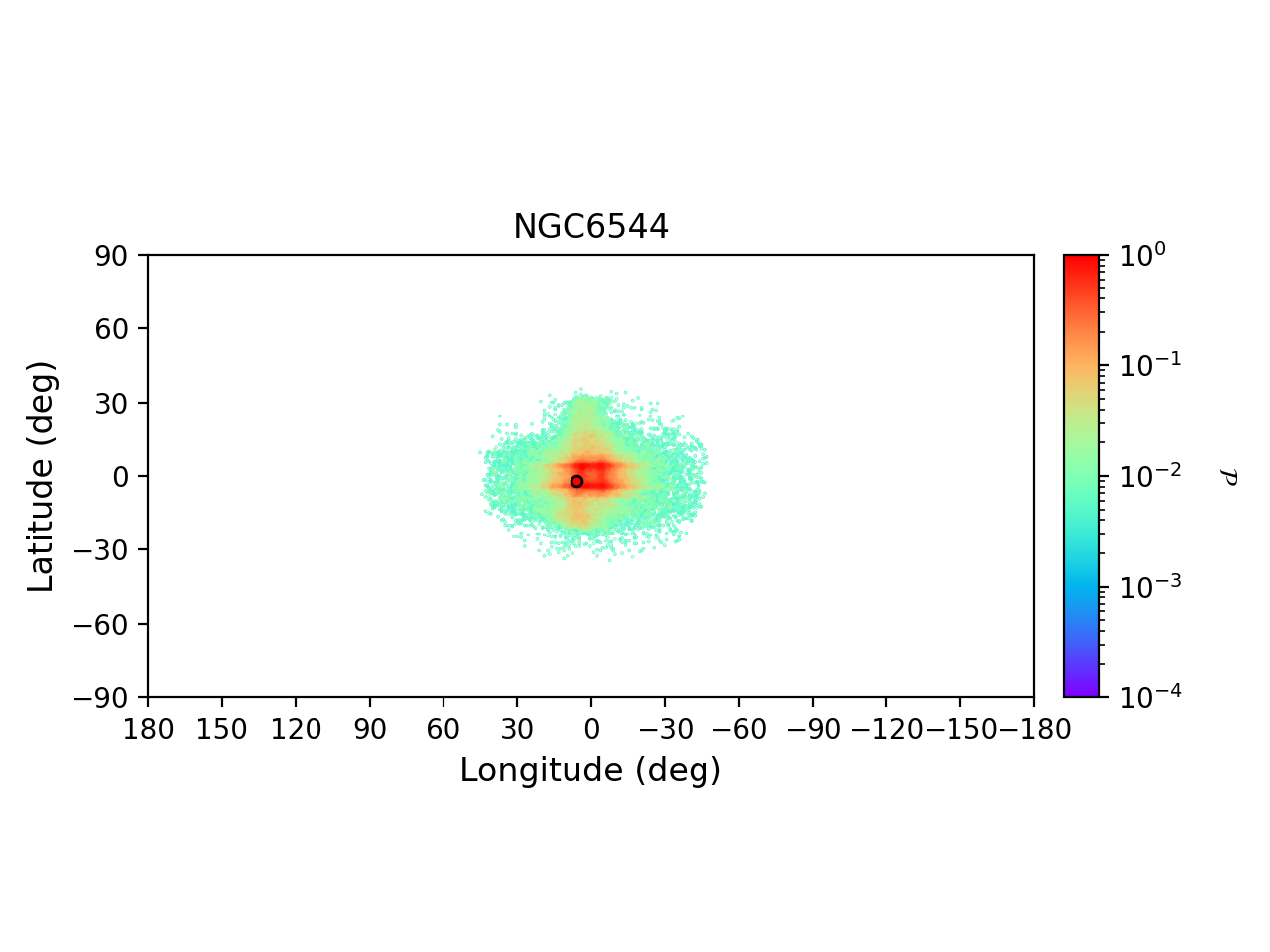}
\includegraphics[clip=true, trim = 0mm 20mm 0mm 10mm, width=1\columnwidth]{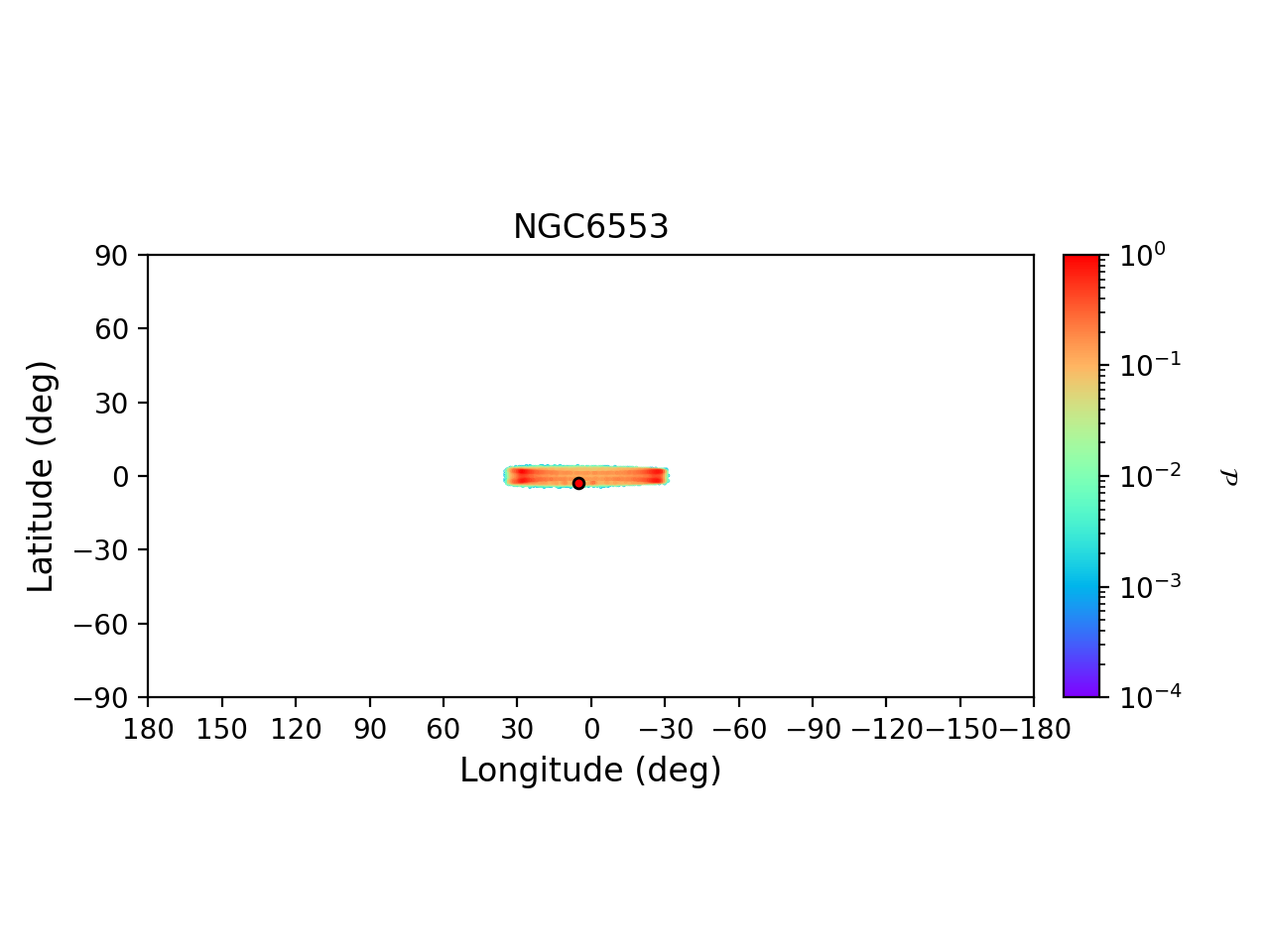}
\includegraphics[clip=true, trim = 0mm 20mm 0mm 10mm, width=1\columnwidth]{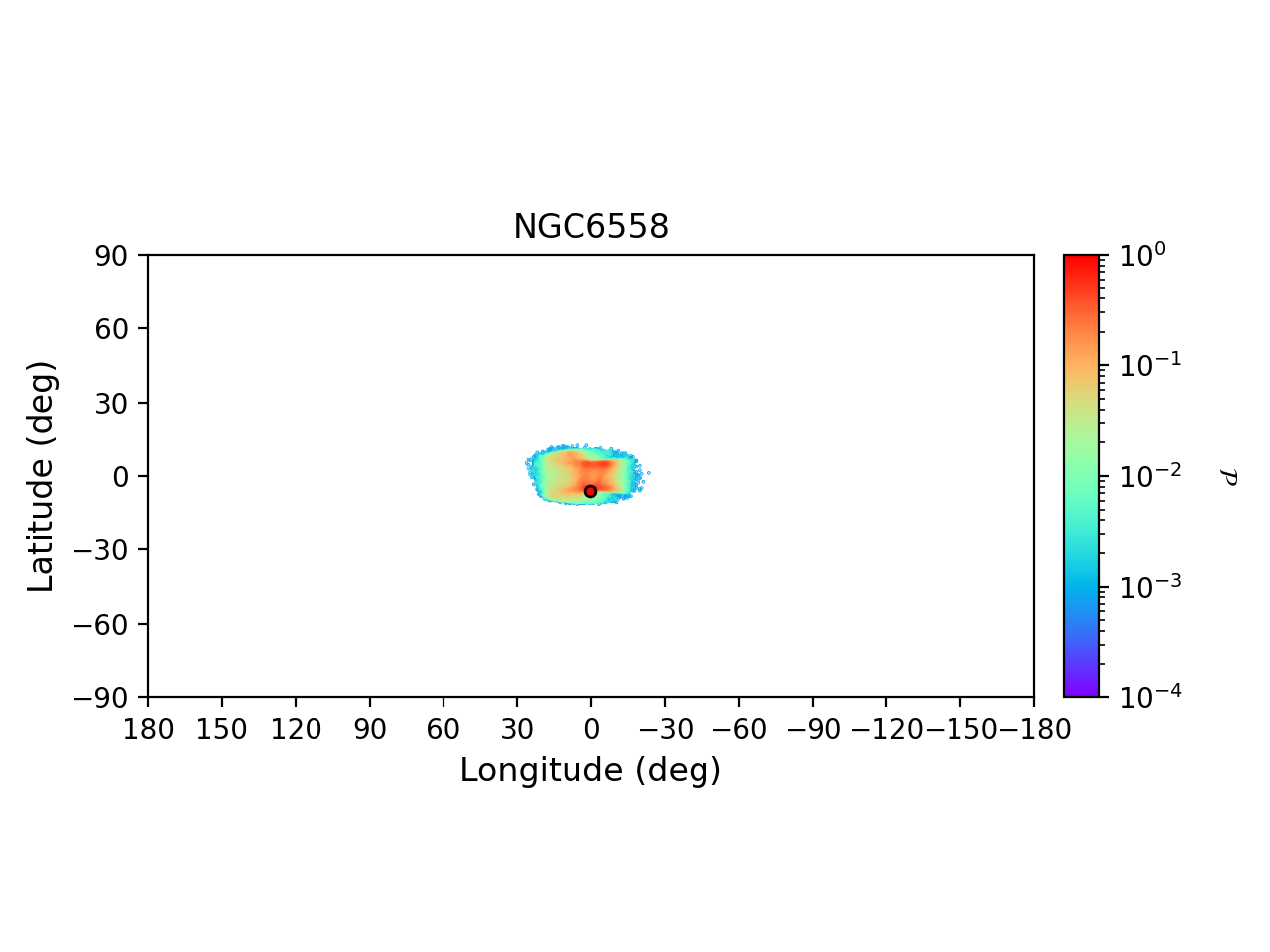}
\includegraphics[clip=true, trim = 0mm 20mm 0mm 10mm, width=1\columnwidth]{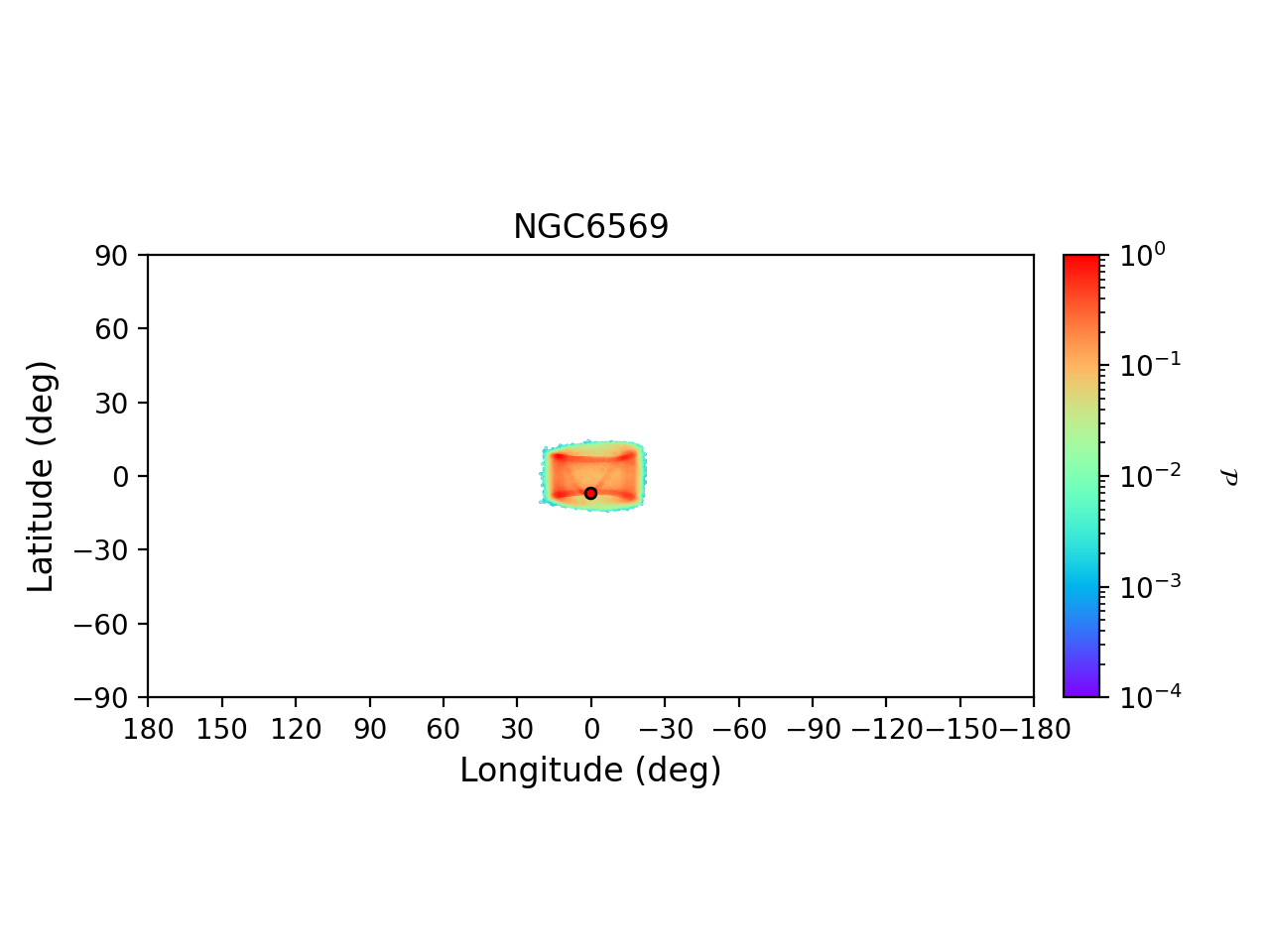}
\includegraphics[clip=true, trim = 0mm 20mm 0mm 10mm, width=1\columnwidth]{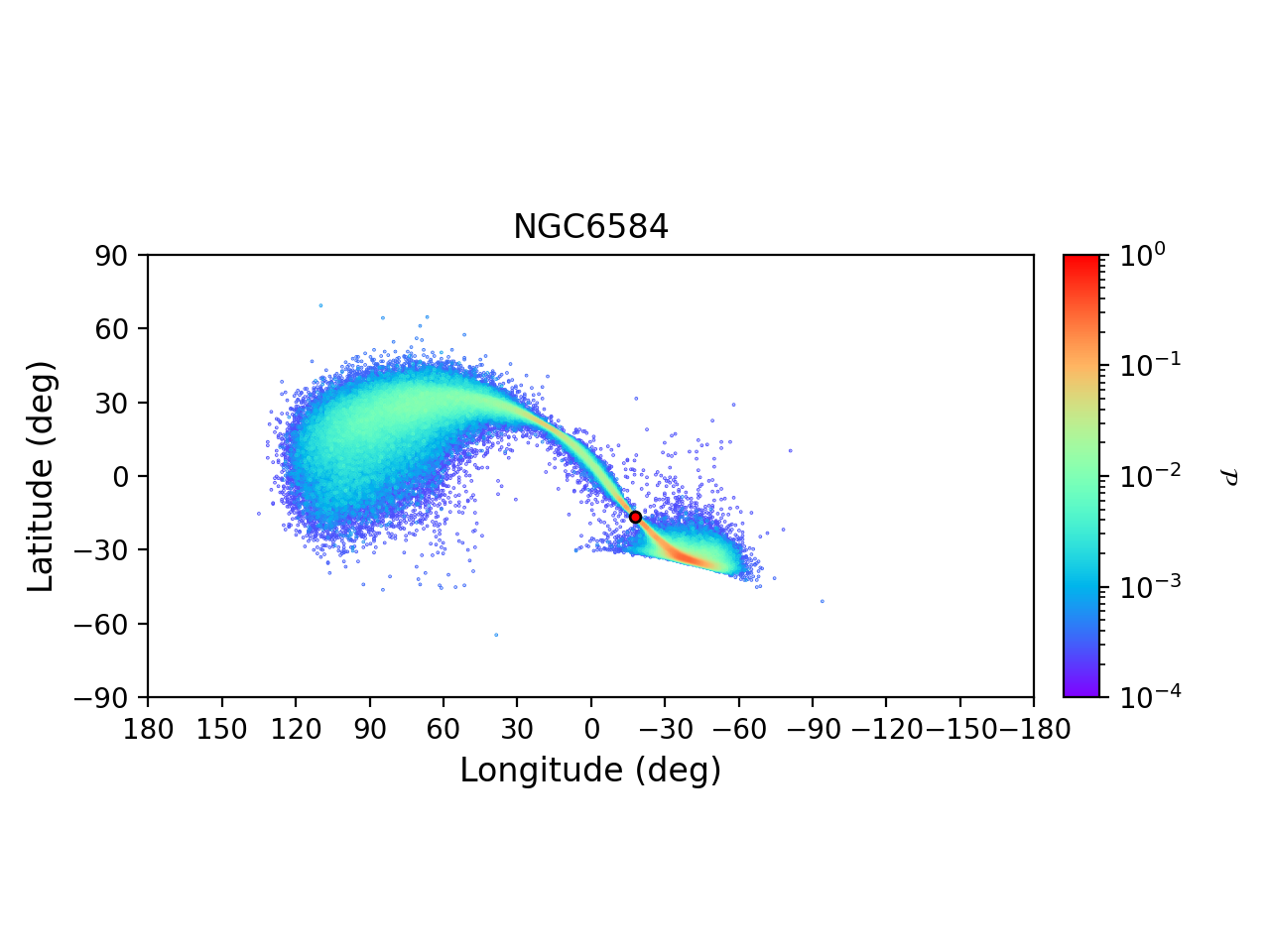}
\includegraphics[clip=true, trim = 0mm 20mm 0mm 10mm, width=1\columnwidth]{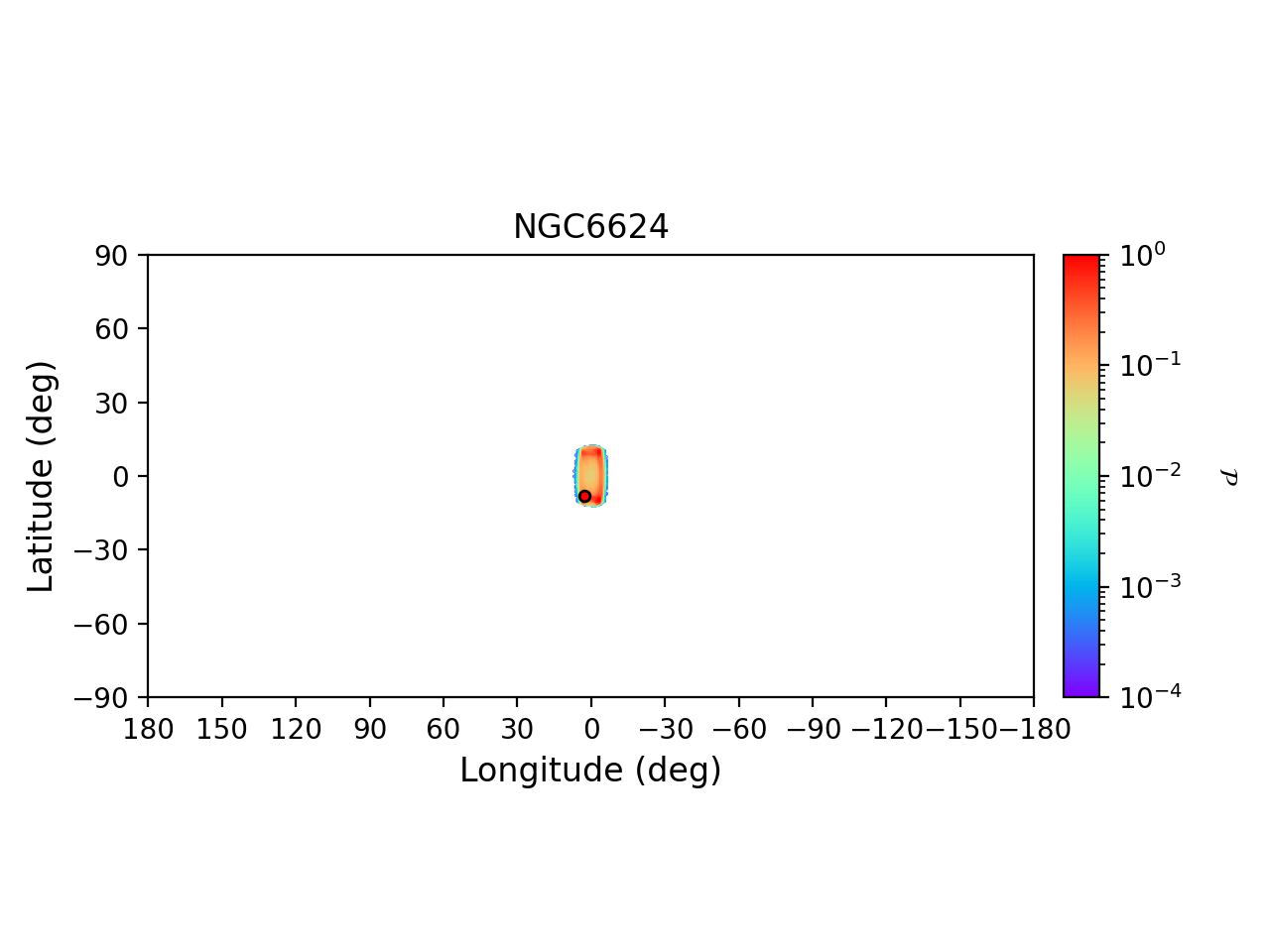}
\includegraphics[clip=true, trim = 0mm 20mm 0mm 10mm, width=1\columnwidth]{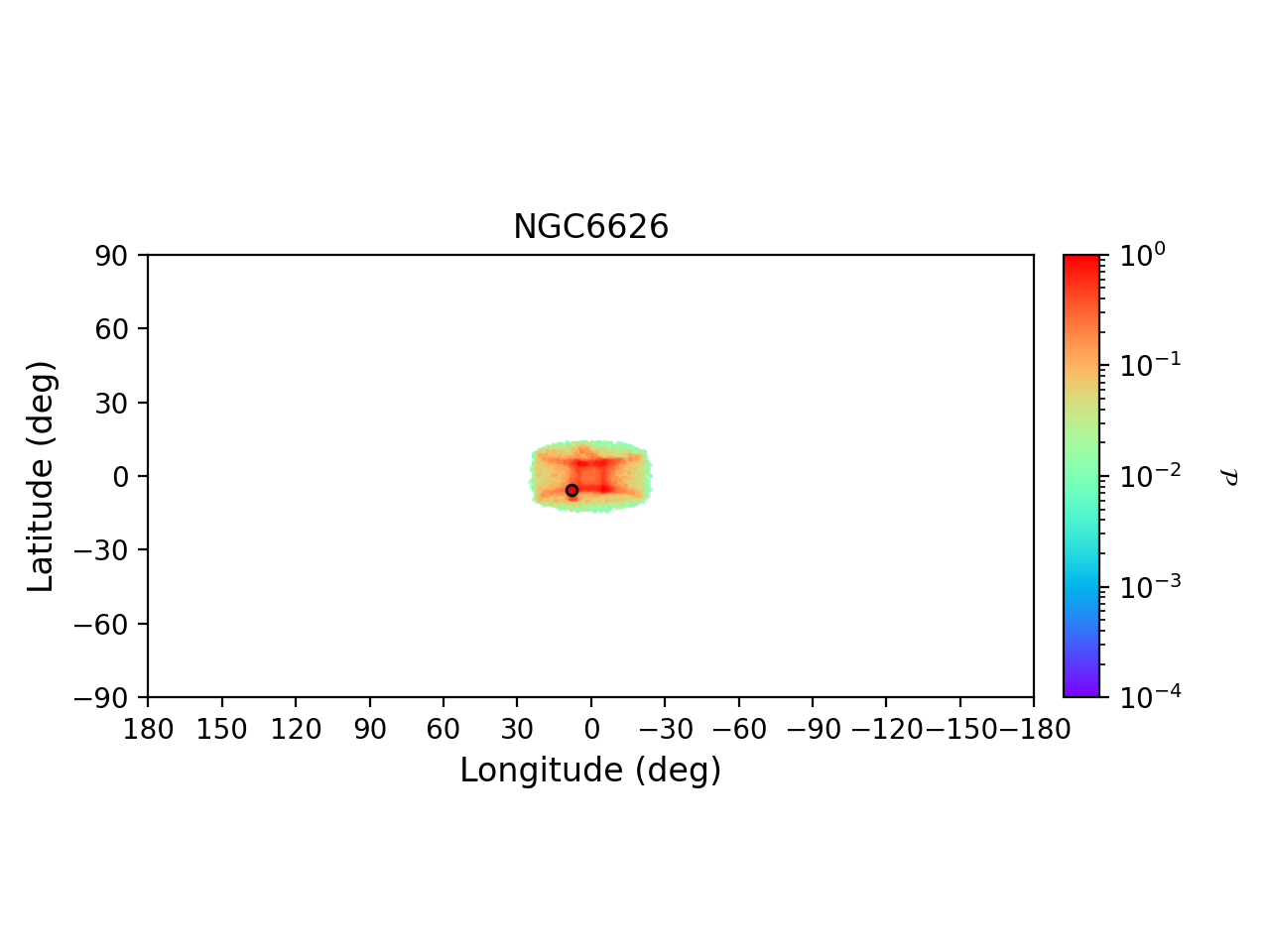}
\caption{Projected density distribution in the $(\ell, b)$ plane of a subset of simulated globular clusters, as indicated at the top of each panel. In each panel, the red circle indicates the current position of the cluster. The densities have been normalized to their maximum value.}\label{stream13}
\end{figure*}

\begin{figure*}
\includegraphics[clip=true, trim = 0mm 20mm 0mm 10mm, width=1\columnwidth]{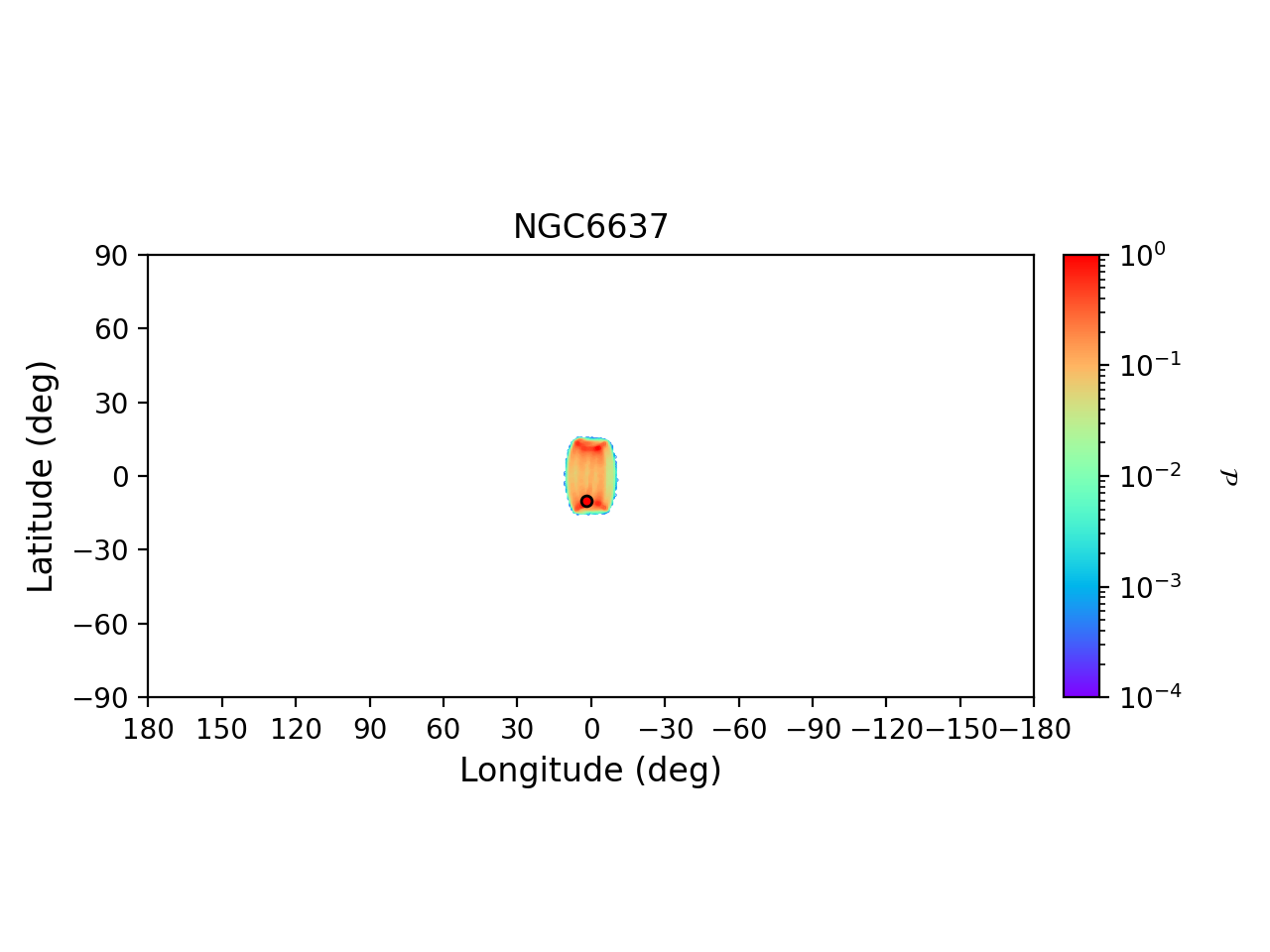}
\includegraphics[clip=true, trim = 0mm 20mm 0mm 10mm, width=1\columnwidth]{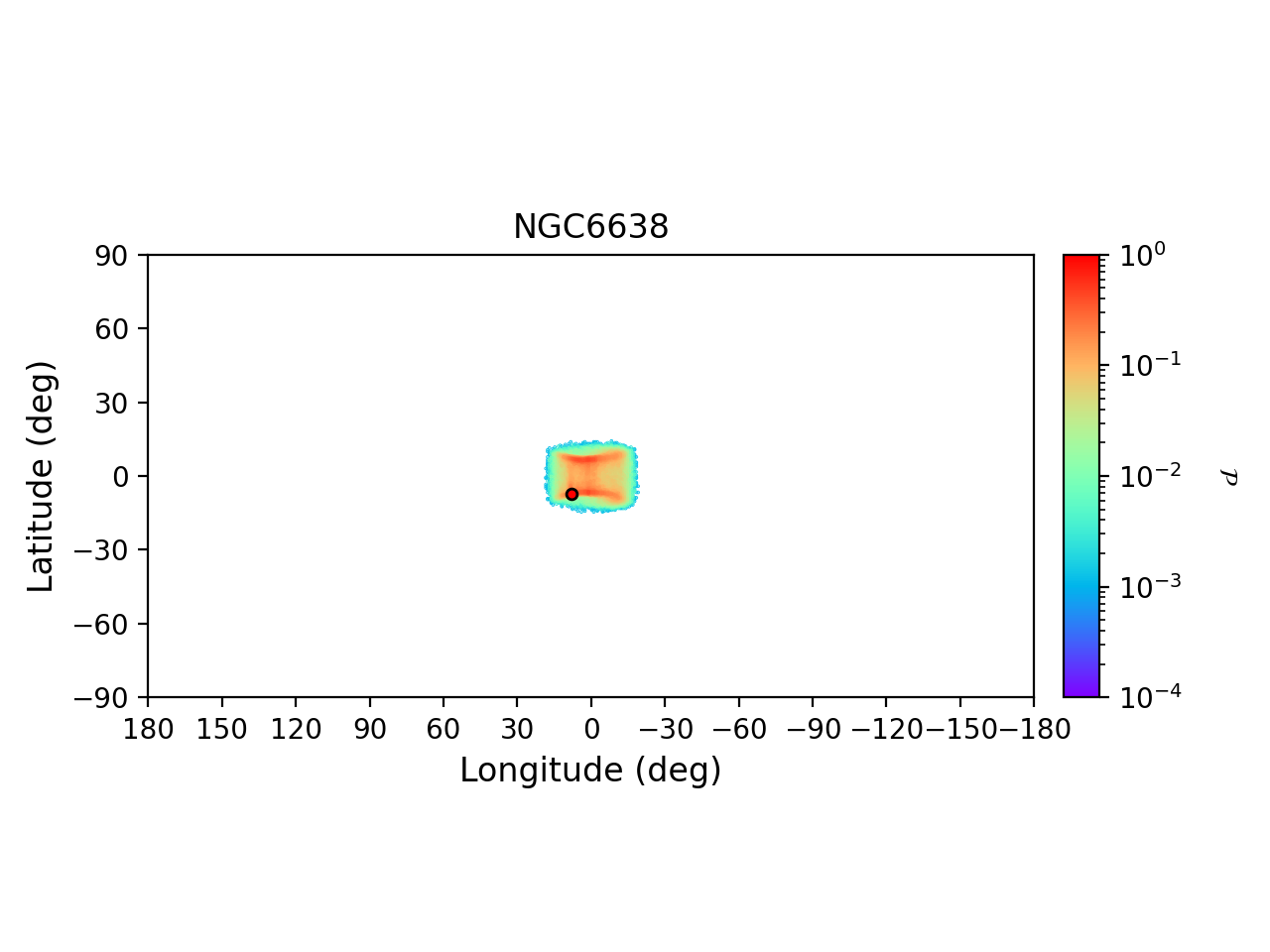}
\includegraphics[clip=true, trim = 0mm 20mm 0mm 10mm, width=1\columnwidth]{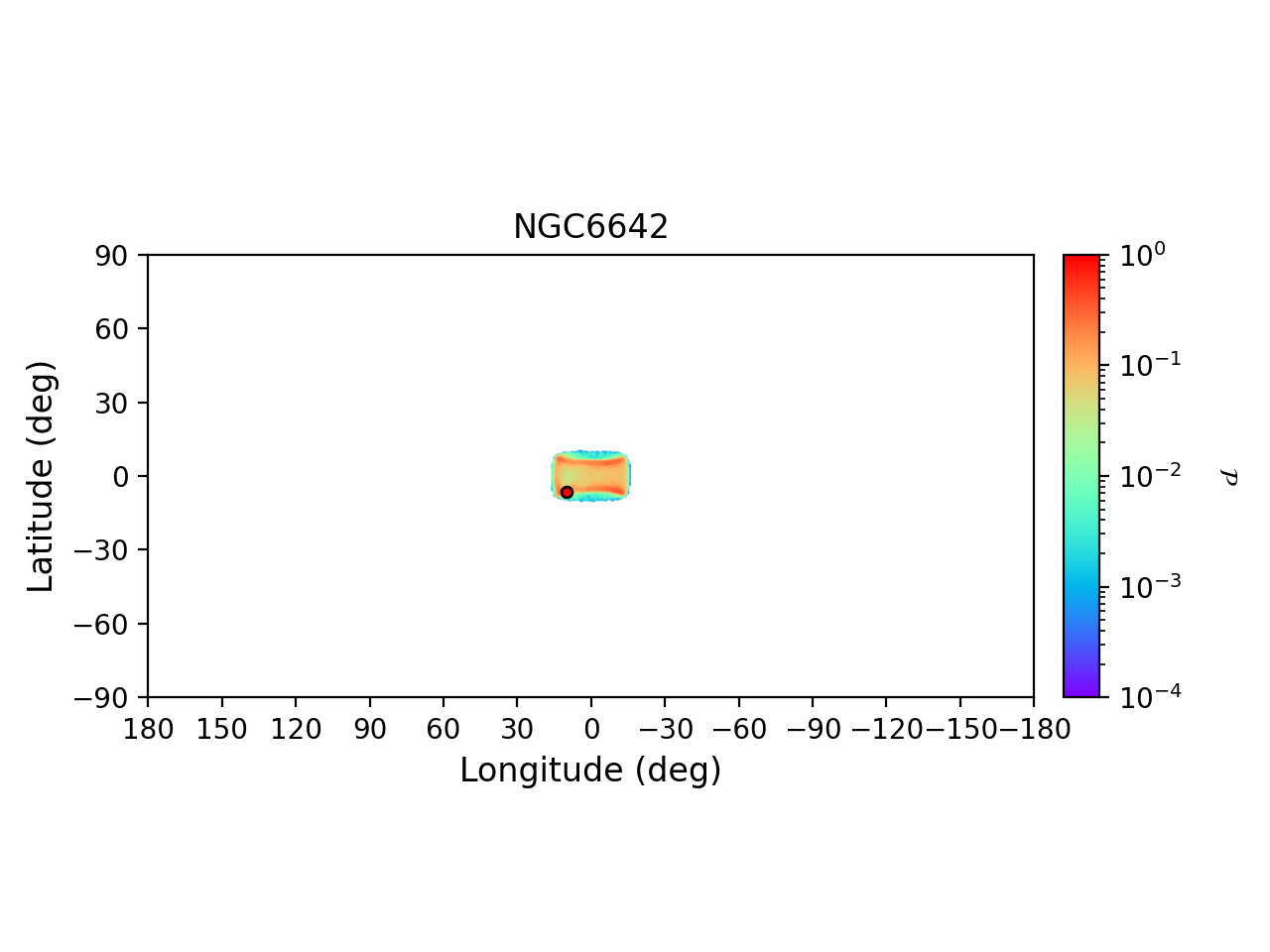}
\includegraphics[clip=true, trim = 0mm 20mm 0mm 10mm, width=1\columnwidth]{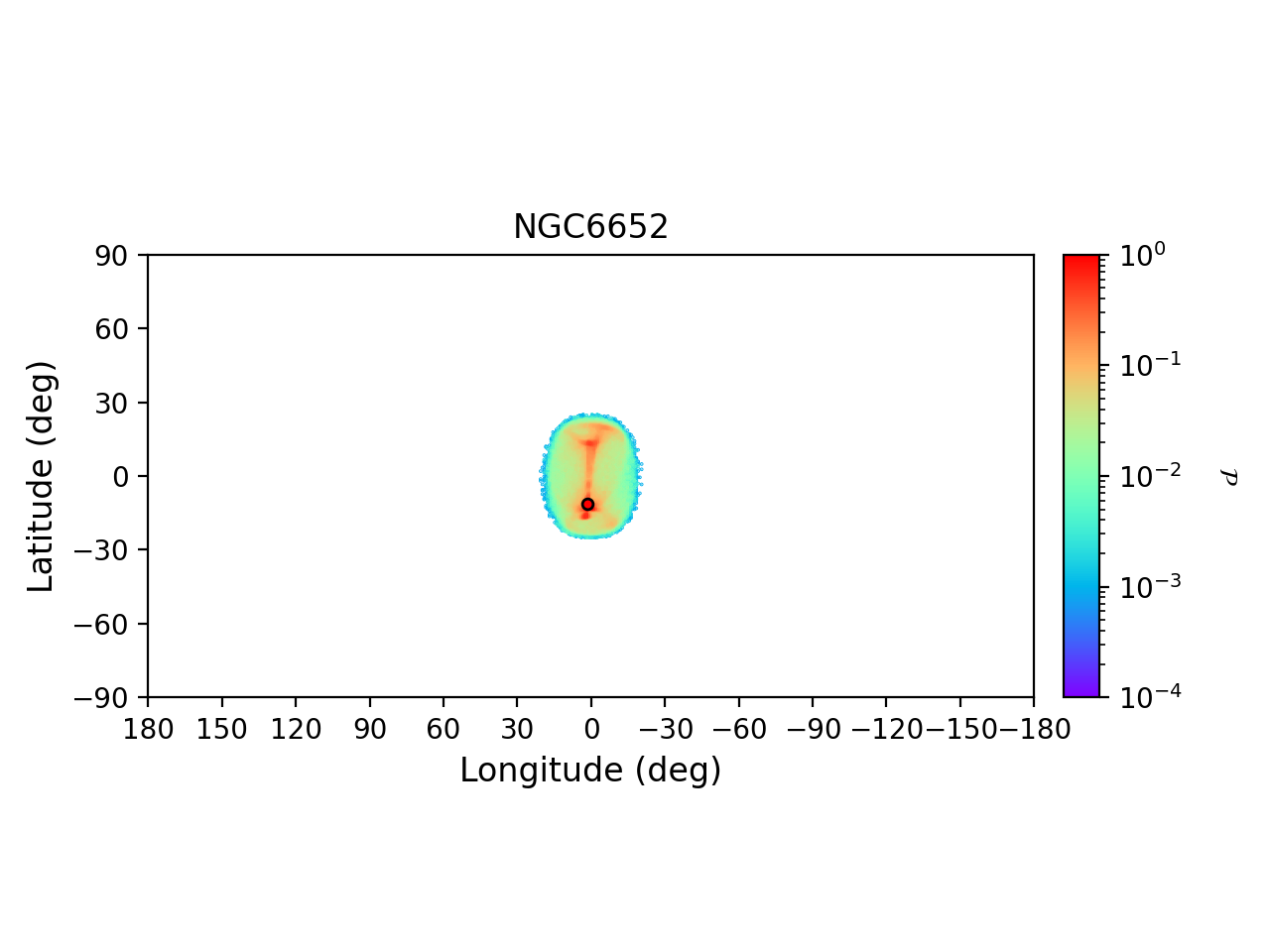}
\includegraphics[clip=true, trim = 0mm 20mm 0mm 10mm, width=1\columnwidth]{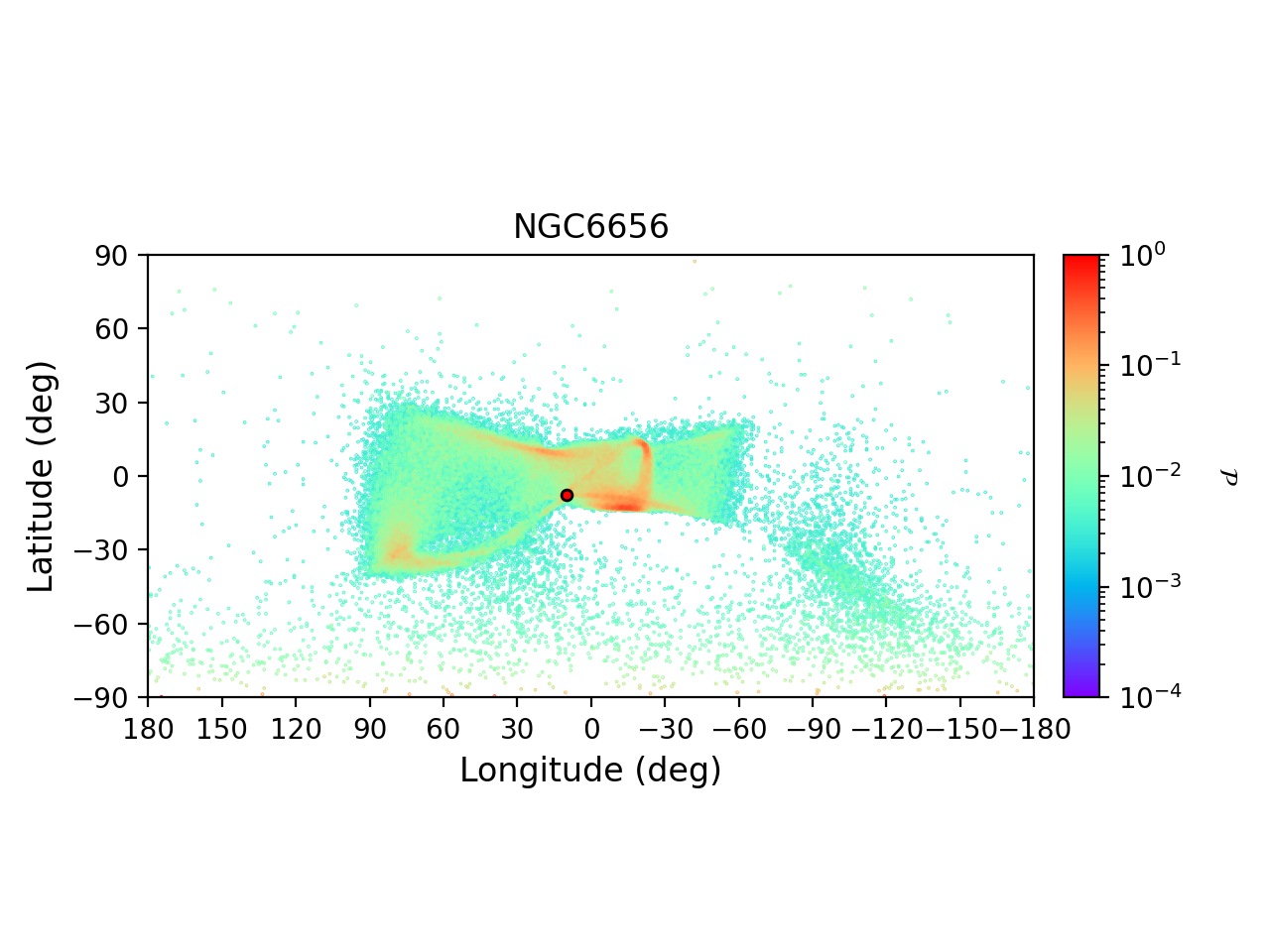}
\includegraphics[clip=true, trim = 0mm 20mm 0mm 10mm, width=1\columnwidth]{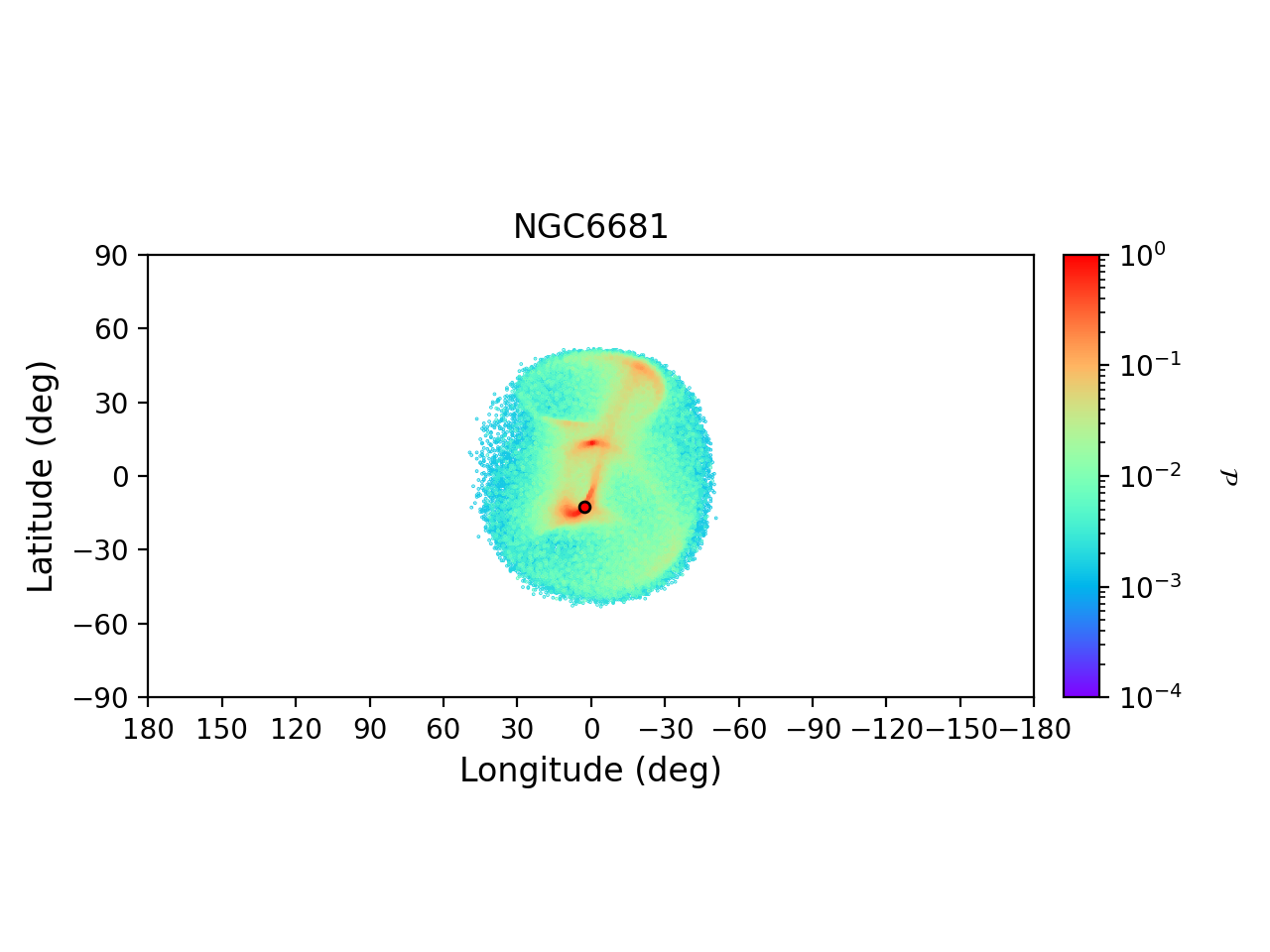}
\includegraphics[clip=true, trim = 0mm 20mm 0mm 10mm, width=1\columnwidth]{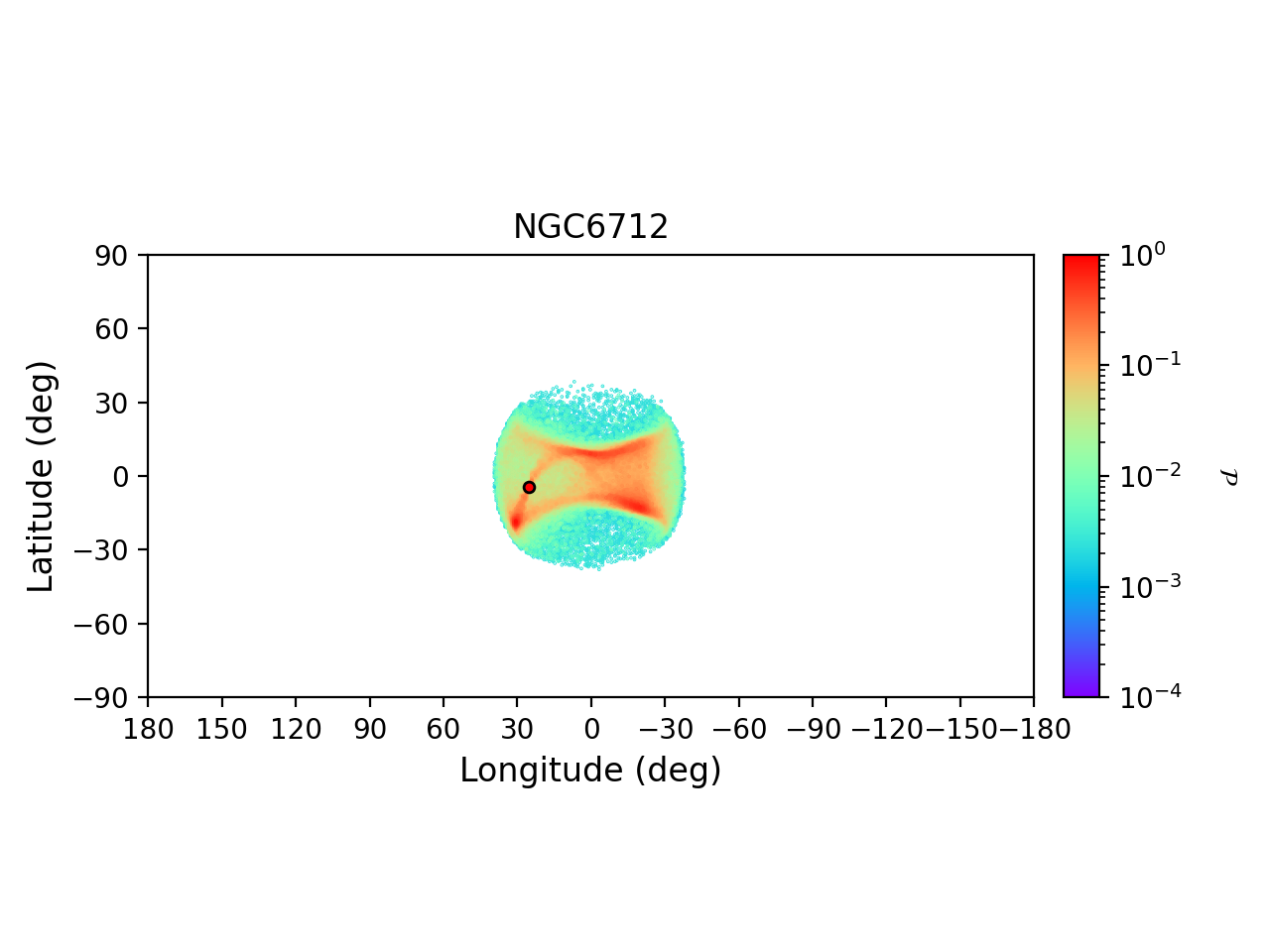}
\includegraphics[clip=true, trim = 0mm 20mm 0mm 10mm, width=1\columnwidth]{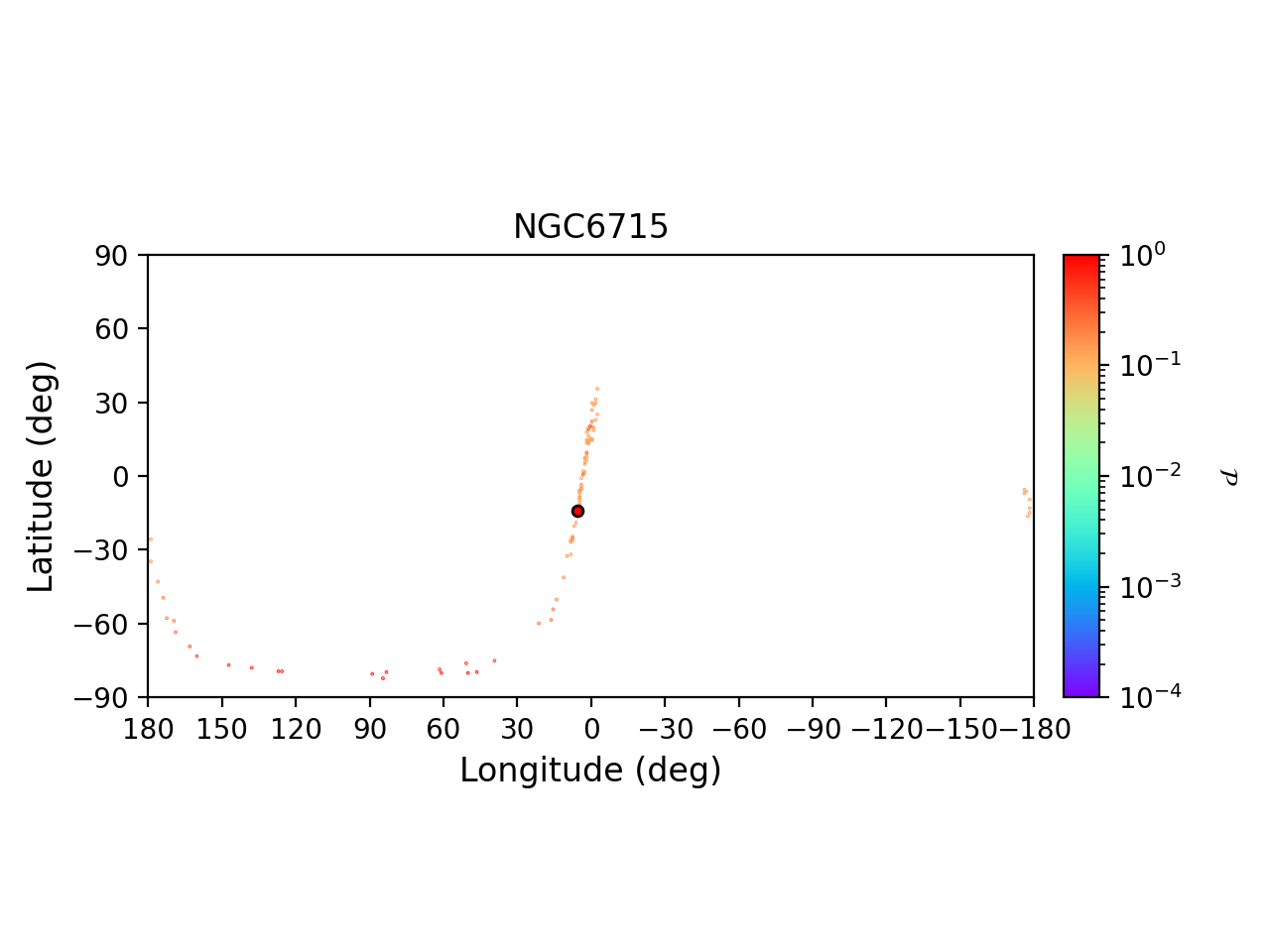}
\caption{Projected density distribution in the $(\ell, b)$ plane of a subset of simulated globular clusters, as indicated at the top of each panel. In each panel, the red circle indicates the current position of the cluster. The densities have been normalized to their maximum value.}\label{stream14}
\end{figure*}

\begin{figure*}
\includegraphics[clip=true, trim = 0mm 20mm 0mm 10mm, width=1\columnwidth]{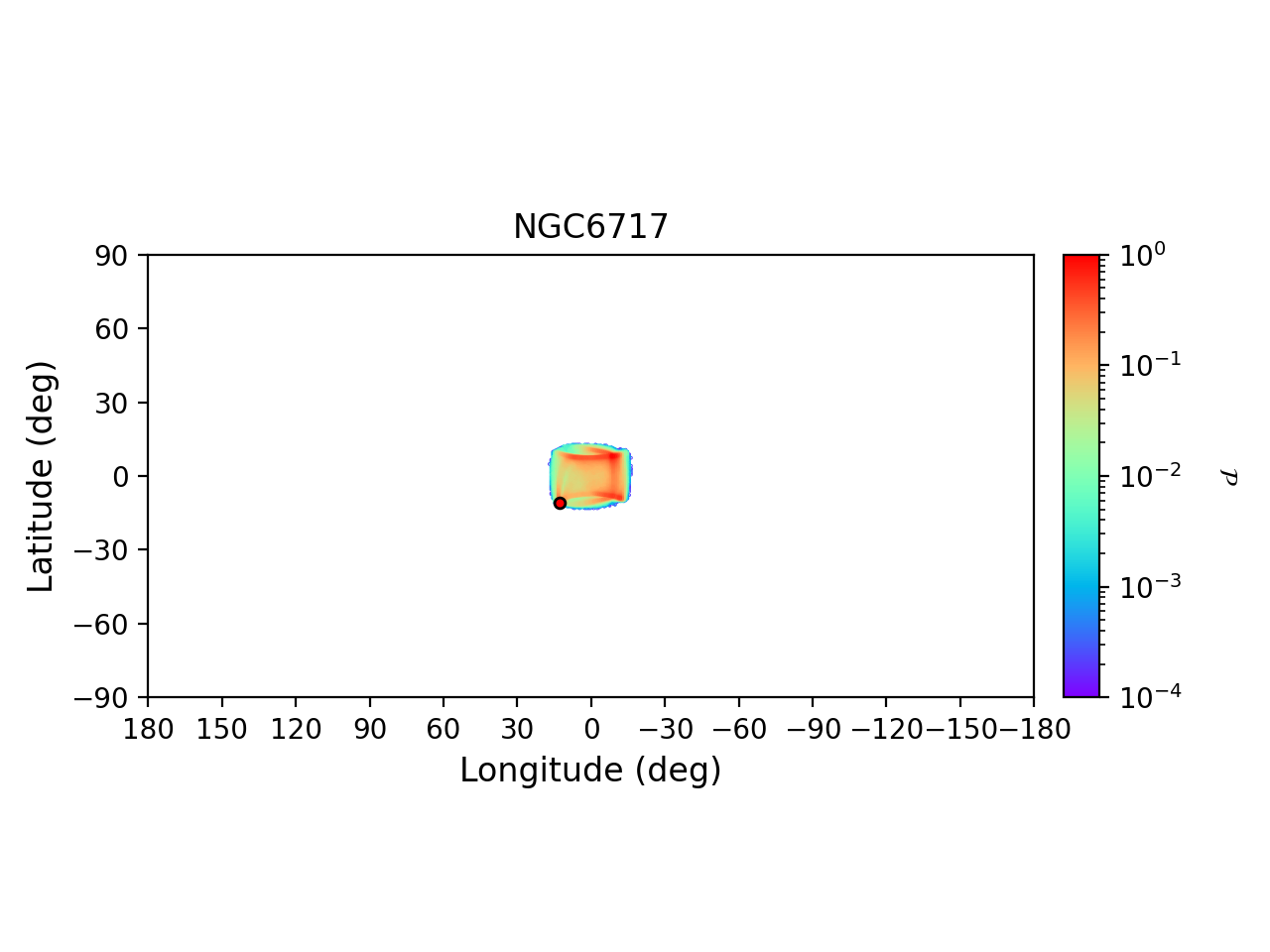}
\includegraphics[clip=true, trim = 0mm 20mm 0mm 10mm, width=1\columnwidth]{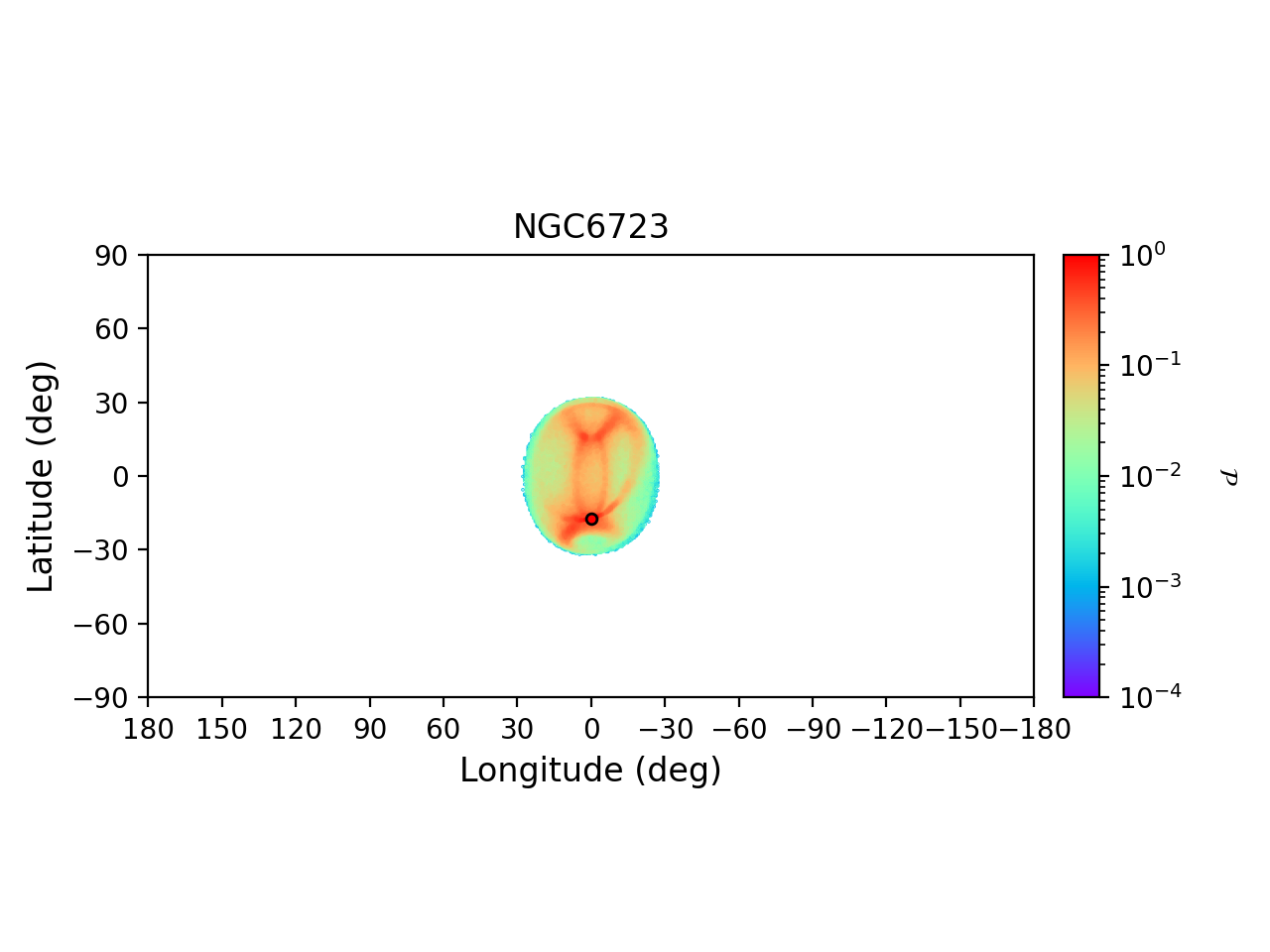}
\includegraphics[clip=true, trim = 0mm 20mm 0mm 10mm, width=1\columnwidth]{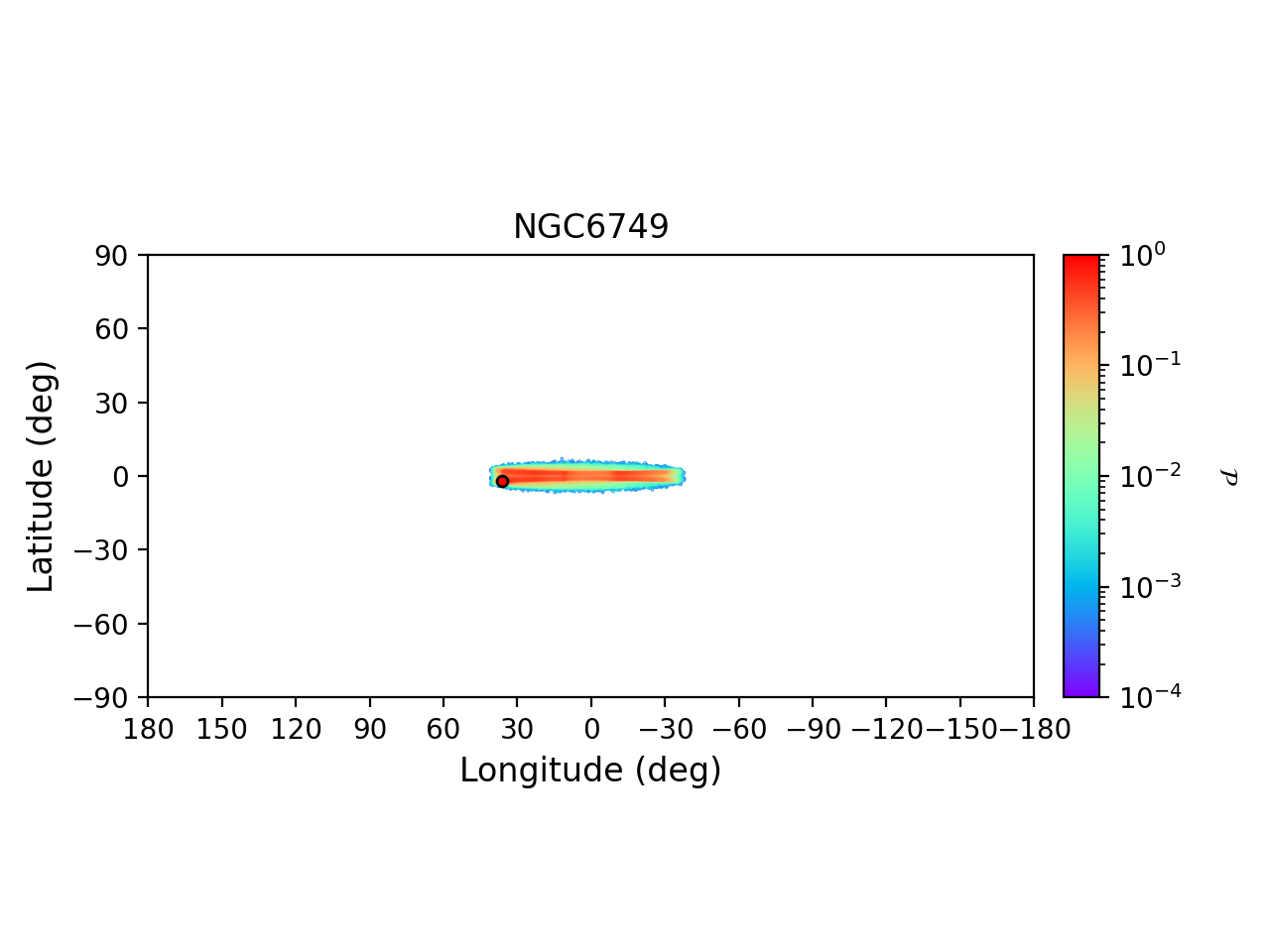}
\includegraphics[clip=true, trim = 0mm 20mm 0mm 10mm, width=1\columnwidth]{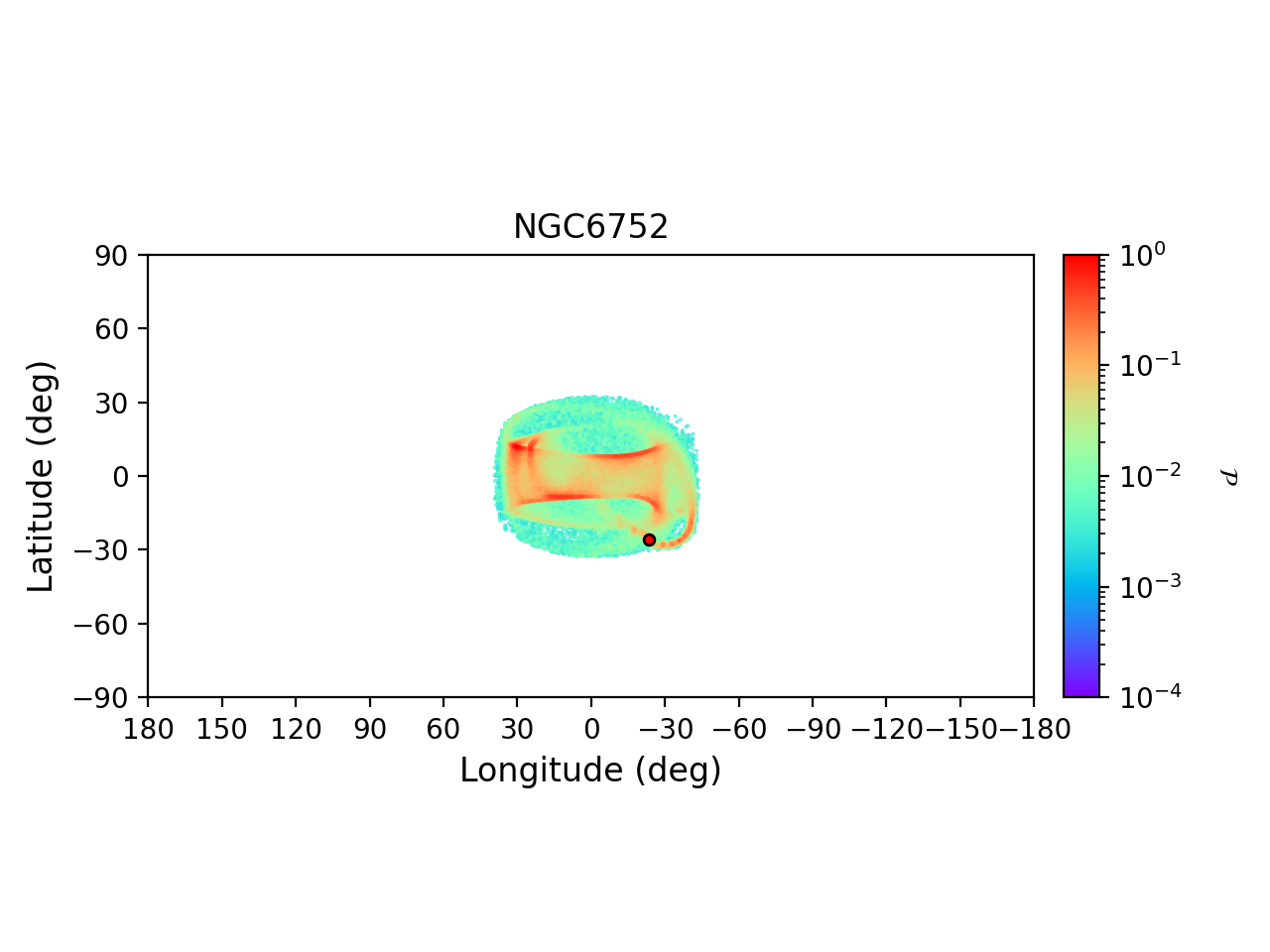}
\includegraphics[clip=true, trim = 0mm 20mm 0mm 10mm, width=1\columnwidth]{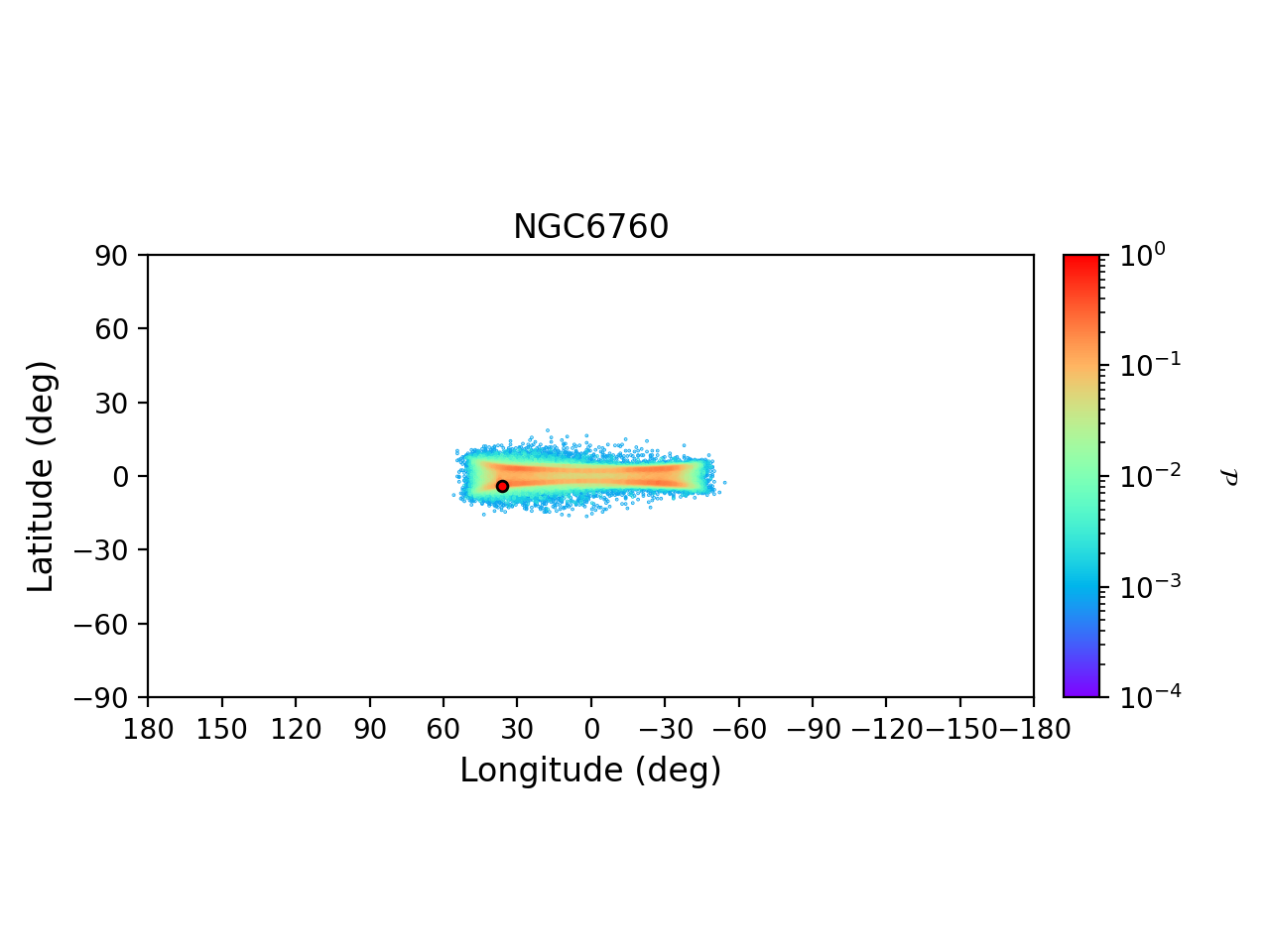}
\includegraphics[clip=true, trim = 0mm 20mm 0mm 10mm, width=1\columnwidth]{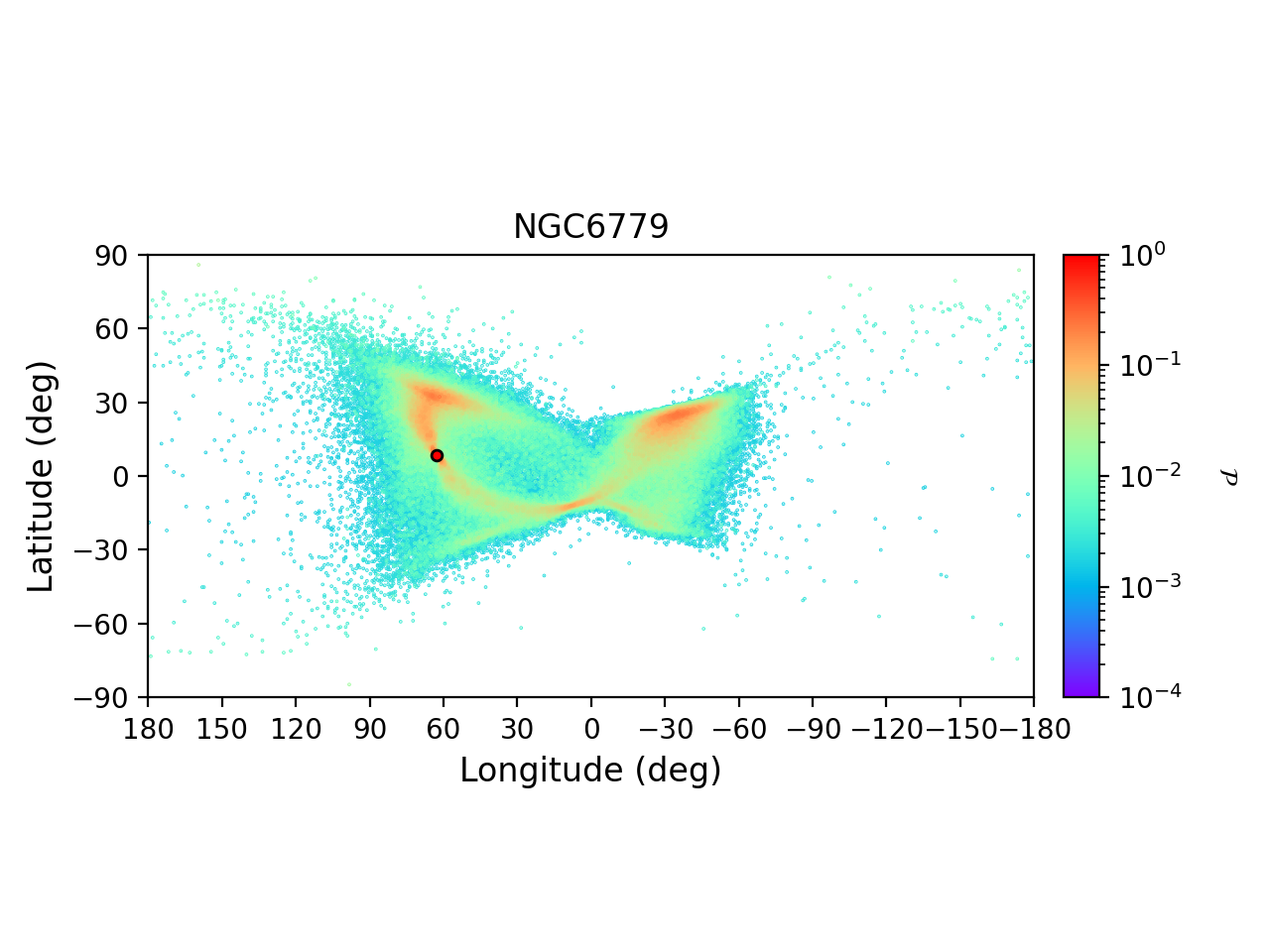}
\includegraphics[clip=true, trim = 0mm 20mm 0mm 10mm, width=1\columnwidth]{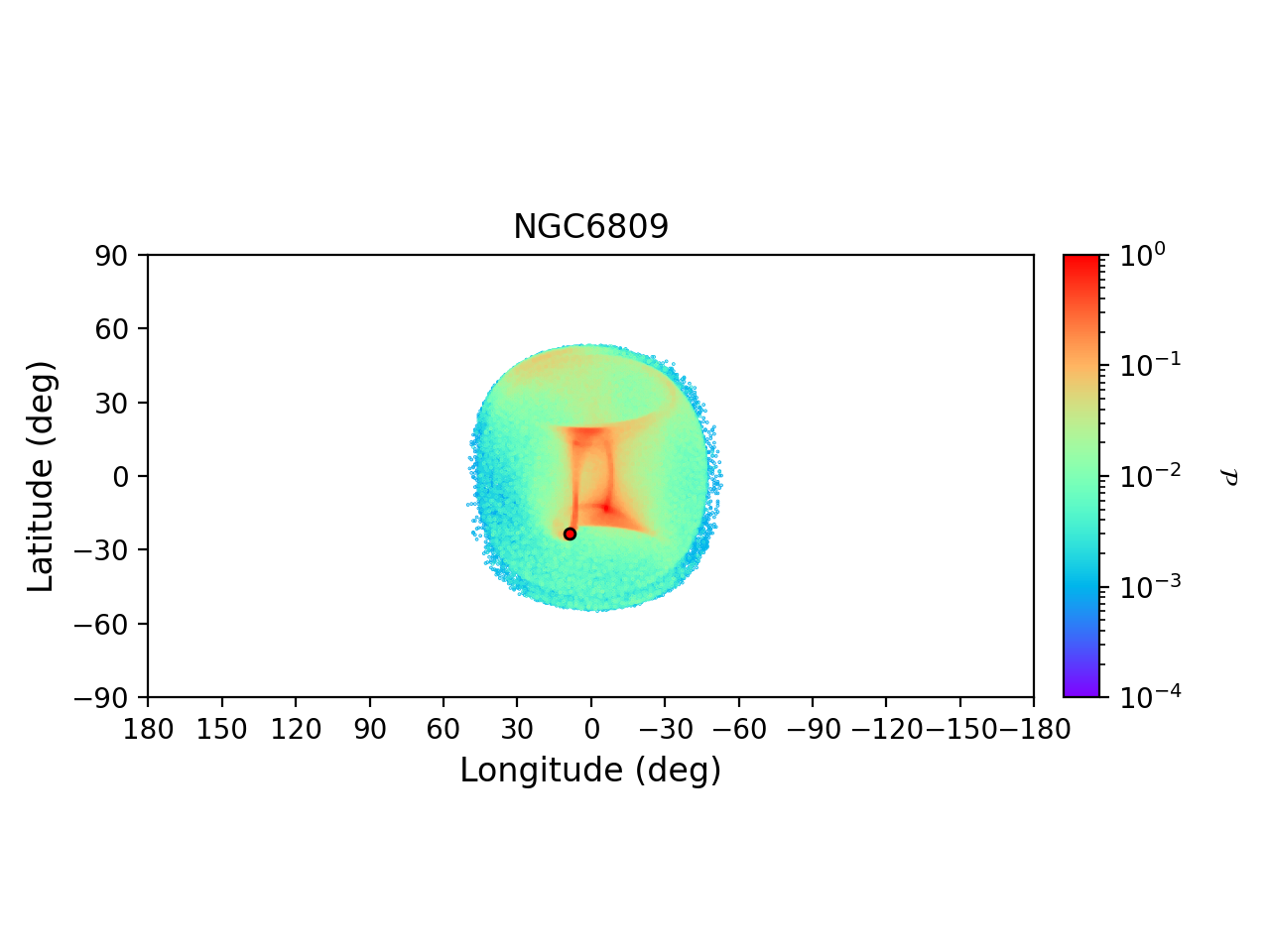}
\includegraphics[clip=true, trim = 0mm 20mm 0mm 10mm, width=1\columnwidth]{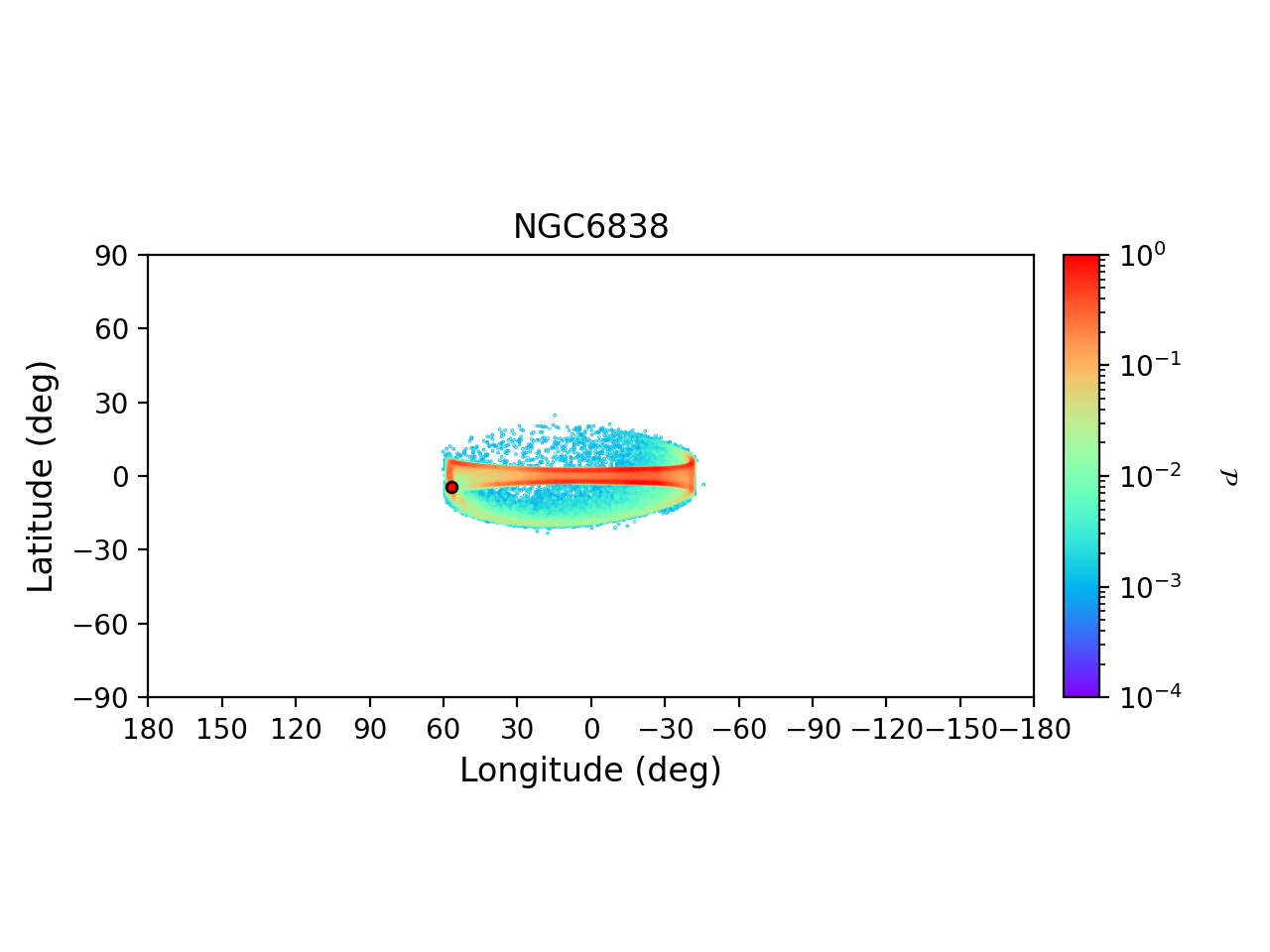}
\caption{Projected density distribution in the $(\ell, b)$ plane of a subset of simulated globular clusters, as indicated at the top of each panel. In each panel, the red circle indicates the current position of the cluster. The densities have been normalized to their maximum value.}\label{stream15}
\end{figure*}

\begin{figure*}
\includegraphics[clip=true, trim = 0mm 20mm 0mm 10mm, width=1\columnwidth]{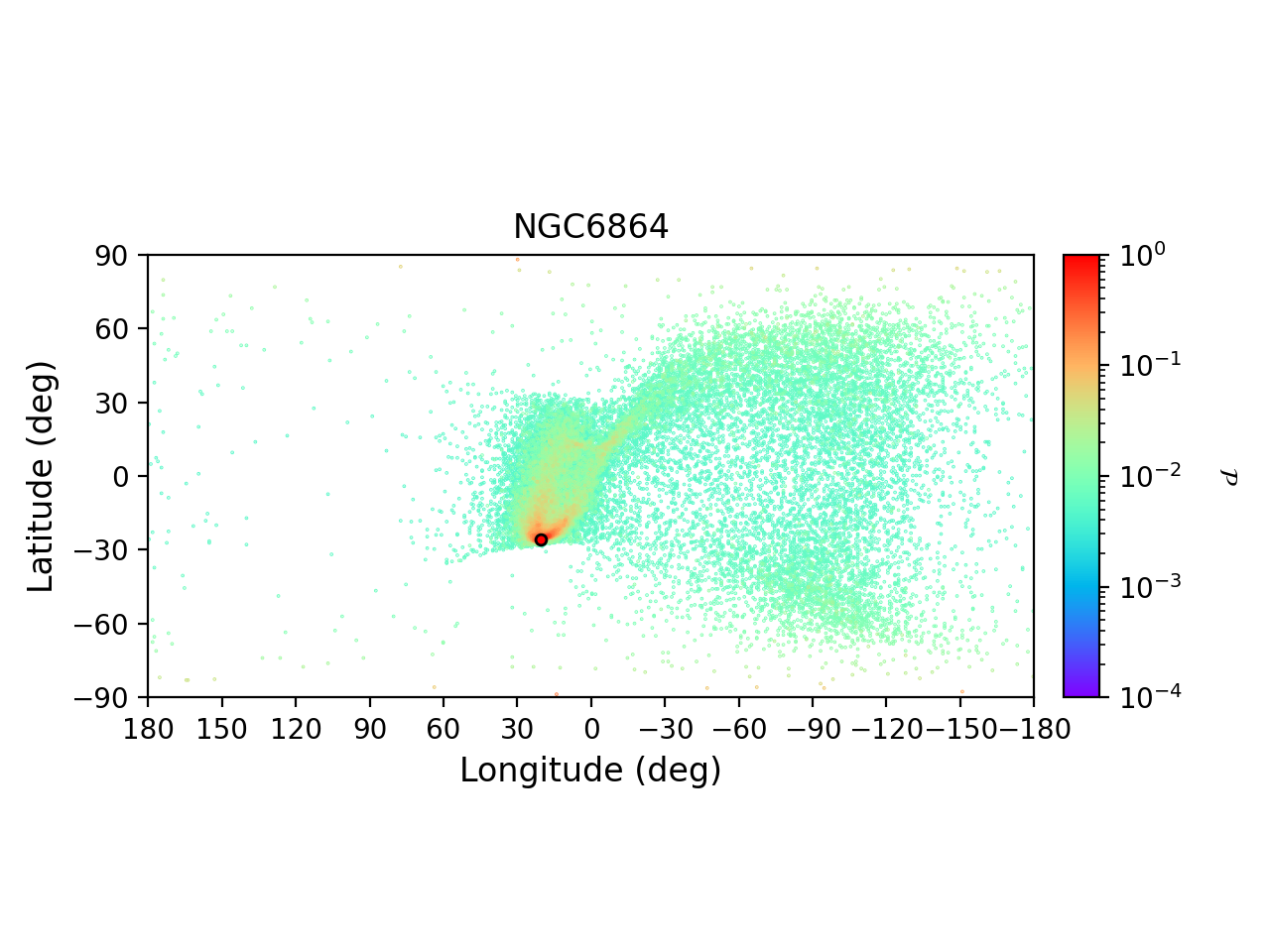}
\includegraphics[clip=true, trim = 0mm 20mm 0mm 10mm, width=1\columnwidth]{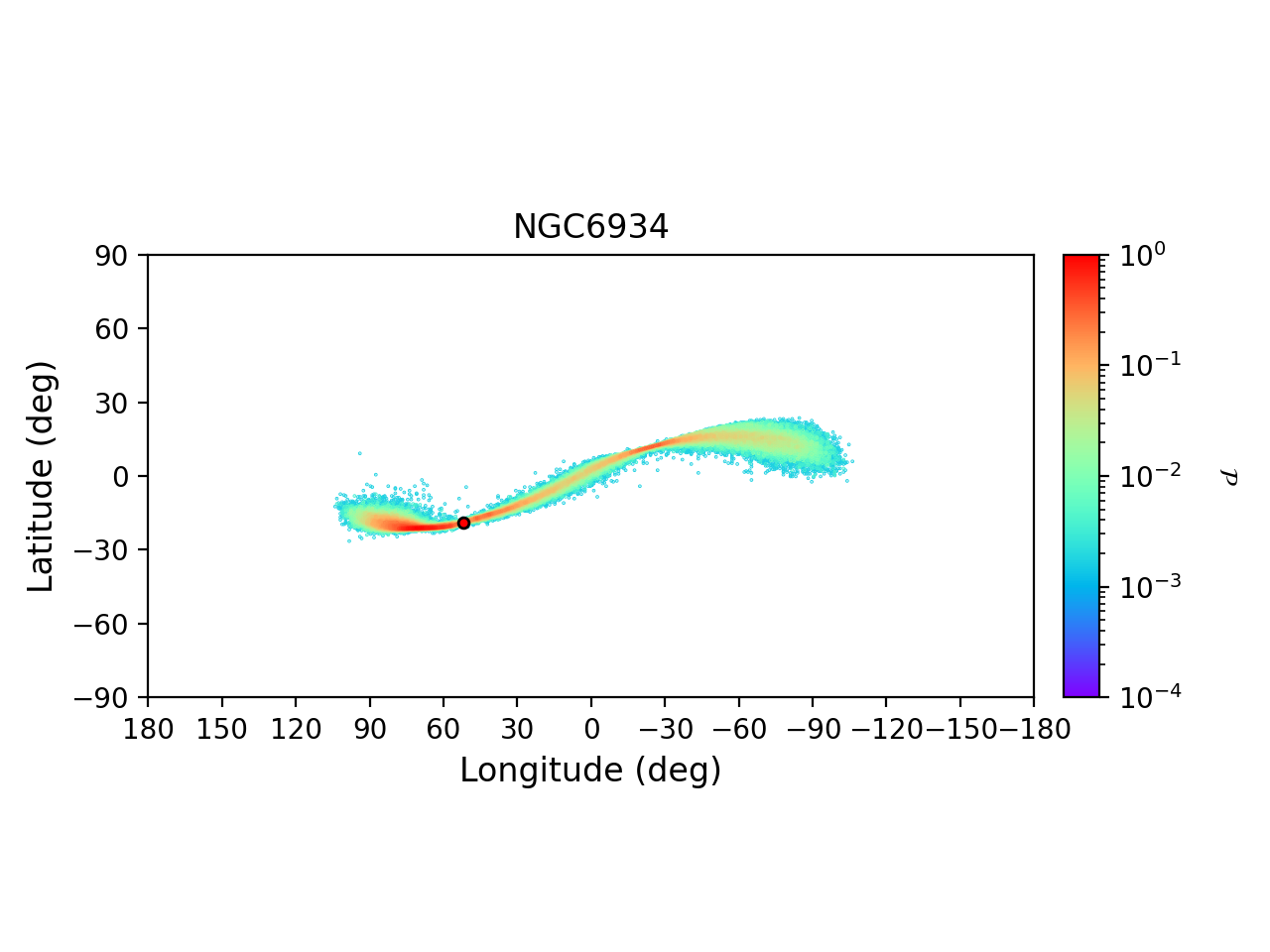}
\includegraphics[clip=true, trim = 0mm 20mm 0mm 10mm, width=1\columnwidth]{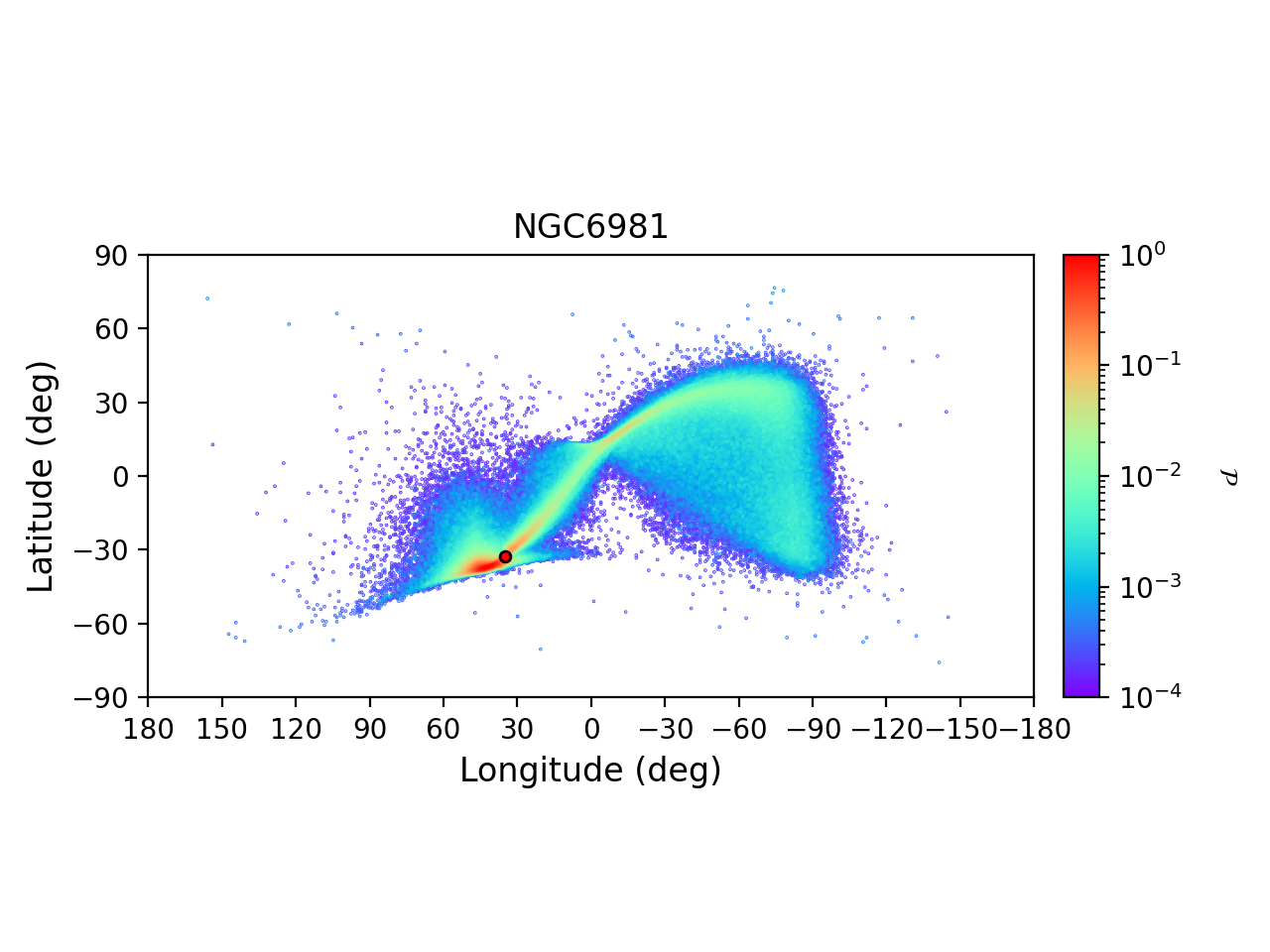}
\includegraphics[clip=true, trim = 0mm 20mm 0mm 10mm, width=1\columnwidth]{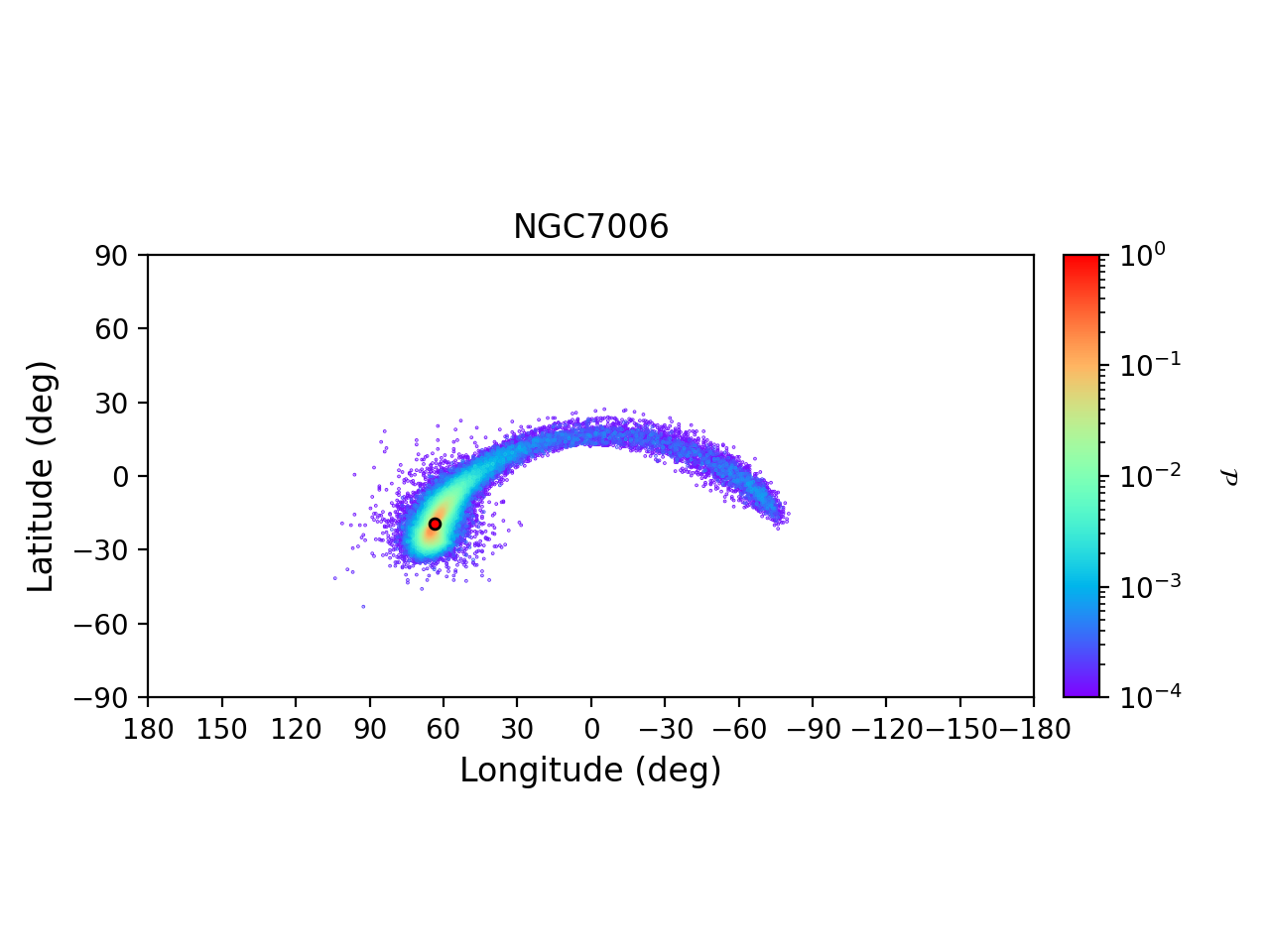}
\includegraphics[clip=true, trim = 0mm 20mm 0mm 10mm, width=1\columnwidth]{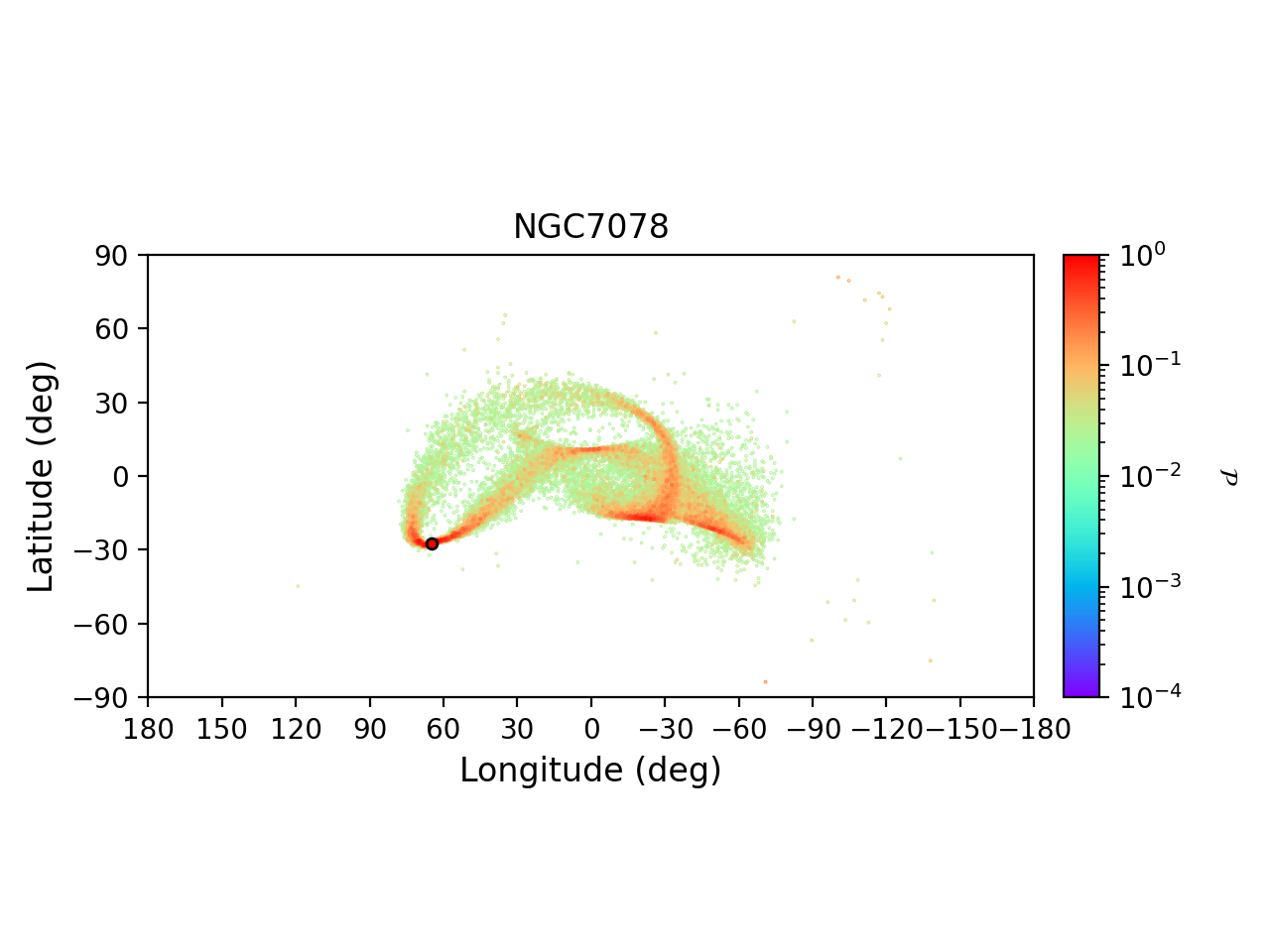}
\includegraphics[clip=true, trim = 0mm 20mm 0mm 10mm, width=1\columnwidth]{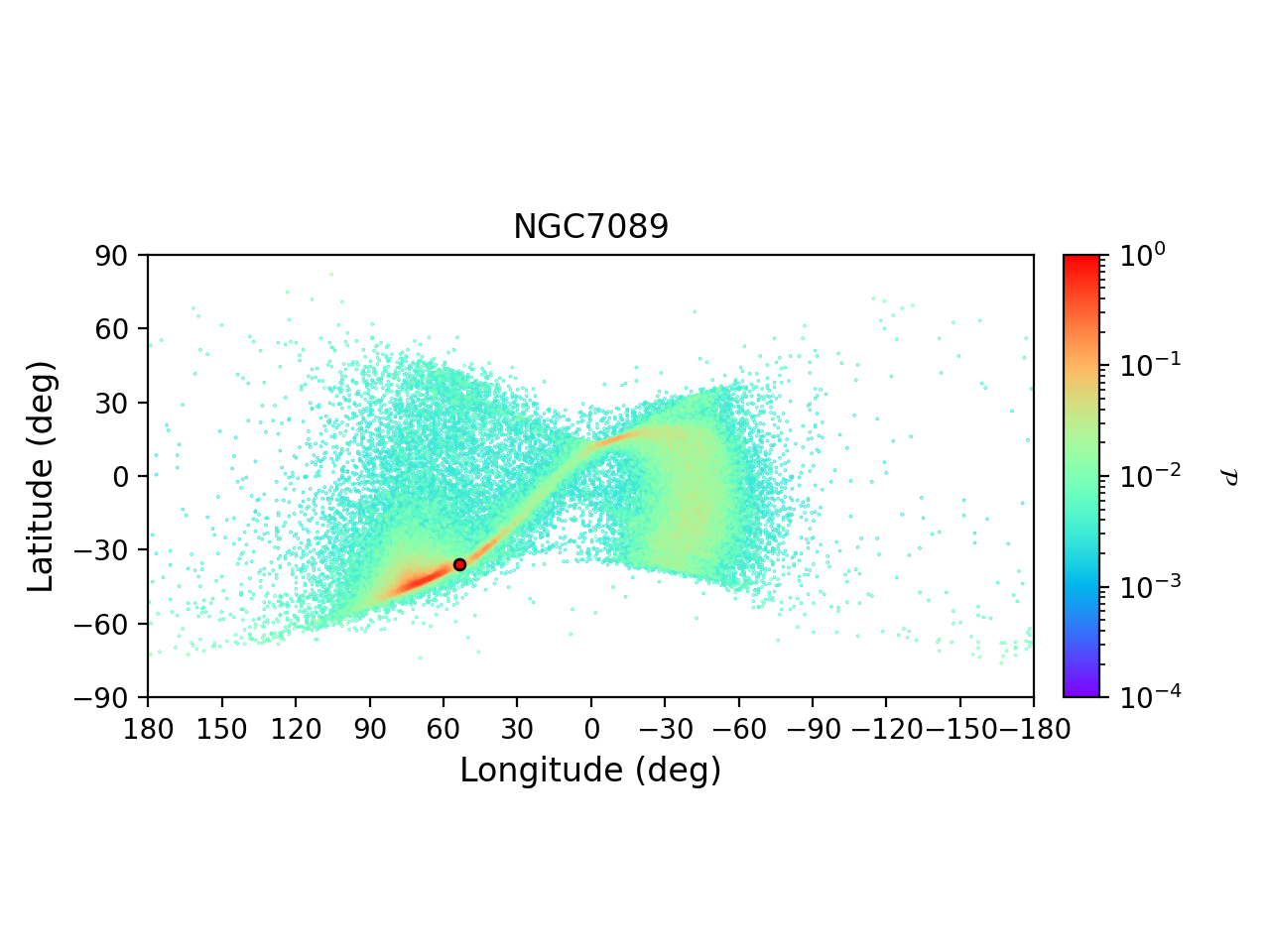}
\includegraphics[clip=true, trim = 0mm 20mm 0mm 10mm, width=1\columnwidth]{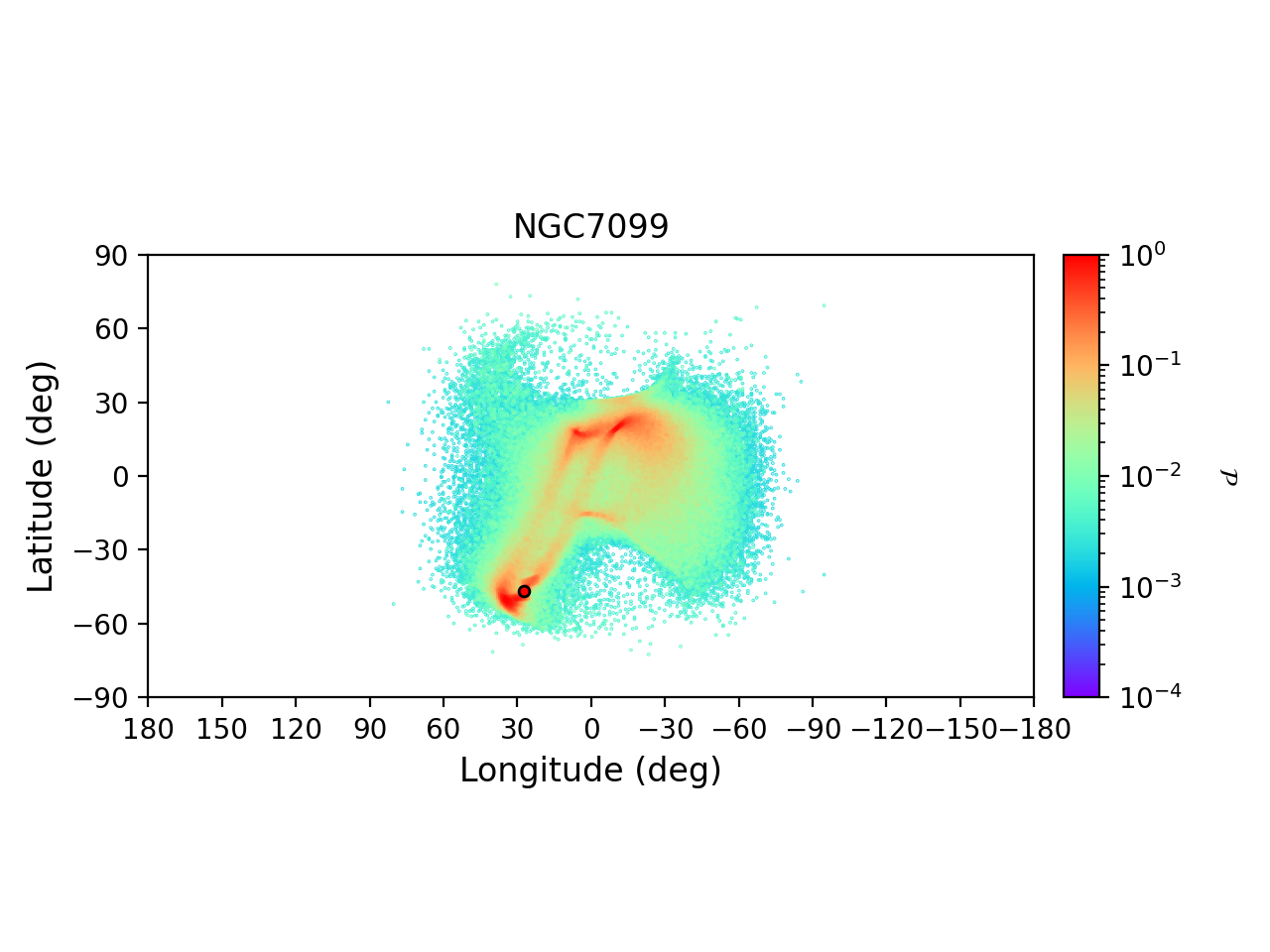}
\includegraphics[clip=true, trim = 0mm 20mm 0mm 10mm, width=1\columnwidth]{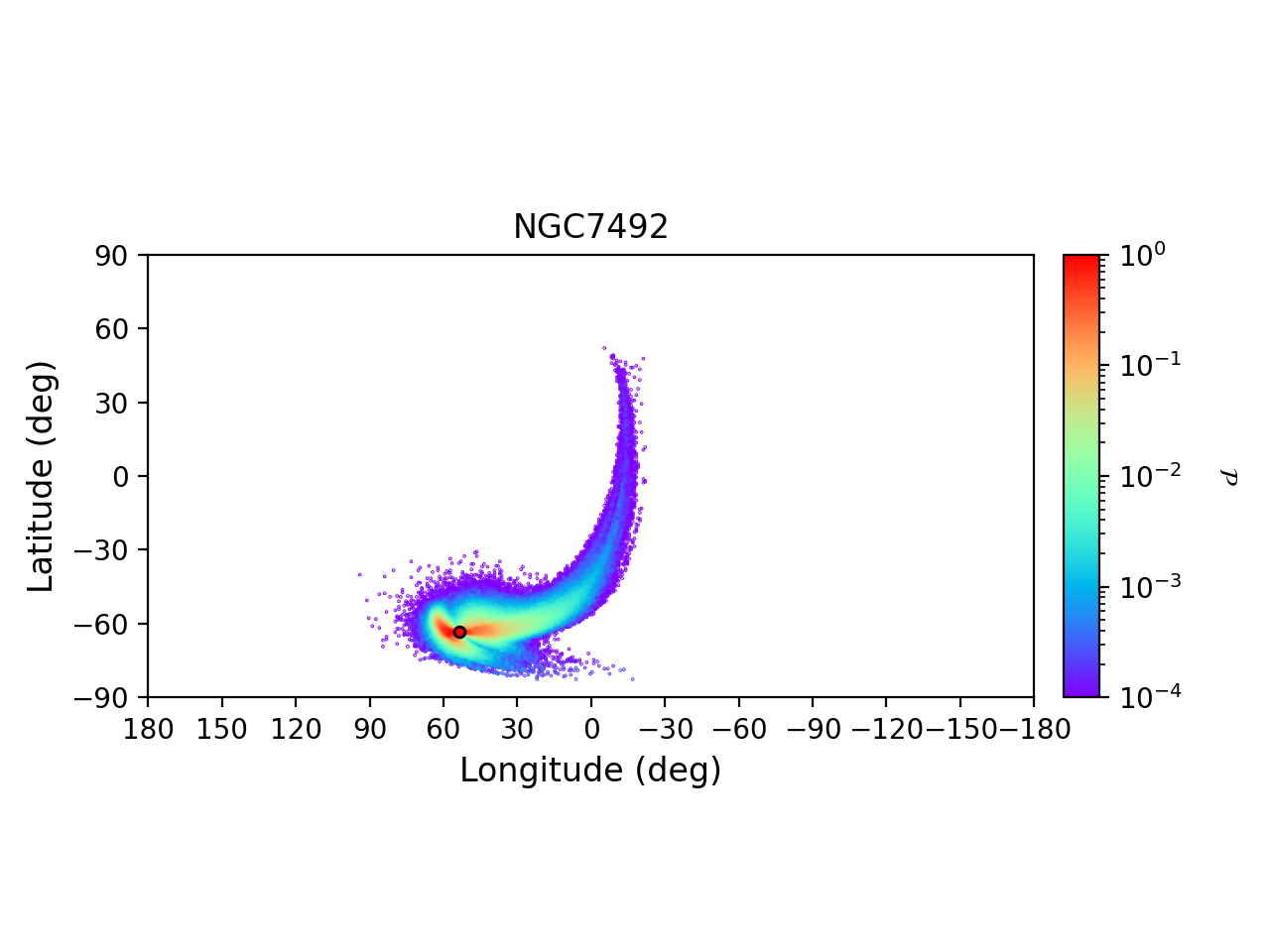}
\caption{Projected density distribution in the $(\ell, b)$ plane of a subset of simulated globular clusters, as indicated at the top of each panel. In each panel, the red circle indicates the current position of the cluster. The densities have been normalized to their maximum value.}\label{stream16}
\end{figure*}

\begin{figure*}
\includegraphics[clip=true, trim = 0mm 20mm 0mm 10mm, width=1\columnwidth]{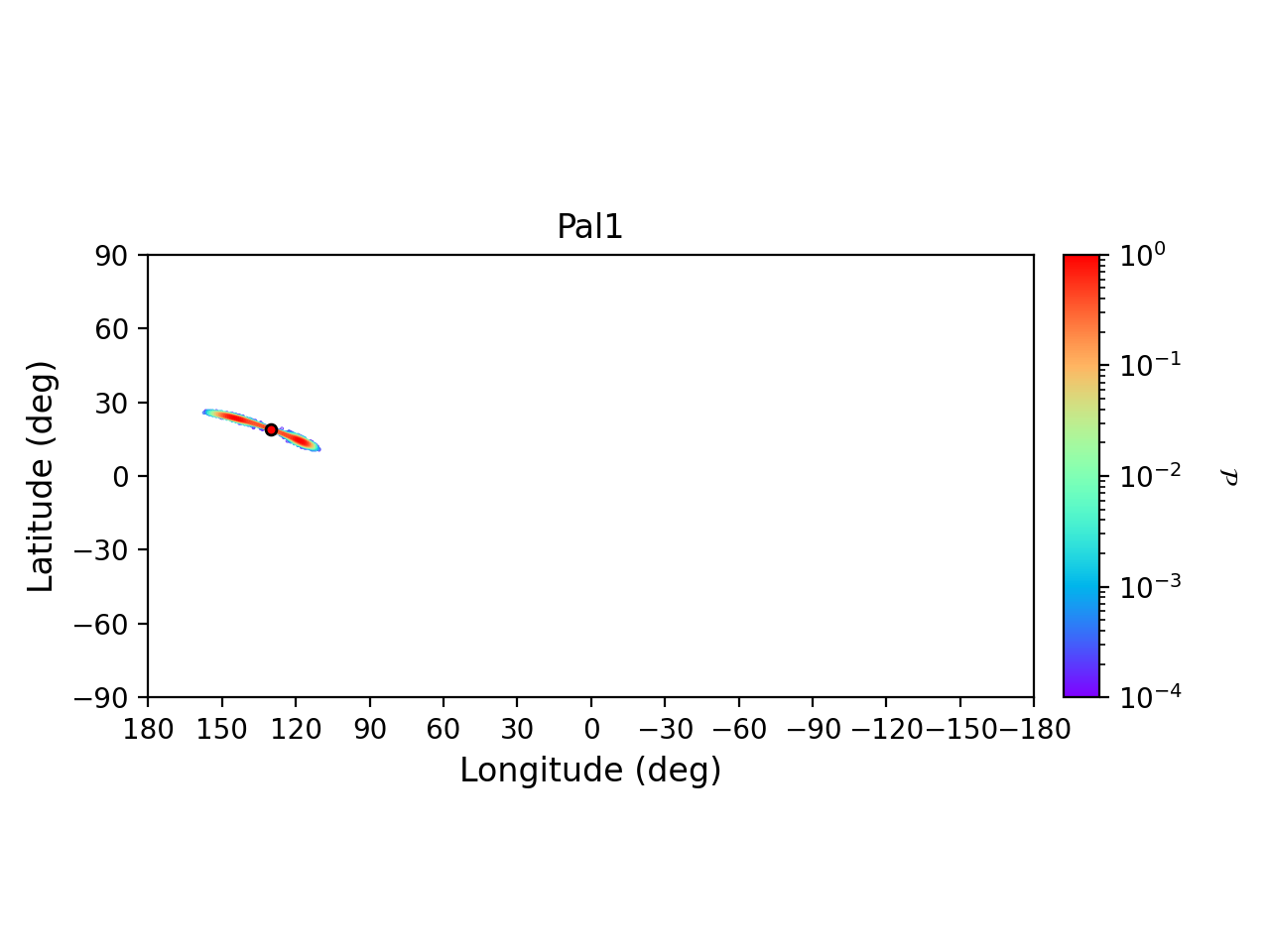}
\includegraphics[clip=true, trim = 0mm 20mm 0mm 10mm, width=1\columnwidth]{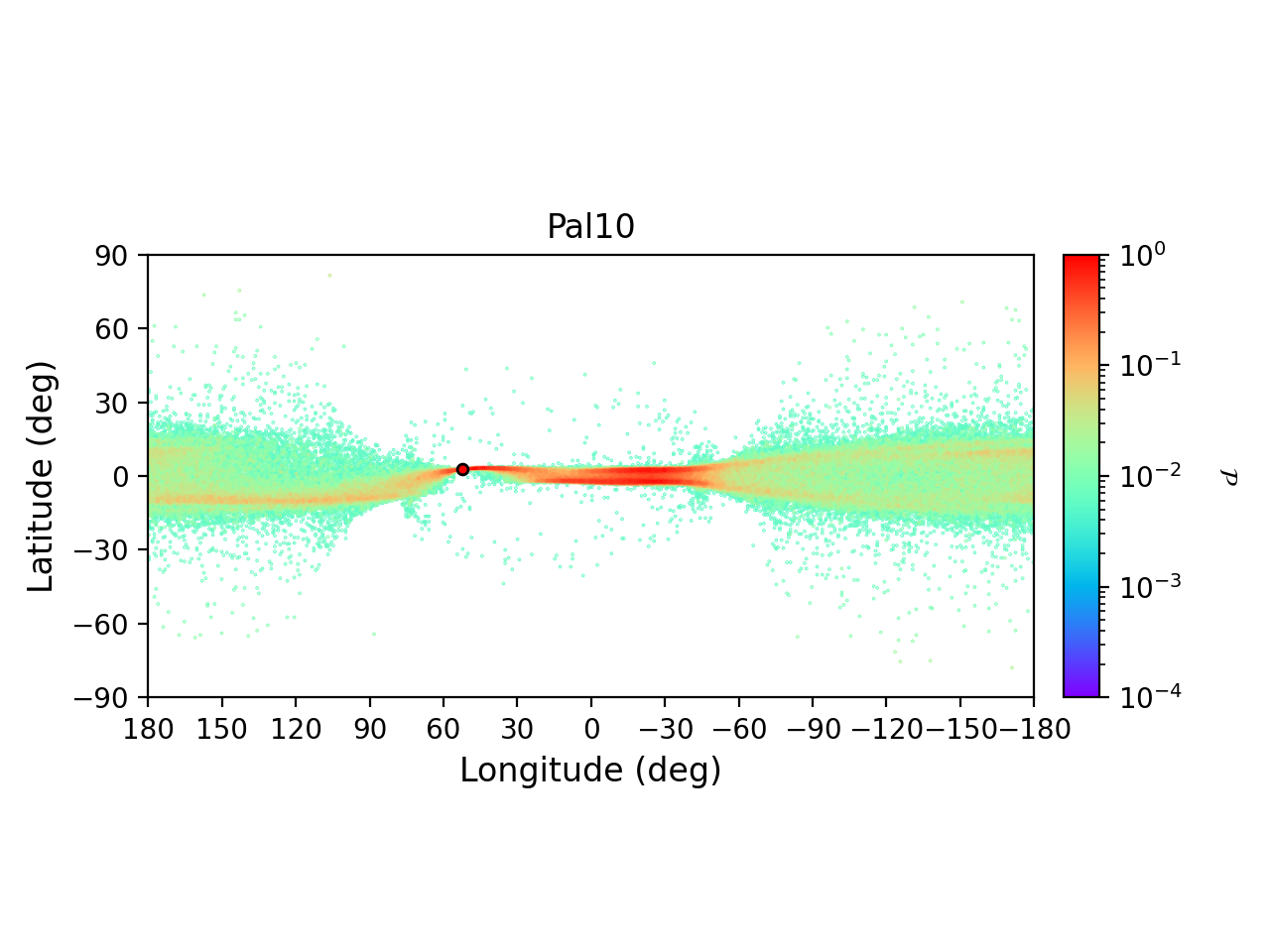}
\includegraphics[clip=true, trim = 0mm 20mm 0mm 10mm, width=1\columnwidth]{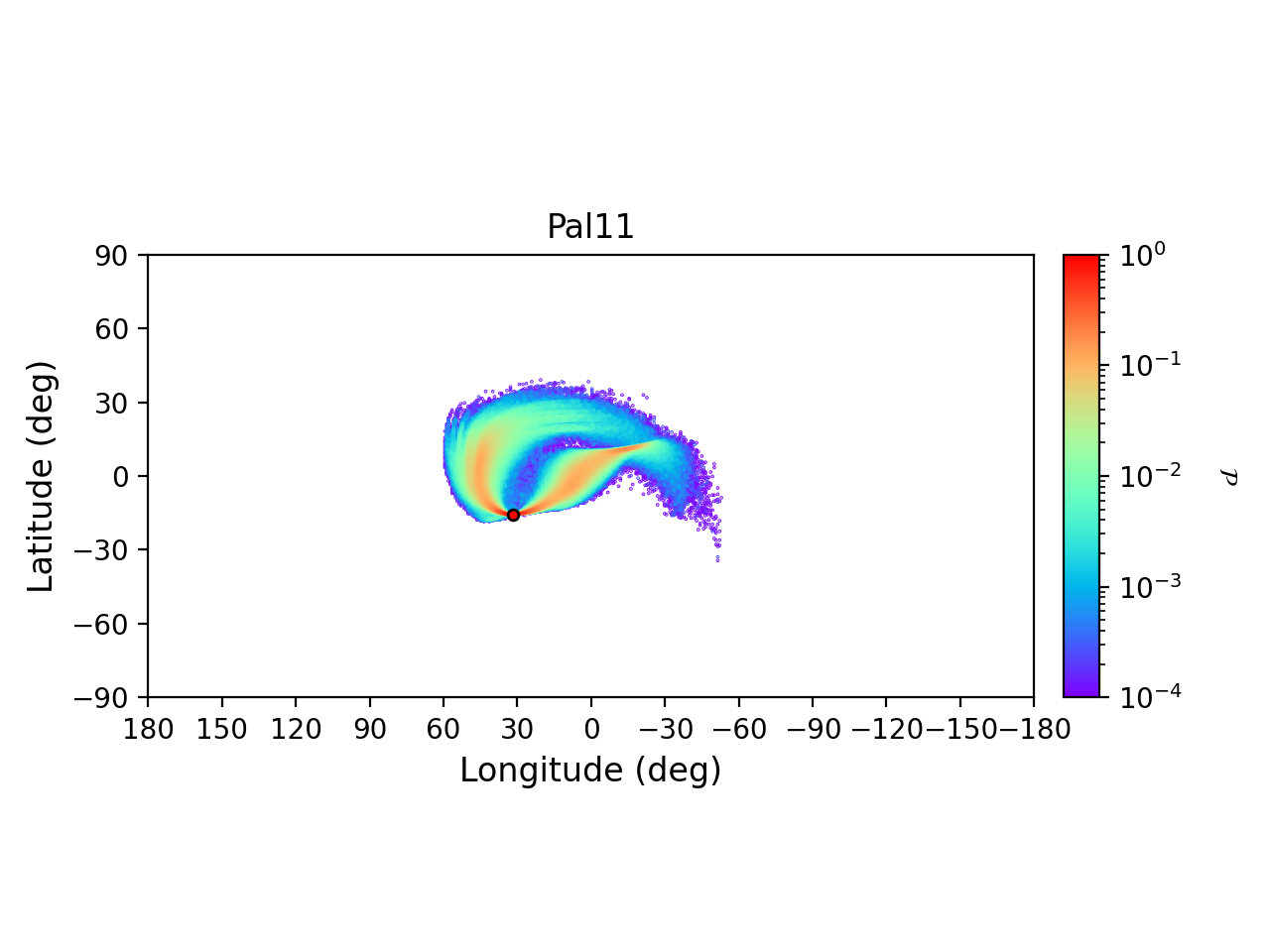}
\includegraphics[clip=true, trim = 0mm 20mm 0mm 10mm, width=1\columnwidth]{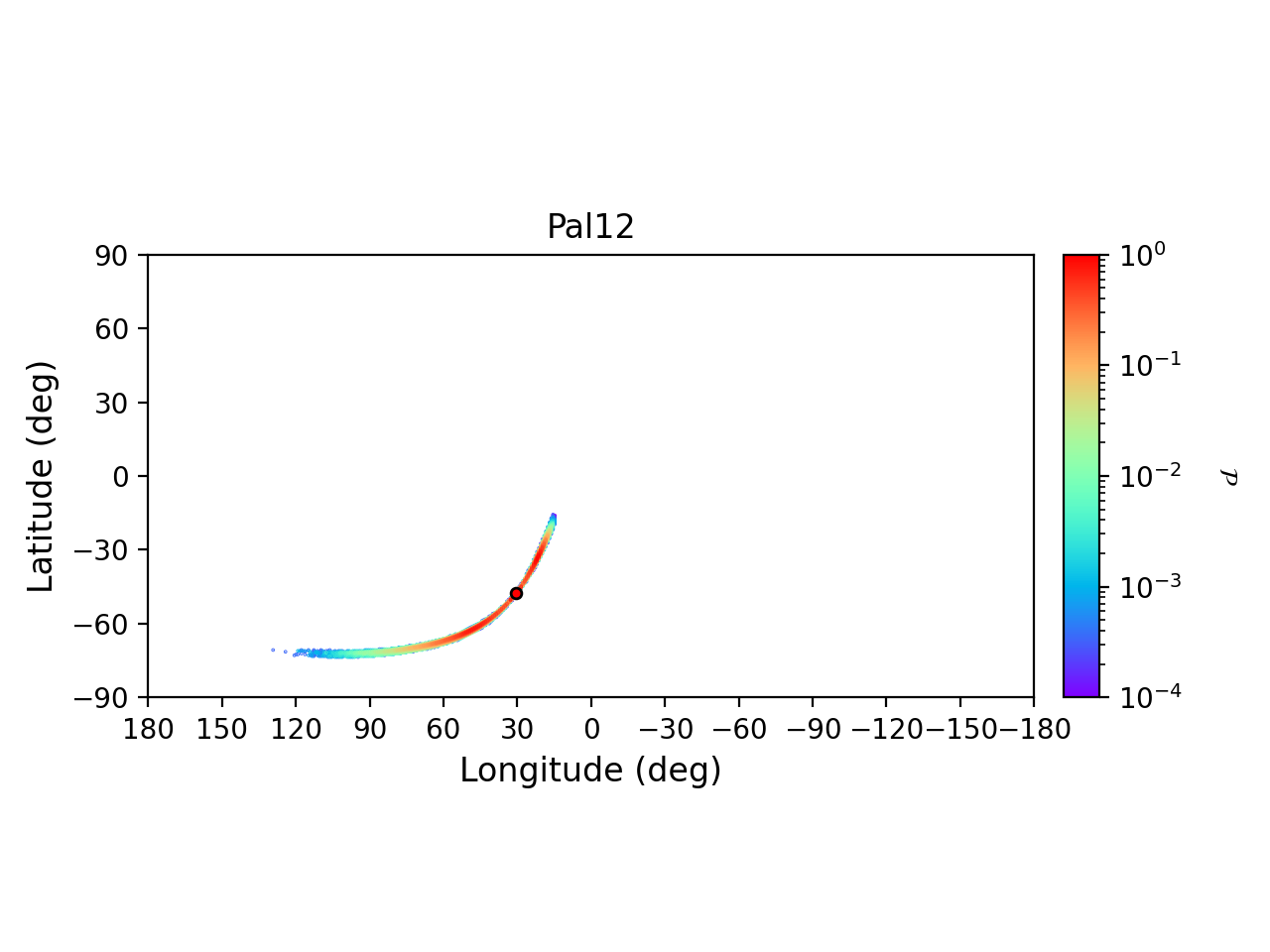}
\includegraphics[clip=true, trim = 0mm 20mm 0mm 10mm, width=1\columnwidth]{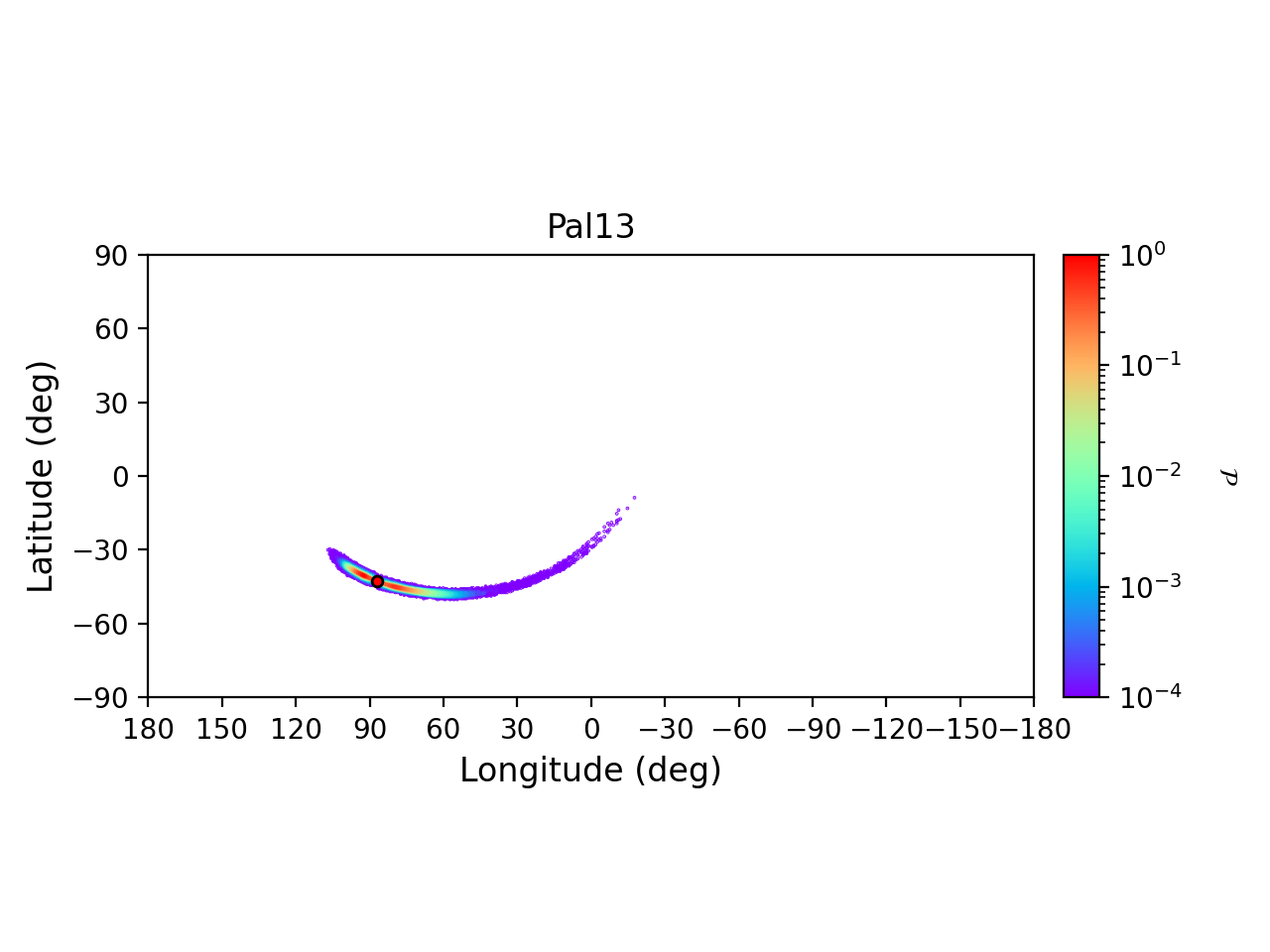}
\includegraphics[clip=true, trim = 0mm 20mm 0mm 10mm, width=1\columnwidth]{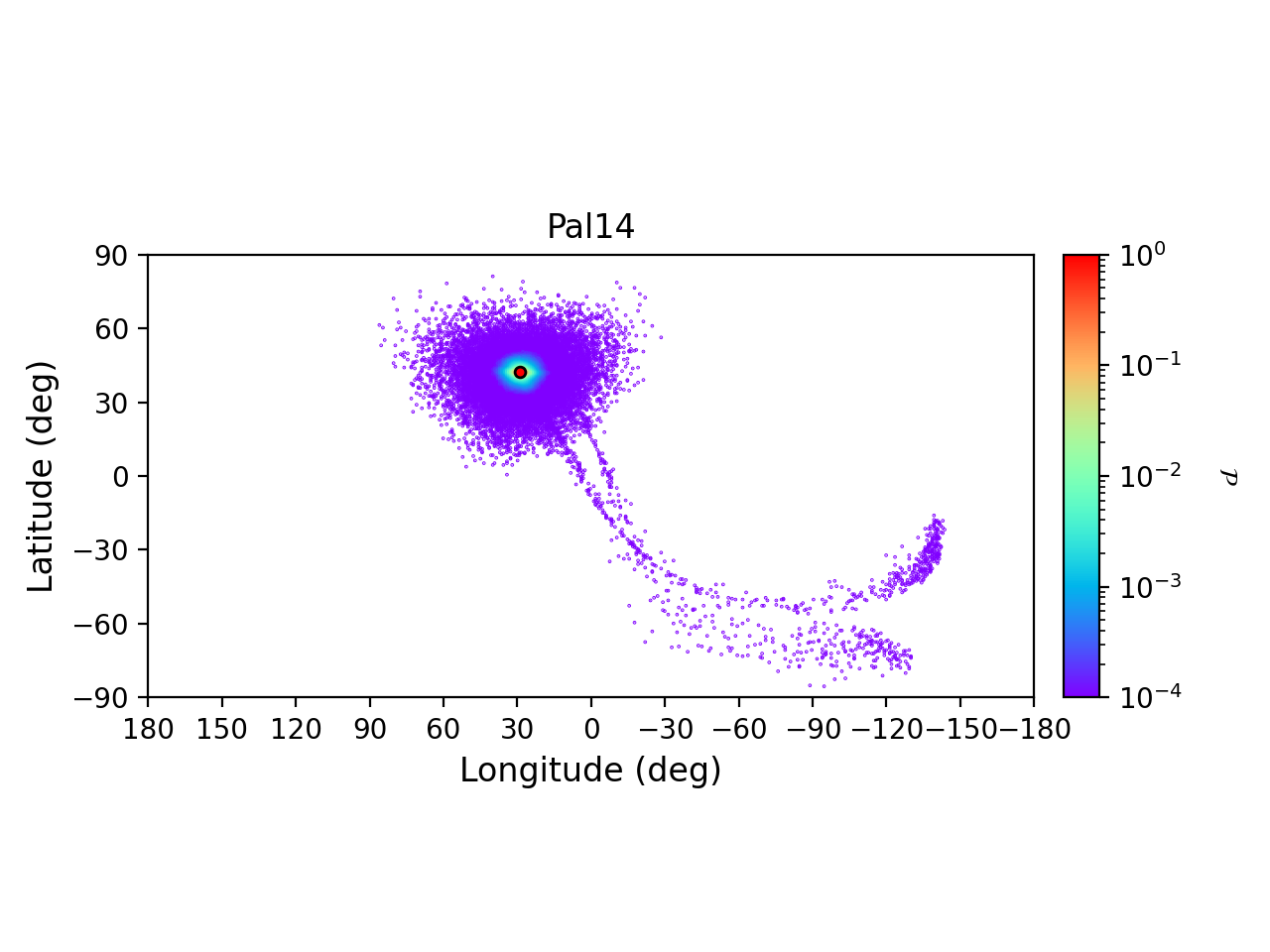}
\includegraphics[clip=true, trim = 0mm 20mm 0mm 10mm, width=1\columnwidth]{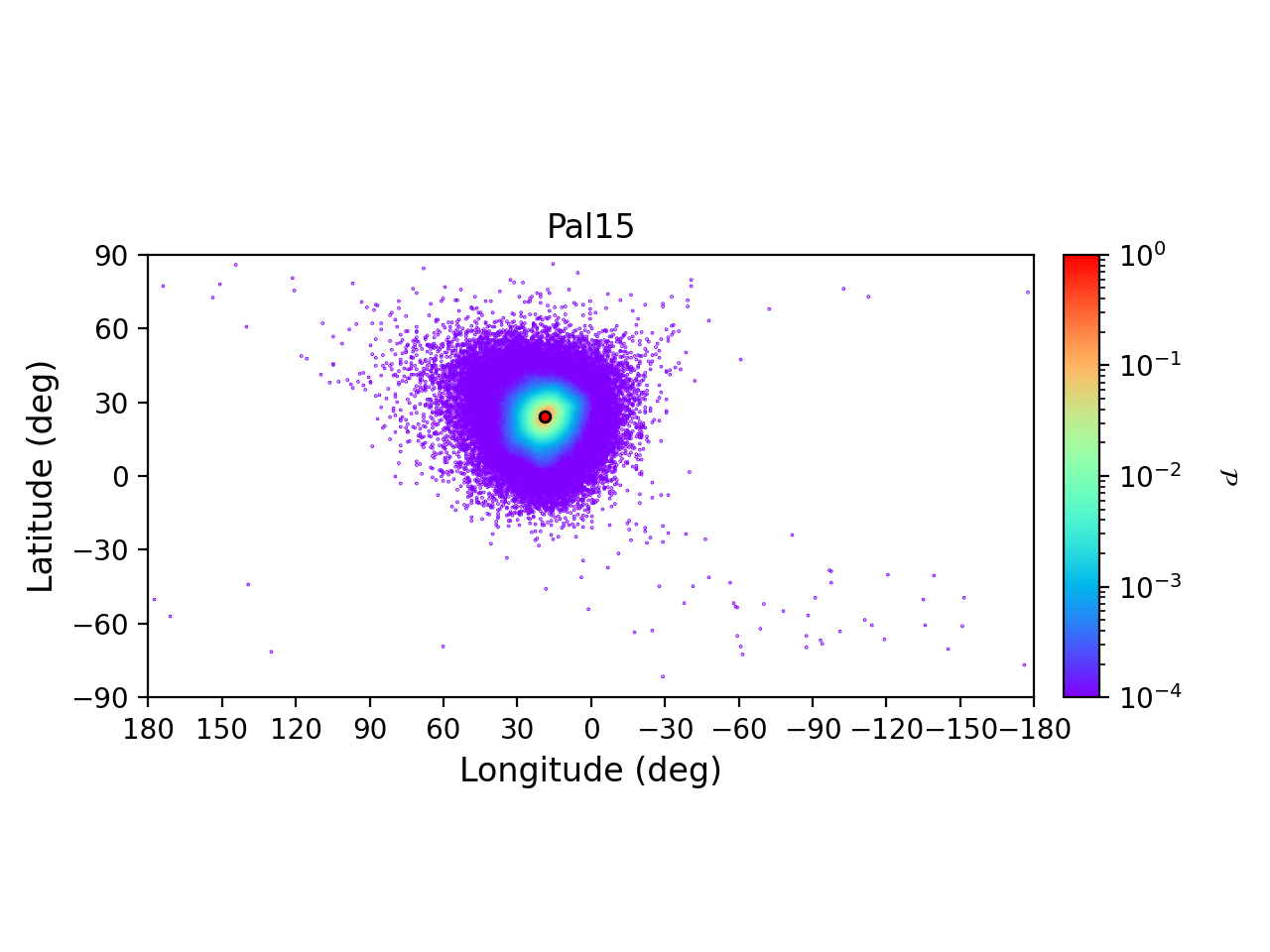}
\includegraphics[clip=true, trim = 0mm 20mm 0mm 10mm, width=1\columnwidth]{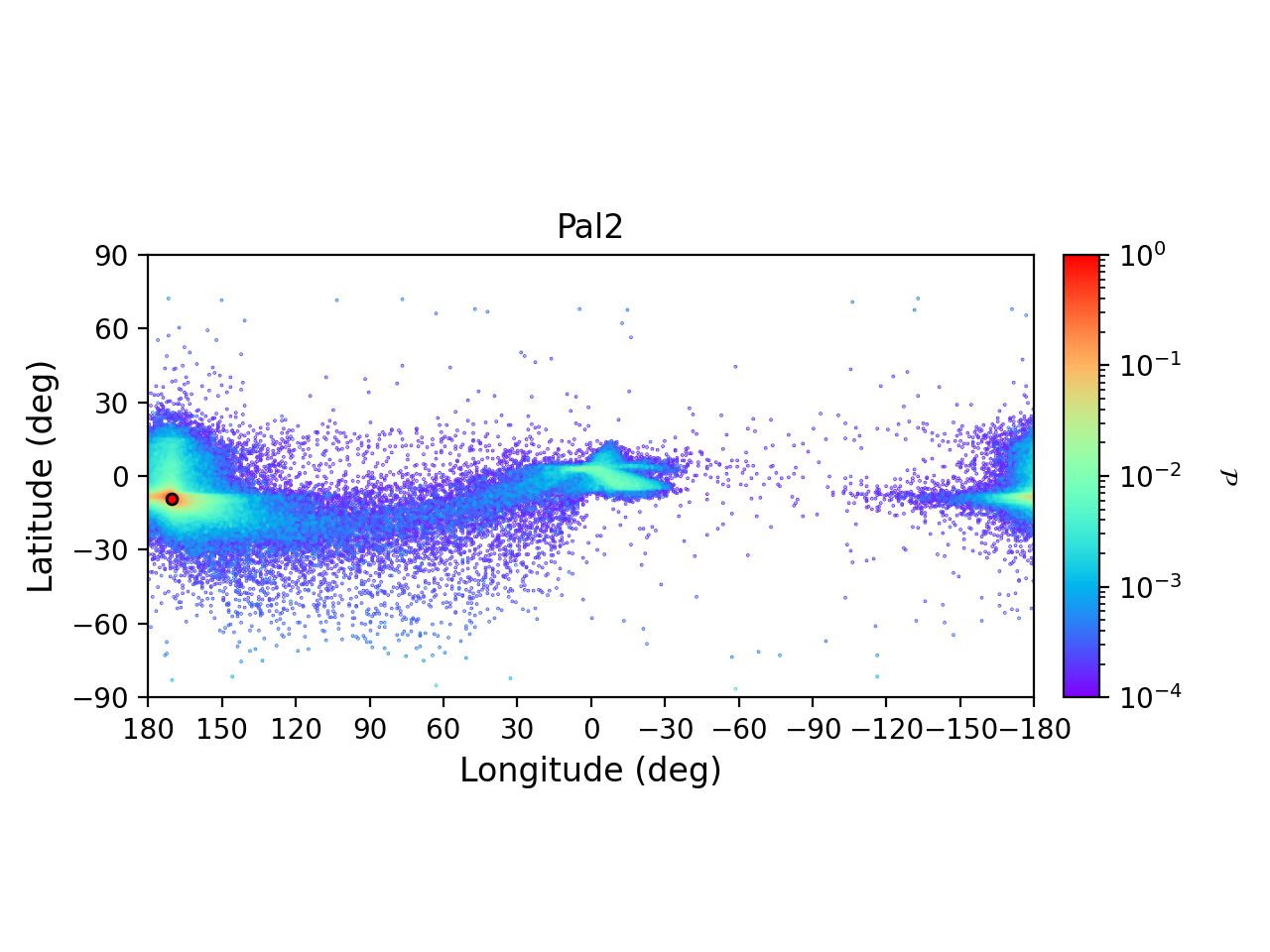}
\caption{Projected density distribution in the $(\ell, b)$ plane of a subset of simulated globular clusters, as indicated at the top of each panel. In each panel, the red circle indicates the current position of the cluster. The densities have been normalized to their maximum value.}\label{stream17}
\end{figure*}

\begin{figure*}
\includegraphics[clip=true, trim = 0mm 20mm 0mm 10mm, width=1\columnwidth]{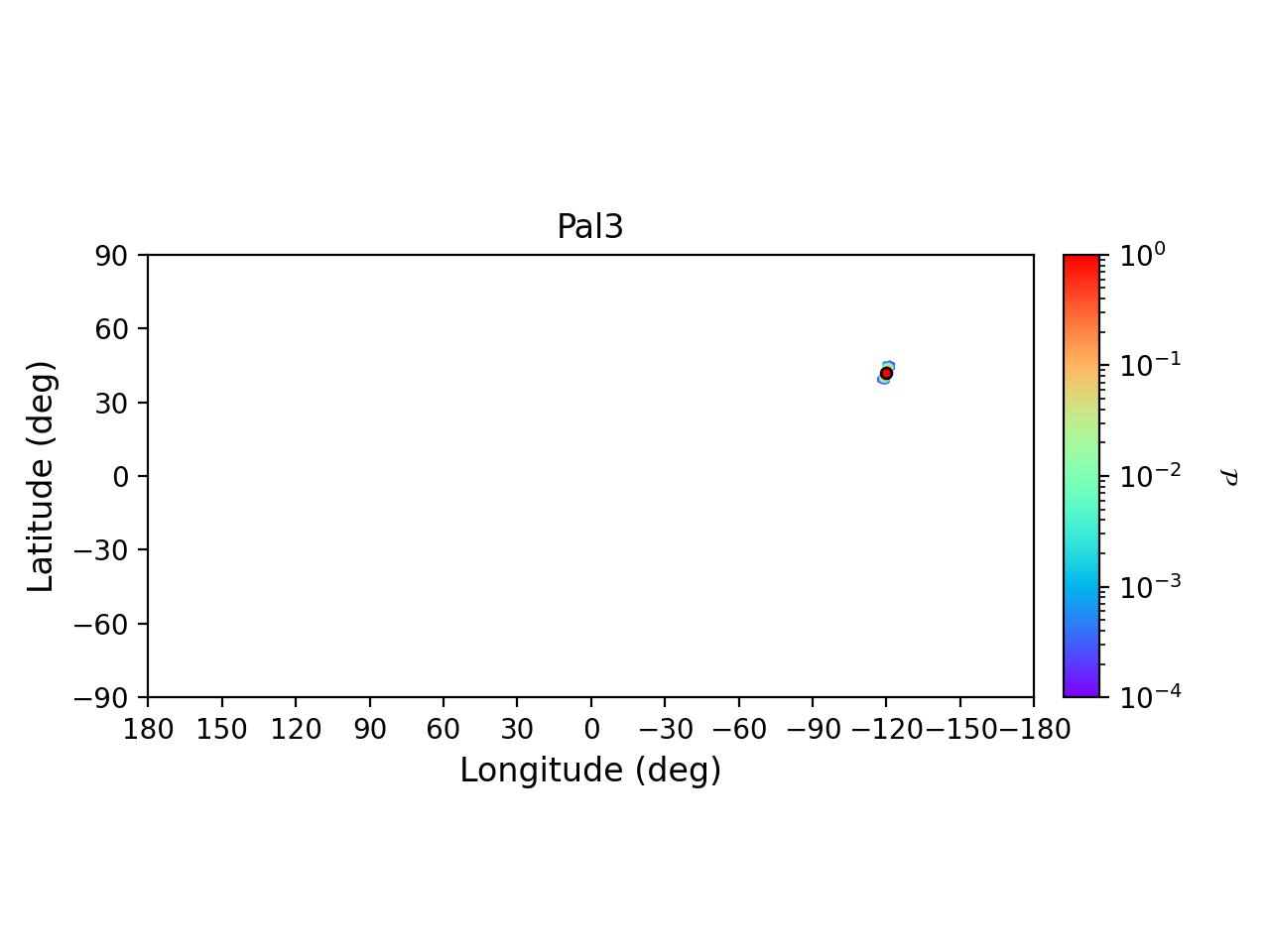}
\includegraphics[clip=true, trim = 0mm 20mm 0mm 10mm, width=1\columnwidth]{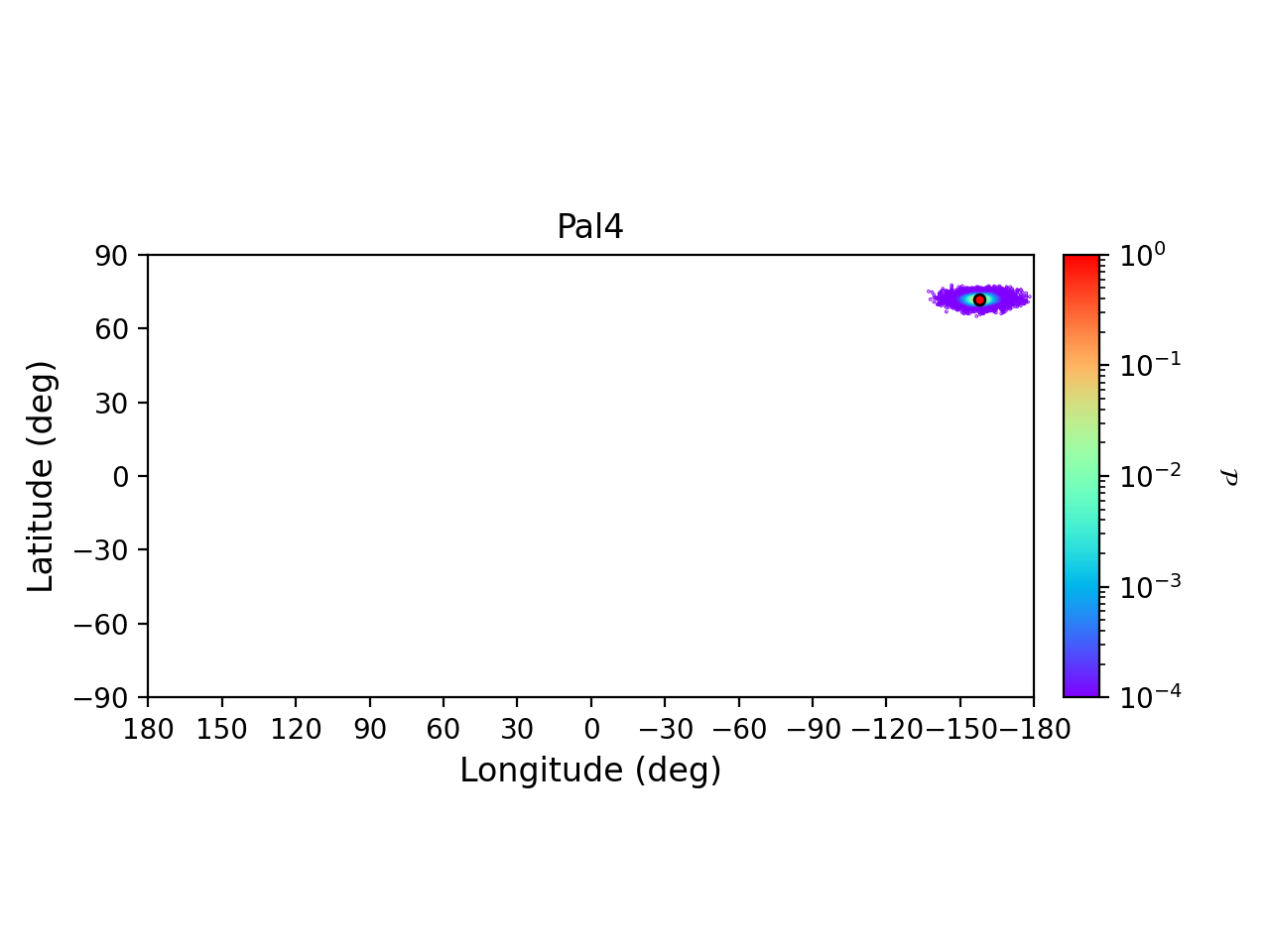}
\includegraphics[clip=true, trim = 0mm 20mm 0mm 10mm, width=1\columnwidth]{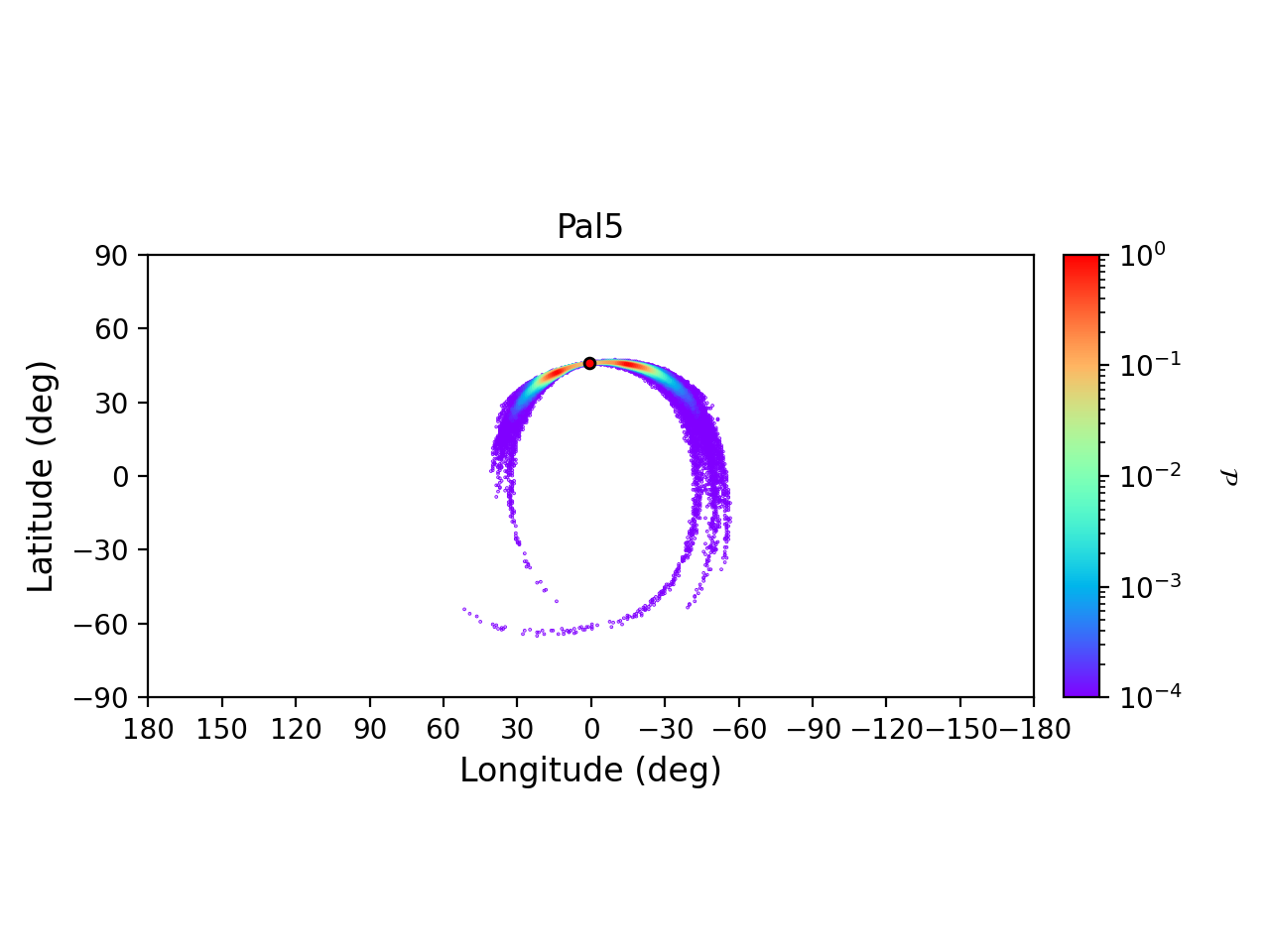}
\includegraphics[clip=true, trim = 0mm 20mm 0mm 10mm, width=1\columnwidth]{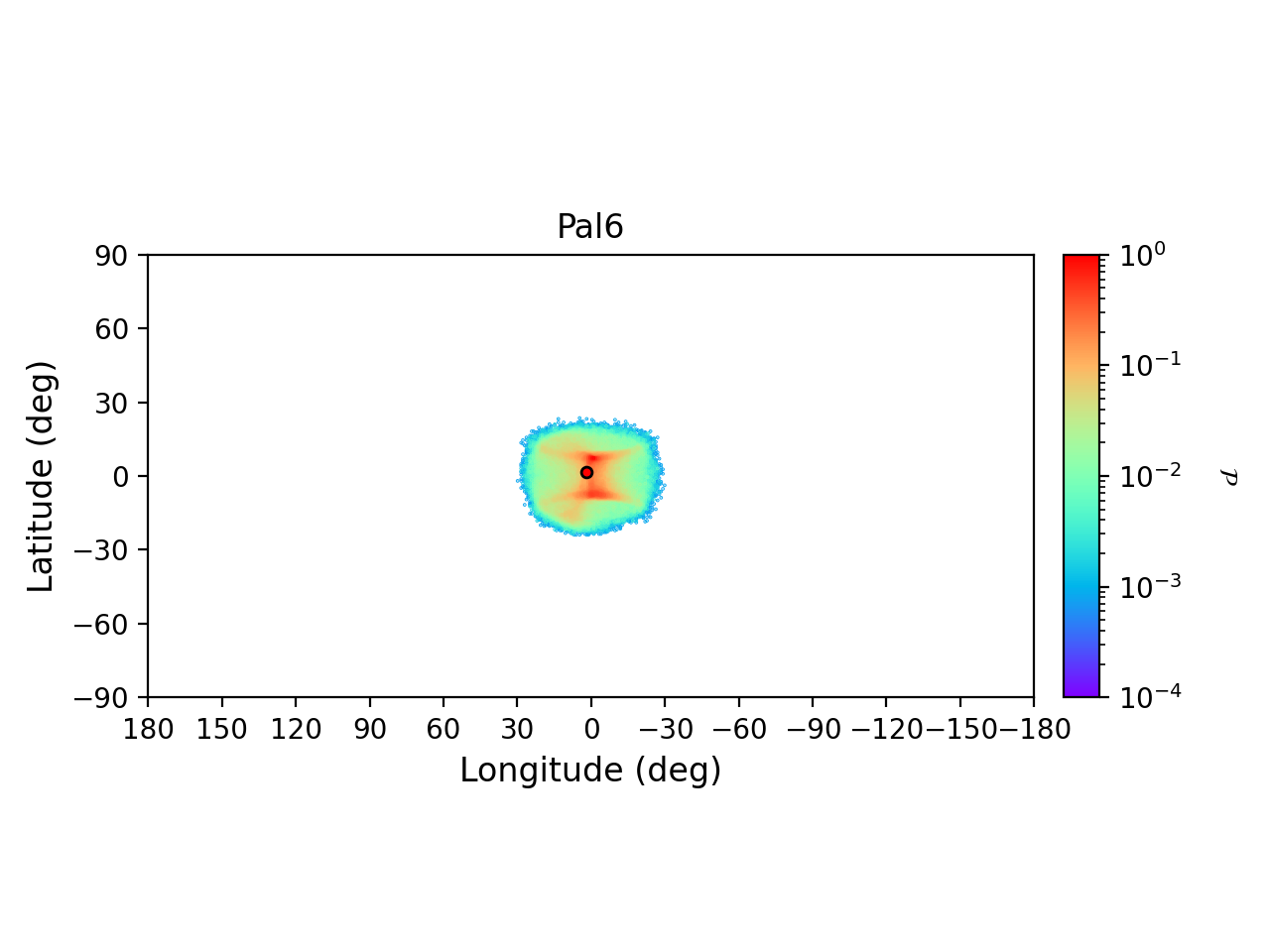}
\includegraphics[clip=true, trim = 0mm 20mm 0mm 10mm, width=1\columnwidth]{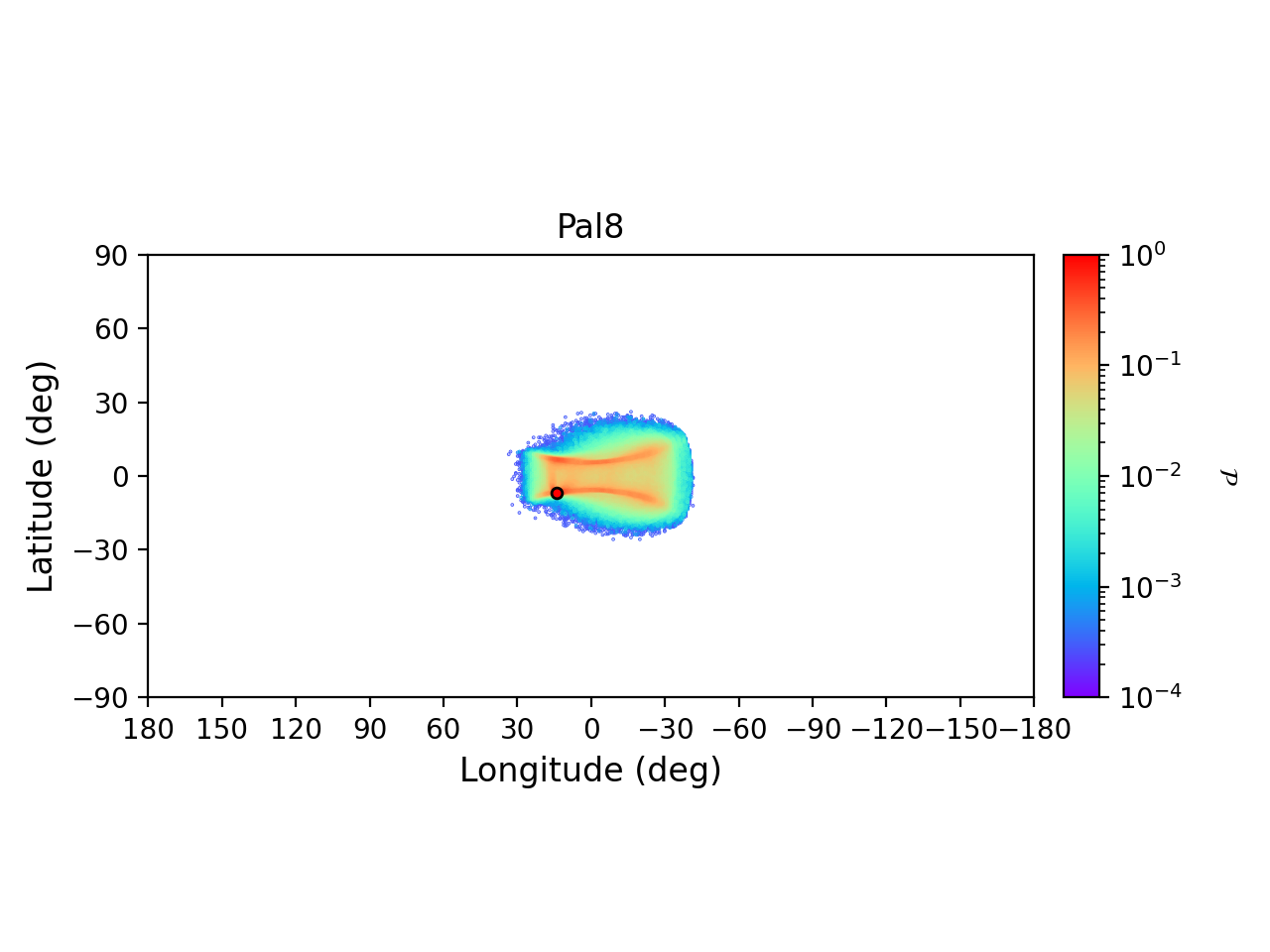}
\includegraphics[clip=true, trim = 0mm 20mm 0mm 10mm, width=1\columnwidth]{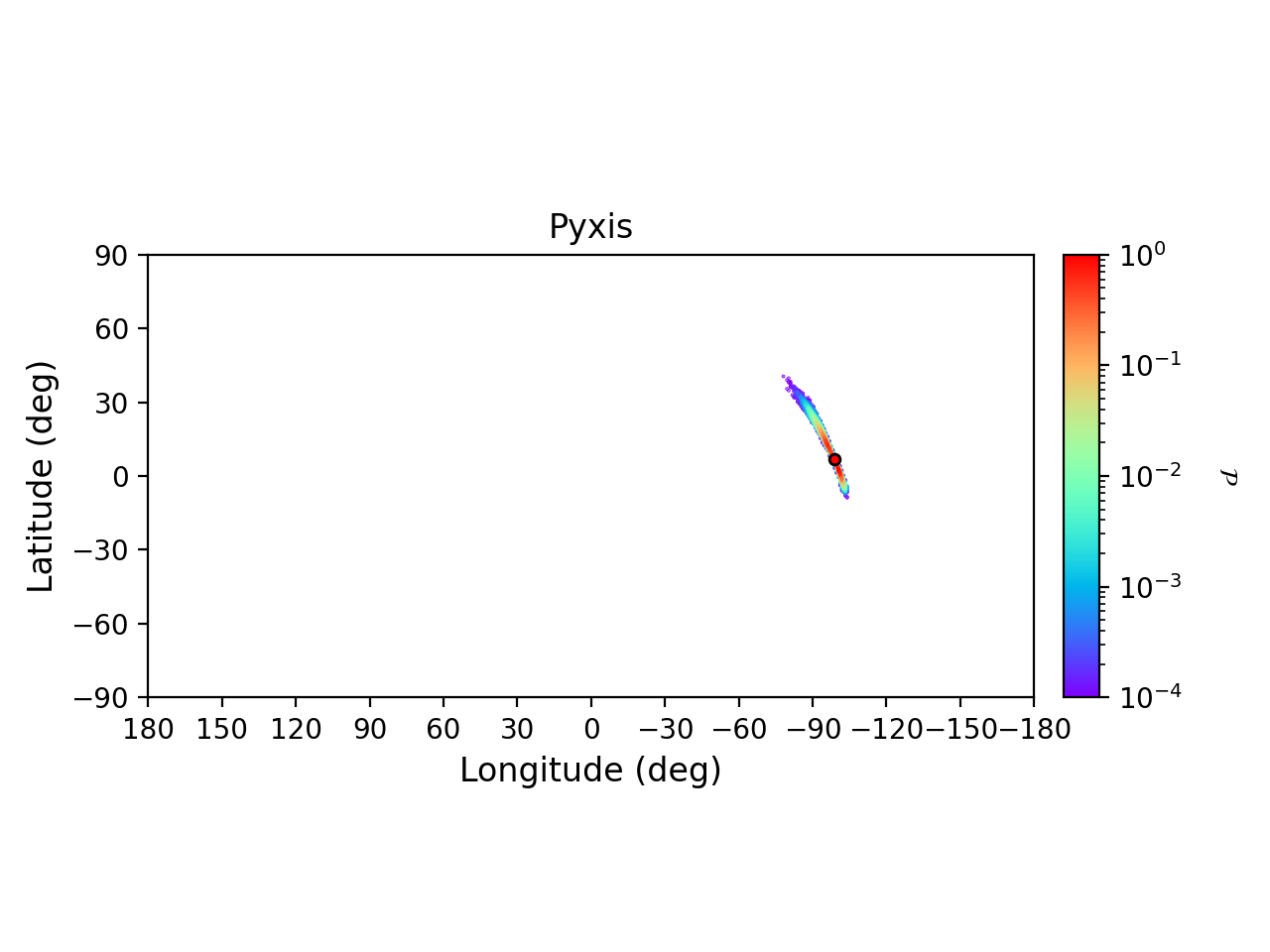}
\includegraphics[clip=true, trim = 0mm 20mm 0mm 10mm, width=1\columnwidth]{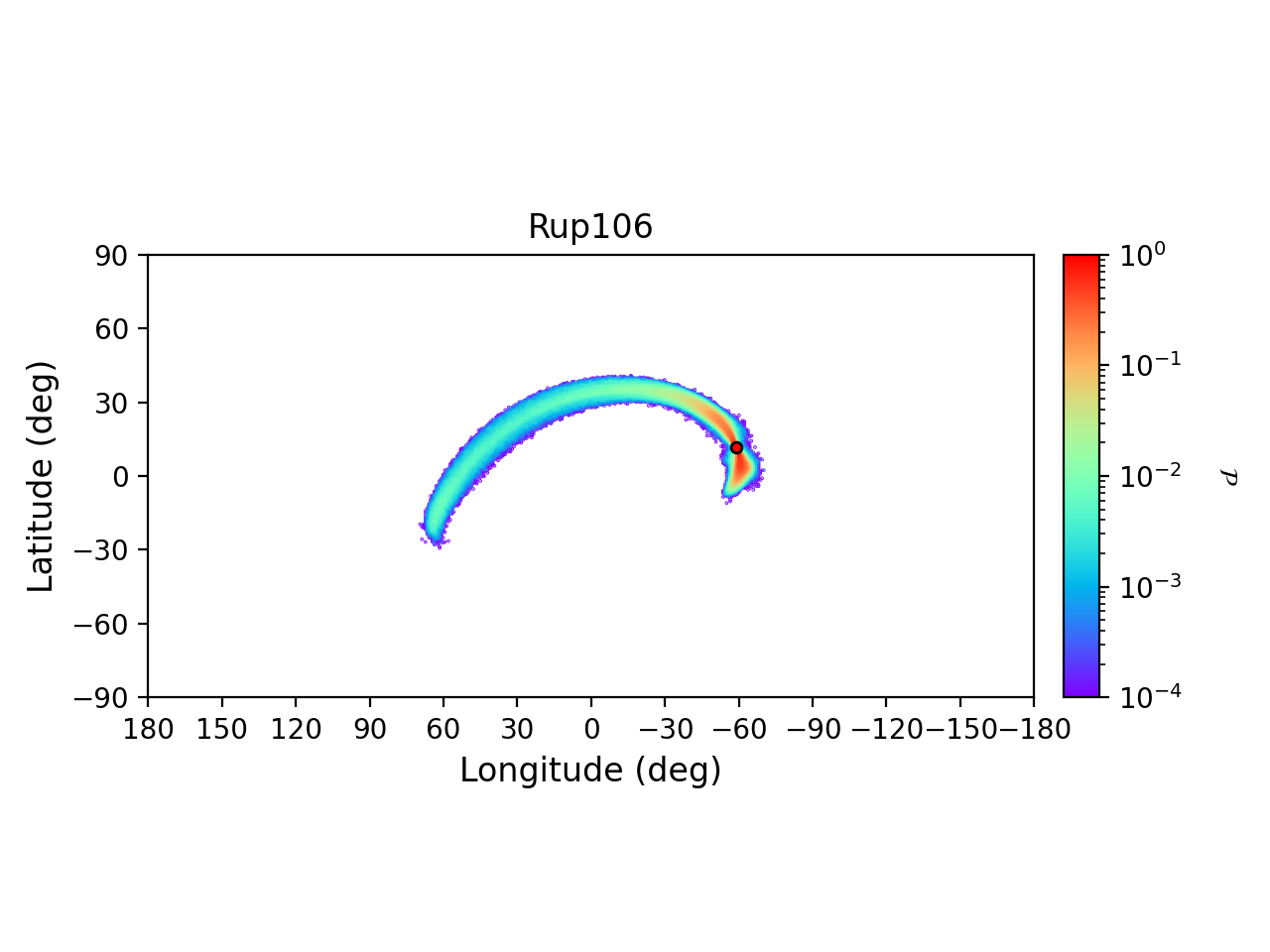}
\includegraphics[clip=true, trim = 0mm 20mm 0mm 10mm, width=1\columnwidth]{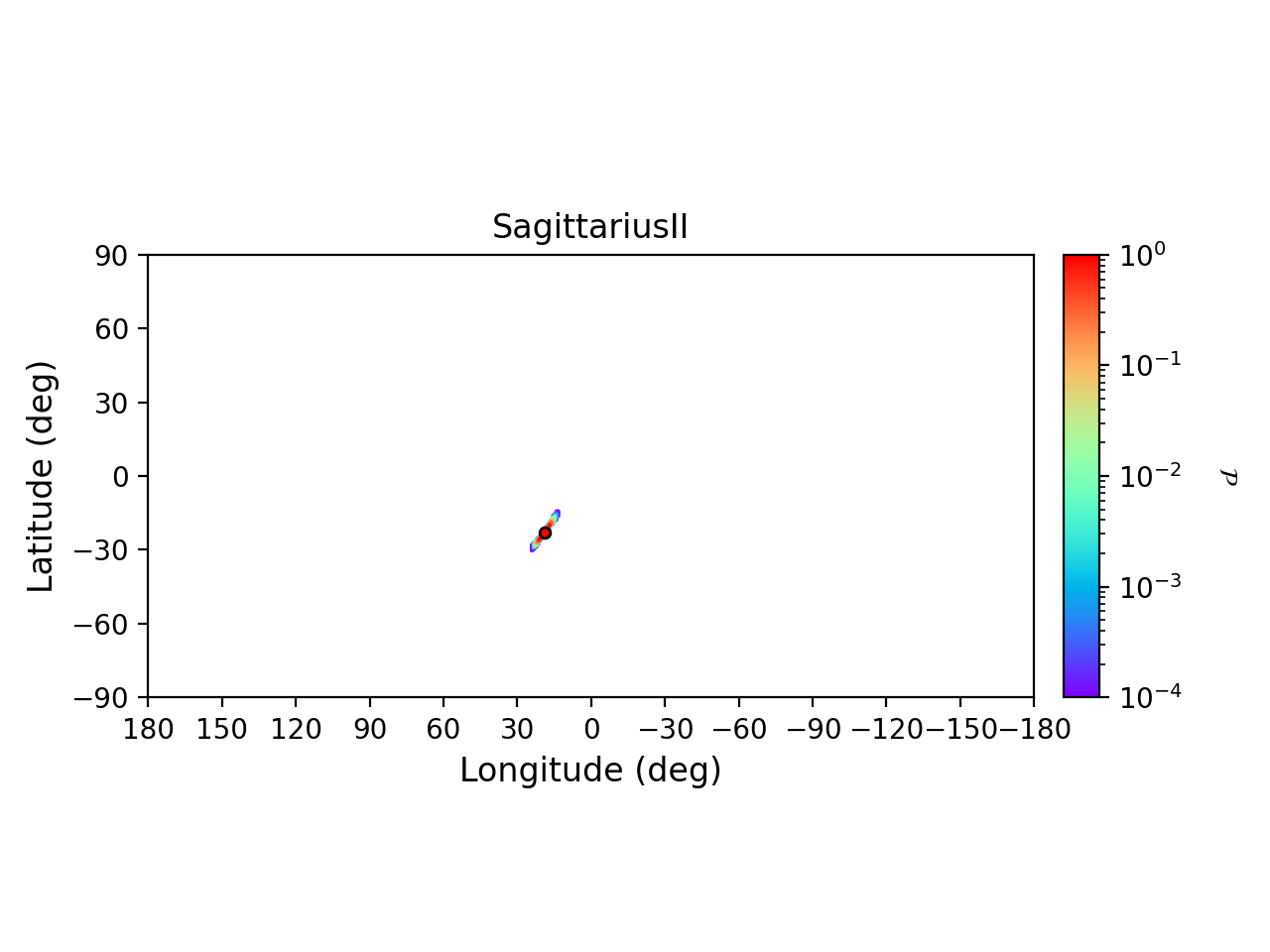}
\caption{Projected density distribution in the $(\ell, b)$ plane of a subset of simulated globular clusters, as indicated at the top of each panel. In each panel, the red circle indicates the current position of the cluster. The densities have been normalized to their maximum value.}\label{stream18}
\end{figure*}

\begin{figure*}
\includegraphics[clip=true, trim = 0mm 20mm 0mm 10mm, width=1\columnwidth]{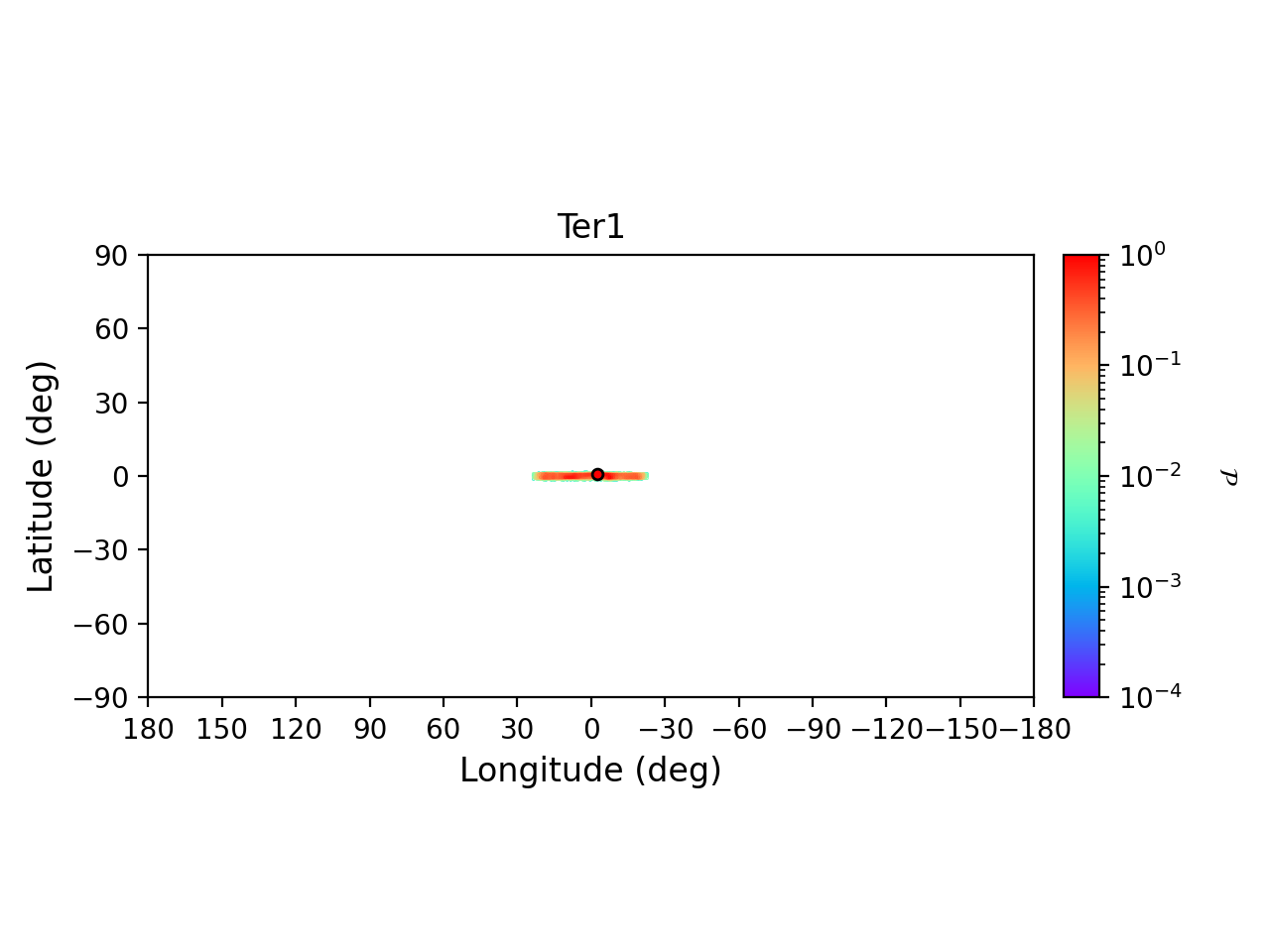}
\includegraphics[clip=true, trim = 0mm 20mm 0mm 10mm, width=1\columnwidth]{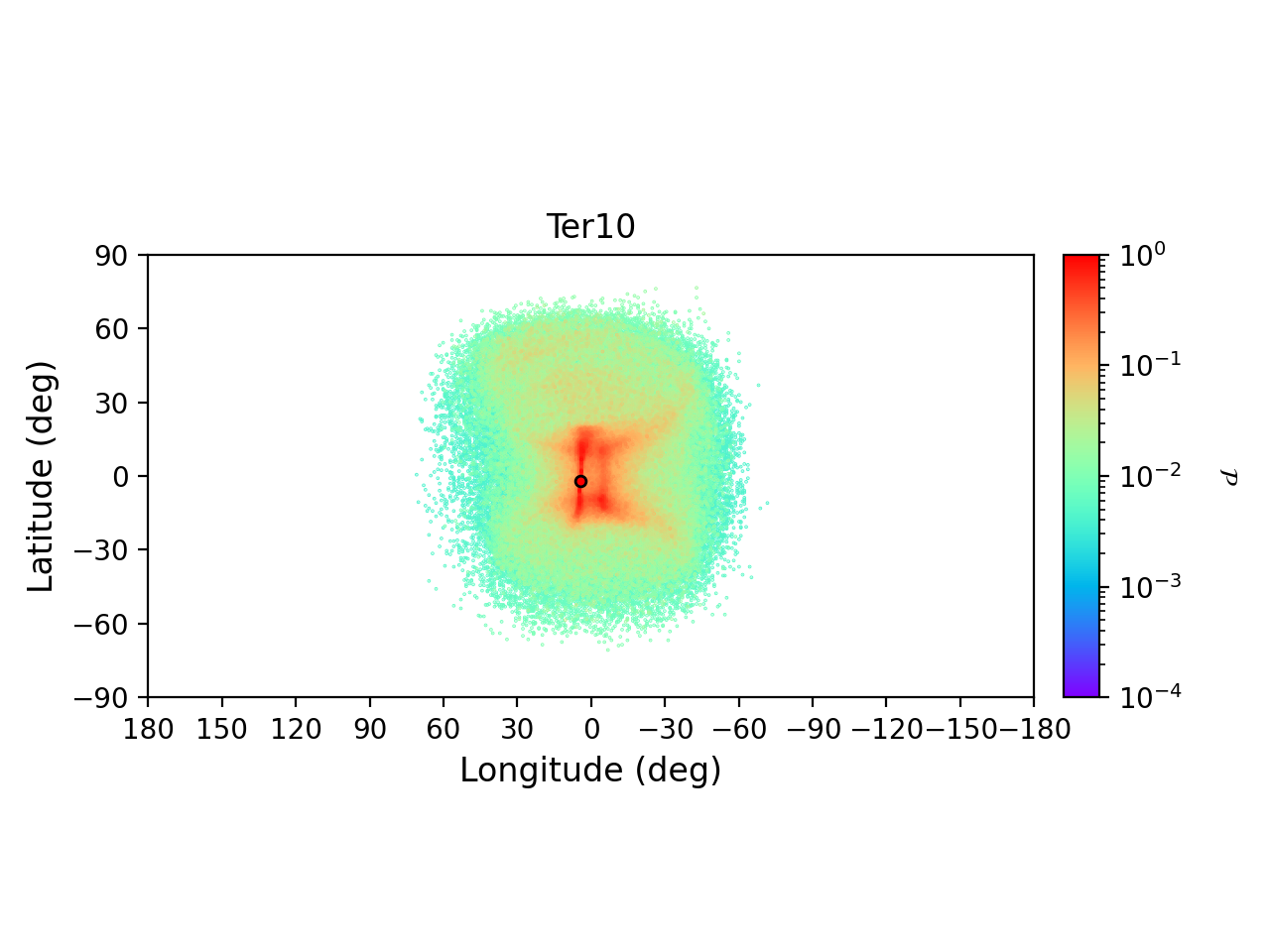}
\includegraphics[clip=true, trim = 0mm 20mm 0mm 10mm, width=1\columnwidth]{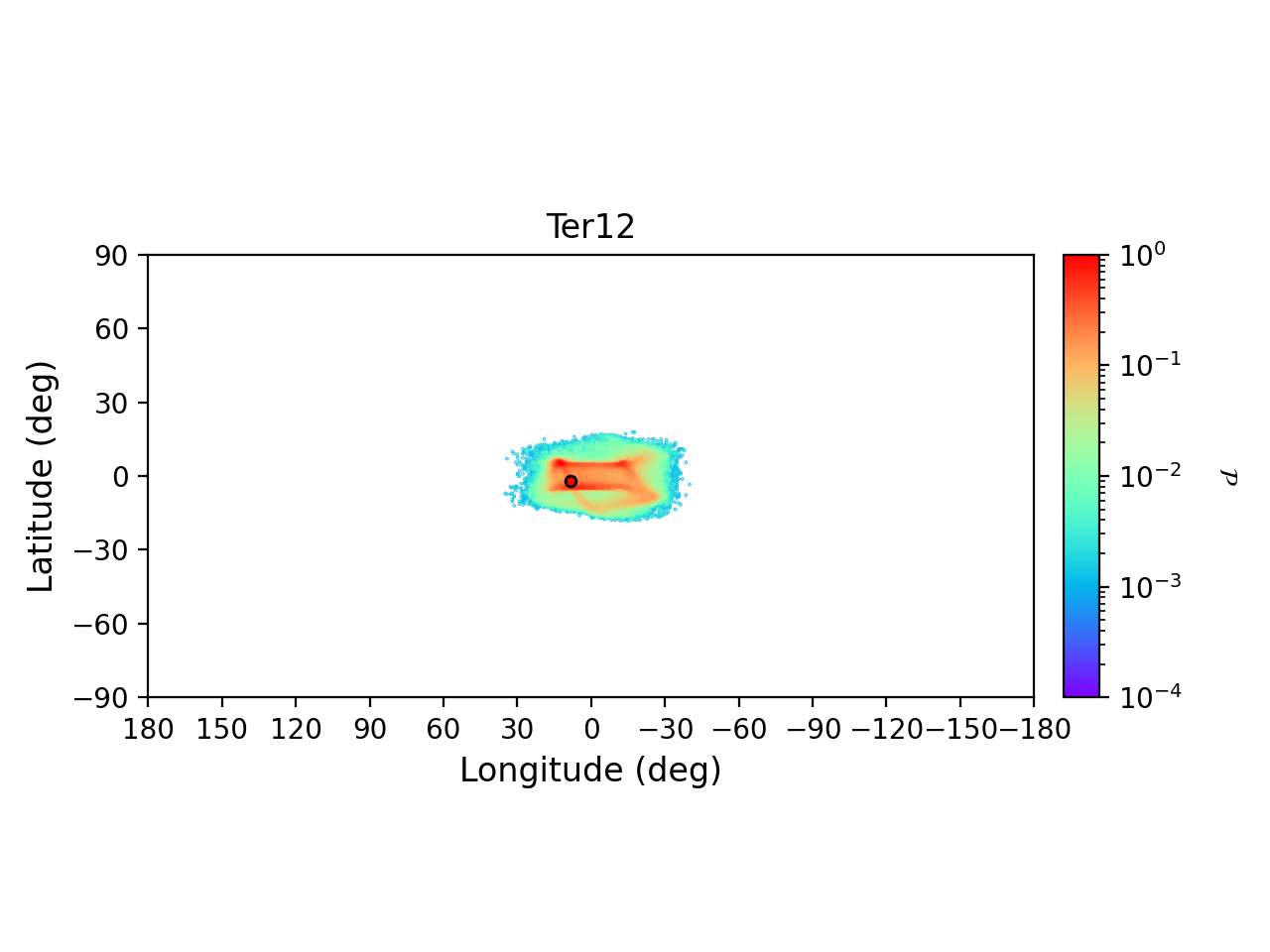}
\includegraphics[clip=true, trim = 0mm 20mm 0mm 10mm, width=1\columnwidth]{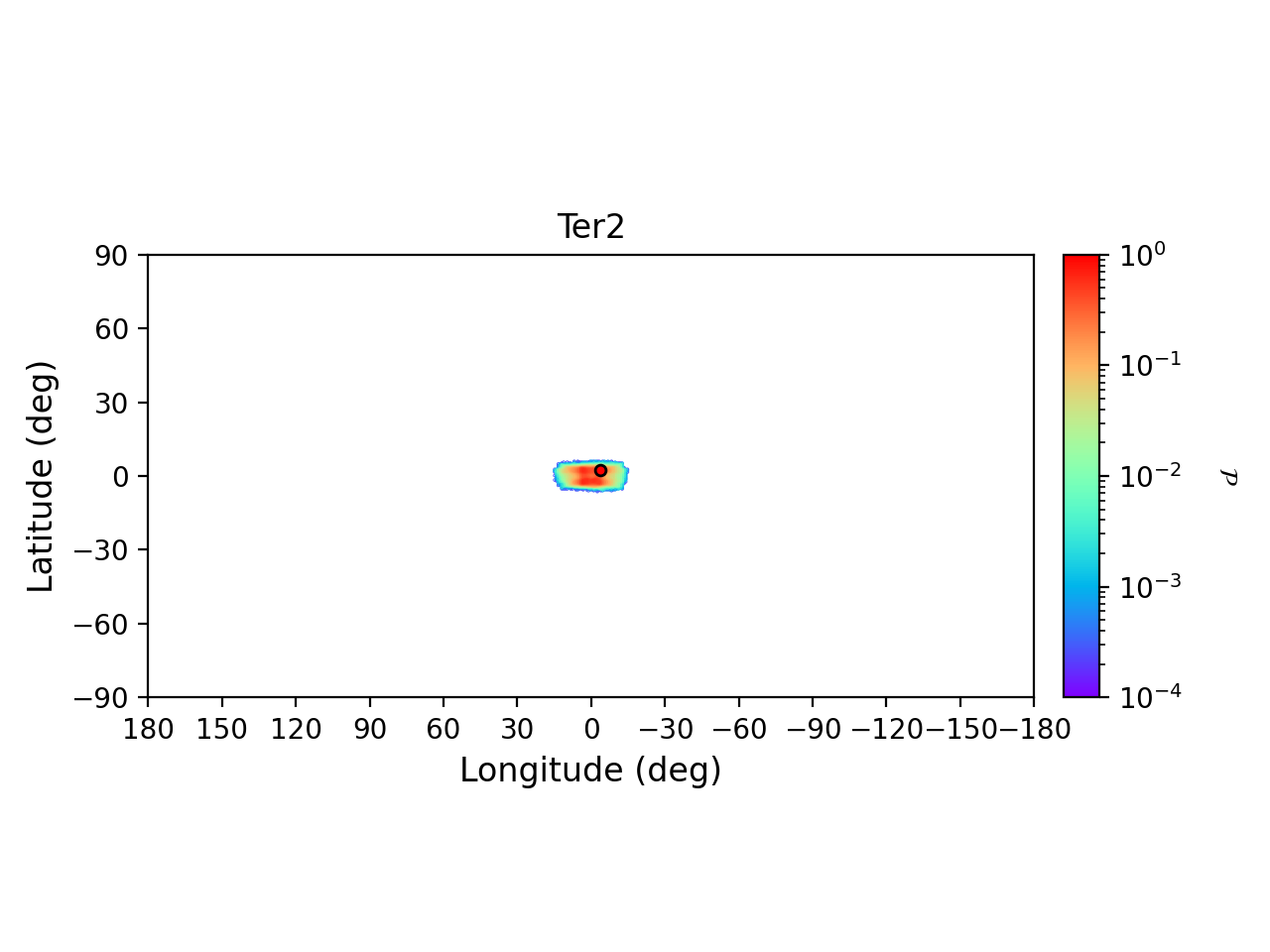}
\includegraphics[clip=true, trim = 0mm 20mm 0mm 10mm, width=1\columnwidth]{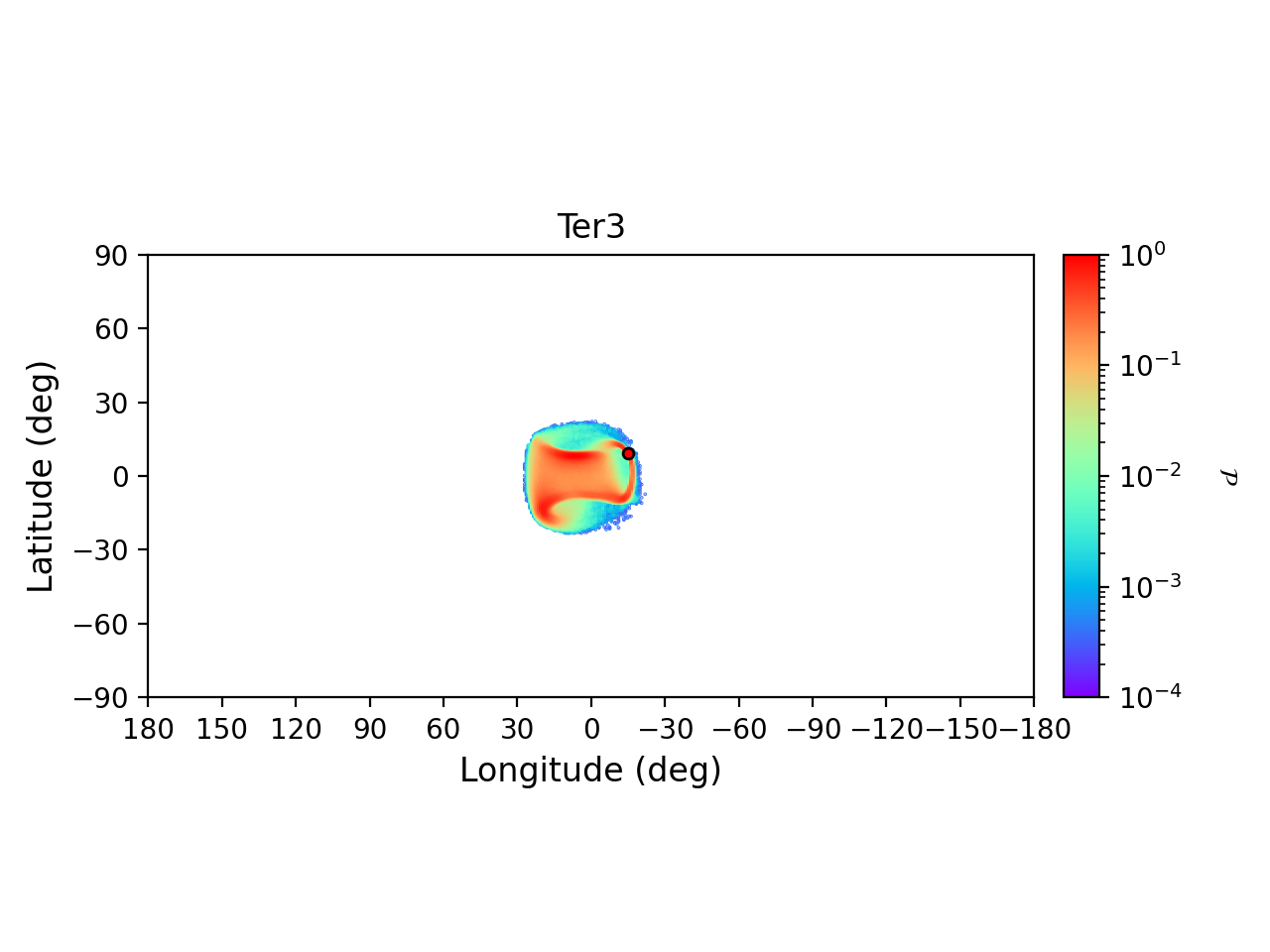}
\includegraphics[clip=true, trim = 0mm 20mm 0mm 10mm, width=1\columnwidth]{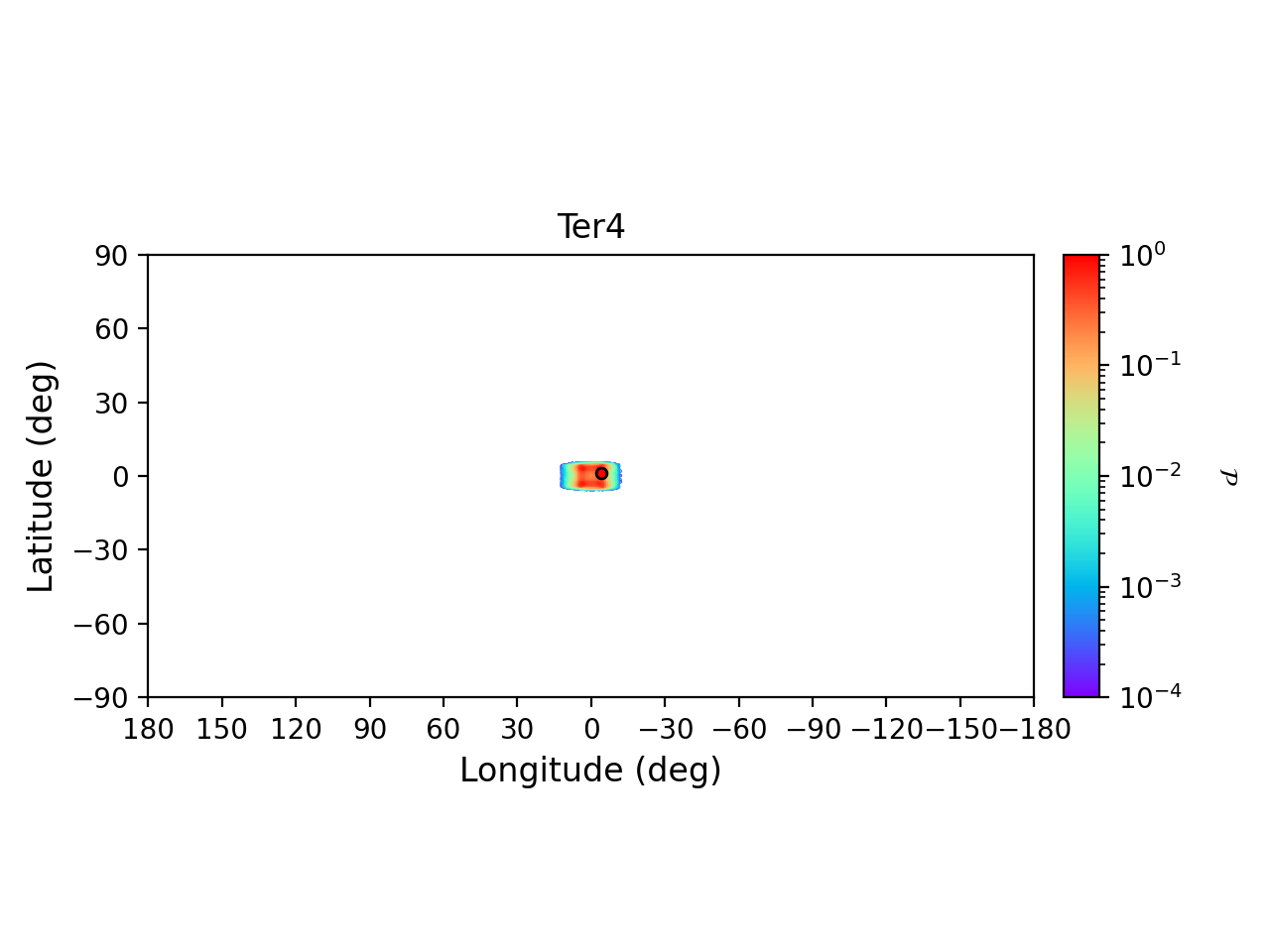}
\includegraphics[clip=true, trim = 0mm 20mm 0mm 10mm, width=1\columnwidth]{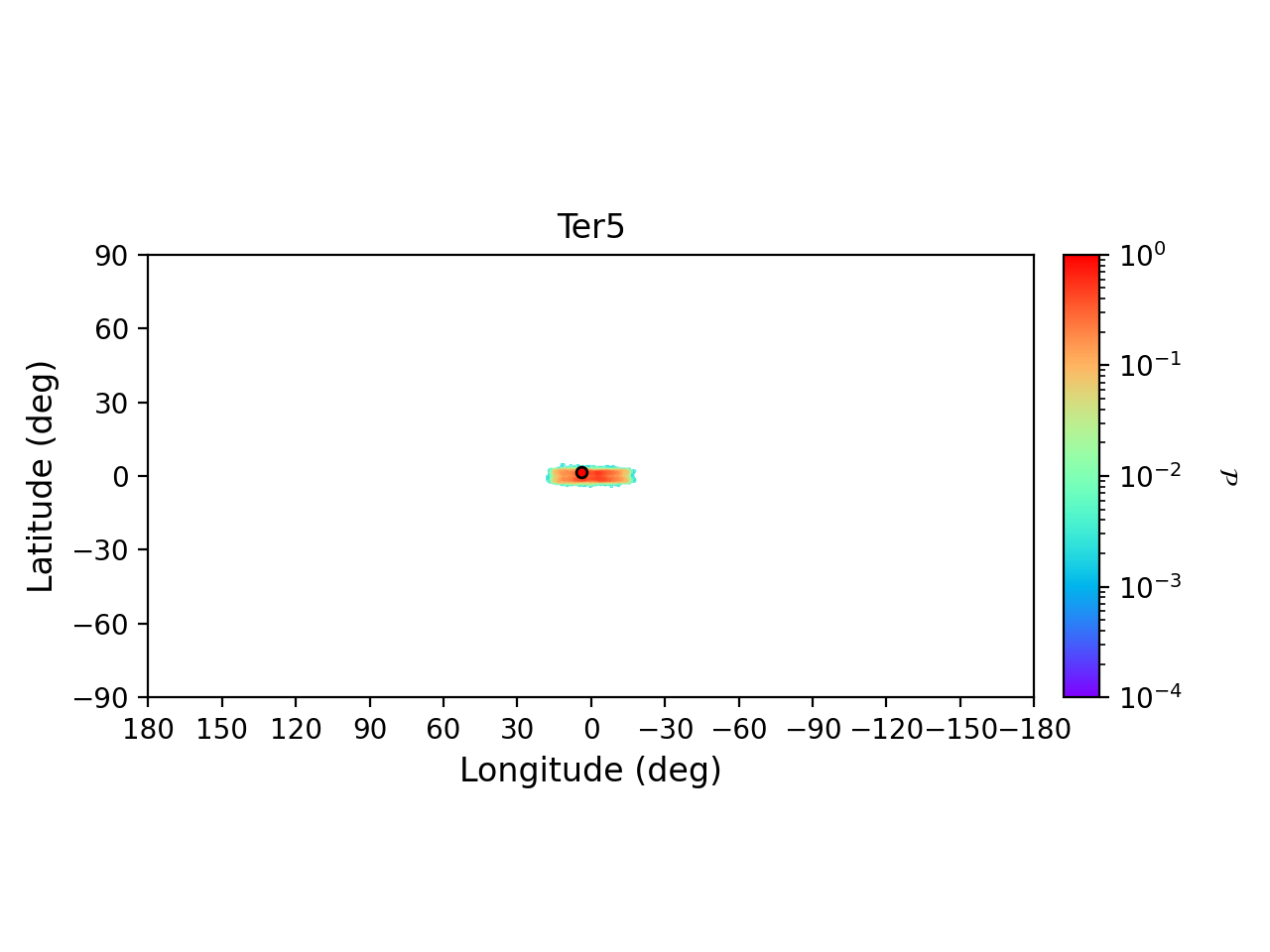}
\includegraphics[clip=true, trim = 0mm 20mm 0mm 10mm, width=1\columnwidth]{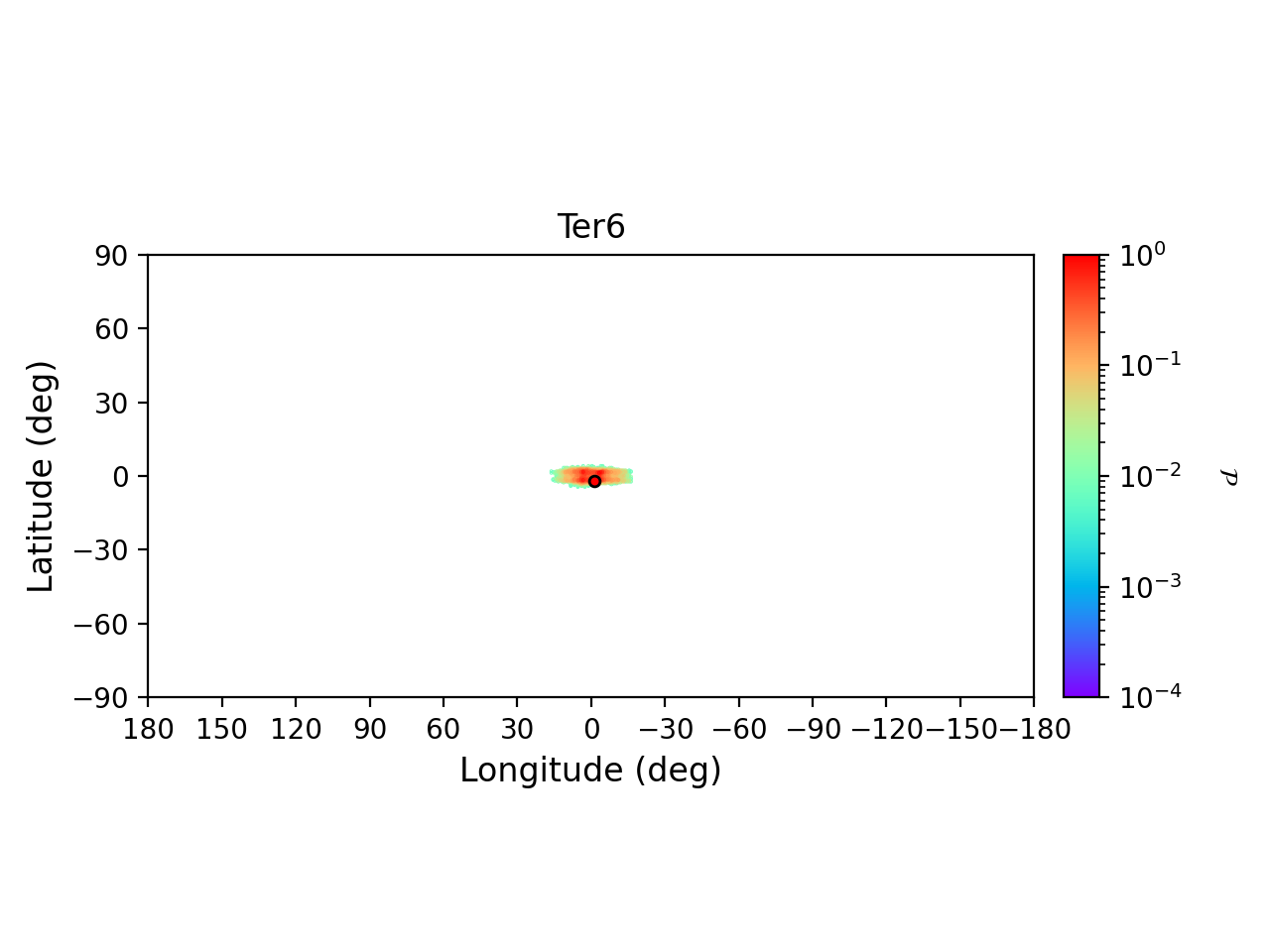}
\caption{Projected density distribution in the $(\ell, b)$ plane of a subset of simulated globular clusters, as indicated at the top of each panel. In each panel, the red circle indicates the current position of the cluster. The densities have been normalized to their maximum value.}\label{stream19}
\end{figure*}

\begin{figure*}
\includegraphics[clip=true, trim = 0mm 20mm 0mm 10mm, width=1\columnwidth]{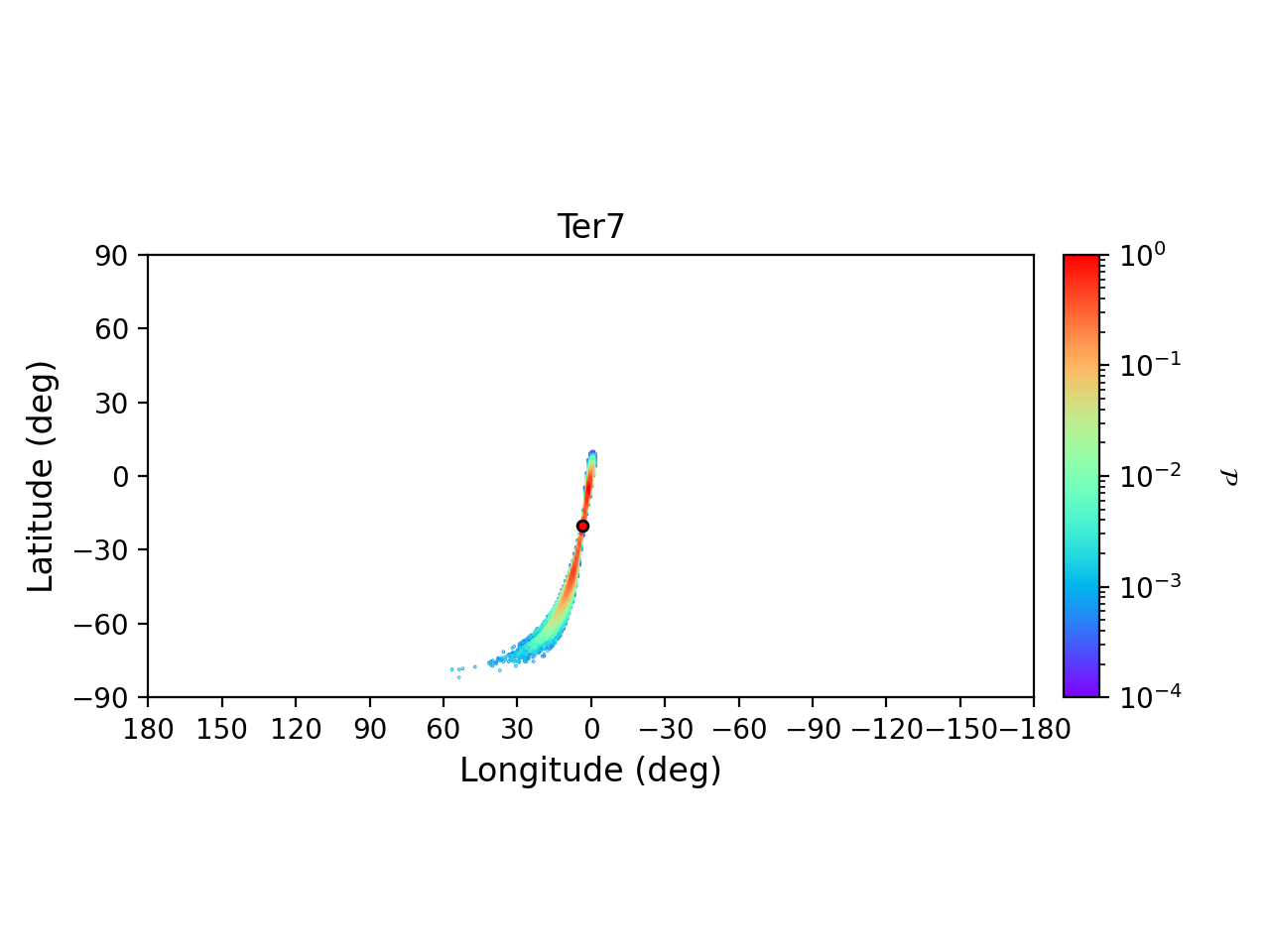}
\includegraphics[clip=true, trim = 0mm 20mm 0mm 10mm, width=1\columnwidth]{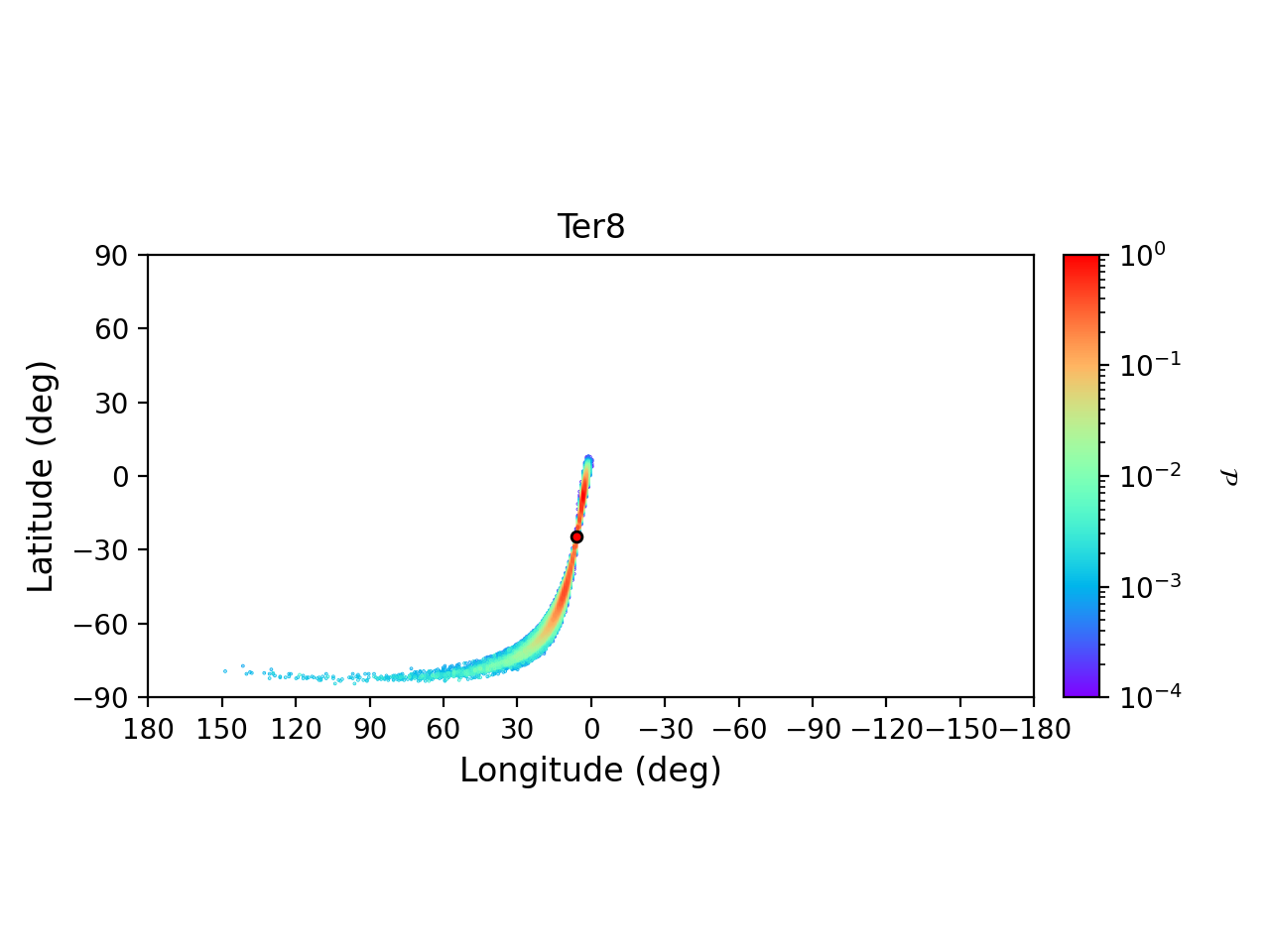}
\includegraphics[clip=true, trim = 0mm 20mm 0mm 10mm, width=1\columnwidth]{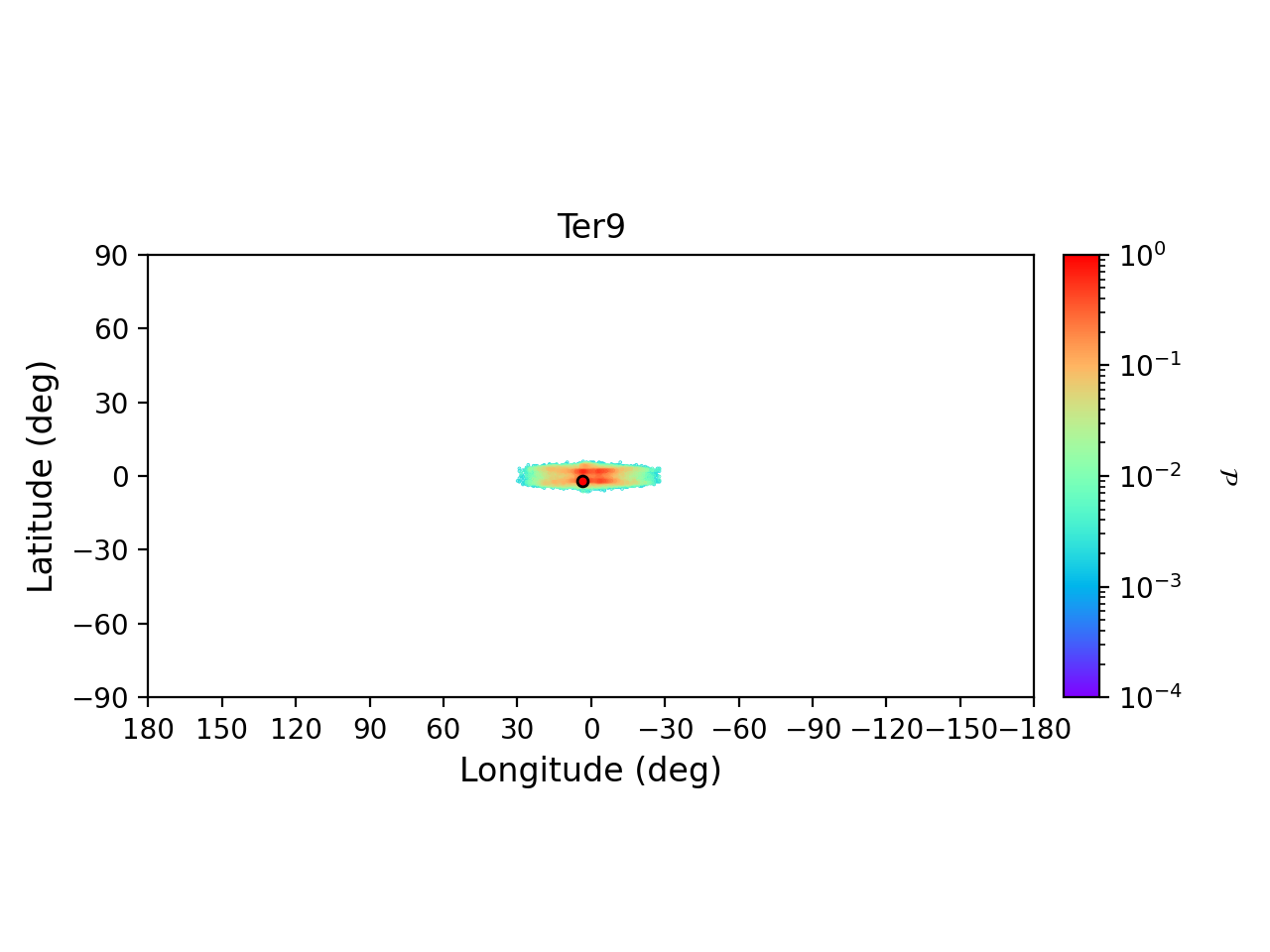}
\includegraphics[clip=true, trim = 0mm 20mm 0mm 10mm, width=1\columnwidth]{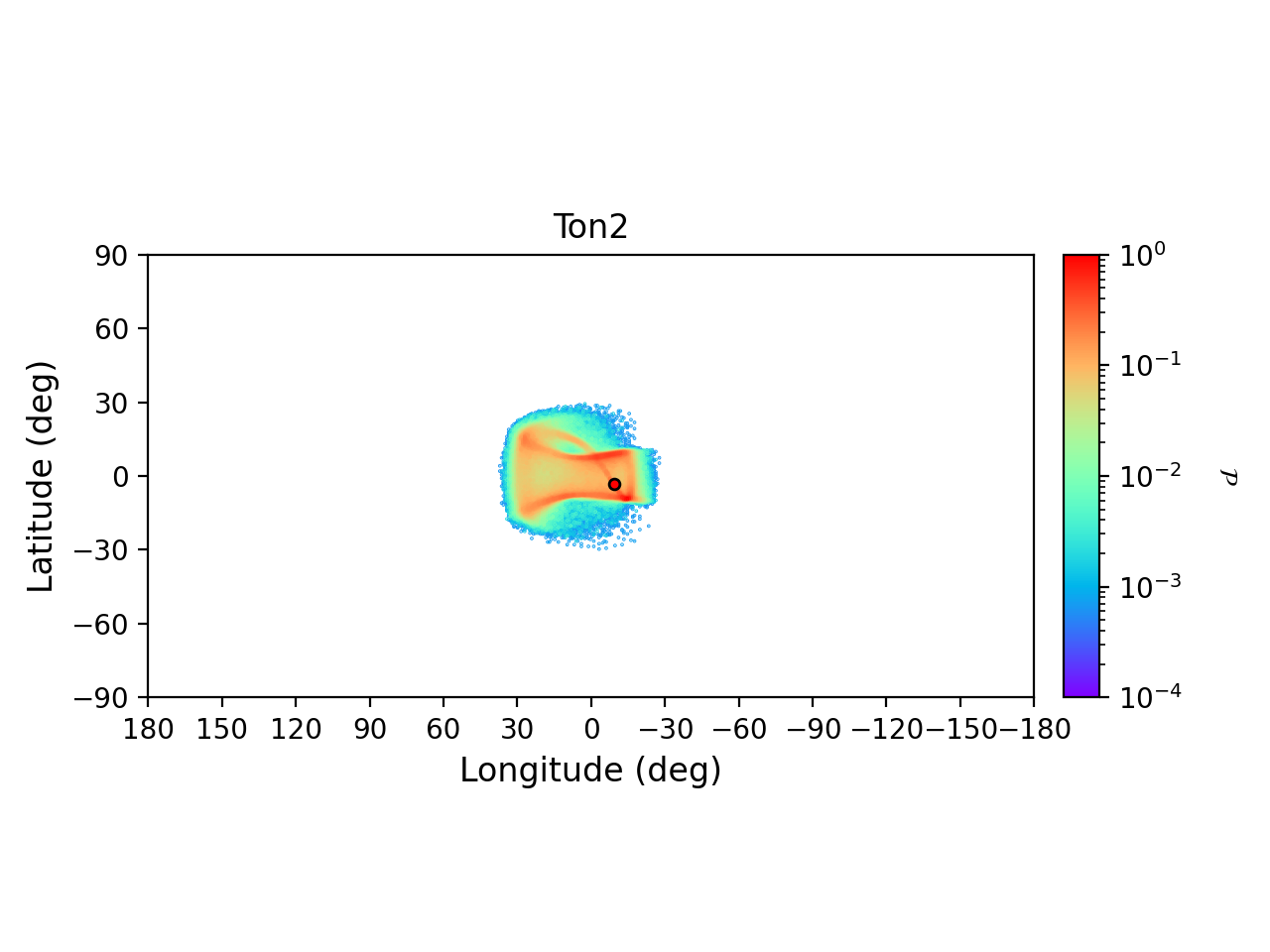}
\includegraphics[clip=true, trim = 0mm 20mm 0mm 10mm, width=1\columnwidth]{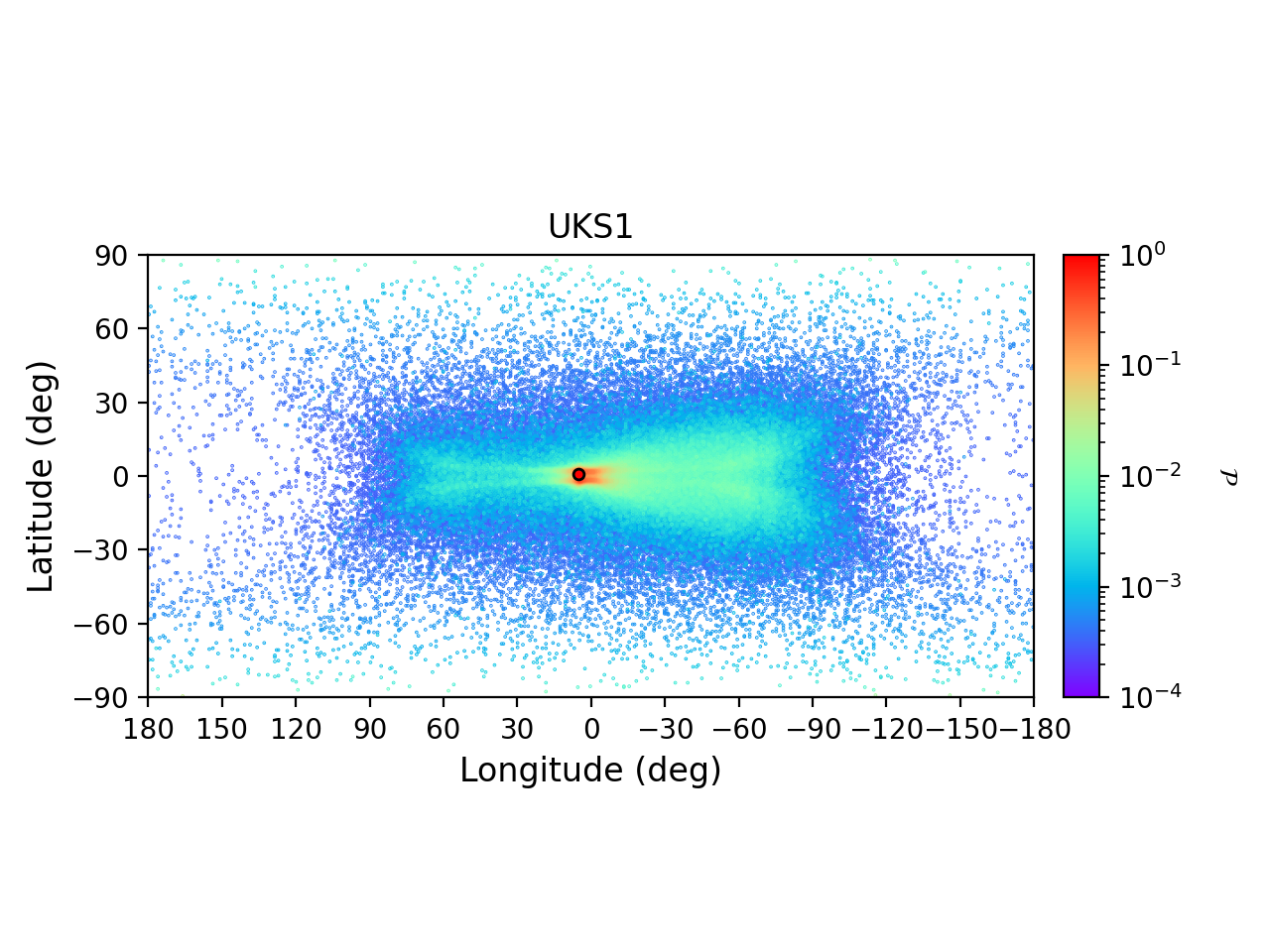}
\includegraphics[clip=true, trim = 0mm 20mm 0mm 10mm, width=1\columnwidth]{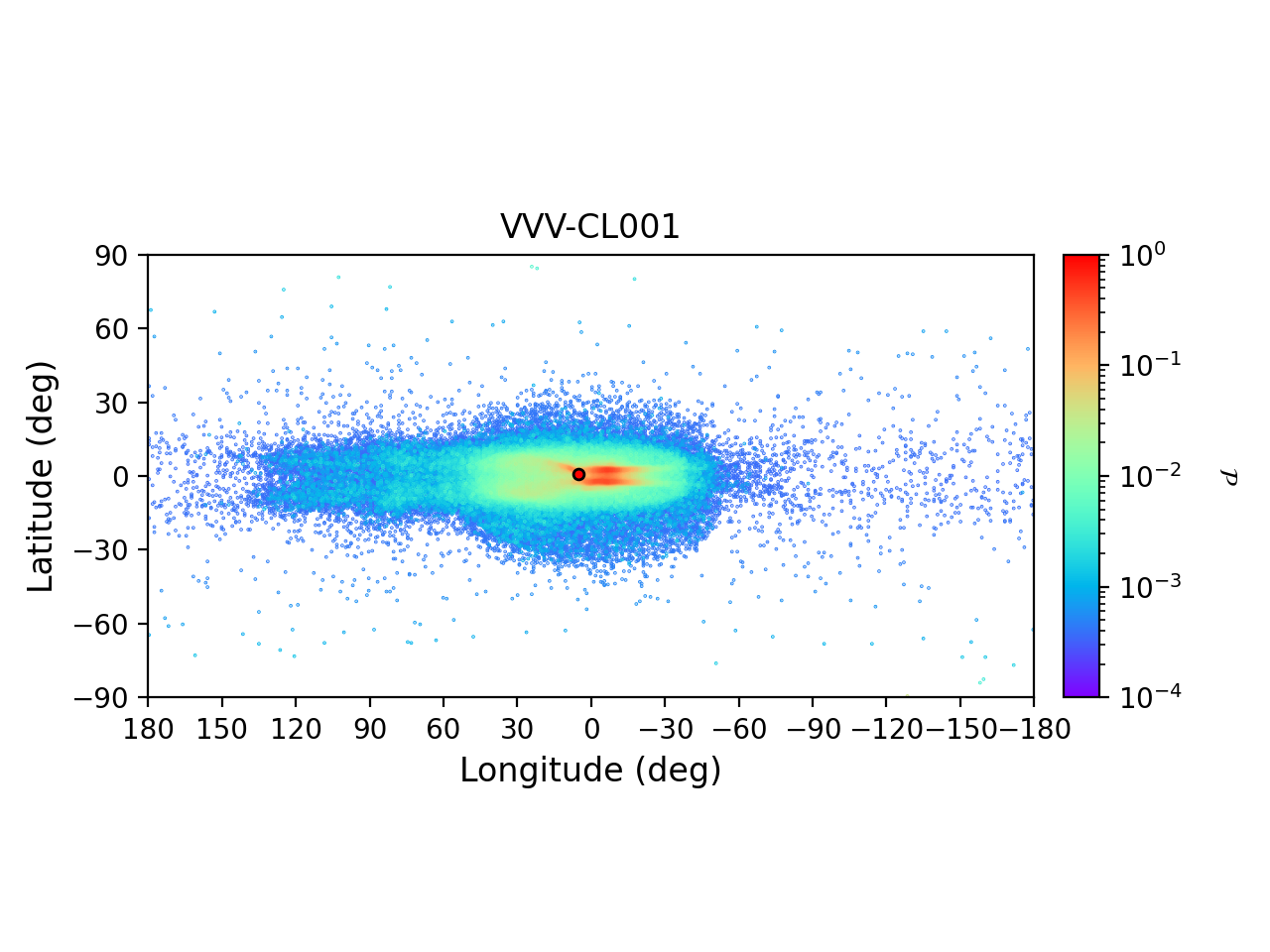}
\includegraphics[clip=true, trim = 0mm 20mm 0mm 10mm, width=1\columnwidth]{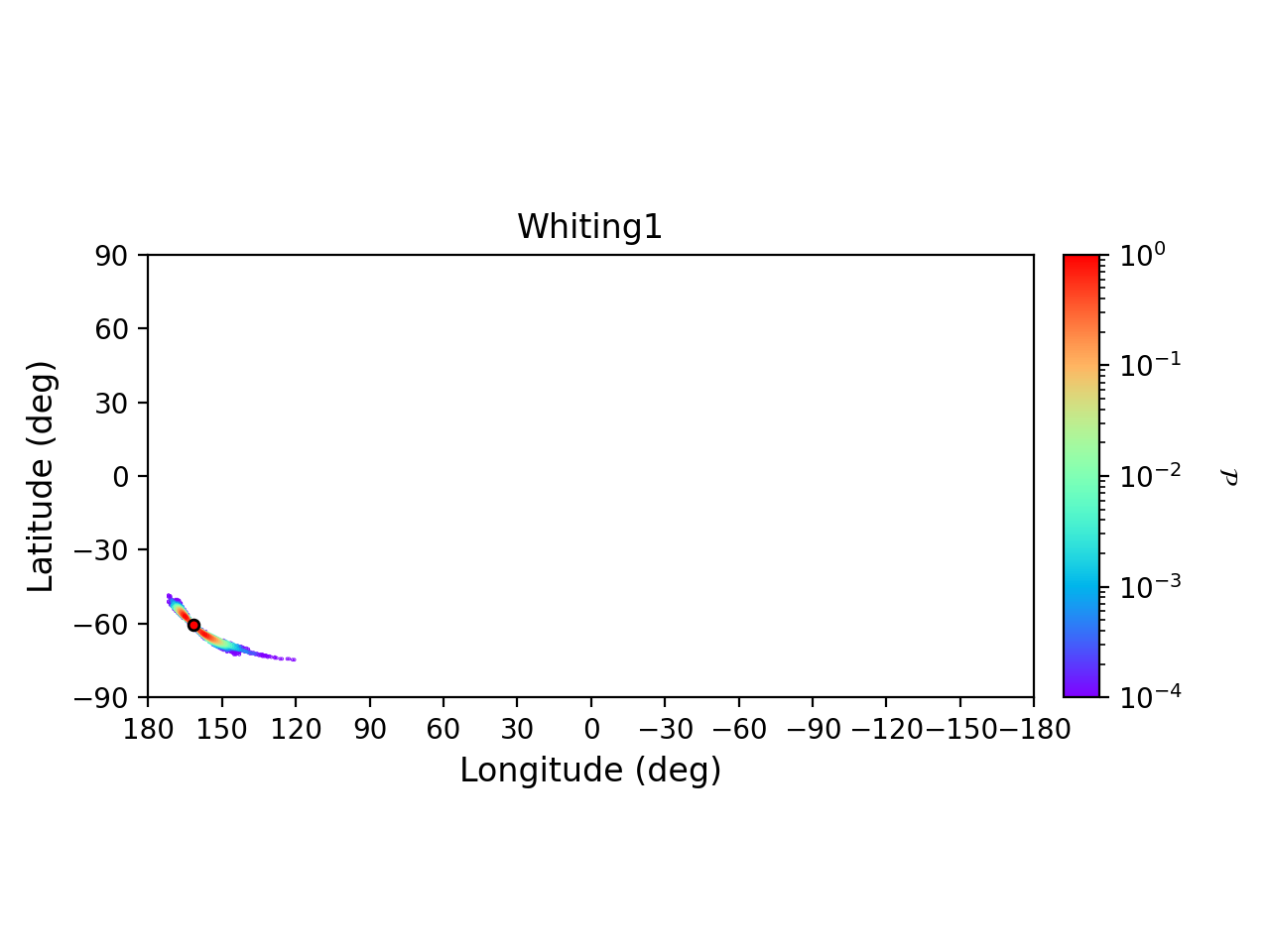}
\caption{Projected density distribution in the $(\ell, b)$ plane of a subset of simulated globular clusters, as indicated at the top of each panel. In each panel, the red circle indicates the current position of the cluster. The densities have been normalized to their maximum value.}\label{stream20}
\end{figure*}

In this section, we report all the extra-tidal features as predicted by our models. For each cluster, we show (from Figs.~\ref{stream1} to \ref{stream20}) the probability density of finding associated extra-tidal features in the sky, by calculating the 2D histogram of the escaped particles. Each particle per cluster is present, that is all 100,000 particles originating each of the 50 Monte-Carlo realizations plus the case with the best values (see Sect.~\ref{numerical}). Thus, each map is a 500x500 histogram that bins 5.1$\times10^6$ particles. The retrieved cumulative 2D histogram is then normalized to its maximum value and plotted in the following figures, in logarithmic scale.


\section{Globular clusters classification}\label{class}
\begin{figure*}
\begin{center}
\includegraphics[clip=true, trim = 0mm 0mm 0mm 0mm, width=0.7\columnwidth]{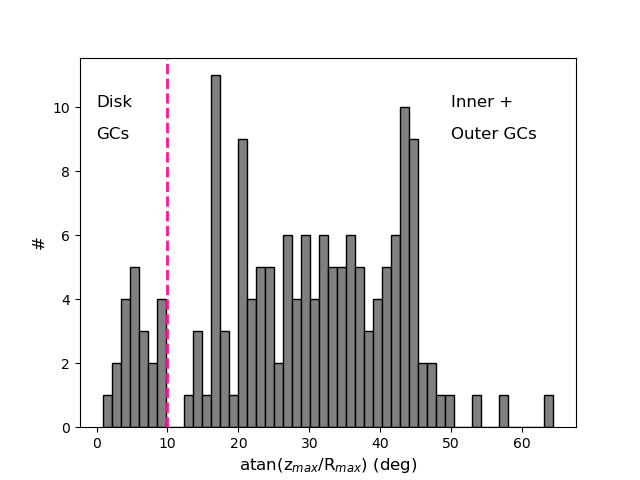}
\includegraphics[clip=true, trim = 0mm 0mm 0mm 0mm, width=0.7\columnwidth]{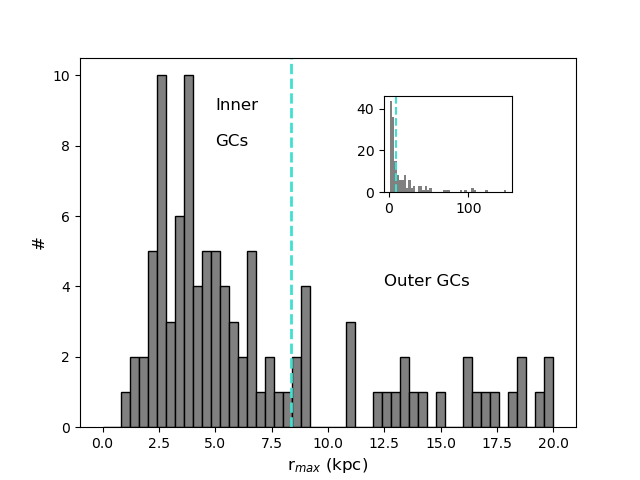}\\
\includegraphics[clip=true, trim = 0mm 0mm 0mm 0mm, width=0.7\columnwidth]{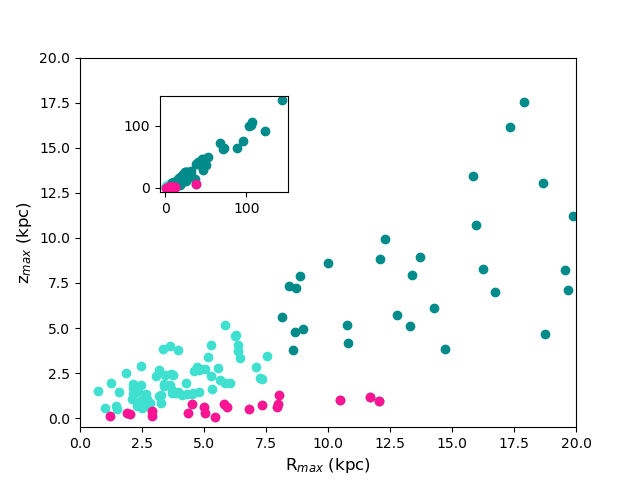}
\includegraphics[clip=true, trim = 0mm 0mm 0mm 0mm, width=0.7\columnwidth]{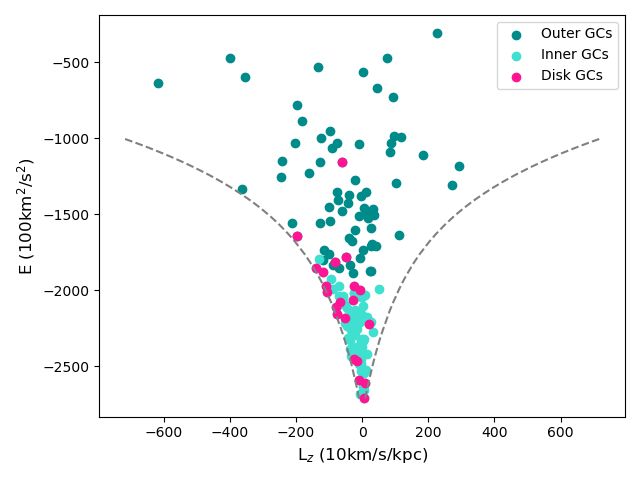}\\
\caption{ \emph{From the top left to the bottom right:} (\emph{First panel}) Distribution of the arctangent of the $z_{max}/R_{max}$ ratio for all simulated GCs. The values are expressed in degrees. The vertical dashed line at $10^\circ$ separates disk clusters (arctan($z_{max}/R_{max}$) $\le 10^\circ$) from the rest of the population of GCs; (\emph{Second panel}) Distribution of the maximum 3D distance, $r_{max}$, from the Galactic center, reached by the GCs orbits in the last 5~Gyr. The main plot shows this distribution for $r_{max} \le 20$~kpc, where as the inset shows the whole distribution, which extends at $r_{max} > 100$~kpc.  In both panels the vertical dashed line, at the solar radius r$_{\odot}$, separates the group of inner GCs from the group of outer GCs. Note that the clusters in one of these two groups which also satisfy the criterion to be disk clusters are classified as disk GCs, and are not in the inner/outer GCs groups. (\emph{Third panel}) Distribution of disk GCs (magenta points), inner GCs (turquoise points), and outer GCs (dark turquoise points) in the $R_{max}-z_{max}$  plane. The main panel shows the distribution of the GCs having $R_{max} \le 20$~kpc, the inset  the distribution of the whole GC sample. (\emph{Fourth panel}) Distribution of the disk GCs (magenta points), inner GCs (turquoise points), and outer GCs (dark turquoise points) in the $E-L_z$ plane. The dashed grey lines correspond, for any given energy E, to the angular momentum of the corresponding circular orbit. Prograde orbits correspond to negative $L_z$ values, retrograde orbits to positive $L_z$.
}\label{orbparam}
\end{center}
\end{figure*}

Fig.~\ref{orbparam} (top panel) shows the distribution of the arctangent of the $z_{max}/R_{max}$ ratio, with $z_{max}$ and $R_{max}$ being respectively the maximum height above or below the Galactic plane and their maximum in-plane distance from the Galactic center, reached in the past 5~Gyr of orbital evolution in the Galactic potential adopted in this paper (see Sect.~\ref{galmod}). As already noticed for field stars \citep[see][]{haywood18}, also the GCs distribution shows a dip at about $10^\circ$, which separates clusters with flattened orbits (arctan($z_{max}/R_{max}$) $\le 10^\circ$) from thicker ones. We thus define a first set of clusters (the \emph{disk GCs}) as that containing all globular clusters with arctan($z_{max}/R_{max}$) $\le 10^\circ$. This first set contains 21 clusters. Of the remaining 138, we distinguish between  a \emph{inner GCs} sample, and a \emph{outer GCs} sample, on the basis of the maximum 3D distance ($r_{max}$) that the cluster reaches from the Galactic center. Inner GCs are those with $r_{max} \le 10$~kpc and outer GCs are those with $r_{max} > 8.34$~kpc, which is the value of the distance of the Sun to the Galactic Center used in this experiment. Such a value allows to discriminate between two classes of tidal debris, for inner clusters are necessarily restricted in latitude and longitude whereas outer cluster can fill the sky. 

Finally the third and bottom panels of Fig.~\ref{orbparam}  show the distribution of these three defined groups in the $R_{max}-z_{max}$ and $E-L_z$ planes. We note that, since disk clusters are uniquely defined on the basis of the ratio between the maximum vertical and in-plane orbital excursion, and not on the circularity of their orbits \citep[as done, for example, by][]{massari19}, some of our disk GCs have elongated (i.e. radial) orbits ($L_z  \sim 0$) or even retrograde ones ($L_z > 0$). Our definition of disk clusters is purely related to a morphological criterium: disk GCs are those whose orbits are confined close to the Galactic plane, independently on their eccentricity.

\begin{table*}
\caption{Classification of the 159 globular clusters studied in this paper in disk clusters ("D"), inner clusters ("I") and outer clusters ("O"). The values of $r_{max}$ and the angle in degrees of the $arctan(z_{max}/R_{max})$ are also given.}\label{classification}
\small
\begin{tabular}{l | c | c | c | l | c | c | c | l | c | c | c}
\hline
      Cluster &    $r_{max}$ &  angle & class & Cluster &    $r_{max}$ &  angle & class & Cluster &    $r_{max}$ &  angle & class \\
            \hline \hline
   2MASS-GC01 &   5.45 &   0.85 &              D &
   2MASS-GC02 &   7.38 &  17.15 &              I &
          AM1 & 123.19 &  36.56 &              O \\
          AM4 &  26.97 &  43.53 &              O &
         Arp2 &  42.44 &  44.54 &              O &
        BH140 &  10.52 &   5.54 &              D \\
        BH261 &   3.83 &  21.02 &              I &
       Crater & 147.37 &  44.46 &              O &
        Djor1 &  11.74 &   5.77 &              D \\
        Djor2 &   2.39 &  16.74 &              I &
           E3 &  12.91 &  24.07 &              O &
  ESO280-SC06 &  13.82 &  33.04 &              O \\
  ESO452-SC11 &   2.53 &  41.47 &              I &
     Eridanus & 109.07 &  44.83 &              O &
      FSR1716 &   5.46 &  17.07 &              I \\
      FSR1735 &   4.43 &  17.10 &              I &
      FSR1758 &  14.35 &  23.13 &              O &
          HP1 &   2.70 &  53.56 &              I \\
       IC1257 &  19.69 &  19.93 &              O &
       IC1276 &   7.98 &   5.71 &              D &
       IC4499 &  27.17 &  42.77 &              O \\
     Laevens3 &  70.74 &  41.35 &              O &
      Liller1 &   1.20 &   5.84 &              D &
       Lynga7 &   4.69 &  17.18 &              I \\
       NGC104 &   7.71 &  24.70 &              I &
      NGC1261 &  20.99 &  36.68 &              O &
      NGC1851 &  19.94 &  29.46 &              O \\
      NGC1904 &  19.57 &  22.81 &              O &
      NGC2298 &  16.69 &  26.93 &              O &
      NGC2419 &  96.76 &  38.13 &              O \\
      NGC2808 &  14.92 &  14.57 &              O &
       NGC288 &  12.50 &  39.00 &              O &
      NGC3201 &  25.53 &  22.62 &              O \\
       NGC362 &  12.27 &  36.03 &              O &
      NGC4147 &  24.90 &  45.18 &              O &
      NGC4372 &   7.36 &  16.30 &              I \\
      NGC4590 &  27.99 &  32.55 &              O &
      NGC4833 &   8.13 &   9.07 &              D &
      NGC5024 &  22.32 &  44.24 &              O \\
      NGC5053 &  18.08 &  44.39 &              O &
      NGC5139 &   7.14 &  21.89 &              I &
      NGC5272 &  16.02 &  40.27 &              O \\
      NGC5286 &  13.32 &  21.11 &              O &
      NGC5466 &  41.16 &  43.51 &              O &
      NGC5634 &  22.18 &  43.13 &              O \\
      NGC5694 &  51.29 &  35.50 &              O &
      NGC5824 &  32.44 &  40.00 &              O &
      NGC5897 &   9.18 &  41.61 &              O \\
      NGC5904 &  24.76 &  42.83 &              O &
      NGC5927 &   5.83 &   7.91 &              D &
      NGC5946 &   5.31 &  23.80 &              I \\
      NGC5986 &   5.00 &  29.36 &              I &
      NGC6093 &   4.04 &  49.12 &              I &
      NGC6101 &  32.24 &  30.56 &              O \\
      NGC6121 &   6.84 &   4.48 &              D &
      NGC6139 &   3.72 &  34.95 &              I &
      NGC6144 &   4.41 &  43.66 &              I \\
      NGC6171 &   3.95 &  33.40 &              I &
      NGC6205 &   8.96 &  39.68 &              O &
      NGC6218 &   4.95 &  31.01 &              I \\
      NGC6229 &  30.26 &  37.44 &              O &
      NGC6235 &   8.37 &  34.55 &              O &
      NGC6254 &   4.98 &  29.32 &              I \\
      NGC6256 &   2.68 &  14.83 &              I &
      NGC6266 &   2.57 &  21.99 &              I &
      NGC6273 &   5.56 &  37.78 &              I \\
      NGC6284 &   6.51 &  36.21 &              I &
      NGC6287 &   6.50 &  30.33 &              I &
      NGC6293 &   3.40 &  37.21 &              I \\
      NGC6304 &   3.38 &  14.75 &              I &
      NGC6316 &   3.80 &  22.98 &              I &
      NGC6325 &   2.57 &  32.34 &              I \\
      NGC6333 &   9.07 &  28.89 &              O &
      NGC6341 &  10.90 &  40.72 &              O &
      NGC6342 &   2.48 &  38.09 &              I \\
      NGC6352 &   4.53 &   9.72 &              D &
      NGC6355 &   3.55 &  29.38 &              I &
      NGC6356 &   8.83 &  28.77 &              O \\
      NGC6362 &   5.41 &  33.16 &              I &
      NGC6366 &   6.04 &  17.80 &              I &
      NGC6380 &   2.35 &  16.74 &              I \\
      NGC6388 &   3.91 &  19.73 &              I &
      NGC6397 &   6.61 &  27.48 &              I &
      NGC6401 &   3.70 &  21.02 &              I \\
      NGC6402 &   3.99 &  32.64 &              I &
      NGC6426 &  16.84 &  22.68 &              O &
      NGC6440 &   1.53 &  25.31 &              I \\
      NGC6441 &   4.67 &  16.48 &              I &
      NGC6453 &   2.71 &  36.68 &              I &
      NGC6496 &   5.71 &  26.65 &              I \\
      NGC6517 &   3.31 &  21.34 &              I &
      NGC6522 &   1.97 &  42.54 &              I &
      NGC6528 &   2.89 &  16.38 &              I \\
      NGC6535 &   4.92 &  16.72 &              I &
      NGC6539 &   3.64 &  39.67 &              I &
      NGC6540 &   2.54 &  12.61 &              I \\
      NGC6541 &   4.78 &  29.39 &              I &
      NGC6544 &   5.93 &  18.30 &              I &
      NGC6553 &   4.36 &   4.14 &              D \\
      NGC6558 &   2.75 &  21.35 &              I &
      NGC6569 &   2.85 &  26.50 &              I &
      NGC6584 &  20.31 &  35.07 &              O \\
      NGC6624 &   1.61 &  64.40 &              I &
      NGC6626 &   3.22 &  20.88 &              I &
      NGC6637 &   2.11 &  56.81 &              I \\
      NGC6638 &   2.34 &  31.98 &              I &
      NGC6642 &   2.20 &  26.61 &              I &
      NGC6652 &   3.15 &  49.58 &              I \\
      NGC6656 &  10.87 &  21.01 &              O &
      NGC6681 &   6.33 &  41.70 &              I &
      NGC6712 &   5.15 &  28.49 &              I \\
      NGC6715 &  38.71 &  45.11 &              O &
      NGC6717 &   2.48 &  33.11 &              I &
      NGC6723 &   4.26 &  47.80 &              I \\
      NGC6749 &   5.05 &   3.24 &              D &
      NGC6752 &   5.72 &  20.72 &              I &
      NGC6760 &   5.95 &   6.19 &              D \\
      NGC6779 &  13.46 &  30.72 &              O &
      NGC6809 &   6.50 &  36.15 &              I &
      NGC6838 &   7.34 &   5.82 &              D \\
      NGC6864 &  16.06 &  33.89 &              O &
      NGC6934 &  37.41 &  20.95 &              O &
      NGC6981 &  21.54 &  35.97 &              O \\
      NGC7006 &  47.26 &  32.50 &              O &
      NGC7078 &  10.86 &  25.72 &              O &
      NGC7089 &  18.77 &  34.91 &              O \\
      NGC7099 &   8.76 &  40.99 &              O &
      NGC7492 &  25.78 &  45.99 &              O &
         Pal1 &  18.77 &  13.96 &              O \\
        Pal10 &  12.06 &   4.56 &              D &
        Pal11 &   8.69 &  23.83 &              O &
        Pal12 &  41.33 &  42.80 &              O \\
        Pal13 &  49.09 &  42.54 &              O &
        Pal14 &  88.95 &  35.97 &              O &
        Pal15 &  46.62 &  44.52 &              O \\
         Pal2 &  38.20 &   8.94 &              D &
         Pal3 & 104.49 &  43.88 &              O &
         Pal4 & 105.89 &  43.58 &              O \\
         Pal5 &  17.54 &  42.94 &              O &
         Pal6 &   3.53 &  27.77 &              I &
         Pal8 &   3.99 &  20.14 &              I \\
        Pyxis &  73.65 &  47.40 &              O &
       Rup106 &  32.08 &  31.98 &              O &
SagittariusII &  75.32 &  41.80 &              O \\
         Ter1 &   2.92 &   2.13 &              D &
        Ter10 &   6.55 &  32.40 &              I &
        Ter12 &   4.23 &  17.31 &              I \\
         Ter2 &   1.49 &  18.30 &              I &
         Ter3 &   3.81 &  26.97 &              I &
         Ter4 &   1.11 &  28.72 &              I \\
         Ter5 &   2.04 &   7.10 &              D &
         Ter6 &   1.93 &   8.77 &              D &
         Ter7 &  41.39 &  45.10 &              O \\
         Ter8 &  47.16 &  45.55 &              O &
         Ter9 &   2.92 &   7.94 &              D &
         Ton2 &   4.39 &  24.81 &              I \\
         UKS1 &   7.95 &   4.51 &              D &
    VVV-CL001 &   5.04 &   7.09 &              D &
     Whiting1 &  52.30 &  43.09 &              O \\               
\hline
\end{tabular}
\end{table*}


\end{appendix}
\end{document}